%% file: main.tex
\def\hidecorrections{}
\documentclass[a4paper,11pt]{book}
\usepackage{amsmath,amsthm,graphicx,xcolor}
\allowdisplaybreaks
\usepackage{amssymb, latexsym}
\usepackage{graphics,color}
\usepackage{braket}
\usepackage{bm}
\usepackage[hang,flushmargin]{footmisc}
\usepackage{textcomp}
\usepackage[inner=20mm, outer=20mm, top=20mm, bottom=20mm, bindingoffset=6mm]{geometry}
\usepackage{wasysym}
\usepackage{cite} 
\usepackage{hyperref}
\usepackage[font=small,labelfont=bf,format=plain,indention=0cm]{caption}
\usepackage{setspace}
\usepackage{fancyhdr}
\usepackage{epigraph}
\usepackage{tikz}
\usepackage{parskip}
\usepackage{subfiles}
\usepackage{subcaption}
\usepackage{stmaryrd}
\usepackage{wrapfig} 
\usepackage{enumitem}
\usepackage{xspace}

\usepackage{amsfonts}
\usepackage{cancel}
\usepackage[normalem]{ulem} 
\usepackage{pifont}
\usepackage{nicematrix}
\usepackage{multirow}
\usepackage{cellspace}

\setlist[itemize,1]{label=--}   

\ifSubfilesClassLoaded{%
  \AtEndDocument{\bibliographystyle{ref_JHEP}\bibliography{ref}}%
}{}
\ifSubfilesClassLoaded{%
  \usepackage{xr-hyper}
  \externaldocument{../build/auxiliary/main}
}{}

\hypersetup{
	colorlinks=true,      
	linkcolor=black,         
	citecolor=black,        
	filecolor=black,        
	urlcolor=black         
}

\renewcommand{\d}{{\rm{d}}}
\newcommand{\cR}{{\cal R}}

\def \rmin r_{\rm min}

\newcommand{\Mpl}{{m}_*}

\newcommand{\aK}{\alpha_K}
\newcommand{\aB}{\alpha_B}
\newcommand{\aH}{\alpha_H}
\newcommand{\aT}{\alpha_T}
\newcommand{\aM}{\alpha_M}
\newcommand{\abar}{\bar{\alpha}}

\newcommand{\R}{{\cal R}}
\newcommand{\cE}{{\cal E}}
\newcommand{\cN}{{\cal N}}
\newcommand{\Src}{{\cal S}}

\newcommand{\stoiic}{$\mathsf{stoiic}\text{-}\mathsf{gr}$\xspace}
\newcommand{\istoriz}{$\mathsf{istoriz}$\xspace}
\newcommand{\vizir}{$\mathsf{vizir}$\xspace}
\newcommand{\grc}{$\mathsf{GRChombo}$\xspace}
\newcommand{\grt}{$\mathsf{GRTeclyn}$\xspace}

\renewcommand{\d}{{\rm{d}}}
\newcommand{\dN}[1]{#1^{\backprime}}

\newif\ifshowcorrections
\ifdefined\hidecorrections \showcorrectionsfalse \else \showcorrectionstrue \fi
\ifshowcorrections
  \newcommand{\ylc}[1]{{\color{blue}\relax#1}}
  \newcommand{\ylg}[1]{{\color{green!50!black}\relax#1}}
  \DeclareRobustCommand{\yld}[1]{\ifmmode{\color{red}\scriptstyle\cancel{#1}}\else{\color{red}\footnotesize\sout{#1}}\fi}
\else
  \newcommand{\ylc}[1]{#1}
  \newcommand{\ylg}[1]{#1}
  \newcommand{\yld}[1]{}
\fi

\AddToHook{cmd/section/before}{\normalcolor\normalfont}
\AddToHook{cmd/chapter/before}{\normalcolor\normalfont}

\begin{document}

\setlength{\parskip}{0.5\baselineskip}
\setlength{\parindent}{15pt}


\pagestyle{plain}
\setcounter{tocdepth}{4}
\begingroup

\input{preword}\endgroup

\cleardoublepage

\pagestyle{headings}
\pagenumbering{arabic}
\setcounter{page}{1}
\pagestyle{fancy}

\begingroup

\input{intro}\endgroup

\chapter*{List of Acronyms}
\markboth{List of Acronyms}{List of Acronyms}
\addcontentsline{toc}{chapter}{List of Acronyms}
\begingroup\input{acronyms}\endgroup


\begingroup

\input{chap_QCl}\endgroup


\begingroup

\input{chap_ICNR}\endgroup
\begingroup

\input{chap_ICsim}\endgroup


\begingroup

\input{chap_SIGR}\endgroup
\begingroup

\input{chap_SINR}\endgroup
\begingroup

\input{chap_SIbeyondGR}\endgroup

\begingroup

\input{conclusion}\endgroup

\clearpage
\phantomsection
\appendix
\addcontentsline{toc}{part}{Appendix}
\begingroup

\input{appendix}\endgroup

\setstretch{1}
\clearpage
\phantomsection
\addcontentsline{toc}{part}{Bibliography}
\bibliographystyle{ref_JHEP}
\bibliography{ref}

\end{document}

%% file: preword.tex

\pagenumbering{gobble}

\vspace*{3cm}

{\centering

{\Large \textbf{Stochastic dynamics of cosmological inflation in general relativity}}
\vspace*{2.5cm}

{\Large
Yoann Levis Launay
}
\vfill
\includegraphics[width=0.2\linewidth]{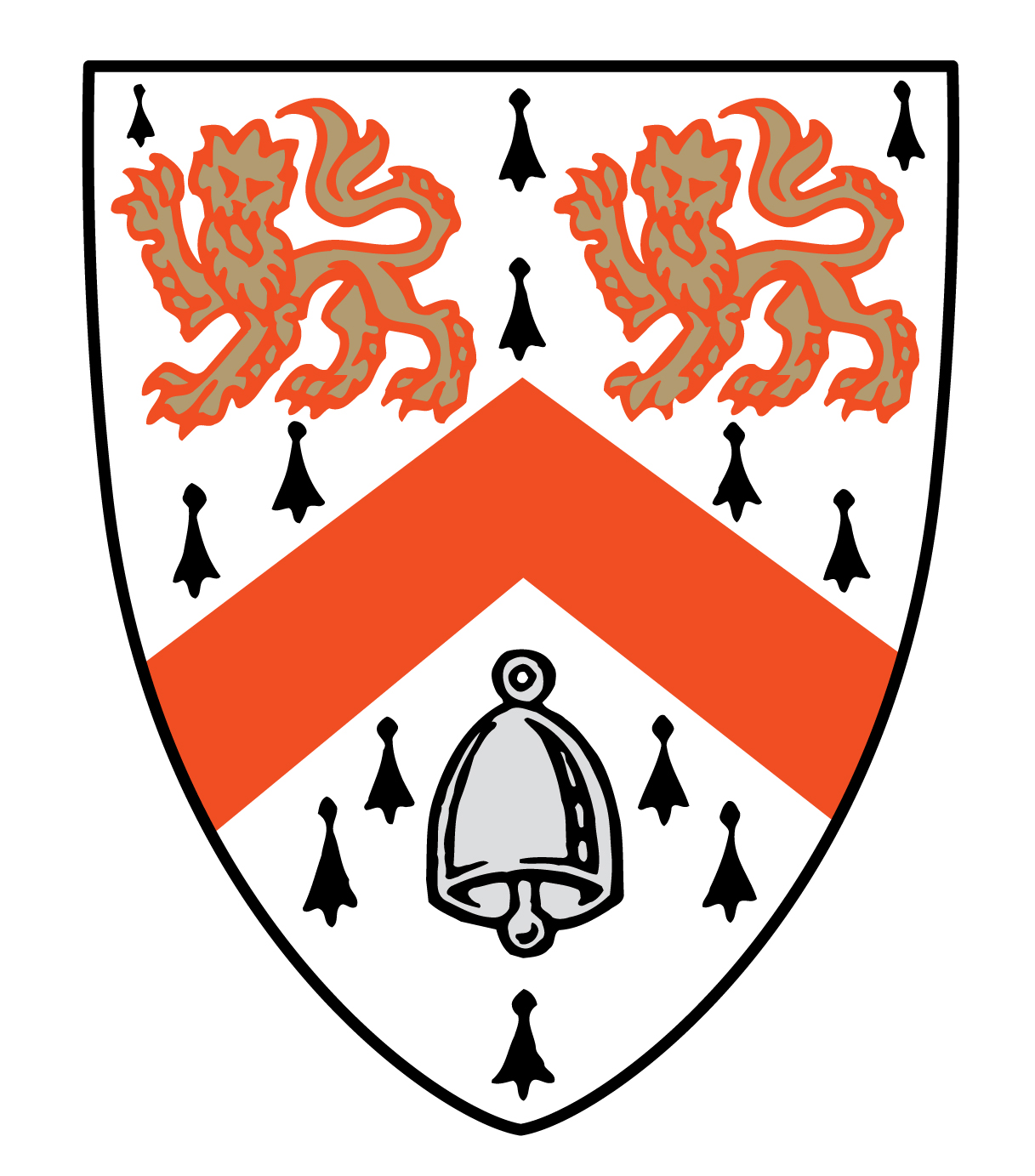}

\vfill
\large{
Wolfson College

\vfill

August 2026
}
\vfill

\setstretch{1.0}
\textit{This thesis is submitted for the degree of \\ Doctor of Philosophy}

\vfill

\includegraphics[width=0.2\linewidth]{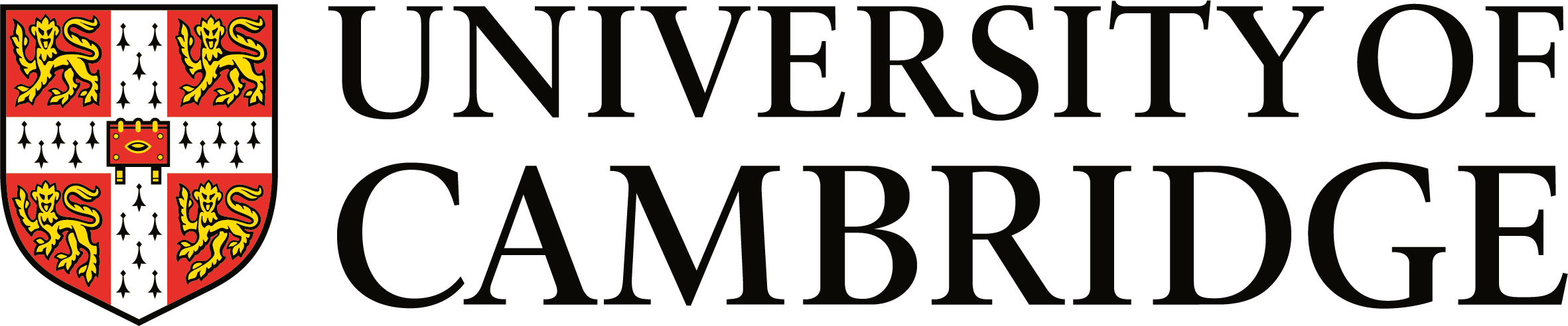}

\setstretch{1.5}
\vspace*{3cm}
}

\newpage

\

\newpage

\pagenumbering{roman}

\begin{center}
    \LARGE{\textbf{Preface}}
\end{center}

\noindent This thesis is the result of my own work and includes nothing which is the outcome of work done in collaboration\ylg{ or content generated by artificial intelligence,} except as declared in the preface and specified in the text. It is not substantially the same as any work that has already been submitted, or is being concurrently submitted, for any degree, diploma or other qualification at the University of Cambridge or any other University or similar institution except as declared in the preface and specified in the text. It does not exceed the prescribed word limit for the relevant Degree Committee.

\vfill

\noindent This thesis is based on results which are presented in the following papers:

\begin{enumerate}

    \item Y.~Launay, G.~Rigopoulos and E.~P.~S.~Shellard (2024),
    \\Stochastic inflation in general relativity,
    \href{https://link.aps.org/doi/10.1103/PhysRevD.109.123523}{\textit{Phys. Rev. D}, \textbf{109}(12), 123523}.

    \item Y.~Launay, G.~Rigopoulos and E.~P.~S.~Shellard (2025),
    \\Quantitative classicality in cosmological interactions during inflation,
    \href{https://doi.org/10.1088/1475-7516/2025/05/071}{\textit{JCAP}, \textbf{2025}(05), 071}.

    \item Y.~Launay, G.~Rigopoulos and E.~P.~S.~Shellard (2025),
    \\Bunch-Davies initial conditions and nonperturbative
    \\inflationary dynamics in numerical relativity,
    \href{https://doi.org/10.1103/qg66-gxh2}{\textit{Phys. Rev. D}, \textbf{112}(4), 043518}.

    \item Y.~Launay, G.~Rigopoulos and E.~P.~S.~Shellard (2026),
    \\Stochastic inflation in numerical relativity,
    \href{https://journals.aps.org/prd/abstract/10.1103/h2cb-q1mt}{\textit{Phys. Rev. D}, \textbf{113}(12), 123515}.

    \item Y.~Launay (2026),
    \\Stochastic scalar-tensor inflation and beyond,
    \ylg{\href{https://doi.org/10.1088/1361-6382/aeb29c}{\textit{Class. Quantum Grav.}, accepted manuscript}}, \href{https://arxiv.org/abs/2606.07467}{arXiv:2606.07467}.

\end{enumerate}

These papers are references \cite{Launay24,Launay2024bis,Launay2025,Launay25ter, Launay2026} in the bibliography and were edited and formatted to form the various chapters of this thesis.
Reference \cite{Launay2024bis} is used in Chapter~\ref{sec:semiclQFT}, \cite{Launay24} in Chapters~\ref{sec:ICNR} and \ref{sec:SIGR}, \cite{Launay2025} in Chapter~\ref{sec:ICsim}, \cite{Launay25ter} in Chapters~\ref{sec:ICNR} and \ref{sec:SIGR}, and \cite{Launay2026} in Chapter~\ref{sec:SIbeyondGR}.
The content of these articles is the fruit of my own theoretical and computational research, based on technical goals set by my supervisors and supported by their regular feedback (with the exception of \cite{Launay2026}).
The key ideas are my own, and the writing was a collaborative effort.
For this thesis, I edited the content of the articles to form a coherent, and accessible narrative, including additional explanations and discussions besides the introduction and conclusion chapters.
Unless specified otherwise, I state here that Artificial Intelligence was solely used for code debugging, grammar, spelling checks, and reference pointing. 
\vfill

\newpage
\
\newpage

{\huge{\textbf{Abstract}}}

{\large \textbf{Stochastic dynamics of cosmological inflation in general relativity}}

{\large \textit{Yoann L. Launay}}

The theory of Inflation, supplemented by fluctuations of the quantum vacuum, is nowadays the leading and most general paradigm to explain the correlations of inhomogeneities at very large distances in the Universe. The study of the inflaton, and the prediction of its self-interactions in the expanding spacetime that it drives are key to constraining high-energy physics models for the origin of the Universe, including the unification of gravity and quantum mechanics.

In the past decades, the application of particle physics methods to the cosmological context has provided perturbative predictions. However, whenever analyticity is lost or when strong non-linearity and non-perturbativity are involved, none of these tools, including semi-analytical methods, are sufficient to provide a complete picture of the underlying physics, leaving a whole landscape of untestable scenarios of inflation. An important advance in this respect has been A. Starobinsky’s stochastic inflation proposal, which can in principle treat nonlinearly and classically inflationary inhomogeneities as the Hubble radius becomes smaller than each of the wavelengths. However, this was never formulated in full general relativity, thus motivating the present work.

After setting this broader context and using the Keldysh formalism of nonequilibrium quantum field theory, this thesis develops, in a first chapter, a validity criterion for the use of classical computations, and thus simulations, to probe inflationary interactions with correlation functions. In general relativity and standard quasi-De Sitter spacetimes, we find analytically that the bispectrum can be extracted from slightly sub-Hubble scales.

This motivates the search for inhomogeneous initial conditions for numerical relativity simulations of inflation in the following chapter, providing the first analytical solutions to the constraints using standard cosmological perturbation theory. These are then used to explore the nonperturbative regime of inflationary dynamics in full numerical relativity, enabling so-called lattice cosmology to evolve the inflating spacetime fully.

This formalism for initial conditions is later used for our main contribution: we show in another chapter how stochastic equations can be obtained for the complete set of equations of the inflaton in general relativity. In addition to incorporating the full nonlinear and backreacting evolution of inflation, this description is found to be gauge-invariant, and free of historical assumptions and associated caveats, such as the gradient-free separate universe approximation.

We then present numerical implementations of these stochastic equations, cast in the BSSN formulation of numerical relativity, demonstrating their correctness in both a slow-roll and an ultra slow-roll scenario, particularly via the satisfaction of the constraints of general relativity. Our first preliminary demonstration within the wider inflationary landscape includes eternal inflation.

In the final chapter, our coarse-graining methods prove to be robust and flexible enough to be extended beyond General Relativity and single-field inflation, offering a comprehensive phenomenological framework with the ambition of serving as a numerical particle collider for the early universe.

\newpage
\
\newpage

{\huge{\textbf{Acknowledgements}}}

More than ever before my PhD, I am very aware of being made of the lucky inputs of the many around me,
whether through mentorship, teamwork, emotional support, sponsorship, laughs and friendship,
sometimes a simple `yes' or an answer to a question, a smile, a favour, an email.
In the following, I hope to name my personal heroes, and to have matched their help and kindness as much as possible,
and can only wish for the forgotten to be assured of my gratitude.

I will start by thanking my supervisors, Paul Shellard and Gerasimos Rigopoulos, for their commitment and constant trust from day one,
in particular with this incredibly ambitious and cherished topic of theirs, which, I'm sure, I will look back on as a very special time in my life.
I am also particularly thankful for them pushing me to high standards of rigour and humility,
as well as for the unmatched career and personal advice, and guidance in what appears to be very challenging and uncertain times.

\ylg{I am also grateful to my examiners, Daniel Baumann and Sébastien Renaux-Petel,
for their careful reading of this thesis, a stimulating viva, and their insightful
comments, which have improved the final version of this work and helped generate some new ideas.}

I was lucky to have additional mentors, advisors and collaborators. Let me thank in particular the Featurespace Innovation Lab team,
notably my supervisor Kamal Parameswaran, and my managers Stuart Burrell and David Sutton, for making my placement a perfect and friendly exposure to industry.
On this note, I should also thank the Department of Applied Mathematics and Theoretical Physics, the University of Cambridge,
and Wolfson College for the privileged services and various support provided to me during my PhD.
In that regard, I would like to specifically thank my advisor Enrico Pajer, my tutor Chris Town, and Paul Murdin for offering their help and making sure of my well-being and success.
On top of this great environment, I would like to thank Will Handley and Miguel Bezarès for their academic support and advice.
While these people have been present for me in the past four years, I would also like to thank those who, before that, helped me become the scholar I wanted to be and feed my curiosity,
but more generally to give back to invested teachers rarely thanked enough.
These include, for my undergraduate studies: Jean-Michel Gillet, Jean-Baptiste Melin, Yannick Alméras, François Sauvageot, Lionel Chaussade, Agnès Desmézières,
and, going back to school: Ms. Gaboreau, Mr. Youssef, Mr. Rouquès, Mr. Jouan, Mr. Le Goëdec, Mr. Adrouji, Ms. Rialland, Ms. Mondolini, Ms. Desmanges and many others.

This project, all its outcomes and my career development would not have been possible without various sponsors.
My PhD and the travelling funds were provided by the UKRI STFC as part of the Cambridge DiS-CDT programme,
as well as the Kavli Institute for Cosmology, Cambridge.
I also acknowledge the resources provided by the DiRAC high-performance computing facility and various entities for covering conference or visit costs: Wolfson College,
Cambridge-LMU collaboration, GRTL Collaboration, UKNR, Newcastle University, Queen Mary University of London, CoBALt conference, Inflation 2025 conference,
ICG Portsmouth, UC Louvain, ICTP Trieste, FZU Prague, Jagiellonian University Kraków, GRAPPA Amsterdam.

I would like to thank those who were part of the great work environment I had at my department, notably Thomas Colas, Ericka Florio, Juliana Kwan, Bowei Zhang among others.
Special thoughts go to my office mates Petar and Ciaran, for their entertaining curiosity and friendship in PhD-hardship, and wishing them the best for the future.

For their unconditional love and support, I want to thank my friends from home: Félix B., Yasmine B., Hugo C., Laura C., Youenn C., Corentin D., Lisenn D., Léo J., Lucas J.,
as well as other dearest from the other side of the Channel, some of whom I hope to reconnect with: Guillaume A., Nathan B., Mathilde L., Antoine T., Elyès Z.
In the extraordinary nature of a Cambridge life, I have felt privileged to have met so many more brilliant, I hope, lifelong partners in laughter and sharing,
some of whom simply made it possible to go through the unexpected challenges of life.
Thank you then to those who have been \textit{my Cambridge}: Aishu A., Ben B., Theja C., Edu C.R., Fanny D. S., Lasse G., Panagiotis G., Charlotte H., Georgia K., Enrica L., James L., Aidan M., Ailie McW. and her family, Andrea P., Viktoria P., Adil S., Emily S., Lars S., Stefan T. and others,
while keeping my most ineffable thoughts for Sophia J..

I will finish by thanking those present in my life since the beginning,
and who will certainly never get back the thanks and attention they deserve given their absolute support.
Thank you to my family, in particular my mum and dad, my grandpa and my grandma, who, at all costs, never said no to my thirst for knowledge.
Thank you to my little brother for making me proud, and teaching me the lessons of life along the way.
Florence L., Pascaline L., Victor L., Fred P. and his family, Julia S. and Enguerrand D, and others, thank you all for this childhood and adulthood.

\newpage
\thispagestyle{empty}
\vspace*{\fill}
\begin{center}
    \begin{tikzpicture}
        \node[inner sep=0] (photo) {\fbox{\includegraphics[width=0.6\linewidth]{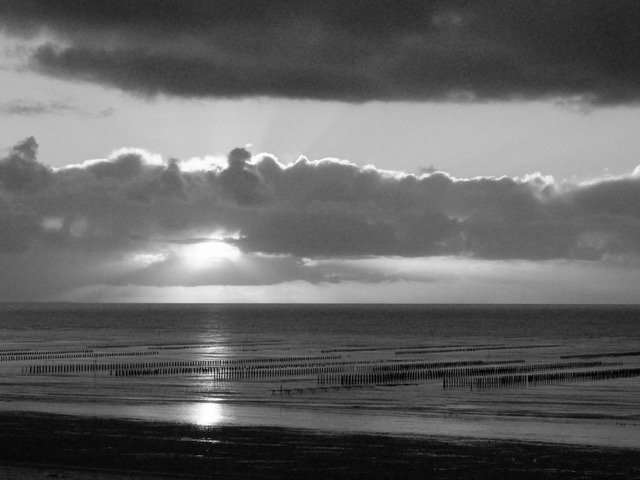}}};
        \node[anchor=east, rotate=-90, xshift=4pt, yshift=3pt, color=gray, font=\tiny] at (photo.south east) {Pascal Launay, Coudeville-Plage, Normandie, France, 2013};
    \end{tikzpicture}

    \vspace{1.5em}
    \textit{À la mémoire de mon oncle, le photographe Pascal Launay (1962--2021).}
\end{center}
\vspace*{\fill}
\newpage

\newpage
\tableofcontents

%% file: intro.tex
\chapter*{Introduction}  \label{sec:intro}
\markboth{Introduction}{Introduction}
\addcontentsline{toc}{chapter}{Introduction}

The present chapter introduces modern developments in cosmology, notably the ones revolving around the concept of inhomogeneity and its origin, and importantly how it birthed the theory of inflation and how this standard model feeds into all observational cosmology today.
Staying at a review level of technicality and assuming basic knowledge of General Relativity, the window of opportunity that justifies the work presented in this thesis is made explicit.
In some of the theoretical overviews, I have relied on various textbooks which should be credited, notably \cite{pajer_fieldtheory, baumann_cosmology_2022, reall_generalrelativity, reall_blackholes, PaulADM}.

\section*{Cosmos meets science}
\addcontentsline{toc}{section}{Cosmos meets science}

In the Cambridge dictionary \cite{cambridge_dictionary}, \textit{Cosmology} refers to ``the study of the nature and origin of the universe''.
Without mention of its modern scientific approach, it is difficult to pinpoint when the understanding of the \textit{cosmos} started being of interest, if not since the dawn of humankind.
Our modern language is rooted in mathematics and luckily shared across the international scientific community.
From this quantitative, relational, object-based, and compositional description, it is easier to identify a clear start to the scientific study of the universe, and to follow its evolution in time according to breakthroughs both in theoretical and observational physics.
In that sense, one would not want to define Cosmology definitively and it is rather safe to state that Cosmology today refers to the study of a certain set of scales of the universe, and a certain set of axioms.
As with other scientific fields and as shown in the present introduction and research, the relaxation of approximations and the wealth of observational data are the main alteration of these in time.

The essence of modern Cosmology was born when the philosopher Giordano Bruno argued in 1585 that there was more beyond Earth and its sky \cite{bruno1584infinito}, building on more observationally and mathematically concrete claims of Copernicus in 1543 \cite{copernicus1543revolutions}, who, probably as a matter of life or death, resorted himself to discarding \textit{geocentrism}.
Bruno was the first explicit \textit{acentric} and seemingly the first to argue that stars were other suns, and that the universe might be infinite.
Later, Galileo quantitatively argued further for \textit{heliocentrism}, despite his understanding through observations that stars were other suns.
This was only confirmed observationally in 1838 by Bessel \cite{Bessel1838}, by finally measuring the distance to a star and thus constraining the luminosity to be solar-like.

Very early on, at least in Greek and Roman thought, the concept of galaxy and the Milky Way as a collection of stars (not suns) was created.
Given our knowledge today, cosmological scales were not yet reached at this point in history.
In the 1700s, technological progress in telescopes allowed astronomers to find new objects, notably what appeared to be faintly glowing clouds, which were called \textit{nebulae}.
The acentric debate was taken to larger scales then, as some argued that these constituted gas or star clusters within the Milky Way, while others argued that these could be other separate galaxies, or as coined by Kant in 1755, other \textit{island universes} \cite{kant1755naturgeschichte}.
This sometimes called ``Great Debate'' lasted until the 1920s, for instance materialising in the famous Shapley–Curtis debate on the 26th of April 1920 at the U.S. National Museum in Washington DC.
Harlow Shapley believed that spiral nebulae were relatively small and lay within the outskirts of the Milky Way Galaxy, while Heber Curtis held that they were independent galaxies, implying that they were exceedingly large and distant, as reported in the ``The Scale of the Universe'' article \cite{shapley1921scale}.
And indeed, the `scale of the universe' was clarified soon after, when Edwin Hubble showed that Andromeda was far outside the Milky Way by measuring Cepheid variable stars \cite{hubble1925cepheids,hubble1925ngc6822,hubble1929m31}.
The modern spatial cosmological scale was then close to the modern one, and the (extended) \textit{Copernican principle} was fully embraced, a term created by Hermann Bondi in 1952 \cite{Bondi1952} in homage to the first acentric discovery.
This means that human observers or observations from Earth are representative of observations from the average position in the universe.

Interestingly, the first mathematical description of the universe on arbitrarily large scales predates an explicit definition of the cosmological scale.
This could be attributed to Newton, whose \textit{Principia} \cite{Newton1687} laid the groundwork for classical mechanics and the understanding of gravitational forces.
Equipped with his theory of the force acting at largest distances, Newton was among the first to apply a mathematical model to a universe filled with gravitationally interacting mass.
This however was not satisfying as a finite Newtonian universe cannot escape gravitational collapse.

\section*{The cosmological spacetime}
\addcontentsline{toc}{section}{The cosmological spacetime}
This problem was barely looked at until 1917, after Albert Einstein published the application of his new and independent theory of general spacetimes, General Relativity (GR) \cite{Einstein1915a,Einstein1915b}, which was then applied to the universe as a whole \cite{Einstein1917}.

For this purpose, let us recall Einstein's modern field equations and postulates of GR, which will be at the core of this thesis, see \cite{reall_generalrelativity} for a first encounter.
We assume that spacetime is described by a 4-dimensional differentiable manifold $\mathcal{M}$, which, for the purpose of this work, will systematically and implicitly be equipped with a chart or coordinate system $\{x^\mu(p)\}$, for any point $p$ in $\mathcal{M}$,
leading to the definition of vector, covector, and tensor fields on the manifold via their covariant and contravariant component indices.
$\mathcal{M}$ is given a metric $g_{\mu\nu}$, which is a symmetric, non-degenerate, rank-2 tensor field used for the raising and lowering of indices, and which defines the geometry of spacetime and allows for the measurement of distances and angles as shown by the elementary proper distance invariant
$
    ds^2 = g_{\mu\nu} dx^\mu dx^\nu.
$
Following the principle of covariance, we assume that the correct metric for the description of a spacetime has a Lorentzian signature $(-,+,+,+)$, which allows for the distinction between timelike, spacelike, and lightlike curves.
The covariant derivative $\nabla_\mu$ is defined as the Levi-Civita connection, or the unique torsion-free connection ($\nabla_a\nabla_b=\nabla_b\nabla_a$ when applied to a scalar field) compatible with the metric ($\nabla g = 0$), or more explicitly
\begin{equation}
    \nabla_\mu V^\nu = \partial_\mu V^\nu + \Gamma^\nu_{\mu\sigma}V^\sigma, \quad
    \nabla_\mu V_\nu = \partial_\mu V_\nu - \Gamma^\sigma_{\mu\nu}V_\sigma,
\end{equation}
where $\Gamma^\rho_{\mu\nu} = \frac{1}{2}g^{\rho\sigma}(\partial_\mu g_{\nu\sigma} + \partial_\nu g_{\mu\sigma} - \partial_\sigma g_{\mu\nu})$ are the Christoffel symbols, while for tensors
\begin{equation}
    \nabla_\mu T^{\nu_1\dots\nu_n}_{\sigma_1\dots\sigma_m} = \partial_\mu T^{\nu_1\dots\nu_n}_{\sigma_1\dots\sigma_m} + \sum_{i=1}^n \Gamma^{\nu_i}_{\mu\rho}T^{\nu_1\dots\nu_{i-1}\rho\nu_{i+1}\dots\nu_n}_{\sigma_1\dots\sigma_m} - \sum_{j=1}^m \Gamma^\rho_{\mu\sigma_j}T^{\nu_1\dots\nu_n}_{\sigma_1\dots\sigma_{j-1}\rho\sigma_{j+1}\dots\sigma_m}.
\end{equation}
The covariant derivative provides a differentiation operator that is compatible with the manifold's geometry, allowing for the definition of parallel transport and geodesics, which are curves that locally extremise the proper distance $ds$.
The curvature of spacetime measures the geometric deviation from a flat partial differentiation along each direction and is described by the Riemann tensor
\begin{equation}
    R^\rho{}_{\sigma\mu\nu}= \partial_\mu \Gamma^\rho_{\nu\sigma} - \partial_\nu \Gamma^\rho_{\mu\sigma} + \Gamma^\rho_{\mu\lambda}\Gamma^\lambda_{\nu\sigma} - \Gamma^\rho_{\nu\lambda}\Gamma^\lambda_{\mu\sigma},
\end{equation}
which can be contracted to obtain the Ricci tensor $R_{\mu\nu}=R^\rho{}_{\mu\rho\nu}$ and the Ricci scalar $R=R^\rho{}_{\rho}$.
In 1917, Einstein wanted to relate this spacetime curvature to the energy content, described here by the stress-energy tensor $T_{\mu\nu}$. The simplest coordinate-independent relation preserving the Newtonian limit and the conservation of energy-momentum is given by his field equations
\begin{equation}
    G_{\mu\nu}\equiv R_{\mu\nu}-\tfrac{1}{2}g_{\mu\nu}R
    = \frac{1}{\Mpl^2}T_{\mu\nu}\equiv -\frac{2}{\Mpl^2}\frac{1}{\sqrt{-g}}\frac{\delta \mathcal{L}_\text{m}}{\delta g_{\mu\nu}},
    \label{eq:Einstein}
\end{equation}
with, if existing, content Lagrangian density $\mathcal{L}_\text{m}$, and the associated geometric Lagrangian giving the l.h.s. was found by David Hilbert at the same time \cite{Hilbert1915} as $\mathcal{L}_\text{g} = \frac{\Mpl^2}{2}\sqrt{-g}R$.
The conservation of energy-momentum is guaranteed by the Bianchi identity $\nabla^\mu G_{\mu\nu}=0$, which implies $\nabla^\mu T_{\mu\nu}=0$.
Tests of Einstein's theory have been very successful so far, see \cite{Will2014,Berti15,Ferreira2019} for reviews, and justify our use as a starting point for the description of the universe.

These equations allowed Einstein to remove the boundary problem of the Newtonian universe, by considering a spatially closed universe: $ds^2 = -dt^2 + R_E^2 d\Omega_3^2$, with $d\Omega_3^2=d\chi^2 +\sin^2\chi(d\theta^2+\sin^2\theta\,d\phi^2)$ the metric of a 3-sphere in spherical coordinates (of radius $R_E$).
Einstein filled the universe with a homogeneous and isotropic perfect fluid, with the stress-energy tensor of a non-relativistic fluid with negligible pressure and velocity, left only with a uniform energy density $\rho$.
At this point, he believed in a static universe and decided to add a term $-\Lambda g_{\mu\nu}$ to the left-hand side of eqn.~\eqref{eq:Einstein}, with $\Lambda$ a constant, which he called the \textit{cosmological constant}, to counteract the gravitational attraction of matter and allow for a static solution in the time-time and space-space equations found from eqn.~\eqref{eq:Einstein}%
\begin{subequations}
    \begin{align}
        \frac{\dot R_E ^2}{R_E} + \frac{1}{R_E^2} - \frac{\Lambda}{3} & = \frac{\rho}{3\Mpl^2},\\
        \frac{\ddot R_E}{R_E} - \frac{\Lambda}{3} & = -\frac{\rho}{6\Mpl^2},
    \end{align}
\end{subequations}
where in fact nulling the time-dependency of $R_E$ only makes sense if $\Lambda\neq 0$, giving $\Lambda = \rho/\Mpl^2$, which is also proportional to the curvature of the sphere $\propto 1/R_E^2$.
This fine-tuning of $\Lambda$ was criticised by de Sitter the same year \cite{deSitter1917a,deSitter1917b}, who also found a solution to Einstein's equations with $\rho=0$ and $\Lambda\neq 0$, which was also static but empty, and was later understood to be, in different coordinates \cite{Lanczos1922}, equivalent to \yld{a}\ylg{the same} time-dependent solution.
The static Einstein universe was later shown to be unstable under inhomogeneities in \cite{Lemaitre1927,Eddington1930}, and so did not address Newton's stability fine-tuning problem. Most importantly, a static universe was incompatible with observations of the recession of galaxies found by Hubble in 1929 \cite{Hubble1929exp}.

Einstein did not write the time-dependent solutions for the radius and used the static assumption directly.
Alexander Friedmann was the first to derive explicitly time-dependent cosmological solutions to Einstein's equations in 1922 and 1924, for a curved isotropic universe \cite{Friedmann1922,Friedmann1924}.
In 1927 and before Hubble's measurements, Georges Lemaître independently \yld{rederived}\ylg{derived} these solutions and predicted the recession of galaxies found later by Hubble, recently renamed as the \textit{Hubble-Lemaître law} \cite{IAU2018}.
While Friedmann and Lemaître found those cosmological solutions from selected approximations,
Howard Robertson \cite{Robertson1935,Robertson1936a,Robertson1936b} and Arthur Walker \cite{Walker1937} separately studied the most general spacetime metric compatible with the cosmological principle of spatial homogeneity and isotropy, and its uniqueness, which is today called the Friedmann–Lemaître–Robertson–Walker (FLRW) metric, which, in various coordinates, can be written as%
\begin{subequations}
    \begin{align}
    ds^2 & = -dt^2+a(t)^2 R_0^2\frac{\delta_{ij}}{(1+k\mathbf{x}^2/4)^2 }dx^idx^j \quad\text{(quasi-cartesian coordinates)},\\
    ds^2 & = -dt^2 + a(t)^2 R_0^2 \left( \frac{dr^2}{1-kr^2} + r^2 d\Omega^2 \right) \quad\text{(spherical coordinates)},\\
    ds^2 & = -dt^2+a(t)^2 R_0^2\bigg(d\chi^2+S(\chi)^2(d\theta^2 + \sin^2(\theta)d\phi^2)\bigg) \quad\text{(spherical comoving coordinates)},
    \end{align}
\end{subequations}
with $t$ the cosmological or \textit{cosmic time}, the proper time measured by fundamental observers. These are observers with zero velocity with respect to the cosmic fluid, including the cosmic microwave background (CMB) radiation.
$a(t)$ is the \textit{scale factor}, and accounts for the isotropic and homogeneous scaling of the distances in the universe in time.
$d\Omega^2 = d\theta^2 + \sin^2\theta\,d\phi^2$ is the metric of a 2-sphere.
$k$ is the spatial curvature parameter and can take the integer values\footnote{$k$ can also take other values but then $\chi$ can be rescaled to be brought back to the case of a $S$-like solid angle; only the sign or nullity matter.} $k = 0, +1, -1$ (respectively $S=\chi, \sin \chi, \sinh \chi$), corresponding to flat, closed\footnote{Triangles' angles add to more than $\pi$. Moving on a 2D surface is locally equivalent to moving on a sphere.}, and open\footnote{Triangles' angles add to less than $\pi$. Moving on a 2D surface is locally equivalent to moving on a saddle surface.} universes respectively.
$(\chi,\theta,\phi)$ are called \textit{comoving coordinates} because observers attached to those coordinates ``comove'' with the expansion. For instance, up to peculiar velocities, galaxies are usually considered at rest in comoving coordinates, and so their comoving distance is constant in time.
The physical distance at a given time (proper distance) is $a(t)\Delta \chi$ for radial instantaneous travel.
Note that we have introduced a fiducial length scale $R_0$ to make $a,k$ and the spatial coordinates dimensionless. This is kept implicit in the following and can be recovered by rescaling $a$ accordingly.

When applied to a perfect fluid $T_{\mu\nu}= (\rho+3p)u_{\mu}u_{\nu}+p g_{\mu\nu}$ in the coordinates of the fluid rest frame where the $4$-velocity is written as $u_{\mu} = (1,0,0,0)$, \yld{and }the Friedmann equations are%
\begin{subequations}
    \begin{align}
        H^2\equiv \left(\frac{\dot{a}}{a}\right)^2 & = \frac{1}{3\Mpl^2}\rho - \frac{k}{a^2} + \frac{\Lambda}{3} \equiv \frac{1}{3\Mpl^2}\sum_i\rho_i,\label{eq:Friedmannvelocity}\\
        \frac{\ddot{a}}{a} = H^2+\dot{H} & = -\frac{1}{6\Mpl^2}(\rho + 3p) + \frac{\Lambda}{3} \equiv -\frac{1}{6\Mpl^2}\sum_i(\rho_i + 3p_i),\label{eq:Friedmannacceleration}
    \end{align}
\end{subequations}
where we have defined the Hubble parameter $H= \frac{\dot{a}}{a}$, coinciding with the Hubble-Lemaître physical recession notation $d_t(a\chi) = H \cdot a\chi$ for an object, at constant $\chi$, such as a galaxy.
Most importantly, we \yld{switched}\ylg{moved} geometrical effects (curvature and cosmological constant) to the r.h.s. of the equations and interpreted them as \textit{effective fluids} with energy densities $\rho_k = -3\Mpl^2 k/a^2$ and $\rho_\Lambda = \Mpl^2\Lambda$ and pressures $p_k = -\rho_k/3$ and $p_\Lambda = -\rho_\Lambda$.

The homogeneous hydrodynamics of the static fluid arise from the conservation of the stress-energy tensor, $\nabla^a T_{ab}=0$, and yields the continuity equation,
\begin{equation}
\dot{\rho} = -3 H (\rho+p),
\label{eq:Friedcontinuity}
\end{equation}
or the scaling of the energy density with the volumetric change rate.
In the non-interacting case\footnote{Beyond that case, see for instance dark sector couplings \cite{Wetterich95}.}, we can write the fluid as a sum of individual densities $\rho = \sum_i \rho_i$ (resp. pressures $p = \sum_i p_i$), all satisfying eqn.~\eqref{eq:Friedcontinuity}.

The interpretation of the Friedmann equations is simple but illuminating, in particular when focusing on the acceleration equation, eqn.~\eqref{eq:Friedmannacceleration}.
Filling the vacuum with \textit{barotropic fluids} ($p=p(\rho)$) such that $\rho + 3p > 0$, \yld{the universe decelerates its expansion}\ylg{the universe's expansion decelerates}, whereas if $\rho + 3p < 0$, \yld{the universe accelerates its expansion}\ylg{it is accelerated}.
In the case of \textit{perfect barotropic fluids} ($p=w\rho$, $w=\text{constant}$), the condition for acceleration is $w < -1/3$. This defines \textit{dark energy} by its equation of state along the way, which is satisfied by the special case of the cosmological constant ($w_\Lambda = -1$).
Spatial curvature effects are rather special as they do not contribute to the acceleration\footnote{Note that Buchert showed that it can via backreaction when considering the averaged Friedmann equations \cite{Buchert00}.} ($w=0$).
Ordinary fluids, such as matter (close to dust with $w\simeq0$) and radiation (with $w=1/3$), decelerate the expansion because they aggregate with positive pressure and so attractive gravity while dark energy acts as a repulsion with its negative pressure.

Modern and observational studies of the Friedmann equations make great use of the \textit{density parameters}, $\Omega = \sum_i\Omega_i = \sum_i \frac{\rho_i}{\rho_c}$,
dimensionless thanks to the normalisation by the critical density $\rho_c= {3H^2\Mpl^2}$, the reference \yld{physical}\ylg{energy} density of a flat universe without dark energy. 
Friedmann equations, eqns.~\eqref{eq:Friedmannvelocity} and \eqref{eq:Friedmannacceleration}, are thus recast as%
\begin{subequations}
    \begin{align}
        1 & = \sum_i \Omega_i \nonumber\\
        & =  \left(\frac{H_0}{H}\right)^2 \sum_i \Omega_{i0}a^{-3(1+w_i)},  \\
        q \equiv -\left(1+\frac{\dot{H}}{H^2}\right) & = \frac{1}{2}\sum_i \Omega_i(1+3w_i) \nonumber\\
        & = \frac{1}{2}\left(\frac{H_0}{H}\right)^2 \sum_i \Omega_{i0}(1+3w_i)a^{-3(1+w_i)},
    \end{align}
\end{subequations}
where all density parameters have been normalised to their present-day values $\Omega_{i0}$, and $H_0$ is the present-day Hubble parameter, with the convention $a_0 = 1$ today, and using
the solution to the continuity equation for a generic dominating $w$-fluid,  {$\rho_i \propto a^{-3(1+w_i)}$}, mapping to $a\propto t^{2/3(1+w_i)}$.

Apart from the \textit{Einstein static} and \textit{De Sitter} cases, very few historical solutions\footnote{
Main solutions include \textit{Einstein-De-Sitter universe} ($\Lambda = k = 0$),
\textit{Friedmann universes} ($\Lambda = 0$),
\textit{flat universes} ($k=0$),
\textit{Milne universes} (curvature only).
} of the Friedmann equations included any cosmological constant.
In their critique of Einstein's universe \cite{Lemaitre1927,Eddington1930}, Lemaître and Eddington were among the first and only ones to credit a nonzero and potentially dominating cosmological constant at some point in time, and the former associated it with the dark energy equation of state.
As an unpopular view at the time, it was suggested instead that the universe could perfectly be decelerating and collapse in the future \cite{Tolman31,Hawking70}.
This notably explains the historical sign convention of the \textit{deceleration parameter} $q$, until the surprise of finding $q_0<0$ and thus an accelerating universe in Supernovae measurements \cite{Riess1998,Perlmutter1999} in 1998.
In fact, before the 1980s, very little was known about the value of all these background parameters.
As already mentioned, the expansion of the universe was discovered by Lemaître in 1927 \cite{Lemaitre1927}, and measured by Hubble in 1929 \cite{Hubble1929exp}, with the value of $H_0$ debated for decades in the range $50-100$ km/s/Mpc (initially found by Hubble as $500$ km/s/Mpc).
Moreover, in the absence of a causal mechanism for inhomogeneities and so structure formation such as the one provided by inflation, constraining the cosmological density parameters $\Omega_i$ from a purely background approach is not straightforward.
One notable approach, built after, but independently from, inflation and structure formation, used the age of the universe as a constraint on the density parameters \cite{Chaboyer95,Chaboyer96}.
As the oldest known objects in the universe, globular clusters were used to provide a lower bound on the age of the universe, thus constraining the background history via $t_0 = \int^{t_0} dt = \int^1_0 (aH)^{-1} da$.
It was found that $t_0 > 12.1 \text{ Gyr}$, which, even if neglecting the structure formation time,
ruled out both a flat matter-dominated universe and a curvature-only (Milne) universe, suggesting the necessity of a cosmological constant in the mix.

\section*{Departing from a homogeneous background}
\addcontentsline{toc}{section}{Departing the background}
As put by Ellis in his editorial note \cite{ellis_editorial_2017} rehabilitating Lifshitz's 1946 pioneering work \cite{Lifshitz46}, it took some time to get cosmological perturbation theory well-established.
Given our knowledge of smaller scales, it was clear that a homogeneous and isotropic theory would need to be extended to include inhomogeneities; however, without a clear understanding of (or proposal for) the origin of these inhomogeneities, it was not clear how and why to do so.
This is certainly why Lifshitz's work was not immediately appreciated.

In \cite{Lifshitz46,Lifshitz63}, many years after the linear treatments of Einstein for the prediction of gravitational waves \cite{Einstein16}, Lifshitz was the first to linearise Einstein's equations around a FLRW background, and to study the evolution of generic inhomogeneities.
As opposed to many other authors later, he handled perturbative gauge-choice subtleties well and notably concluded that such perturbations could not grow fast enough to explain galaxy formation.
This proved to be the wrong conclusion as, according to Ellis, the paper focused on the wrong cosmology; no cosmological constant as commonly done, no baryon-radiation coupling, and $k=\pm 1$. This was fixed later by Sachs and Wolfe \cite{SachsWolfe1967}.
Twenty years later, Hawking presented a coordinate-independent formulation of cosmological perturbation theory (CPT) \cite{Hawking1966}, perturbing the expansion rate, shear, vorticity, and the Weyl tensor in what will later be called the covariant approach \cite{EllisBruni89}.
However, the decisive contribution bridging the more-practical gauge-specific Lifshitz approach and a gauge-agnostic one was made by Bardeen in 1980 \cite{Bardeen1980}, who found gauge-invariant combinations of perturbed quantities, demonstrating along the way the meaning of hypersurface choices.

A notable contribution from Lifshitz is the decomposition of the FLRW metric perturbations into scalar, vector, and tensor modes (\textit{SVT decomposition}), which are usually written in the line element
\begin{equation}
    d s^2=-\alpha_b^2(1+2 \Psi) d t^2+2 \Big(a^2 B_{, i} + V_i\Big) d t d x^i +a^2\Big((1-2 \Phi) \delta_{i j}+2 E_{, i j} +\partial_{(i}C_{j)}+h_{ij}\Big) d x^i d x^j,    
    \label{eq:SVTmetricAll}
\end{equation}
where the four scalar functions $\Phi, \Psi, B$ and $E$ define the scalar perturbations of the geometry (4 d.o.f.s) along that of the field $\delta \phi$, $V_i$ and $C_i$ are two \textit{transverse} (divergenceless) vector fields defining the vector perturbations (2 polarisations each so 4 d.o.f.s), and $h_{ij}$ is the \textit{transverse-traceless} tensor field defining the tensor perturbations (2 polarisations so 2 d.o.f.s), and $a(t)$ and $\alpha_b(t)$ are the background scale factor and lapse function respectively\ylg{ (sometimes implicitly set to $1$ for cosmic time)}.
This decomposition of the 10 d.o.f.s of the metric under rotations into scalars, vectors, and tensors is suited to classifying the breaking of homogeneity and isotropy of FLRW, and when plugged in Einstein's equations, the three types of perturbations decouple at linear order, meaning that they can be studied independently.
Importantly these gauge-dependent objects are called so but do not constitute tensors as they do not transform with the chain-rule under general coordinate transformations. They are only defined with respect to the background FLRW metric.
For an \yld{arbitrary}\ylg{infinitesimal} transformation $x^\mu \to x^\mu + \xi^\mu$, the perturbations transform following the Lie derivative rule instead%
\begin{subequations}
    \begin{align}
        \delta\phi & \to \delta\phi - \xi^\mu\partial_\mu\phi_b,\\
        \delta g_{\mu\nu} & \to \delta g_{\mu\nu} - (g_b)_{\mu\nu,\alpha}\xi^\alpha-(g_b)_{\alpha\nu}\xi^\alpha{}_{,\mu}-(g_b)_{\mu\alpha}\xi^\alpha{}_{,\nu},
    \end{align}
    \label{eq:gaugetransform}
\end{subequations}
or in our notations of eqn.~\eqref{eq:SVTmetricAll} for the scalar perturbations, with $\xi^\mu = (\xi^0, \partial^i\lambda)$,%
\begin{subequations}
    \begin{align}
        \delta \phi & \to \delta\phi - \dot{\phi}_b\xi^0,\label{eq:dphitrans}\\
        \Psi & \to \Psi - \frac{\dot{\alpha}_b}{\alpha_b}\xi^0 - \dot{\xi}^0,\\
        B & \to B + \left(\frac{\alpha_b}{a}\right)^2\xi^0 -\dot{\lambda},\\
                \Phi & \to \Phi + H\alpha_b\xi^0,\label{eq:Phitrans}\\
        E & \to E -\lambda.         
    \end{align}
\end{subequations}
This is what Bardeen used to build gauge-invariant combinations of the perturbations, such as that of eqns.~\eqref{eq:dphitrans} and \eqref{eq:Phitrans} via the \textit{curvature perturbation on comoving hypersurfaces}
\begin{equation}
    {\cal R}\equiv\Phi+\frac{H\alpha_b}{\dot{\phi}_b}\delta\phi.
    \label{eq:defR}
\end{equation}
Nowadays these techniques are standard, and can be pushed very far, as will be shown in Chapter~\ref{sec:ICNR} in the context of inflation, when comparing inhomogeneity as seen from various observers.

With such perturbative techniques at hand came the concrete need for inhomogeneous initial conditions.
Interestingly, the first guesses did not need a concrete scenario for the origin \yld{but rather}\ylg{as} arguments of naturalness gave very good hints.
From Lifshitz's work, the notion of a growing Hubble radius or (causal) horizon $R_H= (aH)^{-1}$ was understood, delimiting the small-scale gravitational fluid dynamics and frozen superhorizon physics.
In particular, because all scales of interest seemed to have been superhorizon at some point in the past, it appeared natural to set initial conditions in this regime.
In the early 1970s, Harrison, Zel'dovich, and Peebles and Yu \cite{Harrison1970,Zeldovich1972,PeeblesYu1970} independently argued that the shape of the initial conditions should be scale-invariant, i.e. that the amplitude of perturbations should be the same for all Fourier modes (constant dimensionless power spectrum), to reflect that Einstein's vacuum equations are scale-free and purely geometric.
In \cite{PeeblesYu1970}, perturbations were also assumed \textit{adiabatic}, i.e. all species present in the primordial plasma (notably matter and photons) are perturbed in the same way in their number density and so entropy, as opposed to so-called \textit{isocurvature} or \textit{entropy} perturbations.

These ideas came independently from the major discovery of the \textit{cosmic microwave background} (CMB) radiation, or the earliest data point of the universe still up to this day.
As early as 1948, Alpher and Herman \cite{Alpher48,Alpher48b} predicted the existence of a relic radiation field from the hot early universe, with a temperature of a few Kelvin.
This was re-discovered theoretically in 1965 in Princeton as reported in \cite{Dicke65}, about the time its isotropic contribution was accidentally measured by Penzias and Wilson \cite{Penzias65}.
With this discovery, came the opportunity to find an early value for the amplitude of inhomogeneities (or anisotropies), as the CMB is a snapshot of the universe at the time of decoupling, when photons last scattered off baryons.
For many years, instruments failed to detect such anisotropies, only putting an upper bound on the relative amplitude of the CMB temperature, reaching down to about $10^{-4}$ by 1984 \cite{Uson84,Uson84b,Boughn81}.
At least at that time, the CMB provided a confirmation that the universe really had gone through a hot, dense, radiation-dominated early phase, and triggered the investigation of the coupled photon-baryon plasma in detail \cite{PeeblesYu1970}.

Regarding theoretical expectations, the first compatible prediction of the primordial energy density contrast $\delta\rho/\rho_b$ was made by Peebles in 1982 \cite{Peebles82}, assuming a flat universe whose matter budget is dominated by \textit{cold dark matter} (CDM).
In fact by 1983, the community struggled to connect the galaxy and so matter contrast found today with the upper limit on CMB anisotropy, when using only the baryon-photon plasma in a radiation-dominated model of the universe.
Using a typical contrast of galaxy counts as a tracer of the matter density, Peebles found that the amplitude of the CMB anisotropies could be as small as a few $10^{-6}$ and, although still to be measured with better precision, reconciled CMB and galaxy observations.
Cold dark matter is a general class of constituents of the universe. CDM particles have negligible random velocities compared to the speed of light ($w=0$, hence \textit{cold}) and interact only gravitationally (\textit{dark} indicates that it interacts very weakly with ordinary matter and electromagnetic radiation).
Thanks to that, it forms in fact gravitational potential wells before the decoupling of matter and radiation and so allows baryons to clump in time to form galaxies after decoupling.
Weakly interacting massive particles (WIMPs) \cite{Blumenthal84} and axions \cite{Preskill82, Abbott82, Dine82} are two known hypothetical examples of this broader class.

This still-unknown constituent of our universe emerged back then as the bridge between the early universe and the late universe.
However in the early 1980s, this was only one solution of a problem that emerged while the greater absence of a causal mechanism for superhorizon perturbations remained, together with some proof that these should be scale-invariant.

\section*{The inflationary paradigm}
\addcontentsline{toc}{section}{The inflationary paradigm}

By the early 1980s, several inconsistencies of the Hot Big Bang model had accumulated and were calling for a complete change of paradigm.

Four years after the measurement of the CMB, Misner pointed out the issue of having a homogeneous signal \cite{Misner69}.
Using the particle horizon formalism of Rindler \cite{Rindler56}, he showed that the comoving distance travelled by light since the beginning of the Hot Big Bang was smaller than the comoving distance between correlated points in the sky, violating causality.
More explicitly, a photon trajectory is integrated over a background Friedmann universe ($ds^2 = 0 = dt^2-ad\chi^2$) to yield the comoving distance travelled by light between two scales $a_1$ and $a_2$ as
 \begin{equation}
 \Delta \chi (a_1,a_2) = \int_{t_1}^{t_2} \frac{dt}{a} = \int_{a_1}^{a_2} R_H(a) {\d\ln a},
 \end{equation}
which gives the \textit{particle horizon}, the comoving distance travelled by a photon since the origin ($a_1=0$), for a $(w\neq -\frac{1}{3})$-fluid as $\chi_p(a) = 2 R_H(a) (3w+1)^{-1}$, and so the ratio of the distance travelled since the CMB by a photon to the typical size of a causal patch at emission
\begin{equation}
    \frac{\Delta \chi(a,a_0)}{\chi_p(a)} = a^{-(3w+1)/2}-1.
\end{equation}
Assuming a dominating ($w>-\frac{1}{3}$)-fluid, e.g. radiation, this ratio is way greater than $1$ at CMB times ($a\ll 1$), and suggests that two photons travelling in approximately straight lines towards us could not have belonged to the same causal patch at emission, and so could not have been correlated. This is the so-called \textit{horizon problem}.

As highlighted in \cite{Dicke79} and before any strong constraint on $\Omega_{k0}$ back then, it might take a lot of fine-tuning to explain a very small $\Omega_k$ up to order of a few (which was expected as such with tentative constraints at that time).
In fact, $\Omega_k \propto (aH)^{-2}$, and so $\dot{|\Omega|}\propto \ddot{a} \propto (1+3w)$ for a dominating $w$-fluid. Clearly if $w>-\frac{1}{3}$, then $\dot{|\Omega|}>0$ and so if there was primordial curvature, it was way smaller in the past than today and so seemingly extremely fined-tuned close to $0$ in the initial conditions.

Finally, Grand Unified theories (GUTs) had been developed to describe the merger of couplings within the Standard Model of particle physics as high energies are reached backward in time.
These theories were predicting the presence of topological defects, formed from the breaking of larger symmetry groups, that would manifest as cosmic strings and magnetic monopoles \cite{Zeldovich78,Preskill79}, produced efficiently in the hot early universe.
However, no such defects had been found by then or today.

These main issues among others drove the community to find a causal process to precede the Hot (or \textit{old}) Big Bang, which would flatten the universe and make it monopole-free.
Initial efforts were certainly focusing on causality itself.
Through the work of Penrose and Hawking from black holes to cosmological spacetimes \cite{Penrose65,Hawking70}, it became clear that tracing the expansion back in time in GR takes the Hot Big Bang to a singularity.
These theorems rely on assumptions that the community will\ylg{ later} keep trying to break in order to escape such an uncomfortable start to the universe, in particular the assumption that Einstein's equations should hold and that the content of the universe satisfies the energy condition with at most zero cosmological constant.
In particular, Gliner and Brout–Englert–Gunzig \cite{Gliner66,Gliner70,Gliner75,Brout78} embraced a cosmological constant domination or de Sitter-like birth emerging from a quantum state and a phase-transition, where Einstein's equations only hold on shell, while Starobinsky considered his famous $R^2$ extension to Einstein-Hilbert's action \cite{Starobinsky79,starobinsky80}. The latter is usually remembered as the first predictive model of inflation.

\begin{figure}[ht]
  \centering
  \begin{subfigure}[t]{0.45\textwidth}
    \centering
    \includegraphics[width=\textwidth]{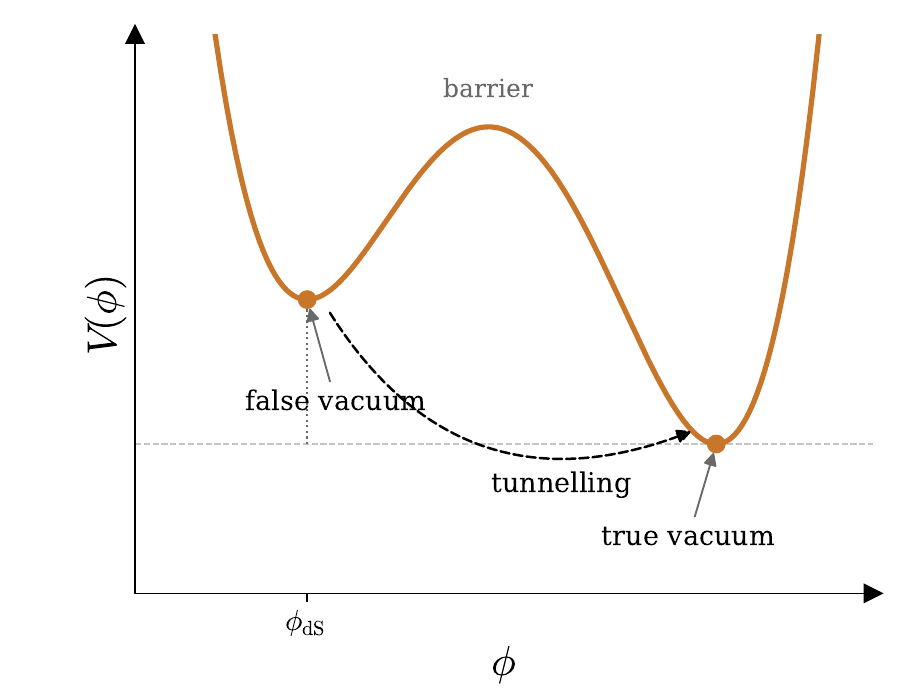}
    \caption{Old inflation (Guth 1981), the inflaton is trapped in a
      false vacuum and escapes via quantum tunnelling through a barrier
      to the true vacuum; the inhomogeneous bubble-nucleation prevents a
      graceful exit.}
    \label{fig:inflation_potentials_old}
  \end{subfigure}
  \hfill
  \begin{subfigure}[t]{0.45\textwidth}
    \centering
    \includegraphics[width=\textwidth]{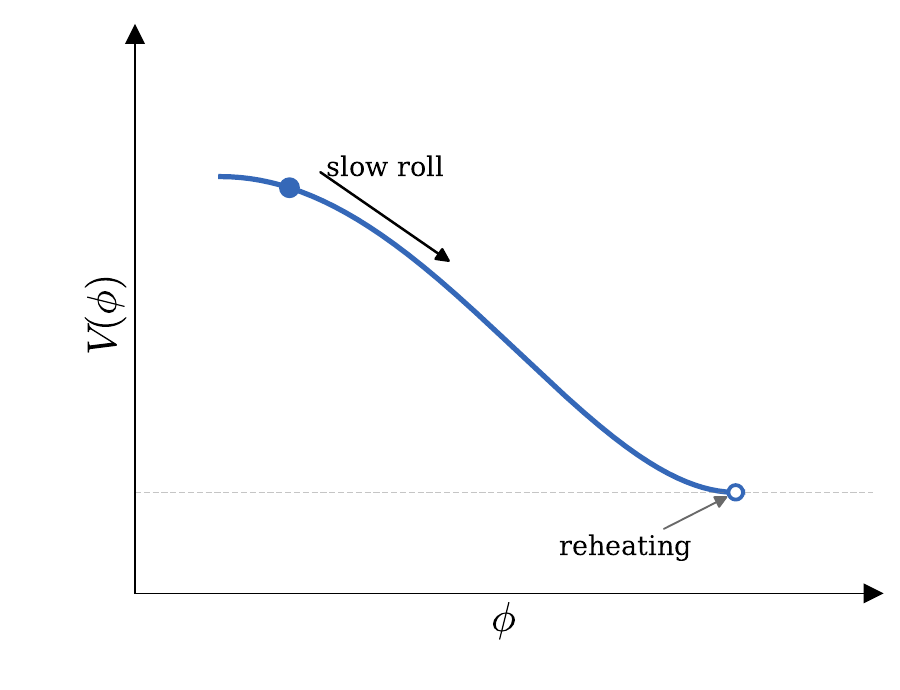}
    \caption{New inflation (Linde 1982; Albrecht \& Steinhardt 1982),
    the inflaton slow-rolls along a flat plateau before reheating at
      the true vacuum, providing a graceful exit.}
    \label{fig:inflation_potentials_new}
  \end{subfigure}
  \caption{Inflationary potentials of the 1980s.}
  \label{fig:inflation_potentials}
\end{figure}

The major breakthrough was that of Guth in 1981 \cite{guth_1981}, who, actively working on cosmological phase transitions \cite{Linde74,Coleman80,Kazanas80,Blumenthal84,Sato81}, emphasised that a GUT-scale phase-transition of a scalar field could lead to a de Sitter false vacuum, and showed how the associated exponential expansion addressed the three aforementioned problems.
Assuming a scalar field $\phi$ on a curved background, with minimally-coupled Lagrangian density
\begin{equation}
{\cal L} = \sqrt{-g}\bigg(\frac{1}{2\Mpl^2}R+X-V(\phi)\bigg),
\label{eq:VanillaAction}
\end{equation}
where $X = -\frac{1}{2}g^{\mu\nu}\partial_{\mu}\phi\partial_{\nu}\phi$ is the canonical kinetic term for this signature, one obtains the associated Friedmann equations applied to the homogeneous field $\phi_b$%
\begin{subequations}
\begin{align}
H^2 & = \frac{1}{3\Mpl^2}\rho_\phi = \frac{1}{3\Mpl^2}\left(\frac{1}{2}{\dot{\phi_b}^2+V(\phi_b)}\right), \\
\dot{H}+H^2 &= -\frac{1}{6\Mpl^2}(\rho_\phi + 3p_\phi) = -\frac{1}{3\Mpl^2}\left(\frac{1}{2}{\dot{\phi_b}^2-V(\phi_b)}\right),
\end{align}
\label{eq:backgroundInflation}
\end{subequations}
while the field undergoes Hubble friction following
\begin{equation}
    {\ddot{\phi_b}}+3H{\dot{\phi_b}}+ \frac{\d V}{\d\phi}(\phi_b) = 0.
\end{equation}
Now assuming that this potential admits a false vacuum at $\phi_{\rm dS}$, as represented in Figure \ref{fig:inflation_potentials_old}, and where the field can stop because of the friction (or via a phase transition changing $V$),
the field acts there as a cosmological constant or de Sitter spacetime where $H(\phi_{\rm dS})=\sqrt{V(\phi_{\rm dS})/3}/\Mpl=\text{cste}$. This clearly confirms that Einstein's term can be understood as a scalar field on a constant potential.

Most importantly, given that the Hubble radius is decreasing in de Sitter $R_H \propto e^{-Ht}$ ($w_{\rm dS}=-1$), the typical correlation scale is allowed to be much larger early on before decaying during inflation and increasing back after, towards CMB scales. 
Taking the maximum length scale between correlated regions observed today as at least the size of the observable universe, i.e. the particle horizon approximated as $R_H(t_0)$,
we require that to be smaller than the particle horizon at the time of inflation $\chi_p (t_i)$, which is at least $R_H(t_i)$ so that we can require $R_H(t_i)>R_H(t_0)$. Assuming de Sitter between $t_i$ and $t_r$ (end of inflation or \textit{reheating}) and radiation domination until $t_0$ (where $H\propto a^{-2}$), one gets a lower bound on the number of e-folds required for causality
\begin{equation}
    \begin{aligned}
 \Delta{\cal N}_{\rm dS}& \equiv\ln\left(\frac{a_r}{a_i}\right)= \ln\left(\frac{R_H(t_i)}{R_H(t_0)}\right)+\ln\left(\frac{H_{\rm dS}}{H_{0}}\right)+\ln\left(\frac{a_r}{a_0}\right)\\
   & >\ln\left(\frac{H_{\rm dS}}{H_{0}}\right)+\ln\left(\frac{a_r}{a_0}\right)\\
   & \simeq \frac{1}{2}\ln\left(\frac{\pi^2g_* T_0^4}{90\Mpl^2H_{0}^2}\right)+\ln\left(\frac{T_r}{T_0}\right),\\
   & > 60-70 + \ln(T_r/10^{15} GeV),             
    \end{aligned}
    \label{eq:60efolds}
  \end{equation}
where we have assumed an instantaneous transition from de Sitter to radiation domination, and $T_r$ is the reheating temperature such that $3\Mpl^2H_{\rm dS}^2 = \frac{\pi^2}{30}g_* T_r^4$, $g_*$ being the effective number of relativistic degrees of freedom at that time.
This matches what Guth required as about $60-70$ e-folds to fix the horizon problem.
The curvature and monopole problem are also fixed naturally as their density parameters are diluted by the exponential expansion, $\Omega_k \propto e^{-2\Delta{\cal N}_{\rm dS}}$ and $\Omega_{mm} \propto e^{-3\Delta{\cal N}_{\rm dS}}$.

In his article, Guth coined the term \textit{inflation} and at the same time exposed its flaws, see also \cite{Guth83}.
In fact, it became clear that ending inflation \yld{requires}\ylg{then required} the field to tunnel out of the false vacuum to reach the true vacuum.
This tunneling, say with nucleation rate per unit volume $\Gamma$ needs to be big enough against $H^4$ to see percolation (i.e. true vacuum bubble collisions), otherwise the de Sitter bubbles keep true vacuum bubbles exponentially apart.
Beyond the obvious need for at least one bubble for us to exist, percolation is essential as bubble walls travelling at the speed of light carry away most of the energy and have a very cold interior, unsuitable for explaining our universe: walls need to collide and dissipate their energy.
$\Gamma$ can also be boosted but this is also a problem as then inflation does not last long enough. The 1981 solution thus brings with it the \textit{graceful exit problem} and is already \textit{old}.

Similarly to fictional Benjamin Button \cite{fitzgerald1922benjamin}, inflation is born old but certainly grows younger quickly.
It took in fact only a year to find a more appropriate potential for the inflaton, and some would even say that Starobinsky's modified action \cite{starobinsky80} was there first, just not conformally transformed into an inflaton and its potential until 1984 \cite{Whitt84}.
With Linde \cite{Linde82}, as well as Albrecht and Steinhardt \cite{Albrecht82}, \textit{new inflation} appeared, as displayed in Figure \ref{fig:inflation_potentials}.
Both papers built on the same Coleman-Weinberg one-loop potential for the SU(5) adjoint Higgs field \cite{ColemanWeinberg73}, represented in Figure \ref{fig:inflation_potentials_new}.
In this paradigm, inflation happens after a phase transition and not before (to escape the graceful exit problem), as long as the inflaton rolls slowly enough (i.e. effectively close enough to de Sitter) down the potential to reach the true vacuum, where reheating happens.
With this, the end of inflation is not a rare, local, and violent quantum event anymore but a smooth, deterministic, and classical process that happens everywhere.
Note that the adjoint Higgs field as the inflaton was ruled out very quickly during the Nuffield Cosmology Workshop in Cambridge in 1982, see for instance \cite{Steinhardt84}. This is not to be confused with the Higgs field of the standard model.
The latter is still a candidate for inflation, see \cite{Bezrukov08}, but in general the inflaton is nowadays often studied as a generic scalar field, with an identity that is not necessarily related to any known particle physics model.

Within this \textit{new inflation} philosophy, a plethora of models have been proposed since, with different potentials and different mechanisms to end inflation, but all sharing the same basic idea of a scalar field driving an accelerated expansion of the universe that is close enough to de Sitter, as well as long enough, to solve the horizon, flatness, and monopole problems.
A quasi-de Sitter behaviour is usually evaluated by the \textit{slow-roll parameters} 
\begin{equation}
\varepsilon_{n+1} = -\frac{1}{H}\frac{\d}{\d t}\ln \varepsilon_n,
\end{equation}
which, apart from $\varepsilon_0 = H$, should be small to not deviate significantly from de Sitter expansion.
When these reach order $1$, the scalar field has reached the end of inflation and reheats the universe.
To this day, inflation still sees new models arising at a faster pace than they are ruled out, see and compare the \textit{Encyclopedia Inflationaris} \cite{encyclopedia_inflationaris} to the number of researchers working on the topic. 
It is also true that, once again as long as $H$ behaves effectively like a quasi-constant for long enough during inflation\footnote{As long as curvature and monopoles are not enhanced.}, an inflationary model does not have to be formulated through a canonical scalar field with a potential, as illustrated by the Starobinsky model \cite{starobinsky80} or other modifications of gravity \cite{Kobayashi10}, as well as noncanonical kinetic terms (moving from $X-V$ in eqn.~\eqref{eq:VanillaAction} to a generic functional $P(X,\phi)$) as found in $k$-inflation \cite{kInflation}, and multiple-field inflation \cite{Kofman85,Starobinsky85,Linde94}.
One can also consider models with non-scalar fields such as non-Abelian gauge fields \cite{GaugeFlation} etc. 

In this climate of infinite possibilities and a potential origin and singularity simply assumed earlier in time\footnote{Taking the example of the singularity problem: Borde et al. \cite{Borde03} showed that inflation cannot be eternal in the past, and so cannot avoid a singularity.},
past years have seen a lot of Effective Field Theory (EFT) approaches to inflation, see for instance \cite{Cheung2008,Senatore10, senatore_effective_2012}, defining a generic symmetries-motivated action for the inflaton and its perturbations, and so a generic model of inflation as a low-energy effective theory of a more fundamental theory of quantum gravity required above some cutoff scale.
This thesis does not look at such agnostic approaches. However, Chapter~\ref{sec:SIbeyondGR} uses the EFT of \yld{Dark Energy (comprising inflation)}\ylg{Inflation} to map a wide class of scalar-tensor theories.

Whichever model or approach is used, models need to fit the data and this brings us to discuss the inhomogeneous framework of inflation, as eqn.~\eqref{eq:60efolds} is a rather crude qualitative constraint from \yld{just }background physics\ylg{ only}.


\section*{A quantum origin of inhomogeneity}
\addcontentsline{toc}{section}{A quantum origin of inhomogeneities}
The quantum treatment of the early universe predates the inflationary monopoly over early universe physics.
As early as the late 1960s, quantum fluctuations were speculated to be the source of inhomogeneity in the universe \cite{Sakharov66,Harrison1970}, and as remarked earlier, quantum fluctuations may allow the avoidance of singularity theorems.
As long as our scalar field (or else) lives below the Planck energy scale, we do not require a theory of quantum gravity.
Beyond those regimes, one could use a path integral over the metric and fields, as was studied by Hawking \cite{Hawking78}.
The quantum field theory limit of this theory on curved spacetime is then that of a saddle-point approximation, when quantum fields are perturbations of the classical curved background.
Other approaches exist, such as Causal Dynamical Triangulation \cite{TriangularQG}, Wheeler-De-Witt quantisation \cite{WdW}, Loop quantum gravity \cite{Rovelli90}, or string theory \cite{polchinski1998}.

As we will see, to explain the origin of inhomogeneities with the current data, it does not seem necessary to solve the full theory of quantum gravity, and one can stick to standard quantum field theory in curved spacetime (QFTCS), which can still accommodate unknown UV-effects via Effective Field Theory (EFT) approaches \cite{polchinski1998,Cheung2008}.
QFTCS was looked at soon after the first developments of QFT, notably with Schrödinger's 1939 precursor paper looking at the creation and annihilation of particles in an expanding universe \cite{Schrodinger39}.
While Schrödinger drew some ``alarming'' conclusions, Parker's work in the late 1960s is usually cited as the first full treatment of the quantisation of fields on curved backgrounds \cite{Parker68,Parker69,Parker71}.
To start a generic QFT quantisation, a \textit{free} scalar field $\ylc{\varphi}$\ylg{, noted differently from the inflaton $\phi$ as it lives on a fixed background geometry with no backreaction on the metric,} and its conjugate momentum $\pi_{\ylc{\varphi}}$ are written as an infinite sum of harmonic oscillators, one per Fourier mode. \ylg{Here we have assumed a homogeneous background, so that Fourier modes decouple, and an isotropic background and vacuum, so that the mode functions depend only on the norm $k=|\mathbf{k}|$}%
\begin{subequations}
    \begin{align}
    \ylc{\hat\varphi}(t,\mathbf{x}) &= \int\!\frac{d^3k}{(2\pi)^{3}}\,
    f_k(t)
    \hat a_{\mathbf{k}}\, e^{ i\mathbf{k}\cdot\mathbf{x}} +\text{h.c},
    \\
    \hat\pi_{\ylc{\varphi}}(t,\mathbf{x}) &= \frac{\partial {\cal L}}{\partial \ylc{\dot\varphi}}= \int\!\frac{d^3k}{(2\pi)^{3}}\,
    \dot{f}_k(t)
    \hat a_{\mathbf{k}}\, e^{ i\mathbf{k}\cdot\mathbf{x}} +\text{h.c}.\label{eq:QFTexpansionpi}
    \end{align}\label{eq:QFTexpansion}
\end{subequations}
\yld{so that the}The commutation relations from the annihilation and creation operators $a_{\mathbf{k}}$ and $a_{\mathbf{k}}^{\dagger}$, $[a_{\mathbf{k}},a_{\mathbf{k}'}^{\dagger}]=(2\pi)^3 \delta^{\ylg{(3)}}(\mathbf{k}-\mathbf{k}')$, and the associated to a quantum vacuum state $|0\rangle$, give the canonical commutation relations of the field and its conjugate momentum
\begin{equation}
\Big[\ylc{\hat\varphi}(t,\mathbf{x}),\,\hat\pi_{\ylc{\varphi}}(t,\mathbf{y})\Big] = i\,\delta^{(3)}(\mathbf{x}-\mathbf{y}),
\label{eq:commutation}
\end{equation}
as long as $f_k \dot{f}_k^*-f_k^*\dot{f}_k=i$.
In this Heisenberg picture for the free theory, we have given the time-dependence to the operators, here through the mode functions $f_k(t)$, which satisfy
\begin{equation}
\ddot{f}_k(t) + \omega_k^2 f_k(t) = 0,
\end{equation}
coming from the free equation of motion for the field (e.g. $\square\ylc{\varphi} + m^2 \ylc{\varphi} = 0$), and where $\omega_k$ is an angular frequency.
In a Minkowski spacetime as commonly found in particle physics, $\omega_k^2 = k^2 + m^2$ and the mode functions are simply plane waves $f_k(t) = e^{-i\omega_k t}/\sqrt{2\omega_k}$.

With the simple example of a massive scalar field on a fixed FLRW spacetime, $\varphi$, the time-dependent scale factor complicates the action, notably giving the conjugate momentum an $a^3$ volumetric factor and the field still satisfies the Klein-Gordon equation but the D'Alembertian operator is now time-dependent, $\Box_g = -\partial_t^2 - 3H\partial_t + a^{-2}\nabla^2$.
The mode functions of $\varphi$, noted $u_k$, now satisfy a rescaled ($u_k\equiv a f_k$) \ylg{mode }equation
\begin{equation}
   u_k'' + w_k^2(t)u_k= 0,
   \label{eq:fixedBGmodes}
\end{equation}
after defining the conformal time $\eta$ derivative $(\cdot)''$ through $d\eta = dt/a$, and the comoving wavenumber $\mathbf{k}$,
and where, with or without mass, the angular frequency is now a time-dependent function
\begin{equation}
w_{{k}}^2(t) = {k}^2+\frac{1}{R_H^2}\Bigg(2-\varepsilon_1+\frac{m^2}{H^2}\Bigg).
\label{eq:wfixedBG}
\end{equation}
As claimed by Schrödinger, this is alarming (even with $\varepsilon_1 = 0$ back then). In particular, we cannot expand over plane-waves anymore and our Hamiltonian and its vacuum are not conserved.
However let us return to the seemingly safe expansion of eqn.~\eqref{eq:QFTexpansion}. 
We observe that in the limit where $kR_H\gg1$ (assuming bounded $\varepsilon_1$ and $m/H$), which we shall name the \textit{far past} of a given mode or \textit{static} region,
we recover a massless equation in a flat spacetime.
In this region of spacetime, we can initialise the field like in Minkowski\footnote{Parker invoked other reasons than\ylg{ the appeal of Minkowski's} symmetry, arguing for an adiabatic activation of the perturbations and interactions, see also \cite{Lippman50,GellMann51,GellMann53}.}
\begin{equation}
        u_k(\eta)\underset{kR_H\gg 1}{=} \frac{e^{-ik\eta}}{\sqrt{2k}},
        \label{eq:BDvac}
\end{equation}
which we will use and refer to as \textit{Bunch-Davies (BD)} initial conditions \cite{BD}.
Note that this prescription was first used on a curved background in de Sitter \cite{Chernikov68}, as it preserves the de Sitter $O(1,4)$ isometries (see also $\alpha$-vacua \cite{Mottola85,Bruce85}) but also because it is the conformal vacuum (reduces to a conformal transform of the Minkowski-space positive frequency function in the conformal past).
At a different time, one could define a new annihilation operator $\hat b_{\mathbf{k}}=\alpha_k \hat a_{\mathbf{k}}+\beta_k^* \hat a_{-\mathbf{k}}^\dagger$ and creation $\hat b_{-\mathbf{k}}^\dagger$ respectively, such that $|\alpha_k|^2-|\beta_k|^2=1$, and thus a new vacuum $|{\tilde 0}\rangle$.
Clearly the number of $b$ particles in $|0\rangle$ is nonzero, $\langle0|b_{\mathbf{k}}^\dagger b_{\mathbf{k}}|0\rangle\neq 0$, essentially confirming that a time-dependent Hamiltonian has different vacua for different times, seemingly justifying Schrödinger's alarm about particle creation.
In Minkowski spacetime the situation is different because there is a preferred vacuum state, coming from a conserved Hamiltonian with a minimal energy state at all times.
In FLRW we need to specify a physical preference for a vacuum, and we choose to recover Minkowski's in the far past, where it is exactly adiabatic, and which is conveniently the only portion of spacetime where frequencies are defined.\footnote{The reason\ylg{ why} the far past is special is that the adiabatic regime effectively restores time-translation symmetry. While a generic FLRW spacetime has no timelike Killing vector and therefore no unique notion of positive frequency, in the limit $k/a\gg H$ the expansion becomes negligible over an oscillation period, the geometry is locally Minkowskian, and an approximate timelike Killing vector exists. This singles out the Minkowski positive-frequency modes, just \yld{as}\ylg{like} asymptotically stationary regions do in black-hole spacetimes.
See historical developments on observer-dependence of the particle concept in general, in Minkowski \cite{Unruh76} and black holes spacetimes \cite{Zeldovich71,Zeldovich72,ZeldovichStarobinsky72,Starobinsky73,Bekenstein72,Bekenstein73,Hawking71,BardeenCarterHawking73,Hawking74,Hawking75,Davies75,Unruh76,Gibbons77}.
}

In 1981, Mukhanov and Chibisov anchored the idea of a quantum origin of all inhomogeneity \cite{Mukhanov1981}, and even before inflation was born,
showed how a quasi-de Sitter framework was to be naturally considered by looking at Starobinsky's $R^2$ model. They provided the first estimate of a red tilted primordial power spectrum.

In 1982 after the fast arrival of inflation in the West and unaware of the work of Mukhanov and Chibisov, Hawking organised the Nuffield workshop in Cambridge \cite{Nuffield83}, allowing him to complete his early computation of the primordial power spectrum \cite{Hawking1982}, and others to publish on the same matter \cite{Starobinsy82,Guth82pert,Bardeen83} after all agreeing on the result.
Making use of eqn.~\eqref{eq:fixedBGmodes} for a massless field in de Sitter ($m=0, \varepsilon_1=0$ in eqn.~\eqref{eq:wfixedBG}) sourced by the BD vacuum in the far past described by eqn.~\eqref{eq:BDvac}, a scale-invariant power spectrum \yld{is}\ylg{was} found as $|\delta \ylc{\varphi}_k|^2 \simeq H^2 /2k^3$, not without debates on which gauge to use and how to relate the field fluctuations to the late-time curvature perturbations.

\begin{figure}[p]
  \centering
  \begin{subfigure}[t]{0.8\textwidth}
    \centering
    \includegraphics[width=\textwidth]{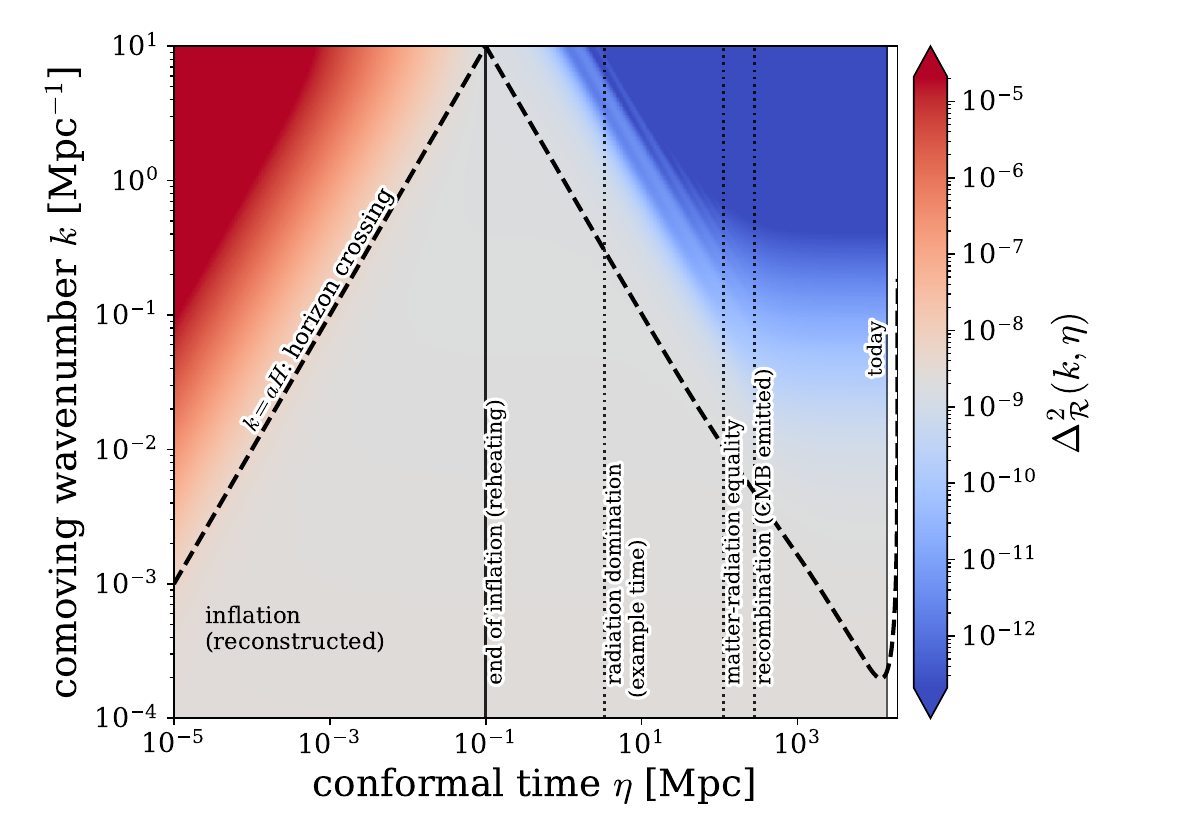}
    \caption{$\Delta^2_{\cal R}(k,\eta)$ across horizon exit, freeze-out,
      and re-entry, with the comoving Hubble radius.\protect\footnotemark}
    \label{fig:ms_spectrum_heatmap}
  \end{subfigure}

  \begin{subfigure}[t]{0.8\textwidth}
    \centering
    \includegraphics[width=\textwidth]{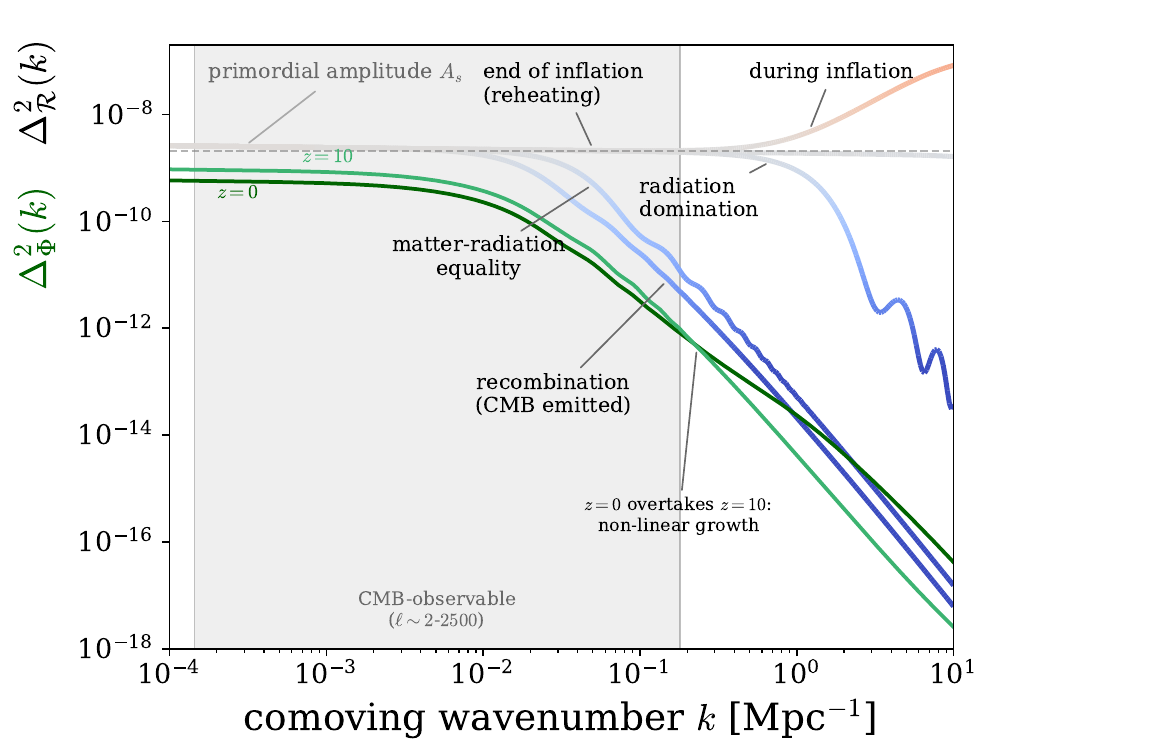}
    \caption{Snapshots of $\Delta^2_{\cal R}(k)$ \yld{at}\ylg{matching} \yld{different}\ylg{the} times highlighted in Figure~\ref{fig:ms_spectrum_heatmap}, with the same colour scheme.
      Green curves are the lensing potential spectra $\Delta^2_\Phi(k)$ at $z=0$ and $10$, non-linearly corrected with halo models.
      The initial curvature perturbation $\cR$ seeds the linear matter density field $\delta_m$ (see eqn.~\eqref{eq:adIC}), for which linear theory breaks down on small scales during matter domination after reaching the critical density for spherical collapse.
      The abundance of collapsed regions for each mass can be convolved with mass-dependent halo density profiles to build the small-scale (\textit{one-halo}) power, which, once added to the large-scale original linear power (called \textit{two-halo}), gives the full nonlinear spectrum.
      The Einstein-Poisson equation can then be used to relate the matter density to the lensing potential, and since
      $\Phi\simeq\tfrac35{\cal R}$ in matter domination, it shares
      $\Delta^2_{\cal R}(k)$'s flat-then-declining shape.
      }
    \label{fig:ms_spectrum_snapshots}
  \end{subfigure}
  \caption{Spectra from inflation to today. Evolution of the primordial curvature power spectrum from
    horizon exit through freeze-out to re-entry. Planck-2018 CAMB output \cite{Lewis:1999bs} is used for the Boltzmann
    evolution from the end of inflation to today, with the initial conditions set by the inflationary power spectrum. 
    The non-linear matter power spectrum is computed with the integrated halo model HMcode-2020 \cite{Mead:2020vgs}.
    Inflationary fluctuations are {reconstructed} by solving the Mukhanov-Sasaki equation backward from Planck-2018 CAMB values for the amplitude and tilt.}
  \label{fig:ms_spectrum}
\end{figure}

In 1985 the full GR linear dynamics were understood better when Mukhanov showed how to account for the propagation and quantisation of the fluctuations through metric scalar degrees of freedom \cite{Mukhanov85}, essential in non-de Sitter inflation.
This was done linearly in the Newtonian gauge and extended to a gauge-invariant treatment by Sasaki a year later \cite{Sasaki86}. With this breakthrough, scalar matter and metric perturbative degrees of freedom became dynamically coupled and gauge-mixed, and inflation became then a precision-calculational framework.
Nowadays this is remembered through the \textit{Mukhanov-Sasaki} (MS) equation for the scalar perturbations in Fourier space
\begin{equation}
   v_k'' + \bigg(k^2-\frac{z''}{z}\bigg)v_k= 0,
   \label{eq:MSvEq}
\end{equation}
after defining carefully the gauge-invariant Mukhanov-Sasaki variable $v \equiv z{\cal R}$, with  $z^2\equiv 2a^2\varepsilon_1$, with ${\cal R}$ defined in eqn.~\eqref{eq:defR}, and initialised with the BD vacuum, see eqn.~\eqref{eq:BDvac}.
The time-dependent effective mass term can be written as
\begin{equation}
    -\frac{z''}{z}= -\frac{1}{R_H^2}\bigg(2-\varepsilon_1-\frac{3}{2}\varepsilon_2+\frac{1}{2}\varepsilon_1\varepsilon_2+\frac{1}{4}\varepsilon_2^2+\frac{1}{2}\varepsilon_2\varepsilon_3\bigg),
\end{equation}
and allows for identifying $R_H$ once again as a boundary between two regimes: the far past or \textit{sub-Hubble} where $kR_H\gg1$ and the \textit{super-Hubble} regime where $kR_H\ll1$ and where clearly $v_k\propto z$ is solution.
This shows that ${\cal R}$ is frozen on super-Hubble scales. This property of single-field inflation is very convenient because it implies that one can ignore what happens to\ylg{ linear} modes crossing the horizon during the end of inflation, after the quasi-de Sitter phase (including the so-called \textit{reheating} phase where the field goes to the bottom of its potential and decays), and still get initial conditions for the post-inflationary evolution of the universe, as illustrated in Figure \ref{fig:ms_spectrum}.
This also explains why the power spectrum of ${\cal R}$ is evaluated at Hubble crossing $k=aH$ at leading order in slow-roll
\begin{equation}
    {\cal P}_{\cal R}(k) \equiv \frac{k^3}{2\pi^2}\bigg|\frac{v_k}{z}\bigg|^2_{k=aH} \simeq \frac{H^2}{8\pi^2\Mpl^2\varepsilon_1}\bigg|_{k=aH},
    \label{eq:PR}
\end{equation}
which, similarly to Mukhanov's and Chibisov's 1981 result, is red tilted because of the full dynamics, as quantified by the \textit{spectral index}
\begin{equation}
n_s - 1 \equiv \frac{d\ln \mathcal{P}_{\mathcal{R}}}{d\ln k}= -2\varepsilon_1 + \varepsilon_2.
\end{equation}
A similar treatment can be performed for the tensor perturbations (historically predating the scalar ones) \cite{Grishchuk74,Starobinsky79, Rubakov82}, or \textit{gravitons} once quantised, which are also frozen on super-Hubble scales (each polarisation following eqn.~\eqref{eq:MSvEq} taking $z = a$ instead), and \yld{whose}\ylg{the} power spectrum\ylg{ of which} is given by
\begin{equation}
    {\cal P}_{\rm t}(k) \equiv \frac{k^3}{2\pi^2}\bigg|\frac{h_k}{a}\bigg|^2_{k=aH} \simeq \frac{2H^2}{\pi^2\Mpl^2}\bigg|_{k=aH},
    \label{eq:Pt}
\end{equation}
with a spectral index $n_t = -2\varepsilon_1$.

In 1992, the COBE satellite finally measured the CMB temperature anisotropy at $\delta T/ T = 10^{-5}$, slightly above what Peebles had proposed with WIMPs. WMAP \cite{Bennett2013} and Planck \cite{PlanckInflation} later confirmed the nearly scale-invariant spectrum with high precision.
As we will see and despite not measuring any tensor modes, this measurement is the most direct evidence of inflation up to this day, and among the most constraining on the inflationary model space.

\footnotetext{The exact
      expansion history immediately after reheating is itself
      model-dependent (e.g.\ kination or an early matter-dominated phase are
      real alternatives to the instantaneous radiation domination assumed
      here and by CAMB).}

\section*{The cosmological collider era}
The MS equation, eqn.~\eqref{eq:MSvEq}, is the inflationary Klein-Gordon equation for the curvature perturbation, and it is linear in the field. This is enough to get the power spectrum, but not enough to get higher-order correlators, which are needed to probe the interactions of the inflaton with itself and with other fields.
Porting the tools from particle physics to cosmology was in some way rather unsuccessful at first.

Let us consider the transition amplitude from a vacuum state $| \Omega_{\text{in}} \rangle$ in the far past $ T\to -\infty$ to $| \Omega_{\text{out}} \rangle$ in the far future $T\to \infty$ in an interacting theory (as opposed to a quadratic free theory).
The transition is described by the evolution operator $U(t,t_1)$, defined as the unique solution of
\begin{equation}
i\,\partial_t\, U(t,t_1) = H(t)\, U(t,t_1).
\qquad U(t_1,t_1) = \boldsymbol{1}
\end{equation}
Picard iteration yields the equivalent \ylg{solution }series%
\begin{subequations}
\begin{align}
U(t_2,t_1) &= {\cal T}\exp\Big(-i\int_{t_1}^{t_2}dt'\,H(t')\Big)\\
&= \sum_{N=0}^{\infty}(-i)^N
\int_{t_1}^{t_2}\!ds_N \int_{t_1}^{s_N}\!ds_{N-1}\cdots\int_{t_1}^{s_2}\!ds_1\;
H(s_N)\,H(s_{N-1})\cdots H(s_1),
\end{align}
\end{subequations}
i.e. operators ordered with the latest time to the left. In the interaction picture, $U$ becomes $U_I$ where $H$ is $H_{int}$ instead.
In the interacting theory and with no equilibrium (time-dependent Hamiltonian),  $| \Omega_{\text{in}} \rangle$ and $| \Omega_{\text{out}} \rangle$ are different and generically do not coincide with the free theory vacua at different times $|0_t\rangle$ (recalling as well that $U_0(t_2,t_1)|0_{t_1}\rangle\neq|0_{t_2}\rangle $ because of particle creation).
However, we assume that sufficiently far in the past and the future, the free theory is stationary so that $|0_{{-\infty/ +\infty}}\rangle=|0_{\text{in/ out}}\rangle$.

In particle physics, computations involve \textit{amplitudes} to go from one state to the other, with some event at $t$ modelled by an operator $\hat{O}_I$ in the interaction picture%
\begin{subequations}
    \begin{align}
\langle \hat O(t) \, \rangle & \equiv  \lim_{T\to\infty} \langle \Omega_T|\,U_I(T,t)\;\hat O_I(t)\;U_I(t,-T)\; \,| \Omega_{-T} \rangle\\
    & = \lim_{T\to\infty(1-i\varepsilon)}
\frac{\langle 0_T |\; U_I(T,t)\;\hat O_I(t)\;U_I(t,-T)\; | 0_{-T} \rangle}
{\langle 0_T |\; U_I(T,-T)\; | 0_{-T} \rangle}.
    \end{align}
    \label{eq:inout}
\end{subequations}
where $U_I$ is the interactive evolution operator.
In this so-called \textit{in-out} formula, we have used the so-called $i\varepsilon$ prescription \cite{Maldacena_2003,Weinberg05,WaldQFTCS,Kaya19} to project the interacting vacuum onto the free ones $|0_{\text{in/ out}}\rangle$, effectively counting bubble diagrams off as seen in the denominator.\footnote{The limit $T\to\infty(1-i\varepsilon)$ and the denominator play
complementary roles. Inserting a complete set of eigenstates of $H$,
$e^{-iHT}\ket{0}=\sum_n e^{-iE_nT}\ket{n}\!\bra{n}0\rangle$, the small
imaginary tilt of $T$ suppresses each excited state by $e^{-\varepsilon E_n T}$,
so only the true (interacting) vacuum $\ket{\Omega}$ projection survives: the
$i\varepsilon$ prescription projects the free vacuum onto $\ket{\Omega}$,
assuming $\braket{\Omega|0}\neq 0$. This projection, however, yields
$\ket{\Omega}$ only up to the divergent factor
$e^{-iE_\Omega T}\braket{\Omega|0}$. Applying the same construction to
$\braket{\Omega|\Omega}=1$ shows that this factor is precisely
$\bra{0}T\{e^{-i\int H_{\mathrm{int}}\,dt}\}\ket{0}$, i.e.\ the sum of vacuum
bubble diagrams; dividing by it therefore normalises the state and, since
disconnected bubbles factorise from any correlator, cancels them
diagram by diagram, leaving only diagrams connected to the external points. This technique is closely related to the \textit{adiabatic switching} of interactions \cite{Lippman50,GellMann51,GellMann53}.}

In cosmology, the experiment is certainly different in the first place as it has already taken place, only once, and we have not chosen or measured the initial and final states, or placed any apparatus of our choice in between.
Following what has been explained previously, one usually relies on the assumption of a state for the fields in the far past to be the vacuum state and then compute statistics of the fields with e.g. $O(t) = \varphi^n$ in the so-called \textit{in-in} formula
\begin{subequations}
    \begin{align}
\langle  \hat O(t) \,\rangle & \equiv  \lim_{T\to\infty} \langle \Omega_{-T}|\,U_I(-T,t)\;\hat O_I(t)\;U_I(t,-T)\; \,| \Omega_{-T} \rangle\\
    & = \lim_{T\to\infty(1-i\varepsilon)}
\frac{\langle 0_{-T} |\; U_I(-T^*,t)\;\hat O_I(t)\;U_I(t,-T)\; | 0_{-T} \rangle}
{\langle 0_{-T} |\; U_I(-T^*,t)\,U_I(t,-T)\; | 0_{-T} \rangle}.
    \end{align}
    \label{eq:inin}
\end{subequations}
In that sense we look at an average value, not a transition matrix anymore.
This was understood in the 60s by Schwinger \cite{Schwinger60} in simple oscillating systems well before applications in cosmology, and was named \textit{Closed-Time-Path} (CTP) formalism as visible from the forward and backward use of the evolution operator.
The point was precisely to compute expectation values for systems out of equilibrium, where \textit{in-out} amplitudes are the wrong object because the future state is unknown.
 Bakshi and Mahanthappa developed the perturbative methods \cite{Bakshi62,Bakshi63}, while Feynman and Vernon (1963) \cite{Feynman63} built the closely related influence functional for a system coupled to (and taken out of equilibrium by) an environment.
Finally Keldysh provided the diagrammatic Green's-function formulation \cite{Keldysh64}. Given this important contribution it is common to refer to the CTP formalism as the \textit{Schwinger–Keldysh} formalism, notably when referring to the diagrammatic use of the \textit{in-in} formula.

Until the 1980s, this formalism lived mostly in nonequilibrium statistical mechanics, not particle physics or gravity.
It became then of interest to the GR community, whose ambition was finding a semiclassical limit or quantum backreaction on the geometry of spacetime. 
Once made accessible \cite{Chou85}, Jordan, and Calzetta and Hu \cite{Jordan86,Calzetta87} set up the CTP formalism in FLRW spacetimes, using notably an equivalent path-integral approach, see \cite{Chen17diag} and Appendix~\ref{app:CTP}.
Many years later, early universe cosmology saw its first \textit{in-in} computations for the inflaton's self-interaction \cite{Falk93,Gangui94}.\footnote{Coincidentally, one of the first computations was done using the formalism of \textit{stochastic inflation}.}
The first complete GR calculation of the leading interactive diagrams, those contributing to the 3-point function or \textit{bispectrum} at the end of inflation $\langle {\cal R}_{\mathbf{k}_1}{\cal R}_{\mathbf{k}_2}{\cal R}_{\mathbf{k}_3}\rangle$, was achieved by Maldacena in 2003 \cite{Maldacena_2003} using eqn.~\eqref{eq:inin},
and later crystallised by Weinberg's treatments \cite{Weinberg05}.

Let us sketch such computations in preparation of Chapter~\ref{sec:semiclQFT}.
One starts from the choice of an action, which is then perturbed to the desired order in its Hamiltonian.\footnote{Note subtleties in transiting from perturbative Lagrangian to Hamiltonian, see e.g. \cite{chen_primordial_2010} for a review.}
We choose to stay generic in our action, and simply assume that we are given bosonic bulk fields with their conjugate momenta, following notations designed in  \cite{dias_numerical_2016-1} for the case of spin-0 only.
Let $\phi^\alpha$, $\alpha=1,\dots,N$, be this arbitrary collection of bosonic fields of integer 
spins $\mathsf s_\alpha$.\footnote{The framework covers arbitrary \emph{integer}-spin bosonic content at tree level. 
Half-integer spin lies outside its scope: Grassmann-valued $X^{\mathsf a}$ require replacing 
$\epsilon^{\mathsf{ab}}$ by a graded symplectic structure and rebuilding the evolution equations 
from anticommutators.}\,\footnote{One could follow treatments suggested in \cite{Deser03,Lee16,Arkani-Hamed15} and decompose these higher-spin modes with rank-$\mathsf s_\alpha$ 
polarisation tensors, built as symmetrised traceless products of the spin-1 ones.
}\,\footnote{Subtleties of the arbitrary spin formalism in curved spacetime seem to still be an active area of research and show very few example calculations in the literature. Here the author is simply noting the large \textit{in-in} literature.}
Typical content include the inflaton on a fixed background $\varphi$ or the curvature 
perturbation ${\cal R}$ ($\mathsf s=0$), isocurvature modes $\sigma^I$ ($\mathsf s=0$), the graviton 
$\gamma^{\pm2}_{\mathbf k}$\ylg{ ($\mathsf s=2$)}, and massive spin-$\mathsf s$ 
fields. Note that the gauge(s) should have been fixed at this point.
We collect the modes and their conjugate momenta into a composite phase-space vector
\begin{equation}
X^{\mathsf a}_{\mathbf k}
\equiv\big(\phi^{\alpha,s}_{\mathbf k}\,,\;\pi^{\,\alpha,s}_{\mathbf k}\big)\,,
\qquad 
\mathsf a \equiv (\alpha,s;\,\phi\ \text{or}\ \pi),
\end{equation}
populated with fields in the interaction picture, that is evolving with the free Hamiltonian.
In this formalism and in addition to BD initial conditions, the quantisation $\hat X^{\mathsf a}_{\mathbf k}(t)
   = f^{\mathsf a}_{k}(t)\,\ylg{\hat a^{\alpha,s}_{\mathbf k}}
   + f^{\mathsf a\,*}_{k}(t)\,\ylg{\hat a^{\alpha,s\,\dagger}_{-\mathbf k}}$\ylg{, with one pair of annihilation and creation operators per degree of freedom $(\alpha,s)$, $\big[\hat a^{\alpha,s}_{\mathbf k},\hat a^{\beta,s'\,\dagger}_{\mathbf k'}\big]=(2\pi)^3\delta^{\alpha\beta}\delta^{ss'}\delta^{(3)}(\mathbf k-\mathbf k')$,} and so the commutation relations of eqn.~\eqref{eq:commutation} translate into
\begin{equation}
\big[\hat X^{\mathsf a},\hat X^{\mathsf b}\big]
= i (2\pi)^3\delta^{(3)}(\mathbf k_a+\mathbf k_b) \begin{pmatrix}
0 & \boldsymbol{1}_{\mathrm{Sp}} \\[2pt]
-\boldsymbol{1}_{\mathrm{Sp}} & 0
\end{pmatrix},
\qquad \boldsymbol{1}_{\mathrm{Sp}}=\text{diag}(\{1_{2\mathsf s_i+1}\}_{i=1}^N).
\label{eq:symplectic}
\end{equation}
A generic cubic interaction\ylg{ arising from a coupling expansion} is a trilinear functional of the phase-space variables which, by 
spatial translation invariance, takes the form
\begin{equation}
\hat H_3 = \sum_{\mathsf{a}_1\mathsf{a}_2\mathsf{a}_3}
\int\!\prod_{r=1}^{3}\frac{\mathrm d^3\mathbf k_r}{(2\pi)^3}\;
(2\pi)^3\delta^{(3)}\!\Big(\textstyle\sum_r\mathbf k_r\Big)\,
\mathcal H_{\mathsf a_1\mathsf a_2\mathsf a_3}\!\big(\mathbf k_1,\mathbf k_2,\mathbf k_3;t\big)\,
\hat X^{\mathsf a_1}_{\mathbf k_1}\hat X^{\mathsf a_2}_{\mathbf k_2}\hat X^{\mathsf a_3}_{\mathbf k_3}\,,
\label{eq:vertex}
\end{equation}
where $\mathcal H_{\mathsf a_1\mathsf a_2\mathsf a_3}$ is a kernel symmetric under simultaneous
exchange of index--momentum pairs. Some further properties can be enforced, such as restricting to rational functions of the momenta using arguments on locality of the underlying Lagrangian.
If we impose invariance under simultaneous rotations to spin-0 fields, $\mathcal H$ is fully specified by magnitudes of the momenta alone,
$\mathcal H_{\mathsf a_1\mathsf a_2\mathsf a_3}(k_1,k_2,k_3;t)$.

Plugging this interaction Hamiltonian in eqn.~\eqref{eq:inin}, expanding up to first order in $\hat H_3$ and, like QFT in flat space, applying Wick's (or Isserlis') theorem to break the Gaussian 6-point function into products of 2-point functions (only nonzero if contracting creation and annihilation operators of the same mode, here internal-external pairs of momenta and their permutations), we get the \textit{tree-level} $({\mathsf a, \mathsf b, \mathsf c})$ 3-point function at time $t$
\begin{equation}
    \begin{aligned}
        \langle \hat X^{\mathsf a}_{\mathbf k_1}\hat X^{\mathsf b}_{\mathbf k_2}\hat X^{\mathsf c}_{\mathbf k_3}\rangle 
        & = -\,i\!\int\limits_{-\infty(1-i\varepsilon)}^{t}\!\mathrm d t'\,
        \big\langle\big[\,X^{\mathsf a}_{\mathbf k_1}(t)\,X^{\mathsf b}_{\mathbf k_2}(t)\,X^{\mathsf c}_{\mathbf k_3}(t)\,,\;
        H_3(t')\,\big]\big\rangle\,\\
        & =  2\cdot 3! \cdot (2\pi)^3\,\delta^{(3)}\!\Big(\textstyle\sum\limits_i \mathbf k_i\Big)\times
   \mathrm{Im}\!\Big[\,
     f^{\mathsf a}_{k_1}(t)\,f^{\mathsf b}_{k_2}(t)\,f^{\mathsf c}_{k_3}(t) \\
    & \int\limits_{-\infty(1-i\varepsilon)}^{t}\!\mathrm dt'
     \sum\limits_{\mathsf a_1\mathsf a_2\mathsf a_3}
     \mathcal H_{\mathsf a_1\mathsf a_2\mathsf a_3}(\mathbf k_1, \mathbf k_2, \mathbf k_3;t')
    \, \times\, f^{\mathsf a_1\,*}_{k_1}(t')\,
     f^{\mathsf a_2\,*}_{k_2}(t')\,
     f^{\mathsf a_3\,*}_{k_3}(t')
   \Big].
    \end{aligned}
\label{eq:ininBispectrum}
\end{equation}
This bispectrum formula is the leading contribution to the fields' non-Gaussian statistics and nonlinear dynamics.
As shown by Maldacena in GR, this formula predicts a nonzero departure from Gaussianity in the primordial curvature perturbation because of the inflaton's self-interactions.

While Maldacena's calculation showed that the quasi-de Sitter bispectra are small (slow-roll suppressed), compatible with still no current detection in the CMB \cite{PlanckNG}, it made the signature of inflation clear: non-Gaussian statistics by the end of inflation should be expected and provide a direct link to the couplings of the underlying theory, similarly to a collider experiment.
In the hope of future detections, the past 20 years have thus seen these methods applied to various actions and interactions of the inflaton, whether analytically \cite{Seery06,chen_observational_2007,Alishahiha04,Cheung2008,senatore_effective_2012,Senatore10b,Chen10} or numerically \cite{chen_large_2007,chen_generation_2008,Bingo,dias_numerical_2016-1,clarke_probing_2021,werth23,Zhang25,Costantini25}.

Recent years have also seen new analytical methods emerge to leave aside the model dependence (such as that of a potential) of those tedious calculations.
This is the so-called \textit{cosmological bootstrap} programme, which aims at fixing the correlator at the end boundary of inflation from first principles rather than computing bulk time integrals vertex by vertex.
Because of sharper analytic properties
than the correlators themselves, the
wavefunction of the universe is used,
\begin{equation}
\Psi[\zeta;\eta_0] = \int^{\zeta} \mathcal D\varphi\, 
e^{iS[\varphi]},
\end{equation}
whose logarithm defines wavefunction coefficients 
$\psi_n(\mathbf k_1,\dots,\mathbf k_n)$ \ylg{when expanded in powers of the field $\varphi$,
}from which equal-time correlators 
follow by Born-rule integration \cite{Maldacena_2003,Anninos15}. For instance, if the theory respects the symmetries of de Sitter space, notably its boost, correlators of light fields 
obey conformal Ward identities at the end of inflation, constraining along the way any exchange before that \cite{Arkani-Hamed18,Baumann20}.
Even in the boostless case, many other properties of the wavefunction coefficients can be used to bootstrap the correlators, such as their singularity structure, unitarity, and manifest locality \cite{Pajer21, goodhew_cosmological_2021-1, Jazayeri21}.
These methods have reproduced and extended the known single-field bispectra without 
reference to a bulk Lagrangian, classifying the allowed shapes directly 
in terms of the singularity structure and symmetries of the boundary observables. 

Beyond self-interactions of the curvature perturbation, one can imagine interactions with other fields, which can be massive or of higher spin.
In recent years, \textit{cosmological collider
physics} \cite{Chen09,Chen10,Arkani-Hamed15,Chen18} has emerged as a way to get special signatures left on the statistics by the end of inflation (including the power spectrum), and so to have a window into the Hubble-scale physics of the early universe.
With a field of a given spin and large-enough mass coupled to the inflaton, the curvature perturbation bispectrum could be imprinted with oscillations modulated by the properties of this field \cite{Arkani-Hamed15, Lee16}, and cannot be mimicked by any local single-field dynamics, making the collider signal a sharp probe of new degrees of freedom during inflation  \cite{Chen18, Pinol23}.
Collider signals have been extensively studied within the wavefunction 
formalism, where unitarity (cutting rules) and boundary differential 
equations determine massive-exchange coefficients for general boost-breaking couplings \cite{Arkani-Hamed18, Cespedes21, Pimentel22, Jazayeri22, Stefanyszyn24}.

Although not central here and beyond the implications a measurement would have, the bispectrum is of particular interest to this thesis as the leading sign of nonlinearity and is used as a proxy for classical dynamics in Chapter~\ref{sec:semiclQFT}.

\section*{Inflation as a seed of all observables}
\addcontentsline{toc}{section}{Inflation as a seed of all observables}

Whether or not inflation is the correct description of the early universe\footnote{See for instance, among many others, the alternative paradigms of \textit{warm inflation} or \textit{cyclic cosmology} \cite{Berera95, Steinhardt02}, all with seemingly various levels of challenges to explain the data as naturally as inflation.}, it is clear that it has been a major step in the understanding of the origin of structure in the universe.
The theoretical foundations of inflation laid in the 1980s have been used extensively for a better understanding of later times \cite{KolbTurner90,Dodelson03,Mukhanov05,kodama84,Mukhanov92}.

Let us briefly review the timeline of the universe flowing from the inflationary era\footnote{\ylg{Before inflation, it is speculated that gravitational, strong and electroweak interactions are unified at the Planck scale, gravity decoupling first and leaving a \textit{Grand Unification} of the remaining forces. The GUT symmetry should then break before or during inflation, so that the resulting monopoles are diluted, and reheating must not restore it, $T_{\rm rh}\lesssim M_{\rm GUT}\sim10^{16}\,$GeV, consistent with current bounds on the tensor-to-scalar ratio, $V^{1/4}\lesssim10^{16}\,$GeV.}}, with some of the key events that have been and are still studied in the literature.

\textbf{Reheating and Preheating}:
At reheating the inflaton reaches the bottom of its potential and converts its potential energy by decaying perturbatively into the particles of the Standard Model with high velocity and temperature. It is expected that the universe could reach a pre-inflation (higher) temperature, before cooling down again with modern expansion.
Additional non-perturbative dynamics could take place between the end of inflation and reheating: resonance phenomena which could lead to exponentially growing solutions for the inflaton before collapsing into e.g. \textit{oscillons}. Preheating is the name given to those non-equilibrium effects.
At this stage bosons should have no mass\ylg{.}\yld{, and it is believed that gravitational, strong and electroweak interactions are \textit{totally unified} at the Planck scale, until a phase transition occurs as the temperature decreases with expansion. Gravity is the first force to decouple, leaving only a \textit{Grand Unification}.}
See e.g. \cite{Kofman94,Allahverdi10,Lozanov19}.

\textbf{EW phase transition}:
When the temperature is small enough ($10^{15}K$), spontaneous symmetry breaking happens via the Higgs potential. Weak bosons are now massive and short-ranged, and become distinct from photons.
Strong interactions are rare and quarks and gluons are massless, almost free.
See e.g. \cite{Kirzhnits72,Morrissey12, Athron23}.

\textbf{QCD phase transition}:
When the temperature is small enough ($10^{12}K$), the coupling becomes strong and short-ranged, confining the latter into hadrons: free quarks bind into protons and neutrons and the universe now contains hydrogen nuclei ${}^1H$.
See e.g. \cite{Witten84,Schwarz03,Guenther20}.

\textbf{Massive decays}: 
Multiple decays occur after particles of the Standard Model acquire mass. When the temperature is smaller than their mass, particles and anti-particles can annihilate.
Hence the decay of the heaviest particles: top quarks, W and Z bosons, the Higgs boson, and b, s quarks.
See e.g. \cite{Laine15,Husdal16,Saikawa18}.

\textbf{Cosmic neutrinos decoupling}: 
Interactions have certain effective ranges; when the expansion is diluting enough, interactions become improbable and the interacting particles start free-streaming through the universe. Neutrinos are among the first to decouple and form a cosmic background (C$\nu$B), when they represent about $40\%$ of the radiative energy.
See e.g. \cite{Alpher53,Dolgov02}.
Only 1 min of cosmological time has passed so far.

\textbf{Big Bang nucleosynthesis (BBN)}:
When the thermal motion of the primordial plasma is small enough (3 min after the end of inflation but 50 000 years before matter-radiation equality), bonds become possible: nontrivial nuclei appear: helium $^4He$ or $^3He$ and a bit of deuterium $^2H$ and lithium $^7Li$. Electrons are still free.
Note that dark matter is also expected to be present. Its production process remains unknown.
See e.g. \cite{Alpher48c,Cyburt16,Fields20}.

\textbf{Recombination}:
As the temperature decreases, electrons become bound to nuclei and photons have a longer mean free path with respect to Thomson scattering.
Recombination is thus both the decoupling time of photons for the Thomson scattering (electromagnetic interaction) and the emission of photons from the de-excitation of the fresh atoms.
The free photons that were emitted here (370 000 y after inflation) constitute the cosmic microwave background (CMB).
See e.g. \cite{Peebles68,Sunyaev09}.

\textbf{Structure formation}:
After radiation-matter equality (and so strictly speaking before recombination), sub-horizon dark matter perturbations start growing steadily and start forming the cosmic web, a network of filaments and voids,
caused by the competition between expansion and gravitational attraction. With time, the cosmic web becomes clumpier, and galaxy clusters and superclusters form. 
See e.g. \cite{Bernardeau02,Angulo22}.

\textbf{Dark ages}:
Between the decoupling and 1 billion years, nothing is known. By cosmological redshift, the CMB quickly becomes below the visible range and so a dark universe is expected.
See e.g. \cite{Madau97,Pritchard12,Liu20}.

\textbf{Reionisation and astrophysical objects}:
Many phenomena occur within these newly-formed dense regions of space; stars notably emerge from the collapse of interstellar media and gas clouds.
The first stars are about $100M_{\odot}$ (population III). Stellar medium can be compared to the pre-recombination plasma. Indeed, those are mainly made of hydrogen re-ionised because of the energy density within stars.
The first heavy atoms are synthesised in stars (\textit{stellar nucleosynthesis}) thanks to collapse-regulating nuclear fusion and its chain reactions.
See e.g. \cite{Gunn65,Barkana01}. Rich astrophysics come out of this, such as binary systems, mergers, gravitational collapse and so supernovae, black holes etc.
Galaxies form notably as gravitationally-bound regions, some of them with a quasar at the centre.

\begin{figure}[ht]
  \centering
  \includegraphics[width=0.95\textwidth]{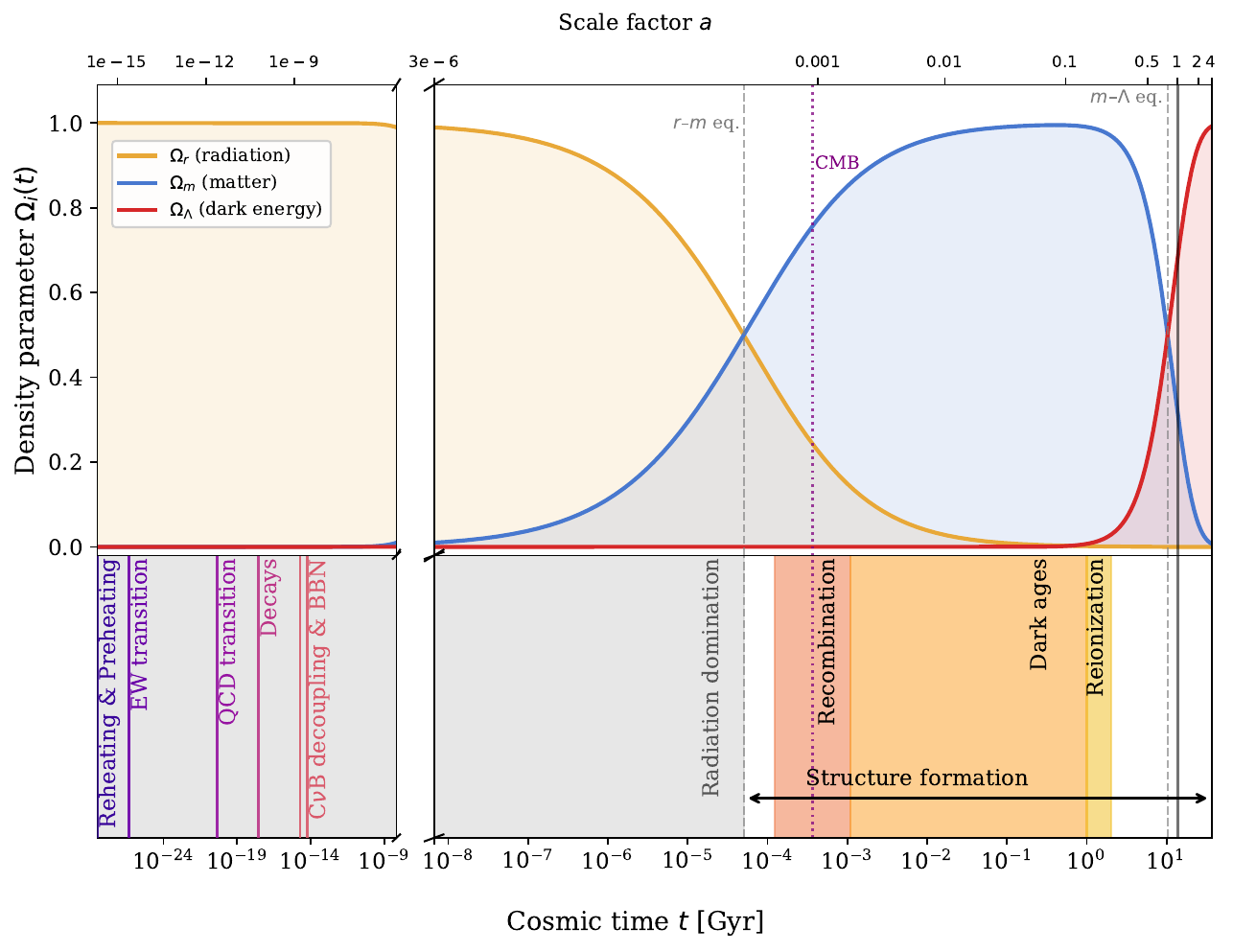}
  \caption{
    Top: evolution of the density parameters $\Omega_i(t)$ for radiation ($\Omega_r$),
    matter ($\Omega_m$), and dark energy ($\Omega_\Lambda$) across cosmic time,
    computed from the Planck 2018 best-fit $\Lambda$CDM parameters.
    Vertical dashed lines mark the radiation--matter and matter--$\Lambda$ equality
    epochs; the dotted line is the CMB emission and the solid line is today
    ($t_0\simeq 13.8\,\mathrm{Gyr}$). Bottom: the corresponding Hot Big Bang
    History eras described \yld{below}\ylg{in the main text}, from reheating and preheating through
    structure formation, on the same cosmic-time axis.
    }
  \label{fig:lcdm_densities}
\end{figure}

These various transitions and epochs are represented on Figure~\ref{fig:lcdm_densities}, together with the evolution of the density parameters $\Omega_i(t)$ for radiation, matter, and dark energy across cosmic time.
These are computed from the Planck CMB experiment's 2018 \cite{Planck2018omega} best-fit to the standard model of cosmology, the so-called $\Lambda$CDM model.

The construction of this model followed directly from breakthroughs on inflation and cosmological perturbation theory described earlier, notably the CDM part from the end of inflation to the CMB emission.
As we mentioned, the starting point is to set adiabatic initial conditions on superhorizon scales for the energy density perturbations of all species in the primordial plasma \cite{PeeblesYu1970,Bardeen83},
meaning perturbed in the same way with respect to their number density. This leads to the superhorizon contrast conditions
\begin{equation}
  \delta_r=\frac{4}{3}\delta_m=-\frac{4}{3}{\cal R}_0.
  \label{eq:adIC}
\end{equation}
These, together with nearly Gaussian and scale-invariant initial perturbations are still standard assumptions.
The second ingredient is the necessary presence, by the time of recombination (or CMB emission), of a cold dark matter component (CDM) to help the clustering of baryons and therefore the formation of large-scale structure \cite{Peebles82,Blumenthal84}, starting from the very small CMB anisotropies.\footnote{The first evidence for a dark matter component came from Zwicky's 1933 study of the Coma cluster \cite{Zwicky33}, who found the dynamical (virial) mass of the cluster far exceeded its luminous mass, later corroborated by Rubin and Ford's galaxy rotation curves in the 1970s \cite{RubinFord70}.}
Between 1984 and 1997, various authors contributed to the modelling of the re-entry of the baryon-photon-CDM perturbations into the horizon, their evolution, and the resulting CMB anisotropies.
In particular in 1984 \cite{Bond84}, supplemented by \yld{the}\ylg{a} statistical analysis in 1987 \cite{Bond87}, Bond and Efstathiou used a relativistic Boltzmann treatment fed by CPT to compute the CMB anisotropies on all angular scales\footnote{As noted by Ellis \cite{ellis_editorial_2017}, Bond and Efstathiou did not refer to inflation back then, only to scale-invariant and adiabatic ICs, but of course made use of established CPT.}, given that the full phase-space distribution of photons is needed to account for directional dynamics in a collisional plasma.
This work combined several effects; the CDM transfer function (CDM particles do not interact with photons and their perturbations grow smoothly and monotonically), the photon-baryon acoustic oscillations \cite{PeeblesYu1970,Sunyaev70}, Silk damping \cite{Silk68} of coupled photons progressively spreading beyond overdense hot regions in the primordial plasma and thus spreading heat or smoothing inhomogeneity on small scales, and massless neutrino free-streaming in a truncated multipole treatment \cite{Stewart72,Bond83}, and the Doppler and Sachs-Wolf effects \cite{SachsWolfe1967}, all together in agreement with the Uson-Wilkinson upper bound on CMB anisotropies at the time \cite{Uson84,Uson84b,Boughn81}.
This was extended later by Ma and Bertschinger in 1995 \cite{ChungPei95} to include or refine the account of massive neutrinos, photon polarisation, and helium recombination, and in 1996 by Seljak's and Zaldarriaga's \textit{line-of-sight integration} or \textsc{CMBFAST} method \cite{Seljak96} to rapidly compute the anisotropies for various configurations of the parameter space.
An increased understanding of the acoustic physics was also brought by Hu et al. \cite{Hu95,Hu95b}, as well as a better understanding of the CMB polarisation by Kamionkowski et al. \cite{Kamionkowski97} and Zaldarriaga and Seljak \cite{Zaldarriaga97}, using the E/B decomposition of the polarisation field to separate the scalar and tensor contributions.
In fact, scalar (density) perturbations generate only E modes at linear order, while tensor perturbations from primordial gravitational waves would generate both E and B.

Without data suggesting it, the $\Lambda$ part of the model came later. Nowadays, it is rather common to see separate explanations for the CMB and for nonlinear structure formation,
however, history taught us better.
Building on their developments of cosmological \textit{N-body simulations} to look at the transition from linear to nonlinear evolution of CDM and baryons \cite{Davis85} to form structures such as galaxies, clusters, and filaments,
Efstathiou et al. \cite{Efstathiou90} were among the first to argue that, on top of CDM, a cosmological constant was needed to reconcile the CMB with large-scale structure.
This was further argued in \cite{Ostriker95}, as one of the ways to reach so-called \textit{cosmic concordance}.

On the experimental side, the supernovae constraints of 1998 \cite{Riess1998,Perlmutter1999} gave the first direct evidence for a cosmological constant late domination, and so for the accelerated expansion of the universe.
CMB together with galaxy surveys were next to push for an accelerating universe at late times.
Through 2dF and SDSS experiments \cite{Colless01,York00}, galaxy surveys reached a large enough volume to measure the matter power spectrum two years after, and so the $h$-scaled matter density parameter $\Omega_m h$ to $0.20$ ($h\equiv H_0/100\,\text{km}\,\text{s}^{-1}\,\text{Mpc}^{-1}$),
independently from the rest of the cosmological parameters \cite{Percival01}. This is possible because the matter power spectrum turnover scale (see green curve of Figure~\ref{fig:ms_spectrum_snapshots}, which is the matter power spectrum at low redshifts\footnote{Note that galaxy power spectra obtained through measurements are biased matter spectra \cite{Desjacques18}.}) is that of the matter-radiation equality comoving horizon and so is related to $\Omega_m h$ only, when no acceleration is present yet.
From the CMB side, and after COBE \cite{Smoot1992} had provided the amplitude of the anisotropy and the confirmed quasi scale-invariance on large scales,
BOOMERanG \cite{BOOMERANG00}, MAXIMA \cite{MAXIMA00}, pre-WMAP \cite{Bennett2003} helped estimate the oscillating peak positions (see Figure~\ref{fig:ms_spectrum_snapshots}) to constrain the curvature and, as flatness was found, thus strengthening the case for the need of a cosmological constant following the old arguments of \cite{Peebles84}.
The degeneracy between $\Omega_m h$ and the other parameters was broken in a 2002 paper \cite{Efstathiou02} using both CMB and galaxy surveys, to finally find $\Omega_\Lambda \approx 0.7$ and $\Omega_m \approx 0.3$ separately from supernovae.
In 2005, 2dFGRS and SDSS galaxy surveys reached the BAO scale (${\simeq 150\text{Mpc}}$),
the typical imprint of the tight photon-baryon coupling before recombination, and measured this most likely galaxy-galaxy separation \cite{Cole05,Eisenstein05}.
This can be seen on the  oscillations of the mid-to-higher $k$ part of the green curve in Figure~\ref{fig:ms_spectrum_snapshots}.
As opposed to CMB experiments, BAO measurements at variant redshifts are sensitive to the expansion history of the universe, and so to the Hubble parameter $H(z)$ and the angular diameter distance $(1+z) D_A(z)=\int dz c {H(z)}^{-1}$ (as opposed to supernovae only accessing the luminosity distance $(1+z)^2D_A(z)$), both functions of $\Omega_m$ and $\Omega_\Lambda$, which can then be probed in time.

Combining all these efforts, notably with the culmination of the latest \textit{Planck} CMB measurements \cite{Planck2015omega,Planck2018omega},
the most stringent constraints on $\Lambda$CDM up to this day are obtained from the CMB and BAO data and can be summarised in the velocity equation fit for the homogeneous part%
 \begin{subequations}
    \begin{align}
        H^2 & = H_0^2 \left(\Omega_{r0}a^{-4} + \Omega_{m0}a^{-3}+\Omega_{k0}a^{-2}+\Omega_{\Lambda 0}\right),\\
        H_0 & = 67.4\pm 0.5 \,\mathrm{km\,s^{-1}\,Mpc^{-1}}, \\
        \Omega_{\Lambda 0}&=0.6847\pm 0.0073,\\
        \Omega_{m0} & = 0.315\pm 0.007,\\
        \Omega_{k0} &= 0.001\pm 0.002,
    \end{align}
 \end{subequations}
with $\Omega_{r0}=1-\Omega_{m0}-\Omega_{k0}-\Omega_{\Lambda 0}\simeq 10^{-5}$,
and the primordial power spectrum of the curvature perturbation for the inhomogeneous part
\begin{equation}
    \mathcal P_{\cal R}(k) = A_s \left(\frac{k}{k_*}\right)^{n_s-1},\quad
    A_s = 2.1\times 10^{-9},\quad n_s = 0.965\pm 0.004,
\end{equation}
with $k_*=0.05\text{Mpc}^{-1}$ the pivot scale, as well as the tensor-to-scalar ratio upper bound from the non-detection of primordial gravitational waves
\begin{equation}
    r \equiv \frac{{\cal P}_{\rm t}}{{\cal P}_{\cal R}} < 0.036 \text{ at } 95\% \text{ C.L.}.
\end{equation}
When one looks at the Planck power spectrum fit, $\Lambda$CDM model shows a remarkable agreement, in line with the best-ever measured blackbody spectrum that the CMB constitutes.
Despite that success, the concordance model is not without observational tensions, in particular when comparing early and late times probes.

\setlength{\tabcolsep}{6pt}
\renewcommand{\arraystretch}{1.15}
\begin{table}[p]
\centering
\resizebox{0.9\textwidth}{!}{%
\begin{tabular}{|Sc|Sc|Sc|Sc|Sc|}
\hline
\multicolumn{2}{|c|}{Signal} & Sources & Experiments (past, current, \textit{future}) & Primary observables \\
\hline
\multirow{16}{*}{Electromagnetic}
& \multirow{8}{*}{Spectrum}
    & CMB-T
    & \shortstack[l]{COBE, WMAP, Planck, ACT, SPT,\\Simons Obs., \textit{CMB-S4}}
    & \shortstack[l]{$A_s,\, n_s,\, \alpha_s \equiv \mathrm{d}n_s/\mathrm{d}\ln k$\\$\Omega_K$, $f_{\rm NL}$ (T bispectrum)} \\
\cline{3-5}
& & \shortstack[l]{CMB spectral\\distortions}
    & \shortstack[l]{COBE-FIRAS, \textit{PIXIE},\\\textit{Voyage-2050 concepts}}
    & \shortstack[l]{$\mu,\, y$ distortions\\$\mathcal{P}_\zeta(k)$ at $k\sim 1$--$10^4~\mathrm{Mpc}^{-1}$} \\
\cline{3-5}
& & PTA
    & \shortstack[l]{NANOGrav, EPTA+InPTA, PPTA,\\CPTA, MPTA, \textit{SKA}}
    & \shortstack[l]{searches for primordial GW backgrounds\\at $f\sim\mathrm{nHz}$: blue-tilted $n_T$, SIGW\\$\to \mathcal{P}_\zeta$ at small scales, strings $G\mu$} \\
\cline{3-5}
& & Lyman-$\alpha$
    & \shortstack[l]{SDSS/BOSS/eBOSS,\\HIRES/UVES, XQ-100\\DESI$^*$, WEAVE-QSO$^*$}
    & \shortstack[l]{$\mathcal{P}_\zeta(k)$ at $k\sim0.1$--$10\,h\,\mathrm{Mpc}^{-1}$:\\extends the CMB lever arm on $n_s,\alpha_s$} \\
\cline{3-5}
& & LSS (spectroscopy)
    & \shortstack[l]{2dF, SDSS/BOSS/eBOSS, DESI$^*$\\($\theta_{1,2}, z$), Euclid$^*$, SPHEREx$^*$\\\textit{Roman}$^*$, \textit{4MOST}$^*$}
    & \shortstack[l]{$H(z), D_A(z)$ (BAO); $f\sigma_8$ (RSD)\\$f_{\rm NL}^{\rm loc}$ (scale-dep.\ bias, $\sigma\!\sim\!1$\\w/ SPHEREx); broadband $P(k)$ shape\\(improves $n_s$ w/ CMB)} \\
\cline{3-5}
& & \shortstack[l]{Line-intensity\\mapping}
    & \shortstack[l]{COMAP, CONCERTO, TIME, CHIME (HI)\\SPHEREx, \textit{CHORD} (HI)\\\textit{CCAT/FYST}, \textit{SKA} (HI IM)}
    & \shortstack[l]{$P(k)$ at high $z$ over large volumes:\\complementary route to $n_s$, $f_{\rm NL}$} \\
\cline{3-5}
& & 21\,cm
    & \shortstack[l]{LOFAR, MWA, HERA,\\EDGES (global)\\\textit{SKA-Low}, \textit{lunar far-side (LuSEE-Night)}}
    & \shortstack[l]{$\mathcal{P}_\zeta(k)$ at $z\sim6$--$30$ (cosmic dawn/\\reionisation; dark ages via LuSEE-Night:\\longest possible $n_s,\alpha_s$ lever arm), PBH accretion} \\
\cline{3-5}
& & Gamma rays
    & \shortstack[l]{EGRET, Fermi-LAT,\\H.E.S.S./MAGIC/VERITAS, \textit{CTAO}}
    & \shortstack[l]{Hawking evaporation: $f_{\rm PBH}(M\lesssim10^{17}\,\mathrm{g})$\\$\to \mathcal{P}_\zeta$ at very small scales\\(model dependent; assumes Gaussian $\zeta$)} \\
\cline{2-5}
& \multirow{5}{*}{Image}
    & LSS (imaging)
    & \shortstack[l]{SDSS, CFHTLenS, DES, KiDS,\\HSC ($e_{1,2},\theta_{1,2},z^*$), Euclid,\\\textit{Rubin-LSST}, \textit{Roman}}
    & \shortstack[l]{$\Omega_m,\ \sigma_8$ / $S_8$ and growth history\\(weak lensing), $3\times2$pt, cluster counts} \\
\cline{3-5}
& & CMB lensing
    & \shortstack[l]{Planck, ACT DR6, SPT-3G,\\Simons Obs., \textit{CMB-S4}}
    & \shortstack[l]{growth of structure at $z\sim2$ ($\sigma_8$),\\$\sum m_\nu$ (w/ BAO), $A_L$ consistency} \\
\cline{3-5}
& & \shortstack[l]{CMB secondaries\\(kSZ/tSZ)}
    & \shortstack[l]{ACT/SPT $\times$ BOSS\\Simons Obs.\ $\times$ DESI\\\textit{CMB-S4} $\times$ \textit{Rubin}/DESI}
    & \shortstack[l]{kSZ tomography: large-scale velocity fields\\\& growth; patchy reionisation; cluster physics (tSZ)} \\
\cline{3-5}
& & \shortstack[l]{Quasars\\(strong lensing)}
    & \shortstack[l]{H0LiCOW/TDCOSMO, STRIDES\\\textit{Rubin-LSST} ($\mathcal{O}(10^3)$ lensed QSOs)}
    & \shortstack[l]{$H_0$ (time-delay distances $D_{\Delta t}$)} \\
\cline{3-5}
& & \shortstack[l]{Microlensing\\(PBH)}
    & \shortstack[l]{EROS/MACHO, Kepler\\OGLE, Subaru-HSC\\\textit{Rubin-LSST}, \textit{Roman}}
    & \shortstack[l]{$f_{\rm PBH}$ for $10^{-10}$--$10^3\,M_\odot$\\$\to \mathcal{P}_\zeta$ at small scales\\(model dependent; assumes Gaussian $\zeta$)} \\
\cline{2-5}
& \multirow{3}{*}{Polarisation}
    & CMB-E
    & \shortstack[l]{WMAP, Planck, ACTPol, SPTpol/3G,\\Simons Obs., \textit{LiteBIRD}}
    & \shortstack[l]{$\tau$ (reionisation), sharpens $n_s$,\\consistency tests} \\
\cline{3-5}
& & CMB-B
    & \shortstack[l]{BICEP2/\textit{Keck}, POLARBEAR, SPIDER\\BICEP Array, SPT-3G, CLASS, Simons Obs.\\\textit{LiteBIRD}, \textit{CMB-S4}}
    & \shortstack[l]{$r<0.036$ (95\%, BK18) $\to \sigma(r)\!\sim\!10^{-3}$\\$n_T$, energy scale of inflation} \\
\cline{3-5}
& & CMB-EB/TB
    & \shortstack[l]{\textit{Planck} reanalyses\\\textit{LiteBIRD}, \textit{CMB-S4}}
    & \shortstack[l]{hints of cosmic birefringence ($\beta\sim0.3^\circ$):\\parity violation (axion--inflaton\\couplings, chiral PGW)} \\
\hline
\multirow{3}{*}{Astroparticles}
& Cosmic rays & EGMF
    & \shortstack[l]{Fermi-LAT + IACTs (blazar cascades), \textit{CTAO}}
    & \shortstack[l]{lower limits on $B_{\rm void}$\\$\to$ inflationary magnetogenesis} \\
\cline{2-5}
& UHECR & \shortstack[l]{Top-down scenarios\\(SHDM decay,\\topological defects)}
    & \shortstack[l]{AGASA, HiRes, Auger(Prime),\\Telescope Array, \textit{GCOS}}
    & \shortstack[l]{superheavy DM decay, topological\\defects (post-inflationary relics)} \\
\cline{2-5}
& Neutrinos & \shortstack[l]{DSNB\\C$\nu$B}
    & \shortstack[l]{Super-K(-Gd), JUNO\\\textit{Hyper-K}, \textit{PTOLEMY (concept)}}
    & \shortstack[l]{DSNB flux; direct detection of the\\relic $\nu$ background ($T_\nu$, thermal history,\\low-$T_{\rm reh}$); complements CMB $N_{\rm eff}$} \\
\hline
\multicolumn{2}{|c|}{\multirow{4}{*}{Gravitational Waves}}
                       & Binaries
    & \shortstack[l]{LVK (O1--O4), \textit{LISA},\\\textit{ET}, \textit{Cosmic Explorer}}
    & \shortstack[l]{$H_0$ (standard sirens), $c_T$ (GW170817),\\sub-solar-mass PBH searches} \\
\cline{3-5}
\multicolumn{2}{|c|}{} & \shortstack[l]{SGWB\\(phase transitions,\\cosmic strings,\\preheating, SIGW)}
    & \shortstack[l]{LVK ($\Omega_{\rm GW}\lesssim3\times10^{-9}$ @ 25\,Hz)\\\textit{LISA}, \textit{Taiji}/\textit{TianQin}\\\textit{DECIGO}/\textit{BBO}, \textit{AION}/\textit{MAGIS}}
    & \shortstack[l]{searches for cosmological backgrounds: FOPT\\parameters ($\alpha,\beta/H$), $G\mu$ (model-dep.,\\$\sim10^{-15}$--$10^{-10}$), $\mathcal{P}_\zeta$ via SIGW} \\
\cline{3-5}
\multicolumn{2}{|c|}{} & PGWB
    & \shortstack[l]{indirect: B-modes (BICEP $\to$ \textit{LiteBIRD}/\textit{S4})\\+ PTA + LVK (spectral shape)\\proposed direct-detection concepts:\\\textit{Ultimate-DECIGO}/\textit{BBO}}
    & \shortstack[l]{$r,\ n_T$ over $\sim21$ decades in $f$,\\reheating equation of state $w_{\rm reh}$} \\
\cline{3-5}
\multicolumn{2}{|c|}{} & Modified GR
    & \shortstack[l]{LVK (polarisation tests), \textit{ET}\\(null stream, extra polarisations),\\\textit{Cosmic Explorer}, \textit{LISA}}
    & \shortstack[l]{extra polarisations, $m_g$, GW dispersion,\\$c_T=c$ $\to$ constraints on\\G-inflation / Horndeski inflaton} \\
\hline
\end{tabular}
}
\caption{Overview of past, present and future observational probes with direct or indirect sensitivity to primordial observables. Rows within each ``Experiments'' cell run past $\to$ current $\to$ future; \textit{italic} marks experiments whose main data are still to come, and $^*$ marks an additional spectroscopic channel providing calibrated redshifts. Additional notations: $e_{1,2}$, $\theta_{1,2}$: galaxy ellipticities and sky coordinates; $D_{\Delta t}$: time-delay distance; $A_L$: lensing-amplitude consistency parameter; $c_T$, $m_g$: GW propagation speed and graviton mass; $T_{\rm reh}$, $w_{\rm reh}$: reheating temperature and equation of state; $\alpha$, $\beta/H$: strength and inverse duration of a first-order phase transition (FOPT); $B_{\rm void}$: void magnetic field; SIGW: scalar-induced GWs; SHDM: superheavy dark matter; BK18: BICEP/\textit{Keck} 2018 analysis. Purely astrophysical or particle-physics constraints are omitted.
See collaboration papers \cite{Smoot1992,Bennett2013,Planck2018omega,ACTDR6spectra,SPT3GEETE,SimonsObservatory,CMBS4ScienceBook,COBEFIRAS,PIXIE,Voyage2050SpectralDistortions,NANOGrav15yrGWB,EPTADR2GWB,PPTADR3GWB,CPTADR1GWB,MPTAGWB,SKAFundamentalPhysics,eBOSSLyaPk,IrsicWDM2017,VielWDM2013,WEAVEQSO,Colless01,eBOSScosmology,DESIscience,DESIDR2,EuclidOverview,SPHEREx,RomanWFIRST,4MOST,COMAPEarlyScience,CONCERTO,TIMEPilot,CHIME,CCATprime,SKARedBook,LOFAREoRlimits,MWAEoRlimits,HERAPhaseIlimits,EDGES,LuSEENight,EGRET3rdCatalog,FermiLAT,CTAScienceBook,CarrPBHconstraints,York00,CFHTLenS,DESY33x2pt,KiDS1000,HSCY3cosmicshear,RubinLSST,Planck2018lensing,ACTDR6lensing,SPT3Glensing,kSZfirstDetection,ACTkSZtSZ,H0LiCOW,TDCOSMOIV,STRIDES,EROS2,MACHO,SubaruHSCPBH,OGLEPBH,BK18,POLARBEARBmodes,SPIDER,LiteBIRD,PlanckBirefringence,NeronovVovkEGMF,FermiLATEGMF,PierreAuger,TelescopeArray,AugerSHDM,GCOS,SuperKDSNB,JUNO,HyperKDesignReport,PTOLEMY,LVKGWTC-3,GW170817,GW170817standardsirenH0,LVKO3isotropicSGWB,LISAdefinitionStudyReport,EinsteinTelescope,CosmicExplorer,Taiji,TianQin,DECIGO,BBO,AION,MAGIS-100}.
Columns $4$ and $5$ were filled with the help of Artificial Intelligence.
}
\label{tab:megaExperiments}
\end{table}

To this day, the most persistent one is the so-called \textit{Hubble tension}: the local distance-ladder measurement of $H_0 \simeq 73\,\mathrm{km\,s^{-1}\,Mpc^{-1}}$ from supernovae \cite{Riess22} has a ${\sim}5\sigma$ gap with the CMB value $H_0 \simeq 67.4\,\mathrm{km\,s^{-1}\,Mpc^{-1}}$ under flat $\Lambda$CDM \cite{Planck2018omega}.
BAO measurements seem to side with CMB ones, and constraints seem to indicate that, if new physics is needed, it would need to be introduced before the modern dark energy domination era.
In contrast, the so-called $S_8$ tension, the ${\sim}2$--$3\sigma$ difference between weak-lensing surveys and the CMB prediction for the clustering amplitude, seems to have dissipated thanks to improved calibrations of the KiDS-Legacy survey \cite{KiDS-Legacy}.
Whether this is settled or not, $\Lambda$CDM has been given a bigger challenge to solve, as DESI DR2 BAO measurements combined with the CMB prefer an evolving dark-energy equation of state, deviating from $\Lambda$CDM by more than $3\sigma$ even without supernovae \cite{DESIDR2}.
The robustness of this result is still highly debated and might reflect an inconsistency between methods rather than genuine dark-energy dynamics \cite{Colgain25}.
In that regard and to mitigate the tensions, it is highly convenient for our community to be reaching a new experimental era, with a new generation of experiments that will provide more precise measurements of the CMB, large-scale structure, as well as new probes such gravitational waves or 21cm cosmology, and more generally cross-correlations in the single experiment that our universe represents.
We report a brief list of past, current, and future experiments in Table~\ref{tab:megaExperiments}, with their main observables, \yld{with }a particular focus on those that are sensitive to primordial physics and inflationary model selection.

The mentioned anomalies might still point to new physics but it is still too early to say. For instance, we know of more pressing physics issues, such as the origin of matter-antimatter asymmetry, the origin's singularity problem that inflation has pushed away, as well as the nature of the inflaton, dark matter, and dark energy.
In raising those concerns and for the sake of this work, it is worth noting that there are many open questions closer in space and time to us than inflation is.
The question is whether they are related to the modelling of the initial conditions (and so to inflation), which seems unlikely given the late-time nature of the tensions.
However, this could be an issue in getting accurate primordial constraints because primordial and late-time parameters are inferred jointly.
For instance, proposed resolutions might introduce pre-recombination physics, notably early dark energy, and might shift the spectral index, giving the tensions indirect but real changes for inflationary model selection \cite{Poulin23}.

Luckily, this thesis focuses on the modelling on the inflationary era, with minimal assumptions on its definition and its constraints.
Its goal is precisely the construction of a generic prediction framework enabled for arbitrary dynamics, and so for a wide range of phenomenology during and after the inflationary paradigm.

\section*{Statement of needs and outline of the thesis}
\addcontentsline{toc}{section}{Statement of needs and outline of the thesis}

We have just highlighted the new observational era that cosmology is entering, together with its challenges.
As a major example, we have mentioned the current inability to measure a nonzero CMB non-Gaussianity within experimental resolution \cite{PlanckNG}.
If one believes that inflation is well described by GR and a slowly-rolling single field, and knows the current constraints on $n_s$, then a detection will take more than the announced and upcoming precisions.
Conversely, if future CMB or LSS experiments such as SO \cite{SimonsObservatory} or CMB-S4\footnote{CMB-S4 is currently cancelled.} \cite{CMBS4ScienceBook} and SPHEREx \cite{SPHEREx} detect non-Gaussianity, then one of these assumptions must be broken.
At the most basic level of inquiry, this could suggest a modified phenomenology (adding other fields), modified dynamics (breaking slow-roll), or going beyond GR.
Making sure that all such theories can be predictive and so testable is the exact gap that this thesis exploits.

More concretely and as we will see in the following chapters, many rather simple scenarios of inflation or possible dynamical outcomes remain untestable because current computational methods, whether theoretical or numerical, rely on numerous assumptions, which were shaped by the historical experiments and early theoretical investigations.
For instance, systematic QFTCS tools \cite{WaldQFTCS, Maldacena_2003, mulryne_pytransport_2017,werth23, Costantini25} are usually limited to tree-level computations and only probe leading-order, and perturbative non-linearity.
On the other hand, commonly used approaches enabling non-perturbative and non-linear studies --- which include stochastic inflation \cite{Starobinsky_stochastic_1988, morikawa_dissipation_1990}, stochastic and standard $\Delta {\cal N}$ formalisms \cite{wands_new_2000, vennin_correlation_2015},  Hamilton-Jacobi approaches \cite{salopek_stochastic_1991, prokopec_ensuremathdeltan_2021}, lattice cosmology \cite{latticeeasy, Cosmolattice, Caravano22thesis}, and numerical relativity for inhomogeneous inflation \cite{East16,clough_robustness_2017,Bloomfield19,joana_inhomogeneous_2021, aurrekoetxea_effects_2020, Corman23,Elley2024, Ijjas22} ---do not fully complement QFTCS as they rely on the long wavelength approximation, or approximate relativistic and quantum-to-classical transitions. 
Beyond standard types of inflation, the landscape of testable scenarios is thus limited to those where strong leading-order approximations are valid.
More than scenarios, some predicted phenomena meant to be originating from inflation, are still a long way from a quantitative and reliable prediction pipeline.
One could take the example of primordial scalar-induced gravitational waves or, in recent years, the example of Primordial Black Holes \cite{Carr74}, which has become the symbol of this need for new inflationary frameworks, with, depending on who is asked, an extreme variety in predictions on their abundance, once formed after inflation from big overdensities produced then.

What this work sees in this context is a lack of a fully nonlinear and nonperturbative treatment of inflationary physics, even in the regime where quantum gravity does not operate yet.
Among complementary methods to the standard quantum field theoretic approaches for the prediction of observables, the \textit{stochastic formalism for inflation}, or \textit{stochastic inflation} \cite{Starobinsky_stochastic_1988}, born not so long after inflation, is the one prioritised in this work.
This formalism makes the most of the semiclassical regime to provide a nonperturbative model of the dynamics, sourced by the quantum vacuum fluctuations' noise. 
Since its early days \cite{Starobinsky_stochastic_1988,salopek_nonlinear_1990,salopek_stochastic_1991,rigopoulos_non-linear_2005,vennin_correlation_2015}, it has been used in the long wavelength limit of General Relativity (GR), yielding numerous developments and predictions on classicalised primordial perturbations
\cite{nakao_stochastic_1988,
nambu_stochastic_1988,
sasaki_classical_1988,
hosoya_stochastic_1989,
habib_stochastic_1992,
bellini_stochastic_1996,
Winitzki00,
rigopoulos_separate_2003,
Liguori04,
tsamis_stochastic_2005-2,
Rigopoulos_multi06,
BattefeldVanTent06,
tolley_stochastic_2008,
gratton_path_2011,
Fujita2013,
levasseur_lagrangian_2013,
levasseur_recursive_2013,
rigopoulos_fluctuation-dissipation_2013,
Garbrecht:2013coa,
burgess_eft_2015,
levasseur_backreaction_2015,
assadullahi_multiple_2016,
burgess_open_2016,
grain_stochastic_2017,
pattison_quantum_2017,
vennin_critical_2017,
prokopec_functional_2018,
pattison_attractive_2018,
noorbala_tunneling_2018,
markkanen_scalar_2019,
pinol_inflationary_2019,
firouzjahi_stochastic_2019,
pattison_stochastic_2019,
ezquiaga_exponential_2020,
Bounakis20,
de_numerically_2020,
pattison_ultra-slow-roll_2021,
ando_power_2021,
cohen_stochastic_2021,
cable_free_2021,
prokopec_ensuremathdeltan_2021,
Jackson22,
Tasinato22,
Figueroa22,
tada_statistics_2022,
cable_second-order_2022,
cruces_stochastic_2022,
mishra_primordial_2023,
wilkins_stochastic_2023,
tomberg_numerical_2023,
Tomberg:2024evi,
tajima_stochastic_2024,
cruces_small_2024,
jackson2024stochasticinflationslowroll,
ballesteros_non-gaussian_2024,
noorbala_classicality_2024,
sharma_stochastic_2024,
Animali25,
tasinato_stochastic_2025,
nassiri-rad_stochastic_2025,
miyamoto_calculating_2025,
Briaud25,
blachier_friction_2025,
barenboim_large_2025,
murata_stochastic_2026,
saha_nonlinear_2026,
beneke_quantum_2026,
ye_nonperturbative_2026,
cruces_consistent_2026,
mishra_eigenvalue_2026,
animali_time-reversed_2026,
Li26b}.
Some of these refinements of Starobinsky's original idea over the years have arisen to describe more and more regimes.
In particular, many works explored the breaking of perturbativity beyond secular growth arising on fixed dS backgrounds.
In general with a full spacetime, whether a theory can produce such secular effects or not \cite{Assassi12} does not protect from nonperturbative dynamics.
For instance, inflation models with slow-roll breaking phases have attracted significant interest in their promise to produce strong overdensities and subsequent primordial black holes, as already mentioned.

The stochastic formalism is not the only candidate for a better description and, within its standard formulation and assumptions, rather limited in its applications and precision.
This is why this work will make the most of its generalisation and, in losing tractable analyticity, will turn to nonlinear simulations inspired by the aforementioned lattice codes, notably Numerical Relativity.
This thesis is organised as follows:
\begin{itemize}
    \item[I)] Chapter~\ref{sec:semiclQFT} builds a semiclassicality criterion from interactive QFT computations.
    \item[II)] Chapter~\ref{sec:ICNR} shows how to link standard cosmological perturbation theory to Numerical Relativity.
    \item[III)] Chapter~\ref{sec:ICsim} follows up by showcasing the generalisation of lattice simulations of inflation, accounting for the geometric dynamics of spacetime, using Numerical Relativity.
    \item[IV)] Chapter~\ref{sec:SIGR} reviews the stochastic approach and demonstrates how to formulate it in full General Relativity\ylc{, with noise computed from linear perturbation theory around a homogeneous background}.
    \item[V)] Chapter~\ref{sec:SINR} reports on our implementation of stochastic Numerical Relativity\ylc{, its validation against linear theory,} and its new conclusions\ylc{, some of which are first looks}.
    \item[VI)] Chapter~\ref{sec:SIbeyondGR} proposes an extension of the formalism beyond single-field General Relativity. 
\end{itemize}

\section*{Notations, Conventions, Units}
\addcontentsline{toc}{section}{Notations, Conventions, Units}

\begin{description}
    \item[Units.] Natural units $\hbar=c=1$, with a mostly-plus metric signature $(-,+,+,+)$. Greek indices $\mu,\nu,\dots$ run over spacetime coordinates, Latin indices $i,j,\dots$ over spatial ones only.

    \item[Planck mass.] The reduced Planck mass $M_{Pl}=(8\pi G)^{-1/2}$ is written $\Mpl$ throughout; it should not be confused with the \textit{running} Planck mass $M(t)$ of Chapter~\ref{sec:SIbeyondGR}, which reduces to $\Mpl$ only in GR. The Numerical Relativity chapters (\ref{sec:ICsim}, \ref{sec:SINR}) instead simulate in dimensionless code units, rescaling $\Mpl$ to a large number (e.g.\ $\Mpl=100$) rather than using physical (GeV) units.

    \item[Time.] $t$ is cosmic time, $\eta$ conformal time, and $\dot{(\,)}$ a derivative with respect to whichever time coordinate is in use. A subscript $b$ marks a background quantity of cosmological perturbation theory\ylg{, kept implicit for the scale factor $a(t)$ and the Hubble parameter $H(t)\equiv \dot a/a$}.

    \item[Fourier convention.] ${\cal F}^{-1}\{\cdot\} = \int \frac{d^3\mathbf{k}}{\sqrt{2\pi}^3} (\cdot)\,e^{-i\mathbf{k}\cdot\mathbf{x}}$. An $n$-point function in Fourier space $\langle\cdot\rangle'$ carries an implicit $\delta^{(3)}(\mathbf{k}_1+\dots+\mathbf{k}_n)$; the 3D space of momenta closing a triangle, $\mathbf{k}_1+\mathbf{k}_2+\mathbf{k}_3=0$, is referred to as the \textit{tetrapyd}.

    \item[Operators.] $\hat{\cdot}$ designates an operator in the QFT sense, its dependence (Heisenberg, interaction picture, \dots) determined by context. A bold operator will usually be a stochastic one.
\end{description}

%% file: acronyms.tex
\vspace{-1.5\baselineskip}
{\Large\textbf{Vocabulary}}
\begin{tabular}{@{\hspace{2.5em}}l@{\hspace{1.5em}}l@{}}
\textbf{AMR}   & Adaptive Mesh Refinement \\[2pt]
\textbf{CCZ4}  & Conformal and Covariant Z4 \\[2pt]
\textbf{CMB}   & Cosmic Microwave Background \\[2pt]
\textbf{CPT}   & Cosmological Perturbation Theory \\[2pt]
\textbf{EFT}   & Effective Field Theory \\[2pt]
\textbf{EOM}   & Equation(s) of Motion \\[2pt]
\textbf{FFT}   & Fast Fourier Transform \\[2pt]
\textbf{FPTA}  & First-Passage-Time Analysis \\[2pt]
\textbf{GR}    & General Relativity \\[2pt]
\textbf{GW}    & Gravitational Wave \\[2pt]
\textbf{IR}    & Infrared \\[2pt]
\textbf{NMC}   & Non-Minimally Coupled \\[2pt]
\textbf{NR}    & Numerical Relativity \\[2pt]
\textbf{PBH}   & Primordial Black Hole \\[2pt]
\textbf{PDF}   & Probability Distribution Function \\[2pt]
\textbf{QFT}   & Quantum Field Theory \\[2pt]
\textbf{QFTCS} & Quantum Field Theory in Curved Spacetime \\[2pt]
\textbf{SGWB}  & Stochastic Gravitational-Wave Background \\[2pt]
\textbf{SI}    & Stochastic Inflation \\[2pt]
\textbf{SIGW}  & Scalar-Induced Gravitational Wave \\[2pt]
\textbf{SR}    & Slow-Roll \\[2pt]
\textbf{SUA}   & Separate Universe Approximation \\[2pt]
\textbf{SVT}   & Scalar--Vector--Tensor \\[2pt]
\textbf{TF}    & Trace-Free \\[2pt]
\textbf{USR}   & Ultra Slow-Roll \\[2pt]
\textbf{UV}    & Ultraviolet \\
\end{tabular}

\clearpage

{\Large\textbf{Names}}
\begin{tabular}{@{\hspace{2.5em}}l@{\hspace{1.5em}}l@{}}
\textbf{ADM}   & Arnowitt--Deser--Misner \\[2pt]
\textbf{BD}    & Bunch--Davies \\[2pt]
\textbf{BSSN}  & Baumgarte--Shapiro--Shibata--Nakamura \\[2pt]
\textbf{DEF}   & Damour--Esposito-Far\`ese \\[2pt]
\textbf{dS}    & De Sitter \\[2pt]
\textbf{EGB}   & Einstein--Gauss--Bonnet \\[2pt]
\textbf{FLRW}  & Friedmann--Lema\^itre--Robertson--Walker \\[2pt]
\textbf{GHY}   & Gibbons--Hawking--York \\[2pt]
\textbf{MS}    & Mukhanov--Sasaki \\[2pt]
\end{tabular}

{\Large\textbf{Codes}}
\begin{tabular}{@{\hspace{2.5em}}l@{\hspace{1.5em}}l@{}}
\textbf{\stoiic} &  Stochastic Initial Conditions for Inflationary Cosmology in GR \\[2pt]
\textbf{\istoriz} & {I}nflationary {Sto}chastic {R}elativity {I}ntegrator for zeta. \\[2pt]
\textbf{\vizir} & {V}isualizer for {I}nflationary {R}elativity
\end{tabular}

\clearpage

%% file: chap_QCl.tex
\chapter{Semiclassical limit of cosmological correlators} \label{sec:semiclQFT}

As highlighted in the introduction, the core assumption of inflation is that the primordial perturbations originate from quantum vacuum fluctuations on small scales, resulting in propagating inhomogeneities in the field and metric. 
However, the focus on the tools of QFTCS may have left a simple question unanswered: how necessary is the quantum treatment for describing cosmological observables, data, and simulations?
This chapter focuses on extracting the semiclassical limit to answer these questions, showcasing the results found in \cite{Launay2024bis}.

As established in previous work \cite{Starobinsky_stochastic_1988, Weinberg05,Weinberg06}, quantum contributions are limited in growth during inflation and leave a universe behaving statistically classically on scales roughly bigger than the Hubble radius.
The vocabulary can be important in such a statement, as (semi-)\textit{classicality} can be ambiguous.
In fact, it is known that the inflationary quantum states do not become classical in the sense of being a coherent state, but rather a squeezed state \cite{Grishchuk90,Albrecht94,Polarski_1996}, the statistics of which behave like a stochastic field, while still showing a high entanglement (notably between modes of opposite momenta) \cite{Martin16}.
More generally, this contributes to explaining why the concepts of classicality and quantumness can have different measures depending on the context, many of which have been applied to cosmology, from Weinberg's criterion to decoherence \cite{burgess_minimal_2022},
quantum purity \cite{Colas24purity} or quantum discord \cite{Ollivier01}, and many other criteria. However, classical behaviour specified according to one approach may or may not be classical according to another.
Most of these studies have focused on scalar fields in a fixed (quasi-)de Sitter background and concentrated on the late-time limit, but it is desirable to explore this behaviour more quantitatively and to obtain precise times for the onset of classicality for a given model.

To understand our choice of what classicality means in this study, it is important to note a growing number of ever more ambitious numerical studies of inhomogeneous inflation such as lattice simulations \cite{latticeeasy,Easther10,Cosmolattice,Caravano22lattice,Caravano2024,caravano2024axionbackreaction, caravano2024usr}, numerical relativity \cite{East16,clough_robustness_2017,Bloomfield19,giblin_jr_preheating_2019,aurrekoetxea_effects_2020, Ijjas2023, Corman23, Elley2024}, and stochastic inflation \cite{Jackson22,cruces_stochastic_2022,tomberg_numerical_2023,Launay24,stolas,Animali25}.
These studies constitute a great leap forward in theoretical predictions because of the possibility to compute correlation functions for any models of inflation and at all orders in perturbation theory. These numerical systems all evolve according to the classical equations of motion (EOM), sourced by initial or stochastic noise, the spectra of which are motivated by the aforementioned quantum fluctuations.
Unfortunately, the precise time from which the non-linear dynamics can be described by the EOM is usually unknown or ignored.
This chapter will show that there can be an answer for each scenario and regime\ylc{, illustrated here for the bispectrum in slow-roll inflation and for a selection of interactions}. With this aim in mind, let us emphasise that in this work, `classical' or `on-shell' refers to any evolution of initial or stochastic sources with the classical EOM.
Strictly speaking, and as we will see, it is thus a form of semi-classicality.

In what follows, Section~\ref{sec:Keldysh_semiclassical} showcases our use of the Keldysh rules, starting from the Dyson \textit{in-in} formula. 
The on-shell and off-shell contributions to the bispectrum are computed explicitly and analysed in Section~\ref{sec:OnOffShell}.
Section~\ref{sec:QCl_timing} focuses on finding scales, regimes and methods which can guarantee the domination of on-shell contributions to the bispectrum.
Section~\ref{sec:QCl_diagsemicl} then examines the resulting bispectrum shape in real space and discusses the implications for stochastic inflation and numerical simulations, before final comments.
The present work relies on the Keldysh formulation of the in-in formalism which we review in Appendix~\ref{app:CTP}, paying attention to the inclusion of an initial state set at a finite past.
Appendix~\ref{app:CSA} reviews the origins of the classical-statistical approximation while Appendix~\ref{app:appRedef} recalls contributions to the bispectrum arising from field redefinitions.   

In this chapter, unless specified otherwise, we focus on a typical choice of early universe scalar cosmological action. It is that of an inflaton $\phi$ in a $P(X,\phi)$ field theory treated in general relativity
\begin{equation}
    S = \frac{1}{2}\Mpl^2\int d^4x \sqrt{-g} R+ \int d^4x \sqrt{-g} P(X,\phi),
\end{equation}
where $X = -\frac{1}{2}\partial^{\mu}\phi\partial_{\mu}\phi$, yielding a speed of sound $c_s = P_{,X}/( P_{,X}+2X P_{,XX})$. This is usually studied in perturbation theory in the right slicing (comoving gauge) to give the perturbation action up to cubic order \cite{chen_observational_2007} 
\begin{eqnarray}
S &=& \int d^4x \,  \frac{1}{2}\frac{az^2}{c_s^2}\Big(\dot{{\cal R}}^2-\frac{c_s^2}{a^2}\left (\partial {\cal R})^2 \right ) \Big) +S^{(3)} +O({\cal R}^4) \nonumber \\
  &\equiv& S^{(2)}+S^{(3)} +O({\cal R}^4), 
\label{eq:linearisedAction}
\end{eqnarray}
where ${\cal R}$ is defined such that $g_{ij} = a^2 \yld{e^{2{\cal R}}}\ylg{(1+2{\cal R})}\delta_{ij}$ in comoving gauge, and the background quantity $z^2(t) = a^2H^{-2}\dot{\phi}^2 = 2a^2\varepsilon_1\Mpl^2$, where $\varepsilon_1 = -H^{-1}d_t\ln H$ is the background slow-roll parameter. This perturbative truncation will be taken as our dynamical action, where the cubic $S^{(3)}$ contribution will be made explicit later. The discussions laid down in this chapter are, moreover, extendable beyond GR, for instance with the EFT of inflation \cite{Cheung2008}.

In this framework, the free theory's equation of motion is ${\cal D}_0{\cal R} = 0$, referred to as the Mukhanov-Sasaki equation introduced previously in eqn.~\eqref{eq:MSvEq}, where the operator $\mathcal{D}_0$ reads in cosmic time
\begin{equation}
\mathcal{D}_0  = -\partial_t\left (\frac{az^2}{c_s^2}\partial_t\cdot\right)+\frac{z^2}{a}\partial^2\,.
\end{equation}
The initial state is chosen as the Bunch-Davies (BD) vacuum \cite{BD}, and,
when looking at the example of a quasi-dS spacetime evolution with such initial conditions,
one can find analytical solutions to the Mukhanov-Sasaki equation using Hankel functions \cite{pajer_fieldtheory,baumann_cosmology_2022}
\begin{equation}
    {\cal R}_k(\eta) = -\frac{H}{\Mpl}\sqrt{\frac{\pi \gamma}{8\varepsilon_1 k^3}}(-k \eta)^{3/2}H^{(1)}_{\nu}[-k\eta],
    \label{eq:quasiDShankel}
\end{equation}
which can be approximated as a quasi massless solution in the slow-roll limit ($\varepsilon_{1,2} \ll 1 \Rightarrow \nu\simeq 3/2^+$) to recover the well-known solution
\begin{equation}
    {\cal R}_k(\eta) = \frac{iH}{\Mpl}\sqrt{\frac{\gamma}{4\varepsilon_1 k^3}}(1+i k \eta)e^{-ik\eta},
    \label{eq:quasiDS}
\end{equation}
where $c_s$ has been set to $1$ for simplicity. This leads to the usual formula for the free two-point spacetime correlation function, and will be essential in our computations in the following.

The parameter $\gamma$ is dimensionless and serves as a book-keeping device to track places where $\hbar$ would appear had we kept SI units.
It will serve as an indicator of contributions that cannot be reproduced by on-shell evolution, i.e.~arising from the classical equations of motion. It should be set to unity when not required, $\gamma\rightarrow 1$.

\section{Semi-classical limit during inflation}\label{sec:Keldysh_semiclassical} 

In the Introduction, we have seen that QFTCS can be studied using various formalisms.
In particular, the use of a path integral generating functional in the CTP formalism \cite{Schwinger60,Bakshi62,Bakshi63,Keldysh64} is often preferred for the study of saddle points and hence semi-classical limits \cite{Keldysh64,Radovskaya2021}.
For a review of these methods in inflationary cosmology, see \cite{Chen17diag} and Appendix \ref{app:CTP}, where we review the propagators involved in the Keldysh-Schwinger formulation of the CTP's diagrammatic expansion, paying attention to how they arise from the imposition of the initial state and the upper boundary of the closed-time-path, a finite and arbitrary time $t_{\rm f}$. 

In this chapter, we retain the \textit{in-in} formalism, while employing the Keldysh basis to extract the semi-classical limit of cosmological correlators. 
Let us apply the \textit{in-in} formula defined in eqn.~\eqref{eq:inin} to the $n$-point correlation function of the comoving curvature perturbation $\hat{\cal R}$ in the interaction picture
\begin{equation}
    \langle \hat{\cal R}(x_1)...\hat{\cal R}(x_n) \rangle = \langle 0 |  {\cal \Bar{T}}\exp\bigg(\frac{i}{\gamma}\int_{\cal C}{\cal H}_{int}[\hat{\cal R}_-^{\cal I}]\bigg) \hat{\cal R}^{\cal I}(x_1)...\hat{\cal R}^{\cal I}(x_n) {\cal T}\exp\bigg({-\frac{i}{\gamma}\int_{\cal C} {\cal H}_{int}[\hat{\cal R}_+^{\cal I}]}\bigg) | 0\rangle,
    \label{eq:ininqcl}
\end{equation}
where the integration is over ${\cal C } = [t_{\rm i},t_{\rm f}]\times \mathbf{R}^3$, and ${\cal T}$ (resp. ${\cal \bar{T}}$) denotes the time ordering operator (resp. anti-time ordering), acting on the exponentiated contributions of the interactions over time.
The evolution operator appears twice, once for the ordered fields and once for the anti-ordered ones, reflecting the ket and bra state vectors involved in quantum-mechanical expectation values.
We have made explicit the two $\pm$ (forwards and backwards temporal evolution) branches of the field, usually kept implicit and encoded in the contours' $\pm i\varepsilon$ prescription \cite{Kaya19}. 

As introduced above, $\gamma$ appears explicitly in the exponent of the evolution operator as it provides one of the first answers to the question `what is quantum?'. 
In this expression $\gamma$ is a critical quantity for the interaction terms as the behaviour of the exponential strongly depends on it. However, dynamics are not the only key to a quantum theory. In particular, one of the key features of eqn.~\eqref{eq:ininqcl} is the cosmological and quantum imprint on the formula: it is assumed that there exists a time (finite or infinite) where the inflaton is in its quantum vacuum state (defined by its free Hamiltonian) and its evolution gives us the state of the inflating universe. The dynamics are highly dependent on how the initial conditions scale compared to $\gamma$. 

A convenient way to distinguish quantum from classically reproducible dynamics lies in the so-called Keldysh formalism of the closed-path integral \cite{Keldysh64}, found with a simple field redefinition
\begin{equation}
\begin{pmatrix}
    {\cal R}_+ \\
    {\cal R}_-
\end{pmatrix}
=
\begin{pmatrix}
    1 & \frac{\gamma}{2} \\
    1 & -\frac{\gamma}{2}
\end{pmatrix}
\begin{pmatrix}
    {\cal R}_{cl} \\
    {\cal R}_{q}
\end{pmatrix}.
\label{eq:redefKeldysh}
\end{equation}
As we hinted above,\textit{ $\gamma$ will be a parameter counting semi-classical orders in our development.} {In particular, as will be seen more explicitly further below, the basis change eqn.~\eqref{eq:redefKeldysh} leads to a re-definition of propagators from ``$++$", ``$--$" and ``$+-$" types, directly arising from eqn.~\eqref{eq:ininqcl}, to ``$\rm{cl}$$-$$\rm{q}$ and ``$\rm{cl}$$-$$\rm{cl}$" propagators - see eqn.~\eqref{eq:defGKGR}. This will allow for a direct counting of powers of $\gamma$ and a corresponding ``classicality distinction" between diagrams with differing numbers of these propagators.} The basis change is usually performed on the path integral functional \cite{Keldysh64,vanderMeulen2007, Radovskaya2021} but can also be performed directly in the \textit{in-in} formula. 

For example, when considering a simple toy $-\lambda a {\cal R}^3\subset {\cal H}_{int}$ interaction, the first-order in-in bispectrum contribution at the ($t=t_f$) boundary is
\begin{equation}
\langle \hat{\cal R}_{\mathbf{k}_1} \hat{\cal R}_{\mathbf{k}_2}\hat{\cal R}_{\mathbf{k}_3}\rangle   
         \simeq -\frac{i}{\gamma}\lambda \int d^4x a \langle 0 |  \hat{\cal R}_-(x)^3  \hat{\cal R}_{\mathbf{k}_1} \hat{\cal R}_{\mathbf{k}_2}\hat{\cal R}_{\mathbf{k}_3}  -  \hat{\cal R}_{\mathbf{k}_1} \hat{\cal R}_{\mathbf{k}_2}\hat{\cal R}_{\mathbf{k}_3}\hat{\cal R}_+(x)^3  | 0\rangle+{\cal O}(\lambda^2),
\label{eq:exampleR3}
\end{equation}
which at this stage would usually be simplified as a commutator and the $\pm$ difference left in the $\pm i\epsilon$ prescription of time integrals. The decomposition of these interaction picture operators in the vacuum's annihilation and creation basis in Fourier space would then allow one to keep only integrated contractions thanks to Wick's theorem. Using the $\pm$ propagators (defined in Appendix \ref{app:CTP}) for these and integrating over time will yield the full quantum correlation. This is the standard \textit{in-in} method.

The Keldysh method instead is adapted for use in the Dyson formula by substituting the decomposition of eqn.~\eqref{eq:redefKeldysh} in eqn.~\eqref{eq:exampleR3}. In particular, at the ($t=t_f$) boundary where $\hat{\cal R} = \hat{\cal R}_{cl}$, the correlator can be expressed as
\begin{equation}
\resizebox{\textwidth}{!}{
$\begin{aligned}
  \langle \hat{\cal R}_{\mathbf{k}_1} \hat{\cal R}_{\mathbf{k}_2}\hat{\cal R}_{\mathbf{k}_3}\rangle = -\frac{i}{\gamma}\lambda \int_{k_a,k_b,k_c}\delta^{(3)}(\mathbf{k}_a+\mathbf{k}_b+\mathbf{k}_c)\int_{t_{\rm i}}^{t_{\rm f}} dt \, a \langle 0 |  \prod_{\mu\in \{ a,b,c\}}\left (\hat{\cal R}_{cl,\mathbf{k}_{\mu}}-\frac{\gamma}{2}\hat{\cal R}_{q,\mathbf{k}_{\mu}}\right)  \hat{\cal R}_{cl,\mathbf{k}_1} \hat{\cal R}_{cl,\mathbf{k}_2}\hat{\cal R}_{cl,\mathbf{k}_3} | 0\rangle\\
  - a \langle 0 | \hat{\cal R}_{cl,\mathbf{k}_1} \hat{\cal R}_{cl,\mathbf{k}_2}\hat{\cal R}_{cl,\mathbf{k}_3}\prod_{\mu\in \{ a,b,c\}}\left (\hat{\cal R}_{cl,\mathbf{k}_{\mu}}+\frac{\gamma}{2}\hat{\cal R}_{q,\mathbf{k}_{\mu}}\right)  | 0\rangle+{\cal O}(\lambda^2),
\end{aligned}$
}
\label{eq:exampleR3bis}
\end{equation}
which greatly simplifies\yld{ and} over the tetrapyd \yld{becomes}\ylg{as}
\begin{equation}
\begin{aligned}
  \langle \hat{\cal R}_{\mathbf{k}_1} \hat{\cal R}_{\mathbf{k}_2}\hat{\cal R}_{\mathbf{k}_3}\rangle' = -2\lambda\int^{t_f}_{t_{\rm in}} dt \, a(t) K_{k_1}(t, t_f)K_{k_2}(t, t_f)G^R_{k_3}(t_f,t)\\
+\frac{1}{4}{\gamma^2}\lambda\int^{t_f}_{t_{\rm in}} dt a(t) G^R_{k_1}(t_f,t)G^R_{k_2}( t_f,t)G^R_{k_3}(t_f,t)+k-perm.+{\cal O}(\lambda^2)\,.\\
\end{aligned}
\label{eq:exampleR3ter}
\end{equation}
Here, we have used Wick's theorem, kept connected diagrams, and simplified terms with even powers of $\hat{\cal R}_{q}$,\footnote{The opposite signs come from having $i$ factors  in $G$ propagators but not $K$.} with the propagators in this basis defined as%
\begin{subequations}
    \begin{align}
        iK(k,t,t') & \equiv \langle \hat{\cal R}_{cl,\mathbf{k}}(t)\hat{\cal R}_{cl,-\mathbf{k}}(t')\rangle \equiv  F(k,t,t'),   \\
        iG^R(k,t,t') & \equiv \langle \hat{\cal R}_{cl,\mathbf{k}}(t)\hat{\cal R}_{q,-\mathbf{k}}(t')\rangle  \equiv -{i}\theta[t-t']G(k,t,t'),  \\
        iG^A(k,t,t') & \equiv iG^R(k,t',t)\,.
    \end{align}
    \label{eq:defGKGR}
\end{subequations}
These can be computed explicitly with the free theory and the initial conditions as shown in Appendix \ref{app:CTP}. Here we have parametrised the magnitudes via the Fourier transforms of the Schwinger (up to $-2i$ factor) and \textit{Hadamard} functions respectively\footnote{Also known as real and imaginary parts of the \textit{Wightman} function.}%
\begin{subequations}
    \begin{align}
        F(k,t,t') &= Re[ {\cal R}_k(t) {\cal R}_k^*(t')],   \\
       G(k,t,t') & = 2Im[{\cal R}_k(t) {\cal R}_k^*(t')].
    \end{align}
    \label{eq:defFG}
\end{subequations}
$K$ is not, strictly speaking, a Green function, but rather a solution of ${\cal D}_0(t) K(t,t') = 0$; it is nonetheless still labelled as a propagator.

The crucial point to note about eqn.~\eqref{eq:exampleR3ter} is that only the $\gamma^0$ terms can be reproduced by classical dynamics; the $\gamma^2$ terms are ``purely quantum'' and cannot be obtained via any sort of classical dynamics {augmented with statistical fluctuations}.

\begin{figure}[ht]
\centering
\includegraphics[width=0.65\textwidth]{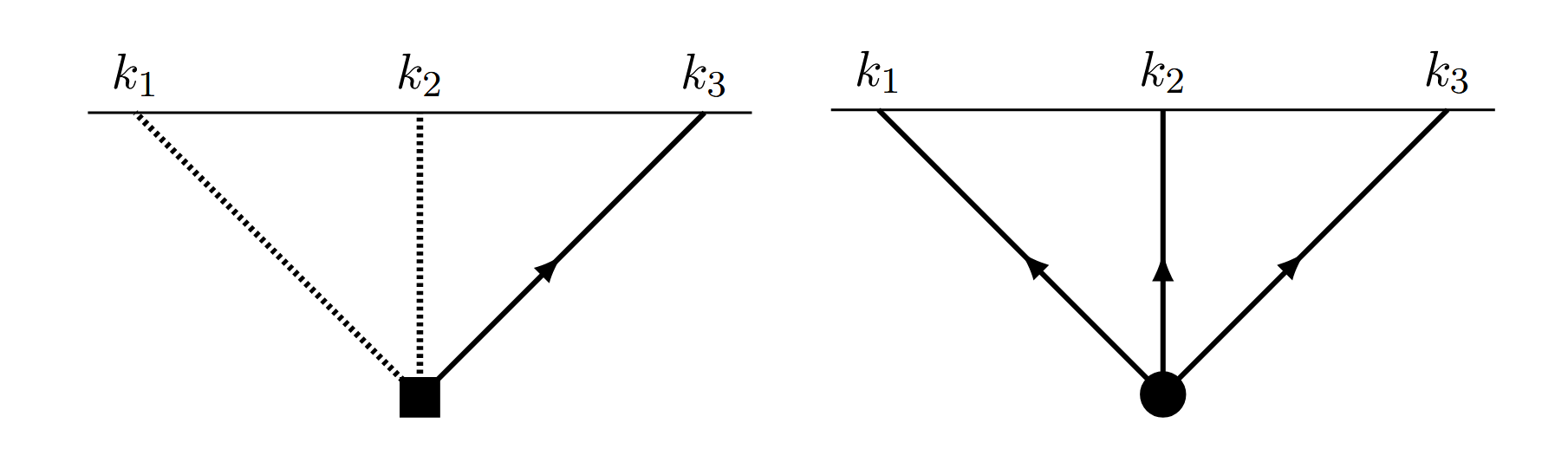}
    \caption{Bulk-to-boundary tree-level diagrams from the $\lambda(t){\cal R}^3$ vertex {decomposed as Keldysh-Feynman `classical' (left) and `quantum' (right) diagrams, up to ($k_1$, $k_2$, $k_3$) permutations. A dashed line stands for a $K$ propagator while an oriented solid one stands for $G^{R/A}$ (see eqn.~\eqref{eq:defGKGR}). Vertex rules are explicit in eqn.~\eqref{eq:exampleR3ter}. The non-oriented solid line is the boundary (time $t_{\rm f}$), intersected at each field's value with momentum $k_i$.}}
    \label{fig:diag1}
\end{figure}

Using $\gamma$ as a counting parameter, one quickly understands that the interaction terms at any order will yield powers of $\gamma$ corresponding to factors of ${\cal R}_q$ in these expressions and in the interaction Hamiltonian or action. Different powers of $\gamma$ yield different vertices. In general, only odd powers of ${\cal R}_q$ appear in the action of a closed system\footnote{An influence functional can arise when integrating out either an environment or a group of scales  \cite{Feynman63} and lead to a theory with interaction terms between $\pm$, and ultimately to non-odd powers of ${\cal R}_q$ in the resulting Keldysh action, see for instance the open EFT of Inflation \cite{Salcedo24}. This will be ignored until the special case of section \ref{sec:QCl_SI}. } and vertices in the Keldysh representation can only include $2n+1$ $\{n=0,1,2,\ldots\}$ factors of ${\cal R}_{q}$. In the ${\cal R}^3$ example studied here two vertices are allowed: one with one quantum ${\cal R}_{q}$ leg  and one with three. In our case, we are interested in the tree-level bulk-to-boundary correlators where ${\cal R}_q(t_{\rm f}) = 0$, which yields only two Feynman diagrams, displayed in Figure~\ref{fig:diag1}. When summing all these contributions with their permutations, one should find the same correlators as the usual in-in approach.

The diagrams with a single $G^{R/A}$ propagator (which have the lowest power in $\gamma$) are usually referred to as \textit{classical} ones, while others are designated as purely \textit{quantum}. A first way to see that is illustrated in Appendix \ref{app:CSA}. It can be summarised in the fact that ${\cal R}_{cl}$ undergoes on-shell dynamics (classical equations of motion) in the $\gamma \rightarrow 0$ limit. The initial conditions, however, remain unchanged. This limit is the semi-classical limit, the leading order of which is called the classical statistical approximation (CSA) \cite{Radovskaya2021} and equates indeed to using only one $G^{R/A}$ propagator per vertex.

Another way to understand the role played by the number of $K$ and $G^{R/A}$ propagators is to look at the commutators and {anti-commutators}%
\begin{subequations}
    \begin{align}
        \bigg\langle\left\{\hat{\cal R}(\mathbf{x}, t), \hat{\cal R}(\mathbf{y}, t^{\prime})\right\} \bigg\rangle & = 2\int d^3 k e^{i \mathbf{k} \cdot(\mathbf{x}-\mathbf{y})} \operatorname{Re}\left[{\cal R}_k(t) {\cal R}_k^*\left(t^{\prime}\right)\right] = \frac{F}{2},\\
        \bigg\langle\left[\hat{\cal R}(\mathbf{x}, t), \hat{\cal R}(\mathbf{y}, t^{\prime})\right]\bigg\rangle & =2\int d^3 k e^{i \mathbf{k} \cdot(\mathbf{x}-\mathbf{y})} \operatorname{Im}\left[{\cal R}_k(t) {\cal R}_k^*\left(t^{\prime}\right)\right] = G,
\end{align}
\label{eq:WeinbergCom}
\end{subequations}
In that sense, discarding multiple $G^{R/A}$ propagators reflects discarding the contributions of commutators to the dynamics. This link is particularly important to relate the Keldysh formalism in cosmology to previous studies of classicality which mostly focus on commutation \cite{Weinberg05, Weinberg06, Assassi12, Lim15, Martin16, burgess_minimal_2022}.
 
For classical dynamics to be an adequate approximation to the full quantum results, vertices with higher powers of $\gamma$ must be negligible with respect to those with no factors of $\gamma$.
For a given $n$-vertex $\lambda(t)\gamma^p{\cal R}_{cl}^{n-p}{\cal R}_q^p$, one will require $p$ $G$ functions ($G/\gamma$) and $n-p$ Keldysh ones ($F$) to calculate any correlation function.
This means that classical dynamics will be an adequate description if 
\begin{equation}
F\gg  G,
\end{equation}
a common property known in non-equilibrium QFT frameworks \cite{berges2015}, which recalls the theorem implying the classicality of the full QFT from the classicality of the free linear theory, proved in \cite{LythSeery08}. 
Counting powers of $G/F$ in calculations is therefore the aim.

\begin{figure}[ht]
\centering
\includegraphics[width=0.65\textwidth]{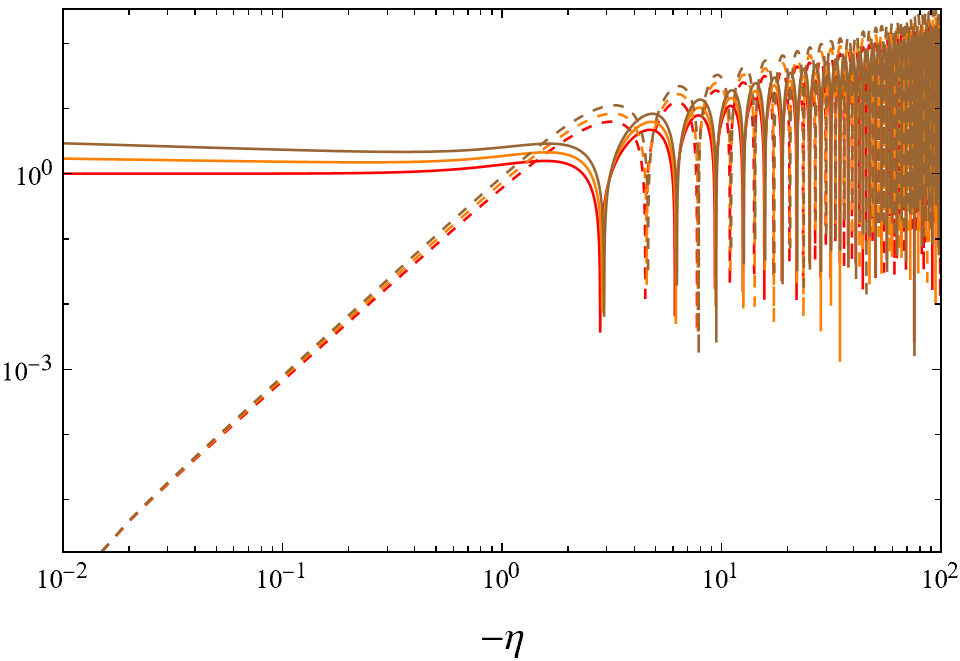}
    \caption{Absolute $F$ (solid lines) and $G$ (dashed lines) in quasi-dS ($\nu = \frac{3}{2}+0$ (red), $+0.05$ (orange), $+0.1$ (brown)) for $\eta$ up to $\eta_f = -0.01 $ and $k=1$, up to constants. Oscillations amplify in the $\eta \rightarrow -\infty$ limit.}
    \label{fig:FG}
\end{figure}

Let us now focus on the present case provided by eqn.~\eqref{eq:quasiDShankel} and its approximation eqn.~\eqref{eq:quasiDS}. One might have noticed that we purposefully did not set $\gamma = 1$ units as opposed to standard treatments. This is to emphasise that both on-shell and off-shell fields can have a $\gamma$ magnitude from their initial conditions. The corresponding kernels of eqn.~\eqref{eq:defFG} in Fourier space and conformal time become  \cite{vanderMeulen2007}%
\begin{subequations}
    \begin{align}
& F\left(k, \eta_1, \eta_2\right)=\frac{\gamma H^2}{4 \varepsilon_1 \Mpl^2k^3}\left[\left(1+k^2 \eta_1 \eta_2\right) \cos k\left(\eta_1-\eta_2\right)+k\left(\eta_1-\eta_2\right) \sin k\left(\eta_1-\eta_2\right)\right], \\
& G\left(k, \eta_1, \eta_2\right)=\frac{\gamma H^2}{ 2\varepsilon_1 \Mpl^2k^3}\left[\left(1+k^2 \eta_1 \eta_2\right) \sin k\left(\eta_1-\eta_2\right)-k\left(\eta_1-\eta_2\right) \cos k\left(\eta_1-\eta_2\right)\right]\,.
\end{align}
\label{eq:FGdS}
\end{subequations}
These two kernels are plotted in Figure \ref{fig:FG}, together with quasi-dS solutions provided by the massive Hankel  solutions of eqn.~\eqref{eq:quasiDShankel}. For a given scale, it takes a certain time for the $F/G$ ratio to become effectively small, partially because UV scales have a divergent 2-point function but mostly because $F$ and $G$ are simply dephased and of comparable magnitude. For this class of solutions, one can show that $G$ goes to $0$ just after horizon crossing, while $F$ does not as shown in Figure \ref{fig:FG}. In that sense, discarding powers of $\gamma$ is allowed in quasi-dS if we focus on super-Hubble scales.

As we have just seen, inflationary semi-classicality can only be applied to a (time-dependent) subset of scales; one can wonder about how to adapt the formalism to the coarse-grained system.
Are the propagators much impacted by the coarse-graining? Let us keep that aside for now as this caveat will be \yld{addressed}\ylg{discussed} in section \ref{sec:QCl_SI}.

\section{On-shell and off-shell correlators analysis}\label{sec:OnOffShell}
In this section we ignore the non-validity of the semi-classical approximation (and so of the CSA) and study the on-shell and off-shell contributions to the bispectrum. We focus on the following interaction vertices:%
\begin{subequations}
    \begin{align}
H_{int} & = \displaystyle\int d^3x \sum_i {\cal H}^{(3)}_i \\
{\cal H}^{(3)}_i & = g_i(t)a(t)^{q_i}f_i[{\cal R}]
\end{align}
\label{eq:eqH3}
\end{subequations}
where the chosen couplings ($g_i$), scaling ($q_i$), and cubic differential functionals ($f_i$) are defined in Table \ref{tab:interactions}. This corresponds to leading cubic terms of an inflationary theory in standard General Relativity \cite{chen_observational_2007}. This means that we need to study vertices like $g_1(t)\dot{{\cal R}}^3$, $g_2(t){\cal R} \dot{{\cal R}}^2$, $g_3(t){\cal R}(\partial{\cal R})^2$, and $g_4(t)\dot{{\cal R}}\partial{\cal R}\partial\chi$, absorbing EOM terms by redefinition \cite{Maldacena_2003} (see Appendix \ref{app:appRedef}).

\renewcommand{\arraystretch}{1.5}
\begin{table}[ht]
\centering
\resizebox{\textwidth}{!}{%
\begin{tabular}{|c|c|c||c|c|}
\hline
$g_i(t)$ & $q_i$ & $f_i[{\cal  R}]$ & $f_i^q[{\cal  R}]$ & $f_i^{cl}[{\cal  R}]$\\
\hline
 $ H^{-3}(\Sigma(1-\frac{1}{c_s^2})+2\lambda)$ & $3$& $\dot{\cal R}_{\pm}^3$ & $\frac{1}{4}\dot{\cal R}_q^3$ & $3\dot{\cal R}_q\dot{\cal R}_{cl}^2$ \\
\hline
 $-\frac{\varepsilon_1}{c_s^4}(\varepsilon_1-3+3c_s^2)$ & $3$& ${\cal R}_{\pm}\dot{\cal R}_{\pm}^2$ & $\frac{1}{4}{\cal R}_q\dot{\cal R}_q^2$ &  $ {\cal R}_q\dot{\cal R}_{cl}^2+2 {\cal R}_{cl}\dot{\cal R}_{cl} \dot{\cal R}_q$  \\
\hline
 $-\frac{\varepsilon_1}{c_s^2}(\varepsilon_1-2s+1-c_s^2)$ & $1$& ${\cal R}_{\pm}(\partial{\cal R}_{\pm})^2$& $\frac{1}{4}{\cal R}_q(\partial{\cal R}_q)^2$ &  $ {\cal R}_q(\partial {\cal R}_{cl})^2+2 {\cal R}_{cl}\partial {\cal R}_{cl} \partial{\cal R}_q$\\
\hline
 $\varepsilon_1^2/c_s^4$ & $3$& $\dot{\cal R}_{\pm}(\partial{\cal R}_{\pm})(\partial\partial^{-2}\dot{\cal R}_{\pm})$ & $\frac{1}{4}\dot{\cal R}_q\partial{\cal R}_q\partial\partial^{-2}\dot{\cal R}_q$ & $\dot{\cal R}_q\partial{\cal R}_q\partial\partial^{-2}\dot{\cal R}_{cl}$ + 5 perm. \\
\hline
\end{tabular}}
\caption{Terms of the slow-roll leading cubic Hamiltonian density of GR coupled to a $P(X,\phi)$ inflaton after redefinition (see eqn. (4.31) and the notations of \cite{chen_observational_2007}) and associated Keldysh vertices after basis change.\protect\footnotemark}
\label{tab:interactions}
\end{table}
\renewcommand{\arraystretch}{1}
\footnotetext{Note the corrected Keldysh factor from $2$ to $3$ in the first classical vertex compared to \cite{Launay2024bis}.}

In \cite{vanderMeulen2007}, a $g \phi^3$ theory is studied in dS to derive the 2-point correlator at second order in $g$, including regulated loops. For a different purpose in another study \cite{Salcedo24}, vertex coefficients of an open EFT in the Keldysh basis have been looked at and bispectrum integral expressions have been found for (quasi-)dS spacetimes, and solved for numerically. In this work, we performed analytical integrations by using the quasi-dS solutions of eqn.~\eqref{eq:quasiDS} in the integrands and simplifying the results using \textit{Mathematica}. In the following, we use the elementary symmetric polynomials to compactify the writing of momenta mixings%
\begin{subequations}
    \begin{align}
        e_1 & = k_1+k_2+k_3, \\
        e_2 & = k_1k_2+k_2k_3+k_1k_3, \\
        e_3 & = k_1k_2k_3.
    \end{align}
\end{subequations}
For the sake of generality, our calculations have not been restricted to correlators at the conformal boundary ($\eta = 0$) but calculated for any future boundary $\eta\leq 0$. Similarly to usual computations in the literature \cite{chen_observational_2007}, we make the following assumptions for these derivations:
\begin{itemize}
    \item The integrations are carried out from the far past ($t_{\rm in}$ or $\eta_i = -\infty$) to the boundary\ylg{ ($\eta\leq0$)}, with the right $\pm i\epsilon$ prescription in any integration of diverging oscillations in the $t_{\rm in}$ limit: the prescription is still necessary in this basis.
    \item Propagators (such as $K$, $G^{A/R}$, or $G^{\pm\pm}$) can be approximated by their quasi-dS form of eqn.~\eqref{eq:FGdS}.
    \item Coupling parameters ($\lambda, \Sigma, \varepsilon_1,\varepsilon_2,H$) are assumed to be slowly varying so that they can be taken out of the time integrals. For these reasons, we will set these almost constants to $1$ and focus on the $k_i$ dependence. For instance eqn.~\eqref{eq:FGdS} will be used with $\frac{\gamma H^2}{ 4\varepsilon_1 \Mpl^2k^3} \rightarrow 1$.
    \item We are not interested in the redefinition contribution one would typically encounter to simplify EOM terms \cite{Maldacena_2003, chen_observational_2007}. These generate products of 2-point functions and are thus essential to finding the consistency relations for instance \cite{Maldacena_2003}. Separating the quantum and classical parts of these at first order in the coupling is straightforward because it is just the free theory. On top of that, evaluating the correlators at the boundary $\eta_f = \eta$ removes the ${\cal R}_q$ field, which means it is a purely classical effect at this order.
\end{itemize}
In practice, our\ylg{ symbolic} code decomposes the integrals into sums of Euler $\Gamma$ functions (regularised by the $\pm i \epsilon$ prescription), which have analytical expressions. We give the final results below, including Fourier space permutations and keeping the tetrapyd's delta functions implicit. In the following, ${\cal B}_{i} =  \langle \hat{\cal R}_{\mathbf{k}_1} \hat{\cal R}_{\mathbf{k}_2}\hat{\cal R}_{\mathbf{k}_3}\rangle'$ for the type $i$ interaction vertex using \textit{in-in} formalism. Inspired by eqn.~\eqref{eq:exampleR3ter}, ${\cal B}^{cl}_{i}$ (resp. ${\cal B}^{q}_{i}$) are the $FFG$ (resp. $GGG$) contributions only, calculated with the Keldysh method and giving the on-shell (resp. off-shell) interactions. 

For $a^3\dot{\cal R}^3$ and at final conformal time $\eta\leq 0$, we get%
\begin{subequations}
    \begin{align}
{\cal B}_{1} & \propto  \frac{6 e_1 e_3 \eta ^4+6 e_2 \eta ^2-3 e_1^2 \left(e_2 \eta ^4+\eta ^2\right)-6}{2 e_1^3 e_3}, \\
 {\cal B}^{q}_{1} & \propto  3 \left(3 e_1^6-18 e_1^4 e_2+24 e_1^3 e_3+24 e_1^2 e_2^2-48 e_1 e_2 e_3+32 e_3^2\right)\nonumber\\
 & \times \frac{ \Big(e_1^4 \eta ^4-4 e_1^2 \left(e_2 \eta ^4+\eta ^2\right)+8 e_1 e_3 \eta ^4+8 e_2 \eta ^2-8\Big)}{2 e_1^3 \left(e_1^3-4 e_1 e_2+8 e_3\right)^3}.
\end{align}
\label{eq:dot3}
\end{subequations}
For $a^3{\cal R} \dot{{\cal R}}^2$ we get%
\begin{subequations}
    \begin{align}
   {\cal B}_{2} & \propto  \frac{e_1^2 \left(3\eta^2 e_2-2\right) e_3+e_2 \left(1-\eta ^2 e_2\right) e_3+e_1e_2 \left(-e_3^2 \eta ^4-e_2^2 \eta ^2+e_2\right)}{2 e_1^2 e_3^3}, \\
    {\cal B}^{q}_{2} & \propto  \frac{\left(e_1^2-2 e_2\right) \Bigg(\eta ^2 \bigg(8 e_2+e_1 \Big(e_1 \big(\eta ^2 (e_1^2-4 e_2)-4\big)+8\eta^2 e_3\Big)\bigg)-8\Bigg)}{2 e_1^2 \left(e_1^3-4e_1 e_2+8 e_3\right){}^2}.
\end{align}
\label{eq:dot2}
\end{subequations}
For $a{\cal R} (\partial {\cal R})^2$ we get%
\begin{subequations}
    \begin{align}
   {\cal B}_{3} & \propto \frac{\left(e_1^2-2 e_2\right) \Big(e_1^3+(e_3^2 \eta ^4+e_2^2 \eta ^2-e_2) e_1-2e_1^2 e_3 \eta ^2+e_3 (e_2 \eta ^2-1)\Big)}{4e_1^2 e_3^3}, \\
{\cal B}^{q}_{3} & \propto -\frac{\left(e_1^2-2 e_2\right) \Big(8+4 (e_1^2-2 e_2) \eta ^2 -e_1 (e_1^3-4e_1 e_2+8 e_3) \eta ^4 \Big)}{2e_1^2 \left(e_1^3-4e_1 e_2+8 e_3\right)^2}. 
    \end{align}
\label{eq:grad2}
\end{subequations}
For $a^3\dot{\cal R} (\partial {\cal R})(\partial \partial^{-2}\dot{\cal R})$ we get%
\begin{subequations}
    \begin{align}
    {\cal B}_{4}& \propto \frac{-2 e_1^5+7 e_1^3 e_2+e_1 e_3 \eta ^4 \Big(e_3 \left(e_1^2+4 e_2\right)+e_1 e_2 \left(e_1^2-4 e_2\right)\Big)-17 e_1^2 e_3}{8 e_1^2 e_3^3} \nonumber\\
    &+\frac{\eta ^2 \left(e_1^5 e_2-e_1^4 e_3-3 e_1^3 e_2^2+13 e_1^2 e_2 e_3-4 e_1 e_2^3+4 e_2^2 e_3\right)+4 e_1 e_2^2-4 e_2 e_3}{8 e_1^2 e_3^3},\\
{\cal B}^{q}_{4} & \propto \Big(-e_1^2 e_2^2 \left(e_1^2-4 e_2\right)+2 e_3^2 \left(7 e_1^2+2 e_2\right)+2 e_1 e_3 \left(e_1^4-4 e_1^2 e_2-4 e_2^2\right)\Big) \nonumber\\
& \times\frac{ \Big(e_1 \eta ^4 \left(e_1^3-4 e_1 e_2+8 e_3\right)-4 \eta ^2 \left(e_1^2-2 e_2\right)-8\Big)}{4 e_1^2 e_3^2 \left(e_1^3-4 e_1 e_2+8 e_3\right)^2}.
    \end{align}
\label{eq:invdot2}
\end{subequations}
Note that the classical bispectra ${\cal B}_i^{cl}$ have also been computed independently with their permutations and all satisfy ${\cal B}_i= {\cal B}_i^{cl}+{\cal B}_i^{q}$.
These are also all in agreement with eqns.~(4.33--4.36) of \cite{chen_observational_2007} in the $\eta = 0$ limit with its dimensionful coefficients. Another check is that of \cite{Salcedo24}, conjecturing the folded singularities we encounter here and similarly in \cite{Green20} where the first interaction's classical vertex contribution was computed in the $\eta = 0$ limit and matches eqn.~\eqref{eq:dot3}.

Several notable features emerge when comparing the off-shell and on-shell contributions.
We start by taking the \textbf{squeezed limit} in those expressions, which makes $e_1$ and $e_2$ functions of two momenta only, while $e_3 \longrightarrow 0$. In all these expressions, we find 
    \begin{equation}
            {\cal B}_{i} \sim {\cal B}^{cl}_{i} \gg
           {\cal B}^{q}_{i}.
        \label{eq:squeezed}
    \end{equation}
{In this limit, the quantum contribution is ${\cal O}(1)$ (constant in $e_1$ and $e_2$), while the classical parts diverge as $e_3^{-3}$ or $e_3^{-1}$ (interaction $1$).
The result is interesting when thinking of the consistency relation verified by the squeezed limit of the single-field 3-point function \cite{Maldacena_2003, Creminelli04, Ganc10,RenauxPetel10}}
{
\begin{equation} \lim_{\substack{k_3 \rightarrow0}} {\cal B}(k_1,k_2,k_3) = (1-n_s)P_{k_3}P_{k_1}, \label{eq:consistency}\end{equation}
where $P_k = (2\pi)^{-3}\langle \hat{\cal R }_{\mathbf{k}}\hat{\cal R }_{-\mathbf{k}}\rangle$ and $n_s-1 = \frac{d \ln [k^3P_k]}{d \ln k}|_{k_1}$ (evaluated at $k_1$, since $k_2\simeq k_1$ in this limit).
This is true for the previous full \textit{in-in} computations in eqns.~\eqref{eq:dot3}, \eqref{eq:dot2}, \eqref{eq:grad2}, and \eqref{eq:invdot2}.  In particular, all of the mentioned derivations rely on the classicality of at least one long wavelength mode (super-Hubble $k_3$) and suggest that the squeezed limit is the result of the free quantum noise evolution on the perturbed classical background.  This explains why, among other things, a more rigorous $\Delta N$ formalism computation can recover the same result \cite{Abolhasani_2018}. The domination of the squeezed limit by the on-shell evolution expressed by our result in eqn.~\eqref{eq:squeezed} adds an element of depth because interactive classicality has been well defined and identified. Of course, one should not forget that redefinition terms in Appendix \ref{app:appRedef} contribute to the consistency relation too. These are much easier to handle in the long wavelength regime where ${\cal R}$ freezes and small gradients cancel terms. In this limit or not, it is appropriate to \yld{assume}\ylg{say} that these are on-shell at leading order in the couplings, which implies that eqn.~\eqref{eq:consistency} is completely explained by on-shell evolution, at least at leading order.}

In the \textbf{equilateral limit}, there is nothing special appearing at first sight which would make either classical or quantum contributions dominate. However, the \textbf{folded limit} (i.e. $k_1 = k_2+k_3$ and cyclic permutations) is more relevant. It is dominated by the folded  singularities of the form 
    $$\left(e_1^3-4e_1 e_2+8 e_3\right)^2 = (k_1+k_2-k_3)^2 (k_1-k_2+k_3)^2(-k_1+k_2+k_3)^2$$
    which only appear when using the classical and quantum Keldysh contributions, as opposed to the total \textit{in-in} results. In such a configuration, the leading order $\left(e_1^3-4e_1 e_2+8 e_3\right)^{-2}$ of those singularities is
    \begin{equation}
        {\cal B}^{cl}_{i}  \sim - {\cal B}^{q}_{i} \gg {\cal B}_{i}.
    \end{equation}
    In the folded limit, leading contributions cancel between the on-shell and off-shell parts, so that the total bispectrum $ {\cal B}_{i}$ is finite. Accounting for dissipation effects can change this conclusion by smoothing out the folded singularities \cite{Salcedo24}. In \cite{Green20}, one can find an interpretation for these emerging singularities. While the fully quantum \textit{in-in} contributions do not exhibit any physical poles because of the uncertainty principle (i.e. no positive energy for virtual particles of the vacuum), the (on-shell) classical evolution does produce physical particles and so physical poles.

    Overall, it is clear that the shape plays a crucial role in deciding whether a 3-point function can be computed via a fully classical evolution, at least from an infinite past. Only sufficiently squeezed shapes seem to be fully reproducible with such classical evolution of initial quantum fluctuations, though later and at late times we uncover some notable exceptions to these expectations numerically.  
    
\section{Timing classicality with quantum interactivity}\label{sec:QCl_timing}
\begin{figure}[!ht]
    \centering
\includegraphics[width=0.87\textwidth]{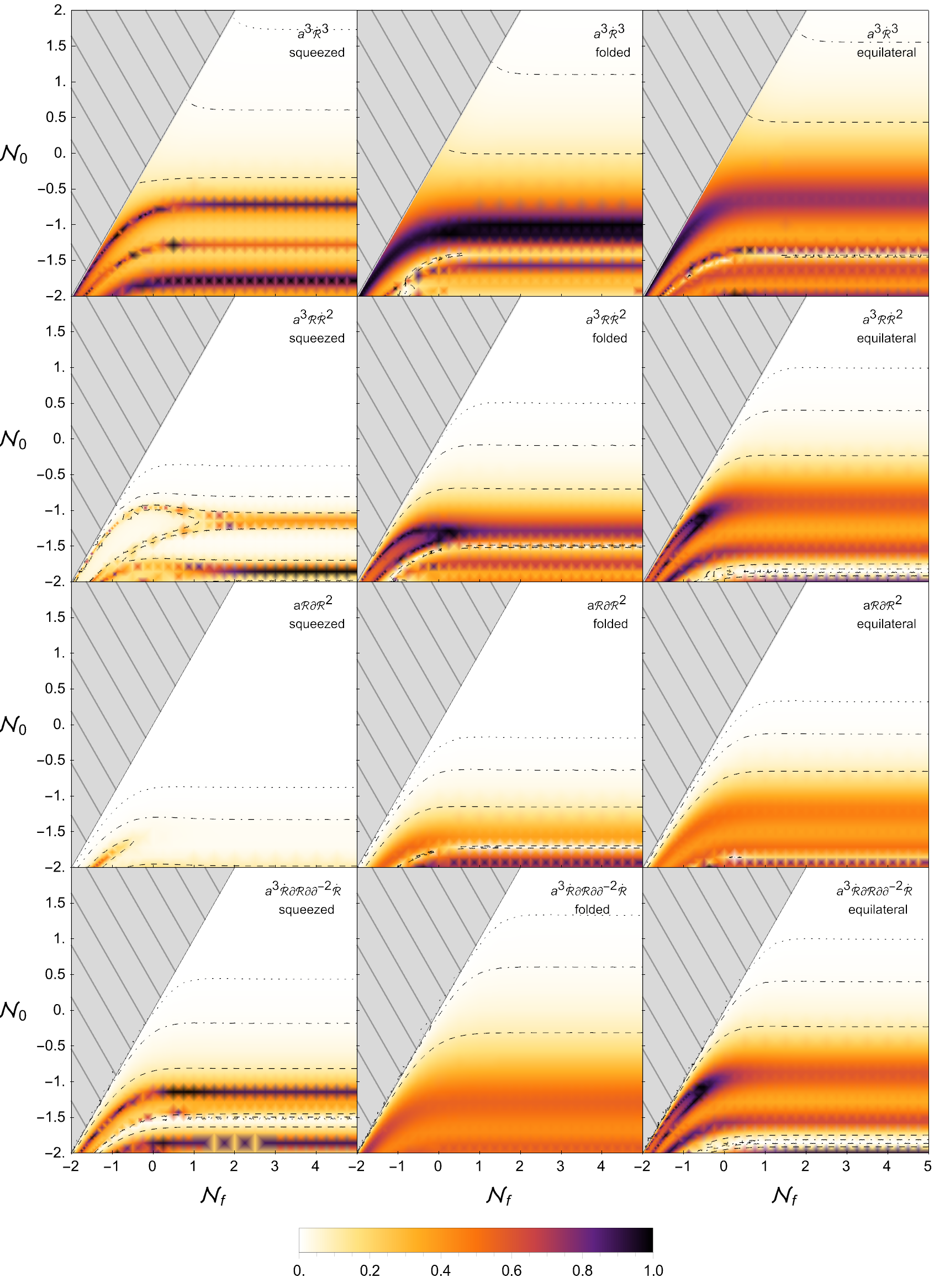}
    \caption{$ {\cal QI}(t_0,t_f)$ for $g_1 a^3\dot{{\cal R}}^3$, $g_2 a^3{\cal R} \dot{{\cal R}}^2$, $g_3 a{\cal R}(\partial{\cal R})^2$, and $g_4a^3\dot{\cal R}(\partial{\cal R})(\partial\partial^{-2}\dot{\cal R})$ in quasi-dS (top to bottom) for {squeezed ($k_1 = k_2 = k_3/0.1 = 1$), folded ($k_1 = k_2/0.55 = k_3/0.5 = 1$), and equilateral ($k_1 = k_2 = k_3 = 1$)} limits (left to right). $k_1$ is set to $1$ and is the last mode to cross the horizon ($k_{2,3}\leq k_1$), time at which we set the time origin ($k_1=aH$ at $\eta = -1$ or ${\cal N} = 0$  equivalently). Hatched grey regions stand for non-causal regions. Dashed contours have values 0.1 (dashed) and 0.01 (dashed dotted) and 0.001 (dotted). 
    }
    \label{fig:RatioTime}
\end{figure}

\begin{figure}[!ht]
    \centering
\includegraphics[width=1.0\textwidth]{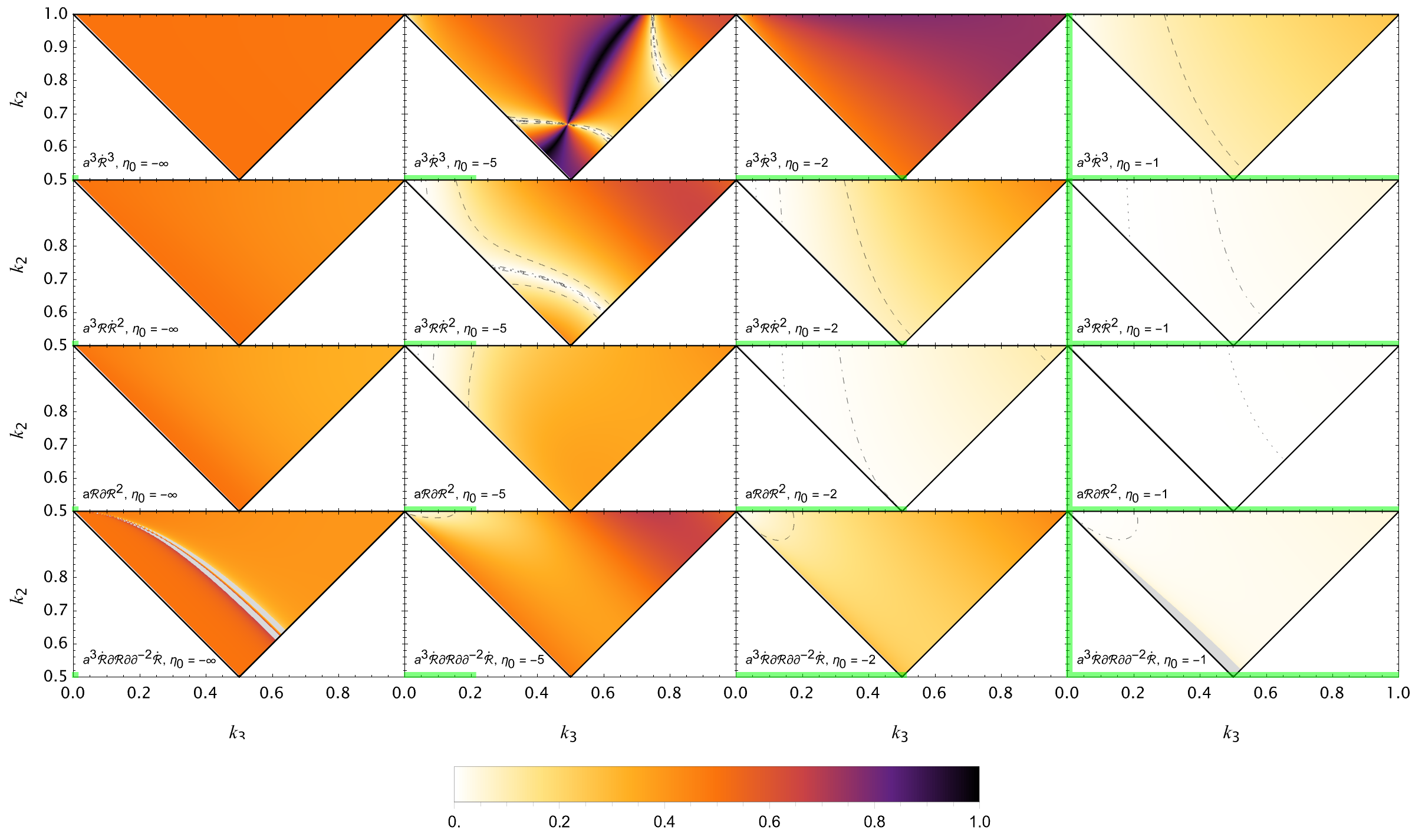}
    \caption{\sloppy ${\cal QI}(1,k_2,k_3,\eta_0, \eta_f = -0.01)$ for $g_1 a^3\dot{{\cal R}}^3$, $g_2 a^3{\cal R} \dot{{\cal R}}^2$, $g_3 a{\cal R}(\partial{\cal R})^2$, and $g_4a^3\dot{\cal R}(\partial{\cal R})(\partial\partial^{-2}\dot{\cal R})$ in quasi-dS (top to bottom) and $\eta_0 =  -\infty^{\pm}, -5, -2, -1$ (left to right, equivalent to ${\cal N}_0 \simeq  -\infty, -1.61, -0.69, 0
    $). Range from $0$ (white) to $1$ (black) before reaching plot threshold or numerically indeterminate (grey). Dashed contours have values 0.1 (dashed) and 0.01 (dashed dotted) and 0.001 (dotted). Green gauges indicated which part of the mode ticks have crossed the horizon. $k_1$ is set to $1$ and is the last mode to cross the horizon ($k_{2,3}\leq k_1$), time at which we set the time origin ($k_1=aH$ at $\eta = -1$ or ${\cal N} = 0$  equivalently).}
    \label{fig:RatioShape}
\end{figure}

We have just argued that we can reliably compute the \yld{stochastic}\ylg{on-shell} evolution of the bispectrum from the Bunch-Davies vacuum in the infinite past and that it is not satisfactory to neglect quantum interactions if QFTCS, together with the BD initialisation, is the appropriate theory describing our universe. We have also seen that classicality can be probed by considering the ratio of the two-point Keldysh and Green propagators. It is true that classicality beyond free theory is implied by that of the free theory itself \cite{LythSeery08} but we are interested in knowing precisely when it can be a good approximation to neglect quantum effects up to a desired accuracy in the $n$-point correlators. As one might expect, the answer is both shape and vertex-dependent. To study this issue, we cut the previous integrals into early and late-time quantum or classical contributions
\begin{equation}
\begin{aligned}
\langle {\cal O}(\mathbf{k}_1,...,\mathbf{k}_n,\eta)\rangle' & = {\cal C}_{early}+{\cal C}_{late} \\
& = {\cal C}_{early}^{q}+{\cal C}_{late}^{q} + {\cal C}_{early}^{cl}+{\cal C}_{late}^{cl},
\end{aligned}
\end{equation}
where the right early-late separation we seek satisfies ${\cal C}_{late}^{q}\ll{\cal C}_{late}^{cl}$ for given $\{\mathbf{k}_i\}$ at a time $\eta$.
At first order, the obvious separation is that of the time integration. One can truncate\ylg{ at $t_0$} the previous integrals (such as eqn.~\eqref{eq:exampleR3ter}) to define schematically for any couplings%
\begin{subequations}
    \begin{align}
        {\cal C}_{early}^{cl} & = \sum_{(a,b,c)\in \text{ perm}(3) }\kappa_{abc}\left (\int_{-\infty}^{t_0} dt\theta_aF_{k_1}\theta_bF_{k_2}\theta_cG_{k_3}|_{(t,t_f)} +k\text{-perm.}\right ),\\
        {\cal C}_{late}^{cl} &  = \sum_{(a,b,c)\in \text{ perm}(3) }\kappa_{abc}\left (\int_{t_0}^{t_f} dt\theta_aF_{k_1}\theta_bF_{k_2}\theta_cG_{k_3}|_{(t,t_f)} +k\text{-perm.}\right ),\\
        {\cal C}_{early}^{q} &  = \sum_{(a,b,c)\in \text{ perm}(3) }-\kappa_{abc}\left (\int_{-\infty}^{t_0} dt\theta_aG_{k_1}\theta_bG_{k_2}\theta_cG_{k_3}|_{(t,t_f)} +k\text{-perm.}\right ),\\
        {\cal C}_{late}^{q} &  = \sum_{(a,b,c)\in \text{ perm}(3) }-\kappa_{abc}\left (\int_{t_0}^{t_f} dt\theta_aG_{k_1}\theta_bG_{k_2}\theta_cG_{k_3}|_{(t,t_f)} +k\text{-perm.}\right ),
        \label{eq:truncateQFT}
    \end{align}
\end{subequations}
{where the $\theta_i$ are linear differential operators from the cubic interaction $\theta_1 {\cal R}\theta_2{\cal R}\theta_3 {\cal R}\subset S^{(3)}$, and $\kappa$ designates the Keldysh coefficients arising from the basis change (vertex-dependent,  see Table \ref{tab:interactions})}.
Early and late contributions are not the same, because of the different integration range. If one wants to calculate explicitly these expressions with the same approximations as in the last section, it is recommended to proceed numerically (e.g.\ with \textit{Mathematica}) because of the many (oscillating) terms arising from such a truncation or shift in the range (which already appear if doing so with the total \textit{in-in} integrals). The cumbersome analytical expressions can be found thanks to our code\yld{, which can be provided on request}. In this chapter, we simply show the outcomes graphically in Figures \ref{fig:RatioTime} and \ref{fig:RatioShape}, showcasing the key ratio which we refer to as the \textit{quantum interactivity} (QI):
\begin{equation}
    {\cal QI}(t_0,t_f,\{k_j\}) = \frac{|{\cal C}^q_{late}(t_0,t_f,\{k_j\}) |}{|{\cal C}^{cl}_{late}(t_0,t_f,\{k_j\}) |+|{\cal C}^q_{late}(t_0,t_f,\{k_j\}) |},
\end{equation}
which can be generalised to any $n$-point function, such as the trispectrum, but also to multiple interactions and to each order in the coupling. It peaks at $1$ when the quantum contributions dominate fully or falls to $0$ in the opposite case. The denominator is not the summed \textit{in-in} late contribution, since we take the absolute value, which prevents cancellation.

In Figure \ref{fig:RatioTime}, we look at the quantum interactivity for each cubic term (or shape) while varying $t_0$ and $t_f\geq t_0$ in the e-folding time ${\cal N}$. The e-folding time always has its origin set at the crossing of the largest $k$ (here $k_1=1$) because this represents the last mode to cross outside the horizon.  We first note the overall trends in Figure \ref{fig:RatioTime}. For each fixed boundary ${\cal N}_f$, looking at a range in ${\cal N}_0$ allows us to see when the interactive quantum contributions are below $10\%$, $1\%$ or $0.1\%$ of the total contribution. 
On the other hand, the range in ${\cal N}_f$ tells us about the stability in time of this behaviour. In particular, all these plots have in common the stability of ${\cal QI}$ after a few e-folds for each given ${\cal N}_0$, which is due to the suppression of the $G$ propagators in the integrands for both types of contributions.

Furthermore, these plots also show a hierarchy in `classicality' between interactions and their associated shapes. Again in Figure \ref{fig:RatioTime}, observe that the lower the dashed contours, the earlier one can start to neglect quantum contributions. From this perspective, it is clear that for the three shapes under consideration, the last interaction to become classical is $g_1(t)\dot{{\cal R}}^3$, preceded by $g_2(t){\cal R} \dot{{\cal R}}^2$ and $g_4(t)\dot{{\cal R}}\partial{\cal R}\partial\chi$, with the earliest being $g_3(t){\cal R}(\partial{\cal R})^2$. It thus appears that the more time derivatives there are, the later the classicalisation of the resulting bispectrum contribution. For each interaction, there is a close correspondence to the shape, from which we infer that there is more time needed for the equilateral shape compared to the folded one and the (mildly) squeezed one. A properly squeezed limit would confirm our conclusion from the previous section, which is that the local shape is mostly evolving classically.

Figure \ref{fig:RatioShape} illustrates the quantum interactivity on a slice through the full tetrapyd over a fixed ${\cal N}_0$--${\cal N}_f$ integration range. In particular, this provides confirmation of the hierarchy in classicality between the different interactions. Most importantly,  it confirms that the equilateral limit is typically the slowest to classicalise and thus constitutes a good diagnostic for when the full tetrapyd will have attained classicality. 
This figure illustrates again that the non-canonical $g_1(t)\dot{\cal R}^3$ interaction receives quantum contributions across a wider range of configurations, persisting until much closer to horizon crossing than for the other terms.

A remarkable observation is that the ${\cal QI}$ diagnostic indicates that sub-Hubble evaluations can provide a good approximation to the final bispectrum for most interactions (with the degree of approximation ranging from slight to considerable depending on shape). The possibility of sub-Hubble `classicality' is certainly not apparent from looking at the ratio $(F/G)^2$ (see Figure \ref{fig:FG}), which clearly takes $1$ to $2$ e-folds after horizon crossing to reach a comparable level. This means that the integration, the vertex and Keldysh coefficients (including $a(t)$) are important for calculating and identifying when the quantum-to-classical transition occurs. Performing the interactive calculus with our approximations is thus useful and quantitative, but one can also obtain excellent diagnostics for classicality at arbitrary accuracy using fully numerical solutions for specific models.\footnote{This is left for future work \cite{LRSZ}, using recent advances in fast bispectral predictions \cite{clarke_probing_2021}.} The question being addressed is thus: how long does it take for the cumulative $K$ propagator contributions to take over those from the $G$ propagators, before the latter closes down any further cumulative additions to the overall integral.

We conclude that, from the point of view of interactions, classicality can be reached earlier than previous criteria had suggested for most interactions. However, we note that the answer will always be interaction and shape dependent. In a way, these results are unprecedented but they are not actually different from what a quantitative study of commutator contributions would lead to when following other studies in the literature, had these been calculated more explicitly. For example, similar to the link established by eqn.~\eqref{eq:WeinbergCom} in free theory, eqn.~(14) given in ref.~\cite{Green20} shows this in interactive theory:
\begin{equation}
\begin{aligned}
     {\cal B}_i(\mathbf{x}_1,\mathbf{x}_2,\mathbf{x}_3,t)-{\cal B}_i^{cl}(\mathbf{x}_1,\mathbf{x}_2,\mathbf{x}_3,t) ~\propto~~ ig_i \int^{t}_{t_0} dtd^3\mathbf{x} a^{q_i}\langle\theta^i_{1\mathbf{x}}\left [ \hat{\cal R}(\mathbf{x}_1,t),\hat{\cal R}(\mathbf{x},t)\right] \\
     \times\,\theta^i_{2\mathbf{x}}\left [ \hat{\cal R}(\mathbf{x}_2,t),\hat{\cal R}(\mathbf{x},t)\right]\theta^i_{3\mathbf{x}}\left [ \hat{\cal R}(\mathbf{x}_3,t),\hat{\cal R}(\mathbf{x},t)\right] \rangle+\theta\text{-perms},
\end{aligned}
\end{equation}
which has been adapted to our notation while decomposing $f_i$ into a product of three differential operators $\theta_i$ acting on $\hat{\cal R}$, i.e.\ reflecting various types of interactions such as those in Table \ref{tab:interactions}\ylg{ and eqn.~\eqref{eq:truncateQFT}}. This makes apparent the high dependence of the quantum part on products of commutators and hence on the $G$ functions. It also emphasises the key contribution of cubic operators $\theta^i_j$ in specifying the time at which these integrals can be neglected: as we have just seen, partial time derivatives are important and can hamper classicalisation close to horizon crossing.

\section{Diagnosing semiclassical approaches}\label{sec:QCl_diagsemicl}
\subsection{The importance of real space and the bispectrum shape}\label{sec:QCl_realspace}
\begin{figure}[!ht]
    \centering    \includegraphics[width=0.85\textwidth]{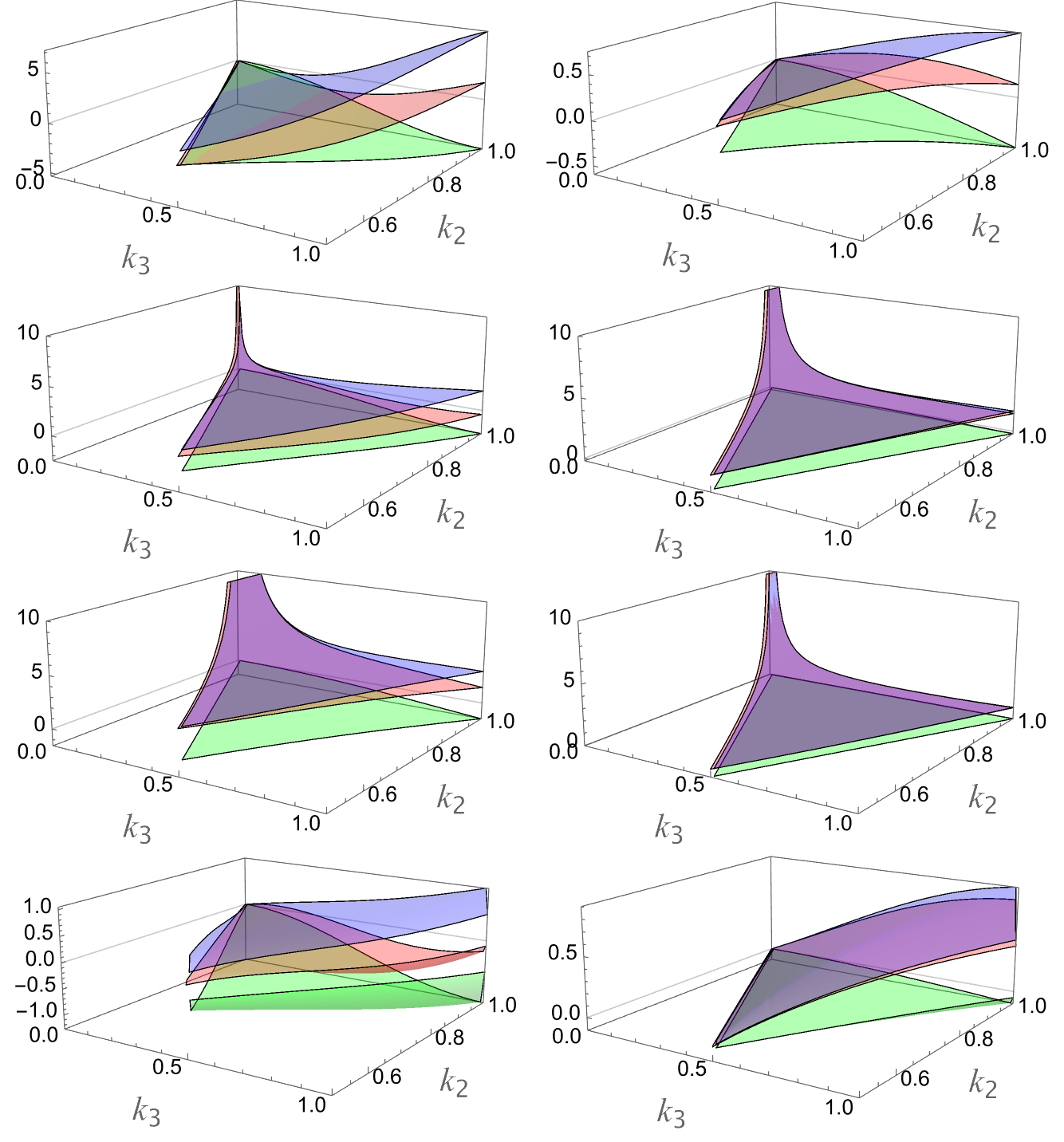}
    \caption{Total (red), classical (blue) and quantum (green) contributions to the shape function of ${\cal C}_{late}$ for ${\cal N}_0 = -1.0$ (left, equivalent to $\eta_0 \simeq -2.72 $) or ${\cal N}_0 = -0.25$ (right, equivalent to $\eta_0 \simeq -1.28 $), to ${\cal N}_f = 4.6$ (equivalent to $\eta_f \simeq -0.01 $) for $g_1 a^3\dot{{\cal R}}^3$, $g_2 a^3{\cal R} \dot{{\cal R}}^2$, $g_3 a{\cal R}(\partial{\cal R})^2$, and $g_4a^3\dot{\cal R}(\partial{\cal R})(\partial\partial^{-2}\dot{\cal R})$ in quasi-dS (top to bottom). $k_1$ is set to $1$ and is the last mode to cross the horizon ($k_{2,3}\leq k_1$), time at which we set the time origin ($k_1=aH$ at $\eta = -1$ or ${\cal N} = 0$  equivalently). Plotting threshold is set at an absolute value of $10$.}
    \label{fig:3Dshape}
\end{figure}

The previous diagnostics have all been defined in Fourier space. However, when returning to real space, the different three-point correlators computed previously will be affected by different tetrapyd regions in Fourier space. In particular, squeezed, folded and equilateral shapes will contribute in distinct ways. The shape function, $e_3^2{\cal B}$, is the actual amplitude of the Fourier transform's integrand, with two late-time contributions plotted on a tetrapyd slice in Figure \ref{fig:3Dshape}, first, well before horizon-crossing for the final mode and, secondly, only just beforehand. The standard gravitational interaction terms from Table \ref{tab:interactions} (i.e., shapes 2--4) have an almost vanishing quantum contribution as horizon-crossing is approached. The most local shape (the third with 
${\cal R}_{\pm}(\partial{\cal R}_{\pm})^2$) indicates that there is only a small quantum contribution even at ${\cal N}_0 = -1.0$. In fact, this small equilateral component is very subdominant relative to the highly peaked squeezed limit where most of the signal resides, which is essentially classical by this time.  This means that from an empirical perspective, when measuring whether this signal is present in CMB or other data, it would still be a good approximation to take an even earlier sub-Hubble prediction of this local shape.  In practical terms for stochastic inflation, this local signal could thus be introduced earlier than naively expected. In other words, the off-shell parts of the tetrapyd could be neglected earlier, which are erased in the integrand back to real space.  We note that these points may also be relevant for multifield inflation models, many of which are expected to exhibit dominant squeezed contributions. 

These observations are in stark contrast to the most equilateral shape (the first EFT shape with $\dot{\cal R}_{\pm}^3$, top of Figure \ref{fig:3Dshape}), which has substantial quantum contributions both at earlier and later times.  On the other hand, the other interactions with a second time derivative (2 and 4) are intermediate between the two extreme cases, though both become essentially classical just prior to horizon-crossing.

Nevertheless, another pertinent issue concerns UV divergences that arise when returning to real space (such as from the coincident limit). This is due to the vacuum energy, which also arises in Minkowski space and might require a regularisation scheme. In particular, when thinking of simulating stochastic initial conditions in general relativity, one cannot go too sub-Hubble because of the risk of introducing divergent non-perturbative noise. One needs to subtract the vacuum energy and only keep its late products. For the same reason, the previous calculations should not extend too far into the sub-Hubble regime at the risk of just investigating  already known oscillating divergences. In a regularisation such as using a cutoff or an adiabatic procedure \cite{Durrer09,pla23}, these far past contributions would be taken out anyway. For this reason and because of the difficult inverse Fourier integration, one should probably be cautious about concluding that classicality in real space occurs earlier than in Fourier space.

\subsection{Stochastic inflation and numerical simulations}\label{sec:QCl_SI}

Let us assume that we can generate initial conditions for a patch of the universe using linear theory, i.e.\ following a 2-point function $F^s$ at a given initial time for all wavenumbers we are interested in within a box of finite size. The simulation then evolves $F^s$ further in time and generates interaction terms, all on-shell, since no quantum simulation exists yet. If the perturbative regime remains, then the previous ${\cal C}^{cl}_{late}$ contributions will be the ones found up to numerical accuracy and up to next order in perturbation theory if $\eta_0$ is the starting time of the simulation. One should of course not forget that such a simulation will not actually separate orders in perturbation theory.

In this framework, the contributions which happened prior to the initial time and are now propagating to the final surface (or those reinteracting for higher orders in perturbation theory) are missing, as is, obviously, any quantum interaction, whether early or late.
For the former, it does not matter because our box is finite anyway and can be chosen so that none of these scales are actually within the spectral range of the simulation box. Only second-order interactions would be missed.
In theory, one could take into account the early contributions by giving next-order and non-Gaussian ICs using the analytical $n$-point correlators and observe these second-order interactions. In practice, introducing these beyond two points with random draws is challenging, though feasible.

The real issue with providing initial conditions for the box's full range of frequencies is the loss of control in the entry time of each mode. In the same way as the previous integrals were computed for some $\eta_0$ chosen for given momenta, the entry of a mode in a simulation must be relative to the mode. If this is not done, each mode is not treated equally in its evolution, thus breaking quasi-scale invariance. This is why stochastic inflation is well-motivated and necessary.

Stochastic Inflation (SI) was introduced in  \cite{Starobinsky_stochastic_1988}, and used as a non-perturbative methodology for light scalars in de Sitter in \cite{starobinsky_equilibrium_1994}. A more thorough introduction to the formalism and its long tradition is given in Chapter~\ref{sec:SIGR}; here we only recall the elements needed for the present discussion.
In stochastic inflation, each  mode that passes a classicality criterion enters the IR system as a stochastic kick \cite{Starobinsky_stochastic_1988, salopek_stochastic_1991,pinol_manifestly_2021, Launay24}. This criterion is usually that the modes be super-Hubble. Linking this formalism with the previous diagrammatics is possible as evoked in \cite{vanderMeulen2007, Garbrecht:2013coa, Garbrecht:2014dca}.

The formalism has undergone various treatments with different underlying assumptions. It is, however, likely that the most rigorous approach extracts the IR sector using a Wilsonian effective field theory, integrating out the UV and the UV-IR mixings to yield an IR action and its path integral. Performing that operation in our framework would mostly entail the appearance of an IR quadratic term $({\cal Q}_q^>)^2$, schematically\ylg{ for the path integral functional}
\begin{equation}
    Z \propto \int {\cal D Q}_{cl}^>{\cal D Q}_{q}^>\exp{\left (\displaystyle -i\int d^4 x {\cal Q}_{q}^>\frac{\delta S}{\delta {\cal Q}}[{\cal Q}_{cl}^>] -\int d^4 x d^4 x' {\cal Q}_q^>(x)D(x,x'){\cal Q}_q^>(x')\right )},
    \label{eq:generalSI}
\end{equation}
where ${\cal Q}$ is a non-perturbative gauge-invariant quantity\footnote{Examples of such quantities can be found in \cite{rigopoulos_separate_2003,langlois_evolution_2005} and are vector fields instead. Note that one would need to deal with removing gauge redundancies in the path integral but we chose to stay simple.} linearising as ${\cal R}$, or a non-perturbative scalar in a given gauge as in \cite{pinol_manifestly_2021}.\footnote{This partition function is fully non-perturbative.} The amplitude $D(x)$ comes from the UV dynamics and the coarse-graining procedure, for which perturbation theory can usually be used. Without any Wilsonian UV integration or specific initial conditions, there is to our knowledge no way to get such a vertex because of the asymmetry in a raw Keldysh basis change such as in Section \ref{sec:OnOffShell}. 

The derivation of eqn.~\eqref{eq:generalSI} has been performed in \cite{morikawa_dissipation_1990,moss_effective_2017,pinol_manifestly_2021} with different notations and typically implies the transfer of initial conditions as implicit boundary conditions to their propagated linear contributions once classical, which are the super-Hubble ones. Eqn.~\eqref{eq:generalSI} can be written via the Hubbard-Stratonovich trick \cite{Stratonovich57,Hubbard59} as
\begin{equation}
\resizebox{\textwidth}{!}{%
$\begin{aligned}
     Z & \propto \int{\cal D }\xi\exp{\left(-\frac{1}{2} \int d^4 x d^4 x' \xi(x) D^{-1}(x,x')\xi(x')\right)}\int {\cal D Q}_{cl}^>{\cal D Q}_{q}^>\exp{\left (\displaystyle -i\int d^4 x {\cal Q}_{q}^>(\frac{\delta S}{\delta {\cal Q}}[{\cal Q}_{cl}^>]-\xi)\right )}\\
     & \propto \int{\cal D }\xi\exp{\left(-\frac{1}{2} \int d^4 x d^4 x' \xi(x) D^{-1}(x,x')\xi(x')\right)}\int {\cal D Q}_{cl}^>\delta\left[\frac{\delta S}{\delta {\cal Q}}[{\cal Q}_{cl}^>]-\xi\right],
\end{aligned}$
}
    \label{eq:HStrick}
\end{equation}
which leads to a similar stochastic equation as for the CSA but with contributions at all times. This feature is visible because of the non-zero  stochastic kernel of the free theory $F^{SI}(x_1,x_2) = \langle {\cal Q}_{cl}^{>,(0)}(x_1){\cal Q}_{cl}^{>,(0)}(x_2)\rangle$.
It can be shown \cite{morikawa_dissipation_1990,moss_effective_2017, pinol_manifestly_2021} that $D$ relates to the full quantum free theory kernel 
\begin{equation}
    D(k,t,t^\prime) = {\cal D}_{k,t}{\cal D}_{k,t^\prime}\omega_k(t)\omega_k(t^\prime)F_{\cal Q}(k,t,t^\prime),
    \label{eq:eqDSI}
\end{equation}
where $\omega_k$ is a time-dependent window function selecting only modes smaller than a certain $k_\sigma(t)=\sigma a(t)H(t)$, ${\cal D}$ is the free operator, and $F_{\cal Q}(k,t,t^\prime) = \langle {\cal Q}_{cl}^{(0)}(\mathbf{k},t){\cal Q}_{cl}^{(0)}(-\mathbf{k},t^\prime)\rangle$. Focusing now on a perturbative framework for the IR too, $F^{SI}$ can be calculated as
\begin{equation}
    F^{SI}(x,y)=\int d^4x_2d^4x_3d^4y_2d^4y_3 G(x_2,x)G(y_2,y){\cal D}_{x_2}{\cal D}_{y_2}W(x_2,x_3)*F_{\cal Q}(x_3,y_3)*W(y_2,y_3),
    \label{eq:eqFSIdef}
\end{equation}
using the convolution operator $*$ in real space. By using integration by parts and the definition of the Green's function, this implies that $F^{SI}$ is related to $F$ through
\begin{equation}
    F^{SI}(k,t,t') = \omega_k(t)\omega_k(t')F_{\cal Q}(k,t,t').
    \label{eq:eqFSI}
\end{equation}
This is basically all one needs to build a theory of stochastic inflation. From that, we can calculate the bispectrum using the Keldysh stochastic rules introduced previously but using this new $F^{SI}$. In particular, in the case where the window is a Heaviside function $\omega_k(t)= \Theta(k_\sigma-k)$, it is completely equivalent to computing late contributions ${\cal C}_{late}^{cl}$ as in section \ref{sec:QCl_timing} where $\eta_0$ is the time at which the mode with greatest $k$ is equal to $k_\sigma$, which is around horizon crossing. In our language, stochastic inflation is the on-shell trajectory of fluctuations which have crossed the horizon.

Consequently, this also means that the previous ratios tell us how accurate stochastic inflation is with respect to the bispectrum by assuming its classicality, as long as the scales still behave perturbatively. The further away from horizon crossing the better for classicality, but, as we just explained, it doesn’t mean that being close to the horizon is not under control. In particular, key interactions happen around horizon crossing and are usually neglected in the usual stochastic treatments which require super-Hubble cutoffs to invoke the separate universe approximation. Furthermore, it was recently pointed out that associated effects, notably non-negligible gradients, can be crucial beyond single phase slow-roll inflation \cite{Figueroa22,cruces_stochastic_2022,jackson2023,Launay24}. At least in a slow-roll phase, the above computations have shown that classicality is not the reason one should invoke to justify a super-Hubble cutoff, but rather a will for analytical simplicity.

We have made explicit which part of perturbation theory stochastic inflation can reproduce. In particular, as found in \cite{vennin_correlation_2015}, the effectively very classical squeezed limit of \cite{Maldacena_2003} is indeed reproduced by stochastic inflation in a perturbative scenario. 

Despite requiring a perturbative entry of UV modes, stochastic inflation is mostly used to probe non-perturbative IR scenarios with no need for quantum gravity (such as ultra slow-roll, secular growth and non-perturbative IR couplings). In that case, there is even less point in waiting for the decay of quantum interactions to let modes enter the IR system: non-perturbative growth will likely enhance the dominating linear theory and consequently non-perturbative $n$-spectra will follow and dominate any non-Gaussian initial conditions with very little influence from perturbative interactions which happened before crossing. This seems to have only been studied in de Sitter space so far \cite{cohen_stochastic_2021}.

\section*{Summary} \label{sec:QClconclusion}

In this chapter we examined when the bispectrum can be derived from classical equations of motion as a proxy for classicality. As highlighted early on, there is a rich but incomplete literature on the subject. In particular, confusion arises quickly when it comes to defining what ‘quantum’ means in cosmology. Following earlier work \cite{vanderMeulen2007, Garbrecht:2013coa, Garbrecht:2014dca, Salcedo24}, we applied the field theory tools built by condensed matter and statistical physics communities to the case of an inflating universe with fluctuations that originated from a quantum vacuum.

In doing so, we clarified the identification of the classical (on-shell) and perturbative evolution of this noise through stochastic PDEs and their diagrammatics. For the convenience of cosmologists, we also adapted the usual \textit{in-in} techniques for this matter. In particular, we derived analytically the classical and purely quantum full slow-roll inflating GR contributions to the three-point correlation function in Fourier space for the first time. These expressions allowed us to imagine what a classical evolution would look like for the bispectrum if performed from the far past, noting the important insight that the consistency relations seem to be due to the on-shell evolution only.

A purely classical evolution of a Bunch-Davies vacuum from the far past is of course unrealistic if the universe is as quantum as QFTCS claims. However, truncating this formalism allowed us to focus on times where the quantum contributions are known to decay and where it becomes realistic to neglect them.
Comparing the Keldysh classical and non-classical interactions in what we defined as quantum interactivity, we managed to show that common interactions are safe to work with classically even before horizon crossing, while more exotic terms, such as those arising from non-canonical sound speeds, can be treated classically slightly after horizon crossing. 
This is the first quantitative study of its kind, as past works focused on general arguments to justify that classicality was roughly and safely reached at late times, rather than the detailed nature of this process. The bispectrum is the lowest-order interactive statistic to compute\yld{ but}\ylc{, and we expect, without having shown it, that} these conclusions can\yld{ surely} be extended to higher $n$-point correlation functions, despite being at higher orders in the coupling’s perturbation theory. Similarly, going beyond the cubic interaction term in GR is also higher order in perturbation theory. Focusing here on the bispectrum is thus a good leading-order probe of the quantum interactivity and leads to quantitative results beyond free theory and without requiring loop calculations.

To tackle questions which motivated the need for this study, we have provided insights for numerical purposes and for empirical measurements. If one wants to simulate an inflating patch of universe and look at correlation functions, it is clear that the analytical comparison at first order lies in the truncation of the \textit{in-in} integrals. This is valid for both evolutions of either an initial range of modes or the equivalent open systems of these modes crossing the horizon sequentially. The latter, called stochastic inflation, is mostly used for investigating the non-perturbative growth of perturbations. Our work mostly focussed on early classicality during perturbative phases, thus obviating the need to go to super-Hubble where the separate universe approximation ultimately struggles to be accurate anyway.

Finally, this chapter has produced clarified methods and new calculations which give a better understanding of statistical signals. The shapes from classical evolution of inflation are now known \ylc{for the slow-roll bispectrum in GR} and can be calculated for many other models.
We hope to do so in the near future using the Keldysh decomposition in numerical estimates of the in-in formalism \cite{clarke_probing_2021,werth23,LRSZ}, so that the simulation community can have easy-to-use tools to probe classicality and quantum interactivity in their models.

%% file: chap_ICNR.tex
\ifSubfilesClassLoaded{\setcounter{chapter}{1}}{}
\chapter{Bridging Cosmology and Numerical Relativity}\label{sec:ICNR}

Within the inflationary paradigm, analytical predictions for different models and, more recently, wide classes of models have been calculated for higher-order correlators over the past two decades \cite{Maldacena_2003, chen_primordial_2010, Arkani-Hamed18,wang_bootstrapping_2022, DuasoPueyo23}.
They all share the key restriction of perturbativity. Model-independent techniques additionally rely on simplifications afforded by assumed symmetries.
Adding to these limitations, loop computations in cosmologically relevant spacetimes are far less comprehensive than those of flat spacetime due to the analytical challenges they pose \cite{Senatore10,Assassi12}.

To move beyond analytically tractable models and symmetries, numerical methods must be employed.
The leading Fourier amplitude of quantum fluctuations can be obtained from the Mukhanov-Sasaki equations for the perturbations \cite{Mukhanov1981,Sasaki86} given (numerical) solutions of the background Friedmann equations for inflation. 
For the computation of higher-order correlation functions of quantum cosmological perturbations several methods have been developed, among which are  the \textit{in-in} formalism computations \cite{ Maldacena_2003, chen_large_2007,clarke_probing_2021} together with transport equations \cite{mulryne_pytransport_2017, werth23}. So far, these have been implemented numerically mostly up to the bispectrum and for cubic-order interactions, though in theory they are not limited to these. However, as mentioned above, the absence of a general treatment of quantum loops on a curved spacetime implies that none of the mentioned numerical techniques seem to tackle generic loop computations, despite the existence of some analytical examples and methods \cite{Senatore10, Assassi12, lee_leading_23, Iacconi24}. In principle, some hope can be found using flow equations up to a certain order
\cite{dias_numerical_2016-1,werth23}, but the computational resources needed would increase drastically together with the difficulty of the task. 

However, for many interesting and realistic models and as hinted in Chapter \ref{sec:semiclQFT}, quantum effects are suppressed in the long-wavelength limit when a mode crosses the Hubble radius \cite{Weinberg05,Weinberg06, burgess_minimal_2022}.
This is noteworthy because it thereby becomes feasible to study the evolution of the relativistic and classical equations of motion alongside non-perturbative scenarios.
In general, these equations can fully account for classical non-linearity but also for all orders in perturbation theory up to numerical precision, the studied range of scales, and simplifying assumptions.
From this context emerged the separate universe approximation \cite{wands_new_2000, salopek_nonlinear_1990, rigopoulos_separate_2003, Rigopoulos_multi06} and
the $\delta N$ approach \cite{wands_new_2000, Lyth03}, but also the stochastic inflation community from extensive analytical use of the super-Hubble limit \cite{Starobinsky_stochastic_1988, morikawa_dissipation_1990, salopek_stochastic_1991, vennin_correlation_2015} together with multiple recent codes \cite{Jackson22,cruces_stochastic_2022,tomberg_numerical_2023,stolas,Animali25}.
Recent advances \cite{Jackson22, cruces_stochastic_2022, Launay24} have demonstrated that this limit and the accompanying analytical simplifications do not, however, describe all scenarios because the long-wavelength approximation neglects gradient terms and associated key phenomena.

Cosmological computations on a lattice initialised with the Bunch-Davies vacuum have recently emerged as an approach that can include spatial gradients \cite{latticeeasy,Caravano22thesis,Cosmolattice},
which in turn neglects the spacetime geometry. 
Recent work \cite{Caravano2024} presented the first non-perturbative non-separate universe simulations of a field from a lattice initialised with full Gaussian initial conditions.
Although these recent advances incorporate non-perturbative global backreaction regimes or local perturbative ones \cite{caravano2024axionbackreaction,caravano2024usr}, the full geometric dynamics and associated subtleties (full backreaction, coordinate choice, etc.) are not accessible and thus keep the testable inflationary landscape rather narrow.

In this context, Numerical Relativity (NR), or full General Relativity on a lattice, appears as the minimal extension to include full geometric dynamics. 
It is worth noting that inhomogeneous inflationary spacetimes have already been successfully simulated using such technologies, building on the expertise and challenges encountered in black hole simulations - see e.g.~\cite{aurrekoetxea24} for a review of cosmology using NR.
Although the focus has been that of post-inflationary evolution and preheating dynamics, inflationary spacetimes have also been investigated, focusing on the study of the robustness of inflation given generic single-mode initial conditions \cite{East16,clough_robustness_2017,Bloomfield19,joana_inhomogeneous_2021, aurrekoetxea_effects_2020, Corman23,Elley2024}, as opposed to the full spectrum of the Bunch-Davies vacuum initial conditions.

As in other physical setups, the primary difficulty in employing NR for cosmology has been the construction of initial conditions which must satisfy the highly non-linear constraints of general relativity.
Starting from simplified analytical solutions \cite{Clough_2015}, the NR community later made use of numerical solvers for the constraints to provide initial inhomogeneities in the inflaton and its momentum together with a 3D geometry \cite{East12,aurrekoetxea_cttk_2022}. 
Because of its assumptions such as conformal flatness and associated caveats in the construction of physical solutions \cite{Baumgarte25}, these methods might only be reliable if the exact form of the perturbations does not matter \cite{aurrekoetxea24}.
Previous applications of these solvers include the study of inflationary exit scenarios, such as recent preheating studies \cite{giblin_jr_preheating_2019,aurrekoetxea_oscillon_2023}.

Beyond exploring the more exotic scenarios of inflation and deepening our understanding of inflationary spacetimes, the next step for these simulations seems to be moving from qualitative and convenient simplifications to quantitative analysis that provides high-precision predictions for direct comparison against observation, including present and future CMB and large-scale structure surveys.
To achieve this, NR still needs to start from the exact initial conditions provided by QFTCS, such as those sourced by the Bunch-Davies vacuum, which involve a complete spectrum. As emphasised in Chapter~\ref{sec:semiclQFT}, this needs to be done in the semiclassical portion of spacetime. 
By doing so, we can evaluate and move beyond previous studies, accounting for all non-linearities, non-perturbative effects, and a broader range of models, as we will explore here.

Reporting on some of the results found in \cite{Launay24, Launay2025}, this Chapter presents and exploits analytical foundations of cosmological perturbation theory (CPT) and the 3+1 decomposition of general relativity.
We start by introducing the latter in Section~\ref{sec:initValprob}, leading to the formulation of the initial value problem in general relativity, central to the initialisation of a numerical relativity simulation.
Section~\ref{sec:ICNRcpt} then introduces the formalism of linear cosmological perturbation theory, whether in a gauge-dependent or gauge-invariant manner.
It provides \ylc{analytical bridges between gauges of interest at linear order}, and notably the construction of initial conditions for inhomogeneous inflationary spacetimes in NR-friendly gauges directly from the Fourier statistics of the gauge-invariant curvature perturbation on comoving hypersurfaces, more familiar to cosmologists.
These constitute\ylc{, to our knowledge,} the first analytical constructions of inhomogeneous inflationary NR initial conditions, consistent with standard CPT\ylc{ at first order}.
This paves the way for simulations with fluctuations following the classicalised Bunch-Davies vacuum, carried out in the following Chapter~\ref{sec:ICsim}.
\ylc{Section~\ref{sec:initValprob} (ADM and BSSN formulations, initial value problem) is standard textbook material (see e.g.\ \cite{baumgarte_shapiro_2010}), recalled for cosmologists in our own notations, while Sections~\ref{sec:secIIB} and \ref{sec:secIIIA} review the linearised ADM equations, with the notations of \cite{PaulADM}, and the standard gauge-invariant variables and Mukhanov-Sasaki equation, collecting the relations expressing them in terms of $\cR$.
The new results \cite{Launay24, Launay2025} are those of Section~\ref{sec:INCRgauges}, which expresses all scalar perturbations in terms of $\cR$ for families of gauges generalising the usual ones (pure metric, fixed curvature, Newtonian, uniform e-folds and generalised synchronous gauges), including the residual gauge freedom of the synchronous gauge (Appendix~\ref{app:dofsyncgauge}), and of Section~\ref{sec:ICNRbssninitdata}, which translates these into a dictionary for BSSN initial data (Appendix~\ref{app:dicBSSN}).
Finally, Section~\ref{sec:classicaldraw} adapts the standard drawing of Gaussian random fields with a given spectrum to the classical limit of Chapter~\ref{sec:semiclQFT}, with a single stochastic grid for all fields.}

\section{Formulating the numerical problem in relativity}\label{sec:initValprob}
In this section, we introduce useful notations while recalling some basics of general relativity not commonly emphasised in a cosmology context, pointing out where the difficulty arises in solving\ylc{. This section is a review of standard material, see e.g.\ \cite{baumgarte_shapiro_2010}}. An expert reader might wish to proceed to the next section.

\subsection{ADM formulation of Einstein's equations}\label{sec:ICNRadm}

For a 4-dimensional metric tensor with components $g_{\mu\nu}$ describing a spacetime with manifold $\cal M$ and Ricci scalar $R$, filled with 'matter' described by a Lagrangian density ${\cal L}_m$, the action is written as
\begin{equation}
S =  \frac{\Mpl^2}{2} \int_{\cal M} d^4 x \sqrt{- g} \, R + \int_{\cal M} d^4 x \, {\cal L}_m + S_{\text{GHY}},
\label{eq:GRaction}
\end{equation}
where the Einstein-Hilbert action has been supplemented with the Gibbons-Hawking-York boundary term $S_{\text{GHY}}$ \cite{GHY_Y,GHY_GH}, an integral over the boundary of the manifold $\partial {\cal M}$.
Unless specified otherwise, this term will not be relevant in the following as we will be working with the equations of motion, which discard such a term built with a total derivative.

When varied with respect to the metric $g_{\mu\nu}$ and applying the least action principle, Einstein's equations are recalled from eqn.~\eqref{eq:Einstein}
\begin{equation}
    \frac{2}{\Mpl^2}\frac{\delta S}{\delta g^{\mu\nu}}= G_{\mu\nu}-\frac{1}{\Mpl^2}\, T_{\mu\nu} = 0,
\end{equation}
where $T_{\mu\nu} = -\frac{2}{\sqrt{-g}}\frac{\delta {\cal L}_m}{\delta g^{\mu\nu}}$ such that $\nabla_{\mu}T^{\mu\nu} = 0$ by construction of the Einstein tensor $G_{\mu\nu}= R_{\mu\nu}-\frac{1}{2}g_{\mu\nu}R$ satisfying the Bianchi identities.
In this work and unless specified otherwise, we will study a single $\phi$ scalar field fluid described by
\begin{equation}
    {\cal L}_m = -\sqrt{-g}\left(\frac{1}{2}\, g^{\mu\nu} \nabla_{\mu}\phi\nabla_{\nu}\phi+ V(\phi)\right).
\end{equation}
Einstein's field equations are not suited for numerical simulations as they are under-determined.
On top of the $10$ equations, $4$ equations constrain the evolution via the Bianchi identity, and, most importantly,
a system of coordinates needs to be chosen to fix the gauge freedom of general relativity, leaving only $6$ independent equations for the $6$ independent components of the metric tensor.
Until a system of coordinates is chosen, the form of the PDEs is simply unknown, and so is the well-posedness of the system.
Moreover, a modern simulation of a spacetime requires a foliation of the manifold into spatial hypersurfaces, which is not guaranteed to exist in general.

In this thesis, we look at inflationary spacetimes admitting a global time function $t: {\cal M} \to \mathbb{R}$,
such that the normal to surfaces of constant $t$,  $n^\mu = -\frac{\partial^\mu t}{\sqrt{-g^{\alpha\beta}\partial_\alpha t \partial_\beta t}}$ (here normalised to $n_\mu n^\mu = -1$), is a timelike and future-directed vector field.
We additionally require the existence of a Cauchy surface, a surface with no closed causal curves through which the entire manifold can be reached and mapped via causal curves.
This latter condition defines a \textit{globally hyperbolic} spacetime \cite{reall_blackholes}, which allows us to perform a 3+1 decomposition of the coordinates, where the field $t$ can be picked constant on each of the hypersurfaces, so that the foliation of spacelike hypersurfaces is written as levels of $t$, $\Sigma(t) = \{p \in {\cal M} | t(p) = t\}$.
The spatial coordinates $x^i$ of a given point $p\in {\cal M}$ can then be defined as the ones chosen on the initial hypersurface $\Sigma(t=t_0)$ at the point intersecting the integral curve of $n^\mu$ passing through $p$.
This is known as the Arnowitt-Deser-Misner (ADM) formalism \cite{ADM}, which leads to the parameterisation of $n$ and the metric%
\begin{subequations}
    \begin{align}
    n_\mu  = (-\alpha,0,0,0), \qquad
    n^\mu  = \left(\frac{1}{\alpha},\frac{\beta^i}{\alpha}\right), \\
    d s^2=-\alpha^2 d t^2+\gamma_{i j}\left(d x^i-\beta^i d t\right)\left(d x^j -\beta^j d t\right),     \label{eq:ADMmetric}
    \end{align}
\end{subequations}
where $\alpha(t,x^k)$ is the \textit{lapse} function, $\beta^i(t,x^k)$ is the \textit{shift} vector, and $\gamma_{ij}=g_{ij}$ is the induced metric on the 3-dimensional spatial hypersurfaces $\Sigma(t)$.
With this parametrisation, we focus on a wide subset of gauges parameterised by the lapse and shift (or $n$), which are not dynamical and can be chosen freely to fix the gauge freedom of general relativity.
The lapse and shift indeed specify which causal curves (and therefore observers, within the light cone) are followed to move from one hypersurface to the next, and thus how the (spatial) coordinates are defined on each hypersurface.
The lapse measures the proper time between two hypersurfaces, while the shift measures the spatial displacement of a point on a hypersurface to the next one along the integral curve of $n^\mu$.

The action in eqn.~\eqref{eq:GRaction} can be recast to reflect this $3+1$ splitting.
This is done by using the Gauss relation between the Ricci scalar $R$ in terms of the 3-dimensional Ricci scalar ${}^3R$ of the spatial metric $\gamma_{ij}$, which, up to implicit boundary terms added to the GHY one, yields
\begin{equation}
S =  \frac{\Mpl^2}{2} \int d t d^3 x \, \alpha \sqrt{\gamma}\Big({ }^{3} R-K^2+K_{i j} K^{i j}\Big)
 +\int d t d^3 x \, \alpha \sqrt{\gamma}\Big(\frac{1}{2} \Pi^2-\frac{1}{2} \partial_i \phi \partial^i \phi-V(\phi)\Big),    \label{eq:ADMaction}
\end{equation}
where we have defined $\Pi$ as the ADM conjugate momentum of the field
\begin{equation}
\Pi= n^\mu \partial_\mu \phi=\frac{1}{\alpha}\left(\dot{\phi}+\beta^i \partial_i \phi\right),
    \label{eq:ADMphi}
\end{equation}
and $K$ is the extrinsic curvature
\begin{equation}
K_{ij} \equiv n_{i|j} \equiv \tilde{K}_{ij}+\frac{1}{3}\gamma_{ij}K = -\frac{1}{2\alpha}(\gamma_{i j,0}+\beta_{i|j}+\beta_{j|i}),
    \label{eq:ADMgamma}
\end{equation}
which is usually split into its trace and traceless components $K$ and $\tilde{K}_{ij}$ respectively, and where 
we adopt the vertical bar notation $()_{|i}$ for a covariant derivative component on the 3-space submanifold.
Writing the extrinsic curvature as a Lie derivative, $K_{ab} = -\frac{1}{2}\mathcal{L}_n \gamma_{ab}$, one sees that it tracks how fast the induced spatial metric changes as it is dragged along the direction normal to the spatial hypersurfaces making the foliation.

Variation of  eqn.~\eqref{eq:ADMaction} with respect to $\phi$  yields the  field evolution equation
\begin{equation}
    \frac{1}{\alpha}\left(\dot{\Pi}+\beta^i \Pi_{\mid i}\right)-K \Pi-\frac{\alpha^{\mid i}}{\alpha} \phi_{\mid i}-\phi_{\mid i}^{\mid i}+\frac{d V}{d \phi}=0,
    \label{eq:ADMPi}
\end{equation}
while variation with respect to $\gamma_{i j}$  yields the gravitational field's evolution equations%
\begin{subequations}
\begin{align}
 \dot{K}+\beta^i K_{, i}+\alpha^{\mid i}{ }_{\mid i}-\alpha\left({ }^{3} R+K^2\right)
 - \Mpl^{-2} \alpha\left(\frac{1}{2} S-\frac{3}{2} \rho\right) & = 0, \label{eq:ADMKeq}\\
  \dot{\tilde{K}}_{ij}+2\alpha\tilde{K}_{il}{\tilde{K}}^l{ }_j+\beta^k \tilde{K}_{ij \mid k}-2\beta_i{ }^{\mid k} \tilde{K}_{jk}+ \alpha_{\mid i\mid j} & \notag\\
  -\frac{1}{3} \alpha^{\mid k}{ }_{\mid k} \delta_{ij} 
-\alpha\left({ }^3\tilde{R}_{ij}+\frac{1}{3}K \tilde{K}_{ij}\right) + \Mpl^{-2} \alpha \tilde{S}_{ij}& =0, \label{eq:ADMKtildeeq}
\end{align}
    \label{eq:ADMK}
\end{subequations}
where the $3+1$ components of the energy-momentum tensor are the energy and momentum densities%
\begin{subequations}
\begin{align}
\rho & \equiv n_\mu n_\mu T^{\mu v}=\alpha^2 T^{00}=\frac{1}{2} \Pi^2+\frac{1}{2} \partial_i \phi \partial^i \phi+V(\phi), \label{eq:ADMrho}\\
{\cal { J}}_i & \equiv -n_{\mu}P_{ \nu i} T_{\mu \nu} = \alpha T^{0}{}_i = -\Pi \partial_i\phi, \label{eq:ADMJi}
\end{align}
\end{subequations}
and the stress tensor $S_{i j}\equiv P_{ \mu i}P_{ \nu j} T^{\mu \nu} = T_{ij}$, decomposed as%
\begin{subequations}
    \begin{align}
        S & \equiv{ }^{3} g^{i j} S_{i j}=\frac{3}{2} \Pi^2-\frac{1}{2} \partial_k \phi \partial^k \phi-3 V(\phi), \label{eq:ADMSscalar} \\
\tilde{S}_{i j}& \equiv S_{ij}-\frac{1}{3}{ }^{3} g^{i j}S =  \frac{1}{2}\left(\partial_i \phi \partial_j \phi-\frac{1}{3} \partial_k \phi \partial^k \phi \,\delta_{i j}\right) . \label{eq:ADMStilde}
    \end{align}
        \label{eq:ADMS}
\end{subequations}
In these expressions, $n$ plays the role of the projector normal to the hypersurfaces, while $P_{\mu\nu} = g_{\mu\nu}+n_\mu n_\nu$ is the projector onto the hypersurfaces.

Variation with respect to the lapse and shift is left aside as they do not yield evolution equations but rather constraints elaborated upon in Section~\ref{sec:ICNRicproblem}.

\subsection{BSSN formulation}\label{sec:ICNRbssn}
\ylg{When it comes to numerical simulations, the ADM equations, although already first order in time, are only weakly hyperbolic: small violations of the constraints, such as numerical errors, can grow without bound, so that long free evolutions are not well-posed.
The BSSN formulation \cite{BSSN_SN,BSSN_BS} cures this, keeping equations first order in time and second order in space.
It evolves the conformal factor and the trace of the extrinsic curvature separately from the conformal metric and the traceless extrinsic curvature, and, crucially, promotes the contracted conformal connection $\tilde{\Gamma}^i$ defined below to an independent variable.
This is where well-posedness shows up in the equations: in the Ricci tensor of the conformal metric,
\begin{equation*}
    \tilde{R}_{ij} = -\frac{1}{2}\tilde{\gamma}^{kl}\partial_k\partial_l\tilde{\gamma}_{ij} + \tilde{\gamma}_{k(i}\partial_{j)}\tilde{\Gamma}^k + \left(\text{terms with first derivatives of } \tilde{\gamma}_{ij}\right),
\end{equation*}
the mixed second derivatives of the metric are traded for first derivatives of $\tilde{\Gamma}^i$, so that the principal part of the $\tilde{\gamma}_{ij}$--$\tilde{A}_{ij}$ system becomes a wave operator.
Using the evolution equation of $\tilde{\Gamma}^i$, in which the momentum constraint has been added, the resulting system is strongly hyperbolic for suitable gauge choices such as the 1+log slicing and Gamma-driver shift commonly used \cite{baumgarte_shapiro_2010}.
The definition of $\tilde{\Gamma}^i$ then becomes an additional constraint, satisfied by the initial data and monitored during the evolution.
BSSN equations are now the common equations for these purposes.}
In this formalism, the metric is written
\begin{equation}
    d s^2=-\alpha^2 d t^2+\gamma_{i j}\left(d x^i+\beta^i d t\right)\left(d x^j+\beta^j d t\right),
        \label{eq:BSSNmetric}
\end{equation}
where $\alpha$ and $\beta^i$ are the lapse and shift, which is the same as our ADM metric but with $\beta^i \longrightarrow-\beta^i$. The induced metric is decomposed thanks to a conformal factor $X$ as
\begin{equation}
\gamma_{i j}=\frac{1}{X} \tilde{\gamma}_{i j},\qquad \operatorname{det} \tilde{\gamma}_{i j}=1, \qquad X=\left(\operatorname{det} \gamma_{i j}\right)^{-\frac{1}{3}},
    \label{eq:ConformalFactor}
\end{equation}
The extrinsic curvature is decomposed into its trace $K=$ $\gamma^{i j} K_{i j}$, and its conformally rescaled traceless part $\tilde{\gamma}^{i j} \tilde{A}_{i j}=0$ as
\begin{equation}
K_{i j}=\frac{1}{X}\left(\tilde{A}_{i j}+\frac{1}{3} K \tilde{\gamma}_{i j}\right).    
    \label{eq:ConformalK}
\end{equation}
Finally, \ylg{the independent variable discussed above is} the conformal connections defined as  $\tilde{\Gamma}^i=\tilde{\gamma}^{j k} \tilde{\Gamma}_{j k}^i$ where $\tilde{\Gamma}_{j k}^i$ are the Christoffel symbols associated with the conformal metric $\tilde{\gamma}_{i j}$. 
As a summary, NR consists of evolving the 7 quantities $X$, $\tilde{\gamma}_{ij}$, $K$, $\tilde{A}_{ij}$, $\tilde{\Gamma}^i$ and $\phi$, $\Pi$ according to%
\begin{subequations}
\begin{align}
 \partial_t \phi-\alpha \Pi-\beta^i \partial_i \phi & = 0, \label{eq:BSSNphi}\\
 \partial_t \Pi-\beta^i \partial_i \Pi-\alpha \partial_i \partial^i \phi-\partial_i \phi \partial^i \alpha
-\alpha\left(K \Pi-\gamma^{i j} \Gamma_{i j}^k \partial_k \phi-\frac{d V}{d \phi}\right)& = 0,   \label{eq:BSSNPi}\\
 \partial_t X -\frac{2}{3} X \alpha K+\frac{2}{3} X \partial_k \beta^k-\beta^k \partial_k X & = 0,  \label{eq:BSSNX}\\
 \partial_t \tilde{\gamma}_{i j}+2 \alpha \tilde{A}_{i j}-\tilde{\gamma}_{i k} \partial_j \beta^k-\tilde{\gamma}_{j k} \partial_i \beta^k 
 \quad+\frac{2}{3} \tilde{\gamma}_{i j} \partial_k \beta^k-\beta^k \partial_k \tilde{\gamma}_{i j} & =0,  \label{eq:BSSNgammatilde}\\
  \partial_t K+\yld{\gamma^{i j} D_i D_j \alpha}\ylg{\alpha^{\mid i}{ }_{\mid i}}-\alpha\left(\tilde{A}_{i j} \yld{A^{i j}}\ylg{\tilde{A}^{i j}}+\frac{1}{3} K^2\right)
-\beta^i \partial_i K-\yld{4 \pi}\ylg{\frac{1}{2}\Mpl^{-2}} \alpha(\rho+S) & = 0,  \label{eq:BSSNK}\\
 \partial_t \tilde{A}_{i j}-X\left[-\yld{D_i D_j \alpha}\ylg{\alpha_{\mid i\mid j}}+\alpha\left(R_{i j}- \Mpl^{-2}\yld{\alpha S_{i j}}\ylg{S_{i j}}\right)\right]^{{TF}} 
-\alpha\left(K \tilde{A}_{i j}-2 \tilde{A}_{i l} \tilde{A}_j^l\right) & \notag \\
-\tilde{A}_{i k} \partial_j \beta^k-\tilde{A}_{j k} \partial_i \beta^k 
+\frac{2}{3} \tilde{A}_{i j} \partial_k \beta^k-\beta^k \partial_k \tilde{A}_{i j} & = 0, \label{eq:BSSNAtilde}\\
 \partial_t \tilde{\Gamma}^i-2 \alpha\left(\tilde{\Gamma}_{j k}^i \ylg{\tilde{A}^{j k}}-\frac{2}{3} \tilde{\gamma}^{i j} \partial_j K-\frac{3}{2} \ylg{\tilde{A}^{i j}} \frac{\partial_j X}{X}\right) 
+2 \ylg{\tilde{A}^{i j}} \partial_j \alpha-\beta^k \partial_k \tilde{\Gamma}^i & \notag \\
-\tilde{\gamma}^{j k} \partial_j \partial_k \beta^i-\frac{1}{3} \tilde{\gamma}^{i j} \partial_j \partial_k \beta^k 
-\frac{2}{3} \tilde{\Gamma}^i \partial_k \beta^k+\tilde{\Gamma}^k \partial_k \beta^i+ 2 \Mpl^{-2}\alpha \tilde{\gamma}^{i j} {\cal J}_j & = 0, \label{eq:BSSNGammatilde} 
    \end{align}
        \label{eq:BSSN}
\end{subequations}
\ylg{Here, $()_{\mid i}$ is again the covariant derivative associated with the induced metric $\gamma_{ij}$, $R_{ij}$ and $\Gamma^k_{ij}$ are the Ricci tensor and Christoffel symbols of $\gamma_{ij}$, indices of $\tilde{A}_{ij}$ are raised with the conformal metric $\tilde{\gamma}^{ij}$, and $[\,]^{TF}$ denotes the trace-free part with respect to $\gamma_{ij}$.}

In this thesis, whenever these equations are solved starting from initial data, we will make use of the latter formulation, implemented in the various solvers of the \textsc{GRTL} collaboration, including \grc and most recently \grt, details of which will be given in the following chapters.

\subsection{The initial condition problem}\label{sec:ICNRicproblem}

Given these systems of equations, evolution could be started from initial data for the aforementioned tensors and scalars on a spatial hypersurface $\Sigma$. 
However, we have seen in Section~\ref{sec:ICNRadm} that a lapse and shift need to be specified, as each or most of the previous quantities {involve a $3+1$ split of tensor components} and so are basis-dependent objects. 
For that reason, the coordinates need to be specified by four algebraic constraints in a four-dimensional spacetime. 
As an example, one could impose the {reference frame} of a stationary observer with respect to a source. 
Writing the evolution equations in the ADM form \eqref{eq:ADMPi}, \eqref{eq:ADMK} is more convenient in such scenarios as the non-dynamical lapse $\alpha$ and the shift $\beta^i$ can be evolved to impose the coordinate system we want. 

In any case, following the fundamental principle of general covariance, the physical result needs to be independent of the coordinate system. Defining the \textit{Hamiltonian and Momentum constraint densities} in both ADM and BSSN formalism%
\begin{subequations}
\begin{align}
\mathcal{H} &\equiv { }^{3}R+\frac{2}{3} K^2-\tilde{K}_{i j} \tilde{K}^{i j}-2 \Mpl^{-2} \rho = { }^{3}R+\frac{2}{3}K^2-\tilde{A}_{i j} \tilde{A}^{i j}- 2 \Mpl^{-2}\rho ,  \label{eq:Ham}\\
 \mathcal{M}_i &\equiv  \tilde{K}^j{}_{i \mid j}-\frac{2}{3} K_{\mid i}- \Mpl^{-2} {\cal { J}}_i =\tilde{\gamma}^{k l}\left(\partial_k \tilde{A}_{l i}-2 \tilde{\Gamma}_{l(i}^m \tilde{A}_{k) m}-3 \tilde{A}_{i k} \frac{\partial_l X}{2 X}\right) 
        -\frac{2}{3} K_{\mid i}- \Mpl^{-2}{\cal J}_i,   \label{eq:Mom}
\end{align}\label{eq:constraints}
\end{subequations}
variation with respect to $\alpha$ and $\beta^i$ yields the Hamiltonian and Momentum constraints $\mathcal{H}=0$ and $\mathcal{M}_i=0$ respectively, which are the four algebraic constraints on the gauge freedom.\footnote{It is also possible to write the ADM Hamiltonian \ylg{density }to see how shift and lapse are simply Lagrange multipliers for the dynamics.}
Indeed, if these are satisfied, any coordinate change will leave the action invariant, and, more precisely, ${\cal H}$ and ${\cal M}_i$ are the generators of the time and spatial reparametrisations respectively.
An appropriate physical interpretation of the constraints is that they ensure the balance between the spacetime curvature's and its contents' energy and momentum densities. 
In addition, finding initial data $K,h$ on a spacelike hypersurface $\Sigma$ that satisfies the constraints is beneficial for global hyperbolicity.
In fact, such data determine a unique maximal globally hyperbolic development \cite{ChoquetBruhatGeroch},
which in turn admits a foliation by spacelike Cauchy hypersurfaces \cite{BernalSanchez2003}, as opposed to assuming such a setup as we did earlier.
This is precisely the structure required for a successful numerical implementation.

These equations are highly non-linear and must be satisfied at all times over the entire spatial hypersurface.
Both analytical and numerical solutions usually require assumptions on the fields' symmetries for feasibility \cite{baumgarte_shapiro_2010}, especially for generic distributions of the spacetime's energy-momentum content. 
For these reasons, it is often said that the real challenge of GR is the construction of initial data, whether these are physical or not yet.
Once initially satisfied, the constraints are fully preserved by the evolution equations, although this property sometimes needs to be enforced in numerical evolution schemes (see constraint damping \cite{CCZ41, CCZ42}), to counter numerical error accumulating at each time step.

\section{Physical cosmological initial data}\label{sec:ICNRcpt}
In this section, we present the first method constructing initial data for inhomogeneous inflationary spacetimes consistent with both standard cosmological perturbation theory and numerical relativity.

\subsection{Scalar linear ADM equations}\label{sec:secIIB}

This section reviews the linearised version of the ADM formalism of scalar perturbation theory without making any gauge choice.\footnote{Notations are those of \cite{PaulADM}.} 
The linear equations presented below match exactly the non-linear left-hand sides (l.h.s.) of eqns. \eqref{eq:ADMphi}---\eqref{eq:ADMS} and \eqref{eq:constraints}. Since the standard Einstein and ADM systems of equations are equivalent up to substitutions and factors, the equations below are also equivalent to the more commonly adopted formulations of cosmological perturbation theory.

Discarding decoupled linear vector and tensor perturbations in eqn.~\eqref{eq:SVTmetricAll}, we incorporate general scalar perturbations to the metric through the line element
\begin{equation}
    d s^2=-\alpha_b^2(1+2 \Psi) d t^2+2 a^2 B_{, i} d t d x^i +a^2\Big((1-2 \Phi) \delta_{i j}+2 E_{, i j}\Big) d x^i d x^j,    
    \label{eq:SVTmetric}
\end{equation}
where the four scalar functions $\Phi, \Psi, B$ and $E$ define the perturbations at a spacetime coordinate $(t, x^i)$, while $a(t)$ and $\alpha_b(t)$ are the background scale factor and lapse function, respectively.
The dictionary from ADM to the first-order scalar decomposition follows from the metrics
\begin{equation}
        \alpha  =\alpha_b(1+\Psi), \qquad
         \beta_i =-a^2 B_{, i}, \qquad 
         \gamma_{i j} =a^2\Big((1-2 \Phi) \delta_{i j}+2 E_{, i j}\Big),
        \label{eq:dictionnarySVT}
\end{equation}
which leads to the linear extrinsic curvature
\begin{equation}
 K^i{}_j =-H \delta_j^i+\frac{1}{3} \kappa \delta_j^i -\left(\partial^i \partial_j-\frac{1}{3} \Delta_b \delta_j^i\right) \chi, \label{eq:linK}
\end{equation}
with background trace relating to the Hubble rate as $H \equiv -\frac{1}{3} K_b = \frac{1}{\alpha_b}\frac{\dot{a}}{a} $, and linear trace and traceless parts respectively given by
\begin{equation}
        \kappa  \equiv 3\left(\frac{\dot{\Phi}}{\alpha_b}+H \Psi\right)-\Delta_b \chi,  \qquad  \chi \equiv-\frac{a^2}{\alpha_b}(B-\dot{E}).
        \label{eq:defkappa}
\end{equation}
Here $\Delta_b=g_b^{ij}\partial_i \partial_j=a^{-2}\nabla^2$ is the background Laplacian. The energy-momentum first-order densities are given by%
\begin{subequations}
\begin{align}
        \delta\rho & = \rho_b+\frac{\dot{\phi_b}}{\alpha_b}\frac{\delta\dot{\phi}}{\alpha_b}+\frac{dV}{d\phi}(\phi_b)\delta\phi-\frac{{\dot{\phi_b}}^2}{\alpha_b^2}\Psi, \label{eq:linrho}\\
        \delta{\cal J}_i & =  -\frac{\dot{\phi_b}}{\alpha_b}\partial_i\delta \phi,\label{eq:linJi}\\
        \delta S_{i j} & = (P_b+\delta P) \delta_{i j}+\left(\partial_i \partial_j-\frac{1}{3} \delta_{i j} \nabla^2\right) \sigma. \label{eq:linSij}
\end{align}
\end{subequations}
In this work, and unless specified otherwise, no anisotropic shear is considered so that $\sigma = 0$. At linear order, this is a gauge-invariant assumption and implies $\delta\tilde{S}_{i j} =0$.
Substituting these expressions into the full ADM equations, eqns.~\eqref{eq:ADMPi}, \eqref{eq:ADMK}, and \eqref{eq:constraints}, yields a set of zeroth and first-order perturbation equations which, importantly, do not yet incorporate a choice of gauge.
Matching the term ordering of the ADM equations, the background terms yield%
\begin{subequations}
\begin{align}
6 H^2 - 2\Mpl^{-2}\left(V(\phi_b)+\frac{1}{2}\frac{\dot{\phi_b}^2}{\alpha_b^2}\right) & =  0, \label{eq:backgroundEqHam}\\
3\Mpl^{-2}V(\phi_b)-9 H^2 - 3\frac{\dot{H}}{\alpha_b} & = 0, \label{eq:backgroundEqK}\\
\frac{\ddot{\phi_b}}{\alpha_b^2}+\left(3H-\frac{\dot{\alpha_b}}{\alpha_b^2}\right)\frac{\dot{\phi_b}}{\alpha_b}+ \frac{dV}{d\phi}(\phi_b)  & = 0,\label{eq:backgrounfEqfield}
\end{align}
    \label{eq:backgroundEq}
\end{subequations}
from the Hamiltonian constraint, the extrinsic curvature, and the field evolution equations respectively, while other equations are trivially satisfied.
These are of course the equations of homogeneous and isotropic FLRW cosmology. At first order, the gravitational evolution equations become%
\begin{subequations}
\begin{align}
        \frac{\dot{\kappa}}{\alpha_b}+2 H \kappa+\left(\Delta_b+3 \frac{\dot{H}}{ \alpha_b} \right)\Psi -\frac{1}{2}\Mpl^{-2}(\delta \rho+3 \delta P) = 0, &\\
         \left(\frac{1}{3} \delta_j^i \nabla^2-\partial_i \partial_j\right)\left(\frac{\dot{\chi}}{ \alpha_b}+H \chi-\Psi+\Phi- \Mpl^{-2} \sigma\right)=0, &
\end{align}
    \label{eq:Kcpt_eq}
\end{subequations}
while for the field equation 
\begin{equation}
    \frac{\delta\ddot{\phi}}{\alpha_b^2}+\left(3 H-\frac{\dot{\alpha_b}} { \alpha_b^2}\right) \frac{\delta \dot{\phi}}{\alpha_b}+\left(\frac{d^2 V}{d \phi^2}(\phi_b) -\Delta_b \right)\delta \phi
    -\frac{\dot{\phi_b}}{\alpha_b}\left(\kappa+\frac{\dot{\Psi}} { \alpha_b}-3 H \Psi\right)+2 \Psi \frac{d V}{d \phi}(\phi_b)=0. 
        \label{eq:fieldCPT}
\end{equation}
Finally, the first-order constraint equations are%
\begin{subequations}
\begin{align}
        4\Delta_b \Phi-4H \kappa-2\Mpl^{-2}\delta \rho = 0, &\\
         -2\partial_i\Big(\Delta_b \chi+\kappa-\frac{3}{2}\Mpl^{-2}\frac{\dot{\phi_b}}{\alpha_b}\delta \phi \Big)= 0. &
\end{align}
    \label{eq:ConstraintsCPT}
\end{subequations}
Hence we have five equations for the five unknowns $\delta \phi, \Psi, \Phi, E$ and $B$. 
In general, we can use the gauge freedom to fix two of these variables. The two constraints then further reduce the dynamical scalar degrees of freedom down to one, usually encapsulated in a variable that is also gauge invariant. 

\subsection{Scalar perturbations, all-in-one}\label{sec:secIIIA}
In the following and unless stated differently, $\alpha_b = 1$ without loss of generality, which implies that $\dot{()}$ is a cosmological time derivative.

Gauge-invariant quantities present many advantages beyond their invariance under coordinate transformations: they naturally reduce the dynamics to the single dynamical scalar degree of freedom and their dynamical equations guarantee that the constraints of General Relativity are automatically satisfied.
Many linear combinations of $\delta \phi, \Psi, E, B$ and $\Phi$ are known to be gauge-invariant at first order, in particular, the Bardeen potentials $\Phi_B$ and $\Psi_B$ \cite{Bardeen1980}, the gauge-invariant field perturbation $\delta\phi_{gi}$, the curvature perturbation on uniform density hypersurfaces $\zeta_{gi}$ or the curvature perturbation on comoving hypersurfaces $\mathcal{R}$. They are given by%
\begin{subequations}
\begin{align}
         \Phi_B & \equiv \Phi + H\chi, \\
         \Psi_B & \equiv \Psi - \dot{\chi},\\
         \delta\phi_{gi} & \equiv \delta\phi-{\dot{\phi_b}}\chi,\\
         \zeta_{gi} & \equiv -\Phi+ \frac{\delta\rho}{3(\rho_b+ P_b)}, \\
         \cR & \equiv \Phi + \frac{H}{\dot{\phi_b}}\delta \phi.
    \end{align}
    \label{eq:GIquantities}
\end{subequations}
With the help of the coordinate transformation of the perturbations given in eqn.~\eqref{eq:gaugetransform}, one can check that these quantities are indeed perfectly invariant.

Those are convenient for the community as they appear to have simple gauge-invariant evolution equations, remembering the Mukhanov-Sasaki (MS) equation for ${\cal R}$, eqn.~\eqref{eq:MSvEq}, or in cosmic time
\begin{equation}
\ddot{\cR} + H\left(3-\varepsilon_2\right)\dot{\cR} +\frac{k^2}{a^2} \cR = 0.
\label{eq:MScosmic}
\end{equation}
Inversely, by knowing the background and a universal description of the\ylg{ scalar} perturbations such as $\cR$, one should be able to write all fields in any gauge, including ones employed in Numerical Relativity, which we will exploit extensively in the next sections.
However, this constitutes a gap in the literature, and we will provide examples in the next subsections.

We use here $\mathcal{R}$ as our master gauge-invariant variable. Before picking a gauge, it is possible to express all other gauge-invariant variables in terms of $\mathcal{R}$ as%
\begin{subequations}
    \begin{align}
         \Phi_{B}  & = -\varepsilon_1 H a^2 k^{-2}\dot{\cR}, \label{eq:PhiB} \\
         \Psi_{B}  & = \varepsilon_1\cR  + \varepsilon_1a^2k^{-2}\Big(\ddot{\cR} +H(2-\varepsilon_2)\dot{\cR} \Big), \label{eq:PsiB} \\
         \delta\phi_{gi}  & =  \sqrt{2\varepsilon_1}\Mpl\Big(\cR +\varepsilon_1a^2 H k^{-2}\dot{\cR} \Big),\label{eq:deltaPhiGI} \\
         \zeta_{gi}  &= -\cR +\frac{1}{3} H^{-1}\dot{\cR}. \label{eq:ZetaGI}
    \end{align}
    \label{eq:GaugeInvExplicit}
\end{subequations}
These relations are most easily obtained in the comoving gauge (see below) but, since they are relations between gauge-invariant variables, they hold in \emph{all} gauges.
Eqn.~\eqref{eq:GaugeInvExplicit} will be more than useful in some of the gauges that we will examine below when the constraint equations will not be directly solvable.
Note also that by using eqn.~\eqref{eq:MScosmic} one easily shows that $\Phi_B=\Psi_B$, which is known to hold for a scalar field as there is no anisotropic stress at linear order.
\yld{We}\ylg{In Chapter~\ref{sec:SIGR}, we} will see the importance of keeping these two variables distinct before using the MS equation\yld{ in Chapter~\ref{sec:SIGR}}.

\subsection{Spacetime foliations using different gauges}\label{sec:INCRgauges}
In this section, we present the decomposition of the perturbation variables in different gauges, all expressed in terms of $\cR$ and the background\ylc{. These decompositions are new and were first given in \cite{Launay24, Launay2025}}.
\subsubsection{Pure metric gauge}\label{sec:INCRcomgauge} A gauge will be called \textit{pure metric} if it sets $\delta\phi$ and $E$ to small arbitrary spacetime functions $\delta\phi^{\rm PM}_G$ and $E^{\rm PM}_G$, their smallness being of the order of other perturbative quantities such as $\cR $.
The null case $\delta\phi^{\rm PM}_G=E^{\rm PM}_G=0$ is commonly used in cosmology and is referred to as the \textit{comoving gauge}.
Such a gauge is among the simplest ones to deduce a whole three-dimensional hypersurface\ylg{ in terms of $\cR$}.
Indeed, $\Phi $ is given directly via the definition of $\cR $ together with the chosen $\delta\phi^{\rm PM}_G$.
Substituting it in the linearised energy and momentum constraints eqn.~\eqref{eq:ConstraintsCPT} yields the result for $\Psi $ and $B $%
\begin{subequations}
\begin{align}
 E^{\rm PM}   & = E^{\rm PM}_G, \label{eq:ComE}\\
     \delta\phi^{\rm PM}    & = \delta\phi^{\rm PM}_G, \label{eq:Comphi}\\
    \Phi^{\rm PM}    & = \cR  -(2 \varepsilon_1\Mpl^2)^{-1/2}\delta\phi^{\rm PM}_G, \label{eq:ComPhi}\\
    \Psi^{\rm PM}  & = (2 \varepsilon_1\Mpl^2)^{-1/2}\Big(\varepsilon_1+\frac{1}{2}\varepsilon_2\Big)\delta\phi^{\rm PM}_G
    -H^{-1}\dot{\cR}  + (2 \varepsilon_1\Mpl^2H^2)^{-1/2}\dot{\delta\phi}^{\rm PM}_G,\label{eq:ComPsi}\\
    B^{\rm PM}  & = \dot{E}^{\rm PM}_G+k^{-2}\varepsilon_1\dot\cR  +(a^2H)^{-1}\Big(\cR  -(2 \varepsilon_1\Mpl^2)^{-1/2}\delta\phi^{\rm PM}_G\Big). \label{eq:ComB}
\end{align}
    \label{eq:PureMetricGaugeDecomp}
\end{subequations}
In particular, the comoving gauge is written%
\begin{subequations}
\begin{align}
 E^{\rm C}   & = 0,\\
     \delta\phi^{\rm C}    & = 0,\\
    \Phi^{\rm C}    & = \cR, \\
    \Psi^{\rm C}  & = -H^{-1}\dot{\cR},  \\
    B^{\rm C}  & = k^{-2}\varepsilon_1\dot\cR  +(a^2H)^{-1}\cR,  
\end{align}
    \label{eq:ComGauge}
\end{subequations}
so that the whole perturbative state is known from $\cR$ and the background.
Alternatively, we could have used eqn.~(\ref{eq:GaugeInvExplicit}) directly to obtain the same relations.
Either way, all quantities have been expressed in terms of $\cR $, and the constraints are satisfied by construction.

\subsubsection{Fixed curvature gauge}\label{sec:ICNRfixedcurvaturegauge} A gauge will have \textit{fixed curvature} if it sets $\Phi$ and $E$ to small arbitrary spacetime functions $\Phi^*$ and $E^*$. The null case $\Phi^*=E^*=0$ corresponds to the well-known spatially flat gauge of cosmological perturbation theory.
This family of gauges is as straightforward to handle as the pure metric family above. Noting that $\cR $ is proportional to one of the perturbation quantities, here $\delta\phi$, the Momentum constraint can be used to obtain $\Psi$ while the Hamiltonian constraint gives $B$%
\begin{subequations}
\begin{align}
 E^{\rm FC}   & = E^{\rm FC}_G,\\
  \Phi^{\rm FC}   & = \Phi^{\rm FC}_G,\\
    \delta\phi^{\rm FC}    & = \sqrt{2\varepsilon_1}\Mpl\Big(\cR -\Phi^{\rm FC}_G\Big),\\
    \Psi^{\rm FC}    & =\varepsilon_1\Big(\cR -\Phi^{\rm FC}_G\Big)-H^{-1} \dot{\Phi}^{\rm FC}_G,\\
    B^{\rm FC}   & = \dot{E}^{\rm FC}_G+\varepsilon_1k^{-2}\dot\cR +(Ha^2)^{-1}\Phi^{\rm FC}_G.
\end{align}
    \label{eq:FixedRGaugeDecomp}
\end{subequations}
Again, we could have used eqn.~(\ref{eq:GaugeInvExplicit}) directly to obtain the same relations, and we obtained the field perturbation in terms of the only degree of freedom for the evolution of the perturbations.

\subsubsection{Newtonian gauge}\label{sec:ICNRnewtoniangauge} A gauge will be called \textit{Newtonian} if it sets $B$ and $E$ to small arbitrary spacetime functions $B^{\rm N}_G$ and $E^{\rm N}_G$.
Unlike the two previous gauge families, using the constraints alone is not enough to express all quantities in terms of $\mathcal{R}$.
We can, however, resort to eqn.~(\ref{eq:GaugeInvExplicit}). Starting from $ \chi^{\rm N}_G   = -a^2(B^{\rm N}_G-\dot{E}^{\rm N}_G)$, one gets $\Phi $ from $\Phi_B $ and then $\Psi $ is obtained from either the Momentum constraint or $\Psi_B $%
\begin{subequations}
\begin{align}
     E^{\rm N}   & = E^{\rm N}_G, \\
     B^{\rm N}   & = B^{\rm N}_G, \\
    \Phi^{\rm N}    & = \Phi_{B}  -H \chi^{\rm N}_G =   -\varepsilon_1 Ha^2 k^{-2}\dot{\cR}-H \chi^{\rm N}_G, \\
    \delta\phi^{\rm N}    & = \sqrt{2\varepsilon_1 }\Mpl\Bigg(\cR  +  \frac{\varepsilon_1 Ha^2}{k^2}\dot{\cR} -a^2\Big(B^{\rm N}_G-\dot{E}^{\rm N}_G\Big) \Bigg),\\
    \Psi^{\rm N}  & = -\varepsilon_1 a^2 H k^{-2} \dot\cR  
            -2a^2H\Big(B^{\rm N}_G-\dot{E}^{\rm N}_G\Big)-a^2\Big(\dot{B}^{\rm N}_G-\ddot{E}^{\rm N}_G\Big).
\end{align}
    \label{eq:NewtGaugeDecomp}
\end{subequations}
Thanks to analytical work in the previous gauge families, the Bardeen potentials $\Phi_B$ and $\Psi_B$ are indeed known as functionals of $\cR$.
The standard Newtonian gauge found in cosmology is of course obtained for $B^{\rm N}=E^{\rm N}=0$.  

\subsubsection{Uniform e-folds gauge}\label{sec:ICNRuniformN} 
If one defines $\cal{N}$ to be the number of e-folds elapsed from an arbitrary time $t_\circledast$ via%
\begin{subequations}
\begin{align}
    \partial_0{\cal N} \equiv -\frac{1}{3}\alpha K =  \frac{1}{6} \partial_0 \ln {^{(3)}g},\\
    {\cal N}(t_\circledast,\mathbf{x}) = 0,
\end{align}
\end{subequations}
then a gauge will be of \textit{uniform} (or \emph{fixed}) ${\cal N}$ if $\delta{\cal N}$ is fixed to a small arbitrary spacetime function $\delta{\cal N}^{\rm UE}_G$ while $B^{\rm UE}_G = 0$.
As will be explained in Chapter \ref{sec:SIGR}, this type of gauge with $\delta{\cal N}^{\rm UE}_G = 0$ is used in the standard formulation of Stochastic Inflation.

As ${\cal N} \equiv -\frac{1}{3}\int_{t_0} K \alpha dt $, one gets ${\cal N}_b = \int_{t_0} H dt$ at zeroth order and  $\delta{\cal N} = \Phi+\frac{1}{3}k^2E$ at first order \cite{pattison_stochastic_2019}. To calculate the whole hypersurface, one first needs to solve for $E $ by using $\Phi_{B }$ to obtain
\begin{equation}
    \dot{E}^{\rm UE}  -\frac{1}{3}a^{-2}H^{-1}k^2 E^{\rm UE}  = -a^{-2}H^{-1}\delta{\cal N}^{\rm UE}_G - \varepsilon_1 k^{-2} \dot\cR, 
    \label{eq:prepUniNGauge}
\end{equation}
the solution of which provides $E$ as
\begin{equation}
    E^{\rm UE}= e^{\int^{t}_{t_\circledast}\!\frac{k^2}{3a^2H}}\left[E_\circledast(\mathbf{x})-\int^t_{t_\circledast}e^{-\int^{s}_{t_0}\!\frac{k^2}{3a^2H}dt'}\left( \frac{\delta{\cal N}^{\rm UE}_G}{a^2H} +\frac{\varepsilon_1  \dot\cR}{k^2}  \right)ds\right],
\end{equation}
with $E(t_\circledast,\mathbf{x})\equiv E_\circledast(\mathbf{x})$. Note that there is still some gauge freedom left in the choice of the initial value of $E$ in space. Once $E $ is known, one gets%
\begin{subequations}
\begin{align}
  \Phi^{\rm UE}  & = -\frac{1}{3}k^2E^{\rm UE} +\delta{\cal N}^{\rm UE}_G,\\
    \delta\phi^{\rm UE}    & = \sqrt{2\varepsilon_1}\Mpl\Big(\cR +\frac{1}{3}k^2E^{\rm UE} -\delta{\cal N}_G^{\rm UE}\Big),\\
    \Psi^{\rm UE}    & =\varepsilon_1\Big(\cR +\frac{1}{3}k^2E^{\rm UE} -\delta{\cal N}^{UE}_G\Big)+H^{-1} \Big(\frac{1}{3}k^2 \dot{E}^{\rm UE} -\delta\dot{{\cal N}}_G^{\rm UE}\Big).
\end{align}
    \label{eq:UniNGaugeDecomp}
\end{subequations}

\subsubsection{Generalised synchronous gauges}\label{sec:secIIIB5}
A gauge will be called \textit{generalised synchronous} or \textit{NR} if it sets $\Psi$ and $B$ to small arbitrary spacetime functions $\Psi^{\rm NR}_G$ and $B^{\rm NR}_G$ to stay within perturbation theory.
The null case corresponds to the synchronous gauge in cosmology. Similarly to the Newtonian case, constraints are not enough to express all perturbations in terms of $\cR $, which is why we use eqn.~\eqref{eq:GaugeInvExplicit} on top of it.
This limitation is a first indication of the gauge gap between numerical relativity and cosmology.

Using the definitions of the gauge-invariant $\Psi_B $ and $\Phi_B$ and the latter's relation to $\cal{R}$, see (\ref{eq:GaugeInvExplicit}), we can obtain the following cascade%
\begin{subequations}
\begin{align}
 \chi^{\rm NR}    & = \displaystyle\int^t_{t^{\circledast}} (\Psi^{\rm NR}_G  -\Psi_B  ) dt' + \chi_{\circledast}^{\rm NR},\\
     E^{\rm NR}   & = \displaystyle\int^t_{t^{\circledast}}  [B_G^{\rm NR}  +a^{-2}\chi^{\rm NR}   ]dt' +E_{\circledast}^{\rm NR},   \\
    \Phi^{\rm NR}    & = \Phi_{B}  -H \chi^{\rm NR},    \\
    \delta\phi^{\rm NR}    & = \sqrt{2\varepsilon_1 }\Mpl[\cR   - \Phi^{\rm NR}  ]\,,
\end{align}
    \label{eq:ADMgaugeDecomp}
\end{subequations}
which are all functions of $\cR $ and its first- and second-order time derivatives and their integrals.

In this gauge both $\chi $ and $E $ need to be initialised at a given time by space-only functions $\chi_{\circledast}(\mathbf{x})$ and $E_{\circledast}(\mathbf{x})$.
It is indeed a well-known problem that the synchronous gauge does not fix all gauge degrees of freedom and that there is still spatial dependence \cite{baumgarte_shapiro_2010}, even though this can sometimes be confused with physical choices in the literature; see Appendix \ref{app:dofsyncgauge} for proof.
In this work, we will choose gauges such that $\chi_{\circledast}=E_{\circledast}=0$  at an arbitrary time, hidden in an implicit boundary of the time integrals.

\subsection{From cosmological perturbations to BSSN initial data}\label{sec:ICNRbssninitdata}
In Section~\ref{sec:ICNRbssn}, the fields used in NR were given.
The initial hypersurface needs to specify the dynamical quantities  $X$, $\tilde{\gamma}_{ij}$, $K$, $\tilde{A}_{ij}$, and $\phi$, $\Pi$.\footnote{$\tilde{\Gamma}^i$ being an intermediary quantity, a function of the others.}
The previous eqn. \eqref{eq:ADMgaugeDecomp} can thus be translated into its BSSN version, using the linearisation of the latter in terms of the cosmological perturbations we dealt with so far\ylc{. This dictionary is new \cite{Launay2025}}.
This is provided in Appendix \ref{app:dicBSSN} and leads to the following dictionary%
\begin{subequations}
\begin{align}
 \delta{\Pi}^{\rm NR} &  = \delta\dot{ \phi}^{\rm NR}-\dot{{ \phi}_b}\Psi_G^{\rm NR},\\
      \delta X^{\rm NR} & = a^{-2} \bigg(2\Phi^{\rm NR} -\frac{2}{3}\nabla^2 E^{\rm NR}\bigg), \\
    \delta\tilde{\gamma}_{ij}^{\rm NR} & =  2\bigg(\partial_i\partial_j-\frac{1}{3}\delta_{ij}\nabla^2\bigg)E^{\rm NR},\\ 
    \delta K^{\rm NR} &  =  3\bigg(\dot{\Phi}^{\rm NR}+H\Psi^{\rm NR}\bigg)-\Delta_b \chi^{\rm NR},\\
    \delta\tilde{A}_{ij}^{\rm NR} &   = \bigg(\partial_i\partial_j-\frac{1}{3}\delta_{ij}\nabla^2\bigg)\bigg(B^{\rm NR}_G-\dot{E}^{\rm NR}\bigg).
\end{align}
\label{eq:eqDictionary}
\end{subequations}
Substituting the previous perturbations eqn.~\eqref{eq:ADMgaugeDecomp} and evaluating at $t^{\circledast}$ yields the perturbative BSSN initial data in NR gauge as functionals of $\cR$%
\begin{subequations}
\begin{align}
    \delta\phi^{\rm NR,\circledast}_{k} & = \sqrt{2\varepsilon_1^{\circledast}}\Mpl(\cR_k^{\circledast}  - \Phi_{Bk}^{\circledast}), \\
    \delta\Pi^{\rm NR,\circledast}_{k} & = \ylg{\sqrt{2\varepsilon_1^{\circledast}}\Mpl\Big[-H^{\circledast}\Psi_G^{\rm NR,\circledast} +\dot{\cR}_k^{\circledast} - \dot{\Phi}_{Bk}^{\circledast}
     -\frac{\varepsilon_2^{\circledast}}{2}H^{\circledast}(\cR_k^{\circledast}  - \Phi_{Bk}^{\circledast})\Big]},\\
     \delta X^{\rm NR,\circledast}_{k} & = 2\Phi_{Bk}^{\circledast} ,\\
    (\delta\tilde{\gamma}_{ij})^{\rm NR,\circledast}_{k} &  = 0,\\ 
    \delta K^{\rm NR,\circledast}_{k} & = 3(\dot{\Phi}_{Bk}^{\circledast}+H^{\circledast}\Psi_{Bk}^{\circledast}),\\
    (\delta\tilde{A}_{ij})^{\rm NR,\circledast}_{k} & =  \bigg(\partial_i\partial_j-\frac{1}{3}\delta_{ij}\nabla^2\bigg)B^{\rm NR, \circledast}_G,
\end{align}
\label{eq:eqInitialConditions}
\end{subequations}
where the Bardeen potentials are expressed in terms of $\dot{\mathcal{R}}$ via eqns.~\eqref{eq:PhiB} and~\eqref{eq:PsiB}, and the gauge initialisations $(\Psi^{\rm NR, \circledast}_G,B^{\rm NR, \circledast}_G)$ are left free to choose but are usually set to zero.
The background initial data is straightforwardly set as
\begin{equation}
    X_b^{\circledast} = a^{-2}(t^{\circledast}) = 1, \qquad
(\tilde{\gamma}_{b})_{ij}^{\circledast} = \delta_{ij}, \qquad
 K_b^{\circledast}=-3H^{\circledast}, \qquad
  (\tilde{A}_b)_{ij}^{\circledast} = 0.
\end{equation}
One should appreciate that, in trying to motivate initial data without a numerical solver, we proved that initial conformal flatness ($\tilde{\gamma} = \delta$) is valid in this gauge at first order.
However, looking at eqn.~\eqref{eq:eqDictionary} suggests that $\delta X$ cannot be $0$ initially simultaneously and vice-versa, making the solver used in previous cosmological studies less reliable \cite{aurrekoetxea_cttk_2022}.

Incidentally, none of the previous relations use the equation of motion of $\cR$.
However, the constraint equations are imprinted by the underlying relativistic theory and so link to the equation of motion's content.
In that sense, any choice or transformation of $\cR$ keeps the decompositions found previously for each gauge (including the initial data in eqn.~\eqref{eq:eqInitialConditions}) satisfying the linear constraints as written in eqn.~\eqref{eq:ConstraintsCPT}.
Importantly for later purposes, that is the case \yld{for}\ylg{when} \yld{a windowed}\ylg{filtering} $\cR$.

\subsection{Classical \yld{drawing}\ylg{draw}}\label{sec:classicaldraw}
The introduction justified that scalar perturbations should be treated using a quantum theory.
However, numerical relativity solves the equations of motion, which are classical (or on-shell), and thus requires classicalised initial data.
Quantisation is usually written down explicitly as an inverse Fourier transform ($\mathcal{F}^{-1}$) expansion of an operator such as
 \begin{equation}
  \hat{\cal O} = {\cal F}^{-1}\{ {\cal O}_{\mathbf{k}}\hat{a}_{\mathbf{k}}\} + h.c,
         \label{eq:ModesExpGeneral}
 \end{equation}
 using each mode's harmonic annihilation and creation operators such that $[\hat{a}^{}_{\mathbf{k}},\hat{a}_{\mathbf{k}'}^\dagger ] =  \delta^{(3)}(\mathbf{k}-\mathbf{k}')$ and any other commutator being $0$.
 The operators, the Fourier amplitudes of which are written in eqn.~\eqref{eq:eqInitialConditions} or that of $\cR$, are constructed this way together with the canonical commutation of $\cR$ and its conjugate momentum; see eqn.~\eqref{eq:commutation}.
 However, we do not (and cannot) input a quantum operator into a classical simulation and thus work within the classical regime of the theory, as demonstrated in Chapter~\ref{sec:semiclQFT}.

Following the aforementioned chapter, we will assume that the operators become random numbers and drop their non-commuting parts, for modes above the comoving Hubble radius $k \leq aH$.
It was also argued that inputting slightly sub-Hubble scales could also be used for examining the signatures of (cubic) GR interactions. 
In this classical limit, eqn.~\eqref{eq:ModesExpGeneral} can be summarised by its spacetime correlation functions from which random realisations can be drawn.
The statistics of such a harmonic mix are Gaussian and thus are completely described by the two-point functions. This is equivalent to writing an expansion of stochastic modes for each operator $i$
\begin{equation}
  \boldsymbol{{\cal O}}_i = {\cal F}^{-1}\{ {\cal O}_{i,\mathbf{k}}\boldsymbol{\alpha}_{\mathbf{k}}\},
         \label{eq:StoModesExpGeneral}
\end{equation}
where $\boldsymbol{\alpha_{\mathbf{k}}}$ are random variables such that
\begin{equation}
           \langle \boldsymbol{\alpha_{\mathbf{k}}}\boldsymbol{\alpha_{\mathbf{k}'}}^* \rangle_{\mathbb{P}}  = C \delta^{(3)}(\mathbf{k}-\mathbf{k}'),
\label{eq:stocomm}
\end{equation}
where $C$ is a normalisation constant (see section \ref{sec:randomdraw} and Appendix \ref{app:random} for determination). It is important to remember that there is only one dynamical scalar degree of freedom, as seen from the Mukhanov-Sasaki equation, and so a single stochastic grid.
This is why {all quantities on the initial} hypersurface can be fully expressed in terms of $\cR_k$ {and $\dot{\cal{R}}_k$ as} in eqn.~\eqref{eq:eqInitialConditions} in Fourier amplitude space, but also in operator space in terms of $\boldsymbol{{\cal R}}$: there is only one set of $\{\boldsymbol{\alpha}_{\mathbf{k}}\}$ for all operators.

\section*{Summary}
In summary, we have solved the linear constraints of GR by expressing all quantities on the initial hypersurface in terms of a variable very well known to cosmologists, the linear, gauge-invariant curvature perturbation $\mathcal{R}$ and its time derivatives--- see equation \eqref{eq:eqInitialConditions}.
Without a solver, this allows us to draw physically-motivated initial data in the classical limit of a given scenario of inflation. The resulting arrays will satisfy the constraints at first order in perturbation theory for a common gauge: the geodesic (or synchronous, or NR) gauge.
We emphasise that the method itself is not restricted to this particular gauge choice and could be adapted to other choices too\ylg{, as well as beyond $\cR$}.

This analytical material is essentially all that is needed to start a full GR simulation and learn about full non-linearities and non-perturbative features during inflation, as done in the next chapter\ylc{, with initial data that satisfy the constraints at first order only. As the amplitude of the fluctuations or the range of scales grows, this is no longer sufficient and the constraints must be solved nonperturbatively (Appendix~\ref{app:UVdep})}.
Note that completely equivalent reasoning can be made for graviton perturbations (tensor modes) sourced by the vacuum and has been used in parallel work \cite{Florio2024}, although it involves two additional degrees of freedom and the constraints are trivially satisfied at first order, which does not necessitate the reasoning of the present work. 
Stochastic gravitons will be discussed in Chapters~\ref{sec:SIGR} and \ref{sec:SIbeyondGR}.

%% file: chap_ICsim.tex
\ifSubfilesClassLoaded{\setcounter{chapter}{2}}{}
\chapter{Nonlinear simulations from inflationary perturbations} \label{sec:ICsim}

In this chapter, we present the methods and results of our earlier unification of the classical initial data of inflationary cosmology with numerical relativity, by performing nonlinear simulations in full numerical relativity.

We organise this chapter following the findings of \cite{Launay2025}. 
Sections~\ref{sec:ICsimNumericalInput} and \ref{NR} start by showcasing the full numerics pipeline and apply it to a vanilla slow-roll type of inflation in Section \ref{sec:correct}, using the stochastic initial conditions introduced in the previous chapter as the starting point of a full GR simulation.
As perturbations remain small in this case, linear perturbation theory results are fully recovered from our simulations, thereby offering a useful validation check. 
Section \ref{sec:ICsimNPinf} is a proof of concept that \ylc{non-perturbative dynamics of the scalar field and the geometry can be probed, starting from initial data that satisfy the constraints at linear order, and within the limits on constraint satisfaction and resolution discussed there}.
We showcase this by studying  inhomogeneous inflation with an inflection point potential that exhibits ultra slow-roll, and a strong resonance model leading to eternally inflating spatial regions. Appendix \ref{app:UVdep} shows why deep UV modes cannot be investigated in this framework. 
\begin{figure}[ht]
    \centering
\includegraphics[width=0.9\linewidth]{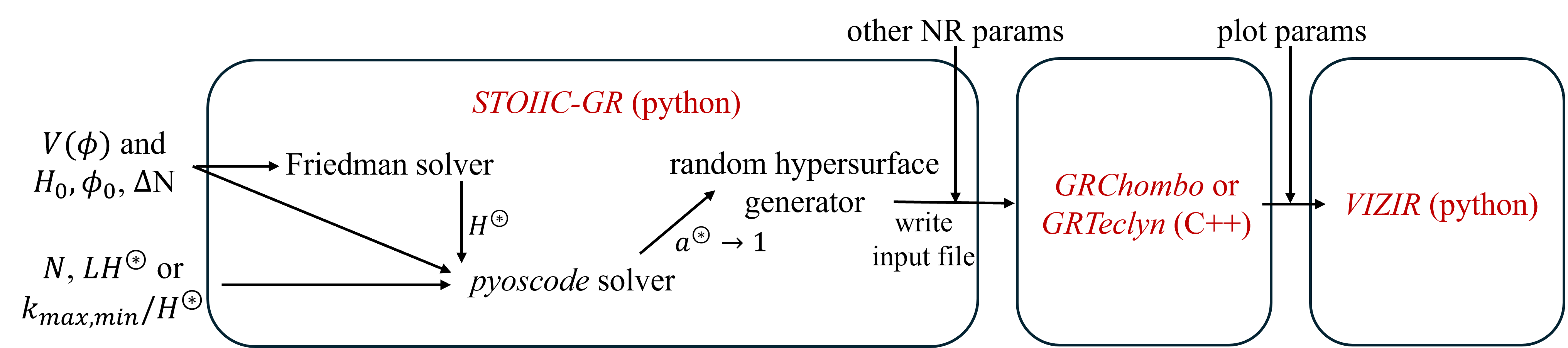}
    \caption{Summary of the full numerical pipeline. The user specifies in chosen units a potential $V$ either entirely or as a functional of parameters, a Hubble constant $H_0$ to start with or an initial field value $\phi_0$ (see \ref{subsec:background}), a number of background e-folds $\Delta \mathcal{N}$ between the Bunch-Davies vacuum initial conditions and the starting time of the GR simulation, $N$ the number of pixels per dimension to generate and get simulated, the length $LH^{\circledast}$ or the Fourier range $k_{max,min}/H^{\circledast}$ of the simulated lattice in terms of the Hubble radius at $t^{\circledast}$. The user can also specify parameters related to NR evolution (stepping, damping, adaptive mesh refinement, etc), or plotting requests from the final extractions. 
         \textbf{Sequence:} 
             i) Friedmann and Mukhanov-Sasaki equations are evolved from the imposition of the Bunch-Davies vacuum to the start of the GR simulation.
             ii) A single 3D random lattice is drawn and given the right spectrum for each of the BSSN quantities of eqn.~\eqref{eq:eqInitialConditions}. The constraints are accurately satisfied at first order in perturbation theory (violation only at second order and beyond). The included wavenumbers go from slightly sub-Hubble to a few times the Hubble radius.
             iii)  This stochastic lattice is then evolved using a full GR simulation. The main software used is \textit{\grc} \cite{Andrade2021} but the results have been successfully recovered using the GPU-enabled ported software \textit{\grt} \cite{GRTeclyn}.
             iv) Quantities of interest are extracted after a few e-folds.}
    \label{fig:pipeline}
\end{figure}

\section{Numerical input}\label{sec:ICsimNumericalInput} 

This section introduces the practical strategy employed to generate the initial data for the full GR simulation,
laying the foundations behind our code \stoiic, to be released in the near future.

\subsection{Scaling the inflationary patch}\label{subsec:designpatch}
The parameter space for a simulation with fixed evolution mostly reduces to dimensionless physical initial conditions and resolutions in space and time. However, these choices can influence or constrain each other.

Our goal is to observe the inflaton's self-interaction in GR. In a perturbative framework, from which we always start, these interactions are key for the scales crossing the Hubble radius defined previously because the horizon crossing entails a spectral turnover from a sub-Hubble to a super-Hubble slope. This means that our simulation needs to showcase at least one mode crossing this horizon: the lattice resolution and the input analytical spectrum need to be such that $k_{max}^{box},k_{max}^{\cal R}\geq(aH)^{\circledast} = (R^{\circledast}_H)^{-1}$ but also such that discrete gradients are accurate enough to resolve the perturbations $k_{max}^{box}> k_{max}^{\cal R}$. Written that way, wavenumbers $k$ will always be taken in comoving Fourier space and so will our spatial coordinates be in the GR simulation, those of eqn.~\eqref{eq:SVTmetric}. 
Our simulations will thus see the Hubble radius $R_H$ shrinking on the lattice until it disappears beyond the discrete resolution.

For a fixed $k_{max}^{\cal R}$ and the gradients-friendly choice of $k_{max}^{box}$, the computational resources will determine the maximal size of the lattice (or $k_{min}$) through the maximum number $N$ of pixels per dimension we can run the code with. Beyond that size, we assume that the simulated universe is made of periodic patches such as this one. This is equivalent to giving a null spectrum to that range of superscales or a known spectrum de-correlated from those smaller scales.  This is a very good approximation in the case of inflation if the patch is of size a few times $R^{\circledast}_H$ because of the separate universe behaviour, where separate patches barely influence each other and can be regarded as different random draws of the same stochastic lattice of finite size \cite{salopek_nonlinear_1990, salopek_stochastic_1991}. 

If $N$ is large enough to have both a super-Hubble-sized box and sub-Hubble scales, it is tempting to take $k_{max}^{box},k_{max}^{\cal R}\gg (aH)^{\circledast}$ similarly to what was achieved in the Friedmann lattice code of \cite{Caravano2024}. From a numerical point of view, nothing hampers the treatment of that regime.
From a physical point of view, we {claim} that it is not possible to go too deep sub-Hubble without encountering well-known UV divergences of the MS solutions.
We believe that the choice of gauge can also change how sub-Hubble one can go. The key difference with \cite{Caravano2024} is that of the full GR treatment versus the Friedmann local one, which in practice, and in the right gauge, \yld{translates}\ylg{could translate} to using the MS linear equation of the Fourier modes with a lattice-averaged Hubble parameter.
It is fine to leave perturbation theory and use the Mukhanov-Sasaki evolution to recover the horizon-crossing perturbations, as long as no back-reaction is investigated too deep in the horizon.
In our case, going too deep would probably drive the evolution equations away from the background and linear solutions when using the full GR action and equations, but most importantly perturbation theory breaks down and so does our initialisation - see Appendix \ref{app:UVdep}.
This problem is fundamental and links to the debate on UV renormalisation (see \cite{pla23} for a recent treatment); the vacuum energy might need to be removed as only its late products are gravity-friendly. 
Until this topic is completely addressed with a {satisfactory treatment of the gravitational effects of deep UV modes}, this work will simply be cautious by applying a simple sharp momentum cutoff imposed by our lattice resolution (see $\sigma_{max}$ below), {to stay within perturbation theory and at the initial time}.\footnote{In the future, we plan to address this inconvenience with the help of an adiabatic switch, as mentioned in the introduction and also implemented in numerical correlator codes \cite{werth23}.}
Most importantly and as argued in the past two chapters, one should not look at modes that are too deep sub-Hubble at the initial time because they are not classicalised and so cannot be treated as a classical stochastic draw.
For instance, we expect that lattice Friedmann simulations would not be able to recover the full QFT bispectrum beyond the squeezed limit.

We summarise all our recommendations as
\begin{equation}
         k_{max}^{\cal R}  = \sigma  /R^{\circledast}_H, \qquad
         dx  \leq \frac{2\pi R^{\circledast}_H }{n\sigma},  \qquad
         1 \leq \sigma \leq \sigma_{max},
    \label{eq:designbox}
\end{equation}
where $n$ is the number of pixels to cover the associated smallest wavelength (usually $n\geq 4$ in our work, reasonable with $6^{\rm th}$ order stencils \cite{radia_lessons_2022}) and ensures that initial data is smooth enough.
$\sigma_{max}$ is the scenario-dependent perturbativity and classicality bound above which we cannot guarantee that our simulation will handle {the corresponding modes perturbatively initially and with valid classicality}.
We also assumed that the IR sector is such that $k_{min}^{\cal R} = k_{min}^{box} = \frac{2\pi}{Ndx}$. This setup also automatically imposes at least a Hubble-sized initial box as it only takes $N\geq  \frac{2\pi}{dx}R^{\circledast}_H\geq 1$ to have $k_{min}\leq 1/R^{\circledast}_H$.

\subsection{Background solutions}\label{subsec:background}
The first choice to make is that of the background quantities of inflation and in particular of the background initial conditions. There is not always a unique way to impose certain constraints on the background dynamics. In this work and unless stated differently, we usually choose an initial $H_0$ and a potential $V_m(\phi)$ parametrised by an unknown parameter $m$ first, then we determine $\phi_0,m$ to match certain values of a slow-roll attractor. For instance, in a $V(\phi)=\lambda \phi^p $ inflation ($p\geq 0$ even), one can use the Friedmann equations to set%
\begin{subequations}
\begin{align}
    \phi_0 & = -\left|\frac{ (3 - \varepsilon_1)p\Mpl}{ \sqrt{2  \varepsilon_1}  (\varepsilon_2 / 2 + \varepsilon_1 - 3)}\right|, \\
           \lambda & =  \left |\frac{\sqrt{2 \varepsilon_1}  (\varepsilon_2/ 2 + \varepsilon_1 - 3) }{ (3 - \varepsilon_1) p\Mpl}\right|^p \Mpl^2  H_0^2 (3 - \varepsilon_1),
\end{align}
\label{eq:potentialposition}
\end{subequations}
for a given set of slow-roll parameters $\varepsilon_{1,2}$, notably $\varepsilon_2 = -2\varepsilon_1$ for the attractor. 

Another method we employed was to fix entirely $V(\phi)$ and pick $\phi_0$ at a point of interest (for instance before a feature).
The field's velocity would then be fixed using the slow-roll limit expression $\dot{\phi_0}=-\frac{V_{,\phi}}{V}\displaystyle |_{\phi_0}\Mpl^2$, thus fixing $H_0$ through the Friedmann constraint, the last quantity required for solving.
Both methods guarantee a slow-roll evolution, and a perfect satisfaction of the zeroth-order constraints eqns.~\eqref{eq:backgroundEqHam} and \eqref{eq:backgroundEqK} until the start of the GR simulation.

We will refer to these two methods as  \textit{potential-position-fixing} and \textit{$H_0$-fixing} respectively.  The former might be the most appropriate to relate to observational constraints, usually made on the shape of the potential, the e-folding duration, and the spectral index.
Note that we do not claim that any of our models are appropriately scaled to match e.g. Planck data \cite{PlanckInflation}. The potentials of section \ref{sec:ICsimNPinf} though should still have portions consistent with data.

From the background ICs, the pre-NR pipeline evolves the Friedmann and field equations \eqref{eq:backgroundEq} recast as \cite{Agocs20}%
\begin{subequations}
\begin{align}
        \frac{\d\ln{\Omega_k}}{\d{\cal N}_b} & = 4 -2a^2 \Omega_kV(\phi),\\
        \Big(\frac{\d\phi}{\d{\cal N}_b}\Big)^2&  = 6 -2a^2\Omega_kV(\phi),
    \end{align}
    \label{eq:rewriteFriedman}
\end{subequations} 
where $\Omega_k = (aH)^{-2}= R_H^2$ is the curvature density in a flat universe.
This is evolved for a number of e-folds $\Delta{\cal N}$ up to the chosen start of NR evolution,
which is useful to stabilise into a potential attractor but mostly, as we are about to explain, for modes to have enough time to transit from sub-Hubble to super-Hubble and classicalise before the start of the GR simulation.
\clearpage
\subsection{Linear solutions}
Once the background dynamics have been solved, it is possible to solve eqn.~\eqref{eq:MScosmic}, the solution of which might be highly oscillatory at high wavenumbers.
This can be numerically problematic and requires an adequate solver such as \textit{pyoscode} introduced in \cite{Agocs20,agocs20bis}.
The code was primarily developed for the Mukhanov-Sasaki equation and this part of our pipeline adapts some of the example codes to our needs.

The evolution of eqn.~\eqref{eq:rewriteFriedman} for  $\Delta{\cal N}$ e-folds lands on realistic values of the Hubble parameter, scale factor, field, and its momentum at $t^{\circledast}$, the GR simulation starting time:  $(H^{\circledast}, a^{\circledast}, \phi^{\circledast},\dot{\phi}^{\circledast})$.
From that, our choice for the range of modes with respect to the Hubble radius at that time can be made explicit.
In particular, if following the previous recommendations in eqn.~\eqref{eq:designbox} and if $\Delta{\cal N}$ is large enough, these should all be deep sub-Hubble back at $t_0$ when the Bunch-Davies vacuum will be imposed. 
This is why we usually recommend
\begin{equation}
\Delta{\cal N}\geq \ln\Bigg(\frac{ k_{max}^{\cal R}}{k_{min}^{\cal R}}\Bigg).
\end{equation}
The discrete set of $\{k_{min}^{\cal R},...,k_{max}^{\cal R}\} $ gives as many Mukhanov-Sasaki equations to solve using this code.
After that, we have all the values needed for ${\cal R}_k$ to impose the associated spectrum on our discrete grid.

\subsection{Random hypersurface}\label{sec:randomdraw}
After solving the MS equation for all modes of our grid up to $t^{\circledast}$, we can perform the classical draw of the initial data for the GR simulation.
One way to impose a spectrum ${\cal P}_k = |f_k|^2$ with Fourier amplitude $f_k$ on a Gaussian and independent random 3D draw in real space is to multiply the Fourier amplitude of the latter by $\sqrt{{\cal P}_k} = |f_k|$.\footnote{Equivalently in the large $N$ limit, the random draw part can be done directly in Fourier space using complex normal draws $\mathbb{C}{\cal N}(0,1)$ and imposing Hermitian symmetry \cite{peligrad_central_2019}.} 
This work imposes several Fourier amplitudes $f_k$ defined in eqn.~\eqref{eq:eqInitialConditions}, all functions of ${\cal R}_k$ and its time derivative. To reach the perfect satisfaction of the linear constraints \eqref{eq:ConstraintsCPT}, an adaptation of the usual method is needed because the sign of perturbations for every quantity matters against others. Hence, our amplitudes will be $f_k$ as opposed to $|f_k|$, which does not impact the validity of the method.
In this sense, we are applying the formula directly \eqref{eq:StoModesExpGeneral}.
It is also worth recalling from section \ref{sec:classicaldraw} that there is only one three-dimensional random draw and dynamical d.o.f. despite having multiple stochastic quantities. All these quantities are therefore cross-correlated accordingly.
To respect those shared statistics, we draw the $\boldsymbol{\alpha}_{\mathbf{k}}$ first, then build curvature perturbation ${\cal R}$ and its time derivative $\dot{\cal R}$, and build the other quantities by real space combination as defined in eqn.~\eqref{eq:eqInitialConditions}.

When working with discrete Fourier transforms, power spectra, and different conventions, reaching the right normalisations and units-invariance can quickly become challenging.
Appendix \ref{app:random} makes transparent the numerical construction of our perturbations using the method described above and the link with QFT.
In particular, one should remember that cosmological power spectra from QFT have surprising units because of the ones implicitly carried by annihilation and creation operators.
Our generator of perturbations added to an inflating background has been tested successfully with respect to the linear and full GR constraints. It has shown on multiple occasions that switching a sign or any slight modification of the constraint formulae would show a violation of these (see initial time in Figures~\ref{fig:quadviolations}, \ref{fig:NPconstraintsa} and \ref{fig:NPconstraintsb}).
The satisfaction of the constraints is the result of the highly precise and cautious perturbative method developed in Chapter~\ref{sec:ICNR}.

The transition to the GR simulation is the next step, once constraints and spectra have been checked. This is usually done by producing an input parameters file for the GR simulation, together with a grid file it can start from.
Before switching to NR, our pipeline also performs a rescaling of the scale factor, the Planck mass, and their dependencies to reach a range of numbers within double precision, valid for the entire duration of the simulation.
Typical choices include $a^{\circledast}=1$ and $\Mpl=100$ (equivalent to having quantities in $0.01\Mpl$ units of masses). 
The latter choice will reveal itself as particularly useful to get second-order perturbative violations of the constraints above the numerical accuracy (see applications in section \ref{sec:ICsimNPinf}).

\section{Numerical relativity}\label{NR}

\subsection{Implementation}\label{NRimplem}
\grc and most recently \grt are two open-source codes from the GRTL collaboration \cite{GRTL,radia_lessons_2022,GRTeclyn} for evolving the BSSN equations of numerical relativity, eqns.~\eqref{eq:BSSN}.
 Both were built to make use of hybrid MPI and OpenMP parallelism and vector intrinsics to achieve the best performance on the latest HPC architectures while offering adaptive mesh refinement (AMR) technology.
The latest software was motivated to enable the use of GPUs thanks to the AMReX library \cite{amrex} and has thus allowed consequent speed-ups.
In this work, the pipeline shown in Figure~\ref{fig:pipeline} can use either of them and leads to the same results.
Nonetheless, as it is more established so far and as \grt was still undergoing refinements at the time of these experiments, \grc is behind every plot of this chapter.

Evolving full non-linear inhomogeneous inflation seems challenging in cosmology, but it is a rather light problem for NR, up to complications before and after evolving mentioned earlier (see also \cite{aurrekoetxea24}).
From this full NR library, we will not make use of the AMR or the CCZ4 damping to eliminate gauge artefacts \cite{CCZ41, CCZ42} but simply the BSSN equations \cite{BSSN_BS,BSSN_SN} (see eqns.~\ref{eq:BSSN}) with our initial conditions and periodic boundary conditions.

In BSSN codes, the numerical gauge is usually a 1+log-slicing evolving together with Gamma drivers \cite{baumgarte_shapiro_2010,radia_lessons_2022}, but our work only uses the sub-case of a geodesic observer, which,  in the cosmological language, is the synchronous gauge presented earlier in Section~\ref{sec:ICNRbssninitdata}, with $\Psi_G^{\rm NR} = B_G^{\rm NR} = 0$.
We will see that this gauge comes with some caveats that \yld{can be}\ylg{are} kept under control. 

Regarding the time resolution in these coordinates, this work sets the Courant factor value as $\Delta t/\Delta x = 0.025 $ while extracting $15$ to $50$ checkpoint lattices for post-processing.
The typical size of our box is $256^3$. This size usually requires less than a few hours to run on $64$ to $152$ processors and still provides insightful results.\footnote{Note that tests have also been successfully run with $1024^3$ in \grt but are computationally and memory-wise more demanding to run and post-process (see in-situ visualisation solutions of the GRTL collaboration, and future work).}
Convergence tests are available in Section~\ref{sec:correct}.

\subsection{The final values problem}\label{sec:finalvalues}
The previous software provides three-dimensional raw fields at multiple time slices and requires efficient\footnote{We made the choice of using the \textit{python} \textit{YT} library \cite{YT}.} and careful extraction of summary statistics, including e.g. spectra.
Most importantly, a cosmologist will appreciate conversions back from BSSN to SVT formalism, as highlighted in Appendices \ref{app:dicBSSN} and \ref{app:extract}.

One of the key problems in physical extractions from numerical relativity is that of the impeded interpretation due to gauge ambiguities. If one is interested in a specific observer's point of view, it is not always guaranteed to be simulation-friendly given previous considerations on well-behaved numerical gauges.
The usual gauges of NR are also rarely anywhere close to observational ones. For these reasons, it is essential to extract gauge-invariant quantities such as ones built in perturbation theory (${\cal R}$, Bardeen quantities, etc ). However, that is only valid if the simulation stays perturbative and is not our purpose.
In previous studies of non-perturbative cosmological frameworks, the key diagnostic extraction was that of the gauge-invariant matter energy density $\rho$ or more recently the Weyl curvature \cite{Ijjas2023,Elley2024} to include geometric contributions.
We want to emphasise in our work the existence of a generalisation for the usual linear gauge-invariants such as ${\cal R}$.
The full GR comoving curvature on comoving hypersurfaces was first defined as  \cite{rigopoulos_non-linear_2005}
\begin{equation}
    {\cal R}_i = -\frac{1}{6}\partial_i\ln |\gamma|+\frac{1}{6}\frac{\partial_0\ln |\gamma|}{\partial_0 \phi} \partial_i \phi,
    \label{eq:NLgaugeinv}
\end{equation}
and constitutes a gauge-invariant quantity on long-wavelength scales.
Furthermore, it was shown in \cite{langlois_evolution_2005} that a covariant tensorial (covector) generalisation can be defined using a timelike vector field $n^a$ (\textit{frame vector}, see 3+1 formalism introduced in Section~\ref{sec:ICNRadm}) as
\begin{equation}
    \tilde{\cal R}_a  = -\frac{1}{6}D_a\ln |\gamma|+\frac{1}{6}\frac{n^b\partial_b\ln|\gamma|}{n^b\partial_b{\phi}} D_a \phi
                = -\frac{1}{6}\partial_a\ln|\gamma|+\frac{1}{6}\frac{n^b\partial_b\ln|\gamma|}{n^b\partial_b{\phi}}\partial_a\phi,
\end{equation}
where the nontrivial term of the induced covariant 3-derivative $D_a f= \partial_a f+n_an^b\partial_b f$ cancels perfectly and conveniently in the second equality.
When the foliation is defined by the observers comoving with the fluid ($D_a\phi = 0$), $\tilde{\cal R}_a$ is minus the covariant 3-derivative of the local number of e-folds, suggesting that this is the right generalisation of the curvature perturbation on comoving hypersurfaces, up to a gradient.
Interestingly, the frame vector is simply $n^a = (1,0,0,0)$ in the synchronous gauge, and so the spatial components of $\tilde{\cR}_a$ are exactly ${\cR}_i$. Using \cite{rigopoulos_non-linear_2005} we thus expect $\tilde{\cR}$ to also be unchanged on large scales.\footnote{For instance, using the comoving foliation with $u^a\propto\nabla^a\phi$, we find  $\frac{n^b\partial_b\ln|\gamma|}{n^b\partial_b{\phi}}= \frac{(\partial^0\phi\partial_0+\partial^i\phi\partial_i) \ln|\gamma|}{(\partial^0\phi\partial_0+\partial^i\phi\partial_i) \phi}$, which in the long-wavelength limit reduces to $\frac{\partial_0\ln|\gamma|}{\partial_0 \phi}$, as expected.}

The link of these generalisations with ${\cal R}$ becomes apparent when linearised\footnote{Note the correction compared to \cite{Launay2025} with no impact on the extractions.}
\begin{equation}
     {\cal R}_i^{(1)}  = \Big(\Phi +\frac{H}{\dot{\phi}_b}\delta\phi - \frac{1}{3}\nabla^2E \Big)_{,i} = \Big(\cR - \frac{1}{3}\nabla^2E \Big)_{,i}.
   \label{eq:RNL}
\end{equation}
When running simulations so that all modes cross the horizon, terms other than ${\cal R}_{,i}$  are gradient-suppressed.\footnote{The closest nonlinear generalisation of the curvature perturbation on comoving hypersurfaces (removing those extra gradient-suppressed terms in the linear limit) was given in \cite{UzanPitrou2007}, in the context of covariant quantisation.}
This means that we now have a good proxy to evaluate non-perturbativity and non-linearity by comparing $\cR$ with
\begin{equation}
    {\cal R}^{\rm NL} \equiv \nabla^{-2} \partial_i \cR_i.
\end{equation}
This is extracted from a simulation for the first time in Section~\ref{sec:ICsimNPinf}.\footnote{\ylc{In comoving gauge, $\delta\phi = 0$ and the spatial metric takes Maldacena's form $\gamma_{ij} = a^2 e^{2\zeta}\left(e^{h}\right)_{ij}$ with $h_{ij}$ traceless, so that $\det e^{h} = 1$ \cite{Maldacena_2003}.
The second term of eqn.~\eqref{eq:NLgaugeinv} then vanishes and $\ln|\gamma| = 6(\ln a + \zeta)$, so that ${\cal R}_i = -\partial_i \zeta$ and ${\cal R}^{\rm NL} = -\zeta$ exactly, at all orders in perturbation theory. ${\cal R}^{\rm NL}$ is therefore minus the nonlinear curvature perturbation $\zeta$ of Maldacena; the sign is the one already present at linear order with our convention $\gamma_{ij}\supset a^2(1-2\Phi)\delta_{ij}$, for which ${\cal R} = -\zeta$.
In any other slicing, such as the synchronous gauge of our simulations, let us write $|\gamma|^{1/6} \equiv e^{{\cal N}(t,\mathbf{x})}$, with ${\cal N}$ the local number of e-folds.
Eqn.~\eqref{eq:NLgaugeinv} reads ${\cal R}_i = -\partial_i{\cal N} + \frac{\partial_0{\cal N}}{\partial_0\phi}\partial_i\phi$, which, at leading order in gradients, is minus the gradient of ${\cal N}$ evaluated on the hypersurface of uniform $\phi$ through each point, i.e.\ on the comoving hypersurface, which coincides with the uniform-density one on super-Hubble scales.
This is the nonlinear $\zeta$ of the $\delta N$ formalism \cite{SasakiDNformula, Lyth03, langlois_evolution_2005}, so that ${\cal R}^{\rm NL} \simeq -\zeta_{\delta N}$ up to gradient corrections. This is also why the e-folds of Chapter~\ref{sec:SIGR} can be used to describe ${\cal R}^{\rm NL}$.}}
\ylc{Since ${\cal R}^{\rm NL}$ and ${\cal R}$ share the same linear limit, up to the gradient-suppressed $\nabla^2E$ term of eqn.~\eqref{eq:RNL}, their difference is quadratic in the fluctuations, and far below the statistical scatter of our spectra in perturbative cases.
Their agreement therefore only tests the extraction (gauge and gradient terms, perturbative amplitude), not the linearity of the dynamics: the latter would affect both alike, and is instead tested against the Mukhanov-Sasaki prediction ${\cal R}^{\rm MS}$.
Conversely, a visible difference means that the linear extraction fails and that only ${\cal R}^{\rm NL}$ should be used.}

\subsection{Perturbative correctness and convergence}\label{sec:correct}
To evaluate the correctness of our simulations, a reference case or regime is useful. For comparison, theoretical results from QFTCS are mostly available in slow-rolling perturbative scenarios of inflation.
In that respect, the background solution to the Friedmann equations and the linear prediction for the power spectrum from eqn.~\eqref{eq:MScosmic} will constitute a good diagnostic of correctness. For this comparison, \stoiic thus solves for the background and ${\cal R}$ fields beyond $t^{\circledast}$ for the entire duration of the GR simulation.

Our vanilla case of inflation is the fully perturbative quadratic potential one $V(\phi) = \frac{1}{2} m^2\phi^2$ with $\phi_0$ and $m$ set using the \textit{potential-position fixing} method described by eqn.~\eqref{eq:potentialposition} and by imposing a slow-roll $\varepsilon_1 = -0.5\varepsilon_2 = 0.01$ and an initial Hubble rate $H_0 = 10^{-5}\Mpl$. These values correspond to the BD vacuum fixing time and then are evolved for $\Delta {\cal N} = 5$ e-folds until the start of the GR simulation.

Figure \ref{fig:meanquad} shows the background fields of interest from both the Friedmann equations and the GR simulation.
The energy density is rendered at different times in Figures \ref{fig:3Da}, \ref{fig:3Db}, and \ref{fig:3Dc}.
Figure \ref{fig:specquad} shows the initial and final dimensionless power spectra, cut at the value of $k_{max}^{\cal R}$, which is also where the initial spectrum is windowed with a Heaviside function.
For this simulation, and following notations and requirements of eqn.~\eqref{eq:designbox}, $\sigma = 3$ and $L = 64/H^{\circledast}$. We are looking at slightly sub-Hubble to super-Hubble scale dynamics. 

\begin{figure}[ht]
    \begin{subfigure}{0.38\textwidth}
        \centering
        \includegraphics[width=\linewidth]{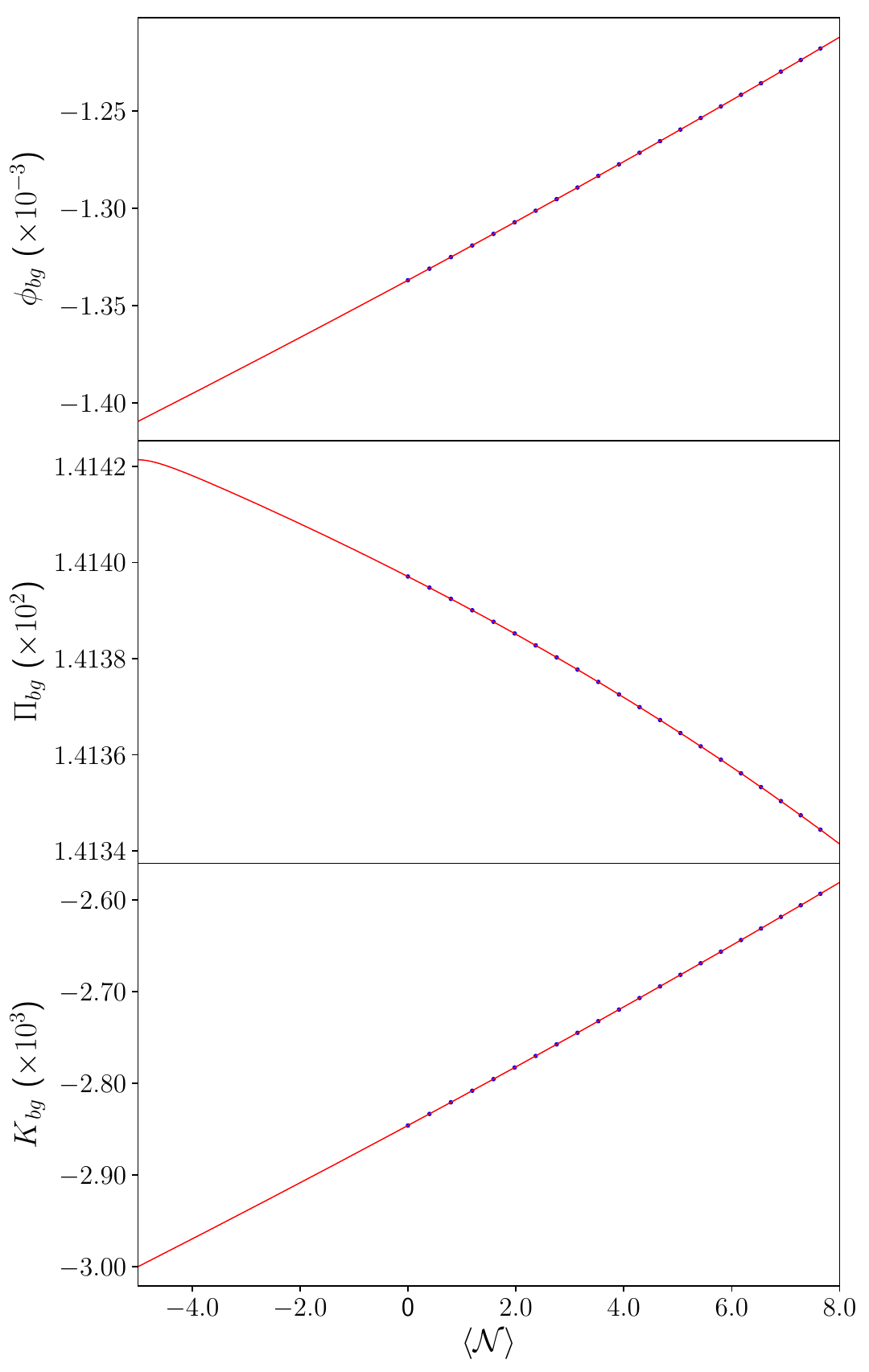}
        \caption{Comparison of the Friedmann solution (solid red) and the mean inflaton (top), conjugate momentum (middle), and extrinsic curvature (bottom) fields extracted from the GR simulation (blue dots).}
        \label{fig:meanquad}
     \end{subfigure}
    \hfill
    \begin{subfigure}{0.60\textwidth}
        \includegraphics[width=\linewidth]{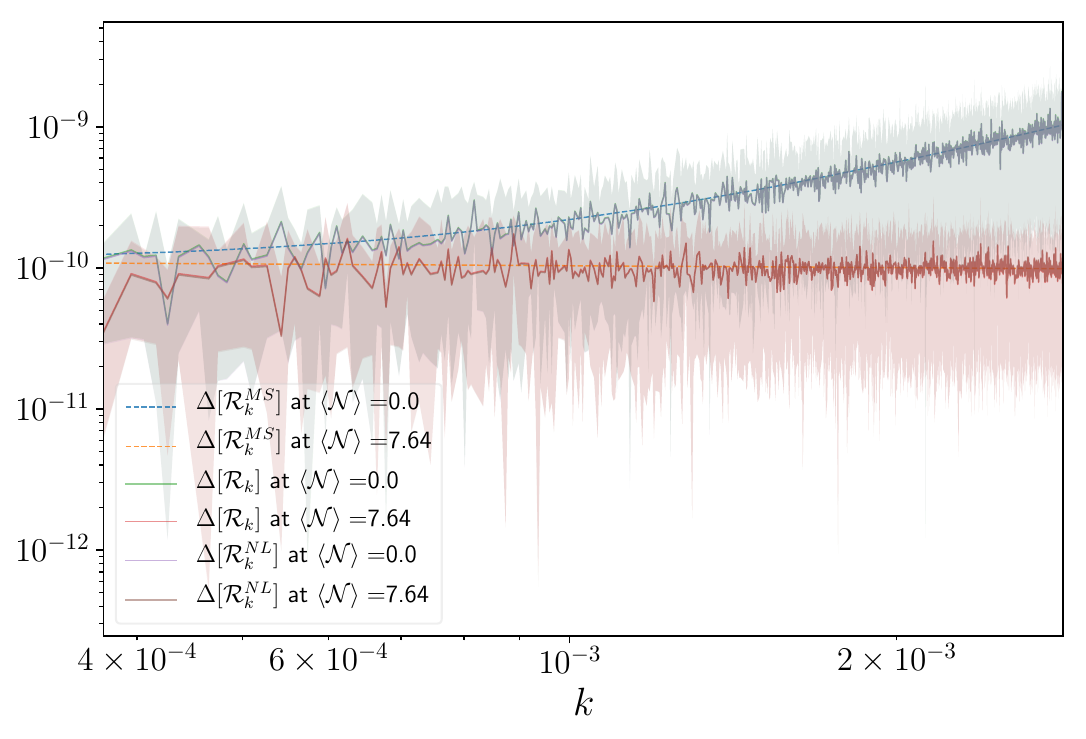}
            \caption{ Linear (dashed, ${\cal R}^{MS}$) and extracted (solid, ${\cal R}$) binned dimensionless spectra ($\Delta[{\cal R}]= \frac{k^3}{2\pi^2} |{\cal R}_k|^2$) compared to extracted ${\cal R}^{\rm NL}$, a scalar built from the fully non-linear generalisation ${\cal R}_i${: perfect superimposition at initial (upper half) and final (lower half) times ({\textcolor[HTML]{95d095}{\rule{1em}{0.5em}}} $+$ {\textcolor[HTML]{cab3de}{\rule{1em}{0.5em}}} $=$ \textcolor[HTML]{608475}{\rule{1em}{0.5em}}, {\textcolor[HTML]{ea9393}{\rule{1em}{0.5em}}} $+$ {\textcolor[HTML]{c6aba5}{\rule{1em}{0.5em}}} $=$ \textcolor[HTML]{b13e3a}{\rule{1em}{0.5em}})}. Contours account for the $1\sigma$-range of points in each bin. Displayed Fourier bins are those of the simulation mesh which have more than 10 realisations and which are smaller than the spectral cutoff of the input ($k_{max}^{\cal R} = 3\times 10^{-3}$).}
        \label{fig:specquad}
    \end{subfigure}
\caption{Background and fluctuations in Fourier space of the quadratic inflation case in $\Mpl=100$ units.}
\end{figure}

The fit between the GR lattice averages and the Friedmann backgrounds is perfect and shows that inflation is proceeding correctly. 
On the inhomogeneous side, the super-Hubble freeze is visible for the density from Figure \ref{fig:3Db} to \ref{fig:3Dc}.
At all times, the extracted power spectra appear to be in close agreement with the linear prediction\ylc{, and ${\cal R}^{\rm NL}$ is very close to ${\cal R}$, as expected for such small amplitudes: this confirms the consistency of the extraction rather than the linearity of the dynamics, which is shown by the agreement with ${\cal R}^{\rm MS}$}.
The spectral turnover is also visible by eye in real space from Figure \ref{fig:3Da} to \ref{fig:3Db}, where the clustering transits to smaller scales.
These observations are of course all subject to both cosmic and the estimator's variance, but also to higher-order interactions, which, in this vanilla and barely sub-Hubble case, are essentially negligible. 
\begin{figure}[ht]
    \centering
        \begin{subfigure}{0.32\textwidth}
        \centering
\includegraphics[width=\linewidth]{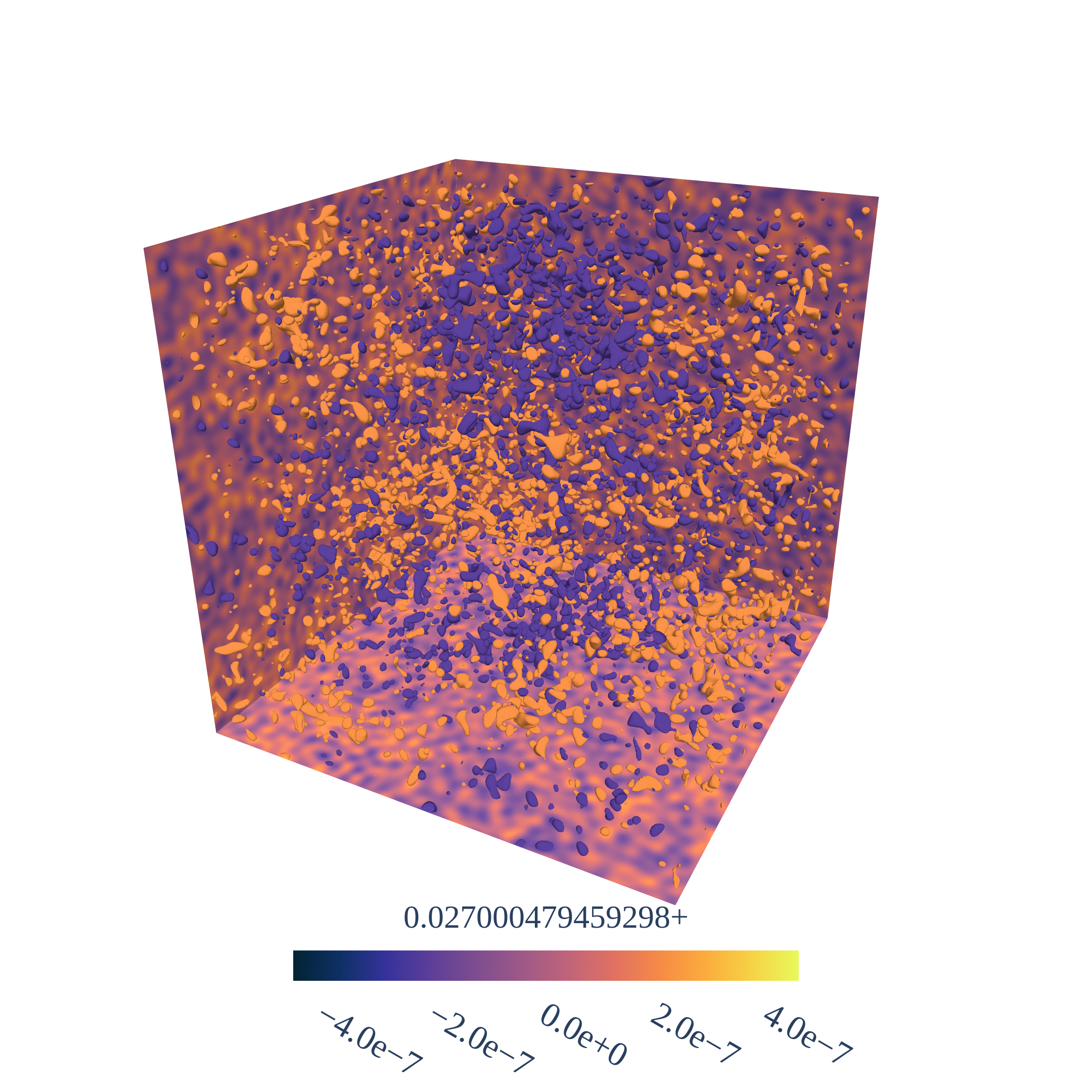}
    \caption{$\langle {\cal N}\rangle=0$}
    \label{fig:3Da}
    \end{subfigure}
    \hfill
\begin{subfigure}{0.32\textwidth}
        \centering        \includegraphics[width=\linewidth]{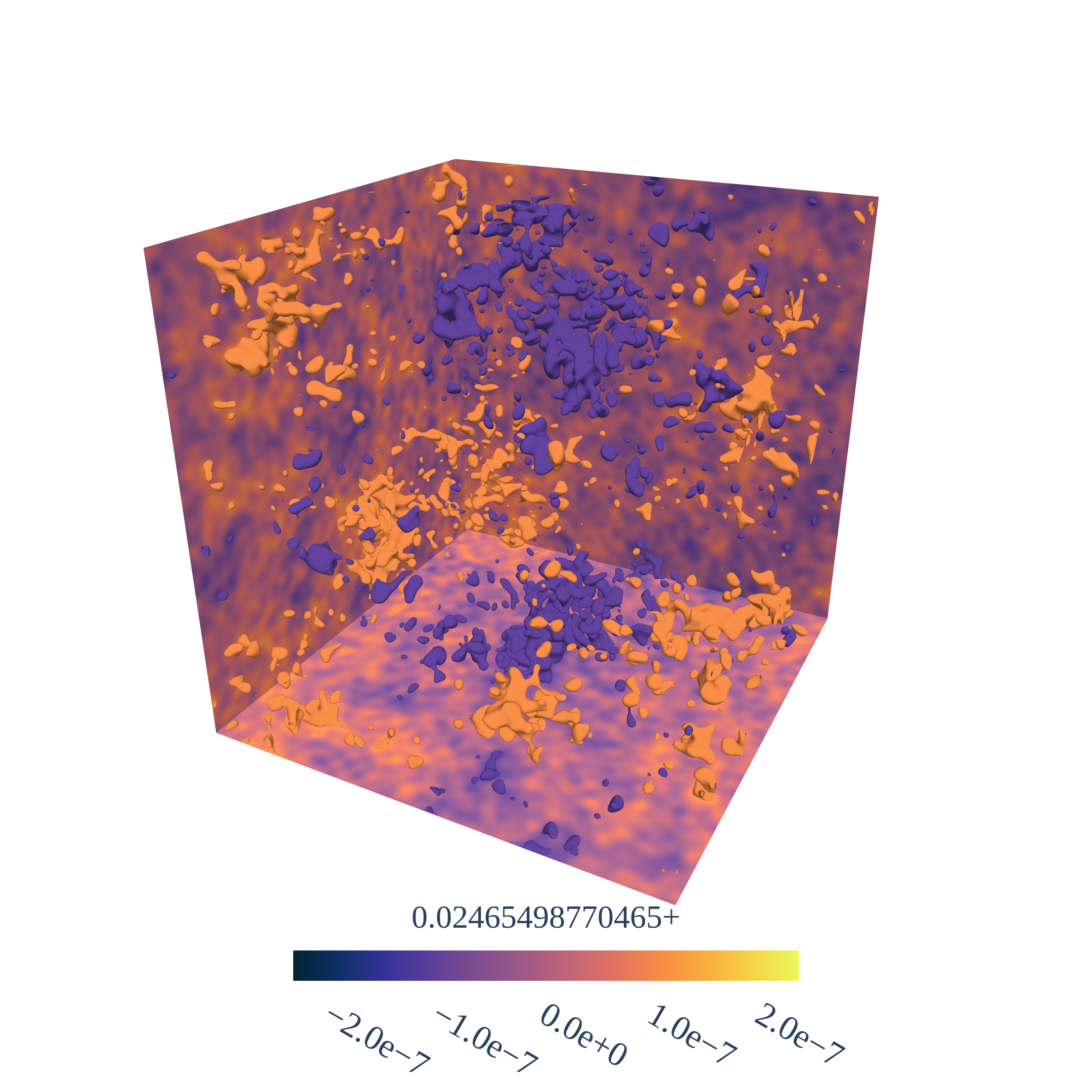}
        \caption{$\langle {\cal N}\rangle=3.91$}
        \label{fig:3Db}
    \end{subfigure}
\hfill
\begin{subfigure}{0.32\textwidth}
        \centering
        \includegraphics[width=\linewidth]{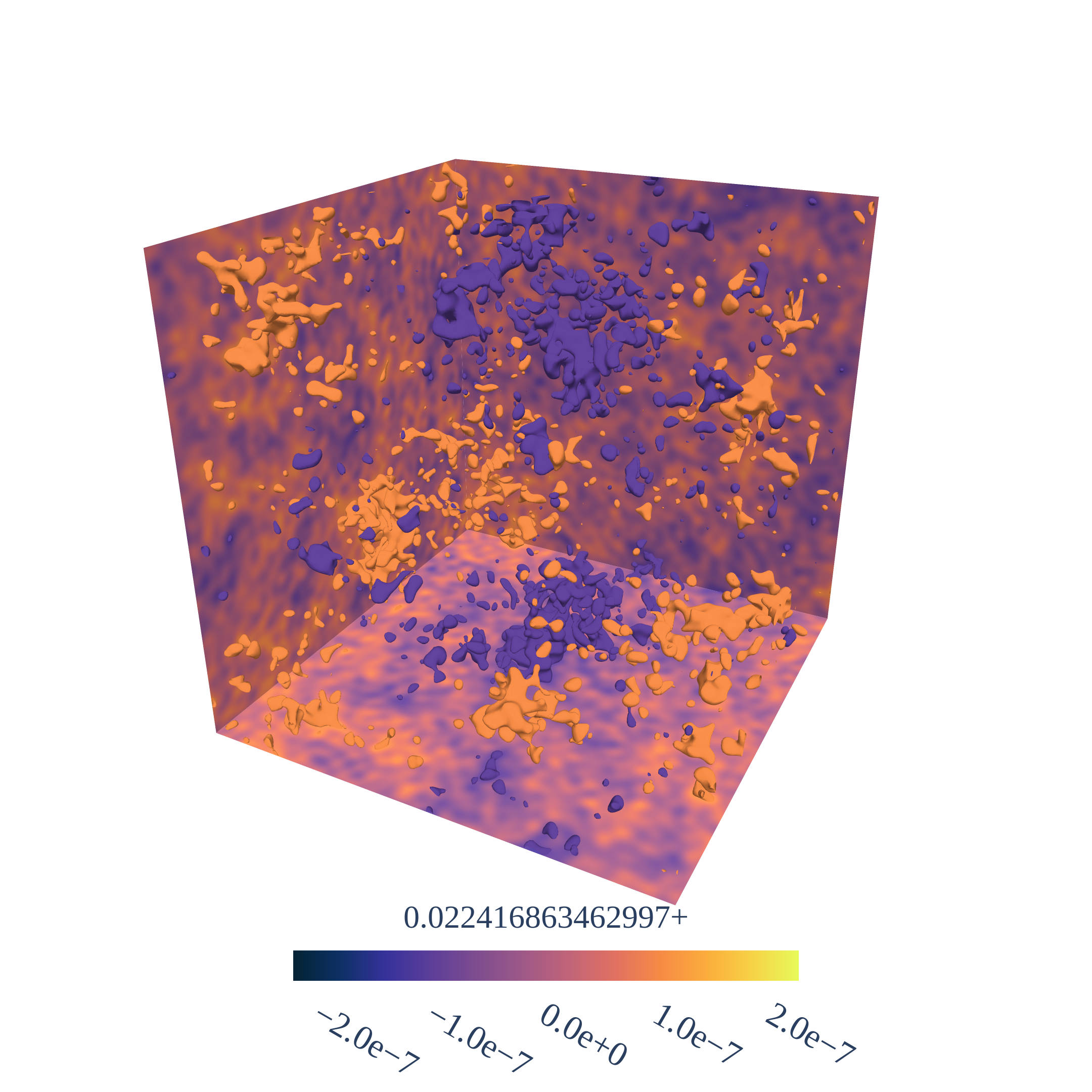}
        \caption{$\langle {\cal N}\rangle=7.64$}
        \label{fig:3Dc}
    \end{subfigure}
\caption{  Energy density evolution (left to right) of the simulated patches for quadratic inflation. 2D boundary heat maps and volumetric isosurfaces share the same colour bar, but not across plots. Isosurfaces are at $\pm 2.5$std. $\Mpl=100$ units. }
\label{fig:3Dplots0}
\end{figure}

\begin{figure}[ht]
\begin{subfigure}{0.43\textwidth}
    \centering
    \includegraphics[width=\linewidth]{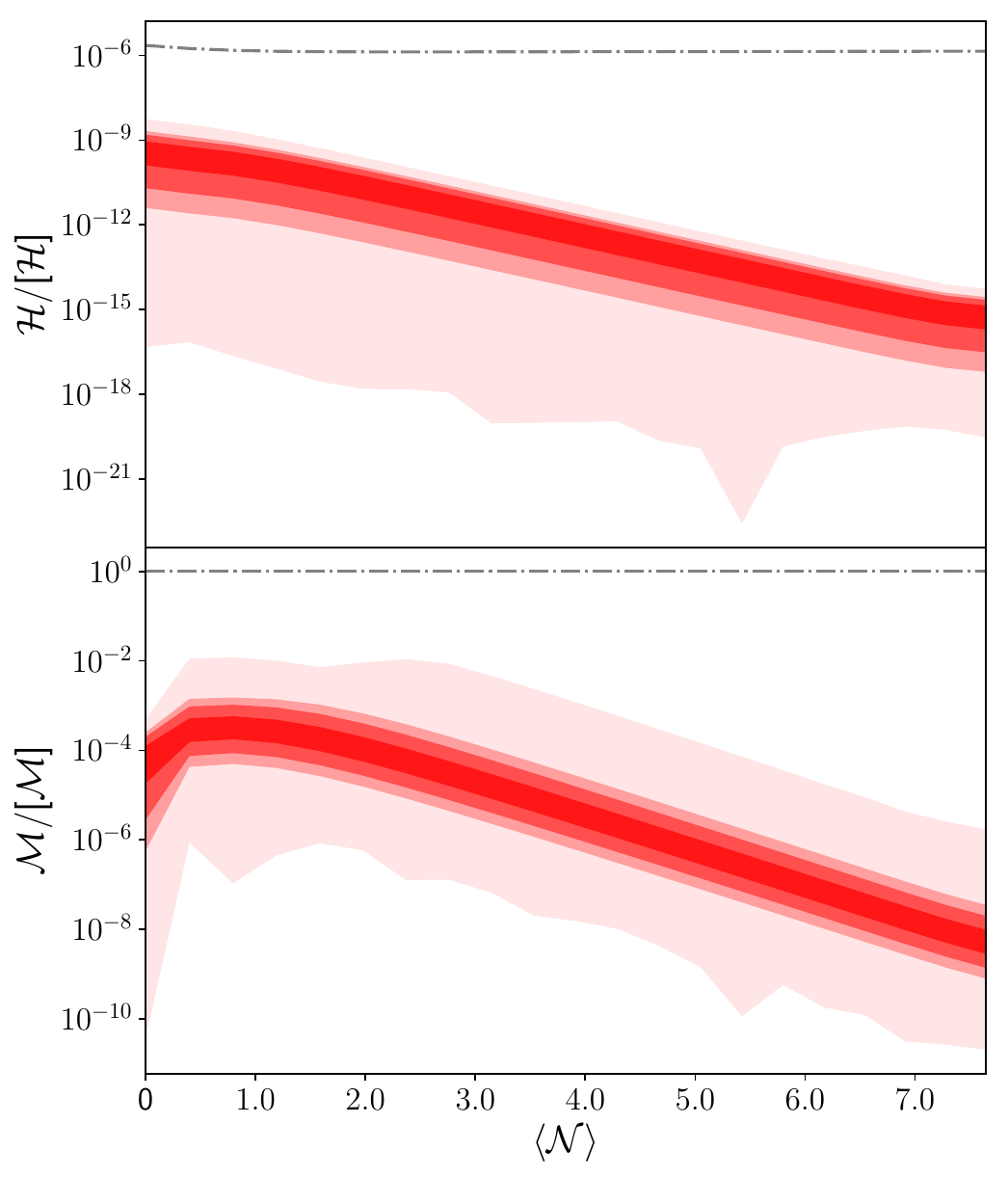}
    \caption{\sloppy Distribution of the relative constraint violations over the lattice at each time step for quadratic inflation, using $68$, $95$, $99$ and $100$ percentile contours (dense to transparent red). The dash-dotted line displays the average first-order Hamiltonian and momentum constraint magnitudes $\langle[{\cal H}^{(1)}]\rangle/\langle[{\cal H}]\rangle$ and $\langle[{\cal M}^{(1)}]\rangle/\langle[{\cal M}]\rangle = 1$.}
    \label{fig:quadviolations}
\end{subfigure}
\hfill
\begin{subfigure}{0.55\textwidth}
    \centering
    \includegraphics[width=\linewidth]{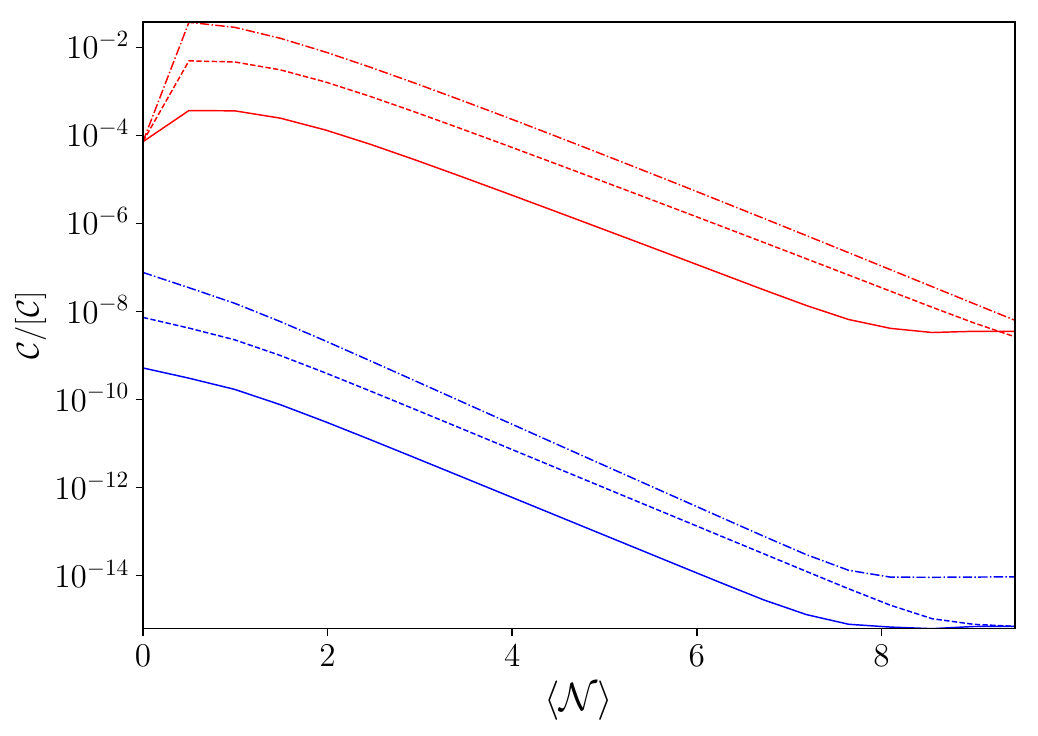}
    \caption{  Mean Hamiltonian (blue) and momentum (red) relative constraints ($\langle|{\cal C}|/[{\cal C}]\rangle$, ${\cal C }= {\cal H}, {\cal M}$) across multiple runs with $N=64$ (dash-dotted), $N=128$ (dashed), and $N = 256$ (solid). Note that the stochastic initial conditions are the same ($\sigma = 3$).}
    \label{fig:ConvTests}
\end{subfigure}
\caption{Evaluation of correctness (left) and convergence (right) of the quadratic inflation simulations. $[{\cal H}]$ and $[{\cal M}]$ are the usual sums of respective absolute terms in the constraints, see eqn.~\eqref{eq:abscons}.}
\end{figure}

As Section~\ref{sec:ICNRicproblem} emphasised, the constraints need to be perfectly satisfied for correctness.
To measure this, we define  the \textit{absolute constraint magnitudes} $[{\cal H}]$ and $[{\cal M}]$, calculated by summing the absolute value of the terms in the BSSN version of the constraints in eqn.~\eqref{eq:constraints}
\begin{subequations}
    \begin{align}
[{\cal H}] & \equiv  \big|{ }^{3}R\big|+\big|\frac{2}{3} K^2\big|+\big|\tilde{A}_{i j} {A}^{i j}\big|+\big|2 \Mpl^{-2} \rho\big|, \\
[{\cal M}_i] & \equiv \big|\tilde{\gamma}^{kl}  \partial_k\tilde{A}_{il} + 2\tilde{\gamma}^{kl}\tilde{\Gamma}^m_{l(i}\tilde{A}_{k)m}\big| 
 +\big| 3\tilde{\gamma}^{kl}\tilde{A}_{ik}\frac{\partial_l\chi}{2\chi}\big| +\big|\frac{2}{3} K_{\mid i}\big|+\big|\Mpl^{-2} {\cal { J}}_i\big|,\\
[{\cal M}] & \equiv \sqrt{\sum_i [{\cal M}_i]^2}.
    \end{align}
    \label{eq:abscons}
\end{subequations}
{These will typically show the magnitude of the different individual terms before algebraic cancellation.
We require our simulations to have $|{\cal H}|/[{\cal H}]\ll 1$ and $|{\cal H}|/[{\cal H}]\ll 1$ as a measure of how well the constraints are satisfied.}
Note that these magnitudes are different in the case of inhomogeneous inflation: $[{\cal H}]$ is usually of order $H^2$ coming from the background, while $[{\cal M}]$ has a leading term at linear order of the perturbations since the Friedmann homogeneous universe trivially satisfies the momentum constraint.
This implies that if our constraints are satisfied at linear order, $|{\cal H}|/[{\cal H}]$ should show the ratio of two orders of perturbation theory while only one for $|{\cal M}|/[{\cal M}]$.\footnote{Note that it is still possible to look at the gap between our Hamiltonian violation and the first-order magnitude by computing $[{\cal H}^{(1)}]/[{\cal H}]$.}

Figure \ref{fig:quadviolations} shows the lattice distribution of these two ratios at each time step.
As a key achievement of Chapter~\ref{sec:ICNR} in the construction of realistic initial conditions for NR, we observe a satisfaction much below the linear magnitude, hence a violation at second order in perturbation theory initially.
If the sign of any quantity is changed in eqn.~\eqref{eq:eqInitialConditions}, we observe a linear-order violation of the constraints, thus confirming the robustness of our analytical result and its implementation.
After the initial time, it is worth noting that a natural damping of the constraint violation occurs, owing to the expansion scaling.
This means that interactions at higher-order in perturbation theory can be safely investigated on our lattice.

Finally, to check that the error built during the evolution is numerical and that it reduces appropriately when increasing resolution while physical scales (and so cutoffs $k_{\rm min,max}^{\cal R}$) are fixed, one performs so-called convergence tests.
In NR, the constraints are once again useful here because of their direct link to the physicality of the solution. In this case, the scaling one should observe for the constraint ${\cal C}(N)$ from a 3D discrete simulation with $N\gg 1$ is \cite{radia_lessons_2022}
\begin{equation}
    \frac{{\cal C}(N_1)}{{\cal C}(N_2)} = \left ( \frac{N_2}{N_1}\right )^n,
\end{equation}
where $n$ is the spatial order of convergence of the stencils used, $n=4$ in our \grc example.
Figure~\ref{fig:ConvTests} gives the right scaling ($\simeq ({128}/{256})^4\simeq 0.06$) in the time evolution for the $N = 128$ to $N=256$ transition.
Initial constraints can be overlooked for convergence as they are built from our linear solutions, as long as they match our theoretical expectations detailed above. 
Final scalings can also become uninformative as the violation stagnates at the bottom of numerical accuracy.
\section{Nonperturbative Inflations} \label{sec:ICsimNPinf}

In the following, we study models known for their non-perturbative dynamics using full GR, providing tests of our pipeline against {more extreme scenarios compared to the slow-roll case studied above}.
These have been studied either theoretically or numerically in the past, but never in full general relativity.
\begin{figure}[ht]
    \centering
         \begin{subfigure}{0.32\textwidth}
        \centering
\includegraphics[width=\linewidth]{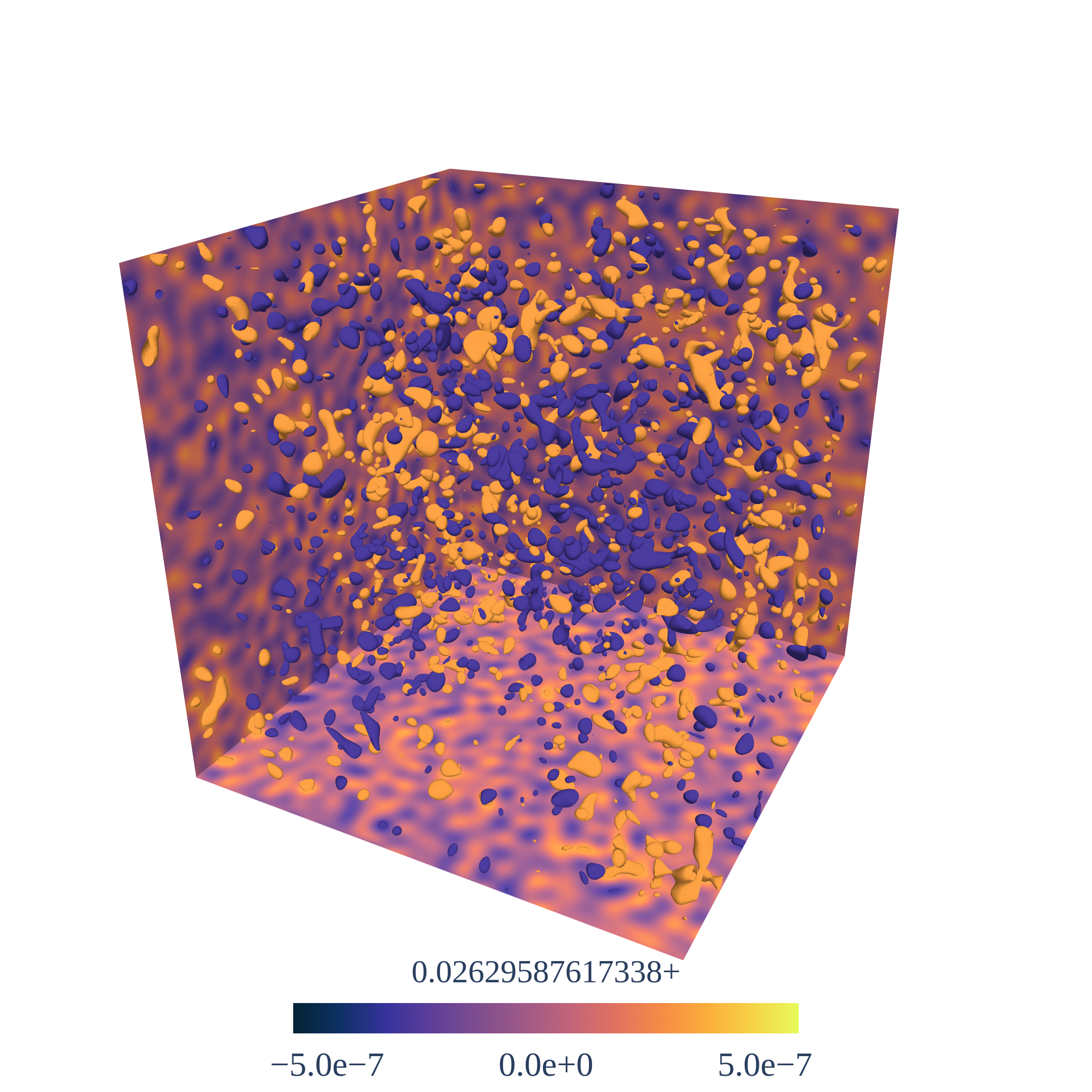}
    \caption{Inflection at $\langle {\cal N}\rangle=0$}
    \label{fig:3Dd}
    \end{subfigure}
    \hfill
\begin{subfigure}{0.32\textwidth}
        \centering        \includegraphics[width=\linewidth]{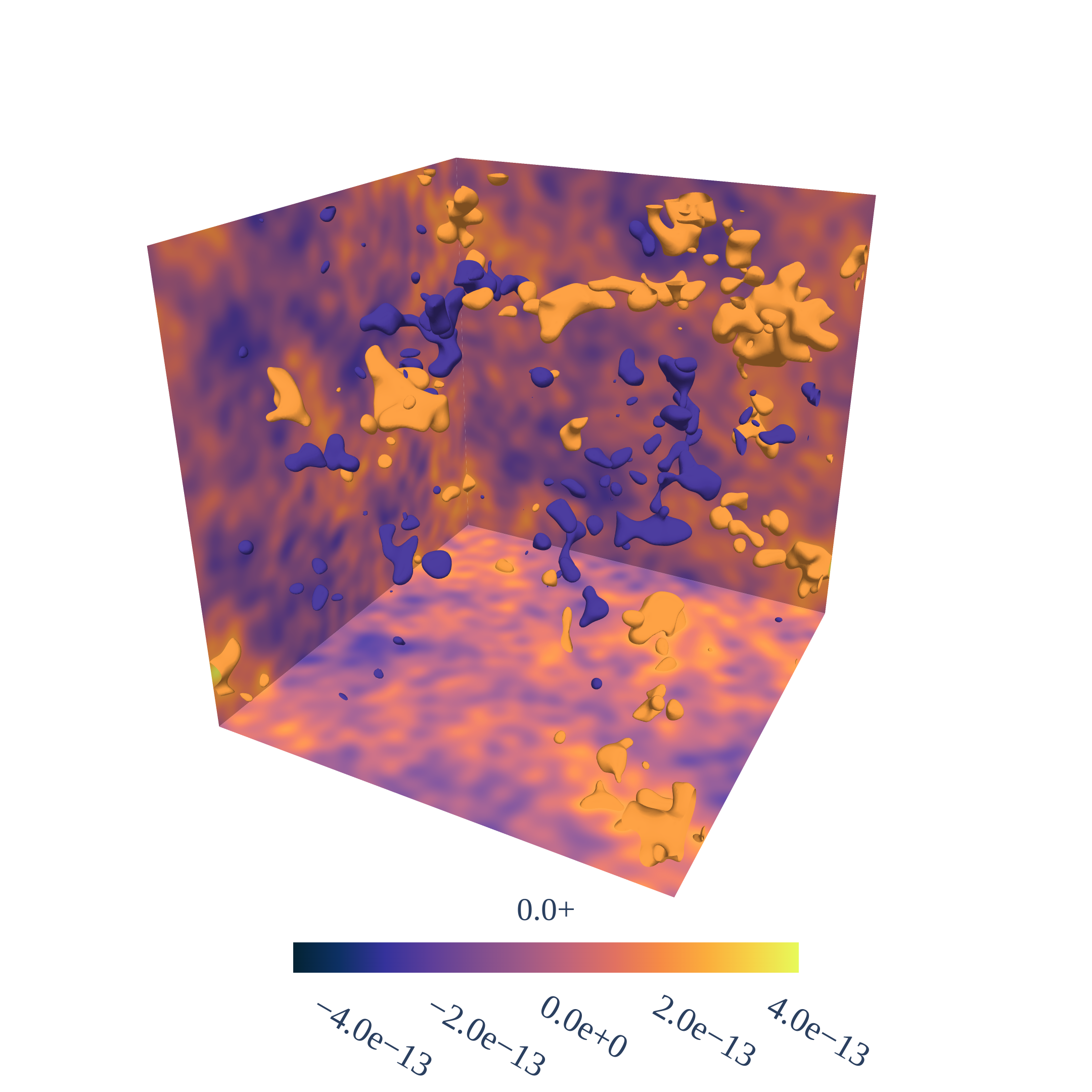}
        \caption{Inflection at $\langle {\cal N}\rangle=4.82$}
        \label{fig:3De}
    \end{subfigure}
\hfill
\begin{subfigure}{0.32\textwidth}
        \centering
        \includegraphics[width=\linewidth]{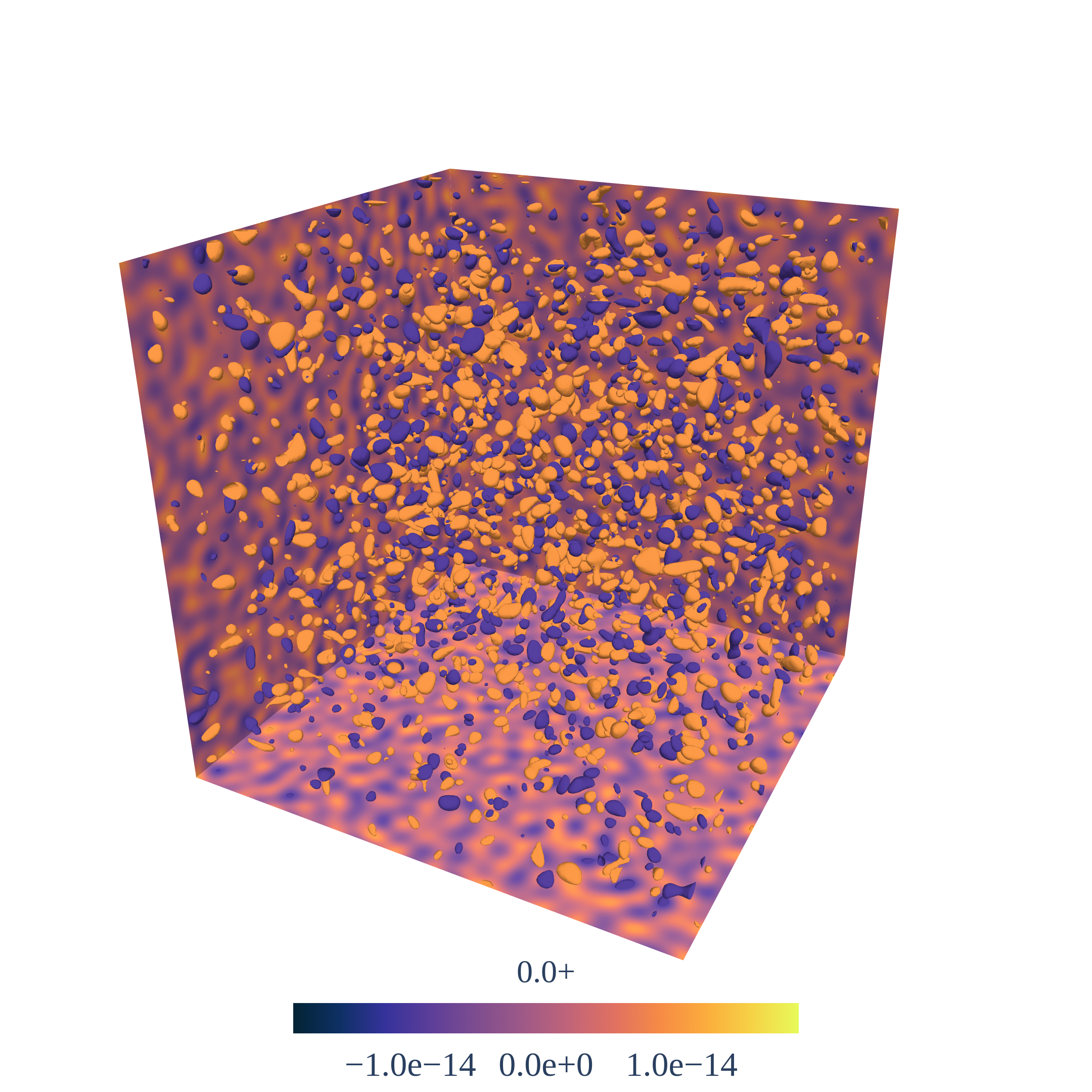}
        \caption{Inflection at $\langle {\cal N}\rangle=9.58$}
        \label{fig:3Df}
    \end{subfigure}

    \centering
        \begin{subfigure}{0.32\textwidth}
        \centering
\includegraphics[width=\linewidth]{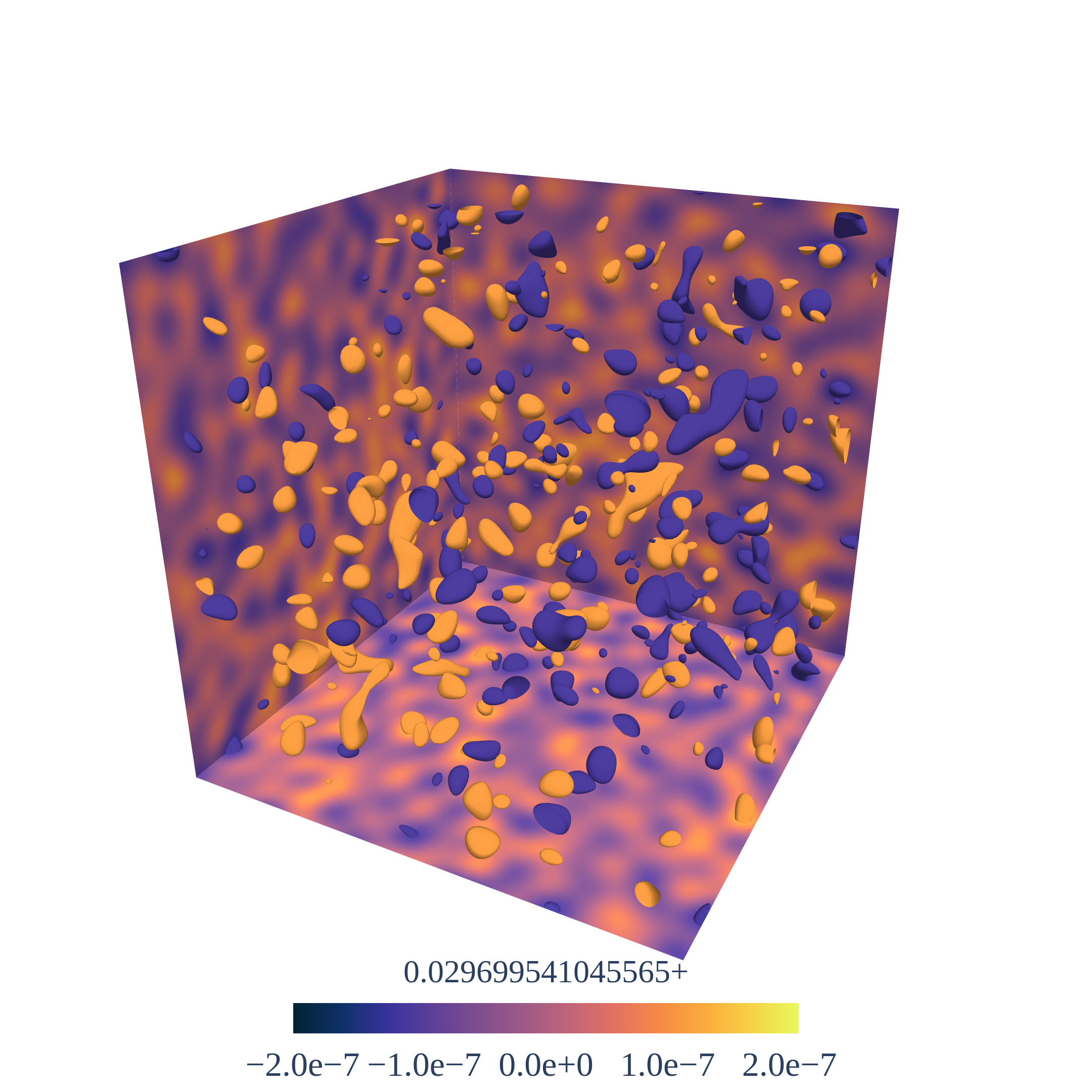}
    \caption{Strong resonance at $\langle {\cal N}\rangle=0$}
    \label{fig:3Dg}
    \end{subfigure}
    \hfill
\begin{subfigure}{0.32\textwidth}
        \centering        \includegraphics[width=\linewidth]{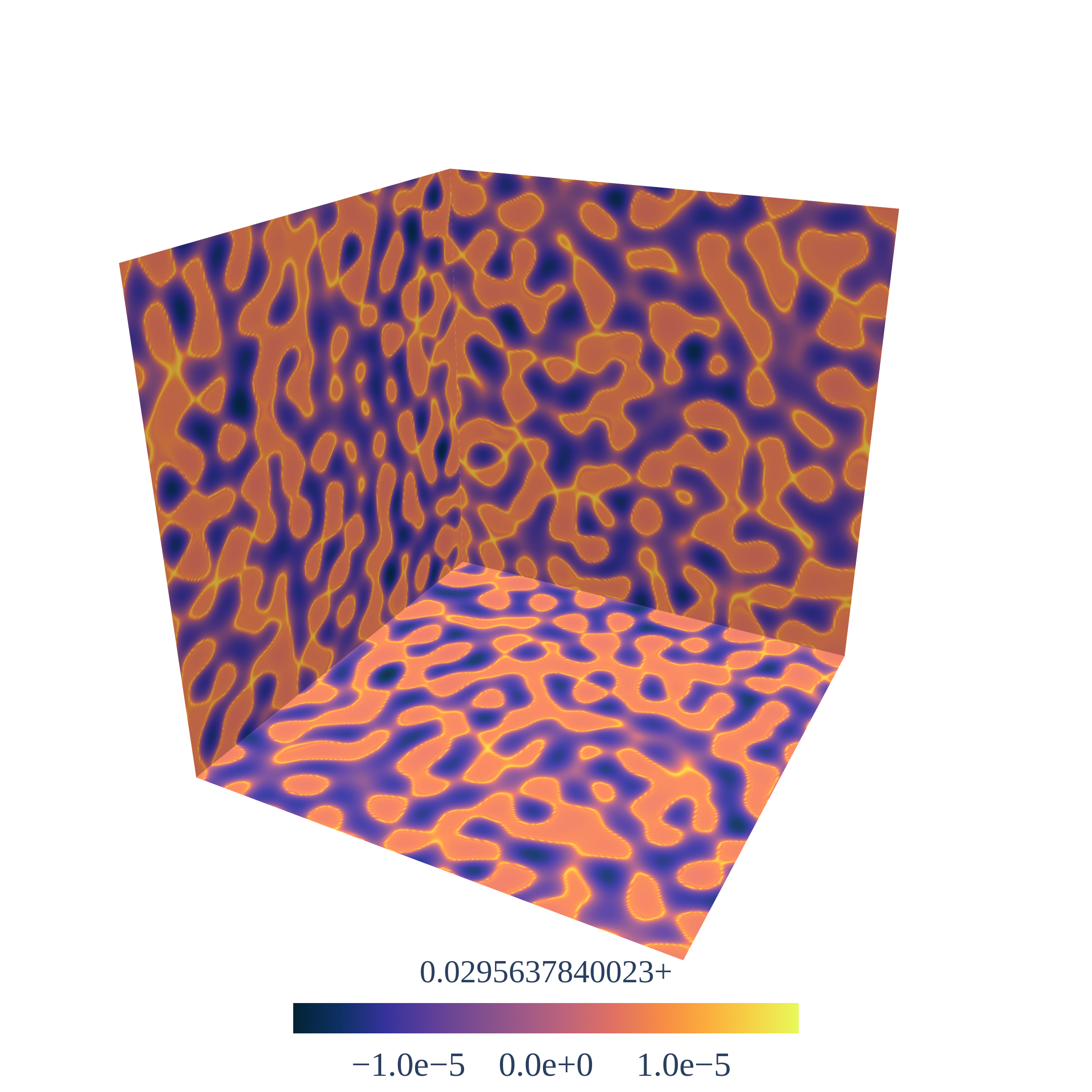}
        \caption{Strong resonance at $\langle {\cal N}\rangle=3.83$}
        \label{fig:3Dh}
    \end{subfigure}
\hfill
\begin{subfigure}{0.32\textwidth}
        \centering
        \includegraphics[width=\linewidth]{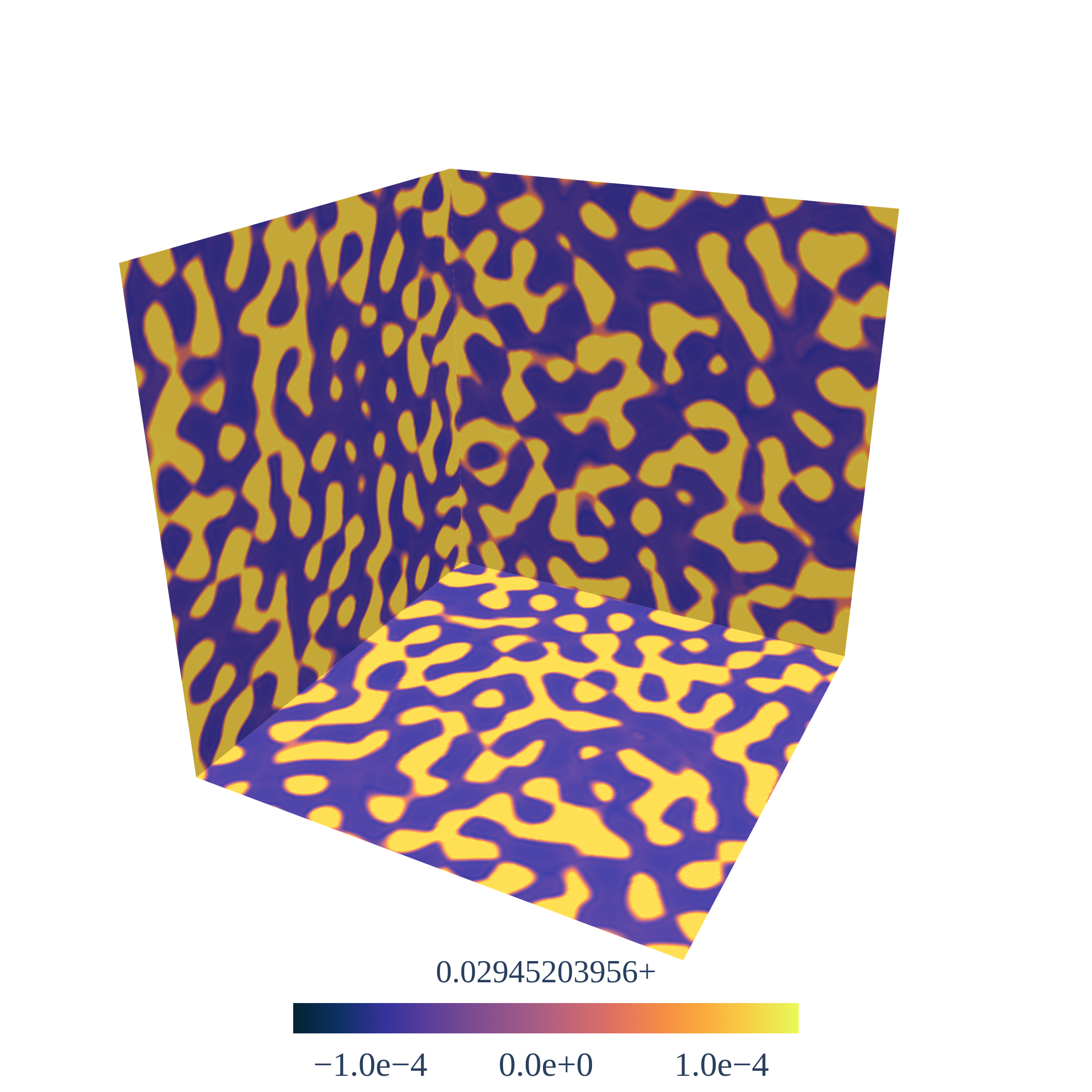}
        \caption{Strong resonance at $\langle {\cal N}\rangle=6.98$}
        \label{fig:3Di}
    \end{subfigure}
\caption{ Energy density evolution (left to right) of the simulated patches for the two non-perturbative studied models (top to bottom). 2D boundary heat maps and volumetric isosurfaces share the same colour bar, but not across plots. Isosurfaces are at $\pm 2.5$std for the inflection example. Note that the resonance is not shown with any isosurface for better visibility. $\Mpl=100$ units. }
\label{fig:3Dplots}
\end{figure}
\subsection{Inflection point inflation}\label{sec:inflection}

We study a single field rolling on the following potential
\begin{equation}
V(\phi) = \Lambda v^4 \frac{\phi^2}{3} \frac{3\phi^2 + 2\sqrt{2}\phi v + 6v^2}{(3\phi^2 + 2v^2)^2},
\label{eq:potProkopec}
\end{equation}
inspired by Higgs inflation \cite{Hamada14,Bezrukov14} and intended to be tuned for the production of primordial black holes (PBHs) in \cite{GarciaBellido17,germani17}.
It produces three phases around the inflection point at $\phi_{inf}\simeq 0.28\Mpl$: ultra slow-roll (USR) sandwiched by slow-roll phases.
The USR phase is key because it provokes an enhancement of the adiabatic power spectrum.
In \cite{germani17}, parameters are tuned to have that peak on sub-CMB scales but we change slightly $v$ to see interesting dynamics within $10$ e-folds.
Our parameters are $\Lambda = 1.86\times 10^{-6}$, $v = 0.198\Mpl$ and we start slow-rolling before the feature with $\phi_0 = -1.2\Mpl$ using the \textit{$H_0$-fixing} method described in Section~\ref{subsec:background}.
These parameters give a decent duration of USR ($\varepsilon_1\simeq 0$, $\varepsilon_2\simeq 6$, yielding a field that almost stops) as Figures \ref{fig:usrrolls} and \ref{fig:meanUSR} showcase. 

\begin{figure}[ht]
    \begin{subfigure}{0.49\textwidth}
        \centering
        \includegraphics[width=\linewidth]{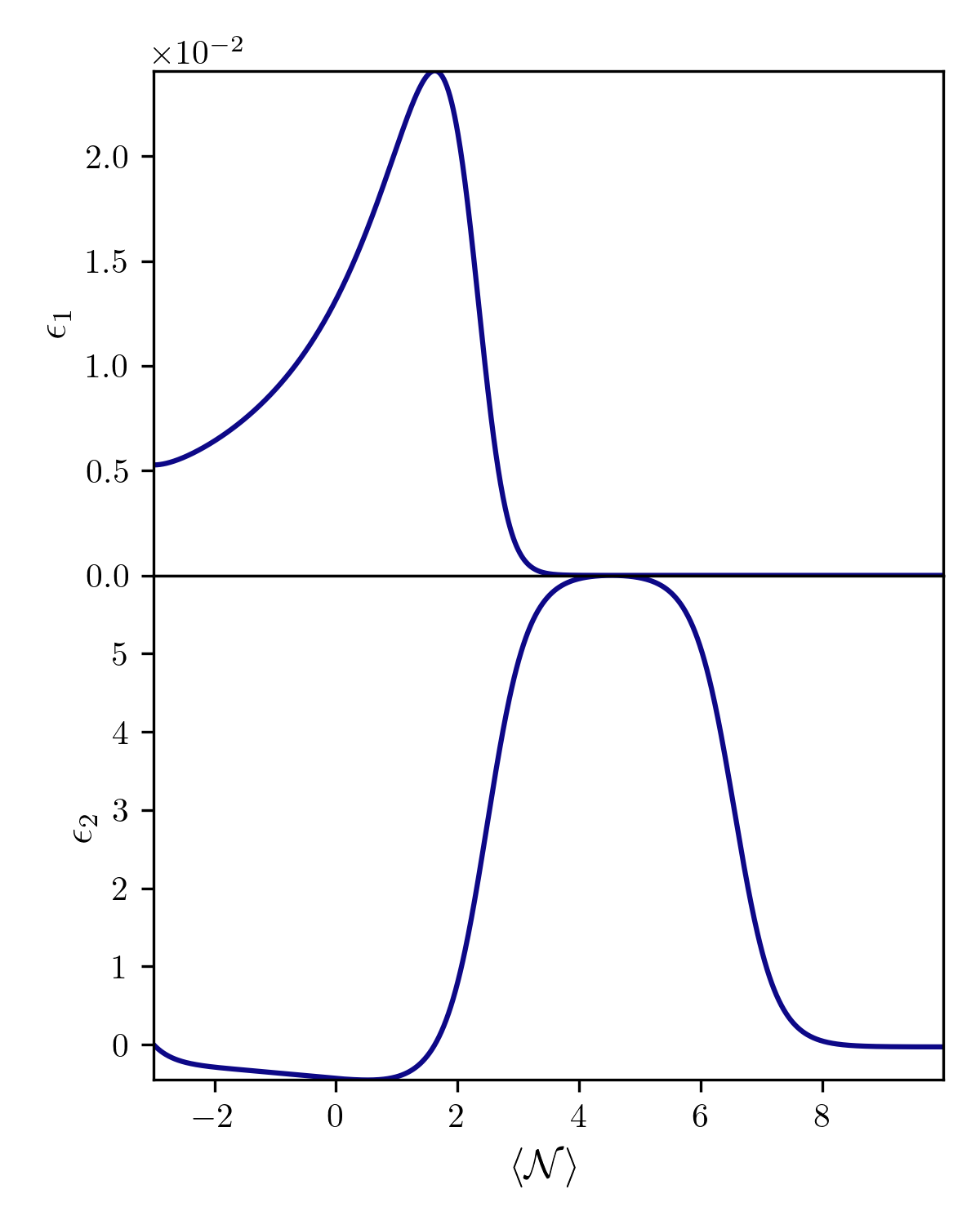}
        \caption{ First (top) and second (bottom) slow-roll parameters.}
        \label{fig:usrrolls}
    \end{subfigure}
    \hfill
    \begin{subfigure}{0.49\textwidth}
            \includegraphics[width=\linewidth]{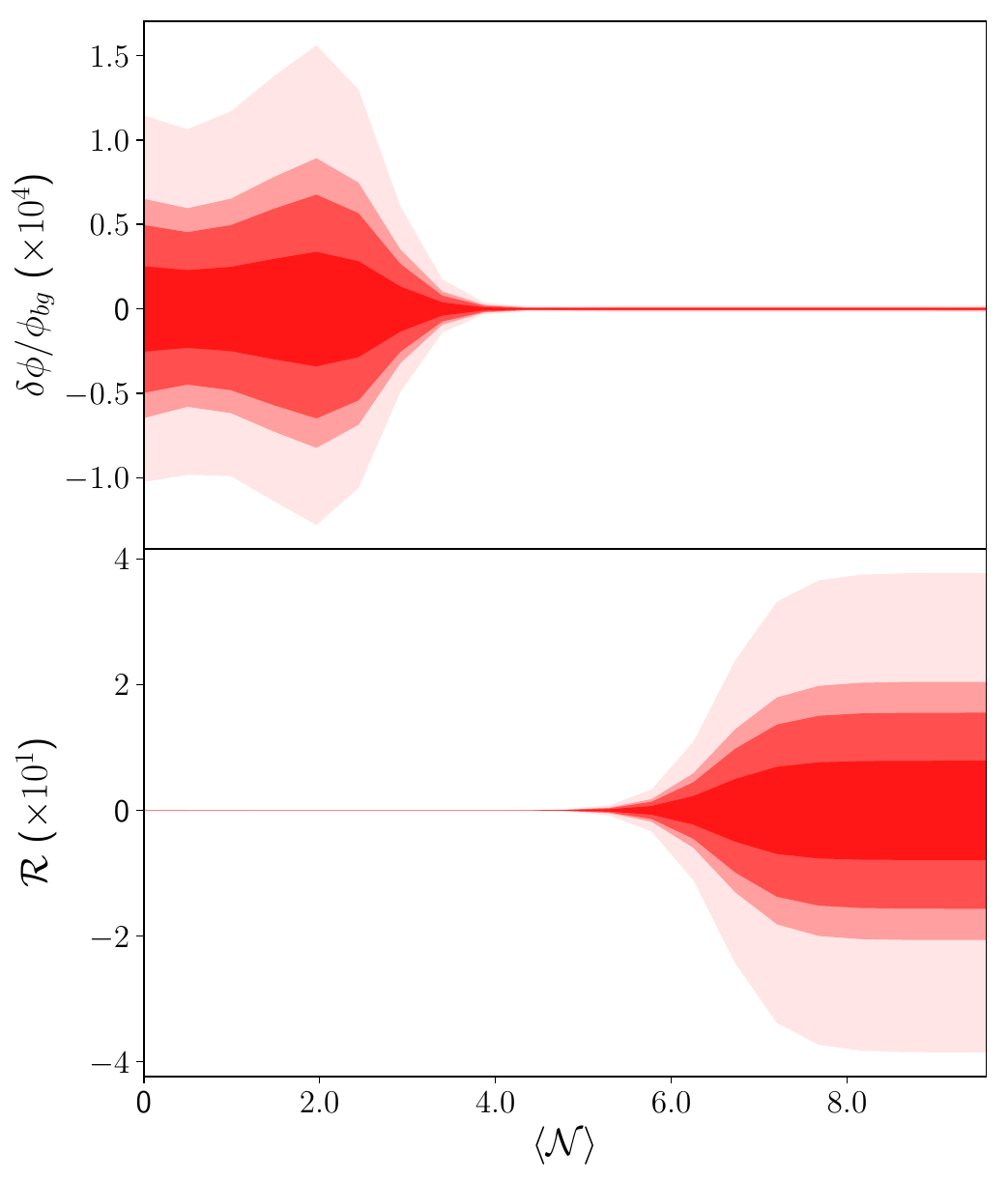}
        \caption{ Contrast of $\phi$ and ${\cal R}$ using $68$, $95$, $99$ and $100$ percentile contours (dense to transparent red).}
        \label{fig:usr_rlin}
    \end{subfigure}
\caption{Background (left) and inhomogeneities (right) of the inflection inflation case in $\Mpl=100$ units.}
\end{figure}

The patch is designed as follows: modes up to $\sigma = 4$ are drawn and accounted for in a patch of size $L=32/H^{\circledast}$.
The resulting lattice average of the simulation in Figure \ref{fig:meanUSR} matches the Friedmann prediction perfectly.

This example is the first to showcase a non-perturbative scenario. In fact, the contrasts can be summarised in that of ${\cal R}$ and show a real space departure from perturbativity in Figure \ref{fig:usr_rlin}: up to a $30\%$ difference from the background.
One would thus expect a departure from linear theory and perturbation theory in general.
The spectrum is plotted in Figure \ref{fig:specUSR}. The simulation shows indeed {a small deviation from} linear theory compared to the quadratic case {at low $k$}.
We expect this gap to be due to starting modes at different times compared to their crossing, as this gap will disappear in Chapter~\ref{sec:SINR}.
We find no visible difference between ${\cal R}$ and ${\cal R}^{\rm NL}$\ylc{. As discussed in Section~\ref{sec:finalvalues}, this only shows that the amplitude of the fluctuations remains small enough for the linear extraction to hold. It does not mean that the complicated and less perturbative interactions occurring at the inflection's phase transition are absent, since those would affect both quantities alike}.
\begin{figure}[ht]
    \begin{subfigure}{0.38\textwidth}
        \centering
        \includegraphics[width=\linewidth]{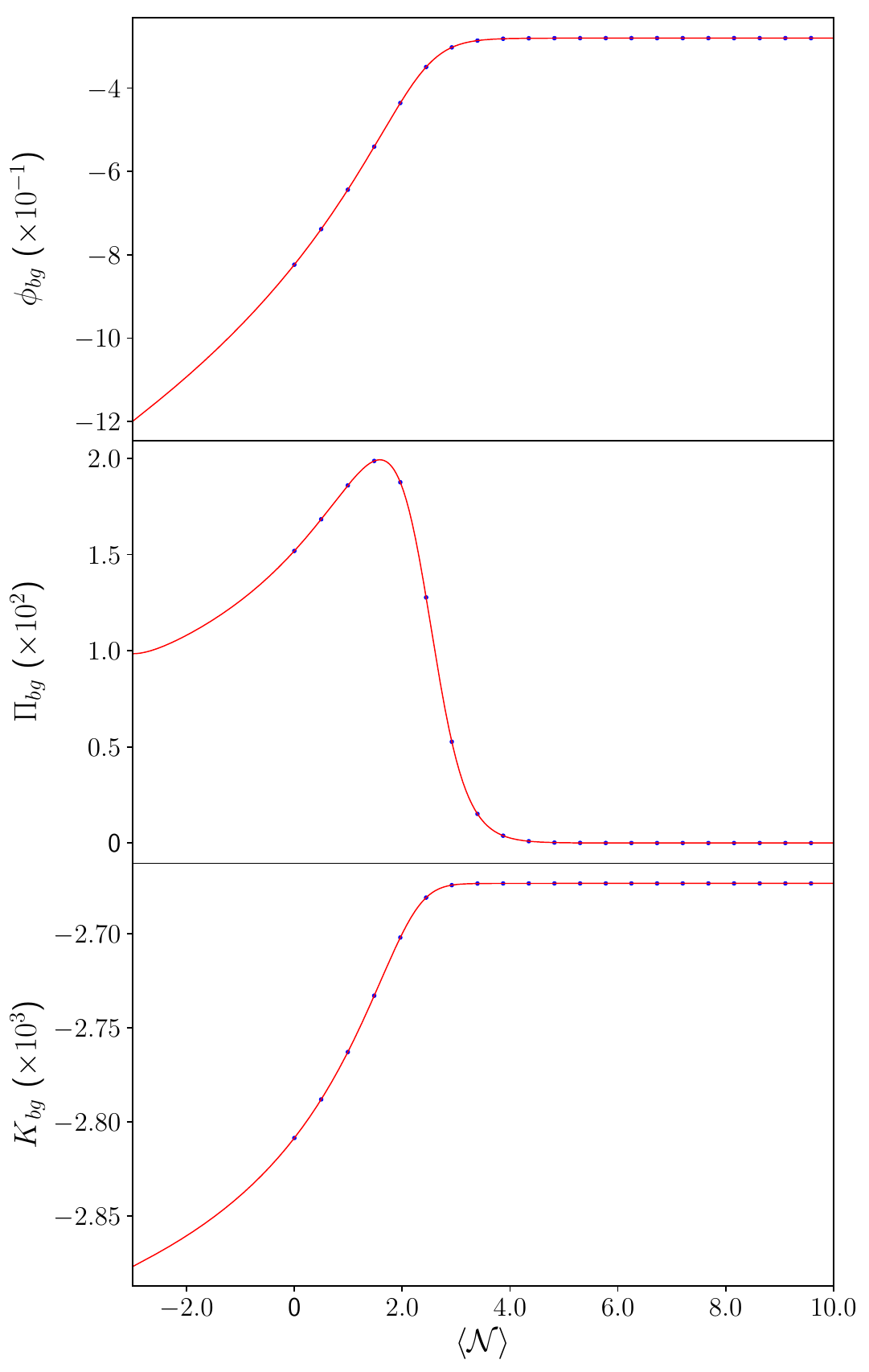}
        \caption{Comparison of the Friedmann solution (solid red) and the mean inflaton  (top), conjugate momentum (middle), and extrinsic curvature (bottom) fields extracted from the GR simulation (blue dots).}
        \label{fig:meanUSR}
     \end{subfigure}
    \hfill
    \begin{subfigure}{0.60\textwidth}
        \includegraphics[width=\linewidth]{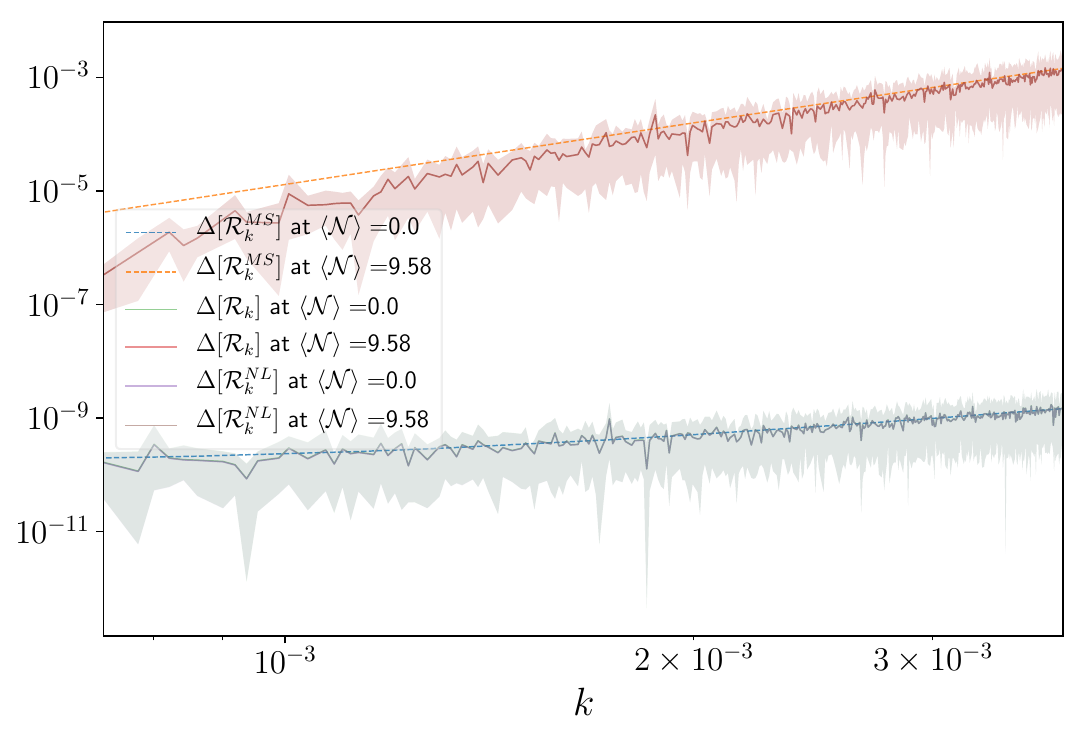}
        \caption{ Linear (dashed, ${\cal R}^{MS}$) and extracted (solid, ${\cal R}$) binned dimensionless spectra ($\Delta[{\cal R}]= \frac{k^3}{2\pi^2} |{\cal R}_k|^2$) compared to extracted ${\cal R}^{\rm NL}$, a scalar built from the fully non-linear generalisation ${\cal R}_i${: perfect superimposition at initial (lower half) and final (upper half) times ({\textcolor[HTML]{95d095}{\rule{1em}{0.5em}}} $+$ {\textcolor[HTML]{cab3de}{\rule{1em}{0.5em}}} $=$ \textcolor[HTML]{608475}{\rule{1em}{0.5em}}, {\textcolor[HTML]{ea9393}{\rule{1em}{0.5em}}} $+$ {\textcolor[HTML]{c6aba5}{\rule{1em}{0.5em}}} $=$ \textcolor[HTML]{b13e3a}{\rule{1em}{0.5em}})}. Contours account for the $3\sigma$-range of points in each bin. Displayed Fourier bins are those of the simulation mesh which have more than 10 realisations and which are smaller than the spectral cutoff of the input ($k_{max}^{\cal R} = 4\times 10^{-3}$).}
        \label{fig:specUSR}
    \end{subfigure}
\caption{Background and fluctuations in Fourier space of the inflection inflation case in $\Mpl=100$ units.}
\end{figure}
Interestingly, the {field}'s contrast shrinks greatly during the USR phase, which in turn benefits that of ${\cal R}$ through the geometry, see Figure \ref{fig:usr_rlin}.
The drop in $\delta\phi$ appears analogous to what the long wavelength and separate universe Hamilton-Jacobi theory predicts for motion on a plateau \cite{prokopec_ensuremathdeltan_2021}, showing consistency with that idealised scenario, as well as the similar decay shown in \cite{Prokopec:2025uvz}.
The evolution of the energy density is also illustrated in Figures \ref{fig:3Dd}, \ref{fig:3De}, and \ref{fig:3Df}, which show that clustering has been altered during the phase transition as opposed to the quadratic case.

Most importantly, such scenarios are of interest to those studying primordial non-Gaussianity. In Figure \ref{fig:cumulUSR} we report the influence of the transition dynamics on the skewness and kurtosis. Leaving a thorough study for future work, we refer to \cite{Florio2024} for another example in the tensorial sector.

\subsection{Strong resonance}\label{subsec:strongRes}
\begin{wrapfigure}{r}{0.45\linewidth}
    \centering
    \includegraphics[width=\linewidth]{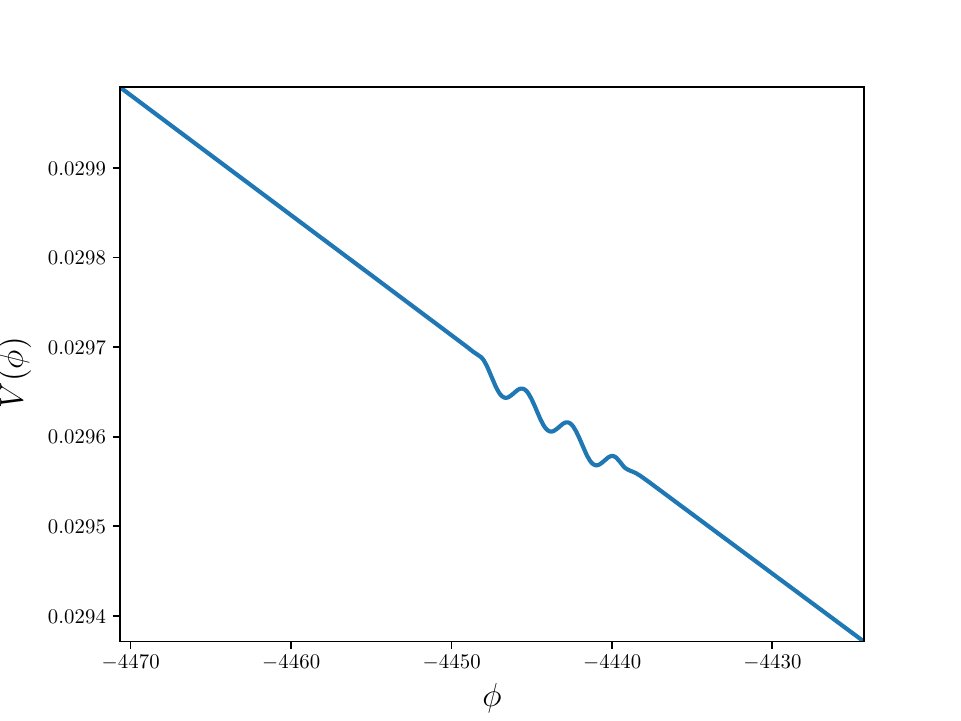}
    \caption{\sloppy Monodromic feature on the quadratic potential. The range in $\phi$ is the one explored during the simulations. $\Mpl=100$ units.}
    \label{fig:potmono}
    \vspace{-2.0\baselineskip}
\end{wrapfigure}
Resonance models have received significant attention as inflationary models with non-trivial features \cite{chen_generation_2008,Flauger10}.
They often consist of a vanilla slow-roll potential, on top of which an oscillatory feature is added. Some models were tested against Planck \cite{PlanckInflation,martin_best_2014-1, Sohn24} without decisive conclusions.
These were still perturbative as, once again, it is our only way to make predictions in QFTCS.
More recently, non-perturbative simulations of inflation gave their first results using lattice cosmology \cite{Caravano2024}, building up from reheating studies \cite{latticeeasy}.

In the cited study, the non-perturbative regime was reached using a strong resonance of the perturbations with the oscillatory feature, inspired by axion monodromy models \cite{Monodromy10}.
Despite offering important insights, these simulations are not fully relativistic and for instance, approximate backreaction and geometry, although some recent improvements clarified the validity of such regimes \cite{caravano2024axionbackreaction,caravano2024usr}.
We believe that our framework can {illuminate} a complete relativistic picture of these processes.
We use a potential similar to that of \cite{Caravano2024}, which is a standard vanilla inflation supplemented by a monodromic feature of width $\Delta\phi$ and frequency $f$
\begin{equation}
\begin{aligned}
   V(\phi)= &  \frac{1}{2}m^2\phi^2+\Lambda^4{\cal W}(\phi-\phi_c)\left(\cos\left(\frac{\phi-\phi_c}{f}\right)-1\right), \\
   \text{where }&{\cal W}(\phi) =  \frac{1}{4}\left(1+\tanh \left(\frac{\phi}{f}\right)\right)\left(1+\tanh \left(\frac{\Delta \phi-\phi}{f}\right)\right),
   \end{aligned}
\end{equation}
the only difference being that our choice of slow-rolling background is taken to be the same as our vanilla example.\footnote{We thank the authors for clarifying their initialisation methods. Here, we choose a different, easier-to-implement set of initial conditions.}
Our oscillatory feature is located such that $\phi_c$ is the value of $\phi^{\circledast}$ in the quadratic example.
Finally, we set our parameters to $\alpha = \dot{\phi}_c/fH_c = 10$, $b = 2\Lambda^4/\dot{\phi}_c^2 = 1.32$, and $\beta = \Delta\phi/\alpha f = 2$, respectively fixing frequency, amplitude, and duration of the oscillations.
The resulting potential is shown in Figure \ref{fig:potmono}.
\begin{figure}[ht]
\begin{subfigure}{0.49\textwidth}
    \centering
\includegraphics[width=\linewidth]{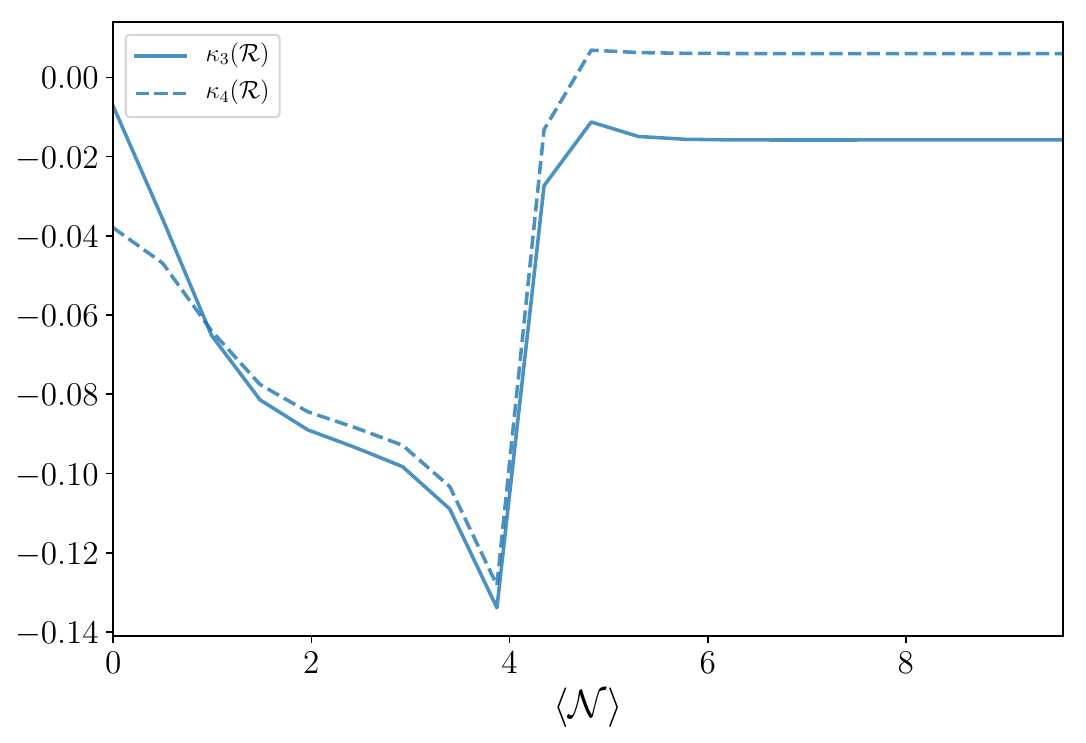}
    \caption{ Cumulants of ${\cal R}$: sample skewness ($\kappa_3$, solid) and kurtosis ($3+\kappa_4$, dashed).}
    \label{fig:cumulUSR}
\end{subfigure}
\hfill
\begin{subfigure}{0.49\textwidth}
    \centering
\includegraphics[width=\linewidth]{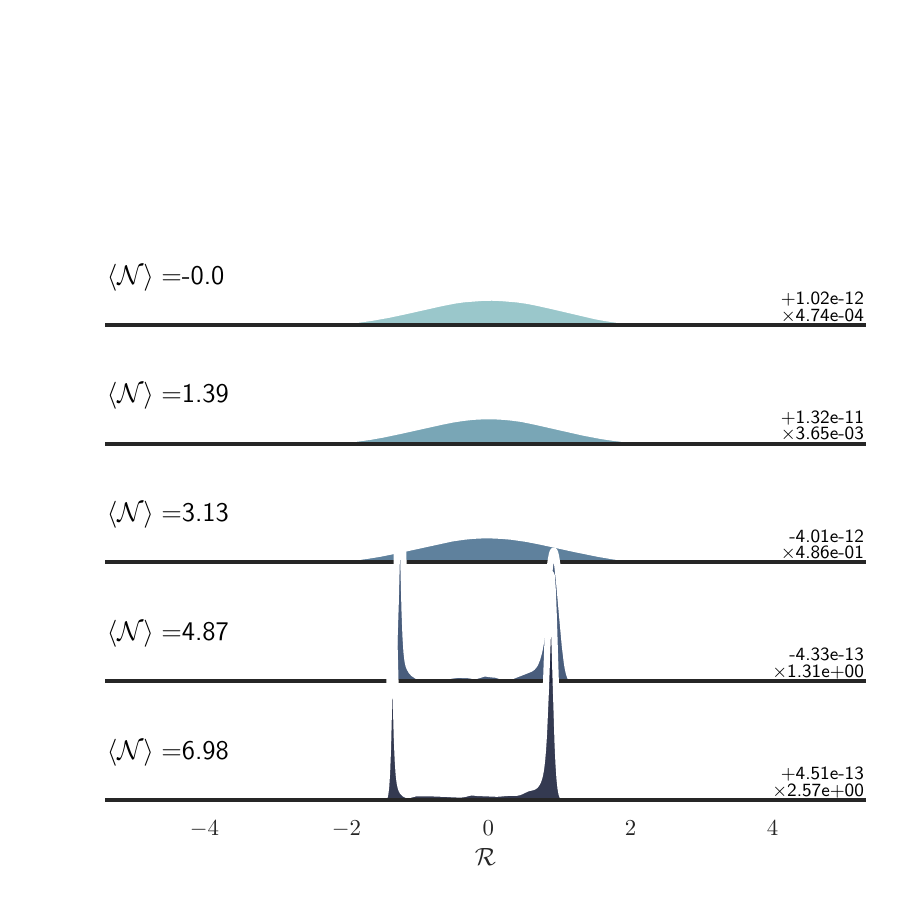}
    \caption{ PDFs of ${\cal R}$: from Gaussianity to bimodal fluctuations.}
    \label{fig:monoPDFs}
\end{subfigure}
\caption{A first look at non-Gaussianity extracted from the GR simulations in the two non-perturbative examples: inflection (left) and strong resonance (right).}
\end{figure}

We start with a range of scales up to $\sigma = 4.5$ in a box of physical size $L = 16/H^{\circledast}$, necessary to probe slightly sub-Hubble scales to observe the resonance with this choice of parameters. In this highly non-perturbative scenario, the background loses its meaning as shown in Figure \ref{fig:meanmono}: compared to the two previous examples, measurable deviations from the Friedmann solution occur after a few e-folds in the oscillating features, with those of $\Pi$ particularly pronounced. In particular, looking at the evolution of the energy density contrasts in Figure \ref{fig:3Dg}, \ref{fig:3Dh}, and \ref{fig:3Di} or at the PDFs in the Joy plot of Figure \ref{fig:monoPDFs} provides a clear picture of the dynamics.
It shows that the perturbations become bimodal and non-perturbative in the gap between the two groups.
A more acute look at the values of the fields shows that some perturbations stay stuck in some of the bumps of the potential: these will undergo eternal inflation in a false vacuum while others proceed successfully en route to the bottom of the potential.
These are the same qualitative conclusions as in \cite{Caravano2024}.

Moreover, we also found an encouraging resemblance between the cited study and our linear spectra extracted in Figure \ref{fig:specmono}.
The departure from linear theory is extreme, but this mostly motivates one to stop looking at a perturbative quantity such as ${\cal R}$.
${\cal R}^{\rm NL}$ shows indeed great independence from the former, thus showing that\ylc{ the amplitude of the fluctuations is large enough for} the linear-order extraction \ylc{to be} no longer valid. This suggests the need for a geometric extraction\ylc{, such as ${\cal R}^{\rm NL}$ itself}.
\begin{figure}[ht]
    \begin{subfigure}{0.38\textwidth}
        \centering
\includegraphics[width=\linewidth]{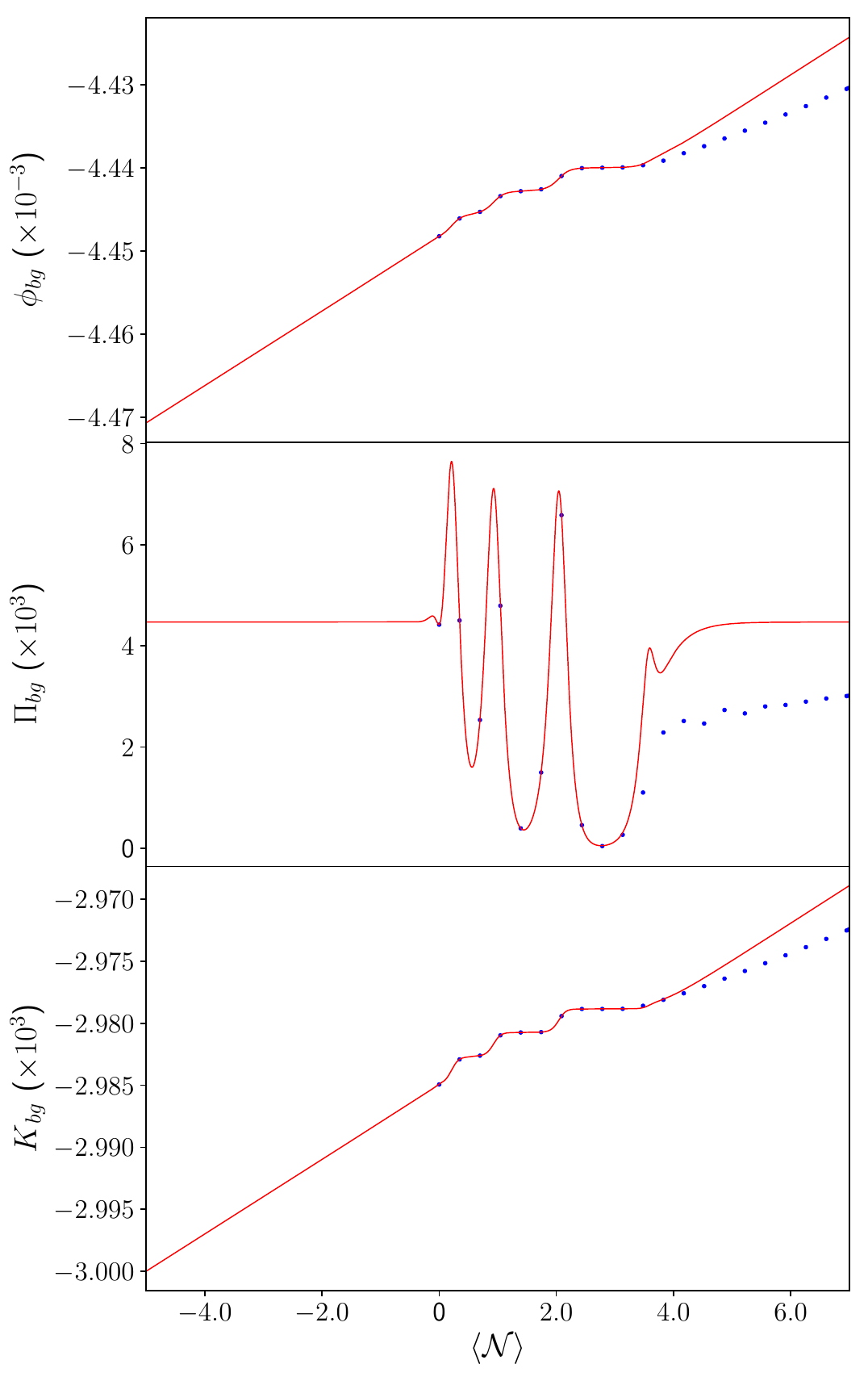}
\caption{ Comparison of the Friedmann solution (solid red) and the mean inflaton (top), conjugate momentum (middle), and extrinsic curvature (bottom) fields extracted from the GR simulation (blue dots). }
 \label{fig:meanmono}
     \end{subfigure}
    \hfill
    \begin{subfigure}{0.60\textwidth}
        \includegraphics[width=\linewidth]{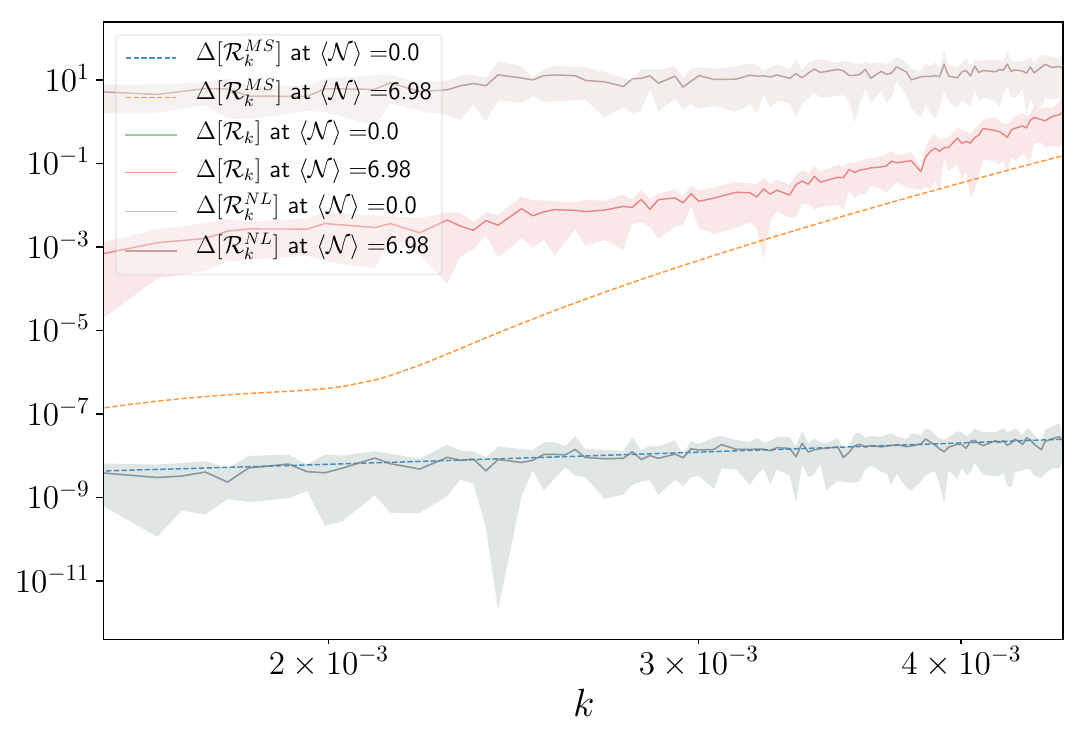}
        \caption{  Linear (dashed, ${\cal R}^{MS}$) and extracted (solid, ${\cal R}$) binned dimensionless spectra ($\Delta[{\cal R}]= \frac{k^3}{2\pi^2} |{\cal R}_k|^2$) compared to extracted ${\cal R}^{\rm NL}$, a scalar built from the fully non-linear generalisation ${\cal R}_i${: perfect superimposition at initial (lower half) time only ({\textcolor[HTML]{95d095}{\rule{1em}{0.5em}}} $+$ {\textcolor[HTML]{cab3de}{\rule{1em}{0.5em}}} $=$ \textcolor[HTML]{608475}{\rule{1em}{0.5em}})}. Contours account for the $3\sigma$-range of points in each bin. Displayed Fourier bins are those of the simulation mesh which have more than 10 realisations and which are smaller than the spectral cutoff of the input ($k_{max}^{\cal R} = 4.5\times 10^{-3}$).}
        \label{fig:specmono}
    \end{subfigure}
\caption{Background and fluctuations in Fourier space of the strong resonance inflation case in $\Mpl=100$ units.}
\end{figure}

\begin{figure}[ht]
    \centering
    \begin{subfigure}{0.49\textwidth}
\includegraphics[width=\linewidth]{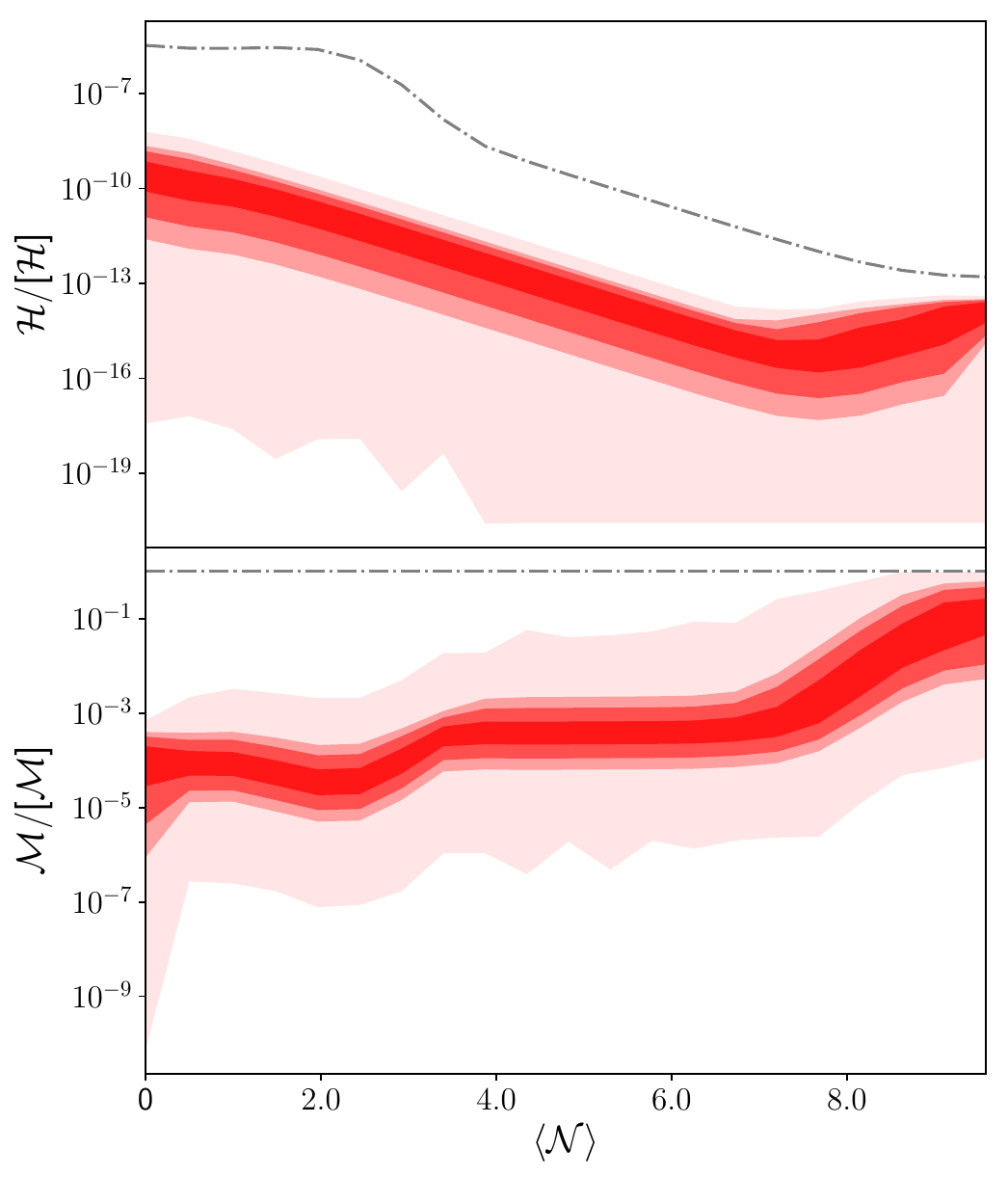}
\caption{Inflection inflation}
 \label{fig:NPconstraintsa}
    \end{subfigure}
    \begin{subfigure}{0.49\textwidth}
\includegraphics[width=\linewidth]{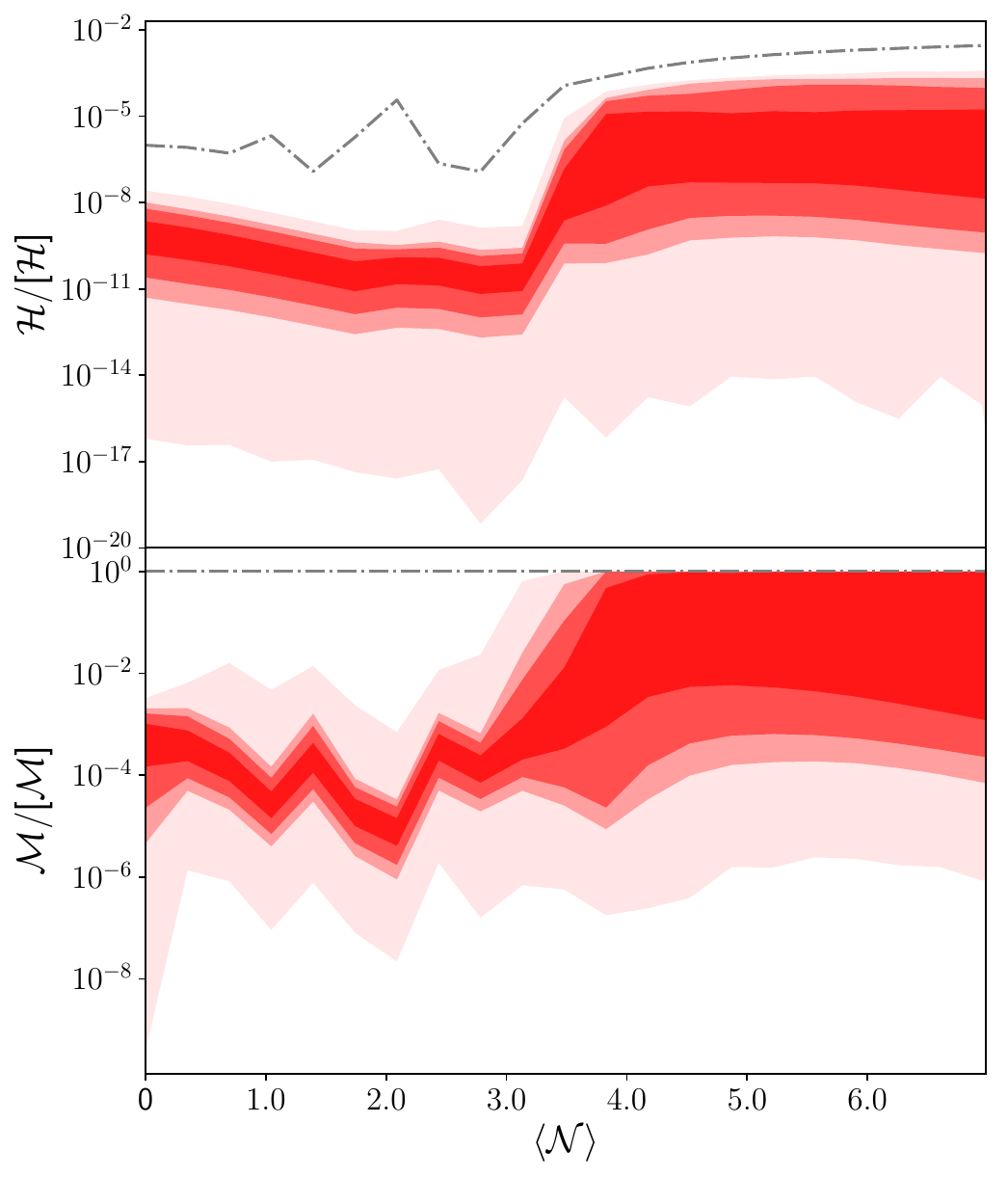}
\caption{Strong resonance }
 \label{fig:NPconstraintsb}
    \end{subfigure}
    \caption{ Evolution of the rescaled constraint violation percentiles for our non-perturbative models. For each time step's box, $68$, $95$, $99$ and $100$ percentile contours are represented (dense to transparent red). The dash-dotted line displays the average first-order Hamiltonian and momentum constraint magnitudes $\langle[{\cal H}^{(1)}]\rangle/\langle[{\cal H}]\rangle$ and $\langle[{\cal M}^{(1)}]\rangle/\langle[{\cal M}]\rangle = 1$.}
    \label{fig:NPconstraints}
\end{figure}

\clearpage
\subsection{Discussion}\label{sec:ICsimDiscussion}

\begin{wrapfigure}{r}{0.55\linewidth}
    \vspace{-2\baselineskip}
    \centering
    \includegraphics[width=\linewidth]{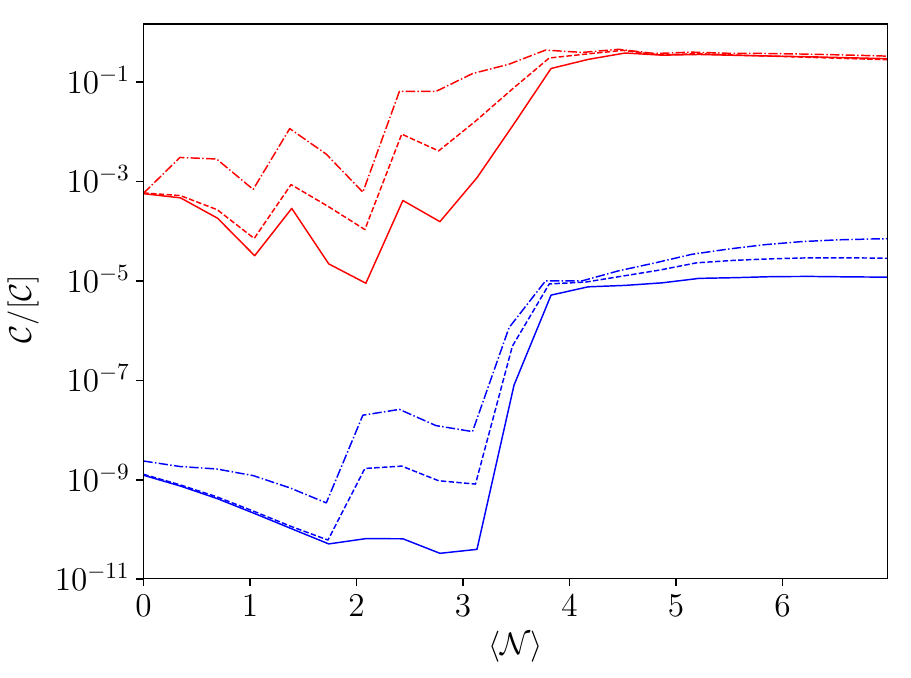}
    \caption{Strong resonance convergence test for the mean Hamiltonian (blue) and momentum (red) relative constraints  ($\langle|{\cal C}|/[{\cal C}]\rangle$, ${\cal C }= {\cal H}, {\cal M}$) across multiple runs with $N=64$ (dash-dotted), $N=128$ (dashed), and $N = 256$ (solid).  With the same range of scales initially, the perturbative phase gets smaller constraints with better resolution, as opposed to the non-perturbative last time steps of the momentum constraint.}
    \label{fig:mono_convergence}
\end{wrapfigure}
The two non-perturbative examples studied above should be of great interest not only to cosmologists but also to the numerical relativity community.
In one case the field slows down to a barely resolvable speed (ultra slow-roll reached during inflection) and, in the other case, domain walls form during the strong resonance of monodromy inflation. 
These nonlinear multi-scale effects provide challenging tests of the present implementation of our methods, at the same time demonstrating their robustness, while also delineating the limits of their applicability.

The degree of accuracy of the numerical evolution can be observed from the relative Hamiltonian ${\cal H}/[{\cal H}]$ and momentum ${\cal M}/[{\cal M}]$ constraints shown in Figure \ref{fig:NPconstraints}.
Both ratios ${\cal H}/[{\cal H}]$ and  ${\cal M}/[{\cal M}]$ should remain much smaller than $1$ for correctness, whether the simulation is perturbative or not.
Both scenarios show good constraint preservation at the initial linear level during at least the first half of the simulations, that is, for over 7 e-folds and 3 e-folds respectively for ultra slow-roll and monodromy inflation. 
{However, a problem develops} in the momentum constraint, while the Hamiltonian constraint (upper panels of Figures \ref{fig:NPconstraintsa} and \ref{fig:NPconstraintsb}) remains satisfied within a  margin $\lesssim 10^{-3}\ll 1$ at all times and {grid points}.
One could think that increasing the resolution would help reduce the relative momentum constraints to values smaller than $1$ {but, as might be guessed from the convergence test provided in Figure \ref{fig:mono_convergence}, increasing the resolution with the same input does not {however} improve the momentum constraint at late non-perturbative times}. 

It is useful to understand which scales {are responsible for such constraint violation} using the Fourier transform of the constraints. This is highlighted in Figure \ref{fig:specCons}: with initial UV mode cutoffs at $k = 4\times 10^{-3}$ and $ 4.5 \times 10^{-3}$ for inflection and resonance cases respectively, it is evident that the constraints and their absolute magnitudes only overlap at late times for modes on scales smaller than the ones included in the initial spectrum, which are scales where gradient computations fail. This momentum constraint breakdown in the axion monodromy case is primarily due to numerical resolution effects after the formation of domain walls, which is a well-known problem for the numerical  evolution of topological defects in an expanding universe. This is evidenced by the localised breakdown of the constraints {along the domain walls (about 15\% of the grid points)}, rather than a generic problem throughout the simulation (e.g.\ it is not due to non-perturbative effects from strong gravity in this case). When these later-emerging, constraint-violating UV scales (we choose conservatively $k>10^{-2}$) are filtered out from the Fourier transform of the various fields, the deviations from the FLRW background shown in Figure~\ref{fig:meanmono} are not affected.
Furthermore, we note that if the simulation of the monodromy potential is stopped at $3.8$ e-folds, just as the bimodal distribution starts forming, and while the majority of points still respect the momentum constraint, the formation of bubbles of trapped fields is still observed.
Hence, the non-perturbative effects of the monodromy potential observed in our simulations are robust against brute-force removal of constraint-violating Fourier modes.\footnote{A more sophisticated dynamical investigation of such mode filtering is beyond the scope of this work.}
\begin{figure}[ht]
    \begin{subfigure}{\textwidth}
        \centering
\includegraphics[width=0.85\linewidth]{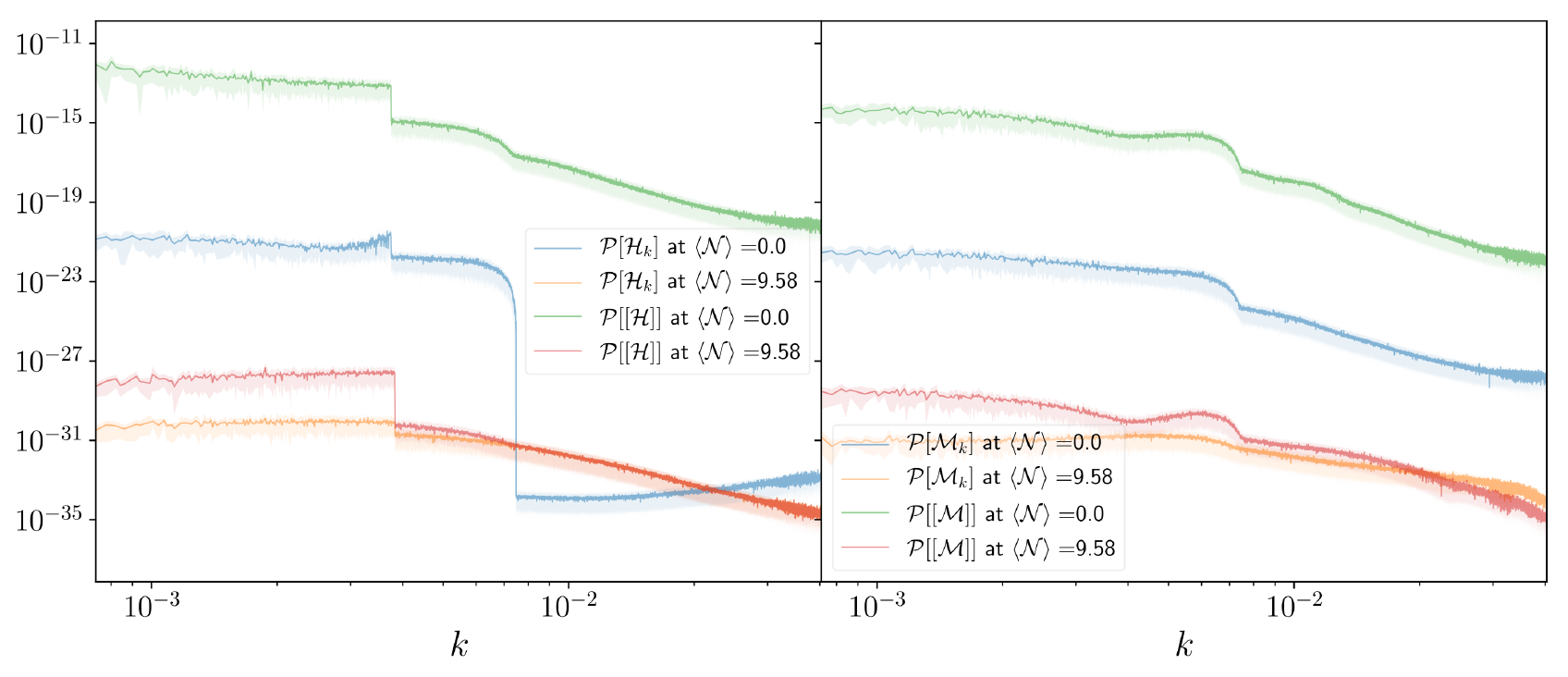}
\caption{Inflection inflation}
    \end{subfigure}
    \begin{subfigure}{\textwidth}
        \centering
\includegraphics[width=0.85\linewidth]{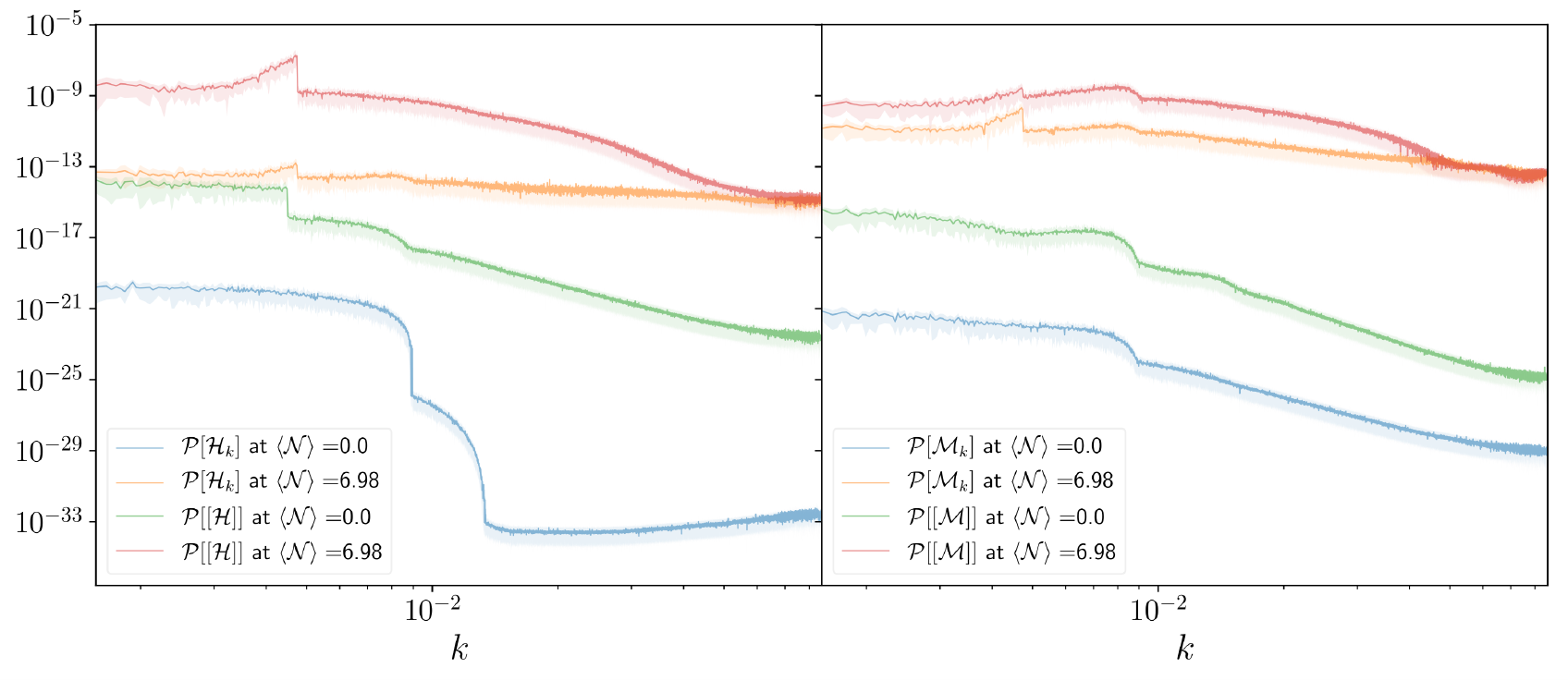}
\caption{Strong resonance }
    \end{subfigure}
\caption{ Initial and final spectra of the Hamiltonian (left) and momentum (right) constraint violations (blue and orange) and absolute constraint magnitudes (green and red) for our non-perturbative examples. {The constraints are well satisfied if the orange line lies below the red.} The spectral input was cut at $k_{max}^{\cal R} = 4\times 10^{-3}$ and $k_{max}^{\cal R} = 4.5\times 10^{-3}$ respectively, visible through spectral drops. $\Mpl=100$ units.}
\label{fig:specCons}
\end{figure}

Accurate numerical resolution of the emerging domain walls on which the momentum constraint currently fails in our simulations is a challenging endeavour: such domain walls have a fixed physical width $\delta_{\rm dw} $ that is determined by the inverse mass associated with symmetry breaking in the potential, $\delta_{\rm dw} \sim m^{-1}_{\rm dw}$.
On the other hand, the physical length scale related to the comoving grid scale $\triangle x$ expands rapidly as $\triangle r(t)  = a(t) \triangle x$ (especially during inflation).
So this soon exceeds the defect width $\triangle r(t)  > \delta_{\rm dw}$ meaning that there is insufficient numerical resolution to evolve the domain walls accurately. 
This is a problem for which there have been a number of alternative solutions developed\footnote{One frequently used approach is the `fat defect' algorithm, pioneered to study axion domain walls in an expanding universe \cite{Press:1989yh}; this essentially makes the mass dynamical $m_{\rm dw}\propto a(t)$, so the domain walls have a fixed comoving width.
However, despite the simplicity of this approach, it modifies the underlying physics of the defect network, notably its gravitational effects.}, the best of which appears to be the use of {adaptive} mesh refinement (AMR) (see, for example, ref.~\cite{Drew:2019mzc}).
We note that this work is compatible with \grt, which is built on the AMReX libraries, with AMR as a core capability.
Although complex topological defect evolution is not the primary focus of this work, we highlight this capability which should improve the evolution of any large perturbations, including the ultra slow-roll scenario. 
In any case, despite the emergence of numerical inaccuracies late in the evolution, we note that they do not cause a more serious numerical breakdown beyond the momentum constraint because the small-scale perturbations become superhorizon and as their evolution freezes. 

{We would like to comment further on the importance of independently monitoring the accuracy of constraint evolution during cosmological simulations. Satisfying the momentum constraint often proves to offer an extra layer of challenge due to the spatial gradients involved.
To our knowledge, this has not often been explicitly diagnosed in previous inhomogeneous inflationary simulations, sometimes unmentioned, sometimes relying on an initial trivial satisfaction \cite{clough_robustness_2017} or the long wavelength approximation \cite{Caravano2024}.
However, it is known that the momentum constraint is a key connector of the Hubble-sized universe patches \cite{prokopec_ensuremathdeltan_2021, Artigas25} and should not be overlooked on those scales.
In our own simulations, we have implemented a self-consistent methodology for the initial conditions to ensure that the constraints are satisfied at linear order initially (i.e.\ effectively with violations only appearing at second-order).
In the present simulation we have stress-tested this approach by setting the whole grid initially inside one Hubble volume $L^{\circledast} \lesssim (H^{\circledast})^{-1}$.
This UV depth of our modes is rather large for axion monodromy where the most subhorizon simulations are necessary to observe large resonances.
In general, Bunch-Davies fluctuations deep inside the horizon rise in amplitude and become less and less perturbative as mentioned earlier, at the risk of leaving the regime of validity of classicality.
Our simulations, such as the previous quadratic case, satisfy the constraints accurately at most points in real space (as opposed to previous work diagnosing the average), and will thus allow us to look more closely into these issues. The two non-perturbative examples presented here are certainly some of the most extreme cases, where the satisfaction of the momentum constraint does not improve over time, but at least the regime of validity of the resulting predictions is well defined. }

\begin{figure}[ht]
    \centering
        \begin{subfigure}{0.3\textwidth}
\includegraphics[width=\linewidth]{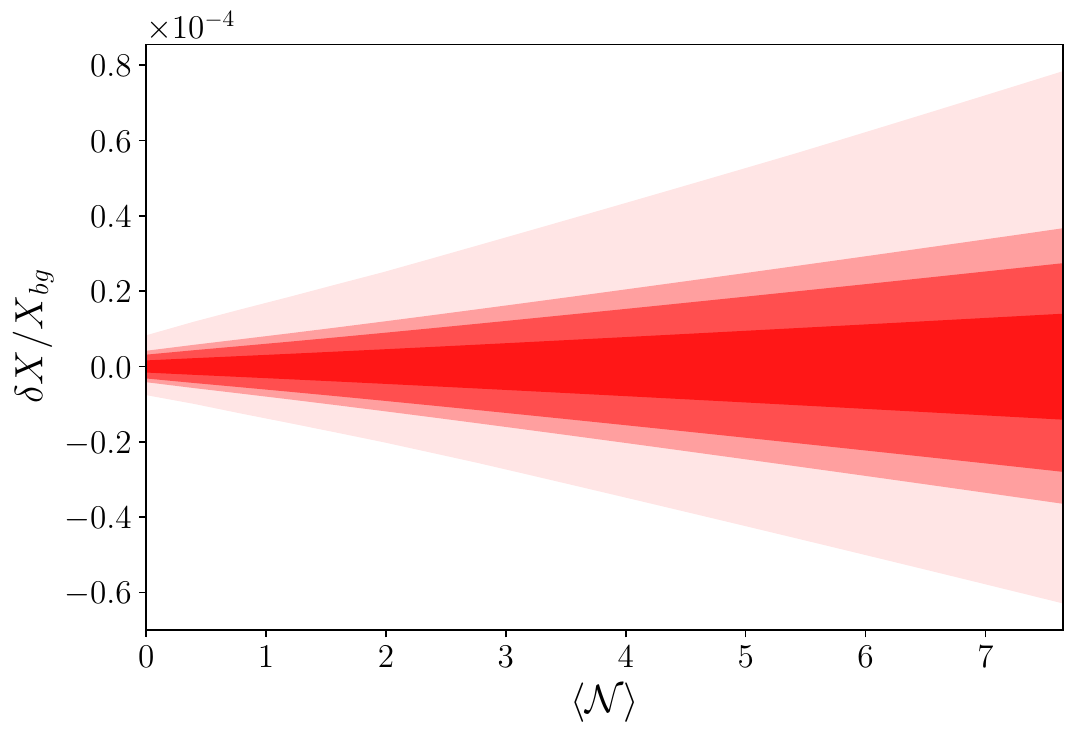}
\caption{Quadratic}
    \end{subfigure}
    \begin{subfigure}{0.3\textwidth}
\includegraphics[width=\linewidth]{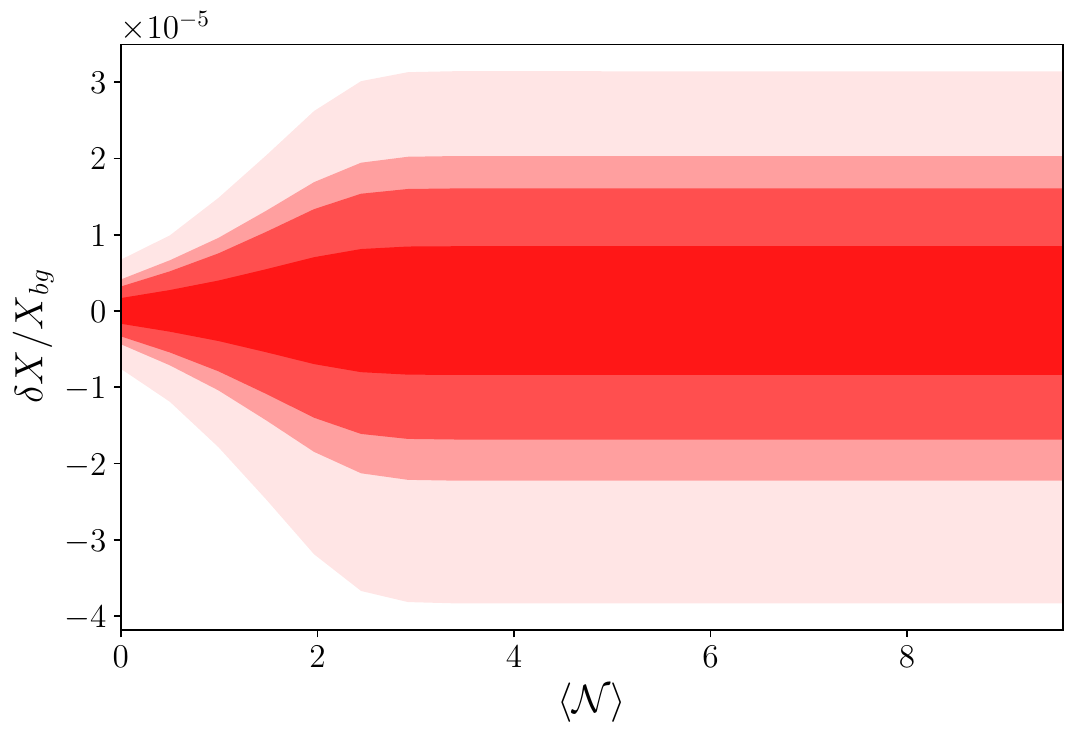}
\caption{Inflection}
    \end{subfigure}
    \begin{subfigure}{0.3\textwidth}
\includegraphics[width=\linewidth]{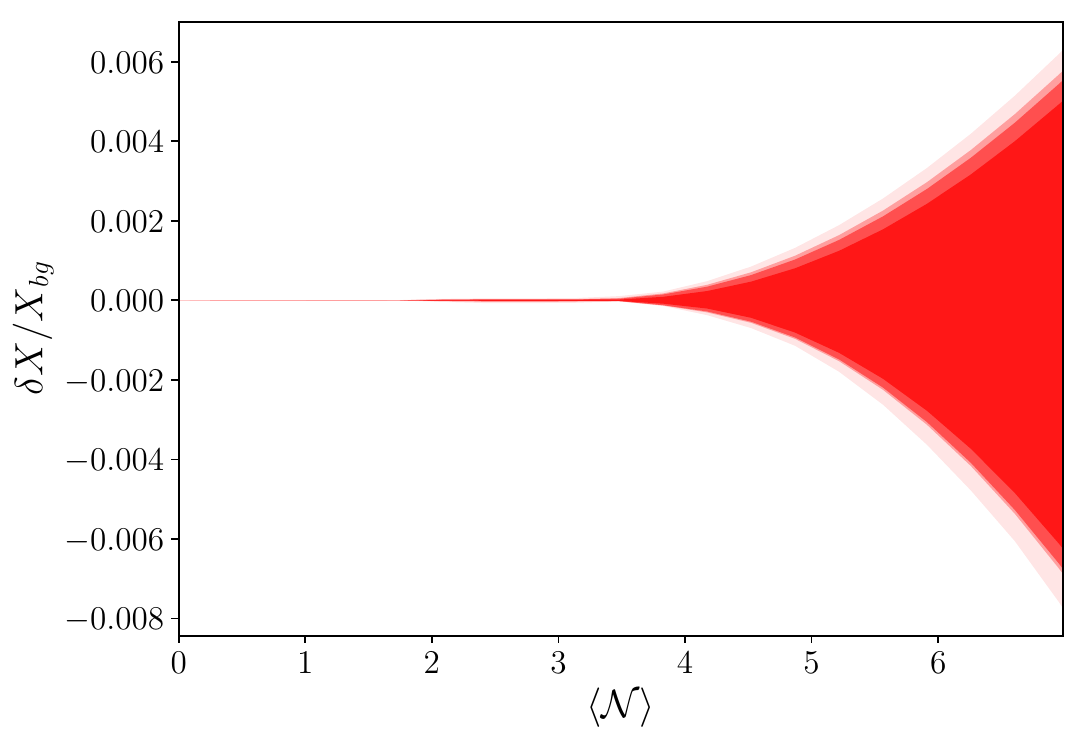}
\caption{Strong resonance}
    \end{subfigure}
    \caption{ $68$, $95$, $99$ and $100$ percentile contours for the contrast of $X$.}
    \label{fig:Xcontrasts}
\end{figure}

Another point of debate, mentioned in \cite{aurrekoetxea24}, is that of\ylg{ the} gauge choice. In principle, the geodesic gauge will end up ill-posed when gravitational collapse becomes possible, because of the crossing of free-falling observers on the grid \cite{baumgarte_shapiro_2010}. In this case, the spatial coordinates can converge, intersect, and so leave a coordinate singularity that prevents further simulation evolution. This is not as visible in inflationary simulations as in strong gravity scenarios, such as black hole simulations, but it is there, especially if we continue to reheating.
There is some evidence of this in the gradual growth of the spatial metric determinant contrast: the ratio $\delta X / X_b $ increases slowly with time according to eqn.~\eqref{eq:ADMgaugeDecomp} and Appendix~\ref{app:dicBSSN}, notably because of the growing $E$ and $\chi$ contributions.
This is shown in Figure \ref{fig:Xcontrasts}, even for the simple slow-rolling phases. As it remains fully perturbative for the whole duration of our simulations, we conclude that using the geodesic gauge is not problematic here.
In the future, we plan to compare this evolution with another successful gauge, adapted from the integrated moving puncture gauge to the case of inhomogeneous inflation \cite{giblin_jr_preheating_2019, Kou21, aurrekoetxea_oscillon_2023}.

\section*{Summary}\label{sec:ICsimConclusion}
In this chapter, we pioneered the full evolution of the Bunch-Davies vacuum initial conditions for inflation in numerical General Relativity.
Three examples were studied, ranging in the level of emerging non-linearity: a very perturbative and linear case of quadratic inflation, a potential with an inflection point and a strongly non-perturbative and non-linear resonance monodromy model.
\ylc{All evolutions, including the most exotic ones, satisfy the constraints for our scales of interest. The Hamiltonian constraint is satisfied at all times and grid points, while the momentum constraint only fails at late non-perturbative times in the two most extreme examples, on the emerging domain walls, i.e.\ on scales smaller than those of the initial spectrum, where the violation does not improve with resolution.}
We \ylc{showed} that looking at non-perturbative scenarios is possible and showcases new phenomena such as highly inhomogeneous geometries and field excursions one can find from a monodromy model\ylc{, with the caveat that the initial data are only constrained at linear order and that the late-time momentum constraint is violated on the domain walls}.
We also commented on the general need to use observer-independent, non-linear variables in order to extract measures of the fluctuations from our numerical boxes, which are not available in Friedmann lattice codes.

This chapter demonstrates that it is possible to \ylc{evolve inflationary initial data from perturbation theory nonperturbatively in full General Relativity. Turning these evolutions into predictions for post-inflation states will require controlling the constraint violations of the most extreme cases and the dependence}.
In future work, we aim to make more refined predictions with bigger simulations and better estimators for the fluctuations and corresponding observables, such as higher-order interactions and correlation functions.
A consequent study will indeed be needed to prove that these simulations can reach very high orders of perturbation theory that have not been achieved analytically (at least quantitatively).

The manner in which the inflationary initial data is input and windowed might significantly impact the predictions. In this chapter's simulations, we are essentially setting a cutoff, and while this should not impact the dynamics, it does mean that modes are not treated equally as they do not enter the simulation at the same stage of their interactivity, as reminded from Chapter~\ref{sec:semiclQFT}.
Stochastic inflation will prove to be a better answer because of the time-dependent cutoff, as described in Chapters~\ref{sec:SIGR} and \ref{sec:SINR}.

%% file: chap_SIGR.tex
\ifSubfilesClassLoaded{\setcounter{chapter}{3}}{}
\chapter{Stochastic Inflation in General Relativity} \label{sec:SIGR}

Not long after the original concept of inflation was introduced, Stochastic Inflation (SI) was formulated in  \cite{Starobinsky_stochastic_1988}, and used as a non-perturbative \yld{methodology}\ylg{method} to probe secular growth arising from light scalars in de Sitter \cite{starobinsky_equilibrium_1994}. 
Concretely, this framework sacrifices the description of negligible non-linearities arising in the sub-Hubble regime in favour of those in the non-perturbative and classical post-Hubble regime, where QFTCS might fail. 
The (semi-)classical limit reached at the crossing of the horizon studied in Chapter~\ref{sec:semiclQFT} is precisely what allows writing stochastic, and in particular nonperturbative, PDEs.
Beyond the historical studies of secular growth, Stochastic Inflation has been used in recent years to study the leading-order departure from perturbation theory provoked by slow-roll\yld{ and dS}-breaking features on the inflaton's potential, including notably ultra slow-roll (USR) phases \cite{pattison_ultra-slow-roll_2021} and its quantum diffusion limit, also found in eternally inflating scenarios \cite{Linde1986eternal}.
These are all of interest for current and future experiments which\ylg{, beyond Planck's CMB scales,} might probe nonlinear and non-perturbative physics, for instance via the production of large overdensities leading to the formation of primordial black holes (PBHs) \cite{pattison_quantum_2017}.

Although able to provide non-perturbative \yld{quantities}\ylg{predictions}, Stochastic Inflation as usually formulated comes with several approximations that simplify the complex dynamics of General Relativity, most notably relying on the Separate Universe Approximation (SUA) \cite{salopek_nonlinear_1990,salopek_stochastic_1991,wands_new_2000} and therefore altogether dropping the dynamics of some degrees of freedom \yld{of}\ylg{in} the gravitational field. To find the spectrum of the stochastic perturbations, it has also been claimed to require a specific choice for the time-slicing to ensure consistency with QFTCS, namely using a uniform e-fold time slicing \cite{Finelli09,Finelli10,pattison_stochastic_2019,artigas_hamiltonian_2022}.
However, there are formulations where such restrictions are not required \cite{rigopoulos_non-linear_2005,cruces_stochastic_2022}, which appears to be linked to debates about the appropriate inclusion of the momentum constraint \cite{prokopec_ensuremathdeltan_2021,cruces_stochastic_2022, Artigas25,Briaud25}.
Furthermore, obtaining the observationally relevant curvature perturbation then requires a first-passage-time analysis (FPTA) to determine a stochastic number of e-folds for a specified point of the scalar potential to be reached \cite{vennin_correlation_2015}.            
This is the so-called stochastic $\Delta {\cal N}$ formalism, used to translate field fluctuations into statistics for the curvature perturbation \cite{Starobinsky_stochastic_1988, Finelli09, Finelli10, pattison_stochastic_2019,Jackson22,figueroa_non-gaussian_2021,Figueroa22,tomberg_numerical_2023,stolas}.
{Recent work within the existing stochastic inflation formalism and $\Delta\mathcal{N}$ \cite{artigas_hamiltonian_2022, jackson2023,Tanaka:2023gul,Tanaka:2024mzw, Briaud25} has aimed to improve on the use of gradient expansions and gauge artefacts, and to address the neglect of anisotropic perturbations but, as we are about to see, stochastic inflation remains firmly rooted in the SUA and does not seem competitive with QFTCS for classical interactions closer to the horizon scale.}

In this thesis, as in most others on the topic, the sub-Hubble scales will be described by linear cosmological perturbation theory (CPT) -- although higher-order perturbation theory can in principle be used for these sub-Hubble scales, see \cite{cohen_stochastic_2021} in de Sitter, but we will not undertake such an endeavour here and stick to this ansatz.
We endeavour to demonstrate here that it is possible to generalise the stochastic inflation equations by making the linear treatment on sub-Hubble scales the \ylc{main} assumption, thereby removing the need for \ylc{the aforementioned approximations on super-Hubble scales. The approximations that remain are the classicality of the IR sector, the computation of the noise from linear CPT around a homogeneous background, the restriction of gauge invariance to gauges close to that background, and a heuristic treatment of backreaction.} More specifically, we formulate equations for Stochastic Inflation within full General Relativity in its ADM formulation, retaining all the variables describing the gravitational field and the freedom to choose the time-slicing and the spatial coordinates within the 3D time-slices. 

To achieve this, this chapter follows our published contribution \cite{Launay24}.
After reviewing the standard approach to Stochastic Inflation in Section~\ref{sec:IntroSI}, our procedure for coarse-graining and obtaining stochastic source terms for the full set of ADM equations is explained in Section~\ref{sec:SIGRderivation}.
These stochastic sources are expressed in terms of $\mathcal{R}$ and are found without fixing a gauge for the perturbations\ylc{, within linear perturbation theory}.
Section~\ref{sec:stogravitons} extends the result to include tensorial mode sources. As a short application, we present and solve stochastic equations for $\Delta{\cal N}$ in Section~\ref{sec:SIGRng} in a toy model exhibiting a short phase of ultra slow-roll, directly obtaining the PDF of the curvature perturbation without the need for a FPTA.

\section{The long tradition of stochastic inflation} \label{sec:IntroSI}

\subsection{Stochastic ansatz}\label{sec:ansatz}

Stochastic Inflation can be thought of as an Effective Field Theory (EFT) of QFTCS \cite{weinberg_development_2021}, valid on scales where quantum correlation functions can be well approximated by classical, stochastic ones.
Its utility lies in treating fluctuations beyond perturbation theory and associated truncations in orders of non-linearity on super-Hubble scales, compared to QFTCS \cite{cable_second-order_2022}. 
By definition, the EFT refers to variables that are coarse-grained beyond a certain length scale, normally commensurate with the Hubble radius, but there is no general method for this coarse-graining in inflationary spacetimes (at least until recently even for simple decoupled spacetimes).
The greatest difficulty of applying known EFT techniques with SI is the nature of its IR-UV split: the cutoff is spacetime-dependent. While building an EFT from a path integral \cite{hosoya_stochastic_1989,prokopec_path_2010,celoria_2021,Cespedes26} is a possible approach in simple cases of test fields in de Sitter (dS), no such approach that also includes the fluctuations of the metric in inflationary spacetimes has been achieved.

Stochastic inflation was\yld{ also} originally designed to tackle the description of IR secular growth of perturbations of a $\lambda\phi^4$ inflaton in de Sitter \cite{Starobinsky_stochastic_1988}.
As IR modes were growing beyond the perturbativity bounds of QFT, it was convenient to observe the (nonperturbative-growth-friendly) classicality of each mode $k$ as the Hubble radius became smaller ($k\leq aH$) during inflation.
Concretely, commutators were found to be decaying from this time-dependent cutoff and quantum operators turned into stochastic numbers, with a quantum-originated spectrum.
The stochastic ansatz is thus simply assuming that, from a statistical standpoint and following our arguments in Chapter~\ref{sec:semiclQFT}, inhomogeneous inflation can be looked at using the classical open IR system of scales, described by stochastic PDEs.\footnote{In principle, this does not require the noise to originate from a quantum theory using quantum initial conditions.}
Beyond that simple ansatz, further nuances appear in diverse treatments, \yld{see}\ylg{with} \yld{for instance}\ylg{various} historical approximations used\yld{ in General Relativity} \cite{salopek_nonlinear_1990,salopek_stochastic_1991}.
The common point of all stochastic equations of motion formulated so far for inflation, which could thus be appended to the stochastic ansatz, is the linear approximation for the UV quantum modes.
In mathematical terms, focusing without loss of generality on a field $X$, with decomposition $X = X^<+X^>$ between its IR classical and nonlinear part $X^>$ and its UV part $X^<$, this can be written with a nonzero r.h.s.
\begin{equation}
  \begin{aligned}
     \cE_{X^>} & \simeq {\cal F}^{-1}\Big\{S_X(k)\boldsymbol{\alpha}_{\mathbf{k}}\Big\}\,,\\
    \Src_X(k) & = \cE^{(1)}_{X}\Big[W_kX_k\Big] \mod{W_k\cE^{(1)}_{X}\Big[X_k\Big]},
  \end{aligned}
  \label{eq:ansatzSI}
\end{equation}
where $\cE_{X^>}$ is the original classical equation of motion applied to the IR field $X^>$, and $\cE^{(1)}_{X}[X^<]$ is the original linearised equation of motion applied to the UV field $X^<$.
The linear operator uses either the pre-computed Friedmann solution as background (\textit{Starobinsky approximation}), or the average of the IR field as its background\footnote{This assumption is commonly used to approximate a Markovian system and at least an additive noise, thus facilitating analytical solutions. 
However, different heuristics have been used in the past for the stochastic update of those r.h.s. and of the background quantities on which the modes evolve.
In the backreacting case where the background and UV spectra noises cannot be pre-computed, solving the equations requires a numerical approach \cite{levasseur_backreaction_2015,Figueroa22,cruces_stochastic_2022}. These studies have all reported a very small impact so far, see Chapter~\ref{sec:SINR} for our own discussion.
}, or the IR field itself depending on the treatment of the stochastic backreaction. We will keep this implicit until needed.

The separation has been made explicit using a window function $W$ selecting the IR modes
\begin{equation}
    X^> = W*X, \qquad X^< = (1-W)*X,
\end{equation}
with typically $W(k) = \Theta(\sigma/R_H-k)$,  $\sigma\leq 1$; however, we will not restrict to a sharp Heaviside window function in the following, as long as it is a smooth function of $k$ and $W(k\to 0)\to 1$ and $W(k\to \infty)\to 0$, with crossing scales of order $\sigma R_H^{-1}$.

We wrote the approximation of eqn.~\eqref{eq:ansatzSI} as an equality up to the term $W*\cE^{(1)}_{X}[X](=0)$ to emphasise that the linear operator is applied to the product of the UV window and the field, which will dramatically simplify this r.h.s..
In practice, the source is computed by solving in parallel for the spectrum of the noise using $\cE^{(1)}_{X}[X_k]=0$. In this manuscript, we refer to this whole algebraic procedure as the \textit{coarse-graining}.

\ylg{We sketch here a simplistic proof for the reader.}
Let $S$ be an action of interest, functional of one field $X$ for simplicity.
\ylg{Let us expand the total action}
\begin{equation}
  S[X^>+X^<] = \underbrace{S[X^>]}_{S^{(0)}}+\underbrace{\int d^4x \frac{\delta S}{\delta X(x)}\Bigg|_{X^>}X^<(x)}_{S^{(1)}}+O\Big((X^<)^2\Big).
  \label{eq:actionExpansion}
\end{equation}
\ylg{$X^>$ contains here a background mode justifying this expansion in amplitude, as opposed to studies of fields in the decoupling limit.
Beyond a naive expansion in the UV field against the IR field, this can be justified in explicit derivations integrating out the UV fields in the Closed-Time-Path formalism by assuming negligible UV interactions\footnote{\ylg{See also first UV loop corrections to the effective potential \cite{moss_effective_2017}.}} \cite{morikawa_dissipation_1990,pinol_manifestly_2021}, or using more implicit Open EFT and Renormalization Group arguments \cite{Cespedes26}.}
This allows us to extract the equation of motion (EOM) of the IR field \yld{up to linear order in UV corrections}\ylg{with linear UV corrections}
\begin{equation}
  \begin{aligned}
      \frac{\delta S}{\delta X^>(x)}\Bigg|_{X^>} & = \frac{\delta S^{(0)}}{\delta X^>(x)}\Bigg|_{X^>} + \frac{\delta S^{(1)}}{\delta X^>(x)}\Bigg|_{X^>}+O\Big((X^<)^2\Big), \\
                                      &  = \frac{\delta S}{\delta X(x)}\Bigg|_{X^>} + \int d^4y \frac{\delta^2 S}{\delta X(\ylg{x})\delta X(y)}\Bigg|_{X^>}X^<(y)+O\Big((X^<)^2\Big),
  \end{aligned}
\end{equation}
after injecting the previous expansion. 
Using now the classicality assumption on $X^>$, we can assume that the evolution is entirely described by
\begin{equation}
    \frac{\delta S}{\delta X^>(x)}\Bigg|_{X^>}=0,
\end{equation}
so that
\begin{equation}
  \frac{\delta S}{\delta X}\Bigg|_{X^>} \ylg{\simeq} - \int d^4y \frac{\delta^2 S}{\delta X(\ylg{x})\delta X(y)}\Bigg|_{X^>}X^<(y).
\end{equation}
This says that the classical EOM of the IR field is the same as that of the original EOM, supplemented by a right-hand side source fuelled by the UV field.
The double functional derivative applied to S is the inverse propagator, which is the linear equation of motion operator. This is of course much more subtle for the metric propagator and more generally when gauge d.o.f.s are involved, which we hope to see clarified in future work.

This short demonstration appears heuristic for two reasons, which can be rigorously answered. First, $X^<$ is technically a quantum operator; however, the operator acting on it will only leave terms with a time derivative acting on the window after simplification of the linear EOM, meaning that the UV modes will only contribute at the time of crossing, where they are classicalised and can be treated as stochastic numbers with \yld{their quantum-originated spectrum}\ylg{the spectrum set by their quantum state at crossing}.
Second, the case of full GR is more subtle because of the gauge-dependence as will be pointed out in this chapter. The procedure developed in the present work successfully tackles that, at least at the level of the EOMs.

Note that this chapter does not focus on the validity of the separation of linear quantum and nonlinear classical scales, and, as advocated in Chapter~\ref{sec:semiclQFT} and most recently in \cite{Ballesteros26},
assumes that every scenario of inflation and every set of scales selected by $W$ should have a classicality criterion checked using QFT-based arguments in their perturbative phase.

This linear source approximation, together with its assumed Gaussianity, is a common feature of the stochastic inflation literature \cite{Starobinsky_stochastic_1988, hosoya_stochastic_1989, grain_stochastic_2017}, but its justification is rarely made explicit. The validity domain of this approximation lies in scenarios where, before horizon-crossing, UV scales have suppressed higher-order statistics compared to the UV tree-level or compared to their IR counterparts generated from the UV tree-level, after horizon-crossing. A quasi-dS universe is probably the safest case in this regard, which is convenient for its believed physical relevance. Some previous work \cite{cohen_stochastic_2021} has even managed to account for suppressed next-order statistics in the noise.
More generally, this stochastic evolution should be reliable for studying any non-linearity and non-Gaussianity generated by the evolution that is larger than the first order in CPT.
However, if no non-perturbative effect is obtained from SI, then one could stick to QFTCS, valid on IR scales and beyond tree-level and before horizon-crossing. However, without a numerical prescription for loops and all Feynman diagrams, capturing the full nonlinearity is practically impossible for QFTCS, as opposed to evolving the full EOMs in the semiclassical regime.

To summarise, stochastic inflation is about reducing QFTCS approximations to UV scales only, allowing for the possibility to study fully non-linear and non-perturbative phenomena on IR scales, i.e. above Hubble scales in our case. 
It is in this ability to provide results beyond QFTCS theoretical and practical capabilities that Stochastic inflation's appeal lies.

\subsection{Current \yld{approaches}\ylg{formulations} \yld{to}\ylg{of} stochastic inflationary dynamics}\label{sec:secIIC2}

So far we have presented a schematic picture of how one might obtain equations describing the dynamics of Stochastic Inflation from coarse-graining the equations of motion.
However, applying the above scheme to the full equations of General Relativity has not been implemented due to its relative complexity, as well as the a priori large number of variables, which can be both dynamical and constrained, and the \ylg{apparent }necessity of choosing a gauge.
As a result and as we are about to review, various approximations have been made, mostly taking the long-wavelength limit and reducing the number of dynamical variables.  

Early work focused on the coarse-graining of the Klein-Gordon field equation only, assuming an unperturbed (initially dS) background and coarse-graining the decoupled field and sometimes its conjugate momentum \cite{Starobinsky_stochastic_1988,sasaki_classical_1988,nambu_stochastic_1988,nakao_stochastic_1988,morikawa_dissipation_1990}.
It was only in \cite{salopek_nonlinear_1990,salopek_stochastic_1991} that the first stochastic equations were formulated fully including metric perturbations and so backreaction. 
It consists in decreasing the interdependence of GR equations by using the long wavelength approximation and judicious gauge choices.
In the full ADM formalism, before any linearisation and coarse-graining, the lowest order gradient expansion of eqns.~\eqref{eq:ADMPi} and \eqref{eq:ADMK} in the convenient $\beta^i = 0$ slicing becomes \cite{salopek_nonlinear_1990,rigopoulos_separate_2003,prokopec_ensuremathdeltan_2021}%
\begin{subequations}
    \label{eq:longGR}
\begin{align}
 \dot{K}-\alpha K^2
 +3\alpha\Mpl^{-2} V = 0, & \\
  \dot{\tilde{K}}^i{}_j-\alpha K \tilde{K}^i{}_j + \Mpl^{-2} \alpha \tilde{S}^i{}_j=0, & \label{eq:longGRKtilde}\\
    \frac{1}{\alpha}\dot{\Pi}-K \Pi+V_{,\phi}=0, & 
\end{align}
\end{subequations}
together with the constraints losing the $3$-Ricci scalar%
\begin{subequations}
    \label{eq:longCons}
\begin{align}
 \frac{2}{3} K^2-\tilde{K}_{i j} \tilde{K}^{i j} -\Mpl^{-2}\Big(\Pi^2+2V\Big) = 0, &\\
 \tilde{K}^j{}_{i \mid j}-\frac{2}{3} K_{\mid i}+\Mpl^{-2}\Pi\partial_i\phi=0. & 
\end{align}
\end{subequations}
These \textit{shift-free gauge IR} equations are the starting point of all recorded stochastic approaches using the coarse-graining at the level of the equations of motion. 

\subsubsection{Hamilton-Jacobi equations}\label{sec:HJ}
Following the original \textit{Salopek} \& \textit{Bond} approach \cite{salopek_stochastic_1991}, some works \cite{prokopec_ensuremathdeltan_2021,Prokopec:2025uvz} chose to neglect anisotropic degrees of freedom (scalar or scalar-induced tensorial).  
In fact, $\tilde{S}^i{}_j$ is usually set to zero in eqn.~\eqref{eq:longGRKtilde} in the absence of anisotropic fluid sources, or by invoking that any possible contributions to it are higher order in the gradient expansion.
This leads to an exponential decay of ${\tilde{K}}^i{}_j$ and results in one fewer equation with only $K$ and $\Pi$ dynamics remaining, namely the \textit{shift-free isotropic $0$th-order gradient expansion} of the ADM equations%
\begin{subequations}
\begin{align}
 \dot{K}-\alpha K^2
 +3\alpha\Mpl^{-2} V = 0, \label{eq:LongIsotropK}& \\
    \frac{1}{\alpha}\dot{\Pi}-K \Pi+V_{,\phi}=0, & \\
      \frac{2}{3} K^2 -\Mpl^{-2}\Big(\Pi^2+2V\Big) = 0, & \label{eq:LongIsotropham}\\
-\frac{2}{3} K_{\mid i}+\Mpl^{-2}\Pi\partial_i\phi=0.  & \label{eq:LongIsotropmom}
\end{align}
\label{eq:LongIsotrop}
\end{subequations}
Defining the local Hubble parameter as $H(x)=-K(x)/3$, eqn.~\eqref{eq:LongIsotropmom} implies that the spatial dependence is purely that of the field $H(x) = H\Big(t,\phi(\mathbf{x},t)\Big)$,
and that its momentum is rewritten with no time derivatives $\Pi(x) = -2\Mpl^2 (\partial_\phi H)_t$ (at constant time). Furthermore, the latter implies 
\begin{equation}
    \frac{1}{\alpha}\partial_t H = \Pi(\partial_\phi H)_t+\frac{1}{\alpha}(\partial_t H)_\phi = -\frac{1}{2\Mpl^2}\Pi^2+\frac{1}{\alpha}(\partial_t H)_\phi,
\end{equation}
which, by injecting eqn.~\eqref{eq:LongIsotropK} and eqn.~\eqref{eq:LongIsotropham}, leads to $(\partial_t H)_\phi = 0$, so that $H$ is purely a function of $\phi$.
Substituting back into the Hamiltonian constraint eqn.~\eqref{eq:LongIsotropham}, one gets a single equation for $\bar{H}=H(\phi)$
\begin{equation}
    \label{eq:HamilJacobi}
    \bar{H}^2=\frac{2}{3}\Mpl^2(\partial_\phi \bar{H})^2+ \frac{1}{3}\Mpl^{-2}V.
\end{equation}
Solving those \textit{Hamilton-Jacobi} dynamics then only requires finding a functional for $\bar{H}(\phi)$, with no time dependence.\footnote{In practice, this can be rather involved if no more approximations are made, see \cite{salopek_stochastic_1991}.}
The idea developed in \cite{salopek_stochastic_1991,prokopec_ensuremathdeltan_2021} is then to apply Starobinsky's stochastic ansatz to get $\Pi(x) = -2\Mpl^2 \partial_\phi \bar{H}+\xi(t)$. Because the chosen functional $\bar{H}$ ensures the satisfaction of the (IR and isotropic) nonlinear constraints, the stochastic configuration satisfies them too; an idea at the core of the Hamilton-Jacobi approach to stochastic inflation.

\subsubsection{Separate Universe Approximation}\label{sec:SUA}

In many later works stemming largely from e.g.~\cite{wands_new_2000,Lyth03}, the similarity of the $\Pi$, $K$ and the Hamiltonian constraint in eqn.~\eqref{eq:LongIsotrop} with Friedmann's background equations of cosmology, eqn.~\eqref{eq:backgroundEq}, is interpreted literally.
The local quantity $-K(x)/3$ replaces again the homogeneous Hubble parameter $H$, so that the whole inhomogeneous universe is described as made up of patches, each of which evolves independently with its own Friedmann equations ($\phi$, $\dot{\phi}$, $H$), and hence the \textit{Separate Universe Approximation} (SUA).

{To allow the application of the coarse-graining using linear perturbation theory and after some criticism \cite{Cruces_failure}, as well as historical debates on the appropriate choice of slicing \cite{Finelli09, Finelli10,vennin_correlation_2015}, the SUA was only recently shown to be fully consistent with first-order CPT in an arbitrary gauge \cite{pattison_stochastic_2019}.
It is in fact possible to rewrite the addition of background and linearised ADM field equations, eqns.~\ref{eq:backgrounfEqfield} and \ref{eq:fieldCPT}, by changing the time variable as $d\tau = \alpha_b(1+\Psi) dt$, yielding
\begin{equation}
    \frac{\partial^2}{\partial \tau^2 }\Big(\phi + \delta \phi\Big) - K \frac{\partial}{\partial \tau }\Big(\phi + \delta \phi\Big) + V_{,\phi}\Big(\phi + \delta \phi\Big) = \frac{\nabla^2}{a^2}(\delta \phi),
\end{equation}
with $K$ given at first order in \ref{eq:linK}, and which clearly looks like the background equation applied to the inhomogeneous fields when taking the long wavelength limit.}
With this principle, the EOM coarse-graining has been applied to the field only, eqn.~\eqref{eq:LongIsotropK}, while enforcing the local Friedmann equations, eqns.~\eqref{eq:LongIsotropK} and \eqref{eq:LongIsotropham}, to be satisfied at all times.
Originally silenced in early works, the satisfaction of the momentum constraint and the account of anisotropic d.o.f.s were recently claimed \cite{Artigas25},  and are further discussed in the next Section~\ref{sec:OtherApproaches}.

Unlike scalars on a fixed background, the inclusion of metric perturbations indeed brings about the issue of the appropriate slicing to be used for both the linearisation and the noise calculation, especially beyond slow-roll scenarios.
The uniform-${\cal N}$ gauge slicing (see Section~\ref{sec:ICNRuniformN}) has emerged as the most natural, with \cite{Finelli09,Finelli10} arguing early on how this slicing recovers CPT equations and thus the long wavelength QFTCS limit when linearising the field equations.
The treatment of \cite{pattison_stochastic_2019} derived above shows how it can be rigorously done in any gauge, at least with respect to CPT.
Beyond the validity of the SUA in such a slicing, this gauge became established and proved convenient given the historical use of the $\Delta{\cal N}$ \textit{formalism}  \cite{wands_new_2000} to derive the statistics of the curvature perturbation by the end of inflation.
In fact, these can be given once one knows the local e-folding probability distribution for a given exit field value, which can be obtained from the Langevin equations via the associated Fokker-Planck equation and its adjoint via a first-passage-time analysis (FPTA), referred to all together as the \textit{stochastic} $\Delta{\cal N} $ \textit{formalism} \cite{vennin_correlation_2015}.
The notable results achieved by this approach are numerous and are predictive \cite{vennin_correlation_2015,pattison_quantum_2017,ezquiaga_exponential_2020}.
These are not investigated in this work as we focus on the Langevin EOMs, and bypass the need for it by directly obtaining the statistics of interest from these, as reported in Section~\ref{sec:SIGRng} and Chapter~\ref{sec:SINR}.

When performing the EOM coarse-graining in this gauge, one gets the SI equations \cite{pattison_stochastic_2019}, which we will refer to as the \textit{SUA Stochastic Inflation} or the `usual' equations\footnote{Note the typo in the constraint of our original publication \cite{Launay24}.}%
\begin{subequations}
\begin{align}
 &\frac{\partial \phi^>}{\partial{\cal N}_b} -\pi^> = \boldsymbol{\Sigma_{\phi}}, \\
 &\frac{\partial \pi^>}{\partial{\cal N}_b}   + (3-\varepsilon_1^>)\pi^>+\frac{1}{(H^>)^2}V_{,\phi}(\phi^>)= \boldsymbol{\Sigma_{\pi}}, \\
& {H^>}^2  = \frac{1}{\Mpl^2}\frac{V(\phi^>)}{3-\yld{(\varepsilon_1^>)^2}\ylg{\varepsilon_1^>}},
\end{align}\label{eq:SIeq}
\end{subequations}
where ${\cal N}_b=\ln (a/a_0)$ is the background number of e-folds, $\varepsilon_1^>=-\partial_{{\cal N}_b}\ln H^> = (\pi^>/\Mpl)^2/2$ the first slow-roll IR parameter, and the correlated stochastic sources given by%
\begin{subequations}
    \label{eq:SInoise}
\begin{align}
\boldsymbol{\Sigma_{\phi}} & = {\cal F}^{-1}\left\{\frac{\partial W_k}{\partial{\cal N}_b}\delta \phi_{{k}}^{\rm UE}\boldsymbol{\alpha_{\mathbf{k}}}\right\}, \\
\boldsymbol{\Sigma_{\pi}} & = {\cal F}^{-1}\left\{ \frac{\partial W_k}{\partial{\cal N}_b}\delta \pi_{{k}}^{\rm UE}\boldsymbol{\alpha_{\mathbf{k}}}\right\},
\end{align}
\end{subequations}
where $\delta\phi_{{k}}^{\rm UE}$, $\delta\pi_{{k}}^{\rm UE}$ are the linear perturbations with quantum initial conditions in the same gauge,
and $\boldsymbol{\alpha_{\mathbf{k}}}$ are the Fourier transform of 3D Gaussian white noise, as previously defined in Chapter~\ref{sec:ICNR} in eqn.~\eqref{eq:stocomm}.
Note that such equations require transformations between gauges; the noise is computed in the flat gauge, then transformed to the uniform-${\cal N}$ gauge for the Langevin equations, and finally -- effectively -- to the uniform-$\phi$ gauge in the \textit{stochastic} $\Delta{\cal N}$ \textit{formalism} after extraction to get to the statistics of $\cR$.
We also note that the window function is almost always chosen to be a Heaviside function making these equations completely decorrelated in space and time to achieve a Markovian system and remove the need for a Fourier transform.
Implementations of those equations and predictions include \cite{Jackson22,tomberg_numerical_2023,stolas,Animali25}.

To get to eqn.~\eqref{eq:SIeq}, the coarse-graining ansatz has been applied as $\delta \phi ^> = W\delta \phi$ and $\delta \pi ^> = W\delta \pi$ in Fourier space \cite{pattison_stochastic_2019}.\footnote{Note that the authors write an opposite sign for the noise contributions when using $\delta X_< = W\delta X$ instead.}
This implies defining $\pi^>$ in terms of the field with a stochastic source, and so not using the true canonical momentum of $\phi^>$, the perturbations of which would be $\partial_0(W_k\delta \phi_{\mathbf{k}})$ instead. We will refer to the former method as the \textit{phase-space} coarse-graining.
Alternatively, and this will be useful for future comparison, one can re-write the previous equations as a unique equation in the IR ADM cosmic time momentum\footnote{Note that it is common to apply the overdamping assumption $|\delta \Pi|^2 \ll |\delta \phi|^2$, thereby reducing the problem to a first-order equation of the field only \cite{Starobinsky_stochastic_1988}. This assumption will not be made in the following.}
\begin{equation}
    \label{eq:SIrewrite}
 \frac{\partial {\Pi}^>}{\partial{\cal N}_b} +3{\Pi}^>+\frac{1}{H^>}V_{,\phi}(\phi^>) = \boldsymbol{\Sigma},
\end{equation}
defining ${{\Pi}^>} \equiv {H^>} {\partial_{{\cal N}_b} \phi^>}$ and $\boldsymbol{\Sigma} \equiv H (\boldsymbol{\Sigma}_{\pi} + \partial_{{\cal N}_b}\boldsymbol{\Sigma}_{\phi}+(3-\varepsilon_1)\boldsymbol{\Sigma}_{\phi})$, proportional to the single stochastic $\boldsymbol{{k}}$-grid $\{\boldsymbol{\alpha_{{k}}}\}$ in Fourier space. 
This noise appears to be exactly the same as the one obtained when coarse-graining the $\Pi$ equation, with the field only prescription  $\delta \phi ^> = W\delta \phi$ so that  $\delta \Pi ^> = W\delta \Pi+ \dot{W} \delta\phi$.
In that sense, {phase-space coarse-graining} is perfectly equivalent to coarse-graining the momentum via the field only, which will be our approach, following the coarse-graining in Wilsonian EFTs and path integrals \cite{moss_effective_2017}.

\subsubsection{Other approaches, extensions and debates}\label{sec:OtherApproaches}
Back to eqn.~\eqref{eq:LongIsotrop}, the role of the momentum constraint has been acknowledged \cite{salopek_nonlinear_1990,kodama_evolution_1998,rigopoulos_separate_2003,prokopec_ensuremathdeltan_2021,cruces_stochastic_2022,cruces_review_2022,Briaud25,Artigas25} as being one of the remaining links between the separate universes in the SUA (model-dependent observation).
Neglecting it by invoking a long-wavelength approximation without care, and particularly beyond attractors, could lead to inconsistencies in CPT \cite{kodama_evolution_1998,Sasaki98,cruces_stochastic_2022}.
This issue has been addressed in recent years by tackling the regime of validity of the SUA by comparing to CPT \cite{pattison_stochastic_2019, artigas_hamiltonian_2022}: it seems that working far enough from the horizon-crossing scale with the uniform-${\cal N}$ gauge is a safe choice to match CPT.

In parallel, other SI approaches using different assumptions have nurtured debates and clarifications. As mentioned previously, SI was formulated with the explicit inclusion of the nonlinear momentum constraint in the \textit{Hamilton-Jacobi formalism}, and the FPTA approach of \cite{vennin_correlation_2015} was applied to this formalism in \cite{prokopec_ensuremathdeltan_2021, Rigopoulos:2021nhv}.
However, this approach was recently criticised \cite{Artigas25} as it discards anisotropic degrees of freedom, notably in the momentum constraint. It was in fact argued that discarding anisotropies in the chosen slicings equates to neglecting the non-adiabatic pressure, $\delta P_{nad}\propto \Psi_B$, as opposed to a SUA-based approach where it is included at 0th order in the gradient expansion in the uniform e-fold gauge.
The non-adiabatic pressure and modes can be key in regimes beyond attractors, such as ultra-slow-roll phases-- of interest as it dramatically enhances perturbations and so PBH production.

Starting from the full set of eqns.~\eqref{eq:LongIsotrop}, another set of stochastic equations for the non-linear generalisation of $\cR$ given in eqn.~\eqref{eq:NLgaugeinv} was developed in
\cite{rigopoulos_non-linear_2005} which by construction reduces to coarse-grained, gauge invariant CPT in all time-slicings with $\beta^i$$\,=\,$$0$.
As shown in \cite{Rigopoulos:2005us, Tzavara:2010ge}, these equations produce the same perturbative results for the bispectrum as the $\Delta{\cal N}$.
The more recent study \cite{cruces_stochastic_2022} also writes stochastic equations in a gauge other than uniform $\mathcal{N}$, the uniform Hubble gauge, and rehabilitates the momentum constraint in the gradient expansion.

Above all issues and debates, the SUA and more generally gradient expansions are not free of remaining issues precisely where gradients are important.
This is particularly visible during \textit{sudden transitions} \cite{jackson2023}, notably those leading to ultra slow-roll from a CMB-like slow-rolling epoch.
For the SUA-based stochastic approach, this means that scales crossing the horizon during the transition simply cannot be looked at.
In \cite{Briaud25}, a heuristic method is built to account for some gradient effects in SUA-based SI in the decoupling limit.
The validity of those fixes beyond the tested scenarios is still questionable, in particular regarding the gradient backreaction, but, most importantly, relies on the long wavelength approximation, and misses considerable nonlinear effects that could be captured by a more complete set of stochastic equations.

The discussion above attempts to provide a flavour of the current status of the SI formalism and the approximations that have been deemed necessary to develop it. In the next section, we demonstrate how one can do away with the long wavelength approximation and the SUA, providing a set of SI equations that retain both the scalar field and all the variables of the gravitational field, the only approximation being scalar linear CPT for the computation of the noise terms. 

\section{Stochastic Inflation in General Relativity}\label{sec:SIGRderivation}

\subsection{Gauge-agnostic coarse-graining}

Applying the stochastic ansatz requires the choice of a target field (and its conjugate momentum if done equivalently in phase-space). 
It can be done blindly by using the dynamical fields for a test-field such as in the original derivation of SI \cite{Starobinsky_stochastic_1988} or in a given gauge \cite{pattison_stochastic_2019}.
However, the latter necessarily attaches an observer to the coarse-graining and leaves the observer rather ill-defined.
In fact, coarse-graining the inflaton's perturbation $\delta\phi \longrightarrow \delta \phi^>$ in the gauge where the number of e-folds is uniform makes sense if these are the only two evolving variables.
When the slow-roll and long-wavelength approximations are not used, and looking back at the expression of $\delta\phi$ as a functional of ${\cal R},\, \dot{\cal R}$ in eqn.~\eqref{eq:UniNGaugeDecomp}, we see that this coarse-graining is disconnected from coarse-graining the fundamental variable receiving \yld{gauge-agnostically }the quantisation, ${\cal R}$\ylg{, or equivalently the hypersurface on which the quantisation is performed for a chosen field}.
The uniform e-fold gauge has thus been a historical and safer choice regarding the SUA, and a meaningful coarse-graining choice in the decoupling limit (where $\delta\phi \propto {\cal R}$) but not beyond.

We now take ${\cal R}$ to be the master dynamical perturbation variable to be coarse-grained, and no gradient or decoupling expansion of any kind is used in the following.
Following the stochastic ansatz described in Section~\ref{sec:ansatz}, the coarse graining corresponds to applying the linear equations of motion to the windowed perturbations.
Stochasticity is simply a 'continuous' resetting of the initial conditions and thus benefits greatly from an understanding of ICs such as the one developed in Chapter~\ref{sec:ICNR}, where we found both gauge-invariant and gauge-dependent perturbations summarised as functionals of ${\cal R}$.
This is particularly useful here, as with ${\cal R}_k^> = W_k {\cal R}_k$, all other gauge-dependent or invariant coarse-grained perturbations can be defined as
\begin{equation}\label{eq:replacement}
\delta Y_k^> \equiv \delta Y_k\big[{\cal R}_k^>\big]. 
\end{equation}
This is really to be understood as a functional, implying that coarse grained time derivatives are to be understood as $\dot{f}^> \equiv d\left(Wf\right)/dt$
in any linearised perturbation variable $f$.

Let us start with the gauge invariants found in \eqref{eq:GaugeInvExplicit} in terms of ${\cal R}$, still using background cosmic time, $\alpha_b =1$.
This defines our coarse-grained linear gauge invariant variables as%
\begin{subequations}
    \begin{align}
         \Phi_{B}^> & \equiv-\varepsilon_1a^2k^{-2}H\dot{{\cal R}}^> ,\\
         \Psi_{B}^>  & \equiv  -\varepsilon_1a^2k^{-2}\bigg(H\dot{\cal R}^>-{\cal S}_{\cal R}\bigg),\\
         \delta\phi_{gi}^>  & \equiv  \sqrt{2\varepsilon_1}{\rm \Mpl}\bigg({\cal R}^> +\varepsilon_1a^2 H k^{-2}\dot{{\cal R}}^> \bigg),\\
         \zeta_{gi}^>  & \equiv -{\cal R}^> +\frac{1}{3H}\dot{{\cal R}}^>. 
    \end{align}
    \label{eq:GaugeInvExplicit>}
\end{subequations}
where we have defined the Fourier amplitude of the source to the coarse-grained MS equation
\begin{equation}
\ddot{{\cal R}}^{>}_k + H(3-\varepsilon_2)\dot{{\cal R}}^{>}_k+ \frac{k^2}{a^2}{\cal R}^{>}_k = {\cal S}_{{\cal R}},
    \label{eq:coarseGrainedReq}
\end{equation}
which simplifies as a functional of ${\cal R}$ and its first time derivative only, after using the Mukhanov-Sasaki equation in ${\cal R}$
\begin{equation}
    {\cal S}_{{\cal R}}  \equiv {\cal R}_k \ddot{W}_k+\big[2 \dot{\cal R}_k + (3-\varepsilon_2) H {\cal R}_k\big]\dot{W}_k.
    \label{eq:SR}
\end{equation} 
One can now understand why it is key to not use the MS equation before coarse-graining, for instance to write $\Psi_B=\Phi_B$.
In other words, the gravitational slip is not zero in the coarse-grained system.

Gauge-dependent perturbations can now be found from these relations, recalling the definitions of each of those invariants\footnote{For IR fields, gauge invariant linear combinations have the same functional form as the non-coarse-grained ones.} from eqn.~\eqref{eq:GIquantities} to define the coarse grained scalar perturbation variables $\Phi^>$, $\Psi^>$, $\chi^>$ and $\delta\phi^>$ in terms of ${\cal R}^>$ and $\dot{{\cal R}}^>$, also explicitly containing ${\cal S}_{\cal R}$%
\begin{subequations}
    \begin{align}
         \Phi^> + H\chi^> & = -\varepsilon_1a^2k^{-2}H\dot{{\cal R}}^>  , \\
         \Psi^> - \dot{\chi}^> & =  -\varepsilon_1a^2k^{-2}\bigg(H\dot{\cal R}^>-{\cal S}_{\cal R}\bigg),\\
        \delta\phi^>-{\dot{\phi_b}}\chi^>  & = \sqrt{2\varepsilon_1}{\rm \Mpl}\bigg({\cal R}^> +\varepsilon_1a^2 H k^{-2}\dot{{\cal R}}^> \bigg),\\
          -\Phi^>+ \frac{\delta\rho^>}{3(\rho_b+ P_b)}  &=  -{\cal R}^> +\frac{1}{3} H^{-1}\dot{{\cal R}}^>, \\
        \Phi^> + \frac{H}{\dot{\phi_b}}\delta \phi^> & = {\cal R}^> .
    \end{align}
    \label{eq:CGvariables}
\end{subequations}
From these relations, and including the case of $W=1$ corresponding to the original closed or non-coarse-grained system of perturbations, there are two ways to proceed:
\begin{itemize}
    \item Gauge-dependently: a gauge is picked, fixing two of the perturbations on the left and thus closing the system. Each gauge ${\rm g}$ thus corresponds to a different functional of ${\cal R}$ and its derivative for each perturbation, $\delta Y^{\rm g} = F_Y^{\rm g}\big[{\cal R}\big]$ for all $\delta Y$ in $\{\delta \phi, \delta g_{\mu\nu}\}$.
    \item Gauge-agnostically: all perturbations can be expressed in terms of ${\cal R}$, $\dot{\cal R}$, and an arbitrary gauge-dependent (but unspecified) one, say $\delta X$, so that $\delta Y = F_Y\big[{\cal R}, \delta X\big]$ instead.
\end{itemize}
The conservation of the $\cR$-functionals or linear combinations in eqn.~\eqref{eq:CGvariables} before and after coarse-graining is precisely what makes the two approaches equivalent.
In other words, we showed that gauge-invariant coarse-graining and gauge-picking commute at the linear level. Note that we have chosen $\cR$ as the master variable to coarse-grain as it is the one associated with quantisation and classicality, but any other gauge-invariant variable could have been chosen instead for other purposes, and the same conclusions would hold\ylg{ (for instance for $\zeta_{gi}$)}.

In both approaches, we now have everything to apply the linear EOM operators (i.e. the l.h.s. of eqns.~\eqref{eq:Kcpt_eq}--\eqref{eq:ConstraintsCPT}) to the coarse-grained perturbations.
In our original work \cite{Launay24}, the first method was used and applied to all gauge families detailed in Section~\ref{sec:ICNR}.
The spectra of the stochastic sources were found to be gauge-independent and so motivated finding the gauge-agnostic method, which is the one reported in the following with the example of the inflaton equation.

\subsection{Example of coarse-graining: ADM field equation}\label{app:CGfield} 

As an explicit example of the procedure described in the previous subsection, we now demonstrate the coarse-graining of the ADM inflaton equation via its corresponding linear equation in Fourier space, both recalled from Chapter~\ref{sec:ICNR}%
\begin{subequations}
    \begin{align}
 \frac{1}{\alpha}\left(\dot{\Pi}+\beta^i \Pi_{\mid i}\right)-K \Pi-\frac{\alpha^{\mid i}}{\alpha} \phi_{\mid i} 
 -\phi_{\mid i}^{\mid i}+V_{,\phi}& =0,  \label{eq:ADMfield2}\\
 \delta\ddot{\phi}+3 H  \delta \dot{\phi}+\frac{k^2}{a^2} \delta \phi
    -\dot{\phi_b}\left(3\dot{\Phi}+\dot{\Psi}+\frac{k^2}{a^2}\chi\right) 
   +V_{,\phi\phi}(\phi_b) \delta \phi +2 \Psi V_{,\phi}(\phi_b)& =0,     \label{eq:fieldCPT2}
    \end{align}
\end{subequations}
where $\alpha_b$ was set to $1$ in the latter.

We want to calculate the spectrum of the $r.h.s.$ of the same equation but for the IR field perturbations $(\delta\phi^>, \Phi^>, \Psi^>, \chi^>)$.
Using eqn.~\eqref{eq:CGvariables} and \emph{without picking a gauge}, it is possible to express all these linear degrees of freedom as functionals of ${\cal R}^>$ and another scalar of our choice.
In the case of the field equation, we choose this scalar to be the inflaton's perturbation $\delta\phi^>$. Eqn.~\eqref{eq:CGvariables} can then be rearranged to extract%
\begin{subequations}
    \begin{align}
        \Phi^>  & = {\cal R}^>-\frac{H}{{\dot{\phi_b}}}\delta \phi^>,\\
        \chi^>  & = -\frac{1}{H}\Phi^> -\varepsilon_1 \frac{a^2}{k^2}\dot{{\cal R}}^> ,\\
         \Psi^>  & =\dot{\chi}^{\rm >}  -\varepsilon_1 \frac{a^2}{k^2} \bigg(H\dot{\cal R}^>-{\cal S}_{\cal R}\bigg).
    \end{align}
    \label{eq:CGvariables1}
\end{subequations}
By making further use of eqn.~\eqref{eq:coarseGrainedReq}, the time derivatives needed in the coarse-graining of \eqref{eq:fieldCPT2} can be expressed as%
\begin{subequations} 
    \begin{align}
         \dot{\Phi}^> & = - \frac{H}{{\dot{\phi_b}}}\bigg( {\delta \dot{\phi}^>}+\frac{\varepsilon_2H}{2} {\delta \phi^>}\bigg)+\dot{\cal R}^> , \\
         \dot{\chi}^>  & = \frac{1}{{\dot{\phi_b}}} \Bigg( {\delta \dot{\phi}^>}+\bigg(\varepsilon_1+\frac{1}{2}\varepsilon_2\bigg) H {\delta\phi^>}\Bigg)-\frac{1}{H}\dot{\cal R}^> 
         +\varepsilon_1\frac{a^2}{k^2}\bigg(H\dot{\cal R}^>-{\cal S}_{\cal R}\bigg), \\
         \dot{\Psi}^> & =  \frac{1}{\dot{\phi}_b} \Bigg( {\delta\ddot{\phi}^>}+\big(2\varepsilon_1+\varepsilon_2\big) H \delta\dot{\phi}^>
        +\bigg(\varepsilon_2-2(\varepsilon_1+\varepsilon_3)\bigg) \frac{\varepsilon_2 H^2}{4}\delta\phi^>\Bigg)\nonumber\\
        &+\big(3-\varepsilon_1-\varepsilon_2\big)\dot{\cal R}^>
         +\frac{1}{H}\bigg(\frac{a^2}{k^2}{\cal R}^>-{\cal S}_{\cal R}\bigg).
    \end{align}
    \label{eq:CGvariables2}
\end{subequations}
Replacing these, together with the previous eqn.~\eqref{eq:CGvariables1} and%
\begin{subequations}
\begin{align}
\displaystyle\frac{\mathrm{d}V}{\mathrm{d}\phi}(\phi_b) & = \bigg(-3+\varepsilon_1+\frac{1}{2}\varepsilon_2\bigg) \yld{\frac{H}{\dot{\phi}_b}}\ylg{H\dot{\phi}_b}, \\
\displaystyle\frac{\mathrm{d}^2V}{\mathrm{d}\phi^2}(\phi_b) & = -\frac{1}{4}\bigg(8\varepsilon_1^2+2\varepsilon_1\big(-12+5\varepsilon_2\big)
+\varepsilon_2\big(-6+\varepsilon_2+2\varepsilon_3\big)\bigg) H^2, 
\end{align}
\end{subequations}
into the l.h.s. of \eqref{eq:fieldCPT2}, all terms perfectly cancel apart from the Mukhanov-Sasaki-Starobinsky term ${\cal S}_{\cal R}$ which survives as a source term. One therefore sees that the IR version of eqn.~\eqref{eq:fieldCPT2} is
\begin{equation}
    \label{eq:fieldCPT-CG}
   \delta\ddot{\phi}^>+3 H  \delta \dot{\phi}^>+\frac{k^2}{a^2} \delta \phi^>
    -\dot{\phi_b}\left(3\dot{\Phi}^>+\dot{\Psi}^>+\frac{k^2}{a^2}\chi^>\right)
    +\frac{\mathrm{d}^2V}{\mathrm{d}\phi^2}(\phi_b) \delta \phi^> +2 \Psi^>\frac{\mathrm{d}V}{\mathrm{d}\phi}(\phi_b)= \frac{\dot{\phi}_b}{H} {\cal S}_{\cal R}.
\end{equation}

Similarly to standard derivations of stochastic inflation, and as recalled in Section~\ref{sec:ansatz}, the obtained linear r.h.s. is now assumed to provide the spectrum of the noise making up the stochastic noise term for the full non-linear version of \eqref{eq:fieldCPT2}, giving the {stochastic} ADM inflaton equation
\begin{equation}
 \frac{1}{\boldsymbol{\alpha}}
 \left(
   \dot{\boldsymbol{\Pi}}
   + \boldsymbol{\beta}^i \boldsymbol{\Pi}_{\mid i}
 \right)
 - \boldsymbol{K}\boldsymbol{\Pi}
 - \frac{\boldsymbol{\alpha}^{\mid i}}{\boldsymbol{\alpha}} \boldsymbol{\phi}_{\mid i}
 - \boldsymbol{\phi}_{\mid i}^{\mid i}
 + \frac{\mathrm{d}V}{\mathrm{d}\phi}(\boldsymbol{\phi})
 = \mathcal{F}^{-1}\{\boldsymbol{\mathcal{S}}_{\Pi}\},
 \label{eq:CoarseGrainedADM_Dyn3}
\end{equation}
where ${\cal F}^{-1}\{\cdot\}$ is the inverse Hermitian Fourier transform {and where
\begin{equation}
    \boldsymbol{{\cal S}}_{\Pi} = \frac{\dot\phi_b}{H}  \boldsymbol{{\cal S}}_{{\cal R}}. 
    \label{eq:SPi}
\end{equation}
Here, we have defined the stochastic noise source with the spectrum given in eqn.~\eqref{eq:SR}
\begin{equation}
\boldsymbol{\mathcal{S}}_{\mathcal{R}} (\mathbf{k})= \boldsymbol{{\cal R}}_{\mathbf{k}} \ddot{W}_k+[2 \dot{\boldsymbol{{\cal R}}}_{\mathbf{k}} + (3-\varepsilon_2) H \boldsymbol{{\cal R}}_{\mathbf{k}}]\dot{W}_k.
\end{equation}
where
\begin{equation}
   \boldsymbol{{\cal R}}_{\mathbf{k}}(t) =   {\cal R}_k(t){}\boldsymbol{\alpha_{\mathbf{k}}}
\end{equation}   
with the usual statistics of the stochastic Gaussian basis $\{\boldsymbol{\alpha}\}$.

\subsection{Full ADM Stochastic Inflation: coarse-graining the geometry}\label{subsec:FASI}

The procedure of inserting the coarse grained perturbation variables in the l.h.s. of the linearised versions of the dynamical equations, and thus obtaining the resulting source term, can also be performed for the other dynamical ADM equations.
In this section, we present the result of the application of this procedure to these remaining equations: a set of stochastic dynamical equations for the extrinsic curvature of the spatial slices that are valid for any choice of coordinates, while satisfying the usual \yld{Energy}\ylg{Hamiltonian} and Momentum constraints.
By construction, standard linear perturbation theory is recovered for the coarse grained variables when this system is linearised. In particular, one can re-derive \eqref{eq:coarseGrainedReq} by working backwards. Clearly, if no time-dependent long-short split is performed $\dot{W}_k=0$, all standard equations are recovered.
All these statements have been found analytically and checked using \textit{Mathematica} \cite{Mathematica}.

Let us recall that, although using linear theory for perturbative UV modes up to their crossing, eqns.~\eqref{eq:CoarseGrainedADM_Dyn3}, \eqref{eq:CoarseGrainedADM_Dyn1}, \eqref{eq:CoarseGrainedADM_Dyn2}, and \eqref{eq:CoarseGrainedADM_Con1}, \eqref{eq:CoarseGrainedADM_Con2}, obtained through this procedure capture any non-linear development once modes have crossed the horizon as set by $W$. 
In the following, the stochastic IR $\boldsymbol{bold}$ notation of the l.h.s. will be kept implicit to avoid notational clutter.
 
The extrinsic curvature {of the spacelike 3D hypersurfaces} is found to obey%
\begin{subequations}
    \begin{align}
    \dot{K}& + \beta^i K_{, i}+\alpha^{\mid i}{ }_{\mid i}-\alpha\left({ }^{3} R+K^2\right)
      - {\rm \Mpl^{-2}} \alpha\left(\frac{1}{2} S-\frac{3}{2} \rho\right) =  {\cal F}^{-1}\{\boldsymbol{{\cal S}}_{K}\}, \label{eq:CoarseGrainedADM_Dyn1} \\
      \dot{\tilde{K}}_{ij}& + 2\alpha\tilde{K}_{il}{\tilde{K}}^l{ }_j+\beta^k \tilde{K}_{ij \mid k}-2\beta_i{ }^{\mid k} \tilde{K}_{jk} 
       + \alpha_{\mid i\mid j}\nonumber\\
     &  -\frac{1}{3} \alpha^{\mid k}{ }_{\mid k}\delta_{ij} 
    -\alpha\left({ }^3\tilde{R}_{ij}+\frac{1}{3}K \tilde{K}_{ij}\right)
    + {\rm \Mpl^{-2}} \alpha \tilde{S}_{ij}={\cal F}^{-1}\{\boldsymbol{{\cal S}}_{\tilde{K}_{ij}}\}, \label{eq:CoarseGrainedADM_Dyn2} 
    \end{align}
    \label{eq:CoarseGrainedADM_K} 
\end{subequations}
where the source terms appearing on the r.h.s. are 
\begin{eqnarray}
 \boldsymbol{{\cal S}}_{K} &=&  -\varepsilon_1\boldsymbol{{\cal S}}_{{\cal R}},\\ 
\boldsymbol{{\cal S}}_{\tilde{K}_{ij}} &=& a^2\varepsilon_1\bigg(\frac{1}{3}\delta_{ij}-\frac{k_ik_j}{k^2} \bigg) \boldsymbol{{\cal S}}_{{\cal R}}.
    \label{eq:r.h.s.ADM}
\end{eqnarray}
This anisotropic evolution equation is meaningful: first-order scalar perturbations source tensorial quantities at higher orders, in agreement with previous work \cite{Tomita67,Matarrese98} and discussed further in Section~\ref{sec:stogravitons}.

The two equations share the same random lattice draw as eqn.~\eqref{eq:SPi}, which is key for the explicit satisfaction of the constraint equations
\begin{eqnarray}
{\cal H} \equiv { }^{3} R+\frac{2}{3} K^2-\tilde{K}_{i j} \tilde{K}^{i j}-2 {\rm \Mpl^{-2}} \rho = 0,  \label{eq:CoarseGrainedADM_Con1}\\
 {\cal M}_i \equiv \tilde{K}^j{}_{i \mid j}-\frac{2}{3} K_{\mid i}- {\rm \Mpl^{-2}} {\cal { J }}_i = 0. 
    \label{eq:CoarseGrainedADM_Con2}
\end{eqnarray}
For the first time, the coarse-graining of the constraints has thus been performed explicitly and shows that no stochastic term on the r.h.s. is to be found.
Indeed, inserting the coarse grained variables in the linearised l.h.s. leads to complete cancellations and no source terms appear.
This is an important property of the stochastic augmentation of the ADM system of equations: the dynamics can be seen as a continuous succession of ``drift and kick'' evolution steps or a {``succession of small time-step initial condition problems''} \cite{salopek_stochastic_1991}.
The l.h.s. of the ADM equations \eqref{eq:CoarseGrainedADM_Dyn3} - \eqref{eq:CoarseGrainedADM_Dyn1} - \eqref{eq:CoarseGrainedADM_Dyn2} preserves the constraints if initial conditions satisfy them.
As we argued in Chapter~\ref{sec:ICNR}, satisfying the constraints is an important consideration in any scheme for setting initial conditions in Numerical Relativity, and we have seen that this has been rather overlooked or debated in the literature on Stochastic Inflation.
Since the stochastic source terms can be interpreted as modifying the `initial conditions' for each successive time step, their addition must not alter the ADM constraints. Our approach for defining these stochastic source terms satisfies the constraints at first order {in perturbation theory} by construction (or up to $\mathcal{O}(\boldsymbol{{\cal S}}_{{\cal R}}^2)$), as a direct consequence of a gauge-agnostic coarse-graining.
If the coarse-graining is not done as we described, e.g. by setting $\Psi_B^> = \Phi_B^>$, one gets a violation of the constraints and a strong gauge-dependence of the r.h.s., which supports our choice.

The stochastic equations derived here rely on linearised perturbations for determining the noise terms and this is an approximation that underlies practically all derivations of stochastic equations for inflation. However, {once a stochastic kick has been given}, the subsequent evolution of the long wavelength configuration is fully non-linear.
The most striking observation is probably the simplicity and similarity of the r.h.s. terms, which contrasts with the nonlinear l.h.s.. This is completely due to the linear framework in the UV and the fact that all dynamics are encoded in one variable. 
Overall, the linear noise approximation together with neglecting the quantum contributions to the interactions (expected to hold from horizon crossing or so, depending on the model, as argued in previous Chapter~\ref{sec:semiclQFT}), and the stochastic backreaction of IR fields on the UV modes that determine the noise via \eqref{eq:ReqISTORIZ1}, are the only approximations left in the above equations. 

Equations \eqref{eq:CoarseGrainedADM_Dyn3} - \eqref{eq:CoarseGrainedADM_Dyn1} - \eqref{eq:CoarseGrainedADM_Dyn2} and \eqref{eq:CoarseGrainedADM_Con1} - \eqref{eq:CoarseGrainedADM_Con2}, solved simultaneously with \eqref{eq:ReqISTORIZ1} which defines the noise sources, constitute our dynamical equations for Stochastic Inflation in General Relativity. 
Note that one can also apply the coarse-graining directly to Einstein's field equations%
\begin{subequations}
\begin{align}
G_{00}-\frac{1}{M_{Pl}^2}T_{00} & = 0,  \\
G_{0i}-\frac{1}{M_{Pl}^2}T_{0i} & =  0 ,\\
G_{ij}-\frac{1}{M_{Pl}^2}T_{ij} & =  {\cal F}^{-1}\{\boldsymbol{{\cal S}}_{ij}\}  ,
\end{align}
    \label{eq:EinsteinCoarseGrained}
\end{subequations}
where 
\begin{equation}
\boldsymbol{{\cal S}}_{ij}  = -a^2\varepsilon_1 \bigg(\delta_{ij}-\frac{k_ik_j}{k^2}\bigg) \boldsymbol{{\cal S}}_{{\cal R}}.
    \label{eq:RHSEinstein}
\end{equation}

\subsection{Discussion}\label{sec:SIGRdiscussion}

In practice, the random variables $\boldsymbol{\alpha_{\mathbf{k}}}$ are drawn once at some fixed early time, which defines a real space random field $\boldsymbol{{\cal R}}(t,\mathbf{x})$ that will be propagated in time by the real space version of the MS equation, written as%
\begin{subequations}
    \label{eq:ReqISTORIZ1}
\begin{align}
        \boldsymbol{\Pi}_{\cal R} &= \frac{9\dot{\phi}_b^2}{\alpha_b K_b^2} \bigg(\partial_0\boldsymbol{{\cal R}}+\beta^i\partial_i\boldsymbol{{\cal R}}\bigg),\\
        \partial_0\boldsymbol{\Pi}_{\cal R}&-\alpha_b K_b\boldsymbol{\Pi}_{\cal R}+\beta_b^i\partial_i\boldsymbol{\Pi}_{\cal R} -\frac{9\dot{\phi}_b^2\alpha_b}{ K_b^2}\yld{\partial_i\partial^i}\ylg{\Delta_b}\boldsymbol{{\cal R}}=0\,. 
    \end{align}
    \end{subequations}
}
Let us recall that those equations' r.h.s. are initially only valid for perturbations around a homogeneous background. However, evolving them in real space has some advantage beyond the software requirement.
As done in the SUA-based SI, it is possible to provide here a heuristic to include the backreaction of the long wavelength inhomogeneous fields on the UV modes used to compute the noise.
This consists in promoting the background parameters in the equation determining the solution for the UV modes into functions of those spatially-dependent long wavelength fields \cite{salopek_stochastic_1991, levasseur_backreaction_2015, figueroa_non-gaussian_2021,Figueroa22}.
Following this general approach, in this work we include the possibility of such long-wavelength backreaction on short wavelength modes by allowing for the coefficients in \eqref{eq:ReqISTORIZ1} to take their local values\footnote{To our knowledge, such a scheme was first proposed but not numerically implemented in \cite{salopek_stochastic_1991}.
More recently, a similar construction for the mode equations was numerically implemented in \cite{figueroa_non-gaussian_2021,Figueroa22,Tomberg:2024evi}.}
$$\phi_b \rightarrow \phi(t,x^i), \quad K_b \rightarrow K(t,x^i), \quad \alpha_b\rightarrow \alpha(t,x^i), \quad \beta^i_b\rightarrow \beta^i(t,x^i),$$
which might make the noise sources non-Gaussian. However, one should probably not stop there as both ${\cal S}_{\cal R}$ and the pre-factors of each ADM equation include background quantities, notably $\varepsilon_1$ and $\varepsilon_2$. These various levels of stochastic backreaction will be discussed in more detail in Chapter~\ref{sec:SINR}.

In a similar manner, the gauge freedom has its own subtlety as the previous equations are valid for any gauge and any background time slicing (the latter being only a matter of variable change for straight time derivatives to get to $\alpha_b'(t') \neq 1$).
The gauge-invariance of the fully non-linear equations is less obvious. In particular, a gauge transformation will leave any non-linear l.h.s. unchanged, but the associated r.h.s. will be consistent only if both the initial and the final gauges are close enough to the homogeneous background one so that we stay within CPT when linearising the r.h.s.. This is why a more accurate property of our equations should be \textit{small gauge invariance}.

One needs to emphasise that the long-wavelength approximation has been removed from the equations.
This is what allowed us to formulate equations in any gauge by bypassing the gauge-mapping issue in this regime \cite{artigas_hamiltonian_2022}, arising before even coarse-graining.
Note that the long wavelength approximation could still be applied to the choice of the window function if one wants to ensure the complete classicalisation of the crossing modes \cite{burgess_minimal_2022}.
However, the window function can now be turned on much closer to the Hubble radius than in the {SUA}, the latter being restricted by the quasi-isotropy assumption \cite{artigas_hamiltonian_2022}.
In particular, one can now safely study regime transitions in SI where gradients are critical, as opposed to the main approach \cite{jackson2023}.
Switching on the window closer to Hubble crossing also gives less interaction time (and so fewer higher-order effects) to the UV modes pre-entry and so strengthens the linear and Gaussian source approximation, as well as empowers SI with the possibility to look at non-Gaussianity as interactions start mostly from the crossing.
In other words, SI can now in principle be more predictive than QFTCS for semiclassical nonlinearities.
Note, however, that we discourage any attempt to apply those equations blindly: QFTCS on curved spacetime is still necessary to define the classicality and interactivity time of the modes for a given lattice and model, as argued in Chapter~\ref{sec:semiclQFT}. 

\subsection{Usual Stochastic Inflation limit}\label{sec:SIGRusuallimit}
In this section, we compare our results with the commonly used eqn.~(\ref{eq:SIrewrite}) of SI \cite{pattison_stochastic_2019}. To achieve this, we need to change the l.h.s. of the $\Pi$ equation to its SUA limit eqn.~(\ref{eq:longGR}) and specify our gauge. 

By fixing $\partial_0{\cal N}=1$, i.e. choosing $x^0$ as the background e-folding ${\cal N}_b$, and $\beta^i=0$, we can write $H^> \equiv -\frac{1}{3}K^> = \frac{1}{\alpha^>}$, and hence $\alpha^>$ is given by the Hamiltonian constraint in the long-wavelength limit.
Via the Hamiltonian constraint, the tractable dependence of $H^>$ on $\Pi^>$ and $\phi^>$ only in the long wavelength limit allows us to drop the stochastic $H$ (or $K$) equation.\footnote{See Chapter~\ref{sec:SINR}, Section~\ref{sec:SINR.QD} to see the link between the satisfaction of the constraints and the stochasticity in both the field and the geometry, as well as the subtlety of the IR $\varepsilon_1$.}
Substituting this in the $\Pi$ equation's l.h.s. yields%
\begin{subequations}
    \label{eq:SImine}
\begin{align}
 \partial_{{\cal N}_b} {\Pi}^>  & = -3{\Pi}^>-\frac{1}{H^>}\frac{dV}{d\phi}(\phi^>)+ \frac{1}{H}{\cal F}^{-1}\{\boldsymbol{{\cal S}}_{\Pi}\},\\
 {\Pi}^> &=  H^> \partial_{{\cal N}_b} \phi^>, \\
 {H^>}^2 & = \frac{V(\phi^>)}{1-\frac{1}{6\Mpl^2}\left(\partial_{{\cal N}_b} \phi^>\right)^2},
\end{align}
\end{subequations}
which is equivalent to the usual eqn.~(\ref{eq:SIrewrite}) but with our own noise term ${\cal F}^{-1}\{\boldsymbol{{\cal S}}_{\Pi}\}$ at first order. Note that since the ${\cal S}_{\cal R}$ has no apparent Laplacian, we leave it unchanged under the long-wavelength approximation.

Our formalism is now comparable to the literature's SI, defined in eqn.~\eqref{eq:SIrewrite}, by looking at the associated noise $\boldsymbol{\Sigma}$ and $\boldsymbol{{\cal S}_{\cal R}}$ only. On the one hand, ${{\cal S}}_{{\cal R}}$ can be re-written in terms of background e-folds slicing ($\alpha_b = H^{-1}$ and $H\partial t = \partial {\cal N}_b$) as
\begin{equation}
    {{\cal S}}_{{\cal R}} = H^2\left ( {\cal R}_k \frac{\partial^2 W_k}{\partial{\cal N}_b^2}+\left[2 \frac{\partial{\cal R}_k}{\partial{\cal N}_b} + (3-\varepsilon_1-\varepsilon_2) {\cal R}_k\right ]\frac{\partial W_k}{\partial{\cal N}_b} \right ).
    \label{eq:Myr.h.s.inEfoldings}
\end{equation}
On the other hand, we can make $\boldsymbol{\Sigma}$ explicit by calculating $\delta \phi$ and $\partial_{{\cal N}_b}\delta \phi$ from the long wavelength limit of our calculus in Section~\ref{sec:ICNRuniformN} where $\delta{\cal N}^* = 0$. By using the long-wavelength limit\footnote{Note that in \cite{pattison_stochastic_2019}, neglecting the $k^2E$ terms is justified only in SR, USR, and Starobinsky models (and it requires fixing the initial value of $E$, i.e. the remaining d.o.f. of the gauge).}%
\begin{subequations}
    \begin{align}
    \delta\phi^{\rm UE}   & \simeq \sqrt{2\varepsilon_1}\Mpl{\cal R},\\
    \delta\pi^{\rm UE}  &  \simeq \sqrt{2\varepsilon_1}\Mpl\left(-\frac{1}{2}\varepsilon_2{\cal R}+\displaystyle\frac{\partial{\cal R}}{\partial{\cal N}_b} \right),
\end{align}
    \label{eq:NoiseUniNlong}
\end{subequations}
which is confirmed to be the same as the spatially flat gauge decomposition in this decoupled and long-wavelength limit.
Finally, by substituting this in $\boldsymbol{\Sigma_{\phi}}$ and $\boldsymbol{\Sigma_{\pi}}$ of eqn.~(\ref{eq:SInoise}), one exactly gets $\boldsymbol{\Sigma} =  {\cal F}^{-1}\{\sqrt{2\varepsilon_1}\Mpl\boldsymbol{{\cal S}}_{\cal R}\} = {\cal F}^{-1}\{\boldsymbol{{\cal S}}_{\Pi}\}$, confirming that our stochastic source is equivalent to the literature's one in this limit.

We should stress that beyond this decoupled and long-wavelength limit, our stochastic source is different from the usual one.
This is because the usual derivation does not coarse-grain $\cR$ effectively, as in general the uniform e-folding gauge is not the same as the flat gauge where the field is precisely proportional to $\cR$.
In theory, the usual coarse-graining is thus inconsistent with respect to the timing of classicality, which should be done with respect to $\cR$.
In practice, this is not an issue as the standard equations are limited to super-Hubble regimes, where classicality should be reached for all perturbative combinations, even without decoupling.

\subsection{Stochastic gravitons}\label{sec:stogravitons}
As already pointed out, the scalar perturbations have an influence on the non-scalar degrees of freedom through the $\tilde{K}_{ij}$ evolution equation of eqn.~\eqref{eq:CoarseGrainedADM_Dyn2}.
At second order and later non-perturbative orders, one thus expects scalar-induced and scalar-coupled tensorial gravitational (or spin-2) waves \cite{Tomita67,Matarrese98}.
In that sense, our equations are the first of their kind to provide a stochastic framework including both scalar and tensorial evolutions.
This is not surprising as previous studies worked in the long-wavelength limit to drop tensorial dynamics.

In previous work \cite{Tasinato22}, it is suggested that we should also include the decohered first order gravitons in the stochastic sources.
This can be simply added if we stay at linear order for the sources. Indeed, we only need to do the same work as for the scalar but for gravitonic perturbations.
This can be done independently and just added to the final scalar result thanks to the scalar-decoupled linear limit.
In practice, this appears to be straightforward as we can build a linear gauge-invariant quantity $h$ by writing the linear tensorial part of the SVT metric, see eqn.~\eqref{eq:SVTmetricAll}, in cosmic time as
\begin{equation}
  ds^2 = -dt^2 + a(t)^2(\delta_{ij}+h_{ij})dx^idx^j,  
\end{equation}
with $h^i{}_i=\partial_ih_{ij} = 0$ (traceless and transverse).
At linear order, $K$, $\tilde{K}_{ij}$, ${^3}R$ and ${^3}\tilde{R}_{ij}$ are the only quantities in the EOM inheriting contributions from $h$.
This leads to adapting the only equation having $h$ terms at linear order, the linearised $\tilde{K}_{ij}$ equation in its exact ordering
\begin{equation}
    -\frac{1}{2}a^2\left[ \partial_t^2 h_{ij}+3H\partial_t h_{ij}-\frac{\nabla^2}{a^2}h_{ij}\right] = 0,
    \label{eq:eqMShreal}
\end{equation}
which in Fourier space appear to be the known Mukhanov-Sasaki equations for the two linear spin-2 components of the gravitons%
\begin{subequations}
    \label{eq:eqMSh}
    \begin{align}
       & h_{ij}(\mathbf{k},t)= \epsilon^+_{ij}(\mathbf{k})h_k^+(t) + \epsilon^\times_{ij}(\mathbf{k}) h_k^\times(t),\\
       & \ddot{h_k^s}+3H\dot{h_k^s}+ k^2h_k^s = 0, \forall s = +,\times,
    \end{align}
\end{subequations}
and which can also be initialised by a Bunch-Davies vacuum. 
We now have everything we need to coarse-grain at linear order with the previous window method using
$$h_{ij}^>(\mathbf{k},t) = W^h_k(t)h_{ij}(\mathbf{k},t),$$
assuming we want the same window for both polarisations. By coarse-graining eqn.~\eqref{eq:eqMSh} to get the spectrum of sources for eqn.~\eqref{eq:eqMShreal}, and promoting the sources to those of the $\tilde{K}_{ij}$ equation, we obtain the updated $\boldsymbol{{\cal S}}_{\tilde{K}_{ij}}$ as%
\begin{subequations}
    \label{eq:eqNewSource}
\begin{align}
     \boldsymbol{{\cal S}}_{\tilde{K}_{ij}} & = \boldsymbol{{\cal S}}_{\tilde{K}_{ij}}^{(\phi)}+\boldsymbol{{\cal S}}_{\tilde{K}_{ij}}^{(h)}, \\
 \boldsymbol{{\cal S}}_{\tilde{K}_{ij}}^{(\phi)}  & = a^2\varepsilon_1\bigg(\frac{1}{3}\delta_{ij}-\frac{k_ik_j}{k^2}\bigg) \boldsymbol{{\cal S}}_{{\cal R}},\\
  \boldsymbol{{\cal S}}_{\tilde{K}_{ij}}^{(h)}  & = -\frac{1}{2}a^2\left(\displaystyle\sum_{s=+,\times}\epsilon^s_{ij}(\mathbf{k}) \boldsymbol{{\cal S}}_{h}^s\right),
  \end{align}
\end{subequations}
  where%
  \begin{subequations}
    \label{eq:eqNewSourceWhere}
\begin{align}
  \boldsymbol{{\cal S}}_{h,\mathbf{k}}^s & = {\cal S}_h(k)\boldsymbol{\alpha}^s_{\mathbf{k}}, \\
  {\cal S}_h(k) & = h_k^s \ddot{W}_k^h + \left[2 \dot{h}_k^s+3Hh_k^s\right] \dot{W}_k^h,\\
  \langle \boldsymbol{\alpha_{\mathbf{k}_1}}^{s_1}\boldsymbol{\alpha_{\mathbf{k}_2}}^{s_2*} \rangle_{\mathbb{P}}  & =  \delta^{(3)}(\mathbf{k}_1-\mathbf{k}_2)\delta^{s_1s_2}.
\end{align}
\end{subequations}
The coarse-graining is now complete and appears to be much easier because of the gauge invariance of the perturbation and the trivial satisfaction of the constraints at linear order.
This extension is not without utility because eqn.~\eqref{eq:eqNewSource} shows a competition between scalar and tensor sources, which seems to be in favour of the latter in slow-roll regimes.
Tensorial degrees of freedom can now be studied in a non-perturbative framework, numerically most likely in the future.\footnote{{For instance, the stochastic backreaction of IR tensorial modes on both tensorial and scalar modes would be involved (credits to Prof. D. Wands and Luiz Felipe Demétrio).}}

\section{A fresh look at non-Gaussianity}\label{sec:SIGRng} 

The stochastic equations of the previous section can be applied to a variety of gauges.
In particular, knowing ${\cal N}$ in uniform-density or uniform-field gauges provides a direct gauge-invariant extraction of quantities such as the nonlinear generalisation ${\cal R}^{\rm NL}$ as defined in Section~\ref{sec:finalvalues}, and this is why the FPTA is required in usual studies where the field is stochastic while the e-fold number is not \cite{vennin_correlation_2015}.
In this context and since a gauge-invariant formulation is being proposed in this work, we propose to apply the uniform field gauge directly to our equations and avoid the FPTA.
Note that this has been attempted recently but starting from the literature's usual equations and assumptions \cite{tomberg_numerical_2023}.

We start from our full ADM equations and set the coordinates such that the field evolves uniformly according to the background dynamics, $\phi(t,\mathbf{x})=\phi_b(t)$, where $\phi_b(t)$ follows eqn.~\eqref{eq:backgroundEq} with $\alpha_b = 1$.
Hence $\phi$ can act as a clock labelling the 3D spatial hypersurfaces. This gauge choice can only be valid if the background field is non-static.
The spatial coordinates on these spatial slices are fixed such that $\beta^i=0$.
In addition, $\phi$ does not now receive stochastic impulses and the stochastic dynamical variables are ${\cal N}$ and $\alpha$ for which Langevin equations can be derived.   

Bypassing its validity for the purpose of an analytical toy and a straightforward numerical stochastic EOM study, this section uses the long-wavelength approximation for the Hamiltonian constraint, re-writing eqn.~\eqref{eq:LongIsotrop}
\begin{equation}
    \frac{6}{\alpha^2}(\partial_t {\cal N})^2 = \frac{2}{\Mpl^2}\left(\frac{{{\dot{\phi_b}}}^2}{2\alpha^2}+V(\phi_b)\right),
\end{equation}
by inserting the definition of ${\cal N}$. From this we immediately obtain
\begin{equation}
  \frac{ \partial{\cal N}}{\partial{\cal N}_b} = \displaystyle\sqrt{1+\bigg(\alpha^2-1\bigg)\bigg(1-\frac{1}{3}\varepsilon_1\bigg)},
  \label{eq:eqUniFieldefolds}
\end{equation}
by making use of the background Friedmann equation. No slow-roll approximation has been made so far.
The second equation we need is given by the field equation which becomes in this gauge an equation for the lapse 
\begin{equation}
  \partial_t \left( \frac{\dot{\phi_b}}{\alpha} \right)= -3\frac{\dot{\phi_b}}{\alpha} \partial_t {\cal N} -\alpha V_{,\phi}(\phi_b) +\alpha\boldsymbol{S}_{\Pi},
\end{equation}
where $\boldsymbol{S}_{\Pi}$ is the Langevin noise found previously. Note again that any multiplier of $\boldsymbol{S}_{\Pi}$ is left undecided concerning the stochastic backreaction.
To get to $\partial_t \alpha = \partial_t f$ where $f(Z_t,t) = \frac{\dot{\phi}_b}{Z_t}$ for  $Z_t = \frac{\dot{\phi}_b}{\alpha}$, the Itô lemma \cite{VanKampen1992, 1994hsmp.book.....G, 2007PhRvE..76a1123L} is used to get 
\begin{equation}
    \label{eq:ito}
    \partial_t \alpha =  \frac{\ddot{\phi_b}}{\dot{\phi_b}}\alpha +\frac{\alpha^3}{\dot{\phi_b}^2}\langle\boldsymbol{S}_{\Pi}({\cal N}_b)^2\rangle
 -\frac{\alpha^2}{\dot{\phi_b}} \left (-3\frac{\dot{\phi_b}}{\alpha} \partial_t {\cal N} -\alpha \frac{dV}{d\phi}(\phi_b) +\alpha\boldsymbol{S}_{\Pi}\right ).
\end{equation}
We have included the Itô correction to the derivative of a function of a random variable as the second term in the above equation.
Note, however, that the effect of this term is minimal (higher order) and the results of the simulations we describe below are practically unaffected by it.
Substituting in the field background equation and the e-folds evolution from eqn.~(\ref{eq:eqUniFieldefolds}), changing the time variable by dividing by $H$  and consequently updating the variance of the Wiener process yields
\begin{equation}
    \label{eq:eqUniFieldlapse}
    \frac{\partial \alpha}{\partial {\cal N}_b}  = 3\alpha\left(\displaystyle\sqrt{1+(\alpha^2-1)(1-\frac{1}{3}\varepsilon_1)}-1\right) +\alpha^3\langle\boldsymbol{S}({\cal N}_b)^2\rangle
    +\bigg(-3+\varepsilon_1+\frac{1}{2}\varepsilon_2\bigg)\alpha(\alpha^2-1)+\alpha^3\boldsymbol{S}({\cal N}_b) ,
\end{equation}
where $\boldsymbol{S} = \boldsymbol{S}_{\Pi}/H\dot{\phi_b}$ is the final stochastic source. In the following, we will keep the coefficient $\alpha^3$ multiplying $\boldsymbol{S}$ as a non-background, stochastic quantity, and will not set it to $1$.
Note that when linearised, these equations still match CPT at first order. Eqns.~\eqref{eq:eqUniFieldefolds} and \eqref{eq:eqUniFieldlapse} together form a coupled system of stochastic PDEs.

In the SUA philosophy, it is common to restrain our case to one patch of the universe for the treatment of the noise, i.e. using the previous equation at one given point $\mathbf{x}_0$ and using only $\boldsymbol{S}({\cal N}_b,\mathbf{x}_0)$. This is completely justified because the window function derivatives yield suppressed correlations beyond the Hubble scale (e.g. a Heaviside window in Fourier space would lead to a cardinal sinus in real space) \cite{Starobinsky_stochastic_1988}. In this framework, it is easier to solve this system of two coupled Langevin equations or stochastic ODEs.

An analytical solution would require specifying the background dynamics and writing the Fokker-Planck equation to get the PDF. To our knowledge, there is no known way to do that analytically without further approximations such as overdamping, model-dependent simplifications, or the neglect of higher-order correlations. As a proof of concept, we decided to provide numerical sample results instead.

The simulations were realised with the Stratonovich evolver of \textit{Mathematica} \cite{Mathematica}.\footnote{\ylg{Since the noise is multiplicative through the $\alpha^3$ factor, the Itô and Stratonovich interpretations of eqn.~\eqref{eq:eqUniFieldlapse} differ by a drift term of order $\alpha^5\langle\boldsymbol{S}^2\rangle$, the same order as the Itô correction of eqn.~\eqref{eq:ito}. As both are negligible here, so is the choice of interpretation, and these subtleties have little influence on our simulations. See \cite{pinol_manifestly_2021} for the discretisation ambiguity in stochastic inflation, and Chapter~\ref{sec:SINR} for the Wong-Zakai argument with smooth windows.}} The following amplitude for $\boldsymbol{S}$, computed in Appendix \ref{app:appD}, is valid for both slow-roll and ultra-slow regimes (USR)
\begin{equation}
\sqrt{\langle\boldsymbol{S}({\cal N}_b)^2\rangle}\simeq \frac{1}{\sqrt{2\varepsilon_1({\cal N}_b)}}\frac{3}{2\pi}\frac{H({\cal N}_b)}{\Mpl}.
\end{equation}
The main difference with usual amplitudes is the $\sqrt{\varepsilon_1}$ in the denominator. This comes from the fact that the lapse equation depends on the amplitude of ${\cal R}$ and not the usual Mukhanov-Sasaki variable ($a\sqrt{2\varepsilon_1}\Mpl\times{\cal R}$) to which field equations are sensitive. Of course, the $\varepsilon_1\rightarrow 0$ limit is problematic, but this is purely a gauge artefact as already noted earlier when defining our gauge and as explained in \cite{Kinney05}.  For the simulation, we choose to evaluate this quantity with the background evolution and so to neglect the stochastic backreaction for simplicity, \yld{as opposed to}\ylg{but keeping} the $\alpha^3$ factor \yld{acting on}\ylg{in front of} $\boldsymbol{S}$.

\begin{figure}[ht]
    \begin{subfigure}{\textwidth}
        \centering
        \includegraphics[width=0.80\linewidth]{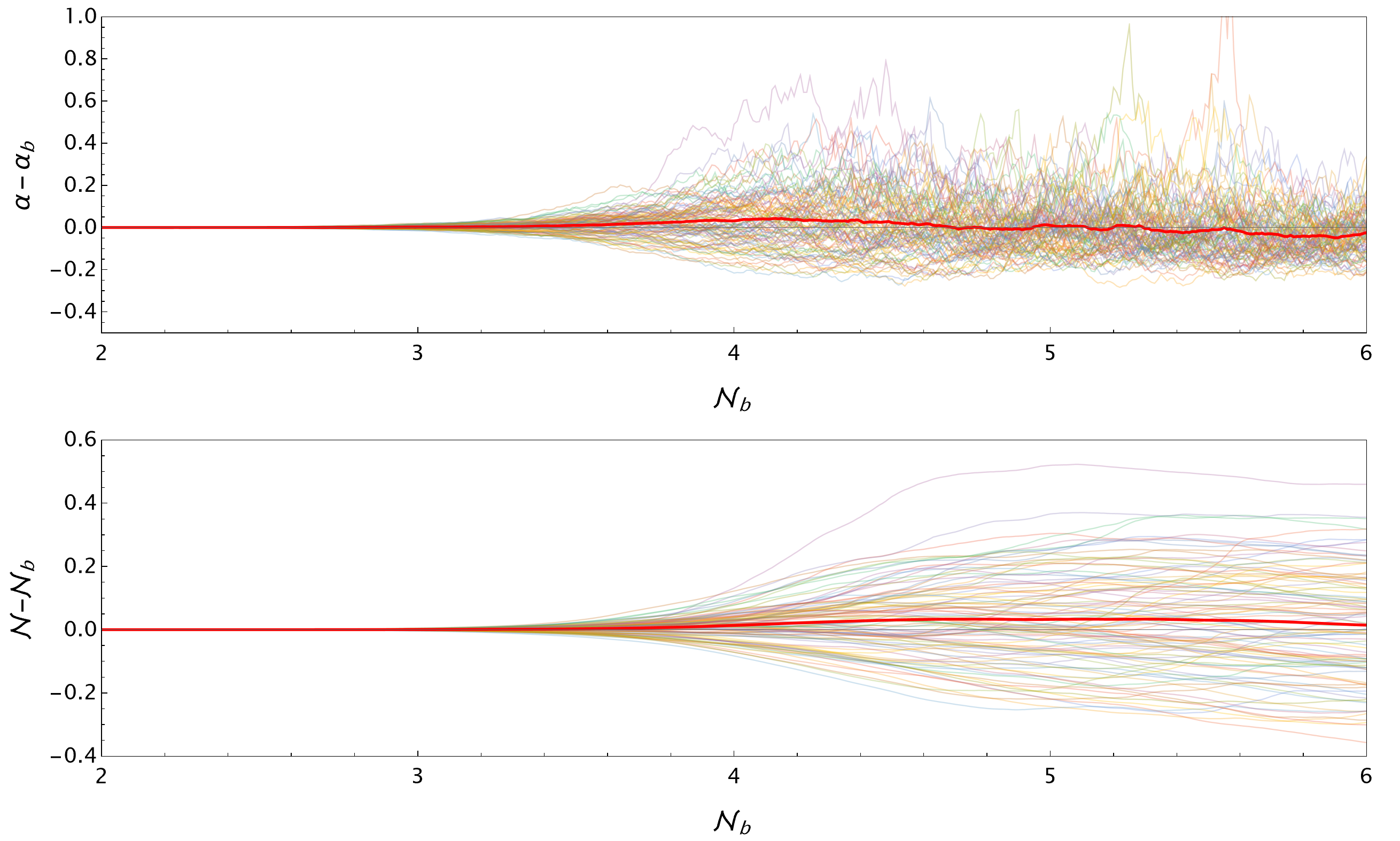}
        \caption{100 of 50000 stochastic paths are shown for the lapse (top) and e-fold (bottom) differences to background. Red curves provide the mean difference over the 100 paths.}
        \label{fig:trajectories}
     \end{subfigure}
    \begin{subfigure}{\textwidth}
        \centering
        \includegraphics[width=0.65\linewidth]{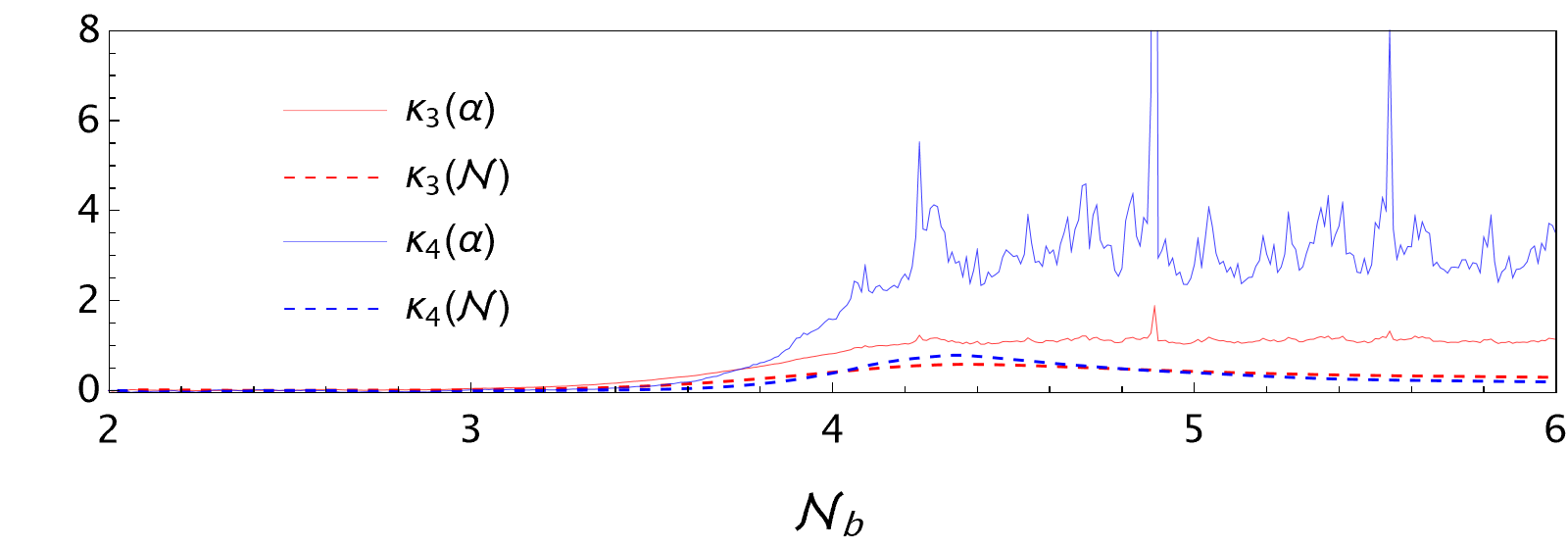}
           \caption{Skewness and kurtosis excess in time for 50000 stochastic paths of the lapse and e-fold differences to background respectively.}
        \label{fig:cumulants}
    \end{subfigure}
\caption{A scalar field undergoing stochastic ultra slow-roll dynamics in uniform field gauge, following eqns.~\eqref{eq:eqUniFieldlapse}, \eqref{eq:eqUniFieldefolds}. The noise is active between $2$ and $6$ e-folds. The field has been initialised at $\phi_b^* = 10$ with an e-folds velocity $\pi_b^*=d\phi_b/d{\cal N}_b^* = -2.0001$ in a potential such that $V(\phi)=V_0$ until $\phi_c = 9.056$, after which $V(\phi)=V_0 [3 (\phi/\phi_c)^2-2(\phi/\phi_c)^3]$, which stops the exponential slowing of the field.  Note that $V_0$ is set to satisfy the Friedmann constraint equation assuming $H^* = 10^{-5}$ initially. Quantities are expressed in $\Mpl= c = 1$ units.}
\label{fig:trajcumul}
\end{figure}
Figures \ref{fig:trajectories} and \ref{fig:cumulants} show 50000 paths starting from uniform lapse and e-folds. $0.01$-e-folds steps were used in the numerical scheme. Without worrying about the realism of our model of Inflation concerning observational constraints, we considered a plateau Inflation followed continuously by a 3rd order polynomial slope of the form $V(\phi)\propto[3 (\phi/\phi_c)^2-2(\phi/\phi_c)^3]$. Initial position and momentum together with the start of the slope are fine-tuned\footnote{Note that this precise tuning depends on our time resolution ($0.01$ e-folds)} for two reasons: we want a decent amount of dispersion and non-Gaussianity by having the field almost stopped on the plateau, but not too much as the gauge definition makes the simulation crash if the velocity is too low. Constraints also need to be satisfied. It appears that USR is reached from 2 e-folds onwards ($\varepsilon_1 \ll1$, $\varepsilon_2 \simeq 6$), which is why the noise is only switched on at this time. When the slope is reached at $\simeq 3.04$ e-folds, the velocity increases extremely slowly, still within the USR regime. 

\begin{figure}[p]
\begin{subfigure}{\textwidth}
    \centering
\includegraphics[width=0.9 \textwidth]{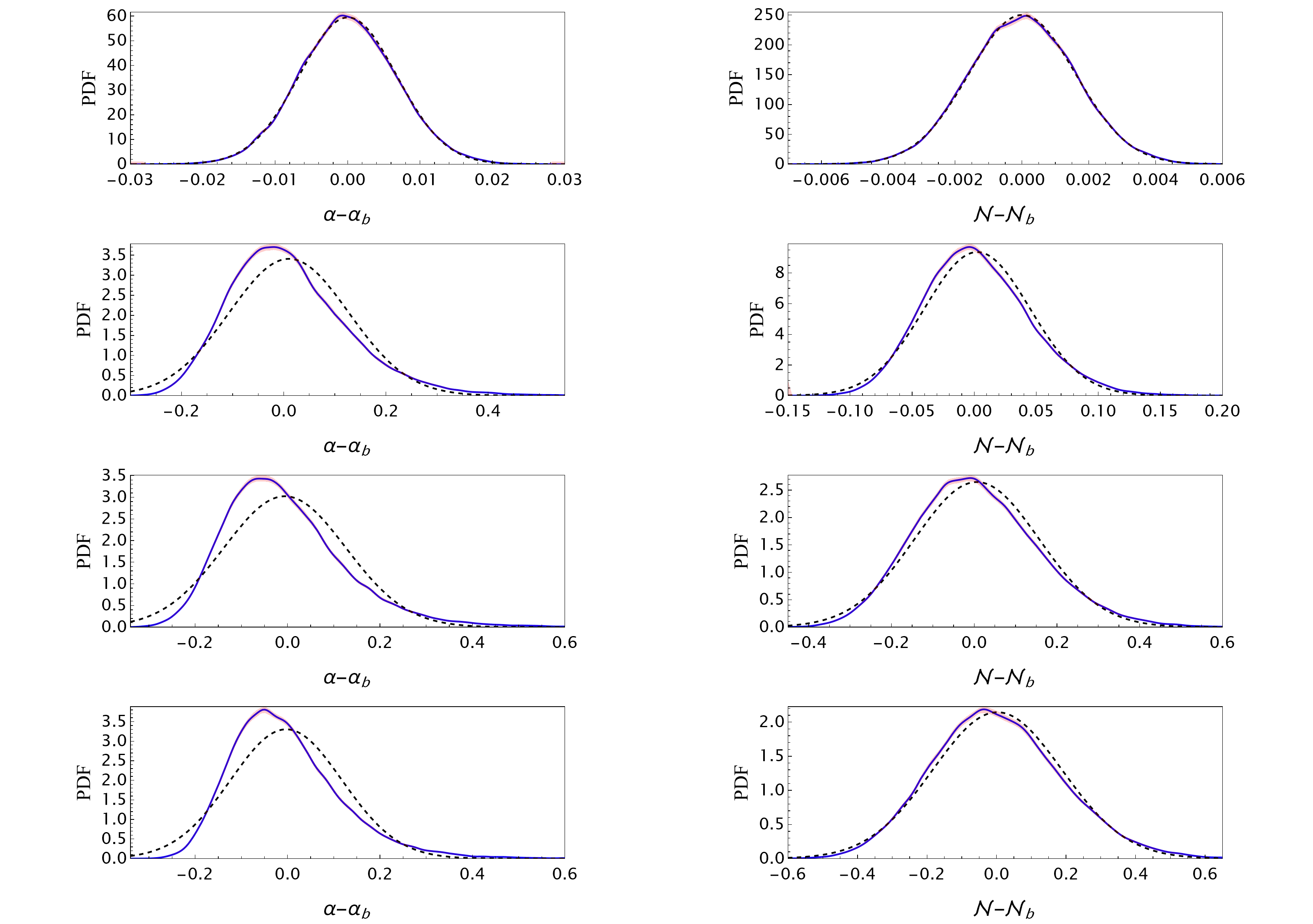}
    \caption{PDFs}
    \label{fig:fig3}
     \end{subfigure}
\begin{subfigure}{\textwidth}
    \centering
\includegraphics[width=0.9 \textwidth]{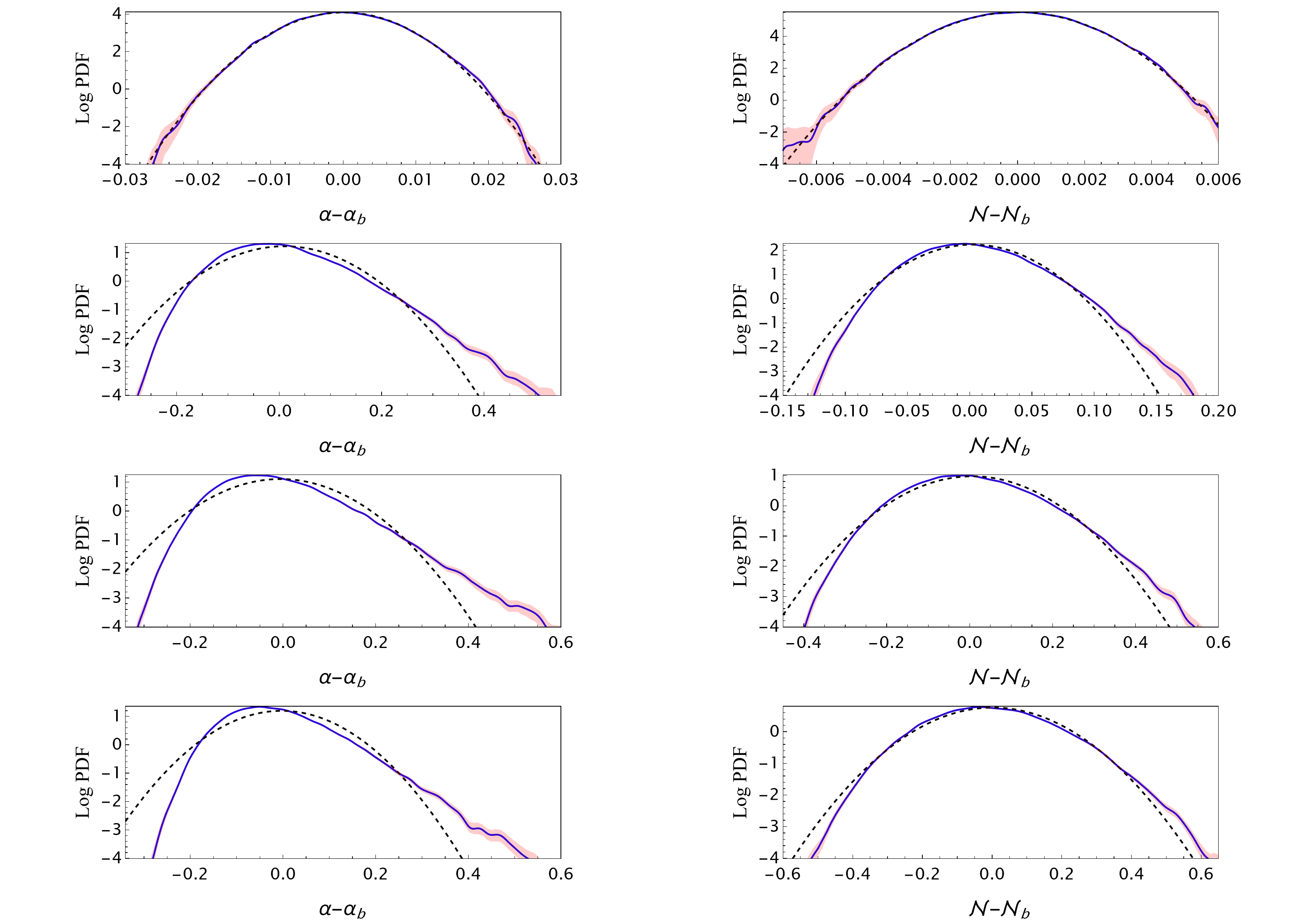}
    \caption{Log PDFs.}
    \label{fig:fig4}
    \end{subfigure}
    \caption{Snapshots of the probability distribution (blue) of a scalar field undergoing stochastic USR dynamics in uniform field gauge, see Figure \ref{fig:trajcumul} for the parameters of the simulation. 50000 paths were used for Kernel Density Estimation to probe the PDFs at four background e-fold times (${\cal N}_b=3,4,5,6$ from top to bottom) for both the lapse (left) and the e-fold (right) differences to the background, against normal PDFs with the same first two moments (black dashed). $95\%$ confidence intervals (red) of these pdf estimators were calculated using bootstrapping.}
\end{figure}

It is well known that a flat potential or a transition can leave strong non-Gaussian imprints on perturbations such as exponential tails \cite{ezquiaga_exponential_2020}. This is actually what we confirm here in Figures \ref{fig:fig3} and \ref{fig:fig4}: both the lapse and the e-folds get an exponential tail on the plateau between $3$ (end of plateau) and $4$ e-folds, which can be confirmed by diverse fittings. This non-Gaussianity can also be tracked in time by looking at the skewness and the kurtosis in Figure \ref{fig:cumulants}.

When it comes to the second phase - and any analytical attempts would probably fail to describe it fully, things eventually stabilise along the slope. From these figures, it becomes clearer that the lapse acts as an extremely non-Gaussian e-folds' momentum. This implies that very strong non-Gaussian changes are given to the e-folds until stopping and eventually the e-folds distribution's non-Gaussianity stabilises later to a lower remnant level\footnote{A study of the PDF's small tail would be possible using the constrained stochastic formalism \cite{Tokeshi23} or importance sampling \cite{Jackson22}.} when all realisations are in the same regime. We have checked that switching the noise terms off at $6$ e-folds makes the lapse come back to its attractor $\alpha_b =1$ and that the e-folds, which are meant to describe ${\cal R}^{NL}$, are indeed conserved. If the noise was frozen later on this potential, one would see that leaving the USR phase makes the lapse and the e-folds come back closer to Gaussianity. This is because the PDF in real space is not a good estimator: adding many Gaussian contributions lowers the relative non-Gaussianity. However, the non-Gaussianity we produced is still imprinted by the end of our simulation, in particular on the statistics of the scales which crossed the horizon before 6 e-folds. This advocates for the necessity to look at quantities such as $n$-spectra or coarse-grained PDFs when looking at data.

It is important to mention that the simulation has also been run by setting $\alpha^3$ to $\alpha_b^3 = 1$. Unusual left-skewed PDFs have been generated and highlight the importance of stochastic backreaction in certain cases. Here it is critical because of the non-perturbative behaviour of the lapse and the e-folds. In particular, this term is a good barrier to reaching an unphysical $\alpha=0$ with stochastic kicks, without adding a prior when solving the Fokker-Planck equation.

What is to be remembered from these equations and simulation is their potential: having coarse-grained but satisfied constraints and an idea of what the stochastic amplitude is can help us skip the tedious and approximate stochastic $\Delta {\cal N}$  and FPT formalisms by writing it all directly in the right gauge. 

However, following the literature, the SUA was used and so the validity is far from the crossing scale. This allows us to reduce to a simple GR framework where we do not need a full Numerical Relativity code, which probably does not exist in such a gauge.  Furthermore, the validity of the long-wavelength approximation is questionable when perturbations become smaller in magnitude than gradient corrections. This could be the case here for the lapse. For these reasons, traceless modes and other terms should be fully accounted for and could stop those perturbations from vanishing completely. We thus emphasise here that it is necessary to run a full NR code even if it takes using an NR gauge. In that sense, this section mainly aims at illustrating how this new framework might compete with the FPT formalism using the same assumptions for the evolution.

\section*{Summary}

In this chapter we have examined the formulation of SI in full General Relativity, dropping most of the approximations made historically such as the long wavelength approximation and the corresponding reduction of gravitational field variables, such as anisotropic degrees of freedom or the momentum constraint, that have been integral to existing versions of the SUA. We have also addressed the issue of time-slicing choice; the choice of lapse and shift in the language of ADM. The \yld{only essential}\ylc{main} approximation kept here is the requirement that CPT is enough to compute the noise source terms, a requirement that could be possibly lifted by retaining higher orders in CPT\ylc{. The other approximations are the classicality of the IR sector, the restriction of gauge invariance to gauges close to the homogeneous background, and a heuristic treatment of backreaction}. Although we have been invariably discussing `stochastic' noise terms, the coarse-graining behind our derivations could have been made at any scale, provided one was content with operator-valued source terms. Of course, the classicalisation of cosmological perturbations is highly convenient, and it is this requirement that sets the IR scale to be placed somewhere slightly above the Hubble radius.        

Returning to the core of 3+1 general relativity, which is essentially the evolution of one temporal 3D-hypersurface to the next, we have proposed a method to coarse-grain the linear theory in a gauge-invariant way using its only degree of freedom, here chosen to be $\mathcal{R}$ the comoving curvature perturbation. We have validated our procedure for several of the most common gauges used in cosmology or numerical relativity, finding that the choice of any of them always results in the same source terms, at least at the linear level. Linking the resulting Langevin terms to their counterpart non-linear equations and adding the treatment of stochastic gravitonic sources, we have provided the first complete set of GR equations for Stochastic Inflation\ylc{ within this linear-noise ansatz}. 

From our equations, we were able to recover the limits where existing SI equations apply. We were also able to go beyond the usual approaches to demonstrate broader applications; our example showed that our formalism could obtain results for the stochastic dynamics of the e-folds $\mathcal{N}$ directly in the uniform field gauge, without shifting 3D-hypersurfaces as required in the stochastic $\Delta {\cal N}$ formalism, for the same SUA assumptions normally made in the existing literature.

We want to highlight the potential of such results. The possible applications are numerous and most notably a key focus of the next chapter will be numerical. The present chapter provides all the tools needed for sourcing a full 3+1 numerical relativity code with stochastic perturbations.
Note that this includes the Langevin terms but also the initial conditions (setting the window to a Dirac delta at initial time, to recover Chapter~\ref{sec:ICNR} by construction) which constitute the main challenge of numerical relativity.
The present systematic treatment gives the opportunity to quantitatively study the nonlinear evolution of inhomogeneities, taking forward previous work which considered inflationary initial conditions in a variety of contexts \cite{aurrekoetxea_oscillon_2023,clough_scalar_2017, aurrekoetxea_cttk_2022, giblin_jr_preheating_2019,joana_cosmic_2022}.
Full GR simulations of super-Hubble dynamics during inflation should provide insights about the nonlinear generation of higher-order correlators. This is important because there currently is no alternative to QFTCS methods except Stochastic Inflation, which has to date traded greater scope on super-Hubble scales in exchange for numerous other approximations. 
With our framework, Stochastic Inflation can also, in \ylc{principle}, resolve \ylc{the nonlinearities generated by the classical evolution} of the semiclassical spacetime region, while (numerical) QFTCS is limited to tree-level and often to the decoupled case. In that sense, it \ylc{could become} a competitive method beyond the simplistic modelling of nonperturbative growth\ylc{, within these approximations}.
Finally, to go beyond the case of a canonical field and its potential in GR, and more generally beyond the Einstein frame, Chapter \ref{sec:SIbeyondGR} will provide the generalisation of the present framework.

%% file: chap_SINR.tex
\ifSubfilesClassLoaded{\setcounter{chapter}{4}}{}
\chapter{Stochastic Inflation in Numerical Relativity}\label{sec:SINR}

In this Chapter, we present a Numerical Relativity scheme where the stochastic inflation equations found in Chapter~\ref{sec:SIGR} are solved for scalar sources in the synchronous gauge\ylc{. Stochastic graviton sources are not included, and the coarse-graining scale is fixed to $\sigma = 1$ throughout}.
Based on our published contribution \cite{Launay25ter}, this consists in implementing stochastic noise kicks in the coupled and nonlinear partial differential equations of general relativity, incarnating the fluctuations induced by UV modes continuously crossing the coarse-graining scale.
Beyond this physical context, this code and its results also constitute the first implementation of stochasticity in numerical relativity.

Section \ref{sec:SINRnumval} introduces the stochastic BSSN equations evolved in our code and its practical implementation, including the design of the lattice and scales involved.
In particular, this leads to the validation of the code against examples well-described by linear theory in Section~\ref{sec:SINRval}, including our usual slow-roll and ultra slow-roll scenarios, with the latter designed to form a nonperturbative curvature perturbation in a linear fashion.
Within those simple scenarios, we then embark on the exploration of newly accessible signals and regimes in Section~\ref{sec:SINRlooks}, where we present the first look at stochastic backreaction in full GR, anisotropic degrees of freedom, and preliminary work on the quantum diffusion domination regime.

\section{Simulations and validation}\label{sec:SINRnumval}

\subsection{Numerical pipeline}
Numerically solving the stochastic ADM equations \eqref{eq:CoarseGrainedADM_Dyn3}, \eqref{eq:CoarseGrainedADM_Dyn1} and \eqref{eq:CoarseGrainedADM_Dyn2} from mostly homogeneous initial data, while satisfying \eqref{eq:CoarseGrainedADM_Con1} and \eqref{eq:CoarseGrainedADM_Con2}, requires the well-posed and stable BSSN formulation of GR, introduced in Chapter~\ref{sec:ICNR}. 
Using the mapping of ADM and CPT variables to BSSN ones (see Chapter~\ref{sec:ICNR} and Appendix \ref{app:dicBSSN}), we can perform the coarse-graining procedure directly on the BSSN equations
and its 7 quantities $X$, $\tilde{\gamma}_{ij}$, $K$, $\tilde{A}_{ij}$, $\tilde{\Gamma}^i$ and $\phi$, $\Pi$. This yields the \textit{stochastic BSSN equations} for inflation%
\begin{subequations}
\begin{align}
 \partial_t \phi-\alpha \Pi-\beta^i \partial_i \phi & = 0, \label{eq:stoBSSNphi}\\
 \partial_t \Pi-\beta^i \partial_i \Pi-\alpha \partial_i \partial^i \phi-\partial_i \phi \partial^i \alpha
-\alpha\left(K \Pi-\gamma^{i j} \Gamma_{i j}^k \partial_k \phi-\frac{d V}{d \phi}\right)& = {\cal F}^{-1}\{\boldsymbol{{\cal S}}_{\Pi}^{BSSN}\},   \label{eq:stoBSSNPi}\\
 \partial_t X -\frac{2}{3} X \alpha K+\frac{2}{3} X \partial_k \beta^k-\beta^k \partial_k X & = 0,  \label{eq:stoBSSNX}\\
 \partial_t \tilde{\gamma}_{i j}+2 \alpha \tilde{A}_{i j}-\tilde{\gamma}_{i k} \partial_j \beta^k-\tilde{\gamma}_{j k} \partial_i \beta^k 
 \quad+\frac{2}{3} \tilde{\gamma}_{i j} \partial_k \beta^k-\beta^k \partial_k \tilde{\gamma}_{i j} & =0,  \label{eq:stoBSSNgammatilde}\\
  \partial_t K+\yld{\gamma^{i j} D_i D_j \alpha}\ylg{\alpha^{\mid i}{ }_{\mid i}}-\alpha\left(\tilde{A}_{i j} \yld{A^{i j}}\ylg{\tilde{A}^{i j}}+\frac{1}{3} K^2\right)
-\beta^i \partial_i K-\yld{4 \pi}\ylg{\frac{1}{2}\Mpl^{-2}} \alpha(\rho+S) & = {\cal F}^{-1}\{\boldsymbol{{\cal S}}_{K}^{BSSN}\},  \label{eq:stoBSSNK}\\
 \partial_t \tilde{A}_{i j}-X\left[-\yld{D_i D_j \alpha}\ylg{\alpha_{\mid i\mid j}}+\alpha\left(R_{i j}- \Mpl^{-2}\yld{\alpha S_{i j}}\ylg{S_{i j}}\right)\right]^{{TF}}
-\alpha\left(K \tilde{A}_{i j}-2 \tilde{A}_{i l} \tilde{A}_j^l\right) & \notag \\
-\tilde{A}_{i k} \partial_j \beta^k-\tilde{A}_{j k} \partial_i \beta^k 
+\frac{2}{3} \tilde{A}_{i j} \partial_k \beta^k-\beta^k \partial_k \tilde{A}_{i j} & =  {\cal F}^{-1}\{\boldsymbol{{\cal S}}_{\tilde A_{ij}}^{BSSN}\}, \label{eq:stoBSSNAtilde}\\
 \partial_t \tilde{\Gamma}^i-2 \alpha\left(\tilde{\Gamma}_{j k}^i \ylg{\tilde{A}^{j k}}-\frac{2}{3} \tilde{\gamma}^{i j} \partial_j K-\frac{3}{2} \ylg{\tilde{A}^{i j}} \frac{\partial_j X}{X}\right)
+2 \ylg{\tilde{A}^{i j}} \partial_j \alpha-\beta^k \partial_k \tilde{\Gamma}^i & \notag \\
-\tilde{\gamma}^{j k} \partial_j \partial_k \beta^i-\frac{1}{3} \tilde{\gamma}^{i j} \partial_j \partial_k \beta^k 
-\frac{2}{3} \tilde{\Gamma}^i \partial_k \beta^k+\tilde{\Gamma}^k \partial_k \beta^i+ 2 \Mpl^{-2}\alpha \tilde{\gamma}^{i j} {\cal J}_j & = 0, \label{eq:stoBSSNGammatilde} 
    \end{align}
        \label{eq:stoBSSN}
\end{subequations}
where the r.h.s. Fourier transforms are found to be very similar to the ADM ones given in eqns.~\eqref{eq:SPi} and \eqref{eq:r.h.s.ADM}, up to overall factors%
\begin{subequations}
\begin{align}
&  \boldsymbol{{\cal S}}_{K}^{BSSN} =-\varepsilon_1\boldsymbol{{\cal S}}_{{\cal R}},\\
&  \boldsymbol{{\cal S}}_{\tilde A_{ij}}^{BSSN}  =\varepsilon_1\bigg(\frac{1}{3}\delta_{ij}-\frac{k_i k_j}{k^2} \bigg) \boldsymbol{{\cal S}}_{{\cal R}},\\
& \boldsymbol{{\cal S}}_{\Pi}^{BSSN} = \sqrt{2\varepsilon_1} \Mpl \boldsymbol{{\cal S}}_{{\cal R}}.
\end{align}
    \label{eq:RHSBSSN}
\end{subequations}
The constraints are still satisfied at first order and intermediary first-order equations eqns.~\eqref{eq:stoBSSNphi}, \eqref{eq:stoBSSNX}, \eqref{eq:stoBSSNgammatilde} and \eqref{eq:stoBSSNGammatilde} have been coarse-grained and found without stochastic sources.
Note that for simplicity and a first implementation, this chapter overlooks the stochastic graviton sources, which would add terms to the $\tilde A_{ij}$ equation. Scalar-induced spin-2 perturbations are however included naturally.
These equations, along with the local Mukhanov-Sasaki defined in eqn.~\eqref{eq:ReqISTORIZ1} for the computation of the noise amplitudes, are the main equations of our stochastic inflation formulation for NR. To our knowledge, {this is the first approach and code of this kind.}

Before incorporating any stochasticity, multiple codes are available to solve the BSSN system.
In this chapter, we use \grt \cite{GRTeclyn} as the core PDE solver of our new code, \istoriz.
In its current version, \istoriz sets NR parameters to fairly simple values (synchronous gauge, no AMR, no CCZ4 damping) similarly to Chapter~\ref{sec:ICsim}, see Section~\ref{NRimplem}.
Given the spectral nature of our stochasticity, using \textsc{AMReX}-based software such as \grt, has additional advantages because of the built-in off-loaded Fast Fourier Transform (FFT) capabilities. Key components of \istoriz are:
\begin{itemize}
    \item[i)] Reading of inflationary background initial conditions ($\phi_b^\circledast$, $\Pi_b^\circledast$, $K_b^\circledast$, $X_b^\circledast$) and spectral initial conditions for the curvature perturbation on comoving hypersurfaces (${\cal R}_k^\circledast$ and $\dot{\cal R}_k^\circledast$ for each $k$ of the discrete Fourier lattice).
    In this work, these are all provided by the \stoiic generator presented in Chapters~\ref{sec:ICNR} and \ref{sec:ICsim}, using \textit{oscode} \cite{Agocs20} to solve the Mukhanov-Sasaki equation from deep sub-Hubble initialisation (time ${\cal N}_0$) to the start of the simulation (time ${\cal N}^{\circledast}$) when the first mode enters the {($\sigma$-dependent) IR sector}.
    \item[ii)]  Random draw of the real space Gaussian discrete fields $\boldsymbol{{\cal R}}^\circledast(\mathbf{x})$ and $\partial_t\boldsymbol{{\cal R}}^\circledast(\mathbf{x})$ from the input spectra (see methodology in Appendix~\ref{app:random}), and, in parallel to the BSSN evolution, evolution of the Mukhanov-Sasaki eqn. \eqref{eq:MScosmic} in real space in its well-posed formulation eqn.~\eqref{eq:ReqISTORIZ1}.
     Stochastic backreaction can be switched on in this evolution, by promoting background quantities such as $\phi_b$, $K_b$ or $X_b$ to their local spatial value. This defines one of three currently implemented levels of backreaction, as discussed and compared in Section~\ref{sec:SINR.Stoback}.
    \item[iii)] Initialisation of the coarse-grained inhomogeneous fields($\phi^{\circledast,>}$, $\Pi^{\circledast,>}$, $K^{\circledast,>}$, $X^{\circledast,>}$) at the initial simulation time $t^{\circledast}$ by applying the procedure used in eqn.~\eqref{eq:eqInitialConditions} to the coarse-grained initial conditions $\big(W*\boldsymbol{{\cal R}}\big)(t^\circledast,\mathbf{x})$ and $\partial_t\big(W*\boldsymbol{{\cal R}}\big)(t^\circledast,\mathbf{x})$.
    This is necessary since these are minor but non-zero inhomogeneous contributions from $W$ being close to $0$ (but not exactly because it is a smooth function) at $t^{\circledast}$ for the modes of interest. If this is not accounted for, we observe a detectable lack of accuracy in the spectrum of the first (low $k$) modes entering the simulations.
    \item[iv)] At each\yld{ RK4} time step and \ylg{Runge-Kutta }substep\ylg{ ($4$)}, the discrete stochasticity is given by
        \begin{itemize}
            \item[a)] Fourier transform of $\boldsymbol{{\cal R}}(t,\mathbf{x})$ and $\partial_t\boldsymbol{{\cal R}}(t,\mathbf{x})$.
            \item[b)] Integration of stochastic sources in Fourier space for ${\cal X} \in \{K,\tilde{A}_{ij},\Pi\}$
            \begin{equation}
                \begin{aligned}
                    \int_t^{t+\Delta t}{\cal O}_{\cal X}(u,\mathbf{k})\boldsymbol{{\cal S}}_{\cal R}(u,\mathbf{k})du
                    & \simeq {\cal O}_{\cal X}(t,\mathbf{k})\Bigg(  {\boldsymbol{{\cal R}}}_{\mathbf{k}}(t)\Big(\dot{W}_k(t+\Delta t)-\dot{W}_k(t)\Big)\\
                    & +\Big(2 \dot{\boldsymbol{{\cal R}}}_{\mathbf{k}}(t)+ (3-\varepsilon_2(t)) H \boldsymbol{{\cal R}}_{\mathbf{k}}(t)\Big)\Big({W}_k(t+\Delta t)-{W}_k(t)\Big)\Bigg),
                \end{aligned}
             \end{equation}
            where the ${\cal O}_{\cal X}$ are the operators in front of $\boldsymbol{{\cal S}}_{\cal R}$ in eqns.~\eqref{eq:RHSBSSN}.
            Window functions are thus integrated exactly\footnote{Note that the sharpness of the window function and its time derivative needs to be discernible after this time step ($\Delta W$ and $\Delta \dot W$ written above need to be within numerical accuracy).} in the limit where $\cR$ and $\dot{\cR}$ are slowly varying over a time step.
            Note that this integral might appear as an Itô integral, but it is in fact a Stratonovich one because our noise has a nonzero correlation time according to the Wong-Zakai theorem \cite{WongZakai65}.
            \item[c)]Inverse Fourier Transform  ${\cal F}^{-1}\{\cdot\}$.
            \item[d)] Addition of the stochastic noise to each r.h.s. of the discretised BSSN. 
        \end{itemize}
        This ensures the satisfaction of the constraints up to ${\cal O}(\text{noise}^2)$.
        The stochastic scheme is detailed in Appendix \ref{app:schemes}.
\end{itemize}

\subsection{Design of the discrete stochastic spacetime}\label{sec:designlattice}
The choice of a geodesic gauge implies that our lattice is a comoving space one. In these coordinates, the Hubble radius $R_H = (aH)^{-1}$ shrinks so that big scales receive stochastic fluctuations first, followed by smaller ones until the lattice's smallest visible scale ($dx$) is reached.
In this work, each mode $k$ evolved by the MS equation before entry, eqn.~\eqref{eq:ReqISTORIZ1}, enters the GR evolution via the stochastic terms when $k_\sigma(t)\equiv\sigma R_H^{-1} (t)$ has increased up to $k$. This is practically the case when the window function is set to the rolling rectangle $\Theta[k_{\sigma}(t)-k]$. However, to demonstrate the generality of our framework and resolve all involved time derivatives of $W$, we pick a smoother window function with activation centred on the crossing \begin{equation}\label{eq:ourWindow}
W_k(t)=\frac{1}{2}\Bigg(1+\tanh\bigg(s\log\bigg(\frac{k_\sigma(t)}{k}\bigg)\bigg)\Bigg),
\end{equation}
where $s=5.0$ is an arbitrary sharpness parameter.\footnote{This should not have an impact on the result in the late-time limit, as long as a window follows requirements such as those defined in \cite{Winitzki00}. For a given discrete numerical scheme, not all windows are suitable. In particular the sharpness should be appropriate for the time stepping and for keeping all terms within perturbation theory.}
The effect and shape of this window on the IR spectra will be visible on the results of Section~\ref{sec:SINRval} and in particular in Figure~\ref{fig:stospec}.
Thanks to a random draw performed in Fourier space and the Fourier transforms at every time step, we effectively input non-Markovian noise when such a non-sharp window function is used: the noise at a given time step depends on the history of the system via its correlation to earlier sources (on top of the stochastic backreaction).
This is seemingly the first time that such a non-Markovian noise is implemented on a lattice with full spacetime correlations, benefiting from a numerical construction of the noise. This should evaporate the fear of \textit{noise-colouring} arising from smooth window functions, as well as the analytical need for the sharp cutoff window function, which has collected a fair amount of criticism \cite{Winitzki00, andersen_stochastic_2022, Mahbub22, Liguori04}.

Let $N$ now be the number of cells per spatial dimension of the simulation's cubic patch of universe and $L = Ndx$ the unitful comoving size of the latter. Assuming real fields and the usual isotropy of the fluctuations, the range of Fourier modes which can be added is thus made of all norms $k$ such that
$ k=dk\sqrt{a^2+b^2+c^2},$ where $(a,b,c) \in \llbracket 0,\lfloor N/2\rfloor \rrbracket^3$ and $dk \equiv k_{\text{min}} = 2\pi L^{-1}$. In this setting, \stoiic picks $L$ so that the first mode (smallest $k$) crosses $k_{\sigma}$ and enters when the simulation starts, which is
$L k_{\sigma}^\circledast = {2\pi}$. This is different from Chapter~\ref{sec:ICsim} where all modes are directly part of the simulation, and where the highest $k$ mode was set to be super-Hubble close to the start of the simulation with all other modes for classicality from the initial time.
{For a good approximation of the gradient stencils ($6$th order in this work), our UV cutoff for the fluctuations is still taken lower than the Nyquist frequency by imposing $k^{\cal R}_{\text{max}} =  N k_{\sigma}^\circledast/4$ as the last and highest-$k$ mode entering the GR simulation.\footnote{Note that Kreiss-Oliger dissipation is also available in \grt to damp out high frequency modes, but not deployed here.}}

Similarly, the design of the time dimension in the lattice or stepping is intrinsically linked to the amount of physical time we are interested in and to the available computational resources constraining the choice of $N$, the number of cells per spatial dimension.
Let $\Delta t$ be the time step, $N_T$ the number of time steps, and $\theta = \Delta t/dx$ the Courant factor, which has an upper bound for stability. We are interested in probing the crossing of all modes in our lattice, from $k_{\text{min}}$ to $k^{\cal R}_{\text{max}}$. Let us assume that our simulation's expansion is roughly that of a de Sitter universe with a constant $H^\circledast$, so that $k_{\sigma}(t) \simeq \sigma H^\circledast e^{\cal N}$, where ${\cal N}= H^\circledast(t-t^\circledast)$ is the number of e-folds since the start of the simulation. This provides a proxy to compute the number of e-folds between the first and the last mode crossing
\begin{equation}
{\cal N}^{\otimes}= \ln \bigg (\frac{k_{\text{max}}^{\cal R}}{k_{\text{min}}}\bigg) = \ln \bigg (\frac{N}{n}\bigg).
\end{equation}
Our simulation needs to last up to that time, together with an amount $\Delta {\cal N}$ of supplementary e-folds for the study of the super-Hubble regime. The feasibility of this study thus depends entirely on our computational resources with respect to the value of $N$.

The choice of the time-to-space ratio $\theta$ is the most constrained one.
Beyond the Courant upper bound and stability in general, the choice of $\theta$ is still key. Given the physical duration ${\cal N}_{tot} = {\cal N}^{\otimes}+\Delta{\cal N}$, the number of time steps is
\begin{equation}
N_T \simeq \frac{{\cal N}_{tot}}{H^\circledast \Delta t} = \frac{N\sigma}{2\pi \theta}({\cal N}^{\otimes}+\Delta {\cal N}).
\end{equation}
In other words, trying to reach an excessively small value for $\theta$ will jeopardise our chances to run the simulation in a reasonable \ylg{compute }time.
Instinctively, the choice of $\theta$\yld{ also} controls the number of modes entering the simulation per time step.
Indeed, when approximating the $k$-width entering the IR between $t$ and $t+\Delta t$ as $\Delta k_{\sigma}(t) \simeq k_\sigma(t)H \Delta t$, we infer that the number of resolved discrete modes entering on the grid is
\begin{equation}
\frac{\Delta k_\sigma(t)}{k_{\text{min}}} \simeq \frac{2\pi\theta}{N\sigma} e^{{\cal N}(t)},
\end{equation}
which essentially says that the window has an exponentially growing range. For a given ${\cal N}^{\otimes}$ and $\sigma$, one can in particular pick $\theta$ such that no more than one discrete mode enters the lattice per time step, by requiring $\Delta k_\sigma^\otimes\leq k_{\text{min}}$, or equivalently
\begin{equation}
\theta < \frac{n\sigma}{2\pi},
\end{equation}
which is more constraining than the stability condition\yld{ only} on super-Hubble cutoffs ($\sigma <1$).

It is important to note that for $\sigma\geq1$, the simulation will not see more than one discrete mode enter at a time if sticking to the Heaviside windowing or a sharp enough window. Using a smooth window function prevents stochastic spikes from happening in that sense.

\subsection{Validation}\label{sec:SINRval}
In the following, we work with the scaling $a^\circledast=1$, units such that $\Mpl = 100$ and lattices of size $N=256$. The coarse-graining is fixed to $\sigma=1$ for all scenarios, implying that modes enter the simulation around their crossing as opposed to traditionally far after.
Similarly to Chapter~\ref{sec:ICNR}, we study both a vanilla example of inflation for the sake of \yld{diagnostics}\ylg{validation}, and a more challenging example of an ultra-slow roll model.
The two scenarios studied are presented in Table~\ref{tab:scenarios} together with their extracted background evolution for the relevant e-fold span in Figure \ref{fig:MeanRolls}. {Note that stochastic backreaction is switched on but its ablation yields minor spectral changes in those two scenarios: we find a detectable but $\lesssim  1$\textpertenthousand\text{} relative change in the power spectrum for all scales, see Section~\ref{sec:SINR.Stoback} for definition and discussion.}

\setlength{\tabcolsep}{6pt}   
\renewcommand{\arraystretch}{1.6}
\begin{table}[]
    \centering
    \resizebox{0.6\linewidth}{!}{
    \begin{tabular}{|c||l|l|}
        \hline
         Potential& Quadratic & Higgs inflection \cite{Hamada14,Bezrukov14} \\
         \hline
         Expression 
         & $V[\phi] = \frac{1}{2}m^2\phi^2$
         & $V[\phi] = \frac{\Lambda v^4\phi^2(3\phi^2 + 2\sqrt{2}\phi v + 6v^2)}{3(3\phi^2 + 2v^2)^2}$ \\
         \hline
         Parameters & $\varepsilon_1 = 0.01$ & $\Lambda = 1.86\times 10^{-6}$ \\
          & $\varepsilon_2 = -0.02$ &  $v = 0.198\Mpl$\\
         & $H_0 = 10^{-5}\Mpl$ &  $\phi_0 = -1.2\Mpl$ \\
         & $m=m(\varepsilon_1,\varepsilon_2,H_0)$
         & $\varepsilon_1=\varepsilon_1(\phi_0)$  \\
         & $\phi_0=\phi_0(\varepsilon_1,\varepsilon_2,H_0)$ 
         & $H_0 = H_0(\varepsilon_1, \phi_0)$\\
         \hline
         Simulation & ${\cal N}^{\circledast}-{\cal N}_0 \sim 5$ & ${\cal N}^{\circledast}-{\cal N}_0 \sim 3$ \\
         & ${\cal N}_f-{\cal N}^{\circledast} \sim 8$& ${\cal N}_f-{\cal N}^{\circledast} \sim 10$ \\
         \hline
    \end{tabular}
    }
    \caption{Potentials used in this study and their parameters, including initial conditions, determined either by \textit{potential-position} or $H_0$ fixing (defined in Chapter~\ref{sec:ICsim}). ${\cal N}_0$, ${\cal N}^{\circledast}$, and ${\cal N}_f$ are the Bunch-Davies initialisation time, the start of the NR simulation, and the final time in e-folds.}
    \label{tab:scenarios}
\end{table}

\begin{figure}[ht]
    \centering
    \begin{subfigure}{0.35\textwidth}
        \centering
\includegraphics[width=\linewidth]{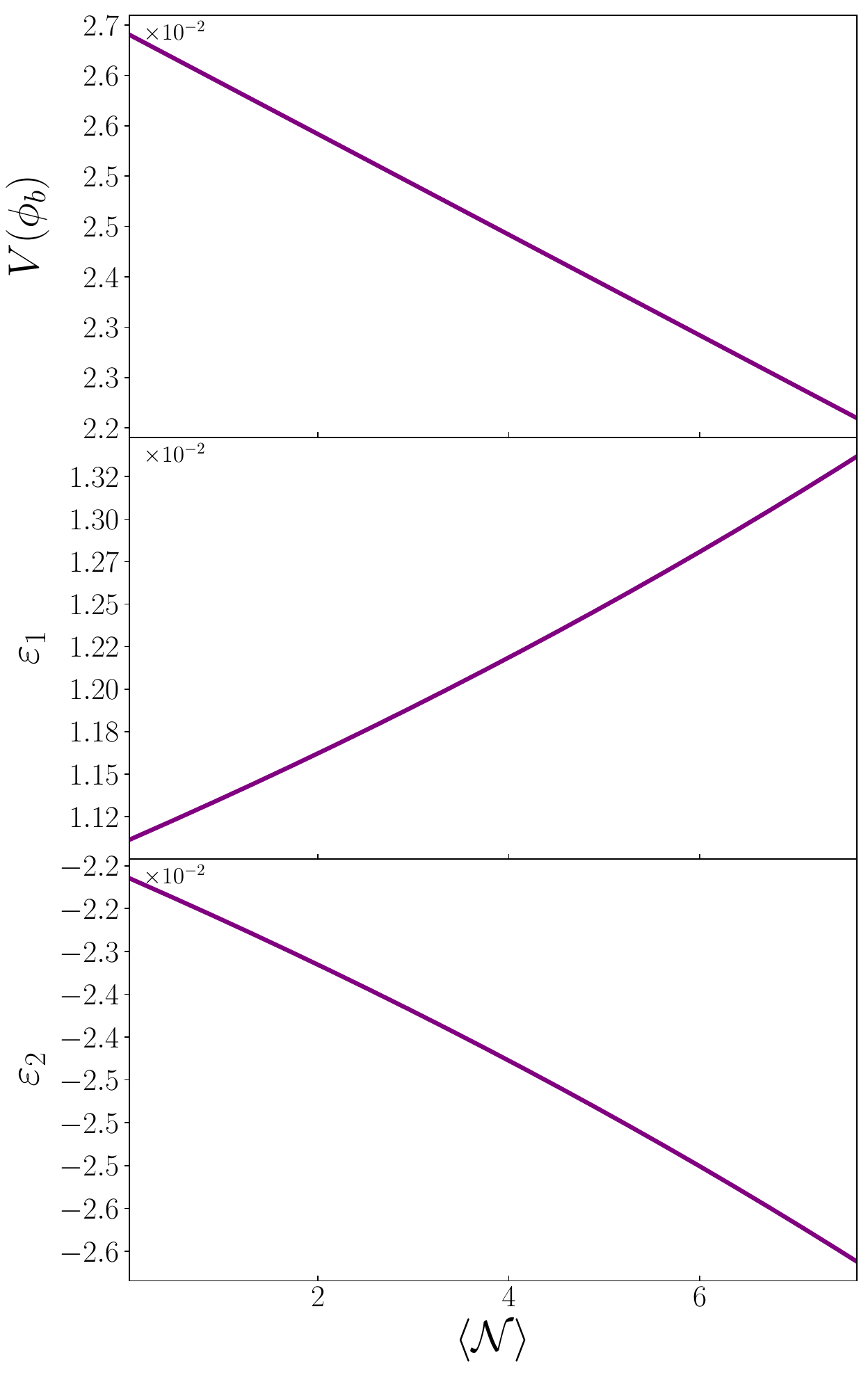}
\caption{Quadratic inflation.}
\label{fig:MeanRollsquad}
    \end{subfigure}
    \begin{subfigure}{0.345\textwidth}
    \centering
\includegraphics[width=\linewidth]{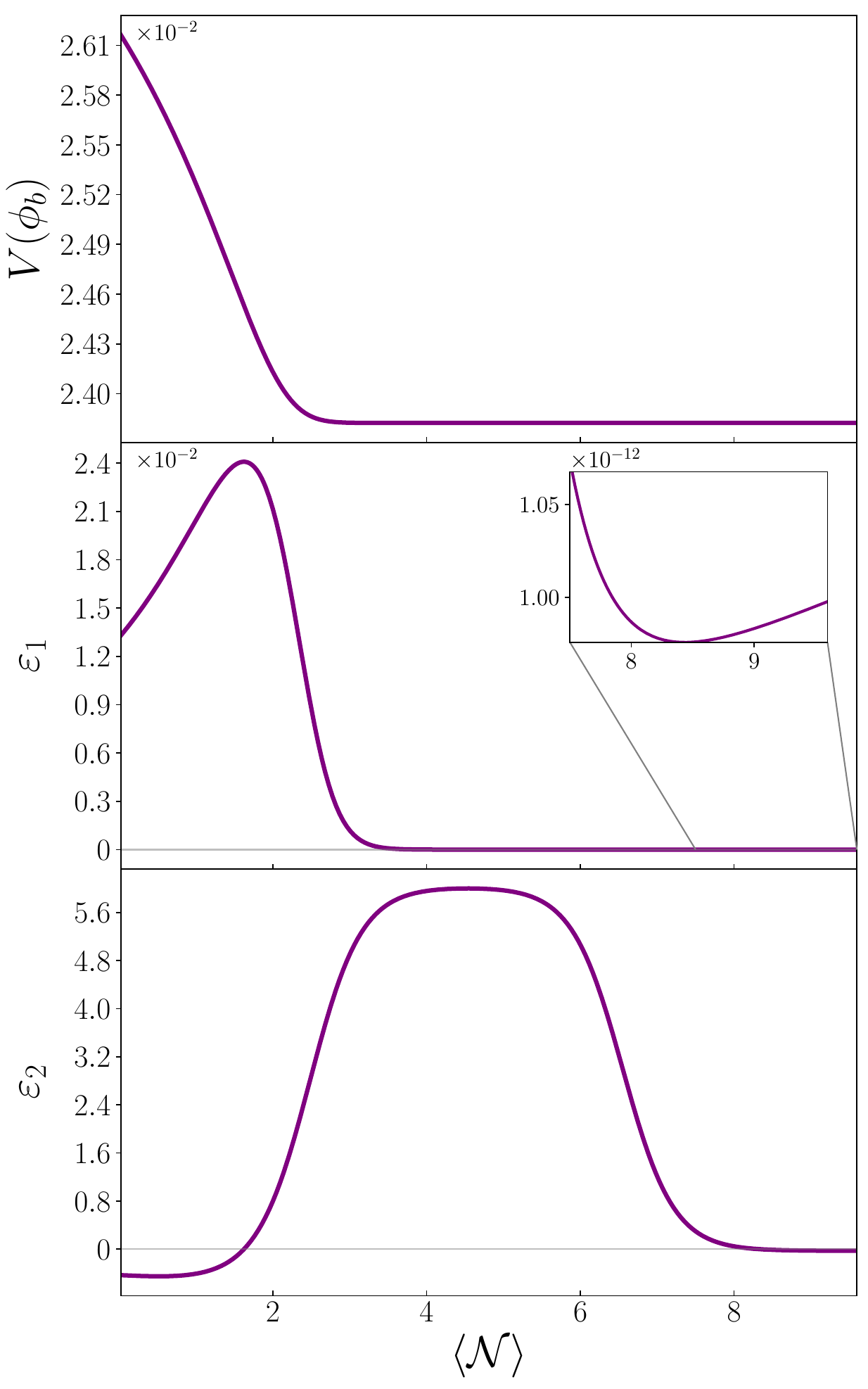}
\caption{Inflection inflation.}
\label{fig:MeanRollsusr}
    \end{subfigure}
\caption{ Time background evolution of the potential value (top), first (middle) and second (bottom) slow-roll parameters for both our validation scenarios. Perfect agreement between Friedmann and NR evolutions. $\Mpl=100$.}
\label{fig:MeanRolls}
\end{figure}

Figures~\ref{fig:quad_contrast} and \ref{fig:quad_spec} report the evolution of a simple quadratic inflation scenario and thus slow-roll (SR) dynamics.
Modes enter the simulation for about $3.5-4$ e-folds as can be seen from the growth of the contrasts in Figure~\ref{fig:quad_contrast}.
Afterwards, quantities such as ${\cal R}$ freeze as expected in the super-Hubble regime. The dynamics of such a vanilla scenario are well-known and a good test that linear physics is recovered\ylg{ at leading order}.
This is particularly visible in the comparison of the spectrum extracted from the NR simulation ${\cal R}^{\rm NR}$ and the Mukhanov-Sasaki solution ${\cal R}^{\rm MS}$ in Figure~\ref{fig:quad_spec}, once the window has fully rolled from the IR to the UV. Note that the extraction of ${\cal R}^{\rm NR}$ from the evolved BSSN variables was detailed in Chapter~\ref{sec:ICsim} and Appendix~\ref{app:extract}. 
Clearly this scenario seems to be well described by both the IC (Chapter~\ref{sec:ICsim}) and the present stochastic NR frameworks.

\begin{figure}[p]
    \centering
    \begin{subfigure}{0.48\textwidth}
        \centering
        \includegraphics[width=0.8\linewidth]{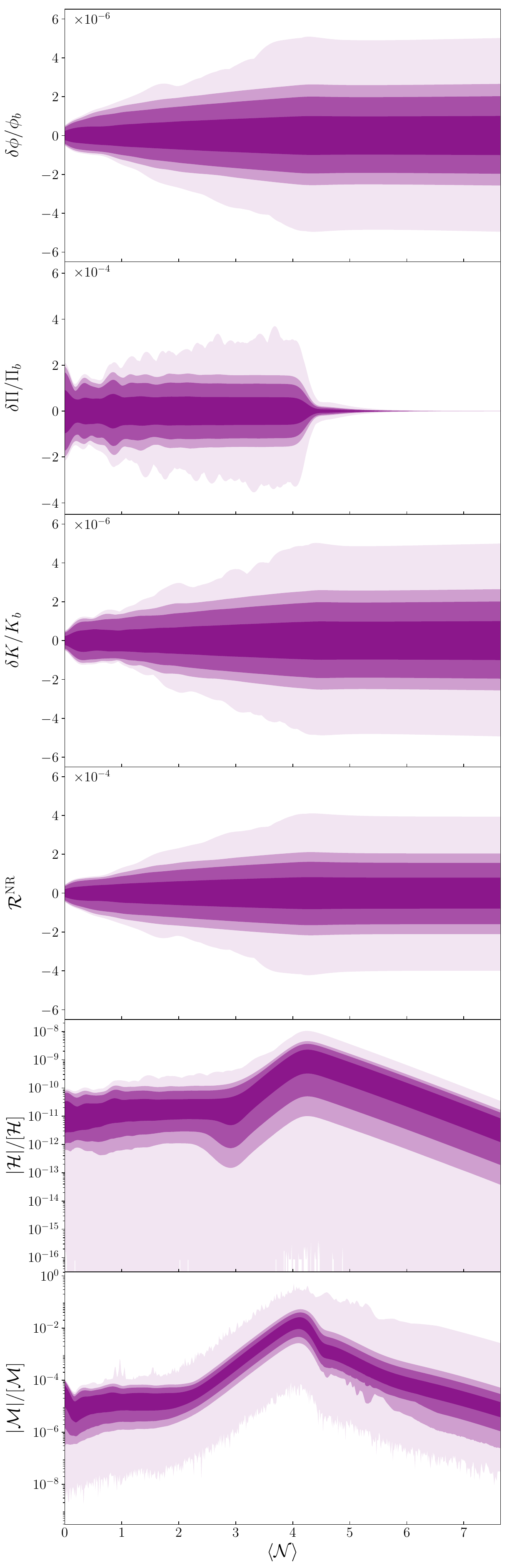}
        \caption{Quadratic Inflation.}
        \label{fig:quad_contrast}
     \end{subfigure}
    \hfill
    \begin{subfigure}{0.48\textwidth}
        \centering
        \includegraphics[width=0.8\linewidth]{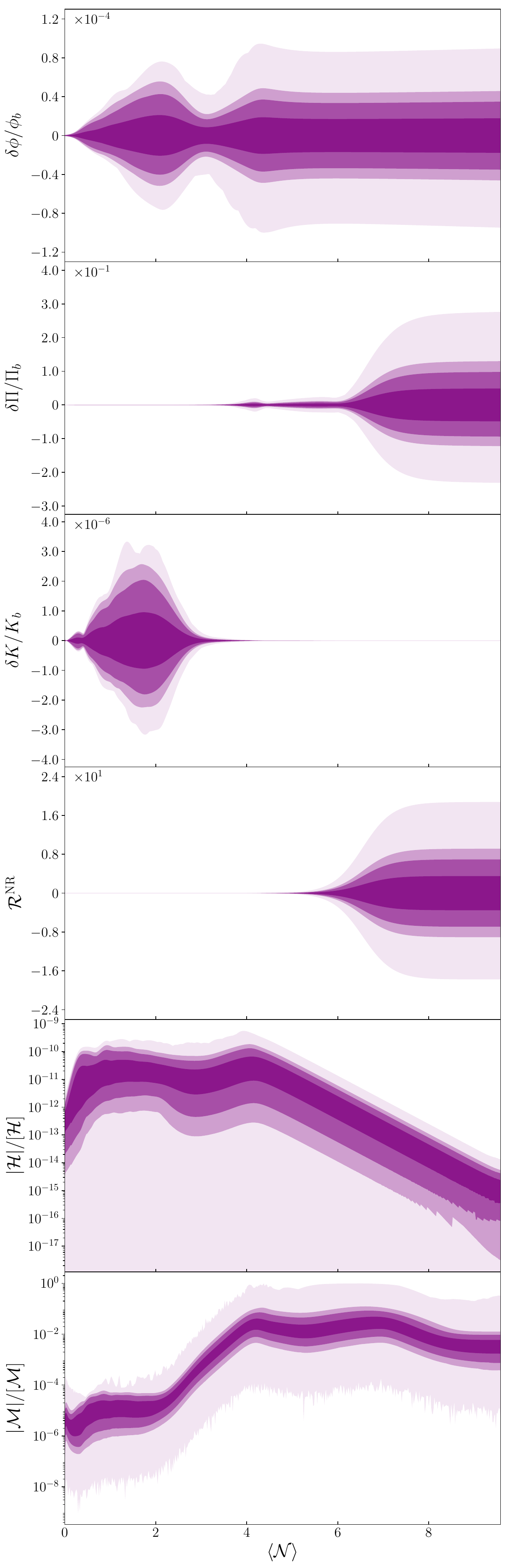}
        \caption{Inflection Inflation.}
        \label{fig:usr_contrast}
    \end{subfigure}
\caption{Contrast of the inflaton (top), its momentum, the extrinsic curvature, the reconstructed ${\cal R}^{\rm NR}$ and the relative Hamiltonian and Momentum constraints (bottom) extracted from the NR simulations using $68$, $95$, $99$ and $100$ percentile contours (dense to light purple). $\Mpl=100$.}
\end{figure}

\begin{figure}[ht]
    \centering
    \begin{subfigure}{0.75\textwidth}
        \centering
    \includegraphics[width=\linewidth]{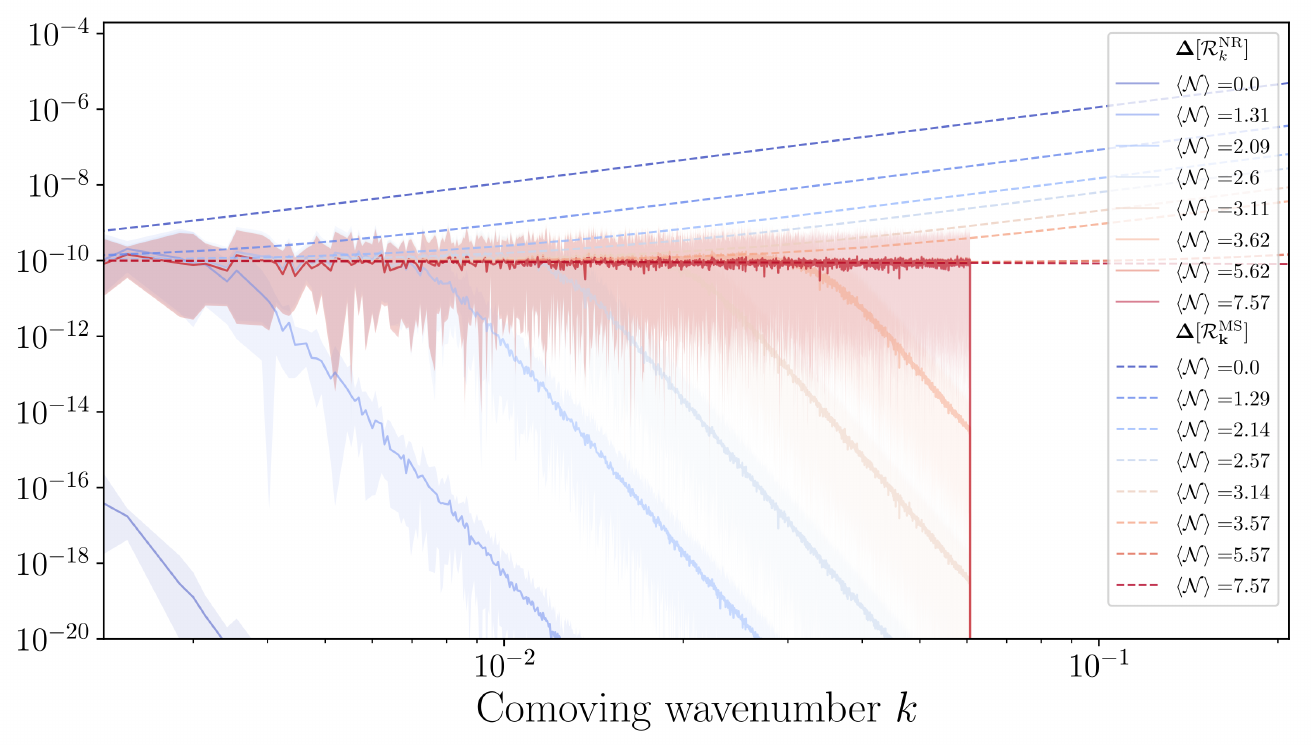}
    \caption{Quadratic Inflation.}
    \label{fig:quad_spec}
     \end{subfigure}
    \hfill
    \begin{subfigure}{0.75\textwidth}
        \centering
\includegraphics[width=\linewidth]{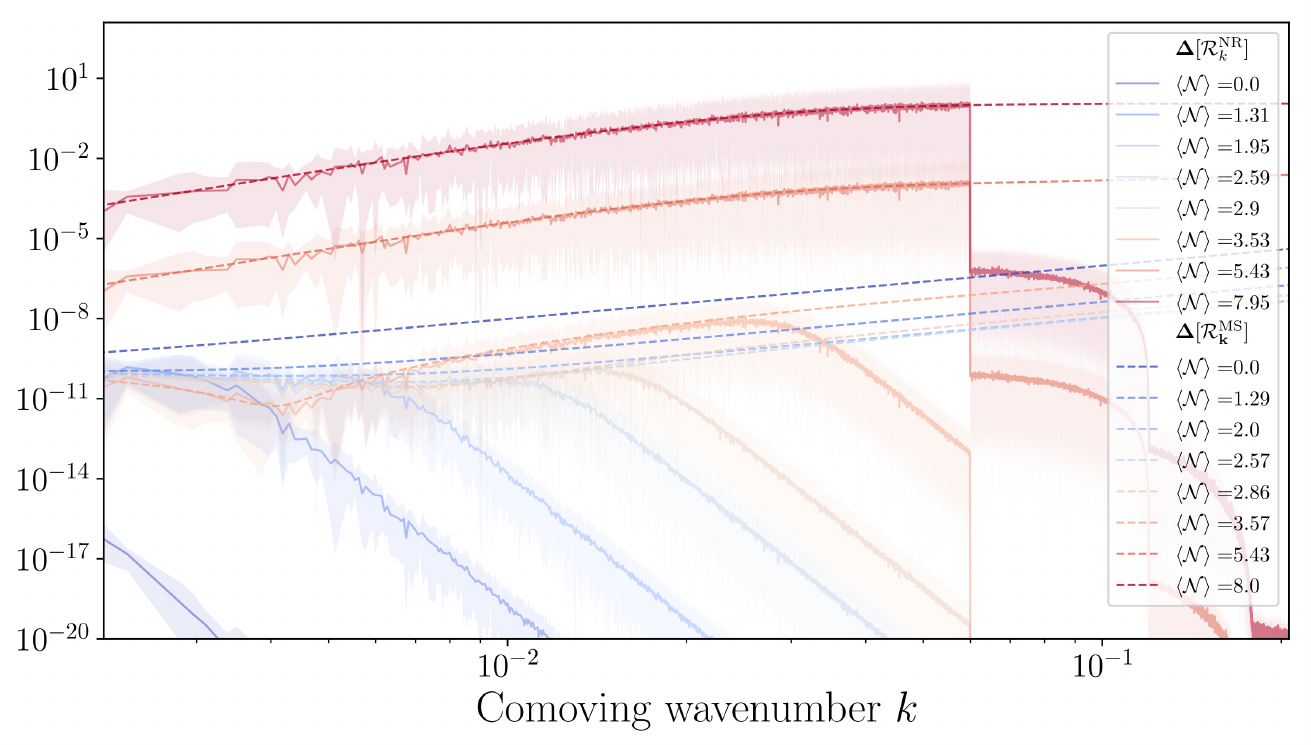}
    \caption{Inflection Inflation.}
    \label{fig:usr_spec}
    \end{subfigure}
\caption{ Linear (dashed, ${\cal R}^{\rm MS}$) and reconstructed (solid, ${\cal R}^{\rm NR}$) binned dimensionless spectra ($\Delta[\cdot]= \frac{k^3}{2\pi^2} |\cdot|^2$) from initial (navy blue) to final (red) times. Contours account for the full range of points in each bin. Fourier bins are those of the lattice with more than 10 realisations. $\Mpl=100$.}
\label{fig:stospec}
\end{figure}

An ultra slow-roll phase (USR), $\varepsilon_1  \sim 0$ and $\varepsilon_2 \sim 6$ in our notations, is famously known for enhancing the power spectrum of modes after crossing the horizon \cite{Leach01}.
In general, slow-roll-breaking regimes can be reached using features of various forms in the inflaton potential, see, for instance, \cite{Mishra19}. Our potential is that of Chapter~\ref{sec:ICsim} \cite{Bezrukov14,Hamada14} and has been once again slightly tuned to show interesting physics within a few e-folds.

In this chapter and as opposed to Chapter~\ref{sec:ICsim}, the lattice can and does contain modes entering the IR system during the USR phase as well as during the transition from SR to USR (while starting in SR before the feature), a numerically challenging regime because of the extreme slow-down of the field and the importance of the gradients in such phase transitions \cite{jackson2023}.
This has been at the heart of issues and attempts to fix them in gradient-agnostic approaches such as the separate universe approximation, alongside $\delta N$ formalisms \cite{Artigas24} or standard stochastic inflation \cite{jackson2023,Briaud25}.
Since our simulations can be run with $\sigma = 1$ and are fully gradient-enabled, this is unprecedented in a stochastic framework, whether it is for USR or not, a feature that was only accessible in lattice Friedmann codes at the expense of losing metric effects \cite{latticeeasy,Cosmolattice}.

One-point distributions for real space inhomogeneities in this USR scenario are displayed in Figure~\ref{fig:usr_contrast}, where we report the contrasts of key quantities. The top three panels show the progressive entry of all modes with their full amplitude within about $4$ e-folds. Note the double feature structure of the $\phi$ contrast evolution in juxtaposition with that of the extrinsic curvature $K$, which lacks the second growth phase. Qualitatively, field fluctuations that were added in the slow roll phase ($\mathcal{N}\lesssim2$) decay as the field slows down during $2\lesssim\mathcal{N}\lesssim 3$, while at the same time more modes are added as they exit the horizon. These latter modes contribute to the growth seen for $3\lesssim\mathcal{N}\lesssim 4$ after which all grid-available modes in the simulation have been used. After $\mathcal{N}\sim 3$ the almost constant part of the potential has been reached and $\phi$ diffuses \ylg{perturbatively }on it without affecting the expansion rate. $K$ therefore does not receive any further perturbations since $\phi$ perturbations no longer change the energy density, leading to the lack of a second growth feature in $K$.
This picture is consistent with previous analytical works \cite{prokopec_ensuremathdeltan_2021, Prokopec:2025uvz}.    

From $\mathcal{N} \sim 3.5$ USR starts and most of the modes making up $\mathcal{R}^{\rm NR}$ are already super-Hubble and undergo an exponential enhancement (expected for USR) until $\mathcal{N}\sim 8$. After that point, a slow roll phase commences with $\left(\varepsilon_1,\varepsilon_2\right)\simeq \left(0,-2.8 \times  10^{-2}\right)$ and  $\mathcal{R}^{\rm NR}$ freezes as expected. This phase occurs after a very shallow maximum is passed and the field starts rolling down the potential with $dV/d\phi<0$.  Note that $\mathcal{R}^{\rm NR}$ reaches non-perturbative values of $\sim 10$ for the furthest outliers in real space. The contrasts $\delta\Pi/\Pi_b$ also become relatively large at late times in sharp contrast to the quadratic case. 

Up to a certain maximum inhomogeneity amplitude (usually defined for the dimensionless power spectrum as $\Delta[{\cal R}]\sim 1$ \cite{Franciolini24}), and due to the $\varepsilon_1\to 0$ limit, USR is usually assumed to be well-described by the linear solution from the Mukhanov-Sasaki equation, despite reaching non-perturbative amplitudes.
This is visible in the spectra displayed in Figure~\ref{fig:usr_spec}, which, at complete overlap of the closed linear and open IR non-linear system, shows excellent agreement\footnote{\ylc{Extracting the non-linear generalisation ${\cal R}^{\rm NL}$, introduced in Chapter~\ref{sec:ICsim}, shows no difference to the linear extraction of $\cR$. As explained in Section~\ref{sec:finalvalues}, this confirms the consistency of the extraction, not the linearity of the dynamics, which follows from the agreement with linear theory.}}
, better than the IC approach of Chapter~\ref{sec:ICsim} (studied on slightly different scales however). 
We find that, despite reaching up to $\Delta[{\cal R}]\sim 0.1-1$, departure from the final linear spectral prediction is seemingly not as strong as reported in lattice Friedmann codes \cite{caravano2024usr}. In the future, a fully quantitative study will make detailed comparisons in this limit.

We note the last two panels of Figures \ref{fig:quad_contrast} and \ref{fig:usr_contrast} as a crucial verification of both our numerical and theoretical pipeline.
In fact, we claimed satisfaction of the constraints at first order in the noise. To evaluate the order of the terms in the constraints, we use the \textit{absolute constraint magnitudes} $[{\cal H}]$ and $[{\cal M}]$ defined in eqn. \eqref{eq:abscons}.
In its entry phase, the noise in ${\cal R}$ has a maximum amplitude of $\sim 10^{-5}$. In both scenarios, the Hamiltonian constraint shows a violation at the next order in perturbation theory ($\sim 10^{-10}$) when compared to the absolute magnitude $[{\cal H}]$. After that, inflation naturally damps the constraint. 
For the Momentum constraint, being null at $0th$ order, the absolute magnitude $[{\cal M}]$ is the first order and we thus observe a second order violation as well in this phase. As a consistency check and similarly to Chapter~\ref{sec:ICsim}, we find that switching any sign in our noise spectra yields a violation at first order (meaning ${\cal H}/[{\cal H}]\sim 10^{-5}$ and  ${\cal M}/[{\cal M}]\sim 1$). In the quadratic case, the last mode crossing makes the $|{\cal M}|/[{\cal M}]$ momentarily a bit higher than second order because of the gradient resolution, which is acceptable given that a smaller $k_{\text{max}}$ would fix that. A similar conclusion can be reached for the inflection case where $|{\cal M}|/[{\cal M}]$ sits later around $10^{-2}$, which is acceptable given an inhomogeneity of order $10$ in ${\cal R}$ at that point. 

Finally, a positive convergence analysis for the constraints and the spectra is reported in Appendix \ref{app:conv}, including subtleties of the stochasticity in the analysis.
This completes the validation and physicality check of our method, making it ready for future work on both fully non-perturbative and non-linear experiments for which no analytical results are known.
\ylg{We provide a first taste of this potential below.}

\section{First looks}\label{sec:SINRlooks}
\subsection{Stochastic backreaction}\label{sec:SINR.Stoback}

\begin{figure}[ht]
    \centering
\includegraphics[width=0.75\linewidth]{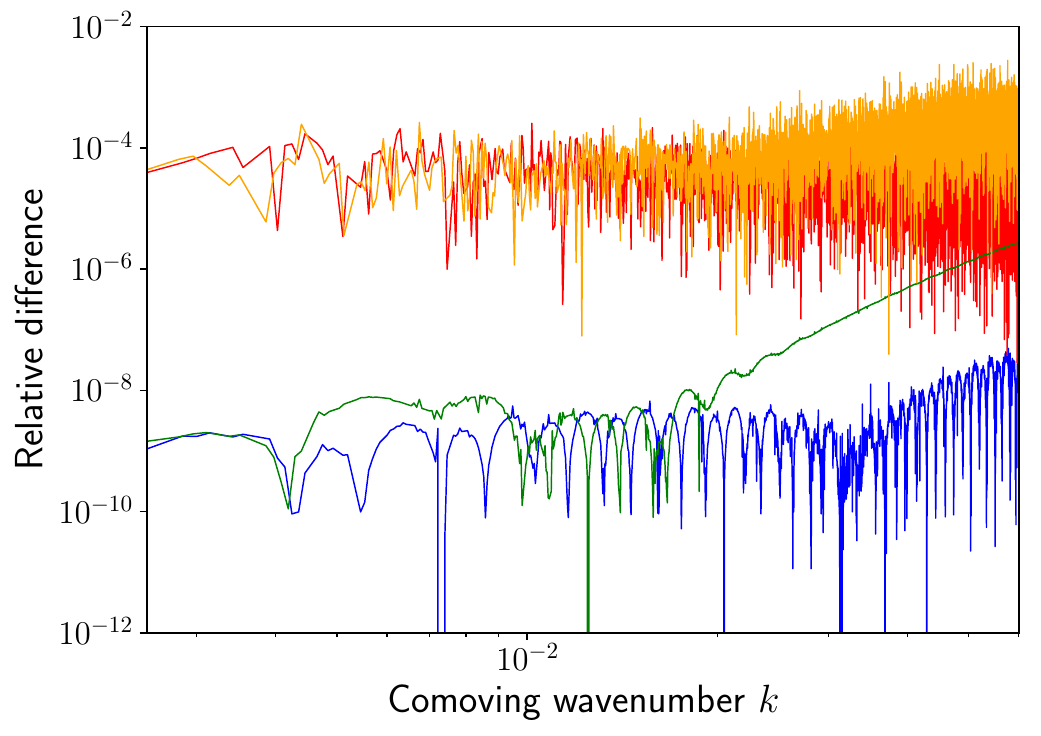}
\caption{ {Relative difference in dimensionless ${\cal R}^{\rm NR}$ power spectra between \textit{level} $2$ and \textit{level} $0$ stochastic backreaction (top 2 curves), or between \textit{level} $1$ and \textit{level} $0$ stochastic backreaction (bottom curves), for quadratic inflation (red, blue) and inflection inflation (orange, green). All the same $N=256$ stochastic seed.}}
\label{fig:ablationback}
\end{figure}

Accounting for stochastic backreaction is still an open problem and thus relies on heuristics in the literature.
The current implementation of \istoriz allows three levels of stochastic backreaction.
The coefficients appearing in the Mukhanov-Sasaki equation \eqref{eq:ReqISTORIZ1} ($K_b, \dot\phi_b, \alpha_b, \beta_b$) can be computed from: 
\begin{itemize}
    \item[i)]\textit{level} $0$ -- The Friedmann equations, evolved separately with the same initial conditions and numerical scheme.
    \item[ii)] \textit{level} $1$ -- Their \ylg{lattice }average values over the stochastically evolving IR lattice. 
    \item[iii)] \textit{level} $2$ --  From the local value of the derived field on the IR lattice.
\end{itemize}
\textit{Level 0} constitutes the original and standard formulation of Stochastic Inflation (so-called \textit{Starobinsky approximation}), and allows one to use pre-computed noise.
In a perturbative regime, any backreaction corrections would be of second order in perturbation theory and so would not matter if the leading behaviour is of interest but would be important for e.g. non-Gaussianity.
Figure \ref{fig:ablationback} compares those corrections in our simulations at a spectral level, indicating the relative importance of level 2 backreaction comparatively to level 0, giving a small but detectable effect on the power spectrum.

\subsection{Anisotropic DOFs}\label{sec:SINR.anis}
\begin{figure*}[!ht]
    \centering
    \begin{subfigure}{\textwidth}
\includegraphics[width=\linewidth]{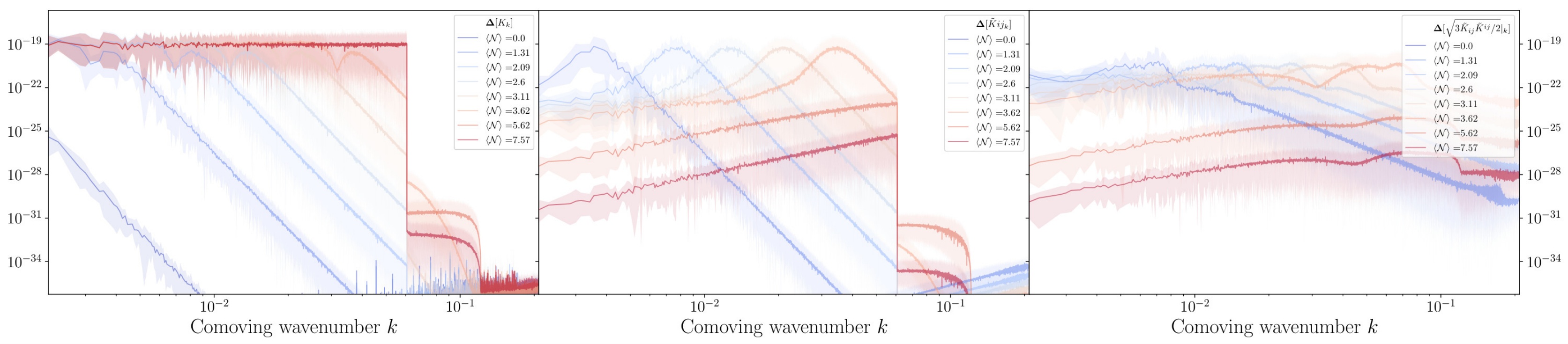}
\caption{Quadratic inflation}
    \end{subfigure}
    \begin{subfigure}{\textwidth}
\includegraphics[width=\linewidth]{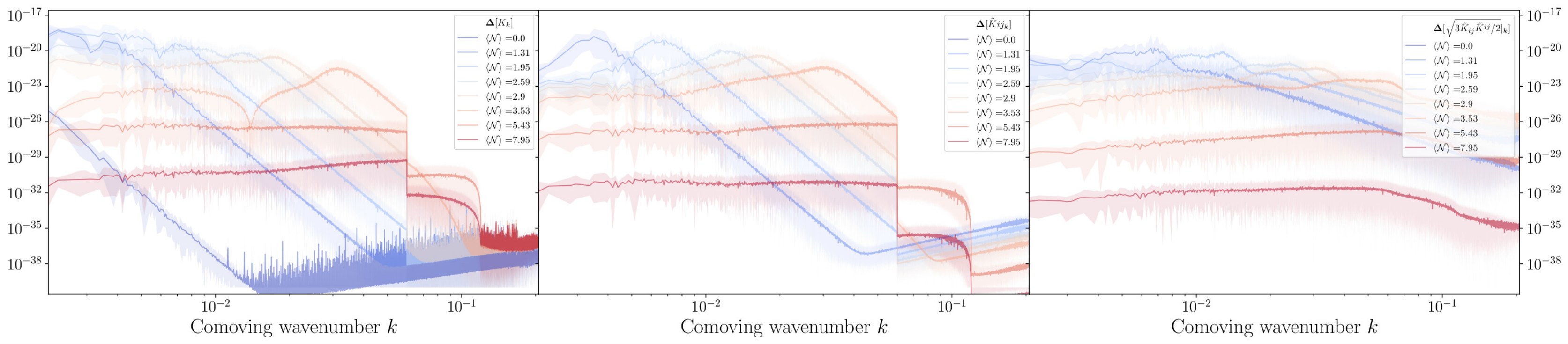}
\caption{Inflection inflation}
    \end{subfigure}
\caption{ Binned dimensionless spectra ($\Delta[\cdot]= \frac{k^3}{2\pi^2} |\cdot_k|^2$) of $K$, $\tilde{K}$, and $\sqrt{3\tilde{K}_{ij}\tilde{K}^{ij}/2}$ in the quadratic (top) and inflection inflation case (bottom) from initial (navy blue) to final (red) times. Contours account for the full range of points in each bin. Fourier bins are those of the lattice with more than 10 realisations. $\Mpl=100$.}
\label{fig:specAK}
\end{figure*}
As a first demonstration of the potential of such simulations beyond standard stochastic inflation, we analyse the evolution of the anisotropic degrees of freedom encoded in $\tilde{K}_{ij}$.
In future work, applying the transverse projector will yield the first extraction of scalar-induced gravitational wave power from a Bunch-Davies vacuum evolved in stochastic NR. We reserve these investigations of PGWs for future work and focus on the general anisotropic amplitude in our two scenarios.

Discarding the anisotropic degrees of freedom of General Relativity is central to the separate universe approximation (SUA) \cite{wands_new_2000} and the Hamilton-Jacobi approach \cite{salopek_nonlinear_1990, rigopoulos_separate_2003, prokopec_ensuremathdeltan_2021} { - but see \cite{Tanaka:2023gul,Tanaka:2024mzw}}. With a fully relativistic evolution of Bunch-Davies fluctuations, we can confirm if this is appropriate and understand the validity of such approximations more quantitatively. We know of course that $K$'s background ($k=0$ mode) is dominating overall given that $\tilde{K}$ has no background order contribution. We are interested in the perturbations at first and second order in perturbation theory, in particular in the case of USR where ${\cal R}$ blows up. For that purpose Figure \ref{fig:specAK} shows the spectra of $K$, $\tilde{K}$ (taken as $\tilde{K}_{ij}(k)\tilde{K}^{ij}(k)^*$ in Fourier space) and $\sqrt{3\tilde{K}_{ij}\tilde{K}^{ij}/2}$. 

It is of interest to compare the anisotropic $(-\tilde{K}_{ij}\tilde{K}^{ij})$ and isotropic ($\frac{2}{3}K^2$) contributions to the Hamiltonian constraint, given that the latter is the only one commonly considered in the SUA. Note that $K$ and $\tilde{K}$ (not squared) do not compete in any of the ADM equations without being squared or contracted respectively, see eqns.~\eqref{eq:CoarseGrainedADM_Dyn3} - \eqref{eq:CoarseGrainedADM_Dyn1} - \eqref{eq:CoarseGrainedADM_Dyn2} 
and
\eqref{eq:CoarseGrainedADM_Con1}, apart from the Momentum constraint \eqref{eq:CoarseGrainedADM_Con2}, usually identified as the culprit hampering such an approximation \cite{cruces_stochastic_2022,Artigas25}. 

Looking at the left panels first, we find that perturbations in $K$ freeze in the quadratic slow-roll case but decay in the USR case. This is also visible in the contrasts present in Figures \ref{fig:quad_contrast} and \ref{fig:usr_contrast}. Switching to the second panels, we see that anisotropic and isotropic powers can be comparable in magnitude in both scenarios initially but the former decays quickly after horizon crossing and is thus negligible in the quadratic case. In the USR case, perturbations in $K$ decay as well, implying that all terms in $K_{ij}$ are of similar magnitude and that the SUA truncation of the momentum constraint could potentially be violated. 

The last panel probes the most relevant terms competing with $K$: these are expected to be of second order given that $\tilde{K}$ has no background order, and, in fact, a careful observation shows anisotropic spectra to be subdominant to the isotropic modes for most, although not all, times. {Clearly, the striking hierarchy between isotropic and anisotropic expansion that characterises the quadratic case is not present in the USR case.} As pointed out in \cite{Artigas25} through the use of the non-adiabatic pressure, the anisotropic sector cannot be neglected a priori for the satisfaction of the momentum constraint in regimes such as this one.  Further quantitative comparison between the isotropic and anisotropic parts of the extrinsic curvature is left for future work.

\subsection{Quantum diffusion domination and eternal inflation}\label{sec:SINR.QD}
\textit{The research presented in this section is preliminary and unpublished.}

There exists another way to break perturbativity, other than going beyond quasi-dS in the background dynamics.
It is a regime where inhomogeneity cannot be injected equivalently from initial conditions: the \textit{diffusion}-dominated regime.
Even if subdominant, diffusion starts from the moment a stochastic source is introduced, but that is not the only way.
Historically, the macroscopic and empirical phenomenology of diffusion was quantified by Fourier in 1822 when he wrote down and studied the heat equation \cite{fourier1822chaleur},
and later applied to matter by Fick \cite{fick1855diffusion}. The search for the microscopic origin of such PDEs was initiated by the kinetic theorists Clausius, Maxwell and Boltzmann \cite{clausius1857waerme,clausius1858weglaenge,maxwell1860illustrations1,maxwell1860illustrations2,maxwell1867dynamical,boltzmann1872weitere}.
In 1905 and 1906, Einstein and Smoluchowski wrote the quantitative bridge between the microscopic and macroscopic worlds, showing that the diffusion strength of a PDE can be related to the mean free path of the underlying particles \cite{einstein1905brownian,smoluchowski1906kinetische}. 
While these treatments were primarily evolving microscopic probability distributions or cumulants, the equivalent stochastic approach to the microscopic equations of motion was only formulated in the early 1900s by Bachelier, for financial speculation \cite{bachelier1900speculation}, and Langevin for physics \cite{langevin1908brownien}, proposing the first stochastic ODEs and re-discovering Einstein's law.
Langevin wrote Newton's equation for a particle undergoing two forces, a systematic drag and a fluctuating random force, or in our case, the Hubble friction sourced by classicalised quantum fluctuations respectively, as seen in eqn.~\eqref{eq:SIeq}.
Depending on the time evolution, this opposition might lead to an equilibrium state described by the \textit{fluctuation-dissipation theorem} \cite{nyquist1928thermal,callen1951irreversibility,kubo1957statistical,kubo1966fluctuation}.
Inflation, however, is an intrinsically non-equilibrium process, and the stochastic EOMs are particularly convenient for this purpose.
Beyond the focus of this work, the equivalent Fokker-Planck treatment for the associated probability distribution is also available, and has been used extensively in the stochastic inflation literature, notably for the analytical extraction of the probability distribution of the curvature perturbation \cite{vennin_correlation_2015,pattison_quantum_2017}.

In the examples presented earlier in this chapter, notably our inflection point scenario, we can see that the stochastic source is subdominant in the dynamics,
as the lattice averages remain close to the deterministic Friedmann background values. 
Although found to be just about nonperturbative in the spectra or field-space spread, the USR example above has been designed to still follow linear theory around its background for validity checks, while still being nonperturbative.
This is achieved by either focusing on modes or jitter entering when the field has not yet slowed down too much or simply a short-enough period in USR (at the risk of reducing the nonperturbative growth in this case), and thus when the stochastic source is still subdominant compared to the drift along the potential.
These conditions define the limit where initial condition codes, such as those of lattice cosmology \cite{caravano2024usr} or our own of Chapter~\ref{sec:ICsim} including the geometry, can safely produce reliable predictions (up to the nonlinear effects of not coarse-graining).

However, in a previous Chapter~\ref{sec:SIGR} (Section~\ref{sec:SIGRng}), long wavelength dynamics were allowed moderate deviations from the background evolution so it could form non-Gaussian features, with noise active during the USR phase (see for instance Figure~\ref{fig:trajcumul}).  
This is an example where the stochastic source is starting to dominate over the drift, and thus where the stochastic formalism is necessary to describe the dynamics of the system.
Given the numerical work developed in this chapter, we now have the ability to explore this diffusion regime in full General Relativity.

Focusing on quasi-dS scenarios to ensure proximity with known physics, there are two specific ways known to reach the diffusion-dominated regime, both relying on the noise being mostly sourced by the overall energy ($H$).
The first one is, as just explained, a USR phase where the field is dramatically slowed down towards an almost flat potential.
There the friction dominates early on to take down the field velocity exponentially fast, as was seen in Figure~\ref{fig:MeanRolls}. Once the field has slowed down enough, the stochastic source can dominate the dynamics.
The other way to reach the diffusion-dominated regime is to increase the energy $H$ by going to high field values, thus enhancing the stochastic source while keeping the same deterministic kinetics.
This is the so-called scenario of \textit{stochastic eternal inflation}, first proposed by Linde in his \textit{chaotic} models of inflation \cite{Linde1986eternal}.

Let us now build a measure of the relative importance of the stochastic source compared to the deterministic drift, in order to progressively explore the diffusion-dominated regime in our simulations.
For the construction of the quantum diffusion proxy, we neglect the l.h.s. gradients of our GR stochastic equations, leading to eqns.~\eqref{eq:SImine}, which together could be expressed as
\begin{equation}
    \dN{Y} \equiv \dN{\begin{pmatrix}
        \phi \\
        {\Pi} \\
        {H} \\
        \end{pmatrix}}   = \begin{pmatrix}
        H^{-1}\Pi \\
        \Delta_c^{(\Pi)} \\
        \Delta_c^{(H)} \\
        \end{pmatrix} 
        + \begin{pmatrix}
            0\\
        \Delta_q^{(\Pi)} \\
        \Delta_q^{(H)} \\
        \end{pmatrix} \xi, \\
    \label{eq:stoY}
\end{equation}
where here $\dN {()}$ denotes the derivative with respect to the number of e-folds ${\cal N}_b$ (uniform e-folds gauge) and $K=-3H$ gives the IR Hubble rate, $\xi$ is a normalised white noise term, and $\Pi$ is the cosmic time derivative of the field.
The classical drift terms are defined by%
\begin{subequations}
\begin{align}
\Big|\Delta_c^{(\Pi)}\Big| &\equiv\bigg|-3\Pi - \frac{V_{,\phi}}{H}\bigg|, \\ 
\Big|\Delta_c^{(H)}\Big| &\equiv \bigg| - 3 H +\frac{V}{H \Mpl^2}\bigg|= \frac{\Pi^2}{2H\Mpl^2}, 
\end{align}
\end{subequations}
\begin{sloppypar}
after substituting the Hamiltonian constraint into the latter.
The quantum noise amplitudes,\allowbreak\ $\Big|\Delta_q^{(\Pi)}\Big|,\Big|\Delta_q^{(H)}\Big|$, could be computed explicitly from the mode functions of the Mukhanov-Sasaki equation, as done in the long wavelength limit ($\sigma \ll 1$), see Appendix~\ref{app:appD}.\footnote{This gives $\Big|\Delta_q^{(\Pi)}\Big| =  \frac{\sqrt{2\varepsilon_{1B}}\Mpl}{H_B} \sqrt{\langle {\cal S}_{\cal R}^2 \rangle}\bigg|_B  \simeq \frac{3H_B^2}{2\pi}$, and
$\Big|\Delta_q^{(H)}\Big| = \frac{\varepsilon_{1B}}{3H_B} \sqrt{\langle {\cal S}_{\cal R}^2 \rangle}\bigg|_B \simeq \sqrt{\frac{\varepsilon_{1B}}{2}} \frac{H_B^2}{2\pi \Mpl}$\ylg{, for some $B$ backreaction prescription}.}
With those equations, we are already in a new \yld{regime}\ylg{setup} as we do not assume overdamping ($\dN{\Pi}\ll|\Delta_c^{(\Pi)}|$ or adiabatic elimination of momentum), and we have allowed stochasticity in the geometry and backreaction
as opposed to the usual treatments.
In the long wavelength limit and in this gauge, as used in Section~\ref{sec:SIGRusuallimit} and standard stochastic inflation, the stochasticity in the geometry ($H$) can be discarded by simply substituting the Hamiltonian (or Friedmann) constraint in the $\Pi$ equation.
As we will see explicitly in eqns.~\eqref{eq:derConstraint} and \eqref{eq:deltaqH}, this is due to the constraint-satisfying construction of the respective noises.
Note, however, that the stochasticity in $\dN{H}$ is still as important as that of the Hubble slow-roll parameter $\varepsilon_1$, creating an offset to the usual relation with the field's kinetics $\varepsilon_1 \neq \Pi^2 / (2H^2\Mpl^2)$.\footnote{However, this is true up to noise, which is why one could still use the equality in the computation of the quantum noise amplitude as the correction is second order in the perturbative noise.}
In this section, we do not simplify out the stochasticity in $H$ because we are interested in its diffusivity and as a proxy of the dynamics beyond the long wavelength limit, in preparation for the full NR results evolving the stochasticity in all d.o.f.s.
\end{sloppypar}

We can define a quantum diffusion (QD) strength, as the ratio of the quantum jump amplitude to the classical drift amplitude for each d.o.f. $X$,
\(\mu_X= |{{\Delta_q X}}|/|\Delta_c X|\).
As is usually done \cite{vennin_correlation_2015}, we can compute it for the field in the overdamped limit
\begin{equation}
\Big|\frac{\dN{\Pi}}{3\Pi}\Big|\to 0, 
\label{eq:SRlimit}
\end{equation}
which is exactly the slow-roll attractor \cite{pattison_attractive_2018} (with $\varepsilon_{n\geq 1}\ll 1$), where the $\dN{\Pi}$ equation becomes
\begin{equation}
\dN{\phi} = \Delta_c^{(\phi)} + \Delta_q^{(\phi)}\xi, \qquad \Big|\Delta_c^{(\phi)} \Big| = {\big|V_{,\phi}\big|}/{3H}, \qquad \Big|\Delta_q^{(\phi)}\Big| = {\Big|\Delta_q^{(\Pi)}\Big|}/{3H}.
\end{equation}
Without computing explicitly the QD ratio $\mu$ for the overdamped $\phi$, this allows us to evaluate the diffusion of the momentum (no overdamping) relative to $\mu$ as
\begin{equation}
\mu_{\Pi} = \frac{\mu}{|g|}, \qquad g \equiv 1+\frac{V_{,\phi}}{3H\Pi}.
 \label{eq:RQDPhiK}
\end{equation}
Without knowing explicitly $\Delta_q^{(H)}$, the case of $\mu_H$ is more subtle.
It requires\yld{ the assumption} that the stochasticity in $H$ makes up for that in the momentum in order to still satisfy the constraints. Let us then differentiate the Hamiltonian constraint, and substitute in eqns.~\eqref{eq:stoY} to get
\begin{equation}
0 = 2 \dN{H}H - \frac{\Pi V_{,\phi}}{3H\Mpl^2} -\frac{\dN{\Pi}\Pi}{3\Mpl^2} = \bigg( 2H\Delta_q^{(H)} -\frac{\Pi}{3\Mpl^2}\Delta_q^{(\Pi)} \bigg)\xi,
\label{eq:derConstraint}
\end{equation}
thus yielding an algebraic relation between the quantum noise amplitudes
\begin{equation}
\Delta_q^{(H)} = \frac{\Pi}{6H\Mpl^2}\Delta_q^{(\Pi)}.
\label{eq:deltaqH}
\end{equation}
This is perfectly verified using the explicit expressions for the noise amplitudes, and we believe that such equality can be expected beyond the long wavelength limit using the full GR formulation (reserved for future work).
The practicality of this result is the fact that it allows us to compute the QD ratio for $H$ in terms of $\mu$ as
\begin{equation}
    \mu_H = \frac{\big|\Delta_q^{(H)}\big|}{\big|\Delta_c^{(H)}\big|} = \frac{\big|\Delta_q^{(\Pi)}\big|}{3\big|\Pi\big|} = \frac{\mu_\Pi\big|\Delta_c^{(\Pi)}\big|}{3\big|\Pi\big|} = |g|\mu_\Pi = \mu.
\end{equation} 
At first sight, this suggests that the diffusion in $H$ follows that of the overdamped field. In the overdamped limit, this feels intuitive as $H$ reflects directly $V$, and thus $\phi$.
Let us mitigate this by writing eqn.~\eqref{eq:stoY} in terms of $\mu$, $\Pi$ and $H$ only,%
\begin{subequations}
    \begin{align}
        \dN{\big(\ln \Pi \big)} & =-3(g+\mu\xi),\\
        \dN{\big(\ln H \big)} & = -\frac{\Pi^2}{2H^2\Mpl^2}(1+\mu\xi),
        \label{eq:muSto}
    \end{align}
\end{subequations}
with inflation ending for a given stochastic path when $\dN{\big(\ln H \big)}$ reaches $-1$. 
These equations are particularly interesting for two reasons. Firstly, the stochasticity in $H$ is directly sourced by that in $\Pi$, and thus the stochasticity in $H$ is not independent of that in $\Pi$.
This means that, whether or not we have stochastic backreaction via $\mu$, or whether or not we have direct stochasticity in $H$, the diffusive dynamics of $\Pi$ can still have an impact via $H$.
Secondly, given the $-\Pi^2$ contribution, this impact is entirely biased towards decreasing $H$ (if $\mu$ is small enough).
The diffusion in $H$ is thus far from being straightforward, and a non-trivial nonlinearity arises in the geometry.
These considerations have been overlooked in the literature, as the slow-roll regime has generally been applied to the Hamiltonian constraint to write $H^2 \simeq V/(3\Mpl^2)$, thus neglecting the $\Pi$-stochasticity in $H$.
However, starting with slow-roll and then switching on stochasticity, it is clear that the impact of the momentum diffusing could dominate before the field itself diffuses as we increase $\mu$.

Leaving the full integration of the stochastic motion for future work,
we can get a grasp of the generic non-Gaussian and nonlinear bias on the dynamics by looking at the proportion of field values going up or down.
For a generic field $X$, Gaussianity of the noise gives
\begin{equation}
\mathcal{P}\Bigg(\dN{X}>0|\dN{X}=\Delta_c^{(X)}(1+\mu\xi)\Bigg)=\frac{1}{2}\Bigg(1+\text{sign}(\Delta_c^{(X)}){\rm erf}\bigg(\frac{1}{\sqrt{2}\mu_X}\bigg)\Bigg),
\end{equation}
so that for $\Pi, H$, with the reasonable assumption $\Delta_c^{(\Pi)}<0$ and since $\Delta_c^{(H)}<0$ we have%
\begin{subequations}
    \begin{align}
\mathcal{P}\Bigg(\dN{\Pi}>0\Bigg) & = \frac{1}{2}\Bigg(1-{\rm erf}\bigg(\frac{|g|}{\sqrt{2}\mu}\bigg)\Bigg),\\
\mathcal{P}\Bigg(\dN{H}>0\Bigg) & = \frac{1}{2}\Bigg(1-{\rm erf}\bigg(\frac{1}{\sqrt{2}\mu}\bigg)\Bigg).
    \end{align}
\end{subequations}
These expressions give the nonlinear bias of deterministic and stochastic kicks together.
At low $\mu$, $\mu \ll |g|$, we recover the deterministic limit where both $\Pi$ and $H$ go down with probability $1$, while at higher $\mu$, $\mu \gg |g|$ for $\Pi$ and $\mu \gg 1$ for $H$, the stochastic domination entails that both go up or down with probability $\frac{1}{2}$.



We now turn to numerics to solve the full GR stochastic dynamics.
We place our pre-stochastic initial conditions in the slow-roll regime of a simplistic potential described by
\begin{align}
    \varepsilon_{1i} & = 5\times 10^{-3}, \qquad V(\phi) = \frac{1}{2}m^2\phi^2, \qquad V'(\phi) = m^2\phi, \qquad  m = 10^{-3}\Mpl,
\end{align}
and choose a large enough initial field value, $\phi_i$, to enforce a starting value of $\mu$, $\mu_i = c w$, where $c$ is a dimensionless parameter that we will vary in the following, and where $w = 0.61$ is chosen so that the probability for the overdamped scalar field to go up after an e-fold is $1/e^3$. Beyond that, eternal inflation is expected to be active \cite{Guth07}.
By writing the slow-roll jumps and the exact Hubble functional as
\begin{equation}
     \Delta^{(\phi_i)}_c \simeq \sqrt{2\varepsilon_{1i}}\, \Mpl, \qquad \Delta^{(\phi_i)}_q \simeq \frac{H_i}{2\pi},  \qquad  H_i = \frac{m}{\Mpl} \frac{|\phi_i|}{\sqrt{2(3 - \varepsilon_{1i})}},
\end{equation}
we can impose the desired $\mu_i$ via
\begin{equation}
     \phi_i = -4\pi \mu_i  \sqrt{\varepsilon_{1i}(3-\varepsilon_{1i})}\bigg(\frac{\Mpl^2}{m} \bigg) .
\end{equation}
Clearly, higher $\mu_i$ requires higher field and $H$ excursions, in agreement with usual realisations of eternal inflation.
Note that, until drastic stochastics bring a change, $\mu$ should be rather constant in the slow-roll regime, as both $H$ and $\varepsilon_{1}$ are slowly varying.

\begin{figure}[p]
    \centering

    \begin{subfigure}{0.48\textwidth}
        \centering
        \includegraphics[width=0.8\linewidth]{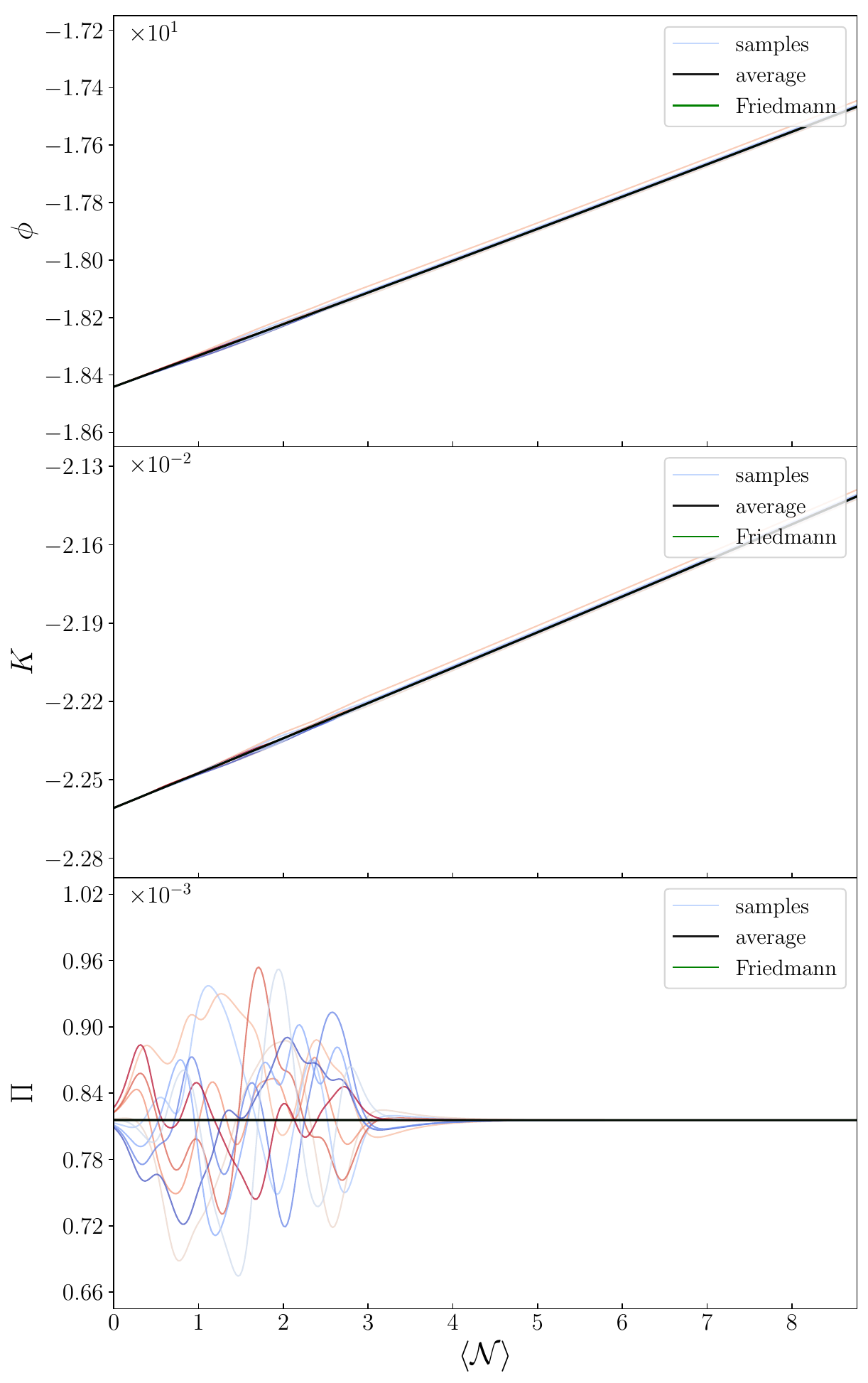}
        \caption{$c=0.02$}
        \label{fig:QD64a}
    \end{subfigure}
    \hfill
    \begin{subfigure}{0.48\textwidth}
        \centering
        \includegraphics[width=0.81\linewidth]{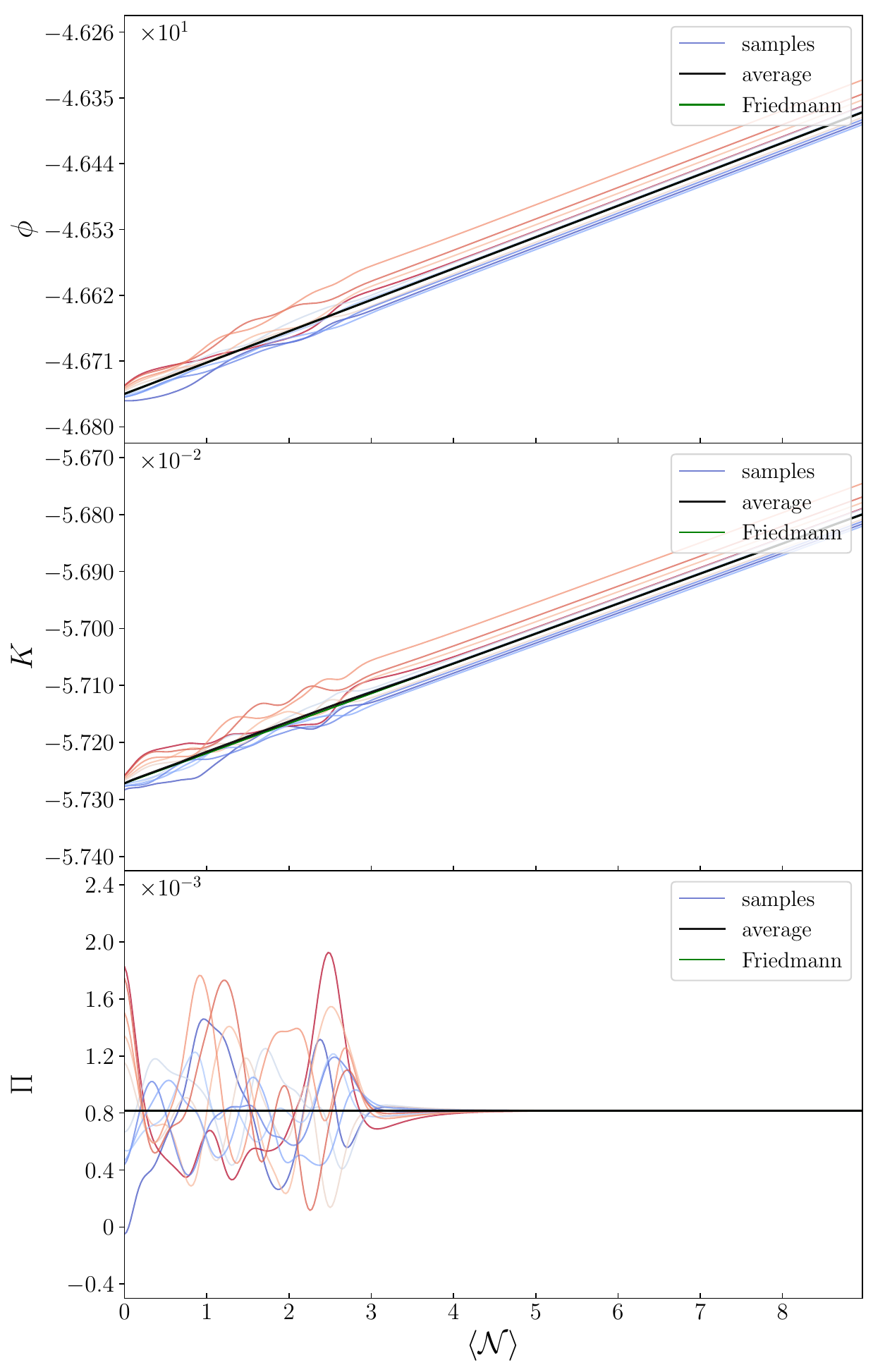}
        \caption{$c=0.05$}
        \label{fig:QD64b}
    \end{subfigure}

    \begin{subfigure}{0.48\textwidth}
        \centering
        \includegraphics[width=0.82\linewidth]{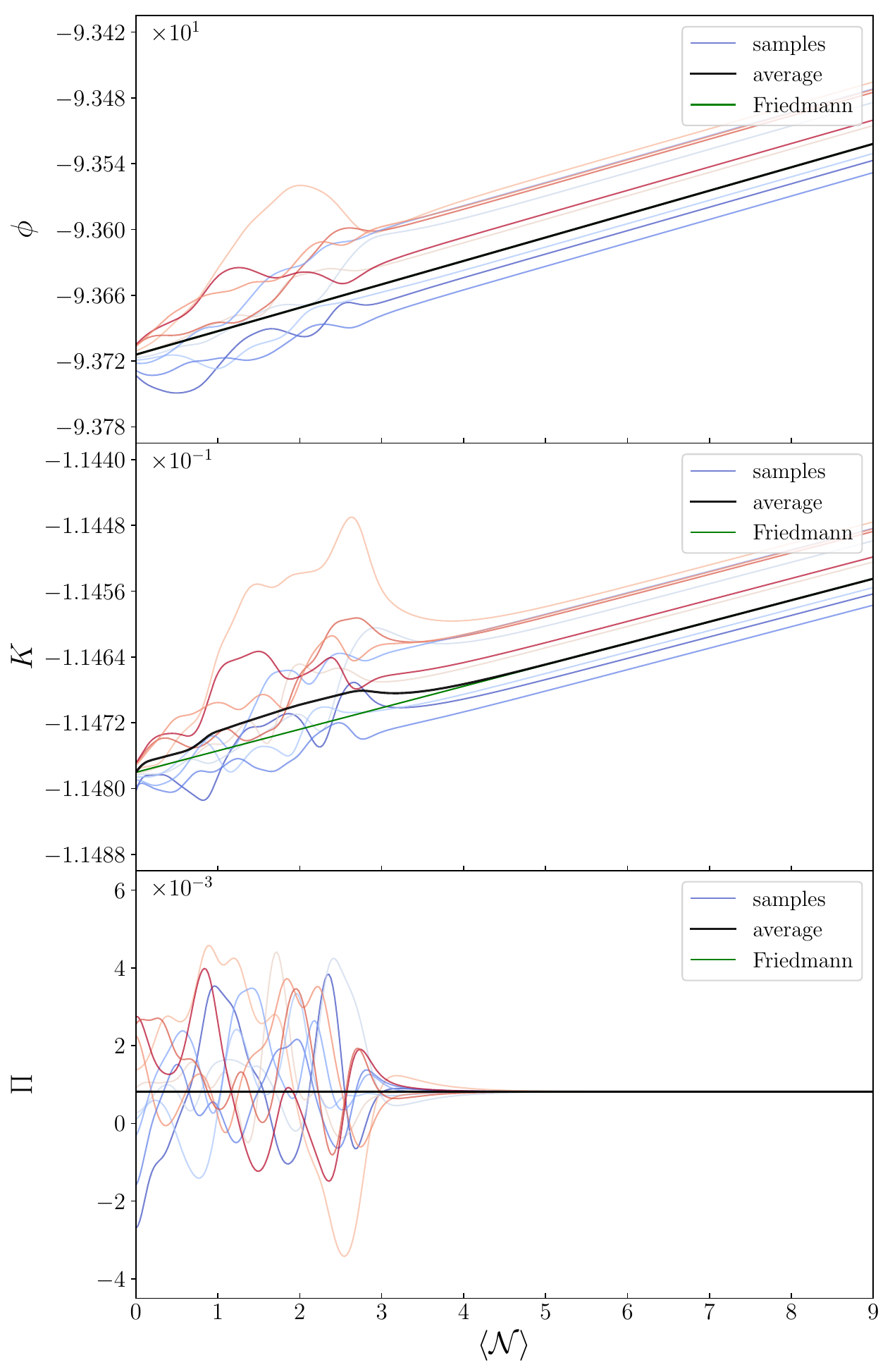}
        \caption{$c=0.1$}
        \label{fig:QD64c}
    \end{subfigure}
    \hfill
    \begin{subfigure}{0.48\textwidth}
        \centering
        \includegraphics[width=0.81\linewidth]{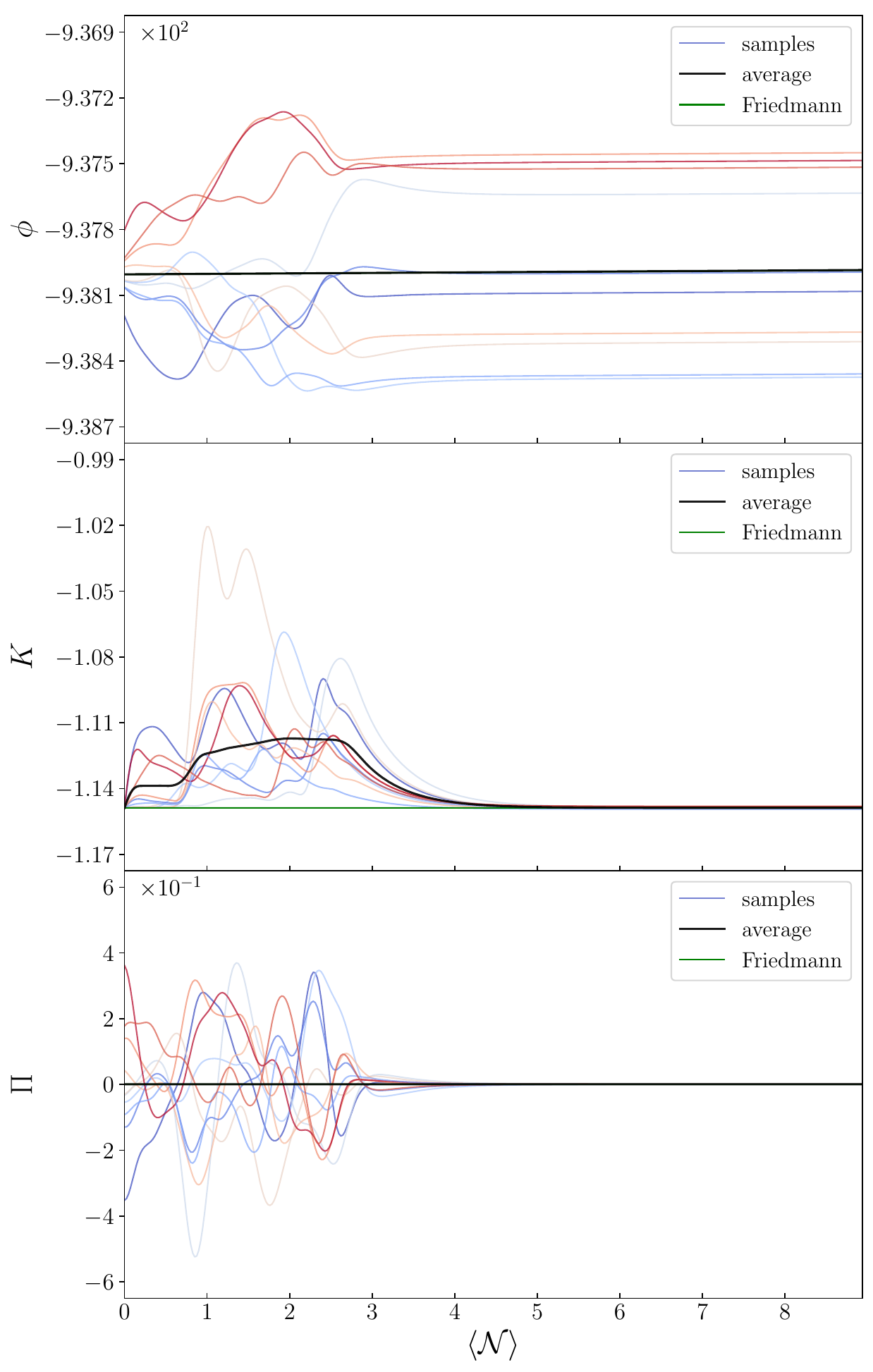}
        \caption{$c=1$}
        \label{fig:QD64d}
    \end{subfigure}

    \caption{Means and perturbations during quadratic inflation as quantum diffusion is boosted (left to right, top to bottom, $c=0.02,\,0.05,\,0.1,\,1$, corresponding to $\mu_i \simeq 0.01, 0.03, 0.06, 0.61$) in full GR, from $N=64$ \istoriz realisations ($\sigma=1$, level $0$ backreaction).
    For the field, momentum, and extrinsic curvature, we show the Friedmann \yld{average}\ylg{background}, the lattice average, and a sample of $10$ stochastic field trajectories. The comoving positions coming from each trajectory are chosen to visualise the representative spread of ${\cal R}$ at the initial time, and coloured accordingly (from positive maximal delta in red to negative maximal delta in blue).
    $\Mpl=100$.}
    \label{fig:QD64}
\end{figure}

\begin{figure}[p]
    \centering

    \begin{subfigure}{0.48\textwidth}
        \centering
        \includegraphics[width=0.8\linewidth]{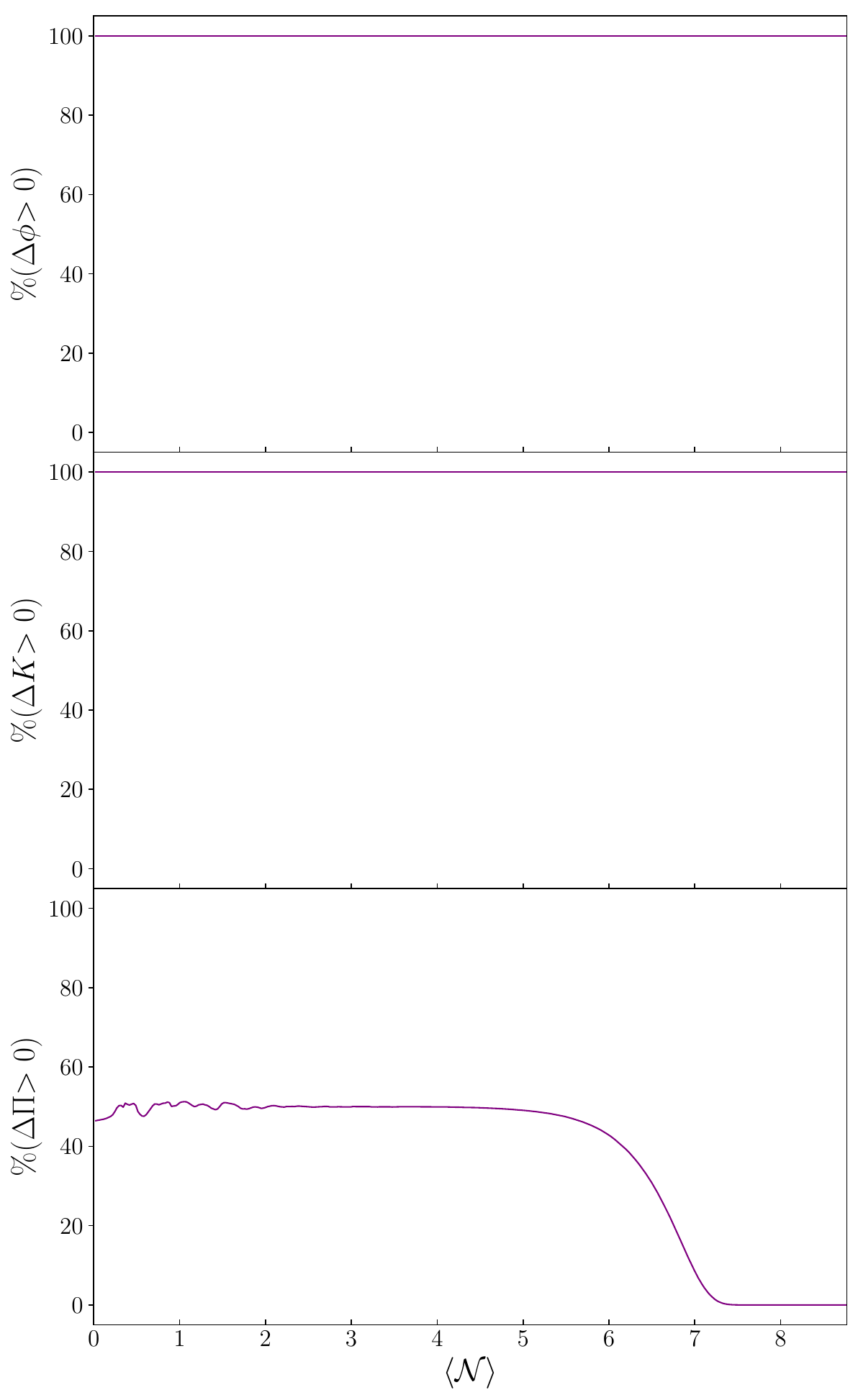}
        \caption{$c=0.02$}
        \label{fig:Propup64a}
    \end{subfigure}
    \hfill
    \begin{subfigure}{0.48\textwidth}
        \centering
        \includegraphics[width=0.8\linewidth]{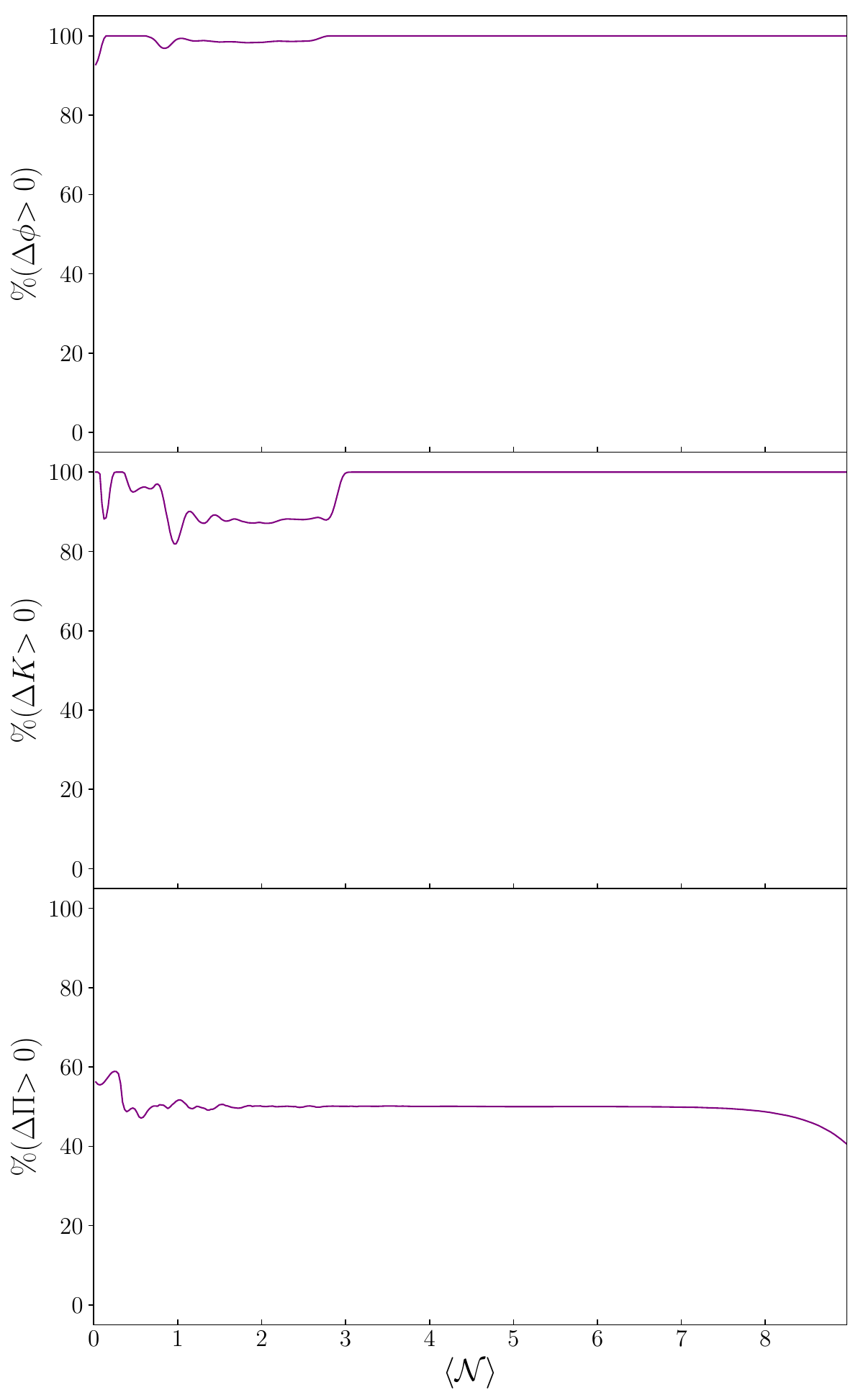}
        \caption{$c=0.05$}
        \label{fig:Propup64b}
    \end{subfigure}

    \begin{subfigure}{0.48\textwidth}
        \centering
        \includegraphics[width=0.8\linewidth]{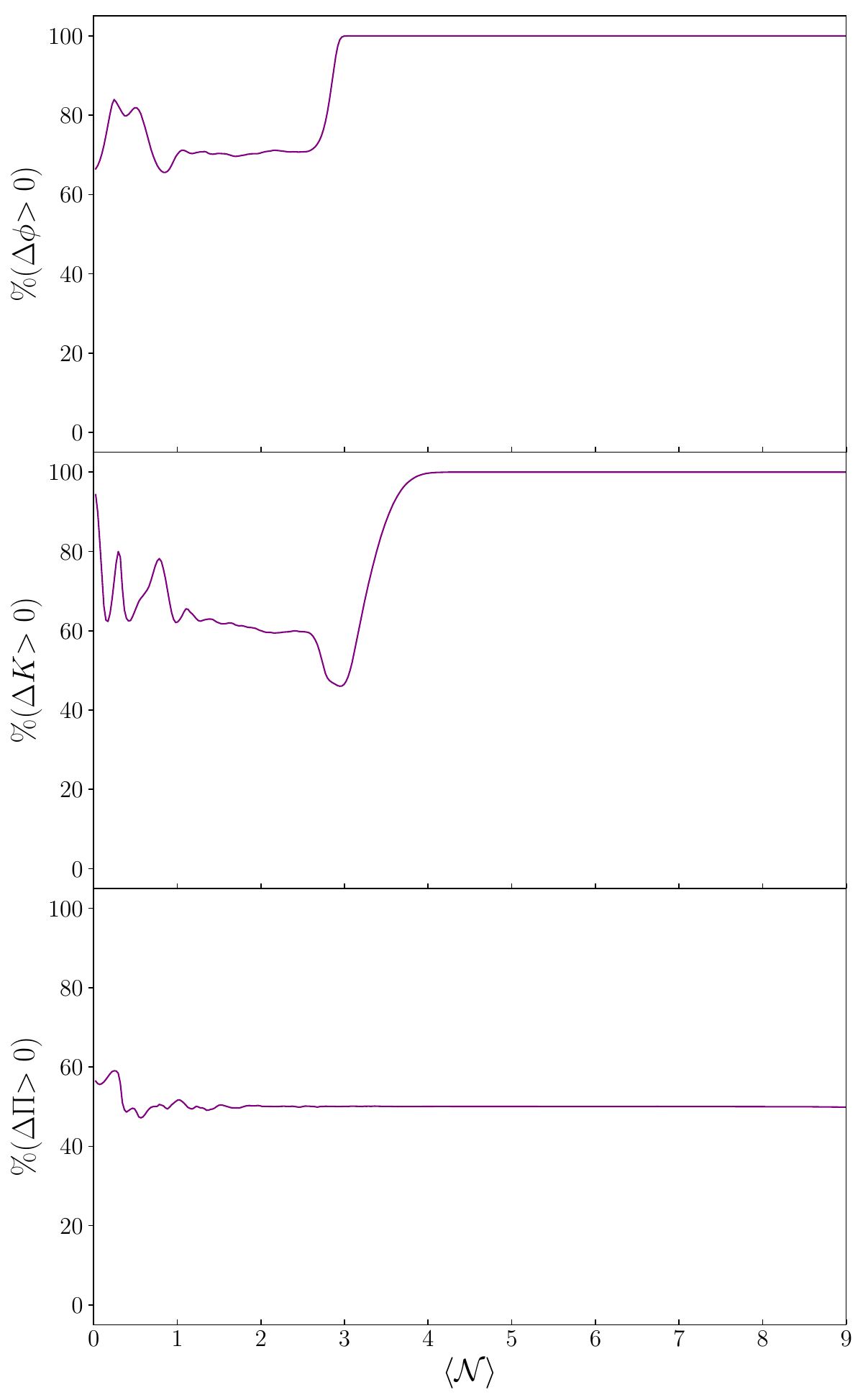}
        \caption{$c=0.1$}
        \label{fig:Propup64c}
    \end{subfigure}
    \hfill
    \begin{subfigure}{0.48\textwidth}
        \centering
        \includegraphics[width=0.8\linewidth]{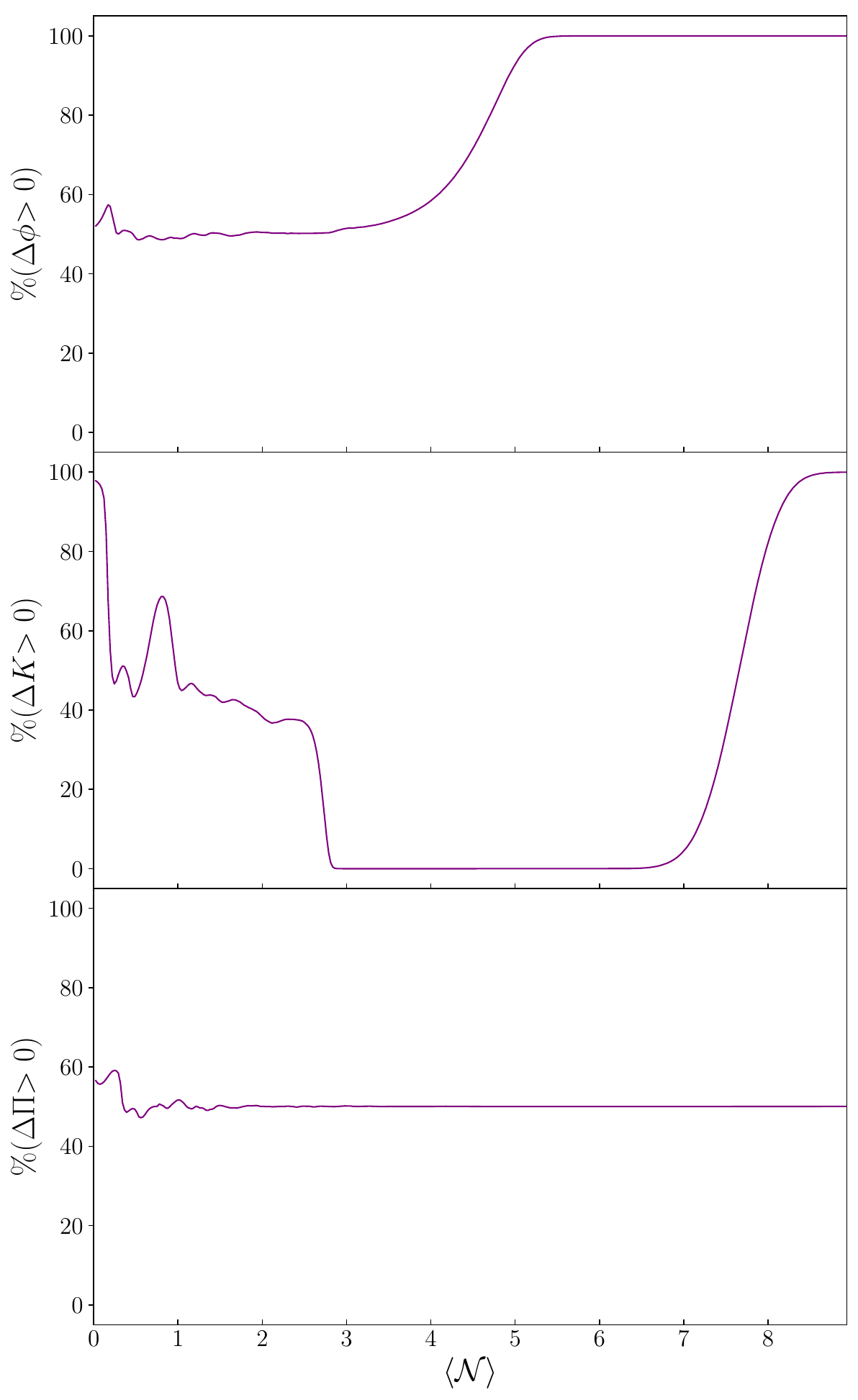}
        \caption{$c=1$}
        \label{fig:Propup64d}
    \end{subfigure}

    \caption{Percentage of pixels going up for each field (top to bottom, within each panel) at each time step, as quantum diffusion is boosted (left to right, top to bottom across panels, $c=0.02,\,0.05,\,0.1,\,1$). The deterministic scalar field rolls from left to right. $\Mpl=100$.}
    \label{fig:Propup64}
\end{figure}

We run \istoriz for $N=64$ realisations (and $\sigma =1$, level $0$ backreaction) with various values of $\mu_i$ and show the results in Figures~\ref{fig:QD64} and \ref{fig:Propup64}.
For each of $X= \phi, \Pi, K$, we show in the former the Friedmann average, the lattice average, and an initially-representative sample of $10$ stochastic field trajectories, while the latter shows the estimation of $\mathcal{P}(\dot{X}>0)$.
Given the limited dynamical range, modes stop entering the simulations at about $3$ e-folds, after which we expect a return to drifting down the potential.  
With as little field-diffusion as $\mu_i \simeq 0.01$ (Figures~\ref{fig:QD64a} and \ref{fig:Propup64a}), the stochasticity is quite noticeable for the momentum ($10\%$ contrast) and quickly reaches complete diffusion from $\mu_i \simeq 0.03$ (Figures~\ref{fig:QD64b} and \ref{fig:Propup64b}), while the field and the extrinsic curvature are simply jittering around their deterministic values, acknowledging some increasing contrast and skewing.
As we increase $\mu_i$ to $0.06$ (Figures~\ref{fig:QD64c} and \ref{fig:Propup64c}), the stochasticity in the momentum is still dominating, but the field and extrinsic curvature are starting to diffuse as well: diffusion domination is there much before what we expected ($\mu_i \simeq 1$).
At this level, we find up to $30\%$ of field values beating the drift up and a noticeable difference between the Hubble rates of the Friedmann background and the lattice average.
Finally, with $\mu_i \simeq 0.61$ (Figures~\ref{fig:QD64d} and \ref{fig:Propup64d}), the stochasticity is dominating all three d.o.f., and the dynamics are fully diffusive: until the extinction of the noise at about $3$ e-folds, half of the points go up and down the potential at each time step.
At this stage, the dynamics of the extrinsic curvature (or $-3H$ on large scales) are of great interest, as seen from eqn.~\eqref{eq:muSto} and its matching realisation in Figure~\ref{fig:Propup64d}. 
Both stochastic kicks in $\Pi$ and $K$ greatly influence the amplitude to the step $K(t+dt)-K(t)\propto +\Pi^2(1+\mu \xi)$.
In particular, the second one hampers the step to be strictly positive. Once the noise is switched off in all equations after about $3$ e-folds, $\Pi$ decreases drastically (including $0$-crossings) to lead to negligible kinetics and thus stagnant $K$.
Once back to the slow-roll attractor, all $K$ values are increasing again ($H$ decreasing).

At this preliminary stage of the study, we can say that the interplay between the stochasticity in $\Pi$ and that in $H$ is non-trivial, even in a regime such as slow-roll where eternal inflation can operate.
In such a regime, a hierarchy seems to be suggested in the emergence of quantum diffusion as one increases a starting value of $\mu$.
It seems to start first with the kinetic diffusion, then in the expansion rate, and finally in the field itself, making previous field-only treatments of stochastic inflation incomplete, or at least questionable.

In the future, we will investigate further the diffusive regime from the stochastic acceleration equations, notably the difference between a slow-roll and ultra-slow roll phase, including effects from and on backreaction, anisotropic d.o.f.s, and other effects.
Most importantly, our current analysis suggests that stochastic eternal inflation is intrinsically problematic for the perturbative noise assumption behind stochastic inflation and the satisfaction of the momentum constraint (verified in our simulation).
In fact, taking $g\to 0$ in eqn.~\eqref{eq:muSto} suggests $\dN{\big(\ln \Pi \big)} =-3\mu\xi$, implying that the kicks' contrast per e-fold is of order $\mu$ and breaks the usual assumptions for perturbative noise as we increase $\mu$.
Our current view is to suggest that this could remain a legitimate approximation as long as the momentum kicks contribute perturbatively to the energy density.

\section*{Summary}

In this chapter, we pioneered the full evolution of the Bunch-Davies vacuum in inflation using numerical relativity and the stochastic formalism.
In the latter, modes do not all enter the simulation at the same initial time as opposed to Chapter~\ref{sec:ICsim} (see also \cite{Florio2024} for tensor modes) but at a given elapsed time relative to their Hubble horizon crossing, thus better accounting for quantum diffusion. By achieving this, we provide a framework generalising stochastic inflation, inflationary numerical relativity and lattice cosmology beyond their standard assumptions\ylc{, while keeping the linear-noise ansatz and the heuristic backreaction of Chapter~\ref{sec:SIGR}}.

Two test scenarios were studied, a very perturbative and linear case of slow-roll inflation and a potential with an inflection point triggering ultra slow-roll and so non-perturbative growth. 
Both scenarios are successfully evolved, matching expectations for the linear spectra and satisfying the Energy and Momentum Constraints \ylc{at the order of the noise} for our scales of interest\ylc{. This validates the pipeline against linear theory, through a slow-roll to ultra-slow-roll transition, for a fixed coarse-graining scale $\sigma = 1$. Its dependence on $\sigma$, which characterises the approximation, remains to be studied.}

{We were additionally able to follow the evolution of anisotropic expansion, normally out of reach for stochastic approaches. Indeed, such simulations \yld{completely }\ylc{remove} the reliance of stochastic inflation on the separate universe approximation and its numerous caveats, the restriction to isotropic expansion, the $\Delta\mathcal{N}$ formalism or the need to pick specific temporal hypersurfaces\ylc{, although the noise itself is still computed from linear theory around a homogeneous background. Backreaction and anisotropies were only given first looks.}}
Finally, a last scenario mixing nonlinear and nonperturbative dynamics was also explored, by looking at the quantum diffusion limit and its realisation in a simplistic but eternally slow-rolling model of inflation.
Accompanied by some analytical proxies and NR, we \ylc{provided a preliminary full GR treatment of quantum diffusion, which suggests that this regime challenges the perturbative noise assumption itself}.
As we will convey in our final conclusions, we hope to continue applying the success of those simulations to relevant non-linear inflationary physics.

%% file: chap_SIbeyondGR.tex
\ifSubfilesClassLoaded{\setcounter{chapter}{5}}{}
\chapter{Stochastic Inflation beyond General Relativity} \label{sec:SIbeyondGR}
\addtocontents{toc}{\protect\setcounter{tocdepth}{3}}


Previous chapters show that, if one is ready to give up analytical solutions, it is possible to build a stochastic formalism in full general relativity, and to solve it numerically, without \ylc{relying on many historical approximations such as the separate universe approximation. What remains is the classicality of the IR sector, the linear noise around a homogeneous background, gauge invariance restricted to gauges close to that background, and heuristic backreaction}.
The procedure employed there is deeply rooted in the mastery of linear perturbation theory in any gauge, and, as argued in this chapter, can be extended beyond single scalar-field inflation in GR.

With the same minimal approximations, this chapter provides the generalisation of stochastic inflation to scalar-tensor theories going beyond General Relativity.
These theories are, in fact, the generic extension of GR with a scalar degree of freedom relevant for inflation and, more generally, dark energy, and encompass a wide class of viable inflationary frameworks while remaining theoretically controlled.
In the perturbative limit, where the sources are derived, these theories are more easily gathered in the same formalism using the Effective Field Theory of Inflation (EFToI)\yld{, as a special case of the EFT of dark energy (EFToDE)}.
Initially stemming from the EFToI, \yld{the latter formalism reached great maturity}\ylg{the EFT of Dark Energy (EFToDE) extended the framework to generic dark energy} in the past two decades, as part of an effort to constrain post-inflation \yld{dark energy }models, and even \yld{rule out some of these}\ylg{saw some of the parameters ruled out} for late eras of cosmology \cite{CreminelliVernizzi17}.
As we will see, the associated notations, in particular those of the so-called $\alpha$ basis \cite{BelliniSawicki14}, are still particularly meaningful for the study of inflation.
The EFToI is today a subset of this scalar-tensor framework, and usually goes further in its approximations.
In particular, mixings with gravity are usually neglected by assuming that the Hubble parameter, or the energy, is above the decoupling scale (equivalent in practice to $\Mpl\to \infty$, and $\dot H \to 0$, at fixed $\Mpl^2\dot H$).
Admittedly, these energies are much higher than the late-time ones, but the freezing of the gauge-invariant fluctuations at Hubble scale crossing makes the imprint at those energies relevant.
The predictability of such an approach has recently culminated thanks to perturbative solvers such as \textit{CosmoFlow} \cite{werth23}.  

For the case of inflation and before even mentioning nonperturbative effects, the perturbative approach is not without generalisations and motivations beyond the decoupling limit.
In particular, computations of the curvature perturbation 3-point function ---as the leading signal of inflationary interactions--- have been performed beyond this limit.
We saw that the pioneering work of Maldacena \cite{Maldacena_2003} gave early on the 3-point function of the curvature perturbation in single-field slow-roll inflation with the mixings with metric perturbations and the leading slow-roll contributions in General Relativity.
This was later extended to more general single-field actions\footnote{As well as various multifield extensions, including the decoupled EFToI treatment of \cite{senatore_effective_2012}.}, such as those with single-field non-canonical kinetic terms \cite{Seery06, chen_primordial_2010, burrage_large_2011}, including braiding \cite{zhang_primordial_2021}, but also various calculations in the general Horndeski classes with those mixings involved \cite{de_felice_inflationary_2011, felice_primordial_2011}, including the gravitonic bispectra \cite{gao_primordial_2011}.
To make such tedious calculations accessible in non-decoupled regimes, numerical pipelines such as \cite{Zhang25} have also been developed.
Importantly, these works emphasise that nontrivial contributions can be found and be missed by the decoupling limit in a generic scalar-tensor theory.
Moreover, even if GR or the decoupling limit are found to be good approximations to certain energy scales, such extended treatments are necessary to quantify the validity of the former.
It is also worth mentioning that a plethora of inflationary models formulated in GR, whether given as interactive actions or simple potentials (e.g. the widespread \textit{scalaron} Starobinsky model \cite{starobinsky80}), have been derived from scalar-tensor theories using transformations to the so-called \textit{Einstein frame}.
However such transformations between frames do not always exist \cite{Zumalacarregui13} or could generate increased complexity beyond the single-field scenario, as a field-space curvature might be generated in the multifield case \cite{Kaiser10}.

In the aforementioned existing calculations, beyond the decoupling limit or not, several approximations are employed on top of perturbativity and the tree-level limitation\footnote{Loop calculations are still challenging today \cite{Senatore10,Assassi12,lee_leading_23} and are thus not part of any numerical pipeline.}, such as the freezing of couplings at the horizon or perturbative expansions on state properties or couplings (including sound speed, slow-roll parameters etc).
More generally, computations of correlators ---even at tree-level--- have never been implemented for a generic scalar-tensor theory, whether using in-in integrations or the aforementioned transport equations.
In this context, one can see that the fully quantum methods are not only unable to handle nonperturbative dynamics at the moment, but also to handle systematically the full variety of perturbative and nonlinear dynamics in the landscape of inflationary models.

At the expense of analytic work and most importantly quantum effects, relying on evolving the equations of motion is usually the easiest way to resolve full nonlinearities and nonperturbative effects, as shown by multiple works in the context of inflation \cite{Cosmolattice,latticeeasy,clough_robustness_2017,Elley2024,Caravano22thesis,Giannadakis25}, including the previous chapters.
However, the stochastic approach is precisely restricted to a semiclassical regime, and so, it is legitimate to relinquish the off-shell part of the system's description, as long as the classicality condition is satisfied.
Providing a theory of stochastic inflation in a generic scalar-tensor theory thus consists in providing a simpler, if not first, access to correlators from generic theories of the origin by exploiting the semiclassicality of the inflationary spacetime.

This chapter is organised as follows, building on the general philosophy and procedure to get stochastic equations for an IR theory introduced in Section~\ref{sec:ansatz}.
In order to derive the stochastic sources of generic scalar-tensor theories via their linear equations without applying the coarse-graining to each theory separately, we introduce the \yld{EFToDE}\ylg{EFT action} and our chosen notations in Section~\ref{sec:EFToDE}.
The stochastic ansatz is then applied to get our main result summarised in Section~\ref{sec:mainresult}: the stochastic sources of the scalar-tensor equations of motion of theories admitting a decomposition in the \yld{EFToDE}\ylg{EFT action} when linearised.
In Section~\ref{sec:EOMs}, we put this result into practice for a variety of theories.
This means finding the \yld{EFToDE}\ylg{EFT action} coefficients that each chosen theory corresponds to, including nonminimally coupled theories, Einstein-Gauss-Bonnet, Horndeski, Braiding, before applying the main result.
Finally, Section~\ref{sec:discussion} concludes by pushing the stochastic ansatz to its edge discussing further possible extensions, inspired by previous works and the generality of our \yld{EFToDE}\ylg{EFT action} notations.
For instance, we \ylc{illustrate} the generality and naturalness of our proposed procedure by \ylc{outlining the coarse-graining of} multifield GR\ylc{, leaving the final equations for future work}. 


In the following, we will use the following tensor for comparison to the Einstein field equations,
$$
  {\cal E}_{\mu\nu} = \frac{2}{M^2}\frac{1}{\sqrt{-g}}\frac{\delta S}{\delta g^{\mu\nu}},
$$
with $M$ a potentially time-dependent Planck mass and $S$ the action of our theory.
When $S$ is the Einstein-Hilbert action with some content described by a stress-energy tensor $T_{\mu\nu}$, we recover ${\cal E}_{\mu\nu} = G_{\mu\nu}-\Mpl^{-2} T_{\mu\nu}$, with the Planck mass $\Mpl = (8\pi G)^{-1/2}$. ${\cal E}_{\mu\nu}$ vanishes in the least action configuration (\textit{on-shell}).
In full generality, we also define 
$${\cal E}_{\ylg{Z}} = \frac{2}{M^2}\frac{1}{\sqrt{-g}}\frac{\delta S}{\delta \ylg{Z}}, $$
useful for instance to look at scalar field equations of motion. \ylg{$Z$ denotes a generic dynamical field or metric variable of the theory.}


\section{Scalar-tensor stochastic sources} \label{sec:procedure}

\textit{In the following, we apply the stochastic ansatz described in Section~\ref{sec:SIGR} to a wide class of \yld{theories of dark energy}\ylg{scalar-tensor theories} beyond GR at the same time (i.e. not one by one specifically) by using an EFT description gathering all linear descriptions at once.}

\ylg{Unless stated otherwise, Sections~\ref{sec:procedure} and \ref{sec:EOMs} assume:
\begin{itemize}[nosep]
  \item a single clock: one scalar degree of freedom, whose background sets the unitary-gauge time slicing;
  \item no matter or radiation sector beyond this scalar and the metric;
  \item linearity of the UV modes on the IR background, as in eqn.~\eqref{eq:actionExpansion};
  \item classicality of the IR modes, which should be checked for each scenario and choice of $W$ \cite{Launay2024bis, Ballesteros26}.
\end{itemize}
They do not assume:
\begin{itemize}[nosep]
  \item a specific theory within the EFT action, its coefficients being left free until Section~\ref{sec:EOMs};
  \item the decoupling limit, mixings with the metric perturbations being kept;
  \item a specific window function $W$;
  \item a Bunch--Davies vacuum: the sources are given by the state of the UV modes at crossing, which need not be quantum (eqn.~\eqref{eq:Qsol} is one example);
  \item the freezing of the UV modes at crossing, their evolution continuing nonlinearly in the IR.
\end{itemize}}

\subsection{EFT primer}\label{sec:EFToDE}

Given the linear origin of the sources, the EFT\yld{ of Dark Energy} is the ideal framework to describe the dynamics of the UV modes and compute the sources of all included theories simultaneously.
It provides indeed a unified way to write down the most general action for a single scalar degree of freedom coupled to gravity, without having to specify the underlying theory.
Later in Section~\ref{sec:EOMs}, we will specify some of the EFT coefficients to match some known higher-order relativistic theories.
The following subsections first introduce the reader to these tools and notations, mostly taken from \cite{Gleyzes13,Gleyzes14}.

\subsubsection{Symmetries and actions}
\yld{The Effective Field Theory of Dark Energy (EFToDE) is a powerful framework to describe the dynamics of dark energy and gravity.}\ylg{The effective field theory of a single scalar degree of freedom spontaneously breaking time diffeomorphisms (the EFT of inflation, EFToI) is a powerful framework to describe the dynamics of the inflaton and its interactions with gravity \cite{Cheung2008}.}
It is based on the idea that \yld{dark energy}\ylg{the dynamics of the scalar field} can be described as a low-energy effective theory, where the high-energy degrees of freedom are integrated out. This allows us to capture the essential features\yld{ of dark energy} without needing to specify its microscopic origin, and thus without needing the fully nonlinear equations of motion.
The action of the \yld{EFToDE}\ylg{EFT} is written in the unitary gauge as \cite{Gleyzes13}
\begin{equation}
    \begin{aligned}
        S  &= \int d^4x \sqrt{-g} \left[ \frac{\Mpl^2}{2} f(t)\,{}^{(4)}R - \Lambda(t) - c(t)g^{00}\right.  \\
       &  \left.  + \frac{M_2^4(t)}{2}(\delta g^{00})^2 - \frac{m_3^3(t)}{2} \delta K \delta g^{00}
       - m_4^2(t) \left(\delta K^2 - \delta K^\mu{}_\nu \delta K^\nu{}_\mu\right) + \frac{\tilde{m}_4^2(t)}{2} R\, \delta g^{00} +\ldots\right],
    \end{aligned}
    \label{eq:EFTaction}
\end{equation}
where $\Mpl$ is the Planck mass, ${}^{(4)}R$ and $R$ are the $4$-dimensional and $3$-dimensional Ricci scalars respectively, and $K_{\mu\nu}$ is the extrinsic curvature. In this form, this is the most general action for a single scalar degree of freedom coupled to gravity, up to second order in perturbations and with at most two derivatives in the quadratic action.
The first line is to be understood as the background action, while the second one represents the quadratic expansion around the FLRW metric.
\ylg{$f$ is the (possibly time-dependent) dimensionless coefficient renormalising the Planck mass in the gravitational sector, while} $\Lambda$ and $c$ play the role of the scalar potential and kinetic energy respectively, \yld{while}\ylg{and} the various time-dependent mass coefficients encode its self and metric interactions.
\ylg{Note that this identification of $\Lambda$ and $c$ with a potential and a kinetic energy is only exact once a background trajectory $\phi(t)$ has been chosen, as in eqn.~\eqref{eq:inflationbgGR} below for inflation: off that limit, $\Lambda(t)$ remains a background quantity, while $V(\phi)$ is a field-space function, and the two coincide only along the chosen trajectory.}
In GR, \yld{the latter}\ylg{the coefficients above} take the values $M_2 = m_3 = m_4 = \tilde{m}_4 = 0$ and $f= 1$.

The least action principle yields the background equations of motion
\begin{align}
    c + \Lambda &= 3\Mpl^2\left(fH^2 + \dot{f}H\right) \equiv \rho_{\ylg{\pi}} \,, \\
    \Lambda - c &= \Mpl^2\left(2f\ddot{H} + 3fH^2 + 2\dot{f}H + \ddot{f}\right)  \equiv p_{\ylg{\pi}} \,,
\end{align}
where $H \equiv \dot{a}/a$ is the Hubble rate and $\rho_{\ylg{\pi}}$, $p_{\ylg{\pi}}$ are the \yld{dark energy}\ylg{boson's energy} density and pressure.

\yld{The EFToI is a special case of the EFToDE corresponding to a data-constrained background evolution.}
In GR inflation\ylg{ with a potential}, \yld{this translates into having}\ylg{the background parameters become}%
 \begin{subequations}
  \begin{align}
    \Lambda&=\Mpl^2 H^2(3-\varepsilon_1)=V(\phi),\\
     c&=\dot{\phi}^2/2 = \Mpl^2 H^2 \varepsilon_1,
  \end{align}
  \label{eq:inflationbgGR}
 \end{subequations}
where $\varepsilon_1 \equiv -\dot{H}/(H^2)$ is the first Hubble slow-roll parameter, small enough for long enough to ensure a sufficient number of e-folds of inflation and a quasi-scale-invariant power spectrum of perturbations \cite{PlanckInflation}.
\ylg{If we were to work with late-time dark energy models, the EFT would have a different background evolution and include matter coupled to gravity.}
For the sake of generality, we will keep general notations.

At this stage, the action suffers from a broken time diffeomorphism invariance, which can be restored by introducing a scalar field $\pi$ that transforms under time diffeomorphisms. 
This scalar field can be thought of as the Goldstone boson associated with the breaking of time diffeomorphisms.
By performing the Stueckelberg trick $t \to t + \pi(x)$, we can restore the full diffeomorphism invariance of the theory and analyse the dynamics of perturbations gauge-agnostically.
The procedure is rather involved as it also requires the modification of the space foliation used in the definition of the extrinsic curvature.
This was performed in past work, including the specific but pioneering case of inflation \cite{Cheung2008}. For our purposes we will only need the resulting linearised equations of motion known from past works.

\yld{As this work is originally motivated for the study of the inflaton, it is worth mentioning that the EFToI predates its generalisation into EFToDE and usually}\ylg{As opposed to late-time dark energy models, it is worth mentioning that the EFToI usually} takes a further step in the approximations by assuming the decoupling limit, which corresponds to neglecting the mixings with gravity and thus the metric perturbations.
This is why a great part of the literature focuses on the interactions of $\pi$ alone but we do not make this approximation here, as we want to capture the full dynamics of the system, including the mixings with gravity, which can be important in some regimes touched by e.g. stochastic inflation.

\textit{For the purpose of generality and interpretability\ylg{ beyond the decoupling limit}, \yld{we will stick to the EFToDE notations without assuming the inflationary limit unless specified otherwise}\ylg{we will use generic dark-energy EFT notations for the study of inflation. Sections~\ref{sec:beyondinflaton} and \ref{sec:beyondinflation} will discuss applications beyond the inflaton}.}

\subsubsection{An interpretable parametrisation}
Rarely used in the context of inflation, another parametrisation of the quadratic mass terms in eqn.~\eqref{eq:EFTaction} is possible and more interpretable.
The so-called $\alpha$-\textit{basis} \cite{BelliniSawicki14} uses five dimensionless and independent functions of time:
\begin{itemize}
  \item $\aK(t)$: \textit{Kineticity}, which quantifies the kinetic energy of the scalar field
  \item $\aB(t)$: \textit{Braiding}, which describes the mixing between the scalar field and the metric
  \item $\aT(t)$: \textit{Tensor speed excess}, which measures the deviation of the speed of gravitational waves from the speed of light
  \item $\aM(t)$: \textit{Planck-mass running rate}, which quantifies the rate of change of the effective Planck mass with time
  \item $\aH(t)$: \textit{Horndeski Planck-mass running}, which captures the time variation of the effective Planck mass in Horndeski theories. When set to zero, the action describes all Horndeski quadratic actions.
\end{itemize}
Analogously to the Hubble slow-roll parameters, we also introduce the notation
$$H\alpha_{X}^{(n+1)} \equiv - d_t{\ln \alpha_X^{(n)}},$$ with $\alpha_X^{(0)} = \alpha_X$ for $X = K, B, H, T, M$.
With \yld{this}\ylg{the} \ylg{$\alpha$-}basis, the quadratic action reads \cite{Gleyzes14}
\begin{equation}
    \begin{aligned}
    S^{(2)} = \int d^3x\, dt\, a^3 \frac{M^2}{2} \Bigg[ \delta K_{ij}\delta K^{ij} - \delta K^2 + (1+\alpha_T)\left(R\frac{\delta\sqrt{h}}{a^3} + \delta_2 R\right) 
+ \alpha_K H^2 \delta N^2  \\
 + 4\alpha_B H\, \delta K\, \delta N + (1+\alpha_H) R\, \delta N \Bigg],    
    \end{aligned}
\end{equation}
where $h_{ij}$ is the spatial metric, $\delta N$ is the perturbation of the lapse function, and the running Planck mass $M$ relates to $\aM$, $\aT$ and $\Mpl$ as%
\begin{subequations}
  \begin{align}
  \aM & =\frac{1}{H} \frac{\d}{\d t} \ln M^2, \\
  M^2 & = \frac{\Mpl^2f}{1+\aT},
  \end{align}
\end{subequations}
with other conversions from eqn.~\eqref{eq:EFTaction} to the $\alpha$-basis given in \cite{Gleyzes14}.
Setting $\aB=\aH=\aT=\aM=0$ and $f=1$, gives back GR, while $\aK=2\varepsilon_1$ is the inflationary limit of the kineticity.

Note that one can also go beyond this framework and include spatial or time derivatives beyond second order, in the latter case as long as they do not give rise to Ostrogradsky instabilities and ghosts.
We will comment on generalisation to those cases in Section~\ref{sec:discussion}.

\subsection{Linear cosmology of the EFT}
Along the $\pi$ boson perturbation, we write the scalar perturbations of the metric with the following convention
\begin{equation}
    \begin{aligned}
        g_{00} & = -1 - 2\Psi, \\
        g_{0i} & = 2 a^2 B_{,i} \\
              & = 2\alpha_{,i}, \\
        g_{ij} & = a^2 \Big((1 - 2\Phi) \delta_{ij}+2E_{,ij}\Big)\\
               & = \yld{a^2 \Big((1 - 2\tilde\Phi) \delta_{ij}+\Theta_{ij}\beta\Big)}\ylg{a^2(1 - 2\tilde\Phi) \delta_{ij}-2{\rm P}_{ij}\Delta\beta},
    \end{aligned}
\end{equation}
where the secondary notations are those of \cite{Gleyzes13}, and where we defined \yld{the operator $a^2\Theta_{ij}=\partial_j\partial_j-\frac{1}{3}g_{ij}\Delta$, with}\ylg{the background Laplacian} $\Delta = \partial^i\partial_i$\ylg{, and the (transverse) projection operator as}
\begin{equation}
  {\rm P}_{ij} \equiv {\partial_i \Delta^{-1} \partial_j} - \frac{1}{3}g_{ij} = a^2\Bigg( \frac{k_ik_j}{k^2}- \frac{1}{3}\delta_{ij}\Bigg).
  \label{eq:defPij}
\end{equation}
To convert the on-shell EOM tensors from the notations of \cite{Gleyzes13,Gleyzes14} to ours, we use in the following
\begin{equation}
  {\cal E}^{(1)} = \frac{1}{2}\frac{\delta S^{(2)}}{\delta \tilde{\Phi}},
  \qquad \tilde{\cal E}^{(1)} = -\frac{1}{2\Delta}\frac{\delta S^{(2)}}{\delta \beta},
  \qquad {\cal E}^{(1)}_{00} = \frac{1}{2}\frac{\delta S^{(2)}}{\delta \Psi},
   \qquad {\cal E}^{(1)}_{0i} = -\frac{a^2}{2\Delta}\partial_i\frac{\delta S^{(2)}}{\delta \alpha},
\end{equation}
where for spatial metric variations we used
\begin{equation}
  \frac{\delta {}^{(1)}g^{ij}}{\delta \tilde{\Phi}}= 2\bar{g}^{ij}, \qquad \frac{\delta {}^{(1)}g^{ij}}{\delta \beta}= -2{\rm P}^{ij}\Delta.
\end{equation}

\subsubsection{Gauge-agnostic equations}\label{sec:linEqns}

The linear equations of motion were derived in \cite{Gleyzes13} in the Newtonian Gauge.
We used those and recovered independently\footnote{
Note that Appendix D of \cite{Gleyzes13} contains typos in the display: 
$H$ factor missing on the $f''$ term of $A_{\pi}$;
the simplification of $E_{\pi}$ should yield $-6\dot{H}c$ as opposed to $+6\dot{H}c$; the $6 \Mpl^2 f'(...)$ term of $E_{\psi}$, the factor $2$ should be on $H^2$ not $H'$
} the $\alpha$ basis linear equations, as displayed in \cite{Gleyzes14} in this gauge for $ {\cal E}^{\mu}{}_{\nu}$ (up-down indices linear Einstein equations).
For use in post-inflationary studies, the literature usually presents either the Newtonian or the synchronous gauge equations. 
This work does not restrict to such choice. It is possible to leave the Newtonian gauge for a gauge-agnostic treatment using the transformation%
\begin{subequations}
    \begin{align}
        \Psi & \rightarrow \Psi - \dot{\chi}, \\
        \Phi & \rightarrow \Phi + H\chi, \\
        \pi & \rightarrow \pi - \chi\,,
    \end{align}
\end{subequations}
where $\chi=-a^2(B-\dot{E})$ rehabilitates arbitrary $B$ and $E$ perturbations of the $g_{0i}$ and $g_{ij}$ metric components.
If one is interested in the linear equations including the boson $\pi$, performing the Stueckelberg trick at this stage is easier than in the EFT action, where the frame vector itself is at stake.
This transformation yields the following gauge-agnostic linear equations of motion and their constraints for the \yld{EFToDE}\ylg{EFT action} in the $\alpha$ basis%
\begin{subequations}
  \begin{align}
    {\cal E}_{00}^{(1)} & =0, \label{eq:linEOO}\\
    {\cal E}_{0i}^{(1)} & = \partial_i Q =0,\label{eq:linEOi}\\
    {\cal E}_{ij}^{(1)} & = \frac{1}{3}{\cal E}^{(1)}a^2\delta_{ij} + \frac{3}{2}{\rm P}_{ij}\tilde{\cal E}^{(1)} = 0, \label{eq:linIJ}\\
    {\cal E}_\pi^{(1)} & = 0,\label{eq:linpi}
  \end{align}
  \label{eq:linEq}
\end{subequations}
where we defined trace and traceless parts with the projector ${\rm P}_{ij}$ defined in eqn.~\eqref{eq:defPij},
and the linear Einstein EOM lhs terms are%
\begin{subequations}
\begin{align}
  {\cal E}_{00}^{(1)} &\equiv
  2\Delta\Big(\Phi(1+\aH)+H\chi(1+\aB)+H\pi(\aH-\aB)\Big) \nonumber\\
  & -6H(1+\aB)\dot\Phi+6 H^3\varepsilon_1(1+\aB)\pi +(6\aB-\aK) H\dot\pi+H(-6-12\aB+\aK)\Psi, \label{eq:E00} \\
   Q &\equiv
  4a^2\Delta\Big(-\dot\Phi+H^2\varepsilon_1\pi+\aB H\dot\pi-(1+\aB)H\Psi\Big), \label{eq:Q} \\
  {\cal E}^{(1)} &\equiv -3\tilde{\cal E}^{(1)}+\ddot\Phi-\aB H \ddot\pi+(1+\aB)H\dot\Psi
    +(-3+2\varepsilon_1+\eta-\aM)\varepsilon_1 H^3\pi+(3+\aM)H\dot\Phi \nonumber\\
  &+\bigg((3-\varepsilon_1+\aM)(1+\aB)-\aB^{(1)}\aB\bigg)H^2\Psi
      +\bigg(\varepsilon_1(\aB-1)-\aB(3+\aM-\aB^{(1)})\bigg)H^2\dot\pi, \label{eq:calE}\\
  \tilde{\cal E}^{(1)} &\equiv
  \frac{2}{3}\Delta\Big(\dot\chi+\aH\dot\pi-(1+\aH)\Psi+(1+\aT)\Phi
    +(1+\aM)H\chi+(\aT-\aM)H\pi\Big).
  \label{eq:tildeE}
\end{align}  \label{eq:lineq2}
\end{subequations}
Note that, by orthogonality, ${\cal E}^{(1)}$ and   $\tilde{\cal E}^{(1)}$ are both null on-shell and the latter could be injected in the expression of the former in eqn.~\eqref{eq:calE}.
However, we keep it this way as the coarse-grained theory will not be on-shell, as explained in Section~\ref{sec:ansatz}.

More involved, the linear EOM of $\pi$ is given by
\begin{align}
 {\cal E}_\pi^{(1)} = &
  -12\varepsilon_1\Big(\varepsilon_1(1+3\aB)+\aB\big(-3+\eta-\aM+\aB^{(1)}\big)\Big)H^4\pi
  \nonumber\\&
  -12\big(3\aB+\aB(\aM-\aB^{(1)})-\varepsilon_1(1+\aB)\big)H^2\dot\Phi -12\aB H\ddot\Phi
  \nonumber\\&
  -2(6\aB-\aK)H^2\dot\Psi -2\aK\Big(3+\aM-2\varepsilon_1-\aK^{(1)}\Big)H^3\dot\pi -2\aK H^2\ddot\pi
 \nonumber \\&
  -2\Big((6\aB-\aK)(3+\aM)-2\varepsilon_1(3+9\aB-\aK)+\aK\aK^{(1)}-6\aB\aB^{(1)}\Big)H^3\Psi
  \nonumber\\&
  +4\Delta\Bigg(\Big(\aB(1+\aM-\aB^{(1)})-\varepsilon_1(1+\aB)\Big)H^2\chi + \aH \dot\Phi\\&\qquad
   \ylg{+}\aB H\dot\chi+(\aH-\aB)H\Psi+ \big(\aM+\aH(1+\aM-\aH^{(1)})-\aT\big)H\Phi \nonumber\\&\quad
  +\Big(\varepsilon_1(1+\aB-\aH)+\aM
    -\aB(1+\aM-\aB^{(1)})+\aH(1+\aM-\aH^{(1)})-\aT\Big)H^2\pi \Bigg).\nonumber
  \label{eq:Epi}
\end{align}
We are not interested in the hydrodynamical equations, and so do not rewrite the $\dot{\rho}$ evolution equation found in \cite{Gleyzes13,Gleyzes14}.
A related discussion on potential extensions is led in Section~\ref{sec:discussion}. The same remark is valid for vector modes.
Spin-2 linear equations are easier to write than the scalar ones as the gauge-invariance of the tensorial perturbations means that the linear contribution to the Einstein equations is proportional to the polarised Mukhanov-Sasaki equations, similarly to GR as seen in Section~\ref{sec:stogravitons}.
We thus focus on scalars until Section~\ref{sec:stochgrav} without loss of generality.

\subsubsection{Curvature perturbation dynamics}
Despite the complexity of eqn.~\eqref{eq:linEq} and \eqref{eq:lineq2}, a single gauge-invariant perturbation can summarise the linear scalar dynamics,
as we recall below using the curvature perturbation on comoving hypersurfaces
\begin{equation}
{\cal R} \equiv \Phi+{H}\pi.
\end{equation} 
Define the field $z$ as 
\begin{equation}
  z^2 \equiv\frac{ a^3 M^2\abar}{(1+\aB)^2} = \frac{a^3 \Mpl^2 f\,\abar}{(1+\aT)(1+\aB)^2}\,,
\end{equation}
and the nonlinear combination $\abar \equiv \aK + 6\aB^2$. The logarithmic cosmic time derivative is
\begin{equation}
 2\dot{\overbrace{\ln{z}}} = H(3+\aM)
  - H\frac{2(6\aB-\aK)\aB\aB^{(1)} + (1+\aB)\aK\aK^{(1)}}{(1+\aB)\abar}.
\end{equation}
A cosmologist would notably find the $-H\varepsilon_2$ term of the GR Mukhanov-Sasaki equation back in $\aK^{(1)}$. 
Define now the sound speed
\begin{equation}
  c_s^2 \equiv -2\frac{(1+\aB)^2}{\bar\alpha}\left(
    1+\aT
    - \frac{1+\aH}{1+\aB}\left(
      1 + \aM +\varepsilon_1
      +\frac{\aB\aB^{(1)}}{1+\aB}-\frac{\aH\aH^{(1)}}{1+\aH}
    \right)
  \right).
\end{equation}
The Mukhanov-Sasaki equation\ylg{ \cite{Mukhanov1981,Sasaki83}} can be found by rewriting the \yld{EFToDE}\ylg{EFT} action or the linear equations in the unitary gauge, giving \cite{Gleyzes13}
\begin{equation}
  \ddot{\R} +  2\dot{\overbrace{\ln{z}}}\,\dot{\R} + \frac{c_s^2 k^2}{a^2}\,\R = 0.
  \label{eq:MS}
\end{equation}
For the case of inflation, and similarly to GR, these modes \yld{can}\ylg{could} be given Bunch-Davies initial conditions in the sub-Hubble regime.
\ylg{This is only one initial-condition choice, and the present formalism remains valid as long as $\R$ remains perturbative; it does not require the modes to freeze, since their evolution simply continues in the IR nonlinearly.}

The value of a ${\cal R}$ mode at (chosen) crossing to the IR has a direct impact on the amplitude of the stochastic sources.
Let us look at its value in the case of inflation. By defining $Q = z{\cal R}$, we can rewrite eqn.~\eqref{eq:MS} in conformal time $\tau$ as
\begin{equation}
  Q_k''+\Big(k^2c_s^2-\frac{z''}{z}\Big)Q_k=0,
\end{equation}
and assume slowly-varying Hubble and EFT rolling parameters at crossing ($-kc_s\tau=1$), so that the solution is analytically known as
\begin{equation}
  Q_k \yld{\propto}\ylg{= A_k}  (-kc_s\tau)^{1/2}H_{\nu_s}^{(1)}(-kc_s\tau)\ylg{+ B_k(-kc_s\tau)^{1/2}H_{\nu_s}^{(2)}(-kc_s\tau)},
  \label{eq:Qsol}
\end{equation}
where $\nu_s^2 = \frac{1}{4}+\frac{2-\varepsilon_1+3\alpha_M/2}{(1-\varepsilon_1)^2}$\yld{, using Bunch-Davies initial conditions in the deep sub-Hubble regime}.
\ylg{Bunch--Davies initial conditions in the deep sub-Hubble regime, $Q_k \to e^{-ic_sk\tau}/\sqrt{2c_sk}$ for $-kc_s\tau\to\infty$, give $B_k=0$ and $A_k = \frac{\sqrt{\pi}}{2}e^{i(\nu_s+1/2)\pi/2}/\sqrt{c_sk}$.
This is where the quantum input enters: quantisation imposes $|A_k|^2-|B_k|^2=\pi/(4c_sk)$, which fixes the amplitude of the sources. Other initial states, e.g. $\alpha$-vacua ($B_k\neq0$) \cite{Mottola85, Bruce85} or classical initial conditions, would only change $A_k$ and $B_k$; the rest of the procedure does not require quantum initial conditions but non-zero Fourier modes for ${\cal R}$.}
\yld{We}\ylg{For the Bunch-Davies example, we} provide a visual comparison of $Q(-kc_s\tau=1)$ with and without running of the Planck mass in Figure~\ref{fig:proxyQ}.
A few percent deviation can be reached with a reasonably perturbative $\aM$ value, translating into a similar difference per e-fold in the amplitude of the stochastics we are about to compute.

\begin{figure}
  \centering
  \includegraphics[width=0.57\textwidth]{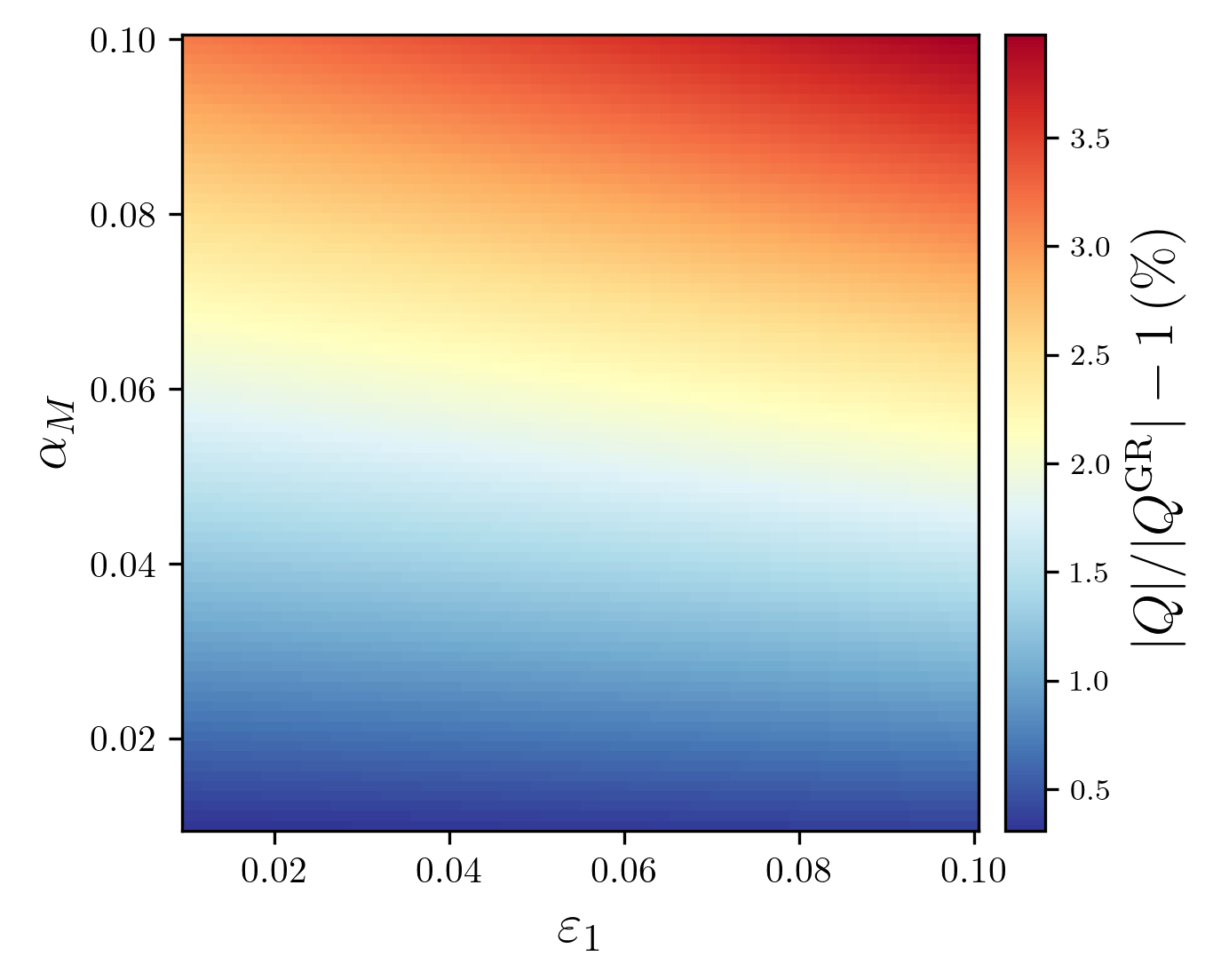}
  \caption{Relative departure of the inflationary Mukhanov-Sasaki variable from its GR value at horizon crossing ($-kc_s\tau=1$) as a function of the Planck mass running $\aM$ and the first Hubble slow-roll parameter $\varepsilon_1$. All Hubble and EFT parameters are assumed slowly-varying.}
  \label{fig:proxyQ}
\end{figure}

\subsubsection{Perturbations as functionals of ${\cal R}$}\label{sec:slicingsOfR}

Following the procedure developed in Chapter~\ref{sec:SIGR} before performing the coarse-graining, we express the gauge invariant combinations in terms of $\R$. 
These are easily obtained in the comoving gauge where $\pi = E = 0$ and where $\Phi = \R$ by definition. 

In the comoving gauge the momentum constraint eqn.~\eqref{eq:linEOi} reduces to solving
$\dot{\R} + \Psi H(1+\aB) = 0$, giving
\begin{equation}
    {\Psi^{\rm co} = -\frac{1}{H(1+\aB)}\dot{\R}\,.}
\end{equation}
Substituting the above into the Hamiltonian constraint eqn.~\eqref{eq:linEOO} yields
\begin{equation}
  {\chi^{\rm co} = -\frac{1}{2(1+\aB)^2}\left(
    \frac{2(1+\aB)(1+\aH)}{H}\R
    + \frac{\, a^2\abar}{k^2}\dot{\R}
  \right),}
\end{equation}
which here equates to $-a^2B^{\rm co}$. The energy density perturbation is also found as $\delta\rho^{\rm co} = -2c \Psi^{\rm co}[\R]$.
We can now evaluate the common gauge invariants used in cosmology, as functionals of $\R$%
\begin{subequations}
\begin{align}
  \Phi_B     &\equiv \Phi + H\chi = \R + H\chi^{\rm co}[\R]\,, \\
  \Psi_B     &\equiv \Psi - \dot\chi = \Psi^{\rm co}[\R] - \dot\chi^{\rm co}[\R]\,, \\
  \delta\phi^{\rm gi} &\equiv \frac{\pi}{\sqrt{2c}} - \sqrt{2c}\chi = -\sqrt{2c}\chi^{\rm co}[\R]\,, \\
  \zeta^{\rm gi}  &\equiv -\Phi + \frac{1}{3}\frac{\delta\rho}{\rho_{\ylg{\pi}}+p_{\ylg{\pi}}} = -\R + \frac{1}{3}\frac{\delta\rho^{\rm co}[\R]}{\rho_{\ylg{\pi}}+p_{\ylg{\pi}}}\,.
\end{align}\label{eq:GIs}
\end{subequations}
where $\rho_{\ylg{\pi}}+p_{\ylg{\pi}} = 2c + M^2(\ddot f - H\dot f)$ is the effective \yld{dark}\ylg{boson} energy density plus pressure, and where we referred to the derived comoving gauge functionals we just derived explicitly.
As shown in Chapter~\ref{sec:SIGR}, knowing these as functionals of $\R$ is particularly useful to get all gauges expressed as functionals of $\R$.
In particular in the relevant gauges, and as Chapters~\ref{sec:ICNR} and \ref{sec:ICsim} achieved in GR, this provides physically-motivated cosmological initial conditions for numerical relativity in modified gravity, without relying on arbitrary initial perturbations and constraint solvers \cite{brady_inflaton_2025}.

Finally, one can compute the gravitational slip as%
\begin{subequations}
  \begin{align}
  \Phi_B - \Psi_B & = \R + H\chi^{\rm co}[\R] - \Psi^{\rm co}[\R] + \dot\chi^{\rm co}[\R]\,. \\
& =  \Bigg( 1 -\frac{1+\aH}{1+\aB}\Big(1+\varepsilon_1+
      \frac{\aB\aB^{(1)}}{1+\aB}-\frac{\aH\aH^{(1)}}{1+\aH}
    \Big) \Bigg) \R \nonumber\\
& +  \Bigg( \frac{ \abar\Big( \aM-2\dot{\overbrace{\ln{z}}} \Big)}{2 (1+\aB)^2}\Big(\frac{k}{aH}\Big)^{-2} - \frac{ \aH}{1+\aB} \Bigg)\frac{\dot \R}{H} - \frac{\abar}{2 (1+\alpha_B)^2} \Big(\frac{k}{aH}\Big)^{-2} \frac{\ddot \R}{H^2}\,\\
& =  \Big(\frac{1+\alpha_H}{1+\alpha_B}\alpha_M-\alpha_T\Big)\R +  \Bigg(
\frac{ \abar\aM}{2 (1+\aB)^2}\Big(\frac{k}{aH}\Big)^{-2}- \frac{\alpha_H}{(1+\alpha_B)}
\Bigg)\frac{\dot \R}{H}\, .
\end{align}
\end{subequations}
In the second line, we injected the Mukhanov-Sasaki equation, eqn.~\eqref{eq:MS}. In GR, this becomes clearly $0$ as expected, but not in the general case.
We have argued in Chapter~\ref{sec:SIGR} that it is important not to inject the Mukhanov-Sasaki equation in the expression of the gravitational slip or any EOMs before coarse-graining.
In fact, the coarse-grained gravitational slip is non-zero, even in GR, which is why it is important to keep track of both expressions above.

\subsection{Sources to all EOMs}\label{sec:stoEOMs}
Now that we have identified the true underlying dynamical field ($\cal R$) in the EOMs, we have everything we need to proceed with the coarse-graining of the linear theory as described in Section~\ref{sec:ansatz}.
We focus first on the scalar sector, which is the most involved, and then on the tensorial one.

\subsubsection{Coarse-graining}
As explained in Section~\ref{sec:ansatz}, the coarse-graining procedure consists in seeing the change of the linear equations of motion of a field $X$ under the transformation
\begin{equation}
  X(t,\mathbf{k}) \;\mapsto\; X(t,\mathbf{k})\,W(t,\mathbf{k})\,,
  \label{eq:coarsegrainmap}
\end{equation}
where $W$ is a window function, before promoting the result to be the spectrum of the stochastic source to the nonlinear EOMs.
However, in full GR, and despite having a single scalar degree of freedom, we saw that multiple scalar perturbations are at play with no clear prescription for their coarse-graining as a group, unless using the coarse-graining $X = {\cal R}$ after having expressed all the gauge-invariant combinations as functionals of $\R$ in the previous section.

In fact, eqns.~\eqref{eq:GIs} allow us to write any field in any gauge as a functional of $\R$. 
If one picks a gauge-dependent perturbation without specifying its value in a given gauge, say $\pi$, then all other gauge-dependent quantities can once again be expressed as functionals of $\pi$ and $\R$%
\begin{subequations}
  \begin{align}
    \chi[\R,\pi]  & = \chi^{\rm co}[\R]+\frac{\pi}{2c},\\
    \Phi[\R,\pi]  & = \R + H\chi^{\rm co}[\R] - H\chi[\R,\pi],\\
    \Psi[\R,\pi]  & = \Psi^{\rm co}[\R]+\dot{\chi}[\R,\pi]-\dot{\chi}^{\rm co}[\R] .
  \end{align}
\end{subequations}
Substituting these three functionals in the linear equations of motion, eqn.~\eqref{eq:linEq} and eqn.~\eqref{eq:linpi}, will \yld{simplify $\pi$ out}\ylg{eliminate $\pi$} and give the Mukhanov-Sasaki equations up to different factors, except for the constraints which will be automatically satisfied by definition and as non-dynamical equations. 
As we are about to see, substituting the coarse-grained functionals
\begin{equation}
  \chi^>\equiv\chi[W\R,\pi^{\ylg{>}}], \qquad \Phi^>\equiv\Phi[W\R,\pi^{\ylg{>}}], \qquad \Psi^>\equiv\Psi[W\R,\pi^{\ylg{>}}],
  \label{eq:CGfuncs}
\end{equation}
\ylg{where we have applied the Fourier coarse-graining defined in eqn.~\eqref{eq:coarsegrainmap} to $X=\R$,} might not give these results, as we are about to see beyond GR.

\paragraph{Mukhanov-Sasaki equation.}
Even as a linear EOM, the Mukhanov-Sasaki equation itself can be coarse-grained and will appear again in the simplification of the sources to the full EOMs.
In Fourier space, ${\cal R}^>(t,\mathbf{k}) = W(t,\mathbf{k}){\cal R}(t,\mathbf{k})$ satisfies the following equation
\begin{equation}
  \ddot{\overbrace{\R^>}} +  2\dot{\overbrace{\ln{z}}}\,\dot{\overbrace{\R^>}}+ \frac{c_s^2 k^2}{a^2}\,\R^> = \Src_{\cal R}^{\rm EFT},
\end{equation}
after simplifying using the MS equation satisfied by ${\cal R}^>$ so that the source Fourier amplitude $ \Src_{\cal R}^{\rm EFT}$ takes the form
\begin{align}
  \Src_{\cal R}^{\rm EFT} &=  \R\ddot W + \Big(2\dot \R+2\dot{\overbrace{\ln{z}}}\,\R\Big)\dot W\,. \label{eq:SReft}
\end{align}
The result in canonical inflation in GR of Chapter~\ref{sec:SIGR}, noted $\Src_{\cal R}^{\rm GR}$, is recovered in this limit where $ 2\dot{\overbrace{\ln{z}}} = H(3-\varepsilon_2)$ and $c_s=1$. This is particularly visible when writing
\begin{equation}
  \Src_{\cal R}^{\rm EFT} = \Src_{\cal R}^{\rm GR}+\gamma\dot W\R,
\end{equation}
with $\gamma$ a function of the $\alpha$ parameters that vanishes in the GR limit and defined as

\begin{equation}
  \gamma \equiv 2\dot{\overbrace{\ln{z}}}-H(3-\varepsilon_2) =  H\Bigg(\varepsilon_2+\aM -\frac{ 2(6\aB-\aK)\aB\aB^{(1)}}{\abar(1+\aB)}-\frac{\aK\aK^{(1)}}{\abar}\Bigg) \,.
\end{equation}

\paragraph{Constraint IR equations.}\label{sec:cons}
We now plug the coarse-grained and gauge-agnostic perturbation functionals $\chi^>,\Phi^>,\Psi^>$ defined in eqn.~\eqref{eq:CGfuncs}, in the Hamiltonian and momentum constraints, eqn.~\eqref{eq:linEOO} and eqn.~\eqref{eq:linEOi}, and call $\Src_{\cal H}$ and $\Src_{\cal M}$ the results in Fourier space, after simplifying using the MS equation satisfied by $\R$.
We find that these source spectra vanish identically
\begin{equation}
  \Src_{\cal H} = 0\,,\qquad \Src_{\cal M} = 0\,.
\end{equation}
As shown in Chapter~\ref{sec:SIGR}, this is the benefit of using the gauge-agnostic coarse-graining.
We should stress once again that no previous work in the stochastic inflation literature, apart from the present one, showed explicit satisfaction of the coarse-grained constraints, which is essential to introduce a physically-balanced stochastic kick at each time step.
At best, the local Friedmann equation corresponding to the Hamiltonian in the separate universe approximation (0th order gradient expansion) would be enforced in most works, and the momentum equation has been used to prepare initial data at the next order in the gradient expansion \cite{prokopec_ensuremathdeltan_2021}.
Some works, including in the context of higher curvature corrections, have found a non-zero source to the constraints (see eqns. (61) and (64) of \cite{aldabergenov_towards_2025}). 
Based on our own previous attempts, we believe that this might be a consequence of coarse-graining in a specific gauge inconsistently (flat gauge in the mentioned work) or simplifying with the MS equation before coarse-graining.

\paragraph{Source to IR scalar.} \yld{Leaving the constraints and the MS equation yields much more complexity at first sight as we get the source Fourier amplitude for eqn.~\eqref{eq:linpi} to be, after substituting the functionals of $W\R$ and $\pi$, simplifying $\pi$ and the MS equation in $\R$,}\ylg{Moving beyond the constraints and the MS equation yields much more complexity at first sight: after substituting the functionals of $W\R$ and $\pi^>$ and simplifying $\pi^>$ and the MS equation in $\R$, we get the source Fourier amplitude for eqn.~\eqref{eq:linpi} to be}
\begin{multline}
  \Src_\pi^{\rm EFT}
  = -\frac{2 H}{(1+\aB)^3}\Bigg(
    \R\ddot W\,(1+\aB)\abar
    + \R\dot W\,H(1+\aB)\abar(3+\aM) \\
    + \dot W\Bigl(2\dot \R\,(1+\aB)\abar
      -H \R\bigl[2(6\aB-\aK)\aB\aB^{(1)} + (1+\aB)\aK\aK^{(1)}\bigr]\Bigr)
  \Bigg)\,.
\end{multline}
However, this compactifies to
\begin{equation}
  \Src_\pi^{\rm EFT}
  = -\frac{2 H \abar}{(1+\aB)^2} \Src_{\cal R}^{\rm EFT},
\end{equation}
after identifying the source Fourier amplitude of the MS equation found in eqn.~\eqref{eq:SReft}.
In the GR limit, this can be written as
\begin{equation}
  \Src_\pi^{\rm GR}
  = -4 H \varepsilon_1 \Src_{\cal R}^{\rm GR}.
\end{equation}
We recover the ADM field equation source Fourier amplitude of Chapter~\ref{sec:SIGR} using $\delta\phi = \sqrt{2\varepsilon_1}H\Mpl\, \pi$ at linear order\footnote{The minus sign comes from having $-\ddot \pi$ in the least action principle as opposed to the ADM equation.}
\begin{equation}
  \Src_\Pi^{\rm GR}
  = \sqrt{2\varepsilon_1} \Mpl \Src_{\cal R}^{\rm GR}.
\end{equation}
Beyond GR, the appropriate field redefinition to recover the same form is $\delta\phi = \sqrt{2c}\, \pi = \dot\phi_b\,\pi$, which can be applied to find the source to ${\cal E}_{\phi}$ as
\begin{equation}
  \Src_\phi^{\rm EFT}
  = -\frac{2 \dot\phi_b H \abar}{(1+\aB)^2} \Src_{\cal R}^{\rm EFT}.
\end{equation}

\paragraph{Source to IR Geometry.} With the same method, the sources to $\cE_{ij}$ are, as observed in the GR case, 
proportional to the MS source too and factorise as
\begin{align}
  \Src_{\cal E}^{\rm EFT}
  & = -\frac{\abar}{(1+\aB)^2}\Src_{\cal R}^{\rm EFT},\\
    \Src_{\tilde{\cal E}}^{\rm EFT}
  & =  \frac{\abar}{3(1+\aB)^2}\Src_{\cal R}^{\rm EFT},
\end{align}
\ylg{which combine, analogously to the on-shell trace-traceless split of eqn.~\eqref{eq:linIJ}, into the spatial source
\begin{equation}
  \Src_{{\cal E}_{ij}}^{\rm EFT} = \frac{1}{3}\Src_{\cal E}^{\rm EFT}\,a^2\delta_{ij} + \frac{3}{2}{\rm P}_{ij}\,\Src_{\tilde{\cal E}}^{\rm EFT}\,,
  \label{eq:EijTraceSplit}
\end{equation}
}meaning\ylg{, after substituting the two expressions above and using $\tilde{\rm P}_{ij}$ defined below,} that the total source to the Einstein tensor is
\begin{align}
  \Src_{{\cal E}_{ij}}^{\rm EFT} =\frac{\abar}{2(1+\aB)^2} \tilde{\rm P}_{ij}\Src_{\cal R}^{\rm EFT},
\end{align}
where we defined the operator
\begin{equation}
\tilde{\rm P}_{ij} = \yld{\Delta\Theta_{ij}=} {\partial_i \Delta^{-1} \partial_j}- \bar{g}_{ij} = a^2\Big( \frac{k_ik_j}{k^2}- \delta_{ij}\Big).
\end{equation}
We thus recover elegantly the GR limit found in Chapter~\ref{sec:SIGR}.

\subsubsection{Stochastic gravitons}\label{sec:stochgrav} 
Linearising in the traceless and divergence-free part of the spatial metric as $g_{ij} = a^2(\delta_{ij} + h_{ij})$, the gravitonic \yld{EFToDE}\ylg{EFT} action is \cite{Gleyzes13,Gleyzes14}
\begin{equation}
  S_h = \frac{1}{2}\int d^4x \frac{a^3 M^2}{4}\left[
    \dot h_{ij}^2 - \frac{c_T^2}{a^2}(\partial_k h_{ij})^2
  \right],
\end{equation}
where the tensor speed can have an excess to celerity $c_T^2 = 1+\aT$.
Hence, using the same approach as for the scalar case and using $z_h^2=\frac{\Mpl^2a^3 f}{4(1+\aT)}$, the effective gravitonic Mukhanov-Sasaki equation can be written for the two Fourier IR polarisation modes $h^>$ as
\begin{equation}
  \ddot{\overbrace{h^>_k}} +  2\dot{\overbrace{\ln{z_h}}}\,\dot{\overbrace{h^>_k}} + \frac{c_T^2 k^2}{a^2}h_k = \Src_{h}^{\rm EFT},
\end{equation}
where the source Fourier amplitude $ \Src_{h}^{\rm EFT}$ takes the form
\begin{align}
  \Src_{h}^{\rm EFT} &=  h\ddot W + \big(2\dot h+2\dot{\overbrace{\ln{z_h}}}\,h\big)\dot W.
\end{align}
The GR result, noted $\Src_{h}^{\rm GR}$, is recovered in the GR limit where $ 2\dot{\overbrace{\ln{z_h}}} = 3H$ and $c_T=1$. This is particularly visible when writing
\begin{equation}
  \Src_{h}^{\rm EFT} = \Src_{h}^{\rm GR}+\gamma_h\dot W h,
\end{equation}
with $\gamma_h$ a function of the $\alpha$ parameters that vanishes in the GR limit and defined as
\begin{equation}
  \gamma_h \equiv 2\dot{\overbrace{\ln{z_h}}}-3H = \frac{\d}{\d t} \ln \Big(\frac{f}{1+\aT}\Big)\,.
\end{equation}
 At linear order, the stochastic gravitons only contribute to the $\cE_{ij}$ equation in its traceless-transverse part.
This $\Src_{h}^{\rm EFT}$ traceless-transverse source is thus to be added to the r.h.s. of the ${\cal E}_{ij}$ equation only, with a $\frac{a^2}{2}$ factor, summed over the two polarisation modes $s=+,\times$.

\subsubsection{Stochastic scalar-tensor equations}\label{sec:mainresult}
  We now collect the previous findings in a shared formalism.
  Within the scope of validity of the stochastic ansatz, the IR relativistic equations of motion of a scalar-tensor theory with action $S$ \yld{and a matching $\{\alpha\}$ to the EFToDE}\ylg{and matching EFT coefficients $\{\alpha\}$} write%
  \begin{subequations}
    \begin{align}
       {\cal E}_{\phi}[g^>, \pi^>] = & - \frac{2\dot\phi_bH\abar}{(1+\aB)^2}  {\cal F}^{-1}\Big\{\Src_{\cal R}^{\rm EFT}\boldsymbol{\xi}^{(0)}_{\mathbf{k}}\Big\},\\
      {\cal E}_{\mu\nu}[g^>, \pi^>]
          = & \frac{\abar}{2(1+\aB)^2}  {\cal F}^{-1}\Bigg\{
          \begin{pmatrix}
            0 & \boldsymbol{0} \\
            \boldsymbol{0} & \tilde{\rm P}
          \end{pmatrix} \Src_{\cal R}^{\rm EFT}\boldsymbol{\xi}^{(0)}_{\mathbf{k}}\Bigg\}\nonumber\\
          & + \frac{1}{2}a^2\displaystyle\sum_{s=+,\times}
             {\cal F}^{-1}\Bigg\{
              \begin{pmatrix}
            0 & \boldsymbol{0} \\
            \boldsymbol{0} & \varepsilon^{(s)}(\mathbf{k})
          \end{pmatrix} \Src_h^{\rm EFT}\boldsymbol{\xi}^{(s)}_{\mathbf{k}}\Bigg\},
    \end{align}\label{eq:main}
  \end{subequations}
  where $\forall s=0,+,\times$,
  \begin{equation}
    \langle \boldsymbol{\xi_{\mathbf{k}_1}}^{s_1}\boldsymbol{\xi_{\mathbf{k}_2}}^{s_2*} \rangle_{\mathbb{P}}  =  \delta^{(3)}(\mathbf{k}_1-\mathbf{k}_2)\delta^{s_1s_2}.
  \end{equation}
It is important to understand that the spin-$0$ stochastic grid $\boldsymbol{\xi}^{(0)}_{\mathbf{k}}$ sees the same realisation in both the scalar field and geometry equations to ensure the displayed satisfaction of the constraints.
After coarse-graining, the original lhs of the (dynamical) EOMs have thus acquired the following power spectra due to their sources in the IR%
\begin{subequations}
\begin{align}
    \langle {\cal E}_{\pi}(\mathbf{k}_1) {\cal E}_{\pi}^*(\mathbf{k}_2) \rangle_{\mathbb{P}} & = \frac{4H^2\abar^2}{(1+\aB)^4}\delta^{(3)}(\mathbf{k}_1-\mathbf{k}_2)|{\cal S}_{\cal R}^{\rm EFT}(k_1)|^2,\\
    \langle {\cal E}_{ij}(\mathbf{k}_1) {\cal E}_{kl}^*(\mathbf{k}_2) \rangle_{\mathbb{P}} & = \frac{ \abar^2}{4(1+\aB)^4}\delta^{(3)}(\mathbf{k}_1-\mathbf{k}_2)\tilde{\rm P}_{ij}\tilde{\rm P}_{kl} |{\cal S}_{\cal R}^{\rm EFT}(k_1)|^2 \notag\\
                                                                                      & +\frac{1}{4}a^4\delta^{(3)}(\mathbf{k}_1-\mathbf{k}_2)\displaystyle\sum_{s=+,\times}\varepsilon^{s}_{ij}(\mathbf{k}_1)\varepsilon^{s}_{kl}(\mathbf{k}_1)|{\cal S}_h^{\rm EFT}(k_1)|^2 \notag \\
                                & = \frac{ \abar^2}{4(1+\aB)^4}\delta^{(3)}(\mathbf{k}_1-\mathbf{k}_2)\tilde{\rm P}_{ij}\tilde{\rm P}_{kl} |{\cal S}_{\cal R}^{\rm EFT}(k_1)|^2\notag\\
                                & + \frac{1}{4}\delta^{(3)}(\mathbf{k}_1-\mathbf{k}_2) \left(\tilde{\rm P}_{ik}\tilde{\rm P}_{jl}+\tilde{\rm P}_{il}\tilde{\rm P}_{jk}-\tilde{\rm P}_{ij}\tilde{\rm P}_{kl} \right)|{\cal S}_h^{\rm EFT}(k_1)|^2.\
\end{align}
\end{subequations}
In the following, we condense the notations of eqn.~\eqref{eq:main} by writing the spin-0 and spin-2 sources
\begin{equation}
  \begin{aligned}
     \boldsymbol{{\cal S}}^{(0)} & = {\cal F}^{-1}\{ \Src_{{\cal R}}^{\rm EFT}\boldsymbol{\xi}^{(0)}_{\mathbf{k}}\}, \\
     \boldsymbol{{\cal S}}_{\mu\nu}^{(2)} & = \frac{1}{2}a^2\displaystyle\sum_{s=+,\times} {\cal F}^{-1}\{\varepsilon^{(s)}_{\mu\nu}(\mathbf{k}) \Src_h^{\rm EFT}\boldsymbol{\xi}^{(s)}_{\mathbf{k}}\},
  \end{aligned}
\end{equation}
together with the convention $\tilde{\rm P}_{00} = \tilde{\rm P}_{0i}= \tilde{\rm P}_{i0} = 0$, and similarly for the polarisation tensors.
Furthermore, we also drop the superscript $^>$ notation of IR fields and keep it implicit.
\section{Examples in motion}\label{sec:EOMs}

To illustrate the potential and generality of what has been shown in the previous section, we pick here historical extensions to General Relativity and find their associated stochastic EOMs for inflation, simply knowing their quadratic action.
For comparison, only a few examples have been incorporated into the stochastic framework to our knowledge \cite{garcia-bellido_jordan-brans-dicke_1994, aldabergenov_towards_2025}, all in the long wavelength limit. We leave the study of this regime and its validity to future work.

A large fraction of these theories can be mapped to the Einstein frame (some of which in the next subsection below), and so, with a singular scalar at stake, can be studied within stochastic inflation in GR, using simply a modified potential.
However, this is not the case for an arbitrary scalar-tensor theory.
For the theories which can be recast to GR, it is still important to master the stochastic formalism as, generically, going beyond single-field phenomenology would not make that possible.

Attached to these examples, we provide Tables~\ref{tab:bgMatch} and \ref{tab:alphaMatch}, which summarise the background and linear matching to the \yld{EFToDE}\ylg{EFT action} parameters, which are the only ingredients needed to find the stochastic EOMs of a theory within our formalism.

\subsection{Stochastic Einstein-Gauss-Bonnet dynamics}\label{sec:EGB}
\textit{The results of this section were obtained using both the procedure specific to the EGB equations and the procedure matching to the \yld{EFToDE}\ylg{EFT action} sources. }

The Einstein-Gauss-Bonnet (EGB) theory is a higher-order scalar-tensor theory that includes a coupling between a scalar field and the Gauss-Bonnet invariant.
It is a well-studied model and has been shown to have interesting phenomenological implications, including in the context of inflation \cite{fernandes_4d_2022,brady_inflaton_2025},
notably in the bias to the effective potential's slope enabling GB-induced ultra slow-roll and so its own Primordial Black Hole production channel \cite{Shinsuke21,aldabergenov_towards_2025}.

Let us write the EGB action as
\begin{equation}
  \Src_{\rm EGB} = \int d^4x\sqrt{-g}\left(
    \frac{\Mpl^2}{2}R
    - \frac{1}{2}\nabla^\mu\phi\nabla_\mu\phi - V(\phi)
    -\frac{1}{8} \lambda(\phi)\,\mathcal{G}
  \right),
\end{equation}
where $\mathcal{G} = R^2 - 4R_{\mu\nu}R^{\mu\nu} + R_{\mu\nu\rho\sigma}R^{\mu\nu\rho\sigma}$
is the Gauss-Bonnet density and $\lambda(\phi)$ is a dimensionless coupling function.
Let us next define the GB-induced energy-momentum tensor, to be compared to the GR one
 \begin{equation} 
\begin{aligned}
T^{\mathrm{GB}}_{\mu\nu} & =
 R_{\mu\rho\nu\sigma} \, \nabla^\rho \nabla^\sigma \lambda
- (\square \lambda) R_{\mu\nu} + 2\, \nabla_\rho \nabla_{(\mu} \lambda \, R^\rho_{\ \nu)} \\
&- \frac{1}{2} (\nabla_\mu \nabla_\nu \lambda)\, R - \frac{1}{2} \left(
2\, \nabla_\rho \nabla_\sigma \lambda \, R^{\rho\sigma}
- (\square \lambda)\, R
\right) g_{\mu\nu}, \\
T_{\mu\nu}^{\phi} & =
\nabla_\mu \phi \, \nabla_\nu \phi - \left(V(\phi)+ \frac{1}{2} \nabla^\rho \phi \, \nabla_\rho \phi \right)g_{\mu\nu}, 
\end{aligned}
\end{equation}
with which we can write the nonlinear EGB equations of motion as
\begin{equation}
  \begin{aligned}
  G_{\mu\nu} - \frac{1}{\Mpl^2}\Big(T^{GB}_{\mu\nu} + T^{\phi}_{\mu\nu}\Big) &=0, \\
    \nabla_\mu\nabla^\mu\phi -\frac{1}{8}\lambda_{,\phi}(\phi){\cal G} - V_{,\phi}(\phi) & = 0. \\
  \end{aligned}
\end{equation}
The 00, ii, and inflaton l.h.s. background equations are respectively given by 
\begin{align}
  3 H^2(1-w) - \frac{1}{\Mpl^2}\bigg(\frac{1}{2}\dot\phi^2_b + V(\phi_b)\bigg)& =0\,, \\
 a^2 H^2\Big(-3+2\varepsilon_1+w(2- \varepsilon_1+\sigma)\Big)   - \frac{a^2}{\Mpl^2}\bigg(\frac{1}{2}\dot\phi^2_b - V(\phi_b)\bigg) & = 0,\\
  -\ddot\phi_b -3H\dot\phi_b -V_{,\phi}(\phi_b) - 3w(1-\varepsilon_1)\frac{H^3\Mpl^2}{\dot\phi_b}&=0,
\end{align}
where we have defined the dimensionless coupling parameters
\begin{equation}
    w  \equiv H \Mpl^{-2} \dot \lambda\,,\qquad
    \sigma  \equiv H^{-1}d_t\ln w\,.
\end{equation}
These notably remind us that the GR relation between the kinetics of the scalar and of the expansion is rather simple compared to here where,
 by addition, \begin{equation}\dot\phi_b^2 = \Big(2\varepsilon_1-w(1+\varepsilon_1-\sigma)\Big)\Mpl^2 H^2.  \end{equation}

The linear theory follows quickly using the \yld{EFToDE}\ylg{EFT action} equations with the following identification of the $\alpha$ parameters to plug in the previous eqn.~\eqref{eq:linEq}
\begin{equation} \Big(\aK, \aH, \aB, \aT, \aM \Big) = \Bigg( \frac{\dot\phi_b^2}{(1-w)\Mpl^2H^2},\, 0,\, \frac{w}{2(w-1)},\, -2 \aB,\, -\sigma \aT  \Bigg), \end{equation}
where the second comes from Horndeski membership of EGB gravity, the following two coming from the matching of $\psi^{\rm co}$ and $\chi^{\rm co}$, $\aT$ using the definition of $z$ with $f=1$ and $\aM$ from its definition.
These are consistent with the literature \cite{Gleyzes13, Gleyzes14} and yield
$$\frac{\abar}{(1+\aB)^2} = \frac{(1-w)}{(1-\frac{3}{2}w)^2}\Bigg( \frac{\dot\phi_b^2}{\Mpl^2 H^2}+\frac{3w^2}{2(1-w)}\Bigg) = \frac{(1-w)^2}{(1-\frac{3}{2}w)^2}\Bigg( \aK+\frac{3w^2}{2(1-w)^2}\Bigg),$$
leading us to write the Stochastic EGB equations of motion as
\begin{equation}
  \begin{aligned}
  G_{\mu\nu} - \frac{1}{\Mpl^2}\Big(T^{GB}_{\mu\nu} + T^{\phi}_{\mu\nu}\Big) & =  \frac{(1-w)^2}{2(1-\frac{3}{2}w)^2}\Bigg( \aK+\frac{3w^2}{2(1-w)^2}\Bigg)
                                   \tilde{\rm P}_{\mu\nu}\boldsymbol{{\cal S}}^{(0)} +\boldsymbol{{\cal S}}_{\mu\nu}^{(2)},\\
    \nabla_\mu\nabla^\mu\phi -\frac{1}{8}\yld{\xi}\ylg{\lambda}_{, \phi}(\phi){\cal G} - V_{, \phi}(\phi) & = -\sqrt{\aK}\Mpl\frac{(1-w)^{\frac{3}{2}}}{2(1-\frac{3}{2}w)^2}\Bigg(1+\frac{3w^2}{2(1-w)^2{\aK}}\Bigg)
                                   \boldsymbol{{\cal S}}^{(0)}, \\
  \end{aligned}
\end{equation}
where we have left implicit that any field $g_{\mu\nu}$ or $\phi$ on the lhs should be taken as its IR part, $g_{\mu\nu}^>$ or $\phi^>$. We will keep this implicit in the rest of the examples for readability.

Let us comment on the result compared to \cite{aldabergenov_towards_2025}. Overall, the structure of the equations is similar in terms of the coefficients but rather different in terms of the sources.
However, as discussed in Section~\ref{sec:cons}, the original constraints are still satisfied in our case, which is not the case in the mentioned work and could point to an inconsistency.

\subsection{Stochastic non-minimally coupled dynamics}\label{sec:NMC}
\yld{Following and introducing historical beyond GR approaches}\ylg{Following and building on historical approaches beyond GR}, we build the stochastic equations for non-minimally coupled theories, starting with subcases such as $f(R)$ theories.

\subsubsection{$f(R)$ dynamics.}
We start with $f(R)$ Inflation. The action is
\begin{equation}
  S =  \frac{\Mpl^2}{2}\int d^4x \sqrt{-g} f(R).
\end{equation}
By introducing an auxiliary field $\varphi$, we can rewrite the action as
\begin{equation}
  S =\frac{\Mpl^2}{2} \int d^4x \sqrt{-g} \Big(  f'(\varphi)(R - \varphi) + f(\varphi) \Big).
\end{equation}
On shell, one has $\varphi = R$ and thus the two actions are equivalent. Finally, by defining $\phi = f'(\varphi)$, we can rewrite the action as that of a dilaton
\begin{equation}
  S =\frac{\Mpl^2}{2}  \int d^4x \sqrt{-g} \Big( \phi R - U(\phi) \Big),
\end{equation}
where the potential $U(\phi)$ is given by $ U(\phi) = f'(\varphi(\phi))\varphi(\phi)- f(\varphi(\phi)) $.
In the unitary gauge and since $M^2 = \Mpl^2 \phi$, we identify the match to the \yld{EFToDE}\ylg{EFT action} parameters as\footnote{Integration by parts is useful to find $\aB$.} 
$$ \Big(\aK, \aH, \aB, \aT, \aM \Big) = \Bigg( 0,\,0,\,- \frac{1}{2}\frac{\dot\phi_b}{H\phi_b},\,0,\, \frac{\dot\phi_b}{H\phi_b} \Bigg), $$
noting the known absence of kinetics in the field. 
Using our main result, the stochastic equations are thus 
\begin{equation}
  \begin{aligned}
  \phi G_{\mu\nu}[g]  + \Big(g_{\mu\nu}\Box - \nabla_\mu\nabla_\nu\Big)\phi+\frac{1}{2}U(\phi)g_{\mu\nu} &= \frac{3\phi_b}{\Big(1-\frac{2H\phi_b}{\dot\phi_b} \Big)^2}\tilde{\rm P}_{\mu\nu}\boldsymbol{{\cal S}}^{(0)} + \phi_b \boldsymbol{{\cal S}}_{\mu\nu}^{(2)}\,,\\
  R[g] - U_{,\phi}(\phi) & = -\frac{12H\phi_b\dot\phi_b}{\Big(1-\frac{2H\phi_b}{\dot\phi_b} \Big)^2}\boldsymbol{{\cal S}}^{(0)},
  \end{aligned}
\end{equation}
which shows the robustness of the method as it can be applied to a theory with no canonical kinetic term.
As we are about to see, this theory is a subcase of Brans-Dicke Inflation, which can be moved to the Einstein frame and can thus be looked at via standard GR stochastic Inflation.

\subsubsection{Damour-Esposito-Farèse dynamics.} Damour-Esposito-Farèse (DEF) gravity and thus inflation adds noncanonical kinetics to $f(R)$ inflation\footnote{Note that this action uses a dimensionless scalar field.}
\begin{equation}
  S = \int d^4x \sqrt{-g} \left( \frac{\Mpl^2}{2} \Big(\phi R - \frac{\omega(\phi)}{2\phi}\nabla^\mu\phi\nabla_\mu\phi\Big) - V(\phi) \right),
\end{equation}
where $\omega$ is the generalised dimensionless Brans-Dicke parameter that controls the strength of the kinetic term, $0$ in the case of $f(R)$ gravity. When $\omega$ is constant, this is the Brans-Dicke theory.
The identification to the \yld{EFToDE}\ylg{EFT action} parameters is
\begin{equation} \Big(\aK, \aH, \aB, \aT, \aM \Big) = \Bigg( {\omega(\phi_b)}\bigg(\frac{\dot\phi_b}{H\phi_b}\bigg)^2,\,0,\,- \frac{1}{2}\frac{\dot\phi_b}{H\phi_b},\,0,\, \frac{\dot\phi_b}{H\phi_b} \Bigg), \end{equation}
leading to the stochastic equations
\begin{equation}
  \begin{aligned}
  \phi G_{\mu\nu}  + \Big(g_{\mu\nu}\Box - \nabla_\mu\nabla_\nu\Big)\phi -\frac{\omega(\phi)}{2\phi} T^{\phi}_{\mu\nu}[\phi]& \\
   +\frac{1}{\Mpl^2}\bigg(1-\frac{\omega(\phi)}{\phi}\bigg)V(\phi)g_{\mu\nu} &= \phi_b\frac{3+2w_b}{\Big(1-\frac{2H \phi_b}{\dot \phi_b} \Big)^2}\tilde{\rm P}_{\mu\nu}\boldsymbol{{\cal S}}^{(0)}+\phi_b\boldsymbol{{\cal S}}_{\mu\nu}^{(2)},\\
\frac{\omega(\phi)}{\phi}\nabla^\mu\nabla^\mu\phi + \frac{1}{2}\left(\frac{\omega(\phi)}{\phi}\right)_{,\phi}\nabla^\mu\phi\nabla_\mu\phi + R - \frac{2}{\Mpl^2}V_{,\phi}(\phi)  &= -4H \phi_b\dot\phi_b\frac{3+2w_b}{\Big(1-\frac{2H \phi_b}{\dot \phi_b} \Big)^2}\boldsymbol{{\cal S}}^{(0)}.
  \end{aligned}
\end{equation}

\subsubsection{Non-minimally coupled dynamics.}
The previous action is equivalent to the non-minimally coupled (NMC) one with coupling functional $F(\phi)>0$,
\begin{equation}
  S = \int d^4x \sqrt{-g} \left[ F(\phi) R - \frac{1}{2}\nabla^\mu\phi\nabla_\mu\phi - V(\phi) \right],
\end{equation}
which has a slightly different identification to the \yld{EFToDE}\ylg{EFT action} parameters as
\begin{equation} \Big(\aK, \aH, \aB, \aT, \aM \Big) = \Bigg( \frac{\dot\phi_b^2}{2FH^2},\,0,\,-\frac{1}{2} \frac{\dot F_b}{H F_b},\,0,\, \frac{\dot F_b}{H F_b} \Bigg), \end{equation}
and directly leading us to the stochastic equations
\begin{equation}
  \begin{aligned}
  F^> G_{\mu\nu}^> + \Big(g_{\mu\nu}^>\Box - \nabla_\mu\nabla_\nu\Big)F^> -\frac{1}{2} T^{\phi}_{\mu\nu}[\phi^>] & = \frac{6F_b+\frac{\dot\phi_b^2F_b^2}{\dot F_b^2}}{\Big(1-\frac{2H F_b}{\dot F_b} \Big)^2}\tilde{\rm P}_{\mu\nu}\boldsymbol{{\cal S}}^{(0)}+ F_b\boldsymbol{{\cal S}}_{\mu\nu}^{(2)},\\
 \nabla_\mu\nabla^\mu\phi^>+ F_{,\phi}(\phi^>)R^> - V_{,\phi}(\phi^>) & = -4H \dot\phi_b\frac{\Big(6F_b+\frac{\dot\phi_b^2F_b^2}{\dot F_b^2}\Big)}{\Big(1-\frac{2H F_b}{\dot F_b} \Big)^2}\boldsymbol{{\cal S}}^{(0)} .
  \end{aligned}
\end{equation}
 
\subsection{Stochastic Horndeski dynamics}\label{sec:horn}
The Horndeski theory is the most general nonlinear scalar-tensor theory with at most second-order derivative equations of motion, which ensures the absence of Ostrogradsky instabilities.
The associated action can be written as
\begin{equation}
S =\int d^4x\sqrt{-g}\Big({\cal L}_2+{\cal L}_3+{\cal L}_4+{\cal L}_5\Big),
\label{eq:action_horndeski}
\end{equation}
where we parametrise the Lagrangian densities as%
\begin{subequations}
  \begin{align}
      {\cal L}_2&=K(\phi, X),\\
{\cal L}_3&=-G_3(\phi, X)\Box\phi,\\
{\cal L}_4&=G_{4}(\phi, X){}^{(4)}R+G_{4,X}\left[
\left(\Box\phi\right)^2-\left(\nabla_\mu\nabla_\nu\phi\right)^2
\right],
\\
{\cal L}_5&=G_5(\phi, X) G_{\mu\nu}\nabla^\mu\nabla^\nu\phi
-\frac{1}{6}G_{5,X}\Bigl[
\left(\Box\phi\right)^3
-3\left(\Box\phi\right)\left(\nabla_\mu\nabla_\nu\phi\right)^2
+2\left(\nabla_\mu\nabla_\nu\phi\right)^3
\Bigr],
  \end{align}
\end{subequations}
with\footnote{Note that sometimes $X$ is defined without the $-1/2$ prefactor \cite{Gleyzes13}.\label{fn:Xnotations}}\edef\fnXnotationNum{\arabic{footnote}} $X:=-\nabla_\mu\phi \nabla^\mu\phi/2$ and $G_{i,Y}=\partial G_i/\partial Y$.
Since all previous examples are special cases of Horndeski theories, they can be matched to a choice of $K$, $G_3$, $G_4$ and $G_5$ functions.
This is given in Table \ref{tab:HornMatch}.

\setlength{\tabcolsep}{4pt}   
\renewcommand{\arraystretch}{1.7}
\begin{table}[ht]
\begin{center}
\resizebox{\textwidth}{!}{%
\begin{tabular}{|l|c|c|c|}
\hline
Theory & $f$ & $c$ & $\Lambda$ \\
\hline
GR & $1$ & $\frac{\dot\phi_b^2}{2}$ & $V(\phi_b)$ \\
\hline
$f(R)$ & $\phi_b$ & $0$ & $\frac{\Mpl^2}{2}U(\phi_b)$ \\
\hline
DEF & $\phi_b$ & $\frac{\Mpl^2\,\omega(\phi_b)\,\dot\phi_b^2}{4\phi_b}$ & $V(\phi_b)$ \\
\hline
NMC & $\frac{2F(\phi_b)}{\Mpl^2}$ & $\frac{\dot\phi_b^2}{2}$ & $V(\phi_b)$ \\
\hline
EGB & $1$ & $\frac{\dot\phi_b^2}{2}$ & $V(\phi_b) + 3w\Mpl^2 H^2$ \\
\hline
Braiding & $1$ & $\frac{1}{2}\dot\phi_b^2\big(P_{,X}+(-\ddot\phi_b+3H\dot\phi_b)G_{,X}-2G_{,\phi}\big)$ & $-P+\frac{1}{2}\dot\phi_b^2\big(-P_{,X}+(\ddot\phi_b+3H\dot\phi_b)G_{,X}\big) $ \\
\hline
General Horndeski & $\frac{2}{\Mpl^2}G_4-\frac{\dot\phi_b^2}{\Mpl^2}G_{5,\phi}-\frac{\dot\phi_b^2\ddot\phi_b}{\Mpl^2}G_{5,X}$ & \multicolumn{2}{c|}{Appendix C of \cite{Gleyzes13}\footnotemark[\fnXnotationNum]} \\
\hline
\end{tabular}
}
\end{center}
\caption{Background EFT coefficient matching for each theory covered in Section~\ref{sec:EOMs}.
$f$ is the gravitational coupling prefactor ($M^2 = \Mpl^2 f/(1+\aT)$),
$c$ the dark energy kinetic coefficient, and $\Lambda$ the potential-like term in the background action~\eqref{eq:EFTaction}.}
\label{tab:bgMatch}
\end{table}

\setlength{\tabcolsep}{2pt}   
\renewcommand{\arraystretch}{1.7}
\begin{table}[ht]
\begin{center}
  \resizebox{0.5\textwidth}{!}{
\begin{tabular}{|c|c|c|c|c|c|}
\hline
Action & $\aK$ & $\aB$ & $\aT$ & $\aM$ & $\aH$ \\
\hline
GR & $2\varepsilon_1$ & $0$ & $0$ & $0$ & $0$ \\
\hline
EGB & $\frac{\dot\phi_b^2}{(1-w)\Mpl^2H^2}$ & $\frac{w}{2(w-1)}$ & $\frac{w}{1-w}$ & $\frac{w\sigma}{w-1}$ & $0$ \\
\hline
$f(R)$ & $0$ & $-\frac{1}{2}\frac{\dot\phi_b}{H\phi_b}$ & $0$ & $\frac{\dot\phi_b}{H\phi_b}$ & $0$ \\
\hline
DEF & $\omega(\phi_b)\!\left(\frac{\dot\phi_b}{H\phi_b}\right)^2$ & $-\frac{1}{2}\frac{\dot\phi_b}{H\phi_b}$ & $0$ & $\frac{\dot\phi_b}{H\phi_b}$ & $0$ \\
\hline
NMC & $\frac{\dot\phi_b^2}{2F_bH^2}$ & $-\frac{\dot F_b}{2HF_b}$ & $0$ & $\frac{\dot F_b}{HF_b}$ & $0$ \\
\hline
Horndeski & \multicolumn{5}{c|}{see \cite{Gleyzes13,Gleyzes14}} \\
\hline
\end{tabular}
  }
\end{center}
\caption{Summary of the $\alpha$ parameters of diverse theories, found through their quadratic action.}
\label{tab:alphaMatch}
\end{table}

\begin{table}[ht]
\begin{center}
  \resizebox{0.75\textwidth}{!}{%
\begin{tabular}{|c|c|c|c|c|}
\hline
Action & $K$ & $G_3$ & $G_4$ & $G_5$ \\
\hline
GR & $X-V(\phi)$ & $0$ & $\frac{\Mpl^2}{2}$ & $0$ \\
\hline
EGB & $X-V+8X^2(3-\ln X)\partial_\phi^4\lambda$ & $\frac{\Mpl^2}{2}+4X(7-3\ln X)\partial_\phi^3\lambda$ & $4X(2-\ln X)\partial_\phi^2\lambda$ & $-4\ln|X|\partial_\phi\lambda$ \\
\hline
$f(R)$ & $-\frac{\Mpl^2}{2}U(\phi)$ & $0$ & $\frac{\Mpl^2}{2}\phi$ & $0$ \\
\hline
DEF & $\frac{\Mpl^2\,\omega(\phi)}{2\phi}X - V(\phi)$ & $0$ & $\frac{\Mpl^2}{2}\phi$ & $0$ \\
\hline
NMC & $X - V(\phi)$ & $0$ & $F(\phi)$ & $0$ \\
\hline
Braiding & $P(\phi, X)$ & $G$ & $\frac{\Mpl^2}{2}$ & $0$ \\
\hline
\end{tabular}
  }
\end{center}
\caption{Matching Horndeski functions to specific theories, see \cite{KYY11,Gleyzes13} for EGB.}
\label{tab:HornMatch}
\end{table}

A correspondence between the $\alpha$ parameters and the $K$, $G_3$, $G_4$ and $G_5$ functions is also known (using Appendix C of \cite{Gleyzes13} and Table 1 of \cite{Gleyzes14}),
which we could use to identify the source factors. 
Instead and for the sake of demonstration, we bypass the full identification of the $\alpha$ parameters using the derivation of the MS action found in \cite{KYY11}, giving us instead%
\begin{subequations}
\begin{align}
  M^2(1+\aT)&=2\left[G_4
-X\left( \ddot\phi G_{5,X}+G_{5,\phi}\right)\right],
\\
M^2&=2\left[G_4-2 XG_{4,X}
-X\left(H\dot\phi G_{5,X} -G_{5,\phi}\right)\right],\\
z^2&=2a^3\bigg(\frac{\Sigma}{\Theta^2}M^4+3M^2\bigg),\\
c^2_s&= \frac{\frac{1}{a}d_t\Big(\frac{\Theta}{a}M^2\Big)-M^2(1+\aT)}{\frac{\Sigma}{\Theta^2}M^4+3M^2},
\end{align}
\label{eq:KobayashiM}
\end{subequations}
where $\Sigma$ and $\Theta$ are defined as%
\begin{subequations}
\begin{align}
\Sigma&:=XK_X+2X^2K_{,XX} \notag\\
& +12H\dot\phi XG_{3,X}+6H\dot\phi X^2G_{3,XX} -2XG_{3,\phi}-2X^2G_{3,\phi X} \notag\\
& -6H^2G_4 +6\Bigl[H^2\left(7XG_{4,X}+16X^2G_{4,XX}+4X^3G_{4,XXX}\right) \notag\\
& -H\dot\phi\left(G_{4,\phi}+5XG_{4,\phi X}+2X^2G_{4,\phi XX}\right)
\Bigr]
\notag\\
& +30H^3\dot\phi XG_{5,X}+26H^3\dot\phi X^2G_{5,XX}
\notag\\&
+4H^3\dot\phi X^3G_{5,XXX}
-6H^2X\bigl(6G_{5,\phi}
+9XG_{5,\phi X}+2 X^2G_{5,\phi XX}\bigr),
\\
\Theta&:=-\dot\phi XG_{3,X}+
2HG_4-8HXG_{4,X}
-8HX^2G_{4,XX}+\dot\phi G_{4,\phi}+2X\dot\phi G_{4,\phi X}
\notag\\&
-H^2\dot\phi\left(5XG_{5,X}+2X^2G_{5,XX}\right)
+2HX\left(3G_{5,\phi}+2XG_{5,\phi X}\right).
\end{align}
\end{subequations}
In the past two blocks of equations, all quantities are implicitly evaluated on the background trajectory, as usual within the stochastic ansatz-- if no heuristic for the stochastic backreaction is used.

Eqn.~\eqref{eq:KobayashiM} is enough to deduce $\Src_{\cal R}^{\rm EFT}$ and $\Src_h^{\rm EFT}$ requiring $(z, c_s,c_T)$ as well as
\begin{equation}
  \frac{\abar}{(1+\aB)^2}= \frac{z^2}{a^3M^2} = 2\bigg(3+\frac{\Sigma}{\Theta^2}M^2\bigg).
\end{equation}
One can check that we recover $2\varepsilon_1$ in the GR limit.

The Einstein and scalar-field equations then follow naturally using the EOMs from \cite{KYY11}
\begin{equation}
\begin{aligned}
\sum_{i=2}^5{\cal G}_{\mu\nu}^i &= \frac{z^2}{4a^3}\tilde{\rm P}_{\mu\nu}\boldsymbol{{\cal S}}^{(0)}+\frac{M^2}{2}\boldsymbol{{\cal S}}_{\mu\nu}^{(2)},\\
\quad \nabla^\mu\left(
\sum_{i=2}^5 J_\mu^i \right)-\sum_{i=2}^5 P^i_\phi&= -\frac{\dot\phi_bHz^2}{a^3}\boldsymbol{{\cal S}}^{(0)},
\end{aligned}
\end{equation}
where the explicit expressions of ${\cal G}_{\mu\nu}^i$, $J_\mu^i$ and $P_\phi^i$ are recalled in Appendix~\ref{app:horndeski}.

\subsection{Stochastic Braiding dynamics} With the general case of stochastic Horndeski theories treated, it is easier to access Braiding dynamics.
Kinetic Braiding is the most general Horndeski theory with nonzero $K=P$ and $G_3=G$ \footnote{Note that we do not use a minus sign.} but GR's $G_4=\Mpl^2/2$ and $G_5=0$, or, in the $\alpha$-basis language, nontrivial $\aK$ and $\aB$ but null $\aT$ and $\aM$. 

Our methods lead directly to the stochastic equations of motion
\begin{equation}
  \begin{aligned}
   G_{\mu\nu} -\frac{1}{\Mpl^2}T_{\mu\nu}^{\rm Braiding}
      & = C_b\tilde{\rm P}_{\mu\nu}\boldsymbol{{\cal S}}^{(0)}+\boldsymbol{{\cal S}}_{\mu\nu}^{(2)},\\
    \nabla^{\mu}\Bigg(\Big((G\Box\phi)_X-P_X+2G_\phi\Big)\nabla_{\mu}\phi +G_X\nabla_{\mu}X \Bigg)-\nabla^{\mu}G_\phi\nabla_{\mu}\phi-P_\phi&= -2\dot\phi_b HC_b \Mpl^2\boldsymbol{{\cal S}}^{(0)},
    \end{aligned}
\end{equation}
where the stress-energy tensor is given by
\begin{equation}
    T_{\mu\nu}^{\rm Braiding} = P_{X} \nabla_{\mu}\phi\nabla_{\nu}\phi 
    + \Big(P +\nabla_{\lambda}G\nabla^{\lambda}\phi\Big) g_{\mu\nu} -G_{,X}\Box\phi \nabla_{\mu}\phi\nabla_{\nu}\phi 
     -2\nabla_{\mu}G\nabla_{\nu}\phi,
\end{equation}
after defining the background prefactor
\begin{equation} C_b \equiv 3+\frac{XP_X+2X^2P_{XX}+12H\dot\phi_b XG_{,X}+6H\dot\phi_b X^2G_{,XX}-2XG_{,\phi}-2X^2G_{,\phi X}-3H^2\Mpl^2}{H-\dot{\phi}_bXG_X/\Mpl^2}.\end{equation}

As identified in the literature \cite{deffayet_imperfect_2010}, the scalar field equation receives a contribution from second-order derivatives of the metric via the term $\nabla^{\mu}\nabla_{\mu}X$ and can thus be simplified using the Einstein equation, here including the stochastic sources. This is a hallmark of Horndeski theories as explained in \cite{KYY11}, as part of the features ensuring that the EOMs stay at most second order in derivatives.
 
\section{Stochastic ansatz at the edge?}\label{sec:discussion}

In the following, we discuss some further insights and extensions.

\subsection{A word on the EFT of Inflation}

The EFT of dark energy, as we have seen through the lens of our notations, completely includes the inflaton, and thus the EFT of inflation.
Recently \ylg{\cite{Choudhury25}}, stochastic Langevin equations were written in a model-independent single-field EFToI framework, using it as a model-agnostic way to look at primordial black hole production
\begin{equation}
  \begin{aligned}
    \frac{\d\hat{\zeta}}{\d\cN} & = \hat{\Pi}_{\zeta} + \hat{\xi}_{\zeta}(\cN), \\
    \frac{\d\hat{\Pi}_{\zeta}}{\d\cN} & = -(3-\epsilon)
    \left(1 - \frac{2\!\left(s - \frac{\eta}{2}\right)}{3-\epsilon}\right)\hat{\Pi}_{\zeta}
  + \hat{\xi}_{\pi\zeta}(\cN), \label{eq:Choudhury}
  \end{aligned}
\end{equation}
where $s = \frac{\d\ln c_s}{\d\cN}$, $\;\epsilon = -\frac{\d\ln H}{\d\cN}$,
$\;\eta = \epsilon - \frac{1}{2}\frac{d\ln\epsilon}{d\cN}$, and $\xi_{\zeta}$ and $\xi_{\pi\zeta}$ are the Gaussian noises from the UV curvature perturbation and its conjugate momentum, respectively.
Compared to our general treatment, this equation focuses on
i) the long wavelength limit,
ii) the terms of the EFT giving $c_s\neq 1$ theories (so that the quadratic action has $f=1$, $\aB=\aM=\aT=\aH=0$),
iii) the linear regime of $\zeta$ on those scales (equating to using eqn.\eqref{eq:MS} in the limit of i) and ii)).

We do not make any claims on the validity of those approximations.
Given the uncertainty on the matter from the cited work, we bring to the reader's attention that
some non-Gaussianity would indeed arise in the tail of the distribution even within this linear framework, as it emerges from the first-passage time structure of the Fokker-Planck problem, or the $\Delta {\cal N}$ formalism (stochastic or not): the distribution of exit e-fold times $\mathcal{N}$ is non-Gaussian even when driven by Gaussian noise and a linear drift, because the question of when a trajectory first reaches the threshold $\zeta_{\rm th}$ is a nonlinear functional of the entire stochastic path.
However, one might want to justify the use of the linear regime, as both the Stueckelberg trick and the conversion from $\pi$ to $\zeta$ are nonlinear operations.
In fact, even if quantum diffusion, i.e. stochastic effects, were to dominate drift and other nonlinear parts of the evolution (such as those enforced by constraints), it is not clear that the linear l.h.s. of eqn.~\eqref{eq:Choudhury} would indeed stay so.
Ultra slow roll dynamics might help simplify dynamics drastically but that is \textit{a priori} not trivial and needs justification.

Besides the quoted study, it is possible to find a fully nonlinear and nonperturbative equation for $\pi$ in a model-agnostic way.
This follows from evaluating the EFT in the decoupling limit \cite{Creminelli25} to get the $\pi$ EOM
\begin{equation}
  \ddot{\pi} = -\frac{\ddot{H}}{2\dot{H}}\Bigg|_{t+\pi}\Bigg(\dot{\pi}^2-\frac{(\partial_i)^2\pi}{a^2}\Bigg).
\end{equation}
Thus if we were to formulate stochastic inflation for $\pi$ in this limit, we would linearise this equation and coarse-grain it.
Interpreting this result could help understand the link between stochastic inflation and nonperturbative methods using the saddlepoint of the decoupled wavefunctional \cite{creminelli_dissipative_2023}, as well as potentially providing an actual stochastic EFT framework--- as opposed to using an EFT as a wide class of theories to apply the stochastic formalism to, as we do in this work.
This is beyond the scope and interests of this work.

\subsection{Numerical evolution}
Of course, one is free to investigate the long wavelength approximation of the equations provided in this work, or the decoupling limit.
Given the recently found caveats of the separate universe approximation regarding e.g. the treatment of gauge and gradient effects and the numerous heuristics it takes to partially fix them \cite{artigas_hamiltonian_2022, jackson2023, Briaud25},
this research advocates for a fully numerical approach once the most 
general equations have been derived, following insights developed 
in Chapters~\ref{sec:SIGR} and \ref{sec:SINR} on general and numerical relativity.

Moreover, deriving the full Horndeski stochastic equations comes at an opportune 
time: the first fully nonlinear NR simulations of inhomogeneous 
inflation beyond GR have recently been carried out using 
\textsc{GRFolres}~\cite{areste_salo_puncture_2023, brady_inflaton_2025}. These simulations rest on 
a series of works establishing the well-posedness of classes of Horndeski theories 
in suitable gauges and variables~\cite{kovacs_well-posed_2020,salo_well-posedness_2022,salo_strong_2024}.\footnote{Well-posedness seems to be currently guaranteed within the weak-coupling regime 
for scalar-tensor theories, though recent (unpublished) work has extended this to 
all polynomial vacuum EFTs beyond weak coupling~\cite{figueras_well-posed_2024}.}
In the future, and given the implementation of \cite{brady_inflaton_2025} within the same software as this work, and given the simplicity of the source conversion to the numerical variables (i.e. going from Einstein's equations to ADM/BSSN ones, see Chapter~\ref{sec:SINR}), stochastic inflation beyond GR is ready to provide its first \yld{assumption-free estimates}\ylc{estimates free of the separate universe approximation} in the classical regime.
Signatures from higher derivatives are rather unclear and limited in expectations, thus reaffirming the motivations of such numerical and general studies.
However, in the proxy built in \cite{Choudhury25} following established derivations \cite{vennin_correlation_2015}, one sees that a nontrivial speed of sound alone should affect the drift in $\zeta$ and therefore the PDF of $\cN$.
As a consequence, the PBH abundance computed from the tail of the PDF will vary greatly, as well as the associated correlators. Conversely, measuring PBHs or not, could help constrain corrections to GR in the early universe.

\subsection{Beyond the EFT action}
Going beyond theories described by the \yld{EFToDE}\ylg{EFT action} seems rather ambitious, at first sight. 
In fact, some of the theories already included in this work showcase rather complicated dynamics, whether we aim at finding the stochastic equations or solving them numerically.
These include beyond Horndeski theories ($\aH\neq 0$), which have higher-order time derivatives in the equations of motion and which we have not looked at \ylg{specifically }in our examples of Section~\ref{sec:EOMs}.
Thanks to recent studies, higher-order time-derivative theories and the presence of ghosts have been disentangled dramatically to go beyond second-order relativity in time \cite{gleyzes_healthy_2015}. 
This made it possible to include $\aH\neq 0$ in the present formalism, but also to look at further classes, including (Unified) Degenerate Higher-Order Scalar-Tensor --- so-called GPLV and (U)DHOST --- theories \cite{LangloisNoui16,DeFelice18} and recent extensions \cite{Gorji26, Michiwaki26}.
One can imagine that the coarse-graining of such theories would be more involved (more time derivatives to handle) but not impossible with our procedure --- in principle.

The point being raised here is indirectly the question of whether or not one needs to know the $\alpha$ coefficients.
Clearly, we have shown in the Horndeski case in Section~\ref{sec:horn} that all one needs to find the sources is to know the prefactors in front of the scalar and gravitonic Mukhanov-Sasaki actions.
Thus, beyond the validity of the stochastic limit, one could imagine that this nonlinear to linear correspondence of the coefficients is all one needs to find the stochastic equations, and so should not be limited to dynamically-friendly dynamics.

\subsection{Beyond the Inflaton: multifields and all}\label{sec:beyondinflaton}
\yld{Our original motivation behind the use of the EFToDE conventions was the clear interpretability it provides through its meaningful $\alpha$ coefficients.
However, in}\ylg{In} general, scalar-tensor theories used for inflation do not have to use that particular scalar as the inflaton,
and more generally, one should be considering the possibility of a much richer phenomenology of all spins surrounding, in the case of inflation, an overall vacuum dark-energy-like period of time.
Importantly, a frame might have a multifield action with field-space curvature and this needs to be accessible.

Starting with a spin-0 landscape, our formalism --- and subsequent NR simulations --- can be generalised to multifields.
On the one hand, its EFT has already been written \cite{senatore_effective_2012}.
On the other hand, Stochastic Inflation already has a consequent literature, though in the long wavelength regime \cite{BattefeldVanTent06, Rigopoulos_multi06, Takahashi26}.
Let us sketch a procedure for the coarse-graining of multifield GR, with field-space metric ${\rm G}_{IJ}(\phi^K)$, potential $V(\phi^I)$, and action\footnote{See \cite{Sasaki98, nibbelink_scalar_2002} for further introduction to the formalism.}
\begin{equation}
    S = \frac{\Mpl^2}{2}\int d^4x \sqrt{-g} R - \int d^4x \sqrt{-g}\Bigg( \frac{1}{2} {\rm G}_{IJ}\nabla_\mu\phi^I\nabla^\mu\phi^J+V(\phi)\Bigg),
\end{equation}
yielding the nonlinear EOMs%
\begin{subequations}
  \begin{align}
    \Box {\phi}^I + \Gamma^{I}_{JK}\nabla_{\mu}\phi^{\yld{I}\ylg{J}}\nabla^\mu\phi^{\yld{J}\ylg{K}} + {\rm G}^{IJ}\partial_J V & = 0, \\
    G_{\mu\nu} - \frac{1}{\Mpl^2}{\rm G}_{IJ}\nabla_\mu\phi^I\nabla_\nu\phi^J + \frac{1}{\Mpl^2} g_{\mu\nu}\left(\frac{1}{2}{\rm G}_{IJ}\nabla^\alpha\phi^I\nabla_\alpha\phi^J + V\right) &= 0,
  \end{align}\label{eq:multifields}
\end{subequations}
where $\Gamma^{A}_{BC}$ are the Christoffel symbols of the field-space metric ${\rm G}$. Compared to GR, all components are simply modified through the effective stress-energy tensor. Our procedure would then, as usual, proceed as follows:
\begin{enumerate}
\item \underline{Identify linear gauge-invariants} such as%
            \begin{subequations}
            \begin{align}
              Q^I \equiv \delta \phi^I +\frac{\dot{\phi}_b^I}{H}\Phi
            \end{align}
            \end{subequations}
            and pick one to coarse-grain (here $Q^I$, following a generalisation of our approach with ${\cal R}$ but that is not compulsory).
            $Q^I$ has a known mass-matrix Mukhanov-Sasaki equation of the form
            \begin{equation}
              \mathcal{D}_t^2 Q^I + 3H\,\mathcal{D}_t Q^I - \frac{\nabla^2}{a^2}Q^I + \mathcal{M}^I{}_J\, Q^J = 0, \label{eq:multiMS}
            \end{equation}
            where $\mathcal{D}_t (\cdot)^I \equiv \partial_t {(\cdot)}^I + \Gamma^I_{JK}\dot\phi^J_b (\cdot)^K$ is the field-space covariant time derivative.
\item \underline{Express all perturbations of one gauge in terms of one of them} ($Q^I$).
      Using the constraints with their new stress-energy contributions%
            \begin{subequations}
              \begin{align}
                \delta\rho &= {\rm G}_{IJ}{\dot{\phi}^I_b} {\delta\dot{\phi}^J}
                + \partial_I V\,\delta\phi^I
                - {\rm G}_{IJ}{\dot{\phi}^I_b \dot{\phi}^J_b}\Psi, \\
                \delta\mathcal{J}_i &= -{\rm G}_{IJ}{\dot{\phi}^I_b}\,\partial_i\delta\phi^J.
              \end{align}
            \end{subequations}
            These terms are noteworthy because they constitute, apart from the $\partial_I V$ contribution, the adiabatic contributions (meaning along the tangent to the background field-space trajectory).
            The multifield comoving gauge thus generalises naturally to nulling the adiabatic sources ${\rm G}_{IJ}{\dot{\phi}^I_b}\,\delta\phi^J=0$ and the usual $E=0$, as done in Section~\ref{sec:slicingsOfR} to get all perturbations in terms of the gauge invariant quantity.

            However, the constraints cannot be solved on their own in this gauge as opposed to the single-field case (entropic contributions of the fields make the Hamiltonian constraints nontrivial),
            which is why the flat gauge ($\Phi=E=0$, $\delta\phi^I = Q^I$) can be used to solve the constraints and extract $\Psi$ from the momentum constraint, and $B$ from the Hamiltonian constraint using that $\Psi$. We find%
            \begin{subequations}
                \begin{align}
                  \delta \phi^{I,{\rm flat}} & = Q^I,\\
                  E^{\rm flat} & = 0,\\
                  \Psi^{\rm flat} & = \frac{1}{\Mpl}\sqrt{\frac{\epsilon}{2}}Q, \\
                  B^{\rm flat} & = \frac{1}{2a^2k^2\Mpl^2}\Big(\sqrt{2\epsilon}H\Mpl\big(\dot{Q}\big)_\parallel+\partial_I V_b Q^I + \sqrt{2\epsilon}(3-\epsilon)\frac{\Mpl}{H}Q_\parallel\Big)\\
                              & = -a^{-2} \chi^{\rm flat},
                \end{align}
            \end{subequations}
          where we defined the multifield slow-roll $\epsilon = G_{IJb}\dot{\phi}^I_b\dot{\phi}^J_b /(2 H^2 \Mpl^2)$ and the field-space adiabatic component projection $X_\parallel \equiv ({\rm G}_{IJ}\dot\phi_b^IX^J)/\sqrt{{\rm G}_{IJ}\dot\phi_b^I\dot\phi_b^J}$.
          This also gives us access to other generalised gauge invariant combinations%
          \begin{subequations}
            \begin{align}
            \Phi_B     &\equiv \Phi + H\chi = H\chi^{\rm flat}[\boldsymbol{Q}]\,, \\
            \Psi_B     &\equiv \Psi - \dot\chi = \Psi^{\rm flat}[\boldsymbol{Q}] - \dot\chi^{\rm flat}[\boldsymbol{Q}]\,, \\
            Q^{I,{\rm gi}} & \equiv \delta\phi^I -  \sqrt{2\epsilon}H\Mpl\chi =Q^I -\sqrt{2\epsilon}H\Mpl\chi^{\rm flat}[\boldsymbol{Q}]\,, \\
            \zeta^{\rm gi}  & \equiv -\Phi + \frac{1}{3}\frac{\delta\rho}{\rho+p} =  \frac{\delta\rho^{\rm flat}[\boldsymbol{Q}]}{6\epsilon H^2\Mpl^2}\,,
          \end{align}
          \end{subequations}
          from which we can express all perturbations in terms of $Q^I$ and another field, or even just $Q^I$ in a specific gauge, see Section~\ref{sec:slicingsOfR}.
\item \underline{Obtain the sources to the nonlinear equations}, eqn.~\eqref{eq:multifields}, by linearising them and applying $Q^I\longrightarrow Q^{J>}\equiv W^I_J Q^J$ in the previously found decompositions,
where $W$ is a windowing matrix allowing one to coarse-grain depending on the crossing of each field.
Given our previous findings, sources should all be found proportional to that of eqn.~\eqref{eq:multiMS}.
The final equations, their extension beyond GR following Section~\ref{sec:procedure}, and their application are kept for future work, e.g. using numerical relativity.\footnote{In fact, there does not seem to be any NR code including field space curvature, only flat field-space approximations \cite{Joana22multi}.}
\end{enumerate} 


Beyond spin-0, we could perfectly imagine a similar construction for multiple spin-2 fields as we did previously or less commonly multiple spin-1 fields\footnote{One could even imagine building a multispin field-space as well --- if that is possible and motivated.},
as long as we understand sufficiently the associated linear theory, their 2-point function, and their regime of classicality or separation of scales.

\subsection{Beyond closed EFTs}
As a final direction for extension, it has become clear in the past few years \cite{Salcedo24, salcedo_open_2024-1, Dufner1, salcedo_phenomenology_2026, Salcedo26sdn} that an open system EFT description in cosmology would be more appropriate if we can only measure some parts of a large system of scales and fields.
The link with stochastic inflation seems rather obvious at first sight, as one focuses on an open system of wavenumbers in the latter.
The so-called \textit{OpenEFT} framework is seemingly more general as it traces out higher-energy but more generally hidden degrees of freedom and is formulated within a fully quantum description, thus containing stochastic inflation as a limit.
Very recently, analytical links have been established between the two approaches \cite{Li26, Cespedes26, Li26b}, although limited to the decoupling limit, similarly to the EFToI framework.
We hope that this could be improved in the next few years before extending our framework. 

\subsection{Beyond inflation}\label{sec:beyondinflation}
\ylg{Although focused on the coarse-graining of scalar-tensor theories of the inflaton,
the purpose of this work is to demonstrate a generic framework for the coarse-graining of fluctuations in a gauge-consistent way, first proposed in Chapter~\ref{sec:SIGR}.
There are quite a few ways to go beyond inflation from the perspective of stochastic or coarse-grained cosmology, which we try to sketch here with no expertise beyond suggestion.

Firstly, one could simply follow the arrow of time starting from any result obtained in an inflationary calculation.
Our methods, if implemented similarly to the GR case (Chapter~\ref{sec:SINR}), enable the creation of nonlinear initial conditions for the post-inflationary universe, which can then be evolved using numerical relativity or other methods in the right regimes.
This includes initial conditions for large-scale structure and $N$-body or GR simulations (e.g.\ $\mathsf{gevolution}$~\cite{1509.01699}, $\mathsf{(G)RAMSES}$~\cite{astro-ph/0111367, 2010.08257}), reheating and preheating lattice codes (e.g.\ $\mathsf{CosmoLattice}$~\cite{2006.15122}), and numerical-relativity collapse codes used for primordial black hole formation (e.g. $\mathsf{GRChombo}$~\cite{2507.19166}).
Bridging stochastic inflation effects and late-time cosmology codes in this way will be a major step in augmenting the testability of inflationary models.

Secondly, the formalism developed in this work is specific neither to inflation nor to
a fundamental field. All the construction needs is a perturbative sector and
a classical nonlinear one, in the UV or the IR, not necessarily reflecting a
quantum-to-classical transition.}

\ylc{In the case of dark energy in scalar-tensor theories, the EFT fixes the sources once a background and the
$\alpha$-functions are chosen, and inflation enters only through that choice with initial conditions.
In particular, the benefit of using the notations of the EFToDE is that it allows us to treat post-inflationary DE fluctuations with the same formalism.
If we keep the Hubble radius as the coarse-graining scale, it is increasing (as opposed to inflation)\ylc{ during the decelerated radiation and matter eras, when modes re-enter it, while the comoving Hubble radius starts shrinking again once dark energy dominates, so that modes then exit it as during inflation,} and separates two mostly linear systems, with no particularly interesting dynamics.
Our stochastic DE scalar-tensor equations would thus be beneficial for the study of the large-scale structure if we set the nonlinearity scale $k_{\rm NL}(t)$ to be that of the window for the study of small scales.
We know, however, that the effect of DE on the growth of structure is mostly linear and that the nonlinearities are mostly due to the matter fluid~\cite{astro-ph/9906174}, since a bigger impact from dark energy fluctuations at the nonlinearity scale would require DE clustering to be significant, found in specific models.}\footnote{\ylc{One way to achieve that is to have a low sound speed~\cite{1101.1026}, which is the case for kinetic braiding~\cite{deffayet_imperfect_2010,1103.5360} or coupled models~\cite{astro-ph/9908023}, as long as the coupling is not screened in dense regions~\cite{astro-ph/0309300}.}}
\ylc{Regarding simulations, using the stochastic EOMs instead of simple stochastic initial conditions is only relevant if entering fluctuations are only accessible dynamically or emerging in time.
During inflation, this is the case because we can only use a classical treatment for modes crossing the dynamical window. For late-time DE, we expect that this could depend on the model.}

\ylg{This work is not restricted to DE fluctuations. In the coarse-graining of multifields in GR carried out in Section~\ref{sec:beyondinflaton}, we have clearly shown that fields (whether or not they include the inflaton) can be coarse-grained and that they can source nonlinear equations of spacetime.
In using different gauge-invariant combinations in the coarse-graining, we could similarly coarse-grain non-scalar fluids with the same procedure, using a windowing on hypersurfaces of constant energy density (as opposed to constant field value) to define the coarse-graining scale.
This would amount to using $ \zeta^{\rm gi}_I  \equiv -\Phi + \frac{1}{3}\frac{\delta\rho_I}{\rho_I+p_I} $ instead of $\R$ (or $Q^I$) and the hydrodynamical equations of the fluid instead of the scalar field equation. This could help refine the construction of the EFT of large-scale structures~\cite{Baumann_2012, carrasco_effective_2012}, and more generally,
the study of the FLRW metric in full nonlinear GR with corrections arising from small scales, and vice versa.
In particular, the backreaction of small-scale inhomogeneities on larger scales, whether tensorial or matter \cite{Isaacson1968a,Isaacson1968b,1011.4920}, could now be studied in a more systematic way with stochastic methods accounting for subscale physics.

The unbroken assumption left in this work is that of an FLRW background for the fluctuations.
If we were to extend the formalism to a more general background, such as compact objects including black holes, we would need to find a new set of gauge-invariant quantities to coarse-grain, which also seems reasonable to find.
Previous work on more generic stochastic gravity includes \cite{verdaguer_stochastic_2007} and could help us in this direction.}

\section*{Summary}

In this chapter, we studied the coarse-graining of perturbations in scalar-tensor theories of dark energy and beyond, including multifields.
Provided a well-defined classical set of scales and a perturbative one, we showed that the stochastic sources for the curvature perturbation and the geometric quantities can be expressed in terms of the Mukhanov-Sasaki source, with modifications arising from the specific features of the theory under consideration.
\ylc{As in GR, these sources are computed from linear theory around a homogeneous background, and the multifield case was only outlined.}
This result provides a unified framework for studying stochastic inflation in a broad class of theories and paves the way for future investigations into the phenomenological implications of these models.

\addtocontents{toc}{\protect\setcounter{tocdepth}{4}}

%% file: conclusion.tex
\chapter*{Conclusion}  \label{sec:Conc}
\markboth{Conclusion}{Conclusion}
\addcontentsline{toc}{chapter}{Conclusion}

When initiated, this thesis had the ambition of finding ways to probe with high precision the full nonlinearities encountered in a generic model of inflation, whether perturbative or not. From the start, it was clear that such an approach would rely in some way on numerical relativity and more generally on the full spacetime equations of motion.

In a fundamentally quantum area of physics, our first concern was that of the validity of using an on-shell description of the inflaton field. This has been known for decades thanks to semi-quantitative arguments for the free theory but never in precise terms for interactions and thus nonlinearity. In Chapter~\ref{sec:semiclQFT}, we defined and computed analytically the \textit{quantum interactivity}, the fraction of nonlinearity of a correlator (here the bispectrum) that cannot be accounted for when neglecting quantum interactions. We adapted it to the finite range of modes a simulation probes and found that these, whether using initial conditions or Langevin sources sampled from the free theory’s Gaussian statistics, make an accurate approximation closer to horizon-crossing than what was previously anticipated, especially for GR interactions previously found to be dynamically classical. We now have a proof of concept\ylc{, demonstrated for the slow-roll bispectrum,} by which to engineer the semiclassical spacetime lattice of any model with respect to a correlator of interest, with the hope of eventually providing a library of numerical tools for anyone planning to investigate such phenomena. In this quest, we also clarified the structure of well-known correlators in quasi-dS scenarios, \ylc{found that, at tree level, Maldacena’s consistency relation arises from the on-shell evolution only}, 
 and clarified the link between initial and dynamical stochasticity with respect to the sourcing of nonlinear effects.

In parallel, our work built on recent but separate advances in lattice cosmology and numerical relativity to extend the former by \yld{incorporating}\ylg{moving to} the latter. Simulations of the inflaton’s Klein-Gordon equation on a lattice are only valid in the decoupling limit and do not give access to the coupled dynamics of the inflating spacetime. On the other hand, the construction of physical and constraint-satisfying initial conditions lies at the heart of the numerical relativity challenge. This is what Chapter~\ref{sec:ICNR} solves for the first time in a general perturbative approach. Cosmological perturbation theory is given a fresh look there to relate the choice of initial data, and more generally perturbed hypersurfaces, in terms of a Mukhanov-Sasaki linear spectrum known to cosmologists.

This methodology was implemented in Chapter~\ref{sec:ICsim} for the study of Bunch-Davies initial conditions evolved nonlinearly on classical geodesic slices ranging across Hubble to super-Hubble lengthscales. In this framework, we confirm the feasibility of cosmological simulations within Numerical Relativity, whether for perturbative, linear nonperturbative, or nonlinear nonperturbative scenarios, where in the latter there should be no known analytical solutions\ylc{. This holds for initial data constrained at linear order, and the momentum constraint is violated at late times in the most extreme examples}. In this process we extended diagnostics of cosmological simulations with relativistic nuance and statistical care in contrast with previous studies. The example of a monodromic resonance added to a slow-roll potential was simulated and found to generate highly nonlinear separately expanding domains in full general relativity.

Phenomenology was not pushed to its edge within this framework, leaving it as a first step of reconciliation between cosmology and nonlinear general relativity. Our focus was on solving the initial conditions problem for every time slice, to enable the continuous entry of each mode as interactivity and classicality switch on dynamically when the Hubble horizon or so is crossed by each mode. With this objective, Chapter~\ref{sec:SIGR} makes the connection between Chapter~\ref{sec:ICNR}’s perturbative solution to the constraints and the stochastic formalism of inflation. Taking such formalism with its minimal assumptions, that of linear UV quantum perturbations assumed with negligible commutator from entry into the on-shell nonlinear IR system, we used the perturbative hypersurface ${\cal R}$-decomposition to build the linear spectra of the Langevin sources to the GR field equations. We demonstrated an excellent satisfaction of the constraints at the order of the noise both explicitly and by construction, an evidence of quantitative precision overlooked in past works. This set of stochastic partial differential equations extends stochastic inflation well beyond reviewed previous approximations, thus removing the gradient expansion, the gauge-dependence in the derivation, and introducing the stochastic gravitonic sources besides the newly included scalar-induced tensorial IR dynamics\ylc{. The noise remains linear around a homogeneous background, so that gauge invariance holds only for gauges close to that background, and the backreaction on the noise relies on heuristics}. Although the sacrifice of previous limiting assumptions leads to an analytically intractable set of equations, it was possible to extract novel analytical insights and new regimes using this more complete non-linear methodology. This was demonstrated by using a new gauge to derive the stochastic $\Delta {\cal N}$ equations, making redundant usual but tedious (adjoint) Fokker-Planck treatments coming after stochastic equations of the field. For instance, ultra-slow-roll non-Gaussian tails were generated and could be understood with respect to previous works.

In Chapter~\ref{sec:SINR}, we reported on the successful implementation of the stochastic sources found for the BSSN equations of Numerical Relativity. At each time step, we managed to transit efficiently to Fourier space for the construction of a given instant’s spectral noise, while retaining algebraic cancellation of the linear kicks in the constraint equations. This is the first code of its kind incorporating full four-dimensional Numerical Relativity. Our benchmark scenarios\ylc{, run with a fixed coarse-graining scale $\sigma=1$ and without stochastic graviton sources,} were in great agreement with linear theory and gave the green light to investigate what has not been looked at before in full General Relativity for inflation. Specifically, we discussed and looked at various levels of scalar stochastic backreaction on the UV and extracted scalar-induced tensorial inhomogeneity to confirm their importance (only recently pointed out) in scenarios of interest for the community such as ultra-slow-roll. 
Leaving for future work further investigation, we provided preliminary proofs of understanding for the diffusion-dominated quasi-dS regimes and \ylc{suggested}, with both analytical and numerical insights, that past treatments of eternally inflating spacetimes \ylc{may be} too simplistic by neglecting what we identified as kinetic diffusion\ylc{. This regime also challenges the perturbative noise assumption itself}.

In the final chapter, Chapter~\ref{sec:SIbeyondGR}, the stochastic formalism ansatz was pushed to its edge by being applied to the widest range of relativistic and high-energy theories to date. The stochastic formalism is highly convenient in its exploitation of linear theory, because it can be parametrised without precise specification of the non-linear counterpart from which it derives. By using the parametrisation given by the EFT of Dark Energy, we found the Langevin spectra of all scalar-tensor theories with at most second-order equations of motion in time, in terms of the unspecified EFT coefficients. These coefficients were then used to map back the stochastic sources to specific nonlinear theories of interest. Stochastic Horndeski inflation is now accessible with our formalism\ylc{ at the level of the equations, yet to be simulated}, still free of historical assumptions such as the gradient expansion. Finally, we also proposed a generalisation of multiple stochastic fields in full General Relativity, leaving a whole new landscape of testable theories in their semiclassical inflationary phase. 
With this method and as we \ylc{argued} in this piece of work, \ylc{we see} no fundamental obstacle to generalising the procedure to any theory\ylg{ with gauge degrees of freedom} as long as it admits a classical set of scales and a valid perturbative one, whether \yld{it}\ylg{the latter} is classical or quantum\ylc{, within the same linear-noise ansatz}.

We believe that this thesis contributed to addressing the objectives stated in the introduction. 
We hope that the reader is now convinced that our formulation of stochastic inflation \ylc{can provide} a complementary and competitive predictive framework to QFT for both nonperturbative and perturbative dynamics. The latter case is to be emphasised, as \ylc{the} on-shell nonlinearities \ylc{generated by the classical evolution} are resolved with the stochastic approach only. We also re-emphasise that our generalisation is essentially bringing together QFT, the standard stochastic approach, lattice and numerical relativity frameworks to remove a long list of approximations in many combinations, related to scale, time, geometry and quantum dependence. These removals imply that we can now investigate \yld{safely }previously out-of-reach or unjustified regimes\ylc{, provided the scales separate into a perturbative and a classical sector, with a linear noise computed around a homogeneous background and a heuristic backreaction}, including scalar-induced tensorial anisotropies, the breaking of quasi-dS and vanilla inflation, the formation of strong over-densities, and the diffusive regime. \ylc{So far, our simulations have validated the construction against linear theory, while its nonlinear regimes have only been given first looks.} Out of the associated numerical simulations capable of scaling on large GPU systems, one will be able to extract high-precision fully multi-messenger and cross-correlated predictions of multiple kinds, such as 3-point functions. In our simulation-based inference era, those kinds of simulated statistics will prove key to getting more out of the same restricted data.

The future looks bright for the field, but one should not forget that some challenges remain. For instance, stochastic backreaction is still to be formulated consistently from first principles{, and the linear noise approximation could fail severely in cases where nonlinearity and classicality become relevant at very different times}. This race towards generality is healthy, but there is certainly discovery potential along the way by identifying unique smoking-gun signatures which may become accessible in the flood of new observational data being obtained about our universe.

\vspace{5\baselineskip}
\begin{center}
    * * * * *
\end{center}

%% file: appendix.tex
\ifSubfilesClassLoaded{\appendix}{}

\makeatletter
\let\oldchapter\chapter
\renewcommand\chapter{\thispagestyle{plain}%
    \global\@topnum\z@
    \@afterindenttrue
    \secdef\@chapter\@schapter}
\makeatother


\chapter{Closed-time path formalism}\label{app:CTP}\

The Bunch-Davies initial conditions can be written as a wavefunctional, useful for the Keldysh formalism often used in the path integral formulation of QFTCS,
\begin{equation}
    \langle{\cal R}^{\rm in}|0\rangle = \mathcal{N}\exp\left(-\frac{1}{2}\int d^3\mathbf{x}d^3\mathbf{y}\, {\cal R}^{\rm in}(\mathbf{x})\frac{\mathcal{E}(\mathbf{x},\mathbf{y})}{\gamma}{\cal R}^{\rm in}(\mathbf{y})\right),
    \label{eq:vacwave}
\end{equation}
where
\begin{equation}
    \frac{\mathcal{E}(t,\mathbf{x},\mathbf{y})}{\gamma} = \frac{z^2(t)}{\gamma}\int \frac{d^3\mathbf{k}}{(2\pi)^3}e^{i\mathbf{k}\cdot (\mathbf{x}-\mathbf{y})}\omega_k,
    \label{eq:BDamp}
\end{equation}
with $\omega_k^2 = k^2$ and ${\cal R}^{\rm in} = {\cal R}(t_{\rm in})$ for some initial time $t_{\rm in}$, often taken $t_{\rm in} \rightarrow -\infty$ for analytical computations.
The above wave functional is derived in the $k\gg aH $ limit, where the theory is free and massless, by following the usual procedures \cite{Weinberg95,Chen17diag, Kaya19}. In that sense, we recover harmonic quantum mechanics for each mode. In this work we will also consider a finite $t_{\rm in}$ sufficiently early such that eqn.~\eqref{eq:BDamp} is an accurate description.

The closed-time path formalism allows us to relate expectation values like eqn.~\eqref{eq:ininqcl} to the path integral generating functional of the theory \cite{Weinberg95,Chen17diag}
\begin{eqnarray}\label{eq:in-in_Z0}
Z[J_+,J_-] &=&  \int {\cal D} {\cal R}^{\rm in}_+ {\cal D} {\cal R}^{\rm in}_-\,\rho_0[{\cal R}^{\rm in}_+,{\cal R}^{\rm in}_-] \nonumber \\
 \!\!\!\!\times && \!\!\!\!\!\!\!\!\!\!\!\int\limits_{{\cal R}_+^{\rm in},\,{\cal R}_-^{\rm in}} \!\!{\cal D} {\cal R}_+ {\cal D} {\cal R}_-e^{\frac{i}{\gamma} \left[S[{\cal R}_+]-S[{\cal R}_-]+\int d^4x J_+(x){\cal R}_+(x) - \int d^4x J_-(x){\cal R}_-(x)\right]}\delta[{\cal R}_+^{\rm f}-{\cal R}_-^{\rm f}],   
\end{eqnarray}
normalised as $Z[J,J] = 1$\footnote{Unlike the often unspecified and ignored normalisation of the in-out generating functional used for scattering amplitudes.} and where $\rho_0$ is the initial condition density matrix, here that of the vacuum $
    \rho_0[{\cal R}^{\rm in}_+,{\cal R}^{\rm in}_-] = \langle {\cal R}^{\rm in}_+|0\rangle\langle 0 | {\cal R}^{\rm in}_-\rangle
$. Expectation values can be retrieved via  
\begin{eqnarray}
   \langle \bar{T}\hat{\cal R}(y_1)...\hat{\cal R}(y_m)T\hat{\cal R}(x_1)...\hat{\cal R}(x_n) \rangle &=& \nonumber \\
  && \hspace{-4cm}\frac{(i\gamma)^m\delta^m}{\delta J_-(y_1)... \delta J_-(y_m)}\frac{(-i\gamma)^n\delta^n}{i^n\delta J_+(x_1)... \delta J_+(x_n)} Z[J_+,J_-]\Big |_{J_+ =J_-= 0}.
  \label{eq:dZdJ}
\end{eqnarray}
In this expression, one understands the role played by the branches: $+$ fields yield the time-ordered contributions, and $-$ the anti-time-ordered ones.\footnote{In fact the path integral only knows about time-ordering and so the two branches are required for expressing real time expectation values.}

\textbf{Temporal boundaries.} The (non-squeezed) vacuum wave functional takes the Gaussian form of free harmonic states
\begin{equation}
    \langle{\cal R}^{\rm in}_{\pm}|0\rangle = \mathcal{N}\exp\left(-\frac{1}{2}\int d^3\mathbf{x}d^3\mathbf{y}\, {\cal R}^{\rm in}_{\pm}(\mathbf{x})\frac{\mathcal{E}(\mathbf{x},\mathbf{y})}{\gamma}{\cal R}^{\rm in}_{\pm}(\mathbf{y})\right),
    \label{eq:vacwaveapp}
\end{equation}
for some appropriate kernel $\mathcal{E}(\mathbf{x},\mathbf{y})$ where we explicitly displayed the dependence on $\gamma$.

At the other end of the two branches, the delta functional imposes the equality of ${\cal R}_+$ and ${\cal R}_-$ at some future final time $t_f$ after all times of interest; it can be left as a free parameter to be set as the largest time of interest or even taken to $+\infty$. This late time $\delta$-functional joining the $\pm$ branches and imposing equality of ${\cal R}_+$ and ${\cal R}_-$ at the final time $t_{\rm f}$ can be represented as
\begin{equation}
    \delta[{\cal R}_+^{\rm f}-{\cal R}_-^{\rm f}] \propto \exp\left[-\int d^3\mathbf{x} \, \frac{\left({\cal R}_+(\mathbf{x},t_{\rm f}) - {\cal R}_-(\mathbf{x},t_{\rm f})\right)^2}{C^2}\right],
\end{equation}
for some infinitesimal $C \rightarrow 0$. 
In most works, only the ${\cal R}_{\pm}$ path integrals are written and these temporal boundaries are confusingly kept implicit in the path ensemble. This is because they are only needed for finding the propagators of the theory, which can be found independently from the path integral, while the path integral is used for interacting calculations.

However, these terms can also be included, despite using only one path integral per branch. This can be achieved by using a Dirac delta function for each temporal boundary%
\begin{subequations}
    \label{eq:in-in_Z}
    \begin{align}
Z[0,0] &= \int {\cal D} {\cal R}_+ {\cal D} {\cal R}_-e^{\frac{i}{\gamma} \int_{\rm in}^{\rm f}dt\int d^3\mathbf{x}d^3\mathbf{y}\left\{{\cal S}[{\cal R}_+(x)]-{\cal S}[{\cal R}_-(x)]+A^{\rm f}[{\cal R}_{\pm},x]\right\} \delta[\mathbf{x}-\mathbf{y}] +A^{\rm in}[{\cal R}_{\pm},x,y]},\\
 A^{\rm in}[{\cal R}_{\pm},x,y] & = \frac{i}{2} \mathcal{E}(t,\mathbf{x},\mathbf{y})\left [{\cal R}_{+}(x){\cal R}_{+}(y)+{\cal R}_{-}(x){\cal R}_{-}(y)\right ]\delta(t-t_{\rm in}),  \\
 A^{\rm f}[{\cal R}_{\pm},x] & = i\gamma C^{-2}\left[{\cal R}_+(x) - {\cal R}_-(x)\right]^2\delta(t-t_{\rm f}).
    \end{align}
\end{subequations}
Under this form, the integral has no constraint on initial and final values as these have been transferred to the integrand. Note that $t_{\rm in}$ and $t_{\rm f}$ need to be finite for the Dirac delta functions to be defined. However, this can be extended using the identity commonly referred to as the $\epsilon$ prescription \cite{Weinberg95, Kaya19}
\begin{equation}
 \lim_{\substack{\epsilon \rightarrow0}} \int^{u_f}_{-\infty}\epsilon e^{\epsilon u}f(u)du = f(-\infty)\,.
 \end{equation}
This means that, for instance, the previous $\delta(t-t_{\rm in})$ can be substituted by $\epsilon e^{\epsilon t }$ and, when all time integrals are performed in any calculation such as in eqn.~\eqref{eq:dZdJ}, the limit $\epsilon \rightarrow 0$ can be taken. This can also be done for the future boundary and is very common in scattering studies where particles come from both past and future infinite boundaries \cite{Weinberg95}. For now, we will keep the finite time expression for simplicity and this will not alter any of our generalities.


\textbf{Propagators.} The propagators of the associated free theory in this $\pm$ basis of fields are usually denoted $G^{++}, G^{--}, G^{+-}$ and $G^{-+}$ and can be found in the literature, for instance in the cosmological case \cite{Fulling89, Chen17diag}. The propagators are obtained from the free theory action which can in general be written as
\begin{equation}
    {\cal S}[{\cal R}_+(x)]-{\cal S}[{\cal R}_-(x)] = \text{boundary terms} + \frac{1}{2}{\cal R}_+{\cal D}_0{\cal R}_+ -\frac{1}{2}{\cal R}_-{\cal D}_0{\cal R}_-,
\end{equation}
in terms of the free differential operator ${\cal D}_0$ and boundary terms arising from temporal integrations by parts.\footnote{In general, it is assumed that \emph{spatial} boundary terms do not contribute, either for topological reasons, e.g.~implying periodic boundary conditions, or because all dynamical fields are assumed to decay sufficiently fast at spatial infinity.} Such temporal boundary terms are usually ignored or hidden in the literature; the propagators are defined via a $2\times 2$ matrix which is the functional and matrix inverse
\begin{equation}
\left(\begin{array}{cc}\mathcal{D}_0& 0\\ 0&-\mathcal{D}_0\end{array}\right)_{\!\!(x)}
   \left(\begin{array}{cc}{ G^{++}}(x;x^\prime)&{ G^{+-}}(x;x^\prime)
              \\
          {G^{-+}}(x;x^\prime) & {G^{--}(x;x^\prime) }\end{array}\right)
    =\begin{pmatrix} \, \delta^{\ylg{(4)}}(x-x^\prime) & 0 \\0 &  \, \delta^{\ylg{(4)}}(x-x^\prime)
    \end{pmatrix}\,,
    \label{eq:green+-}
\end{equation}
and are identified with appropriate 2-point functions of $\hat{\cal R}$
\begin{eqnarray}
iG^{++}(x;x^\prime\,)
  &=& \langle 0| T \hat{\cal {\cal R}}(x)\hat{\cal {\cal R}}(x^\prime)| 0 \rangle
      = \theta(x^0\!-\!x^{\prime 0})iG^{-+}(x;x^\prime)
        + \theta(x^{\prime 0}\!-\!x^0)iG^{+-}(x;x^\prime),
\label{Feynman prop}
\\
iG^{--}(x;x^\prime\,)
  &=& \langle 0|\bar T \hat{\cal {\cal R}}(x)\hat{\cal {\cal R}}(x^\prime)| 0\rangle
      = \theta(x^0\!-\!x^{\prime 0})iG^{+-}(x;x^\prime)
        + \theta(x^{\prime 0}\!-\!x^0)iG^{-+}(x;x^\prime),
\label{antiFeynman prop}
\end{eqnarray}
where the Wightman functions $iG^{+-}$ and $iG^{-+}$ are
\begin{equation}
iG^{+-}(x;x^\prime)=\langle 0 |\hat{\cal {\cal R}}(x^\prime) \hat{\cal {\cal R}} (x)| 0 \rangle
\,,\quad
iG^{-+}(x;x^\prime)=\langle 0|\hat{\cal {\cal R}}(x)\hat{\cal {\cal R}}(x^\prime)| 0 \rangle
\,,
\label{Wightman functions}
\end{equation}
and $|0\rangle$ denotes the physical vacuum state. Note that not all of these in-in propagators are independent since, identically
\begin{equation}
    G^{++}+G^{--}-G^{-+}-G^{+-}=0\,.
    \label{eq:sumG}
\end{equation}

The diagonal elements of the propagator matrix in eqn.~\eqref{eq:green+-} are the time-ordered (Feynman) and anti-time ordered Green's functions of the operator $\mathcal{D}_0$, the first of which appears in usual S-matrix calculations in QFT, the second simply being the Hermitian conjugate of the first. The real difference of the \emph{in-in} formalism compared to the more commonly encountered \emph{in-out} formalism used in computations of scattering amplitudes is the off-diagonal elements in eqn.~\eqref{eq:green+-} which are solutions of the equation of motion:
$\mathcal{D}_0 \, G^{-+}=\mathcal{D}_0 \, G^{+-}=0$. The expressions of eqns.~\eqref{Feynman prop}, \eqref{antiFeynman prop} and \eqref{Wightman functions} do indeed provide a solution to eqn.~\eqref{eq:green+-}, but the necessity of the off-diagonal terms is not clear from this equation. In fact, solving the system in eqn.~\eqref{eq:green+-} requires the temporal boundary conditions coming from $\rho_0$ and the branch closure ($A^{\rm in}$ and $A^{\rm f}$). We include these temporal boundary conditions below as part of the operator to be inverted, assuming that they are set at finite times, and show that they indeed lead to eqns.~\eqref{Feynman prop}, \eqref{antiFeynman prop} and \eqref{Wightman functions}. We do this using the Keldysh fields eqn.~\eqref{eq:redefKeldysh} which make the physical meaning of the initial conditions more transparent and allow for a direct link to the classical-statistical approximation.

\textbf{Solving for Keldysh propagators.}
We assume the action is invariant under translations in 3-space, as is the initial Gaussian density matrix of eqn.~\eqref{eq:in-in_Z}. A Fourier space treatment is therefore ideal to simplify all quadratic forms. At the free theory level, one is therefore looking at simple harmonic quantum mechanics applied to each mode. 

To rewrite the free action, we will exploit the fact that differential operators have spatial symmetries $O(t, \mathbf{x},\mathbf{y}) = O(t, \mathbf{x}-\mathbf{y})$, so that any quadratic term can be written
\begin{equation}
\begin{aligned}
\int d^3\mathbf{x}d^3\mathbf{y}\phi(\mathbf{x})O(t,\mathbf{x},\mathbf{y})\phi(\mathbf{y}) & = \int_{\mathbf{x},\mathbf{y}}\int_{\mathbf{k}_1,\mathbf{k}_2,\mathbf{k}_3}O(t,\mathbf{k}_3)\phi_{\mathbf{k}_1}\phi_{\mathbf{k}_2}e^{i\mathbf{k}_1\cdot \mathbf{x}+i\mathbf{k}_1\cdot \mathbf{y}+i\mathbf{k}_3\cdot (\mathbf{x}-\mathbf{y})} \\
& = \int_{\mathbf{k}_1,\mathbf{k}_2,\mathbf{k}_3}O(t,\mathbf{k}_3)\phi_{\mathbf{k}_1}\phi_{\mathbf{k}_2}\delta[\mathbf{k}_1+\mathbf{k}_3]\delta[\mathbf{k}_2-\mathbf{k}_3] \\
& = \int_{\mathbf{k}}O(\mathbf{k})\phi_{\mathbf{k}}\phi_{-\mathbf{k}}.
\end{aligned}
\end{equation}
In particular, a local $O(t, \mathbf{x},\mathbf{y}) = O(t)\delta^{\ylg{(3)}}(\mathbf{x}-\mathbf{y})\rightarrow O(t)$ in Fourier $\mathbf{k}$-space.

The Keldysh basis redefinitions from eqn.~\eqref{eq:redefKeldysh} take the partition function from eqn.~\eqref{eq:in-in_Z} to the form
\begin{equation}
\begin{aligned}
    Z[0,0] & =  \int {\cal D R}^{\rm in}_{cl}{\cal D R}^{\rm in}_{q} \, \rho_0({\cal R}^{\rm in}_{cl}+\frac{\gamma}{2}{\cal R}^{\rm in}_q,{\cal R}^{\rm in}_{cl}-\frac{\gamma}{2}{\cal R}^{\rm in}_q )\\
    & \times\int_{\rm in}^{\rm f} {\cal D R}_{cl}{\cal D R}_{q}\exp{\left (\int d^4 x\frac{i}{\gamma}S[{\cal R}_{cl}+\frac{\gamma}{2}{\cal R}_{q}]-\frac{i}{\gamma}S[{\cal R}_{cl}-\frac{\gamma}{2}{\cal R}_{q}]\right)} \delta[{\cal R}_q^{\rm f}].
\end{aligned}
    \label{eq:generalcubic}
\end{equation}
and the initial state reads explicitly as
\begin{equation}\label{eq:density-Kelsysh1}
    \rho_0 = \exp\left[-\int d^3xd^3y\left(\frac{1}{\gamma}{\cal R}^{\rm in}_{cl}(\mathbf{x})\mathcal{E}(t_0,\mathbf{x},\mathbf{y}){\cal R}^{\rm in}_{cl}(\mathbf{y}) + \frac{\gamma}{4}{\cal R}^{\rm in}_{q}(\mathbf{x})\mathcal{E}(t_0,\mathbf{x},\mathbf{y}){\cal R}^{\rm in}_{q}(\mathbf{y}) \right)\right], 
\end{equation}
where $\mathcal{E}$ was defined in eqn.~\eqref{eq:BDamp}. To see how the quadratic operator in the exponent of eqn.~\eqref{eq:generalcubic} should be inverted, also taking into account the initial density matrix and final delta functional, we re-write the free partition function ($J_+=J_-=0$) in terms of the Keldysh fields eqn.~\eqref{eq:redefKeldysh} as
\begin{equation}\label{eq:freetotZ}
         Z_0[0,0]  =  \int {\cal D R}_{cl}{\cal D R}_{q}\exp{\left[\frac{i}{2}\int_{t_{\rm in}}^{t_{\rm f}} dt\int_{ \mathbf{k}} 
        \, \begin{pmatrix}
        {\cal R}_{cl}\, , \, 
            {\cal R}_{q}
        \end{pmatrix}_{t,\mathbf{k}} {\mathcal{D}}(t, k)\begin{pmatrix}
        {\cal R}_{cl} \\ 
            {\cal R}_{q}
        \end{pmatrix}_{t,-\mathbf{k}}\right]},  
\end{equation}
where ${\cal D} =  \bar{\mathcal{D}}_0+ \bar{\mathcal{D}}_b$ is the Fourier space sum of the Fourier free propagator matrix
\begin{equation}
    \bar{\mathcal{D}}_0(t,k)  = \begin{pmatrix}
        0 & \mathcal{D}_0  \\ 
             \mathcal{D}_0 & 0 
        \end{pmatrix},
        \label{eq:freetotZk}
\end{equation}
where $\mathcal{D}_0(t,k)  = -\partial_t\left (\frac{az^2}{c_s^2}\partial_t\cdot\right)-\frac{z^2}{a}k^2$, and the Fourier boundary contributions
\begin{equation}
    \bar{\mathcal{D}}_b(t, \mathbf{x},\mathbf{y})  = \begin{pmatrix}
        \frac{2i}{\gamma}\mathcal{E}(t,k) \delta[t-t_{\rm in}]& az^2(\delta[t-t_{\rm f}]-\delta[t-t_{\rm in}])\partial_t \\ 
             az^2(\delta[t-t_{\rm f}]-\delta[t-t_{\rm in}])\partial_t& \frac{i\gamma}{2}\mathcal{E}(t,k)\delta[t-t_{\rm in}] +i C\delta[t-t_{\rm f}]
        \end{pmatrix},
        \label{eq:Dbk}
\end{equation}
where $\mathcal{E}(t,k) = z^2(t)\omega_k$. The 2-point correlators of  ${\cal R}$ are obtained by finding the inverse of $\mathcal{D}(t,k)$ 
\begin{equation}
{\cal D}(t,\mathbf{k})
   \begin{pmatrix}
       K_k(t,t') & G^R_k(t,t^\prime) \\
       G^A_k(t,t^\prime) & H_k(t,t^\prime)
   \end{pmatrix}
    =\begin{pmatrix}  \, \delta[t-t^\prime] & 0 \\0 &   \, \delta[t-t^\prime]
    \end{pmatrix}.
    \label{inversek}
\end{equation}
The 2-point functions are\yld{ then given by}%
\begin{subequations}
    \begin{align}
        \langle {\cal R}_{cl}(t,\mathbf{k}) {\cal R}_{cl}(t',\mathbf{k}^\prime) \rangle = i K_k(t,t')\delta^{\ylg{(3)}}(\mathbf{k}+\mathbf{k}^\prime),\\
        \langle {\cal R}_{cl}(t,\mathbf{k}) {\cal R}_{q}(t',\mathbf{k}^\prime) \rangle = i G^R_k(t,t')\delta^{\ylg{(3)}}(\mathbf{k}+\mathbf{k}^\prime),\\
        \langle {\cal R}_{q}(t,\mathbf{k}) {\cal R}_{cl}(t',\mathbf{k}^\prime) \rangle = i G^A_k(t,t')\delta^{\ylg{(3)}}(\mathbf{k}+\mathbf{k}^\prime),\\
        \langle {\cal R}_{q}(t,\mathbf{k}) {\cal R}_{q}(t',\mathbf{k}^\prime) \rangle = i H_k(t,t')\delta^{\ylg{(3)}}(\mathbf{k}+\mathbf{k}^\prime).
    \end{align}
\end{subequations}
The system \eqref{inversek} involves 12 equations, 4 for each of $t = t_{\rm in}$, $t = t_{\rm f}$ and $t_{\rm in}< t <t_{\rm in}$ portions of the time interval $[t_{\rm in}, t_{\rm f}]$. Starting with the evolution times, eqn.~\eqref{inversek} yields the propagators' evolution%
\begin{subequations}
    \begin{align}
        \mathcal{D}_0(t,k)G^A_k(t,t^\prime) & =  \delta[t-t^\prime], \label{eq:sysa}\\
        \mathcal{D}_0(t,k)H_k(t,t^\prime) & = 0, \label{eq:sysb}\\
        \mathcal{D}_0(t,k)K_k(t,t^\prime) & = 0, \label{eq:sysc}\\
        \mathcal{D}_0(t,k)G^R_k(t,t^\prime) & =  \delta[t-t^\prime].\label{eq:sysd}
    \end{align}
\end{subequations}
Eqns.~\eqref{eq:sysa} and \eqref{eq:sysd} show that $G^A$ and $G^R$ are Green's functions, while $K$ and $H$ are solutions to the e.o.m. At $t=t_{\rm f}$ the boundary terms impose%
\begin{subequations}
    \begin{align}
\partial_tG^A_k(t,t^\prime)\displaystyle|_{t = t_{\rm f}} & = 0,  \\
\partial_tH_k(t,t^\prime)\displaystyle|_{t = t_{\rm in}} & = 0, \\
\partial_t K_k(t,t')\displaystyle|_{t=t_{\rm f}} +i C^{-2}G^A_k(t_{\rm f},t^\prime) & = 0, \\
\partial_t G_k^R(t,t')\displaystyle|_{t=t_{\rm f}} +i C^{-2}H_k(t_{\rm f},t^\prime) & = 0. 
    \end{align}
\end{subequations}
Recalling that $C\rightarrow 0$ to represent the delta functional at $t_{\rm f}$ these become%
\begin{subequations}
    \begin{align}
\partial_tG^A_k(t,t^\prime)\displaystyle|_{t = t_{\rm f}} & = 0, \label{eq:sysfa} \\
\partial_tH_k(t,t^\prime)\displaystyle|_{t = t_{\rm in}} & = 0, \label{eq:sysfb}\\
G^A_k(t_{\rm f},t^\prime) & = 0, \label{eq:sysfc}\\
H_k(t_{\rm f},t^\prime) & = 0. \label{eq:sysfd}
    \end{align}
\end{subequations}
Furthermore, at $t=t_{\rm in}$ we have%
\begin{subequations}
    \begin{align}
    \partial_tG^A_k(t,t^\prime)\displaystyle|_{t = t_{\rm in}} & = \frac{2i}{\gamma a_{\rm in}}\omega_kK_k(t_{\rm in},t^\prime), \label{eq:sysia}\\
\partial_tH_k(t,t^\prime)\displaystyle|_{t = t_{\rm in}} & = \frac{2i}{\gamma a_{\rm in}}\omega_kG^R_k(t_{\rm in},t^\prime), \label{eq:sysib}\\
\partial_tK_k(t,t^\prime)\displaystyle|_{t = t_{\rm in}} & =\frac{i\gamma}{2a_{\rm in}}\omega_kG^A_k(t_{\rm in},t^\prime), \label{eq:sysic}\\
\partial_tG^R_k(t,t^\prime)\displaystyle|_{t = t_{\rm in}} & = \frac{i\gamma}{2a_{\rm in}}\omega_kH_k(t_{\rm in},t^\prime).\label{eq:sysid}
    \end{align}
\end{subequations}
Eqns~\eqref{eq:sysa} - \eqref{eq:sysid} completely determine the 2-point functions, as we now show: Since  $\mathcal{D}_0(t,k)\varphi(t)=0$ is a second-order O.D.E. in time, one sees from eqns.~\eqref{eq:sysb}, \eqref{eq:sysfb} and \eqref{eq:sysfd} that $$H_k(t,t') = 0\,.$$  
Looking now at the boundary conditions in eqns.~\eqref{eq:sysfa} and \eqref{eq:sysfc}, we can see that these initial conditions together with ${\cal D}_0(t,k)G_k^A(t,t') = 0$ for the portion $t>t^\prime$, are enough to find $G_k^A(t,t')=0$ for $t>t'$. Hence: 
\begin{equation}
G_k^A(t,t') = \text{Advanced Green's function}\,.
\end{equation}
Similarly, from the other end of the time interval we learn from eqns.~\eqref{eq:sysib} and \eqref{eq:sysid} (recall that $H_k=0$) that $G_k^R(t,t') = 0$ on the $t<t^\prime$ range and hence
\begin{equation}
G_k^R(t,t') = \text{Retarded Green's function}\,.
\end{equation}
On the other halves of their respective time ranges, these Green's functions do not vanish because of the Dirac delta function at $t=t^\prime$. Integrating eqns.~\eqref{eq:sysa} and \eqref{eq:sysd} from $t=t'-\epsilon$ to $t=t'+\epsilon$ and taking the limit $\epsilon\rightarrow 0$, we find%
\begin{subequations}
    \begin{align}
\frac{az^2}{c_s^2}\partial_t G^A_k\displaystyle|_{t = t^\prime} &  = 1, \label{eq:sysmida} \\
G^A_k(t^\prime,t^\prime) & = 0, \label{eq:sysmidb}\\
\frac{az^2}{c_s^2}\partial_t G^R_k\displaystyle|_{t = t^\prime} & =  -1, \label{eq:sysmidc} \\
G^R_k(t^\prime,t^\prime) & = 0.\label{eq:sysmidd}
    \end{align}
\end{subequations}
Finding the solution of ${\cal D}_0 G^{R/A}=0$ with these conditions then yields $G^R_k(t,t')$ and $G^A_k(t,t')$ for all values of $(t, t') \in [t_{\rm in}, t_{\rm f}]$. For instance, this leads to eqns.~\eqref{eq:defGKGR} and \eqref{eq:FGdS} in a quasi-dS framework.

Note that at no point has the Bunch-Davies vacuum been used so far: the Green's functions encode dynamics only. The Advanced Green's function $G^A(t,t')$ (a propagator from the final temporal point) can then be expressed by including a $\theta[t^\prime-t]$ step function factor, while the Retarded Green's function 
$G^R(t,t')$ (propagating from the initial time) can be expressed by including a $\theta[t-t^\prime]$ step function. We note that $G^R$ and $G^A$ are not independent since $G^R(t,t') = G^A (t',t)$.

Knowledge of the Green's function $G^A$ or $G^R$ along with the initial state (or, more generally, the density matrix) then determines $K_k(t_{\rm in},t')$ and $\partial_tK_k(t,t^\prime)\displaystyle|_{t = t_{\rm in}}$ via 
eqns.~\eqref{eq:sysia} and \eqref{eq:sysic} which for our initial state can be written as%
\begin{subequations}
    \begin{align}
    K_k(t_{\rm in},t^\prime) & =-i \frac{\gamma a_{\rm in}}{2\omega_k}\partial_tG^A_k(t,t^\prime)\displaystyle|_{t = t_{\rm in}} \,, \label{eq:K_ini1}\\
\partial_tK_k(t,t^\prime)\displaystyle|_{t = t_{\rm in}} & =i\frac{\gamma \omega_k}{2a_{\rm in}}G^A_k(t_{\rm in},t^\prime)\,. \label{eq:K_ini2} 
    \end{align}
\end{subequations}
These equations provide the initial conditions at $t_{\rm in}$ which determine $K_k(t,t')$ as the appropriate solution of eqn.~\eqref{eq:sysa} for all values of $(t, t') \in [t_{\rm in}, t_{\rm f}]$. When eqns.~\eqref{eq:K_ini1} and \eqref{eq:K_ini2} are evaluated at $t^\prime = t_{\rm in}$, we can use the previous equations to confirm that $K_k(t_{\rm in},t_{\rm in}) =  -i\frac{\gamma}{2\mathcal{E}(t,k)}$ and $\partial_t K_k(t,t_{\rm in})\displaystyle|_{t=t_{\rm in}} = 0$ as they should for our initial state. Note that, as can be seen from eqns.~\eqref{eq:sysia} and \eqref{eq:sysic}, the cl-cl part of the initial density matrix determines the initial amplitude of $K$ while the q-q part determines the initial velocity of $K$. This fact will also be relevant in the formulation of the classical-statistical approximation below.  

The free propagators $G^R$, $G^A$ and $K$ determined above are used in the Feynman diagrams of perturbation theory. We now discuss vertices and the classical-statistical approximation to the dynamics, both linear and non-linear.    

\clearpage

\chapter{Classical statistical approximation}\label{app:CSA}

If one performs an additional integration by parts in eqn.~\eqref{eq:freetotZ}, the free ${\cal R}_{cl}{\cal D}_0{\cal R}_q$ term can be turned into another ${\cal R}_{q}{\cal D}_0{\cal R}_{cl}$ term, showing that ${\cal R}_{cl}$ is the on-shell part of the field (the one satisfying the equation of motion in the $\gamma \rightarrow 0$ limit). Indeed, in this limit, ${\cal R}_q$ can be integrated out and the exponential becomes a functional Dirac function evaluated at the EOM's left-hand side. This remains valid with interaction terms. This in turn means that ${\cal R}_q$ is the off-shell part of the dynamics.\footnote{Not necessarily quantum effects, just any trajectory violating the strict EOM.} This is why the Keldysh basis is so useful and meaningful.

\textbf{Classical statistical evolution. }
We have seen that our quasi-dS scenario has a semi-classical approximation only valid for the modes which have crossed the Hubble radius. Let us set this aside for now and consider what happens if we perform this approximation regardless.

In this limit, we take $\gamma \rightarrow 0$ up to second order and so only the leading vertex ${\cal R}_q{\cal R}_{cl}^2$ remains. This amounts to on-shell evolution only. 
Overall, the stochastic partition function is written as
\begin{equation}
\begin{aligned}
        Z^{sto}[0,0] & =  \int{\cal D R}_{cl}{\cal D R}_{q}\exp\Bigg(i\int d^4 x d^4 y 
    \, {\cal R}_{q}(x)\left(\mathcal{D}_0{\cal R}_{cl}+3\lambda(t){\cal R}_{cl}^2\right)_{y}\delta[x-y]\Bigg) \\
    & \times \exp\Bigg[- \frac{1}{2}\int d^4 x d^4 y  \, \begin{pmatrix}
        {\cal R}_{cl} 
            {\cal R}_{q}
        \end{pmatrix}_{t,\mathbf{x}} 
        \begin{pmatrix}
        4{\cal A}_{\rm in}(x,y) & 0 \\
        0 & \gamma^2{\cal A}_{\rm in}(x,y) 
        \end{pmatrix}
        \begin{pmatrix}
        {\cal R}_{cl} \\ 
            {\cal R}_{q}
        \end{pmatrix}_{t,\mathbf{y}}\Bigg] ,
\end{aligned}
    \label{eq:Semi-Cl-partitionR}
\end{equation}
where we have kept only the initial conditions for simplicity, and defining 
\begin{equation}
    \mathcal{A}_{\rm in}(x,y) =  \frac{1}{2\gamma}\delta(t_x-t_{\rm in})\delta(t_y-t_{\rm in})\mathcal{E}(t_x,\mathbf{x},\mathbf{y}).
\end{equation}
The Hubbard-Stratonovich transformation \cite{Hubbard59, Stratonovich57} finally allows us to see that this path integral is stochastic
\begin{equation}
   \exp\Bigg( - \frac{1}{2}\int d^4 x d^4 y 
    \, {\cal R}_{q}(x)\mathcal{A}_{\rm in}(x,y){\cal R}_{q}(y) \Bigg) = \int {\cal D} \xi {\rm P}[\xi|\mathcal{A}_{\rm in}] \exp\Bigg(-i\int d^4 x{\cal R}_{q}\xi\Bigg),
\end{equation}
where ${\rm P}[\cdot|\Sigma]$ is the probability density of a Gaussian field with 2-point function $\Sigma$
\begin{equation}
    {\rm P}[\xi|\Sigma] \propto \exp\Bigg( - \frac{1}{2}\int d^4 x d^4 y 
    \, \xi(x)\Sigma^{-1}(x,y)\xi(y) \Bigg).
    \label{eq:probaxi}
\end{equation}
The path integral can now be calculated and, setting $\gamma = 1$ gives
\begin{equation}
 Z^{sto}[0,0] =  \int {\cal D R}_{cl}{\cal D}\xi \,{\rm P}[{\cal R}_{cl}|4{\cal A}_{\rm in}]{\rm P}[\xi|{\cal A}_{\rm in}]\,\delta[\mathcal{D}_0{\cal R}_{cl}+3\lambda(t){\cal R}_{cl}^2-\xi] .
\end{equation}
The above partition function encodes evolution of the classical equations of motion, enforced by the delta functional, of the original initial conditions encoded in $\rho_0$. In fact, $\xi$ encodes the momentum initial condition and similarly another random variable could be used to include the bulk's initial condition from the other Gaussian on ${\cal R}_{cl}$.  This is the classical statistical approximation to quantum dynamics, also described in \cite{Radovskaya2021} for instance.

\textbf{Stochastic EOM and correlators. }
Another way of writing this stochastic path integral is its associated Langevin equation 
\begin{equation}
    \mathcal{D}_0\tilde{\cal R}+\frac{\delta S^{(3)}}{\delta {\cal R}}[\tilde{\cal R}]=\tilde{\xi}.
    \label{eq:langevinlin}
\end{equation}
where $\tilde{\xi}$ has statistics such that, for the free theory, the partition in eqn.~\eqref{eq:probaxi} has the same two-point function as $\tilde{\cal R}$. This corresponds to considering the lowest order in the coupling, which satisfies $ {\cal D}_0\tilde{\cal R}^{(1)}  = \tilde{\xi}$, and assumes that the stochastic two-point function 
\begin{equation}
\begin{aligned}
    K^{stoch}(x, x') & \equiv \langle \tilde{\cal R}^{(1)}(x) \tilde{\cal R}^{(1)}(x') \rangle, \\
    & =\int \int d^4yd^4y' G^R(x,y)G^R(x',y')\langle \tilde{\xi}(y)\tilde{\xi}(y') \rangle,\\
    \end{aligned}
\end{equation}
matches exactly the full quantum theory's $K$ defined earlier.
Note that this $G^R$ is the same as in the full theory with $\gamma = 1$. 

One can check that the diagrammatics of this theory are the same as the in-in theory where the $\gamma^{p\geq 1}$ vertices have been dropped. This has been achieved in \cite{vanderMeulen2007, Garbrecht:2013coa, Garbrecht:2014dca} and we illustrate it below for the case of the bispectrum.  One can get the next order cubic non-linearity from
\begin{equation}
   {\cal D}_0{\tilde{\cal R}^{(2)} = -\frac{\delta S^{(3)}}{\delta {\cal R}}[\tilde{\cal R}^{(1)}}] = -\epsilon^{ijk}\theta_{i}\left[[\theta_{j}\tilde{\cal R}^{(1)}]\theta_{k}[\tilde{\cal R}^{(1)}]\right],
      \label{eq:langevinlin2}
\end{equation}
where the $\theta_i$ are linear differential operators from the cubic interaction $\theta_1 {\cal R}\theta_2{\cal R}\theta_3 {\cal R}\subset S^{(3)}$. $\epsilon^{ijk}$ is a $\pm 1$ tensor summing symmetries ($\epsilon^{i_1jk}=\epsilon^{i_2jk}$) and depending on how many integrations by parts are required to vary $\theta_i{\cal R}$ in the least action principle (e.g. $\nabla^p$ and $\partial_t^p$ require $p$ of them). This stochastic equation is solved using the previous solution
\begin{equation}
\begin{aligned}
\tilde{\cal R}^{(2)}(x) 
& = -\epsilon^{ijk}\int d^4 y G^R(x,y) \theta_{i}\left[\theta_{j}[\tilde{\cal R}^{(1)}]\theta_{k}[\tilde{\cal R}^{(1)}\right](y) \\
& = -\epsilon^{ijk}\int d^4 y\int d^4 z_1\int d^4 z_2 G^R(x,y) \theta_{i,y}\left[\theta_{j,y}[G^R](y,z_1)\theta_{k,y}[G^R](y,z_2)\right] \tilde{\xi}(z_1)\tilde{\xi}(z_2).
\end{aligned}
\end{equation}
For instance, the 3-point function at lowest order (4th order in $\tilde{\xi}$) is given by
\begin{equation}
\resizebox{\textwidth}{!}{
$\begin{aligned}
     \langle \tilde{\cal R}^{(2)}(x_1) \tilde{\cal R}^{(1)}(x_2) \tilde{\cal R}^{(1)}(x_3) \rangle & = - \epsilon^{ijk}\int d^4 y_1 d^4 y_2 d^4 y_3 d^4 z_1 d^4 z_2 \theta_{i,y}\left [\theta_{j,y}[G^R](y,z_1)\theta_{k,y}[G^R](y,z_2) \right ]\\ & \times G^R(x_1,y_1)G^R(x_2,y_2) G^R(x_3,y_3)  \langle\tilde{\xi}(z_1)\tilde{\xi}(z_2)\tilde{\xi}(y_2)\tilde{\xi}(y_3)\rangle \\
     & = - 2\epsilon^{ijk}\int d^4 y G^R(x_1,y)\theta_{i,y}\left[\theta_{i,y}[F](y,x_2)\theta_{j,y}\theta_{k,y}[F](y,x_3)\right]  \\
     &  = -2\int d^4 y \theta_{1,y}[G^R](x_1,y)\theta_{2,y}[F](y,x_2)\theta_{3,y}[F](y,x_3) +\text{perm}.
\end{aligned}$
}
\label{eq:demoSTObi}
\end{equation}
once the Isserlis theorem has been applied and keeping only connected contributions and where the integration by parts has been reversed. Note that the full expression also involves summing over $(x_1, x_2,x_3)$ permutations. This result in real space matches the in-in formalism tree-level formula with $FFG$ vertices only. This will be true at all orders in $\tilde{\xi}$ for any $n$-point correlator \cite{Radovskaya2021,vanderMeulen2007} and can also be generalised to non-cubic vertices with one $G$ or more (if odd) if one wants to look at next orders in $\gamma$.

By now, the notion of what constitutes a 'classical' contribution should be clear: a contribution without 'quantum' features, where 'quantum' is what cannot be generated from the classical equation of motion or any of its stochastic modifications.

This section also gives the opportunity to highlight that a path integral does not have to be quantum. This is justified because Schrödinger-like equations (and so path integrals) can be written for Hilbert spaces of stochastic states \cite{Bounakis20,rigopoulos_fluctuation-dissipation_2013,vastola_stochastic_2019}. Going from the stochastic EOM to the path integral is known as the MSRJ procedure \cite{MSRformula}.

\clearpage

\chapter{Redefinition contributions to the leading bispectrum}\label{app:appRedef}
Eqn.~\eqref{eq:eqH3} actually constitutes the action of the redefined field ${\cal R}_n$ from \cite{Maldacena_2003, chen_observational_2007}. The total bispectrum is

\begin{equation}
    \langle \hat{{\cal R}}_{\mathbf{k}_1}\hat{{\cal R}}_{\mathbf{k}_2}\hat{{\cal R}}_{\mathbf{k}_3}\rangle =     \langle \hat{{\cal R}}_{n,\mathbf{k}_1}\hat{{\cal R}}_{n,\mathbf{k}_2}\hat{{\cal R}}_{n,\mathbf{k}_3}\rangle + \langle f(\hat{{\cal R}}_{n,\mathbf{k}_1})\hat{{\cal R}}_{n,\mathbf{k}_2}\hat{{\cal R}}_{n,\mathbf{k}_3}\rangle+\text{perm},
\end{equation}
where { $f(\hat{{\cal R}}_{\mathbf{k}_1}) =\frac{\varepsilon_2}{4 c_s^2} {\cal R}^2+\frac{1}{c_s^2 H} {\cal R} \dot{\cal R} +\frac{1}{2 a^2 H}\left[(\partial {\cal R})(\partial \chi)-\partial^{-2}\left(\partial_i \partial_j\left(\partial_i {\cal R} \partial_j \chi\right)\right)\right] +...
$ is the redefinition chosen in \cite{chen_observational_2007}.}
Only the first term is usually kept at leading order in slow-roll in the super-Hubble limit where derivatives damp the other terms. However, around horizon crossing, these are perfectly legitimate contributions. We do not consider them because they are fully accounted for in a classical framework, at least at \yld{linear}\ylg{leading} order.

\clearpage

\chapter{Linearised BSSN dictionary}\label{app:dicBSSN}
Using ADM and scalar SVT notations introduced in the introduction, we derive
\begin{itemize}
    \item ${X} = a^{-2} [1+2\Phi -\frac{2}{3}\nabla^2 E]. $\\
            \textit{proof.}  $^{(3)}g = det(^{(3)}g_{ij})= a^6 [1-6\Phi +2(E_{,11}+E_{,22}+E_{,33})] = a^6 [1-6\Phi +2\nabla^2 E]$ \\
            Hence $[^{(3)}g]^{-1/3} = a^{-2} [1+2\Phi -\frac{2}{3}\nabla^2 E].$
            
     \item ${\tilde{\gamma}_{ij}} = \delta_{ij}+2(\partial_i\partial_j-\frac{1}{3}\delta_{ij}\nabla^2)E,$ and ${\tilde{\gamma}^{ij}} = \delta^{ij}-2\delta^{im}\delta^{jn}(\partial_m\partial_n-\frac{1}{3}\delta_{mn}\nabla^2)E.$ \\
      \textit{proof.}  $X{^{(3)}g_{ij}} = a^{-2} [1+2\Phi -\frac{2}{3}\nabla^2 E] a^2 [(1-2\Phi)\delta_{ij}+\partial_{ij}E] = \delta_{ij}+2E_{,ij}-\frac{2}{3}\delta_{ij}\nabla^2E$. The inverse metric is obtained at first order with $\tilde{\gamma}^{ik}\tilde{\gamma}_{kj} = {\delta}{^i{}_j}$.
      
      \item ${K} =  -3H+\kappa$ using the definition from eqn.~\eqref{eq:defkappa}: $\kappa = 3(\frac{\dot{\Phi}}{\alpha_b}+H\Psi)-\Delta \chi$.
      \item ${\tilde{A}_{ij}} =  \frac{1}{\alpha_b}(\partial_i\partial_j-\frac{1}{3}\delta_{ij}\nabla^2)(B-\dot{E}).$ \\
      \textit{proof.} Similarly $K_{ij}^{TF} = -(\partial_i\partial_j-\frac{1}{3}\delta_{ij}\nabla^2)\chi$  \\
      so  $X K_{ij}^{TF} =a^{-2} [1+2\Phi -\frac{2}{3}\nabla^2 E]  \frac{a^2}{\alpha_b}(\partial_i\partial_j-\frac{1}{3}\delta_{ij}\nabla^2)(B-\dot{E}) = \frac{1}{\alpha_b}(\partial_i\partial_j-\frac{1}{3}\delta_{ij}\nabla^2)(B-\dot{E}).$
    \item ${\alpha} = \alpha_b(1+\Psi)$ and ${\beta_{i}} = a^2B_{,i}$.
    \item ${\tilde{\Gamma}^i_{jk}} = \delta^{il} E_{,lkj} - \frac{1}{3}\nabla^2[E_{,j}{\delta}{^i_k}+E_{,k}{\delta}{^i_j}-E_{,l}\delta^{il}\delta_{jk}].$\\
    \textit{proof.} $\tilde{\Gamma}^i_{jk} = \frac{1}{2}[ \delta^{ij}-2\delta^{im}\delta^{jn}(\partial_m\partial_n-\frac{1}{3}\delta_{mn}\nabla^2)E][2\partial_j(\partial_l\partial_k-\frac{1}{3}\delta_{lk}\nabla^2)+2\partial_k(\partial_j\partial_l-\frac{1}{3}\delta_{jl}\nabla^2)-2\partial_l(\partial_j\partial_k-\frac{1}{3}\delta_{jk}\nabla^2)]E 
    = \delta^{ij}[\partial_j(\partial_l\partial_k-\frac{1}{3}\delta_{lk}\nabla^2)+\partial_k(\partial_j\partial_l-\frac{1}{3}\delta_{jl}\nabla^2)-\partial_l(\partial_j\partial_k-\frac{1}{3}\delta_{jk}\nabla^2)]E
    = E_{,lkj}\delta^{il}-\frac{1}{3}\nabla^2[E_{,j}{\delta}{^i_k}+E_{,k}{\delta}{^i_j}-E_{,l}\delta^{il}\delta_{jk}]+o(E).$
    \item ${\tilde{\Gamma}^i} = \frac{4}{3}\delta^{ij}\nabla^2E_{,j}. $\\
       \textit{proof.} Computing $\tilde{\gamma}^{jk} \tilde{\Gamma}^i_{jk} = \delta^{jk} \tilde{\Gamma}^i_{jk},$ as the previous expression is first order.
    \item ${R_{ij}} = (\nabla^2\delta_{ij} + \partial_i\partial_j)\Phi,$ and  ${R^{TF}_{ij}} = (\partial_i\partial_j-\frac{1}{3}\delta_{ij}\nabla^2)\Phi.$\\
    \textit{proof.} One can show that ${^{(3)}{R}{^i{}_j}} = \Delta \Phi \delta^k{}_j+\Phi^{,k}{}_{,j}$ and lower the index.
\end{itemize}

\clearpage
\chapter{Degrees of freedom in small generalised synchronous gauges}\label{app:dofsyncgauge}

To understand this freedom, let us consider the effect of the following gauge transformation
\begin{equation}
        \tilde{x}^i  \longrightarrow x^i+\partial^i\lambda,  \qquad
\tilde{t} \longrightarrow t + \xi^0,
\end{equation}
where $\lambda(t,\mathbf{x}) = f(\mathbf{x})-g(\mathbf{x})(t-t^*)/a(t)^2 $ and $\xi^0(\mathbf{x}) = -g(\mathbf{x})$, consistent with the $\xi^\mu=(\xi^0,\partial^i\lambda)$ notation of eqn.~\eqref{eq:gaugetransform}.
Under this transformation the lapse and the shift perturbations are unchanged%
\begin{subequations}
    \begin{align}
\Psi \longrightarrow &  \Psi-\dot{\xi}^0 = \Psi, \\
B  \longrightarrow & B + \xi^0/a^2-\dot{\lambda} = B -g/a^2+g/a^2 = B.
    \end{align}
\end{subequations}
The conclusion is different for $E$ and $\chi$%
\begin{subequations}
    \begin{align}
E(t, \mathbf{x})\longrightarrow &  E(t,\mathbf{x}) - \lambda(t, \mathbf{x})  =E(t,\mathbf{x})- f(\mathbf{x})+g(\mathbf{x})(t-t^*)/a(t)^2,\\
\chi(t, \mathbf{x}) \longrightarrow &  \chi(t, \mathbf{x})  -\xi^0(t, \mathbf{x}) = \chi(t, \mathbf{x})+g( \mathbf{x}),
    \end{align}
\end{subequations}
which evaluated at $t^*$ shows that $f(\mathbf{x})$ and $g(\mathbf{x})$ can be chosen to set the gauge with any desired value of $\chi(t^*,\mathbf{x})$ and $E(t^*,\mathbf{x})$. Thus, a small 3+1 slicing still has some gauge freedom.

\clearpage
\chapter{Randomness with a power spectrum}\label{app:random}
As a reminder, the Fourier transform ${\cal F}$ and its inverse are defined with a backward normalisation in QFT. 

We define RFFTN (resp. IRFFTN) as a discrete Fourier transform (resp. its inverse) on a positive 3D hemisphere with no normalisations ($RFFTN\{IRFFTN\}=N^3$).

We want to draw a random 3D process with power spectrum $|{f}_{\mathbf{k}}|^2$ defined from QFT as
 \begin{equation}
  \hat{f}(x) = {\cal F}^{-1}\{ {f}_{\mathbf{k}}\hat{a}_{\mathbf{k}}\} + h.c.,
\label{eq:ModesExpApp}
 \end{equation}
There are a few subtleties that must be accounted for.
\begin{itemize}
\item  Assuming that units are such that $[\hat{f}(x)] = M^0$, we understand from the consequently unitless two-point function
\begin{equation}
\langle \hat{f}(x)^2\rangle = {\cal F}^{-1}\{ |{f}_{\mathbf{k}}|^2 \},
\end{equation}
that annihilation and creation operators hide units and that ${f}_{\mathbf{k}}$ is not a simple Fourier density ($[{f}_{\mathbf{k}}] \neq M^{-3}$) but such that $[{f}_{\mathbf{k}}] = M^{-3/2}$.

\item  Building a real stochastic Fourier expansion does not require the addition of the Hermitian conjugate\footnote{If we were doing so, the stochastic approach would require a factor $\sqrt{2}$ due to the commutativity of the latter and not of  QFT.} because we are dealing with numbers (we can find $\boldsymbol{\alpha}$ such that $\boldsymbol{\alpha}_{-\mathbf{k}}= \boldsymbol{\alpha}^*_{\mathbf{k}}$ while $\hat{a}_{-\mathbf{k}}\neq \hat{a}^\dagger_{\mathbf{k}}$) 
\begin{equation}
  \boldsymbol{f}(x) = {\cal F}^{-1}\{ \tilde{f}_{\mathbf{k}}\boldsymbol{\alpha}_{\mathbf{k}}\}, 
\end{equation}
where $\tilde{f}_{\mathbf{k}} = f_k$ (resp. $f_k^*$) if $\mathbf{k}$ is in the $k_z>0$ hemisphere (resp. $k_z<0$ hemisphere). This is essentially saying that we impose a Hermitian symmetry as (I)RFFTN does.
We can compute the coincident two-point function%
\begin{subequations}
    \begin{align}
        \langle \boldsymbol{f}(x)^2 \rangle & = \int\frac{d^3\mathbf{k}_1}{{(2\pi)}^3}\frac{d^3\mathbf{k}_2}{{(2\pi)}^3}\tilde{f}_{\mathbf{k}_1}\tilde{f}_{\mathbf{k}_2}\langle \boldsymbol{\alpha}_{\mathbf{k}_1}\boldsymbol{\alpha}_{\mathbf{k}_2}\rangle\\
        &  = \int\frac{d^3\mathbf{k}_1}{{(2\pi)}^3}\frac{d^3\mathbf{k}_2}{{(2\pi)}^3}\tilde{f}_{\mathbf{k}_1}\tilde{f}_{-\mathbf{k}_2}\langle \boldsymbol{\alpha}_{\mathbf{k}_1}\boldsymbol{\alpha}_{\mathbf{k}_2}^*\rangle \\
 & = \frac{\yld{C^2}\ylg{C}}{(2\pi)^3}\int\frac{d^3\mathbf{k}} {{(2\pi})^3}|f_k|^2 = \frac{\yld{C^2}\ylg{C}}{(2\pi)^3}\langle \hat{f}(x)^2\rangle,
    \end{align}
\end{subequations}
where the integral in $\mathbf{k}_2$ has been reversed and the Hermitian symmetry of $\mathbf{k}$ assumed, before applying eqn.~\eqref{eq:stocomm}.
 The right normalisation is the one that matches the stochastic and quantum correlation functions, thus $C = \yld{(2\pi)^{3/2}}\ylg{(2\pi)^{3}}$.

\item  To generate the $\boldsymbol{\alpha}_{\mathbf{k}}s$, it is possible to use
\begin{equation}
     \boldsymbol{\alpha}_{\mathbf{k}}/\yld{C}\ylg{C^{1/2}} = \boldsymbol{s}_{\mathbf{k}} = \frac{1}{\sqrt{N}^3}\text{RFFTN}\{\Delta(\mathbf{x})\}(\mathbf{k}), 
\end{equation}
when $N$ is big enough and taking each $\Delta(\mathbf{x})$ as independent random normal variables on the grid \cite{peligrad_central_2019}. This will indeed imply the statistics of eqn.~\eqref{eq:stocomm} and their Hermitian symmetry.
\end{itemize}
This leads to the final strategy%
\begin{subequations}
    \begin{align}
   \boldsymbol{f}(x)
 & =  \displaystyle \left(\frac{dk}{2\pi}\right)^{3}\times\text{IRFFTN}\left\{ \left(\frac{dk}{2\pi}\right)^{-\frac{3}{2}}f_{\mathbf{k}}\times (2\pi)^{3/2}\boldsymbol{s}_{\mathbf{k}}\right\} \\
 & = \displaystyle\left(\frac{dk}{\sqrt{2\pi}}\right)^{\frac{3}{2}}\times\text{IRFFTN}\left\{ f_{\mathbf{k}}\boldsymbol{s}_{\mathbf{k}}\right\}.
    \end{align}
\end{subequations}

\clearpage
\chapter{Looking too deep in the UV}\label{app:UVdep}

We demonstrate here the breakdown of our methods when including modes that are too sub-Hubble. When varying $\sigma$, one can study the scaling of the simulation independently from the resolution by rescaling the resolution accordingly, for instance by rescaling $L$. Thanks to this, gradients remain perfectly accurate and cannot violate the constraints with amplitude greater than second order in perturbation theory. Figure \ref{fig:sigdep} showcases such an experiment with the case of quadratic inflation. It shows the ratio of the constraint violation and the typical amplitude of a first-order violation of the constraints $|{\cal H}|/[{\cal H}^{(1)}]$ and $|{\cal M}|/[{\cal M}]$. Given that our method provides violation at second order in perturbation theory, this ratio is a direct comparison of the second and first orders.

Cutting more of the UV makes the contrasts on the lattice decrease, which explains why the constraints do too. Inversely, the contrasts and so the constraints increase when including more and more UV modes, which diverge in this limit. In particular, the momentum's contrast in the geodesic gauge has usually been the first to break perturbation theory in the UV in this work.

When increasing $\sigma$, one has to stop before exiting the perturbative regime. Indeed, the second-order perturbations increase in amplitude, and the further you move in the UV, the bigger the effect. Our constraints are still satisfied to first order but with a smaller gap between each order of perturbation theory. As this gap decreases by including more UV modes, satisfying the constraints at linear order as we do here is no longer sufficient and another approach must be used. However, this does not seem to be the real problem as there is no certainty about the physicality of those deep sub-Hubble modes. While some of us argue that a regularisation should be performed, for instance using the adiabatic method \cite{Durrer09, pla23}, we will simply state that substituting those diverging modes into Einstein's equations and so evolving them in NR would probably lead to unphysical non-perturbative non-linearities. 

\begin{figure}
    \centering
\includegraphics[width=0.65\linewidth]{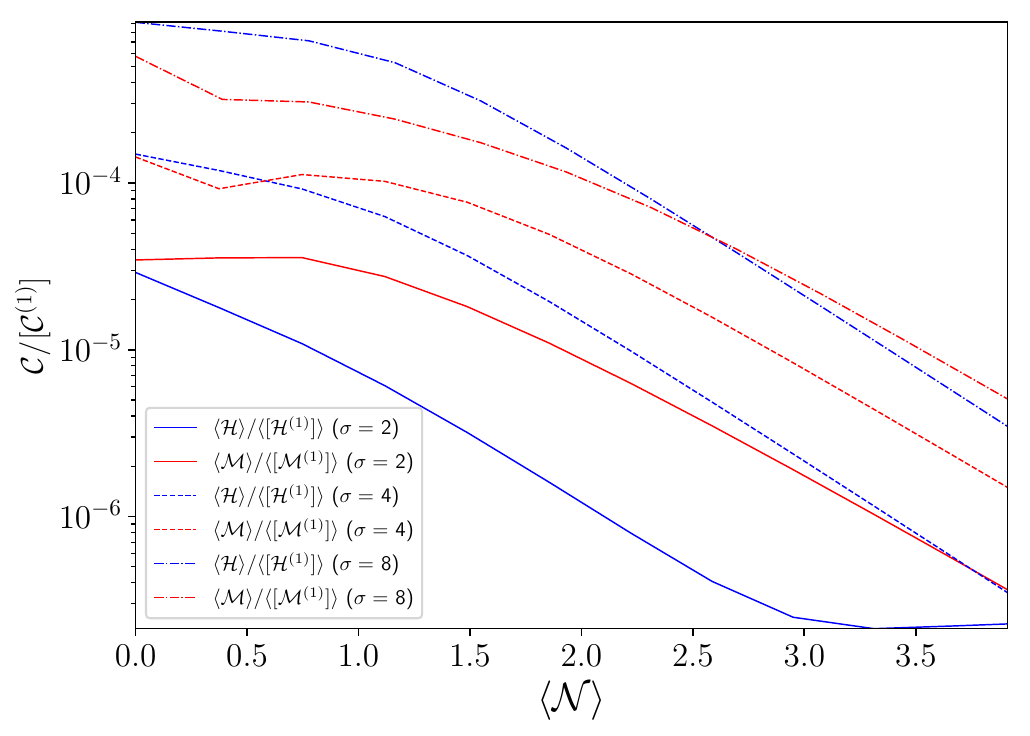}
    \caption{Mean constraint violation to mean first order magnitude ratio  ($\langle|{\cal C}|/[{\cal C}^{(1)}]\rangle$, ${\cal C }= {\cal H}, {\cal M}$) for quadratic inflation with different UV cutoffs. For $\sigma = 1$ we set $L = 128/H^{\circledast}$ and adapt $L$ when changing $\sigma$ to keep their product a constant. \protect\footnotemark}
    \label{fig:sigdep}
\end{figure}

\footnotetext{ Note that the Hamiltonian $\sigma=2$ final ratio flattens because of the constraint violation reaching numerical accuracy (smaller $\sigma$ means smaller perturbations and so violation).}
\clearpage

\chapter{Cosmological extractions}\label{app:extract}
\vizir's postprocessing makes use of diverse relations to extract diagnostics and observables. The most important relations are as follows\footnote{Note the corrected numerical factors compared to \cite{Launay2025}.}, $\langle \rangle_{\scriptscriptstyle \square}$ denoting a box averaging%
\begin{subequations}
    \begin{align}
    \langle {\cal N} \rangle_{\scriptscriptstyle \square} & = -\frac{1}{2}\ln {\langle X \rangle_{\scriptscriptstyle \square}}, \\
    \Phi_k & = \frac{1}{2}(a^2 X_k-1)+\frac{1}{4}\sum_{i,j=1}^3 \frac{k_ik_j}{k^2}(\tilde{\gamma}_{ij}-\delta_{ij}),\\
    E_k & = -\frac{3}{4}\sum_{i,j=1}^3 \frac{k_ik_j}{k^4}(\tilde{\gamma}_{ij}-\delta_{ij}),\\
   {\cal R} &= \Phi -\frac{\langle K \rangle_{\scriptscriptstyle \square}}{3\langle \alpha \rangle_{\scriptscriptstyle \square}\langle \Pi \rangle_{\scriptscriptstyle \square}}(\phi-\langle \phi\rangle_{\scriptscriptstyle \square}).
    \end{align}
\end{subequations}

\clearpage

\chapter{(U)SR noise amplitude}\label{app:appD}

The Fourier amplitude of $\boldsymbol{S}$ is exactly the same as for ${\cal S}_{\cal R}/H^2$ in e-folding time
\begin{equation}
    {{ S}}_k = {\cal R}_k \frac{\partial^2 W_k}{\partial{\cal N}_b^2}+\left[2 \frac{\partial{\cal R}_k}{\partial{\cal N}_b} + (3-\varepsilon_1-\varepsilon_2) {\cal R}_k\right ]\frac{\partial W_k}{\partial{\cal N}_b}.
    \label{eq:eqSk}
\end{equation}
We choose to work \ylg{here} with a Heaviside window $W_k = \Theta(\sigma aH-k)$, far enough from Hubble crossing thanks to $\sigma \ll 1$, the derivative of which is the Dirac distribution $\partial_{{\cal N}_b}W_k = \partial_{{\cal N}_b} (\sigma aH)\,\delta(\sigma aH-k)$.
\ylc{The second derivative of the window function in eqn.~\eqref{eq:eqSk} is not convenient and numerically challenging if we have such a sharp and distributional object.
Instead, we can choose to rewrite eqn.~\eqref{eq:eqSk} generally as}
\begin{equation}
    {{ S}}_k = \ylc{\partial_{{\cal N}_b}\left({\cal R}_k \partial_{{\cal N}_b} W_k\right) + }\left[\partial_{{\cal N}_b}{\cal R}_k + (3-\varepsilon_1-\varepsilon_2) {\cal R}_k\right ] \partial_{{\cal N}_b} W_k.
    \label{eq:eqSk2}
\end{equation}
\ylc{The total derivative is a consequence of not kicking in phase-space but in kinetics only, as explained in Section~\ref{sec:SIGRusuallimit}.
If we were to revert to phase-space kicks and alter the IR momentum $\Pi^>_Y$ relation $\partial_{{\cal N}_b} Y^>=\Pi^>_Y$ with that suggested $\cR_k\,\partial_{{\cal N}_b}W_k$ kick, then the $(3-\varepsilon_1-\varepsilon_2){\cal R}_k$ kick would disappear from the $\partial_{{\cal N}_b} \Pi^>_Y$ equation, leaving only the $\partial_{{\cal N}_b}{\cal R}_k$ kick term in the bracket.

Let us take the example where the field $Y$ is the local efold number ${\cal N}$ with momentum the lapse $\alpha$ (as in Section~\ref{sec:SIGRng} for the uniform field gauge). This redefinition implies that the local efold, and so the Hamiltonian constraint now dynamical, should also be kicked (similarly to the scalar field in a different gauge).}
\ylc{In coarse-grained linear theory, linearising eqns.~\eqref{eq:eqUniFieldefolds} and \eqref{eq:eqUniFieldlapse} around $\alpha=1$ gives $\partial_{{\cal N}_b}\delta{\cal N}^> = (1-\varepsilon_1/3)\,\delta\alpha^>$ and $\partial_{{\cal N}_b}\delta\alpha^> = -(3-\varepsilon_1-\varepsilon_2)\,\delta\alpha^> + \boldsymbol{S}$, so that $\delta{\cal N}^>$ follows the super-Hubble MS friction. The factor $1-\varepsilon_1/3$ comes from the long-wavelength Hamiltonian constraint and is set to $1$ in the following.
Redefining the lapse as $\delta\tilde{\alpha}^> \equiv \delta\alpha^> - \boldsymbol{\Sigma}_{\cal R}$, with $\boldsymbol{\Sigma}_X \equiv {\cal F}^{-1}\{\tilde{X}_{\mathbf{k}}\,\partial_{{\cal N}_b}W_k\,\boldsymbol{\alpha}_{\mathbf{k}}\}$, this becomes the phase-space stochastic system
\begin{equation}
    \partial_{{\cal N}_b}\delta{\cal N}^> =\delta \tilde{\alpha}^>+\boldsymbol{\Sigma}_{\cal R}, \qquad
    \partial_{{\cal N}_b}\delta\tilde{\alpha}^> = -(3-\varepsilon_1-\varepsilon_2)\,\delta \tilde{\alpha}^> + \boldsymbol{\Sigma}_{\partial{\cal R}},
    \label{eq:aux}
\end{equation}
where $\boldsymbol{\Sigma}_{\partial{\cal R}}$ has amplitude $\partial_{{\cal N}_b}{\cal R}_k$: the $(3-\varepsilon_1-\varepsilon_2){\cal R}_k$ term of eqn.~\eqref{eq:eqSk2} has indeed cancelled against the friction acting on $\boldsymbol{\Sigma}_{\cal R}$.}

The solution of ${\cal R}_k$ can be found by solving eqn.~\eqref{eq:MScosmic} in the case of slow-roll and ultra-slow roll. It turns out that the solution is identical despite different equations \cite{Kinney05} and written in conformal time $\tau$ as
\begin{equation}
    {\cal R}_k(\tau) = \frac{H}{\sqrt{4\varepsilon_1\ylc{(\tau)} M_{Pl}^2k^3}}(1+i k \tau)e^{-ik\tau}.
\end{equation}
It is now convenient to verify that\ylc{, keeping the time dependence of $\varepsilon_1$,}
\begin{equation}
    \partial_{{\cal N}_b}{\cal R}_k \delta(\sigma aH-k) = \left(\frac{1}{aH} \partial_{\tau}{\cal R}_k \right )_{-k\tau=\sigma}= \ylc{\frac{\varepsilon_2}{2}{\cal R}_k}-\frac{\ylc{H}\sigma^2}{\sqrt{4\varepsilon_1 M_{Pl}^2k^3}}e^{i\sigma},
\end{equation}
\yld{which is negligible compared to the ${\cal R}_k$ term in eqn.~\eqref{eq:eqSk2} in the $\sigma\ll1$ limit.}\ylc{where the ${\cal O}(\sigma^2)$ term is negligible.
In SR, $\partial_{{\cal N}_b}{\cal R}_k \simeq 0$ at leading order in slow-roll parameters.
In the phase-space formulation of eqn.~\eqref{eq:aux}, this means that the momentum kick $\boldsymbol{\Sigma}_{\partial{\cal R}}$ vanishes and only the e-fold kick $\boldsymbol{\Sigma}_{\cal R}$ remains, which is equivalent to a kick $S_{\rm eff}=3{\cal R}_k$ in the original formulation (kicks in momentum only).

In USR, $\varepsilon_2=6$ gives $\partial_{{\cal N}_b}{\cal R}_k \simeq 3{\cal R}_k$, so that $\boldsymbol{\Sigma}_{\partial{\cal R}}\simeq3\boldsymbol{\Sigma}_{\cal R}$ and eqn.~\eqref{eq:aux}, now with an anti-friction $+3$, can be integrated: a kick at ${\cal N}_b=u$ yields $\delta{\cal N}^>={\cal R}_k\,e^{3({\cal N}_b-u)}$, to which the e-fold kick $\boldsymbol{\Sigma}_{\cal R}$ only contributes the non-growing ${\cal R}_k$.
It can thus be neglected as long as the USR phase lasts more than about an e-fold after the crossing of the modes of interest, $e^{-3({\cal N}_{\rm end}-{\cal N}_k)}\ll1$, leaving the single momentum kick $S_{\rm eff}=3{\cal R}_k$, the same noise as in SR.\footnote{\ylc{In practice, the simulations of our example in Section~\ref{sec:SIGRng} would, as checked numerically, show slightly enhanced non-Gaussianity with the noise added to the e-folds evolution, and thus to the Hamiltonian constraint. We chose not to include it for two reasons. First, we stick to our original publication \cite{Launay24}, with the present warning as an erratum, the point of that section being only a proof of concept in this new gauge. Second, a proper treatment of the Hamiltonian (or phase-space) coarse-graining in full GR is currently in preparation.}}
}

Finally, the amplitude of the Gaussian noise is found by performing the inverse Fourier transform of the spectrum \cite{Starobinsky_stochastic_1988} with spherical invariance
\begin{equation}
    \sqrt{\langle\boldsymbol{S}({\cal N}_b)^2\rangle} = \left(\frac{|\yld{(3-\varepsilon_1-\varepsilon_2) {\cal R}_{\sigma a H}}\ylc{S_{\rm eff}}|^2}{6\pi^2}\frac{d}{d{\cal N}_b}(\sigma a H)^3 \right)^{\frac{1}{2}}
    \simeq \frac{1}{\sqrt{2\varepsilon_1}}\frac{3}{2\pi}\frac{H({\cal N}_b)}{M_{Pl}},
\end{equation}
using $|{\cal R}_{\sigma aH}| \simeq \frac{H}{\sqrt{4\varepsilon_1 M_{Pl}^2k^3}}$\yld{ and in the limit where $\varepsilon_1 \simeq 0 $ and $\varepsilon_2 \simeq 0$ (SR)  or $\varepsilon_2 \simeq 6$ (USR)}.
\clearpage

\chapter{Numerical and stochastic schemes}\label{app:schemes}
\begin{sloppypar}
At each time step and for the vector of fields ${\cal X}$, such that $\partial_t {\cal X}= {\cal G}[{\cal X}(t),t,\mathbf{x}]= {\rm BSSN}[{\cal X}(t),t,\mathbf{x}]+{\cal I}[t,\mathbf{x}]$, then the 4th-order Runge-Kutta (RK4) scheme is written as
\end{sloppypar}
$${\cal X}(t+\Delta t, \mathbf{x})={\cal X}(t, \mathbf{x})+\Bigg(\frac{1}{6}{\cal K}_1+\frac{1}{3}{\cal K}_2+\frac{1}{3}{\cal K}_3+\frac{1}{6}{\cal K}_4\Bigg)\Delta t,$$
where each term is written sequentially as%
\begin{subequations}
    \begin{align}
    {\cal K}_1& ={\cal G}[{\cal X}(t),t,\mathbf{x}],\\
    {\cal K}_2&={\cal G}[{\cal X}(t)+{\cal K}_1\frac{\Delta t}{2},t+\frac{\Delta t}{2},\mathbf{x}],\\
    {\cal K}_3& = {\cal G}[{\cal X}(t)+{\cal K}_2\frac{\Delta t}{2},t+\frac{\Delta t}{2},\mathbf{x}],\\
    {\cal K}_4& ={\cal G}[{\cal X}(t)+{\cal K}_3{\Delta t},t+{\Delta t},\mathbf{x}].
    \end{align}
\end{subequations}
The subtlety here is the stochastic contribution of the r.h.s. ${\cal I}[t,\mathbf{x}] =  {\cal F}^{-1}\{{\cal S}_k(t)\boldsymbol{\alpha_{\mathbf{k}}}\}$ writing generically ${\cal S}_k(t) = a^{\cal X}_k(t)\ddot{W}_k(t)+b^{\cal X}_k(t)\dot{W}_k(t)$.
Note that with stochastic backreaction, ${\cal I}$ becomes a function of the fields as well.
In stochastic PDE language, applying the RK4 method to the stochastic contribution does not give the same convergence for the stochastic part \cite{WarningSRK}, rather that of the Euler–Maruyama method.
In the future, we could use an appropriate stochastic scheme such as those given in \cite{Rossler10} and used in \cite{cruces_stochastic_2022}.
This would require significant edits in the \textsc{AMReX} core, while we expect very little impact if the sources are kept in the perturbative regime.

\clearpage

\chapter{{Stochastic convergence}}\label{app:conv}

The \textit{absolute constraint magnitudes} $[{\cal H}]$ and $[{\cal M}]$ were defined in eqn.~\eqref{eq:abscons}.
These account for the typical magnitude before cancellation. As demonstrated by the lower panels in Figures \ref{fig:quad_contrast} and \ref{fig:usr_contrast}, physicality of the simulations requires $|{\cal H}|/[{\cal H}]\ll 1$ and $|{\cal M}|/[{\cal M}]\ll 1$.
More quantitatively in the case of perturbative inflation, $[{\cal H}]$ is of order $H^2$ from the background, and $[{\cal M}]$ contains linear order gradients. Satisfying the constraints when our self-consistent (linear order) stochastic noise acts therefore means that $|{\cal H}|/[{\cal H}]$ should be of second order and $|{\cal M}|/[{\cal M}]$ first order.

The convergence of the constraints assesses the robustness of the numerical scheme: increasing resolution for the same physics should yield converging accuracy. In Figure \ref{fig:Convergence}, we show that stochastic convergence is attained as we increase $N$: as the sources are first order in the noise, each stochastic kick will violate the constraints at second order as reflected by the amplitudes' early plateau. Once the modes have all entered, the usual inflation-driven exponential decay of the constraints is observed. 
The convergence test loses its meaning later in the evolution as it unfairly disfavours higher resolution runs: either a numerical accuracy floor is reached faster (left) or physics has more small scales to enhance  unphysical noise (right) (see \cite{Launay2025} for the spectral study of the constraints during ultra slow-roll). This does not impact the validity of our simulations as they still satisfy the constraints but rather makes the design of a meaningful and fair convergence test more difficult.

A convergence test can also focus on the extraction of quantities of interest and complete the verification where the constraint convergence is blinded.
We provide here the convergence of the extracted spectra ${\cal R}^{\rm NR}$ to the linear evolution in Figure \ref{fig:specConvergence}. The extreme linearity of quadratic inflation makes the MS solution the reference that a higher resolution should approach, which is what we find.

\begin{figure}[!ht]
    \centering
\includegraphics[width=0.65\linewidth]{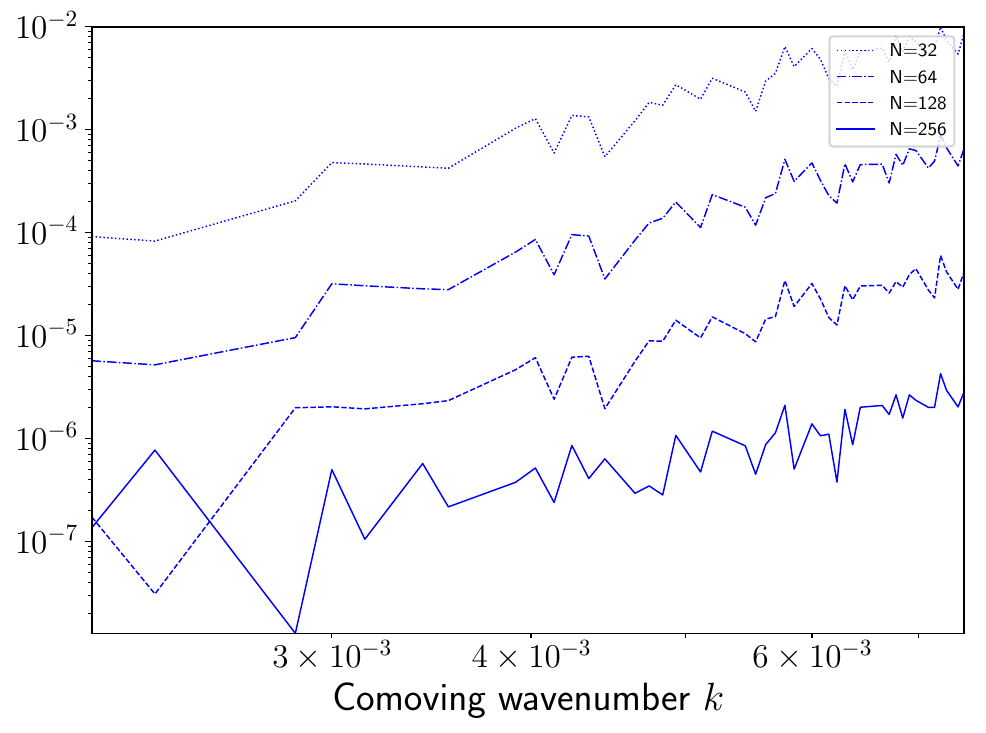}
\caption{ { Convergence test in a linear scenario (quadratic inflation) for the relative difference of the NR curvature perturbation and the Mukhanov-Sasaki equation result $|1-\Delta[{\cal R}^{\rm NR}]/\Delta[{\cal R}]$ across multiple runs with $N=32$ (dotted), $N=64$ (dash-dotted), $N=128$ (dashed), and $N = 256$ (solid). $k_{\text{max}}^{\cal R}/dk$ is set to $8$ across all runs, giving about $\ln 8\simeq 2$ e-folds of stochastic sourcing. }}
\label{fig:specConvergence}
\end{figure}

\begin{figure*}[!ht]
    \centering
    \begin{subfigure}{0.45\textwidth}
\includegraphics[width=\linewidth]{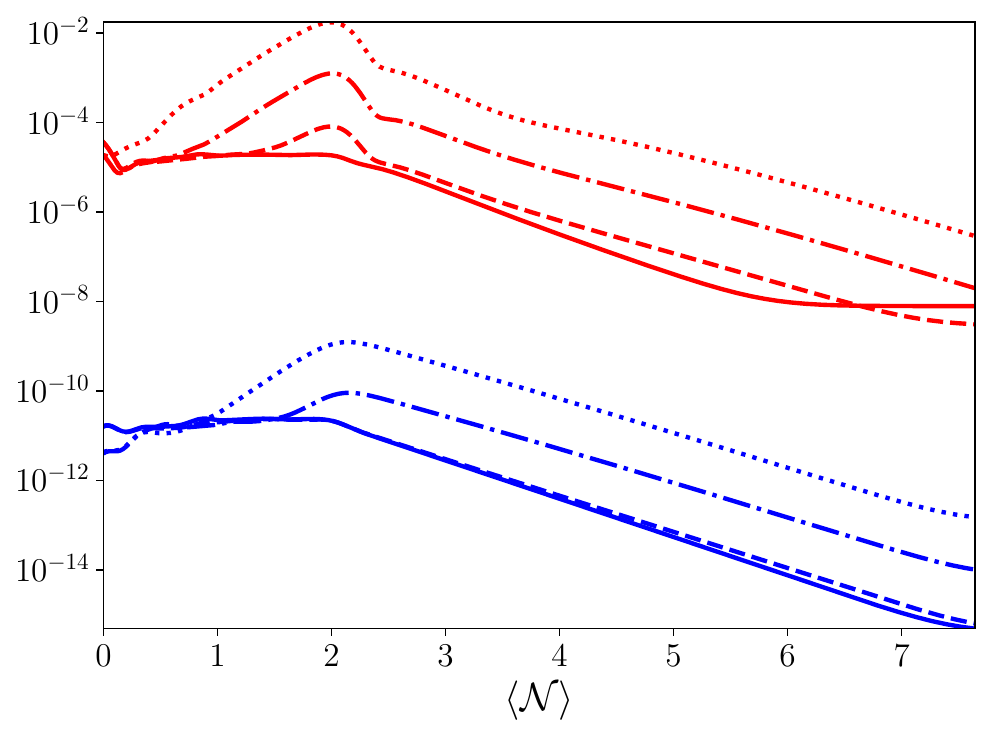}
\caption{Quadratic inflation}
    \end{subfigure}
\begin{subfigure}{0.45\textwidth}
\includegraphics[width=\linewidth]{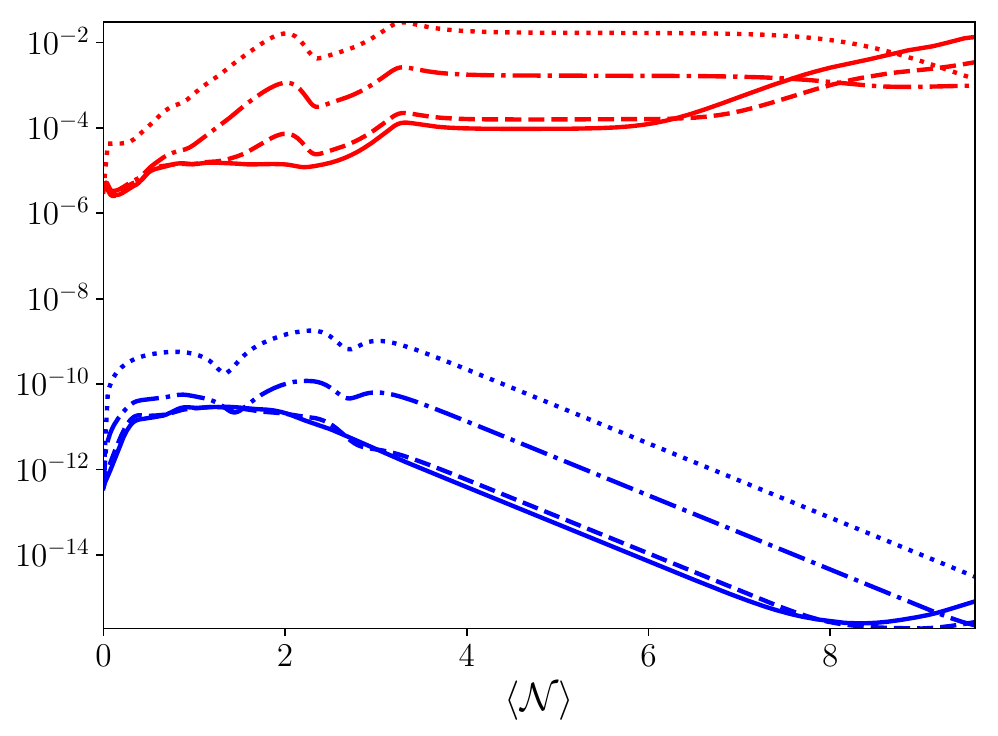}
\caption{Inflection inflation}
    \end{subfigure}
\caption{ { Convergence tests for the mean hamiltonian (blue) and momentum (red) relative constraints ($\langle|{\cal C}|/[{\cal C}]\rangle$, ${\cal C }= {\cal H}, {\cal M}$) across multiple runs with $N=32$ (dotted), $N=64$ (dash-dotted), $N=128$ (dashed), and $N = 256$ (solid). $k_{\text{max}}^{\cal R}/dk$ is set to $8$ across all runs, giving about $\ln 8\simeq 2$ e-folds of stochastic sourcing. See Appendix \ref{app:conv} for interpretation.}}
\label{fig:Convergence}
\end{figure*}

\clearpage

\chapter{Coefficients of the Horndeski EOMs}\label{app:horndeski}
The terms below were calculated in \cite{KYY11, gao_inflation_2011}
\begin{eqnarray}
{\cal G}_{\mu \nu}^2&=&-\frac{1}{2}K_X\nabla_\mu\phi\nabla_\nu\phi-\frac{1}{2}Kg_{\mu\nu},
\\
{\cal G}_{\mu \nu}^3&=&\frac{1}{2}G_{3,X}\Box\phi\nabla_\mu\phi\nabla_\nu\phi
+\nabla_{(\mu}G_3\nabla_{\nu)}\phi-\frac{1}{2}g_{\mu\nu}\nabla_\lambda G_3\nabla^\lambda\phi ,
\\
{\cal G}_{\mu \nu}^4&=&
G_4G_{\mu\nu}-\frac{1}{2}G_{4,X}R\nabla_\mu\phi\nabla_\nu\phi
-\frac{1}{2}G_{4,XX}\left[(\Box\phi)^2-(\nabla_\alpha\nabla_\beta\phi)^2\right]\nabla_\mu
\phi\nabla_\nu\phi
\cr&&
-G_{4,X}\Box\phi\nabla_\mu\nabla_\nu\phi
+G_{4,X}\nabla_\lambda\nabla_\mu\phi\nabla^\lambda\nabla_\nu\phi
+2\nabla_\lambda G_{4,X}\nabla^\lambda\nabla_{(\mu}\phi\nabla_{\nu)}\phi
\cr&&
-\nabla_\lambda G_{4,X}\nabla^\lambda\phi\nabla_\mu\nabla_\nu\phi
+g_{\mu\nu}\left(G_{4,\phi}\Box\phi-2XG_{4,\phi\phi}\right)
\cr&&
+g_{\mu\nu}\biggl\{-2G_{4,\phi X}\nabla_\alpha\nabla_\beta\phi\nabla^\alpha\phi\nabla^\beta\phi
+G_{4,XX}\nabla_\alpha\nabla_\lambda\phi\nabla_\beta\nabla^\lambda\phi\nabla^\alpha\phi\nabla^\beta\phi
\cr&&
+\frac{1}{2}G_{4,X}\left[(\Box\phi)^2-(\nabla_\alpha\nabla_\beta\phi)^2\right]\biggr\}
+2\Bigl[G_{4,X}R_{\lambda(\mu}\nabla_{\nu)}\phi\nabla^\lambda\phi
\cr&&
-\nabla_{(\mu}G_{4,X}\nabla_{\nu)}\phi\Box\phi\Bigr]
-g_{\mu\nu}\left[
G_{4,X}R^{\alpha\beta}\nabla_\alpha\phi\nabla_\beta\phi-\nabla_\lambda G_{4,X}\nabla^\lambda\phi
\Box\phi\right]
\cr&& +G_{4,X}R_{\mu\alpha\nu\beta}\nabla^\alpha\phi\nabla^\beta\phi
-G_{4,\phi}\nabla_\mu\nabla_\nu\phi-G_{4,\phi\phi}\nabla_\mu\phi\nabla_\nu\phi
\cr&&
+2G_{4,\phi X}\nabla^\lambda\phi\nabla_\lambda
\nabla_{(\mu}\phi\nabla_{\nu)}\phi
-G_{4,XX}\nabla^\alpha\phi\nabla_\alpha\nabla_\mu\phi
\nabla^\beta\phi\nabla_\beta\nabla_\nu\phi,
\\
{\cal G}_{\mu \nu}^5&=&
G_{5,X}R_{\alpha\beta}\nabla^\alpha\phi\nabla^\beta\nabla_{(\mu}\phi
\nabla_{\nu)}\phi
-G_{5,X}R_{\alpha(\mu}\nabla_{\nu)}\phi \nabla^\alpha\phi\Box\phi
\cr&&
-\frac{1}{2}G_{5,X}R_{\alpha\beta}\nabla^\alpha\phi\nabla^\beta\phi\nabla_\mu\nabla_\nu\phi
-\frac{1}{2}G_{5,X}R_{\mu\alpha\nu\beta}\nabla^\alpha\phi\nabla^\beta\phi\Box\phi
\cr&&
+G_{5,X}R_{\alpha\lambda\beta(\mu}\nabla_{\nu)}\phi\nabla^\lambda\phi
\nabla^\alpha\nabla^\beta\phi
+G_{5,X}R_{\alpha\lambda\beta(\mu}\nabla_{\nu)}\nabla^\lambda\phi\nabla^\alpha\phi\nabla^\beta\phi
\cr&&
-\frac{1}{2}\nabla_{(\mu}\left[G_{5,X}\nabla^\alpha\phi\right]\nabla_\alpha\nabla_{\nu)}\phi\Box\phi
+\frac{1}{2}\nabla_{(\mu}\left[G_{5,\phi}\nabla_{\nu)}\phi\right]\Box\phi
\cr&&
-\nabla_\lambda\left[G_{5,\phi}\nabla_{(\mu}\phi\right]\nabla_{\nu)}\nabla^\lambda\phi
\cr&&
+\frac{1}{2}\left[
\nabla_\lambda \left(G_{5,\phi}\nabla^\lambda\phi\right)-\nabla_\alpha\left(G_{5,X}\nabla_\beta\phi\right)
\nabla^\alpha\nabla^\beta\phi
\right]\nabla_\mu\nabla_\nu\phi
\cr&&
+\nabla^{\alpha}G_5\nabla^{\beta}\phi R_{\alpha(\mu\nu)\beta}
-\nabla_{(\mu}G_5G_{\nu)\lambda}\nabla^\lambda\phi
\cr&&
+\frac{1}{2}\nabla_{(\mu}G_{5,X}\nabla_{\nu)}\phi\left[
(\Box\phi)^2-(\nabla_\alpha\nabla_\beta\phi)^2\right]
\cr&&
-\nabla^\lambda G_5R_{\lambda(\mu}\nabla_{\nu)}\phi
+\nabla_\alpha\left[G_{5,X}\nabla_\beta\phi\right]\nabla^\alpha\nabla_{(\mu}\phi
\nabla^\beta\nabla_{\nu)}\phi
\cr&&
-\nabla_\beta G_{5,X}\left[
\Box\phi\nabla^\beta\nabla_{(\mu}\phi-\nabla^\alpha\nabla^\beta\phi
\nabla_\alpha\nabla_{(\mu}\phi\right]\nabla_{\nu)}\phi
\cr&&
+\frac{1}{2}\nabla^\alpha\phi\nabla_\alpha G_{5,X}\left[
\Box\phi\nabla_\mu\nabla_\nu\phi -\nabla_\beta\nabla_\mu\phi\nabla^\beta\nabla_\nu\phi\right]
\cr&&
-\frac{1}{2} G_{5,X} G_{\alpha\beta}\nabla^\alpha\nabla^\beta\phi\nabla_\mu\phi\nabla_\nu\phi
-\frac{1}{2}G_{5,X}\Box\phi\nabla_\alpha\nabla_\mu\phi\nabla^\alpha\nabla_\nu\phi
\cr&&
+\frac{1}{2}G_{5,X}(\Box\phi)^2\nabla_\mu\nabla_\nu\phi
+\frac{1}{12}G_{5,XX}\bigl[(\Box\phi)^3-3\Box\phi(\nabla_\alpha\nabla_\beta\phi)^2
\cr&&
+2(\nabla_\alpha\nabla_\beta\phi)^3\bigr]\nabla_\mu\phi\nabla_\nu\phi
+\frac{1}{2}\nabla_\lambda G_5 G_{\mu\nu}\nabla^\lambda\phi
\cr&&
+g_{\mu\nu}\Biggl\{
-\frac{1}{6}G_{5,X}\left[(\Box\phi)^3-3\Box\phi(\nabla_\alpha\nabla_\beta\phi)^2
+2(\nabla_\alpha\nabla_\beta\phi)^3\right]
+\nabla_\alpha G_5R^{\alpha\beta}\nabla_\beta\phi
\cr&&
-\frac{1}{2}\nabla_\alpha\left(G_{5,\phi}\nabla^\alpha\phi\right)\Box\phi
+\frac{1}{2}\nabla_\alpha\left(G_{5,\phi}\nabla_\beta\phi\right)\nabla^\alpha\nabla^\beta\phi
-\frac{1}{2}\nabla_\alpha G_{5,X}\nabla^\alpha X\Box\phi
\cr&&
+\frac{1}{2}
\nabla_\alpha G_{5,X}\nabla_\beta X\nabla^\alpha\nabla^\beta\phi
-\frac{1}{4}\nabla^\lambda G_{5,X}\nabla_\lambda\phi\left[
(\Box\phi)^2-(\nabla_\alpha\nabla_\beta\phi)^2\right]
\cr&&
+\frac{1}{2}G_{5,X}R_{\alpha\beta}\nabla^\alpha\phi\nabla^\beta\phi \Box\phi
-\frac{1}{2}G_{5,X}R_{\alpha\lambda\beta\rho}
\nabla^\alpha\nabla^\beta\phi\nabla^\lambda\phi\nabla^\rho\phi
\Biggr\},
\end{eqnarray}
\begin{eqnarray}
P_\phi^2&=&K_\phi,
\\
P_\phi^3&=&\nabla_\mu G_{3,\phi}\nabla^\mu\phi,
\\
P_\phi^4&=&G_{4,\phi}R+G_{4,\phi X}\left[(\Box\phi)^2-(\nabla_\mu\nabla_\nu\phi)^2\right],
\\
P_\phi^5&=&-\nabla_\mu G_{5,\phi }G^{\mu\nu} \nabla_\nu\phi -
\frac{1}{6}G_{5,\phi X}
\left[
(\Box\phi)^3-3\Box\phi (\nabla_\mu\nabla_\nu \phi)^2+2(\nabla_\mu\nabla_\nu \phi)^3
\right],
\end{eqnarray}
and
\begin{eqnarray}
J_\mu^2&=&-{\cal L}_{2X}\nabla_\mu\phi,
\\
J_\mu^3&=&-{\cal L}_{3X}
\nabla_\mu\phi+ G_{3,X}\nabla_\mu X+2G_{3,\phi}\nabla_\mu\phi,
\\
J_\mu^4&=&-{\cal L}_{4X}\nabla_\mu\phi+
2G_{4,X}R_{\mu\nu}\nabla^\nu\phi
-2G_{4,XX}\left(\Box\phi\nabla_\mu X-\nabla^\nu X\nabla_\mu\nabla_\nu\phi\right)
\cr&&
-2G_{4,\phi X}\left(\Box\phi\nabla_\mu \phi +\nabla_\mu X\right),
\\
J_\mu^5&=&
-{\cal L}_{5X}\nabla_\mu\phi-2G_{5,\phi }G_{\mu\nu}\nabla^\nu\phi
\cr&&
-G_{5,X}\left[G_{\mu\nu}\nabla^\nu X+
R_{\mu\nu}\Box\phi \nabla^\nu\phi-R_{\nu\lambda}\nabla^\nu\phi\nabla^\lambda\nabla_\mu\phi
-R_{\alpha\mu\beta\nu}\nabla^\nu\phi\nabla^\alpha\nabla^\beta\phi
\right]
\cr
&&
+G_{5,XX}\left\{
\frac{1}{2}\nabla_\mu X\left[(\Box\phi)^2-(\nabla_\alpha\nabla_\beta\phi)^2\right]
-\nabla_\nu X\left(\Box\phi \nabla_\mu\nabla^\nu\phi-\nabla_\alpha\nabla_\mu\phi
\nabla^\alpha\nabla^\nu\phi\right)
\right\}
\cr
&&
+G_{5,\phi X}\left\{
\frac{1}{2}\nabla_\mu\phi\left[(\Box\phi)^2-(\nabla_\alpha\nabla_\beta\phi)^2\right]
+\Box\phi\nabla_\mu X-\nabla^\nu X\nabla_\nu\nabla_\mu \phi
\right\}.
\end{eqnarray}
\clearpage

\makeatletter
\let\chapter\oldchapter
\makeatother